%% file: pdficrc.tex
\documentclass[a4paper]{article}
\pdfoutput=1
\usepackage{pdfpages}

\usepackage{icrc2013}

\usepackage{upgreek}
\usepackage{multirow} 
\usepackage{amssymb}
\usepackage{tikz}
\usetikzlibrary{calc,intersections}

\usepackage{amsmath}
\usepackage[english]{babel}

\usepackage{caption}
\usepackage{subcaption}
\usepackage{placeins}
\usepackage{color}
\usepackage{enumitem,blindtext}
\usepackage{multirow}

\title{The JEM-EUSO Mission: Contributions to the ICRC 2013}

\newcommand{\etal}{\MakeLowercase{\textit{et al. }}} 
\shorttitle{JEM-EUSO}

\authors{The JEM-EUSO Collaboration}
\afiliations{}

\abstract{

\vspace*{1cm}

\noindent  \includegraphics[width=0.90\textwidth]{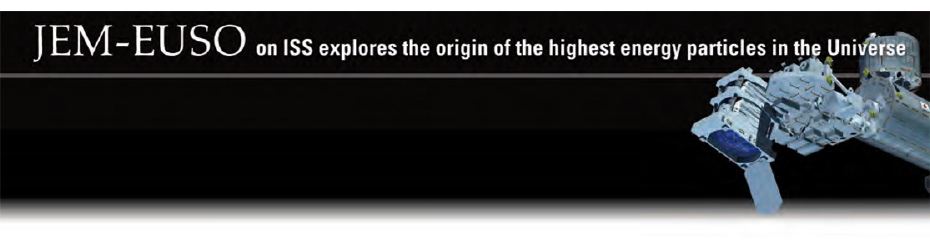}

\vspace*{1cm}

Contributions of the JEM-EUSO Collaboration \\
to the 33$^{rd}$ International Cosmic Ray Conference \\
(The Astroparticle Physics Conference \\
Rio de Janeiro, July, 2013.

\vspace*{1cm}

\includegraphics[width=0.90\textwidth]{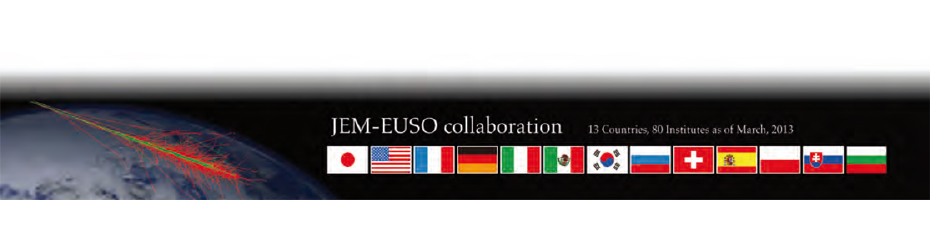}
}


\begin{document}
\maketitle

\pagestyle{plain}
\pagenumbering{arabic}

\clearpage
\newpage

\vspace*{2cm}
\clearpage
\newpage

\onecolumn{
\noindent {\bf The Collaboration:} \\
J.H.~Adams Jr.$^{md}$, 
S.~Ahmad$^{bb}$, 
J.-N.~Albert$^{ba}$, 
D.~Allard$^{bc}$, 
L.~Anchordoqui$^{mf}$,
V.~Andreev$^{me}$, 
A.~Anzalone$^{dh,dn}$, 
Y.~Arai$^{ev}$, 
K.~Asano$^{et}$,
M.~Ave~Pernas$^{kc}$
% ...
\\


{\center
% ...
$^{ba}$ LAL, Univ Paris-Sud, CNRS/IN2P3, Orsay, France\\
$^{bb}$ Omega, Ecole Polytechnique, CNRS/IN2P3, Palaiseau, France\\
$^{bc}$ APC, Univ Paris Diderot, CNRS/IN2P3, CEA/Irfu, Obs de Paris, Sorbonne Paris Cit\'e, France\\
% ...
$^{dh}$ INAF - Istituto di Astrofisica Spaziale e Fisica Cosmica di Palermo, Italy\\
% ...
$^{dn}$ Istituto Nazionale di Fisica Nucleare - Sezione di Catania, Italy\\
% ...
$^{et}$ Interactive Research Center of Science, Tokyo Institute of Technology, Tokyo, Japan\\ 
% ...
$^{ev}$ High Energy Accelerator Research Organization (KEK), Tsukuba, Japan\\ 
% ...
$^{kc}$ Universidad de Alcal\'a (UAH), Madrid, Spain\\
% ...
$^{md}$ University of Alabama in Huntsville, Huntsville, USA\\
$^{me}$ University of California (UCLA), Los Angeles, USA\\
$^{mf}$ University of Wisconsin-Milwaukee, Milwaukee, USA\\
% ...
}
}

\newpage

\onecolumn{
{\bf Content:} \\
\begin{tabular}{lllll}
& & & &  \\
		&					& Overview papers &  \\
& & & &  \\
1. 	&page 7  	&Status of the JEM-EUSO Mission 								&A. Santangelo, P. Picozza, T. Ebisuzaki	&ID0738 \\
& & & &  \\
2. 	&page 11	&The JEM-EUSO Instruments 											&F. Kajino et al. 												&ID1128 \\
& & & &  \\
3. 	&page 15	&JEM-EUSO Science capabilities									&G. Medina-Tanco et al.										&ID0937 \\
& & & &  \\
4. 	&page 19	&EUSO-BALLOON: a pathfinder for observing 			&P. von Ballmoos et al.										&ID1171 \\
							& &UHECRs from space 	 & &  \\
& & & &  \\
5. 	&page 23	&Calibration and testing of a prototype of the 	&M. Casolino et al.												&ID1213 \\
							& &JEM-EUSO telescope on Telescope Array site & &  \\
& & & &  \\
6. 	&page 27	&Atmospheric Monitoring system of 							&A. Neronov et al.												&ID1072 \\
							& &the JEM-EUSO telescope & &  \\
& & & &  \\
		&					& Physics papers &  \\
& & & &  \\
7. 	&page 31  &Estimated exposure of UHECR observation by  		&K. Shinozaki et al. 											&ID1250 \\
							& &the JEM-EUSO mission & &  \\
& & & &  \\
8. 	&page 35	&Identification of extreme energy photons 			&A. D. Supanitzky, G. Medina-Tanco				&ID0481 \\
& &with JEM-EUSO & &  \\
& & & &  \\
9. 	&page 39	&A study on JEM-EUSOs trigger probability 			&A. Guzman et al.													&ID0533 \\
							& &for neutrino-initiated EAS & &  \\
& & & &  \\
10. &page 43	&Sensitivity of orbiting JEM-EUSO to  					&T.J. Weiler et al.												&ID0631 \\
							& &large-scale cosmic-ray anisotropies 	 & &  \\
& & & &  \\
11. &page 47	&Nuclearite observations with JEM-EUSO 					&M. Bertaina et al.												&ID0272 \\
& & & &  \\
		&					& Simulation papers &  \\
& & & &  \\
12. &page 51  &ESAF-Simulation of the EUSO-Balloon 						&T. Mernik et al.													&ID0875 \\
& & & &  \\
13. &page 55	&Simulating the JEM-EUSO Mission:  							&T. Mernik et al. 												&ID0777 \\
							& &Expected Reconstruction Performance & &  \\
& & & &  \\
14. &page 59	&Simulations and the analysis of fake trigger 	&S. Biktemerova et al.										&ID1283 \\
							& &events background in JEM-EUSO experiment & &  \\
& & & &  \\
15. &page 63	&Pattern recognition and direction reconstruction &S. Biktemerova, M. Gonchar, S. Sharakin	&ID1282 \\
							& &for the JEM-EUSO experiment 	 & &  \\
& & & &  \\
16. &page 67	&On-line and off-line data analysis for the  		&L.W. Piotrowski, A. Pesoli								&ID0713 \\
							& &EUSO-TA and EUSO-BALLOON experiments & &  \\
& & & &  \\
		&					& Calibration papers &  \\
& & & &  \\
17. &page 71  &Absolute calibrations of the Focal Surface  		&P. Gorodetzky et al.											&ID0858 \\
							& &of the JEM-EUSO Telescope & &  \\
& & & &  \\
18. &page 75	&Photomultiplier Tube Sorting for  							&C. Blaksley, P. Gorodetzky							&ID0628 \\
							& &JEM-EUSO and EUSO-Balloon  & &  \\
\end{tabular}
}

\newpage

\onecolumn{
\begin{tabular}{lllll}
19. &page 79	&On-board calibration system of the  						&M. Karus et al.													&ID0545 \\
							& &JEM-EUSO mission & &  \\
& & & &  \\
20. &page 83	&Absolute In-flight Calibration of the  				&N. Sakaki et al.												&ID0546 \\
							& &JEM-EUSO Telescope with the Moonlight	 & &  \\
& & & &  \\
21. &page 87	&The JEM-EUSO Global Light System  							&L. Wiencke et al.												&ID0818 \\
& & & &  \\
		&					& Instrumental papers &  \\
& & & &  \\
22. &page 91  &JEM-EUSO Design for Accommodation on the  			&J. Adams, R.M. Young, 								&ID1256 \\
							& &SpaceX Dragon Spacecraft &A. Olinto &  \\
& & & &  \\
23. &page 95	&Multi-Anode Photomultiplier Tube reliability   &H. Prieto et al. 												&ID0343 \\
							& &analysis and radiation hardness assurance & &  \\
							& &for the JEM-EUSO Space mission  & &  \\
& & & &  \\
24. &page 99	&Second level trigger and Cluster Control Board  &J. Bayer et al.													&ID0432 \\
							& &for the JEM-EUSO mission & &  \\
& & & &  \\
25. &page 103	&Performance of the SPACIROC front-end ASIC  		&H. Miyamoto et al.												&ID1089 \\
							& &for JEM-EUSO 	 & &  \\
& & & &  \\
26. &page 107	&The TA-EUSO and EUSO-Balloon optics designs  	&Y. Takizawa et al.		&ID0832 \\
& & & &  \\
27. &page 111	&Manufacturing of the TA-EUSO and the  	&Y. Hachisu et al.						&ID1040 \\
							& &EUSO-Balloon lenses & &  \\
& & & &  \\
28. &page 115	&The Electronics of the EUSO-Balloon UV camera 		&H. Miyamoto et al.											&ID0765 \\
& & & &  \\
29. &page 119	&Global Description of EUSO-Balloon Instrument  	&C. Moretto et al.												&ID0678 \\
& & & &  \\
		&					& AMS papers &  \\
& & & &  \\
30. &page 123 &UV night background estimation in South   	&P. Bobik et al.							&ID0874 \\
							& &Atlantic Anomaly & &  \\
& & & &  \\
31. &page 127	&Retrieving cloud top height in the JEM-EUSO   &A. Anzalone et al. 												&ID0919 \\
							& &cosmic-ray observation system  & &  \\
& & & &  \\
32. &page 131	&Simulations of extensive air showers produced by   &G.~S\'aez~Cano et al.						&ID1281 \\
							& &UHECRs in cloudy sky to be detected by JEM-EUSO & &  \\
& & & &  \\
33. &page 135	&Absolute Fluorescence Spectrum and Yield    		&D. Monnier Ragaigne et al.		&ID0449 \\
							& &Measurements for a wide range of  	 & &  \\
							& &experimental conditions & &  \\
& & & &  \\
34. &page 139	&Towards the Preliminary Design Review of the   	&M.D.~Rodr\'iguez~Fr\'ias et al.		&ID0900 \\
							& &Infrared Camera of the JEMEUSO Space Mission & &  \\
& & & &  \\
35. &page 143	&LIDAR treatment inside the ESAF Simulation  	&S. Toscano et al.						&ID0530\\
							& &Framework for the JEM-EUSO mission & &  \\
& & & &  \\
36. &page 147	&An End to End Simulation code for the IR-Camera  		&J.A. Morales de los Rios et al. &ID0514 \\
							& &of the JEM-EUSO Space Observatory & &  \\
\end{tabular}
}

\newpage
\setcounter{section}{0}
\setcounter{figure}{0}
\setcounter{table}{0}
\setcounter{equation}{0}
 \input{icrc2013-0738.tex}

\newpage
\setcounter{section}{0}
\setcounter{figure}{0}
\setcounter{table}{0}
\setcounter{equation}{0}
 \input{icrc2013-1128.tex}

\newpage
\setcounter{section}{0}
\setcounter{figure}{0}
\setcounter{table}{0}
\setcounter{equation}{0}
 \input{icrc2013-0937.tex}

\newpage
\setcounter{section}{0}
\setcounter{figure}{0}
\setcounter{table}{0}
\setcounter{equation}{0}
 \input{icrc2013-1171.tex}

\newpage
\setcounter{section}{0}
\setcounter{figure}{0}
\setcounter{table}{0}
\setcounter{equation}{0}
 \input{icrc2013-1213.tex}

\newpage
\setcounter{section}{0}
\setcounter{figure}{0}
\setcounter{table}{0}
\setcounter{equation}{0}
 \input{icrc2013-1072.tex}

\newpage
\setcounter{section}{0}
\setcounter{figure}{0}
\setcounter{table}{0}
\setcounter{equation}{0}
 \input{icrc2013-1250.tex}

\newpage
\setcounter{section}{0}
\setcounter{figure}{0}
\setcounter{table}{0}
\setcounter{equation}{0}
 \input{icrc2013-0461.tex}

\newpage
\setcounter{section}{0}
\setcounter{figure}{0}
\setcounter{table}{0}
\setcounter{equation}{0}
 \input{icrc2013-0533.tex}

\newpage
\setcounter{section}{0}
\setcounter{figure}{0}
\setcounter{table}{0}
\setcounter{equation}{0}
 \input{icrc2013-0631.tex}

\newpage
\setcounter{section}{0}
\setcounter{figure}{0}
\setcounter{table}{0}
\setcounter{equation}{0}
 \input{icrc2013-0272.tex}

\newpage
\setcounter{section}{0}
\setcounter{figure}{0}
\setcounter{table}{0}
\setcounter{equation}{0}
 \input{icrc2013-0875.tex}

\newpage
\setcounter{section}{0}
\setcounter{figure}{0}
\setcounter{table}{0}
\setcounter{equation}{0}
 \input{icrc2013-0777.tex}

\newpage
\setcounter{section}{0}
\setcounter{figure}{0}
\setcounter{table}{0}
\setcounter{equation}{0}
 \input{icrc2013-1283.tex}

\newpage
\setcounter{section}{0}
\setcounter{figure}{0}
\setcounter{table}{0}
\setcounter{equation}{0}
 \input{icrc2013-1282.tex}

\newpage
\setcounter{section}{0}
\setcounter{figure}{0}
\setcounter{table}{0}
\setcounter{equation}{0}
 \input{icrc2013-0713.tex}

\newpage
\setcounter{section}{0}
\setcounter{figure}{0}
\setcounter{table}{0}
\setcounter{equation}{0}
 \input{icrc2013-0858.tex}

\newpage
\setcounter{section}{0}
\setcounter{figure}{0}
\setcounter{table}{0}
\setcounter{equation}{0}
 \input{icrc2013-0628.tex}

\newpage
\setcounter{section}{0}
\setcounter{figure}{0}
\setcounter{table}{0}
\setcounter{equation}{0}
 \input{icrc2013-0545.tex}

\newpage
\setcounter{section}{0}
\setcounter{figure}{0}
\setcounter{table}{0}
\setcounter{equation}{0}
 \input{icrc2013-0546.tex}

\newpage
\setcounter{section}{0}
\setcounter{figure}{0}
\setcounter{table}{0}
\setcounter{equation}{0}
 \input{icrc2013-0818.tex}

\newpage
\setcounter{section}{0}
\setcounter{figure}{0}
\setcounter{table}{0}
\setcounter{equation}{0}
 \input{icrc2013-1256.tex}
 \hspace{4cm}
 \newpage

\newpage
\setcounter{section}{0}
\setcounter{figure}{0}
\setcounter{table}{0}
\setcounter{equation}{0}
 \input{icrc2013-0343.tex}

\newpage
\setcounter{section}{0}
\setcounter{figure}{0}
\setcounter{table}{0}
\setcounter{equation}{0}
 \input{icrc2013-0432.tex}

\newpage
\setcounter{section}{0}
\setcounter{figure}{0}
\setcounter{table}{0}
\setcounter{equation}{0}
 \input{icrc2013-1089.tex}

\newpage
\setcounter{section}{0}
\setcounter{figure}{0}
\setcounter{table}{0}
\setcounter{equation}{0}
 \input{icrc2013-0832.tex}

\newpage
\setcounter{section}{0}
\setcounter{figure}{0}
\setcounter{table}{0}
\setcounter{equation}{0}
 \input{icrc2013-1040.tex}

\newpage
\setcounter{section}{0}
\setcounter{figure}{0}
\setcounter{table}{0}
\setcounter{equation}{0}
 \input{icrc2013-0765.tex}

\newpage
\setcounter{section}{0}
\setcounter{figure}{0}
\setcounter{table}{0}
\setcounter{equation}{0}
 \input{icrc2013-0678.tex}

\newpage
\setcounter{section}{0}
\setcounter{figure}{0}
\setcounter{table}{0}
\setcounter{equation}{0}
 \input{icrc2013-0874.tex}

\newpage
\setcounter{section}{0}
\setcounter{figure}{0}
\setcounter{table}{0}
\setcounter{equation}{0}
 \input{icrc2013-0919.tex}

\newpage
\setcounter{section}{0}
\setcounter{figure}{0}
\setcounter{table}{0}
\setcounter{equation}{0}
 \input{icrc2013-1281.tex}

\newpage
\setcounter{section}{0}
\setcounter{figure}{0}
\setcounter{table}{0}
\setcounter{equation}{0}
 \input{icrc2013-0449.tex}

\newpage
\setcounter{section}{0}
\setcounter{figure}{0}
\setcounter{table}{0}
\setcounter{equation}{0}
 \input{icrc2013-0900.tex}

\newpage
\setcounter{section}{0}
\setcounter{figure}{0}
\setcounter{table}{0}
\setcounter{equation}{0}
 \input{icrc2013-0530.tex}

\newpage
\setcounter{section}{0}
\setcounter{figure}{0}
\setcounter{table}{0}
\setcounter{equation}{0}
 \input{icrc2013-0514.tex}

\end{document}

%% file: icrc2013-0738.tex



\title{Status of the JEM-EUSO Mission}

\shorttitle{JEM-EUSO}

\authors{
A. Santangelo$^{1,2}$,
P. Picozza$^{2,3,4}$,
T. Ebisuzaki$^{2}$,
for the JEM-EUSO Collaboration.
}

\afiliations{
$^1$ Institute for Astronomy and Astrophysics, Kepler Center for Astro and Particle Physics, University of T\"ubingen, Germany \\
$^2$ RIKEN Advanced Science Institute, Wako, Japan \\
$^3$ Dipartimento di Fisica, University of Roma Tor Vergata, Italy \\
$^4$ Istituto Nazionale di Fisica Nucleare - Sezione di Roma Tor Vergata, Italy
}

\email{Andrea.Santangelo@uni-tuebingen.de}

\abstract{The Extreme Universe Space Observatory, on-board the Japanese Experimental Module of the ISS (JEM-EUSO), mainly aims at unveiling the origin of the ultra high energy cosmic rays (UHECRs), beyond the suppression due to the Greisen-Zatsepin-Kuz'min effect. JEM-EUSO will also explore fundamental physics at these extreme energies. Designed to measure the arrival directions, the energies, and possibly, the nature of these particles, JEM-EUSO consists of a wide-field of view (60 degrees) telescope, with a diameter of about 2.5m, which points to Nadir from space to detect, during night-time, the UV (290-430 ~nm) tracks generated by extensive air showers propagating in the earth's atmosphere. The high statistics arrival direction map will allow anisotropy studies and, most likely, the identification of individual sources of UHECRs and their association with known nearby astronomical objects. This will shine new light on the understanding of the acceleration mechanisms and, perhaps, will produce new discoveries in astrophysics and/or fundamental physics. The comparison of the energy spectra among the spatially resolved individual sources will eventually confirm the Greisen-Zatsepin-Kuz'min process, validating Lorentz invariance up to $\gamma\sim10^{11}$. In this paper, we will present the current status of the mission, reporting on the most recent technical developments, mission, and programmatic aspects of this challenging space-based observatory.}


\maketitle

\section{Introduction}
The Extreme Universe Space Observatory, on-board the Japanese Experiment Module (JEM-EUSO) of the ISS \cite{cctakahashi,cccasolino}, is a space based mission which aims at studying ultra-high-energy cosmic rays (UHECRs) of the highest energy. These are cosmic particles with energy $E\ge5\times10^{19}$~eV, above the threshold of the Greisen-Zatsepin-Kuz'min suppression of the cosmic ray spectrum \cite{GZK}. Building on the heritage of the Pierre Auger and Telescope Array observatories, JEM-EUSO focuses its science case on the most energetic of those events, at $E\sim 10^{20}$~eV, often referred to as extreme energy cosmic rays (EECRs).

JEM-EUSO is designed to monitor from space, looking towards nadir during night-time, the earth's atmosphere to detect the UV (290-430~nm) tracks generated by the gigantic extensive air showers (EAS) propagating through the atmosphere. By imaging, with a time resolution of the order of $\mu s$, the fluorescence and Cherenkov photons of the EAS, the energy, arrival direction and nature of the primary UHECR particle will be reconstructed. 

Placed at an altitude of $H\sim 400$~km from the earth�s surface, JEM-EUSO orbits the earth, with a speed of $\sim7$~km s$^{-1}$, every $\sim90$~min. with the inclination of the ISS $\pm51.6^{\circ}$. JEM-EUSO can be operated pointing to nadir ("nadir mode"), or slightly tilted ("tilted mode"). 
JEM-EUSO can reach an instantaneous aperture of about $6-7 \times10^{5}$ km$^{2}$~sr \cite{toshi}, beyond the practical limit of any ground-based UHE observatory. An extended study on the expected exposures, obtained taking into account the duty cycle, the cloud coverage of the observed scene, and the trigger efficiencies, is reported in \cite{Adams2013}. An updated study on the expected performance is given in \cite{Shinozaki2013}. JEM-EUSO is expected to reach in nadir mode, at $E\sim 10^{20}$~eV, an annual exposure about 9 times the one of the Pierre Auger Observatory. Operating in tilted mode, it will reach an annual exposure of about 20 times larger than the one of the Pierre Auger Observatory. 

During the lifetime of the mission, JEM-EUSO is expected to observe several hundreds of UHECR events with energies exceeding $5\times10^{19}$~eV, and a few hundreds at $10^{20}$~eV and above. 

In addition to the very large area monitored by space-based UHE observatories, other advantages are the well constrained distances toward showers, the clear and stable atmospheric transmission in the above half of the troposphere, the uniform exposure across both north and south skies. All of these aspects are discussed in detail in \cite{Adams2013}.

\section{Science objectives and requirements}
The science objectives of the mission are divided into one main objective and five exploratory objectives. The main objective of JEM-EUSO is to initiate a new field of astronomy using the extreme energy particle channel, being the first instrument to explore, with high statistics, the energy decade around and beyond $10^{20}$~ eV. At these extreme energies, due to attenuation effects, only a handful of sources must dominate the EECR flux. The main science goals are therefore: 1) The study of the anisotropies of the EE sky; 2) The identification of sources by high-statistics arrival direction analysis and possibly the measurement of the energy spectra in a few sources sources with high event multiplicity; 4) A high statistics measurement of the trans-GZK spectrum.

Besides the prime science objectives, we set five exploratory objectives to which the instrument can contribute, depending on the actual nature of the extreme energy cosmic ray flux: 1.) The study of the UHE neutrino component, by discriminating weakly interacting events via the position of the first interaction point and of the shower maximum; 2) The discovery of UHE Gamma-rays, whose shower maximum is strongly affected by geomagnetic and LPM effect; 3.) The study of the galactic and local extragalactic magnetic fields, through the analysis of the magnetic point spread function.
The science capability of the mission is discussed in detail in \cite{GMTICRC}. 

Among the exploratory objectives, several topics of atmospheric science are included. JEM-EUSO will allow a characterization of the night-glow and of the transient luminous events (TLE) in the UV band. It can also detect the slow UV tracks associated to meteors and meteoroids.

One of the key requirements of the mission is certainly the validation of the observational technique in space operation condition. A proper operation of the main telescope, of the AMS, the measurement of the background and its variability, the determination of the duty cycle, the identification of the atmospheric scene, which is essential to correctly estimate the exposure, are the key requirements to be reached during the first phase of the mission. 

We then aim at a stringent comparison of the energy spectrum with the one measured by the current generation of UHCR observatories around and above $5\times10^{19}$~eV, which implies a comparable annual exposure at lower energies. Since the key goal of the mission is to explore the highest energies, one of the key requirements of JEM-EUSO is to reach an annual exposure of approximately one order of magnitude larger than the one of the ground based observatory at EECRs.

More specifically, from the science objectives, the following scientific requirements have been set: \textit{Statistics of a few hundreds events} above $E>7\times10^{19}$~eV, which implies an exposure (in three years) of $\ge 10^5$~km$^2$~sr~yr; \textit{Angular resolution} $\le 3^\circ$ at $E>8\times10^{19}$~eV (expressed in terms of$ \gamma_{68}$);\textit{ Energy resolution} (expressed in terms of the 68\% of the distribution) $\le 30\%$ for $E>8\times10^{19}$~eV (goal: for  $E>6\times10^{19}$); capability to discriminate between nuclei, gamma ray and neutrinos, which implies \textit{Xmax determination error} $\le 120$ (g/cm$^2$) $E=10^{20}$~eV and zenith angle 60 degrees); \textit{full-sky observation} with $<30\%$ (goal 15\%) non uniformity among hemispheres. We eventually aim at measuring the timing properties of luminous atmospheric events with ms resolution. 

The requirements are currently being validated with end-to-end simulations based on the EUSO Simulation and analysis framework (ESAF) and with the recently developed EUSO-offline framework. More details are found in \cite{Mernik, Berat}.  

\section{The JEM-EUSO instrument}

JEM-EUSO consists of a main telescope, sensitive to near UV, and of an atmosphere monitoring system (AMS). The main telescope is a fast (of the order of $\mu s$) and highly- pixelized $3\times10^5$ pixels) digital camera with a large-aperture wide-Field of View (FoV, 2 $\times$ 30$^{\circ}$) normally operating in \textit{single photon counting} mode but capable of switching to \textit{charge integration} mode in case of strong illumination. 

\begin{figure}[!h]
  \centering
  \includegraphics[width=0.35\textwidth]{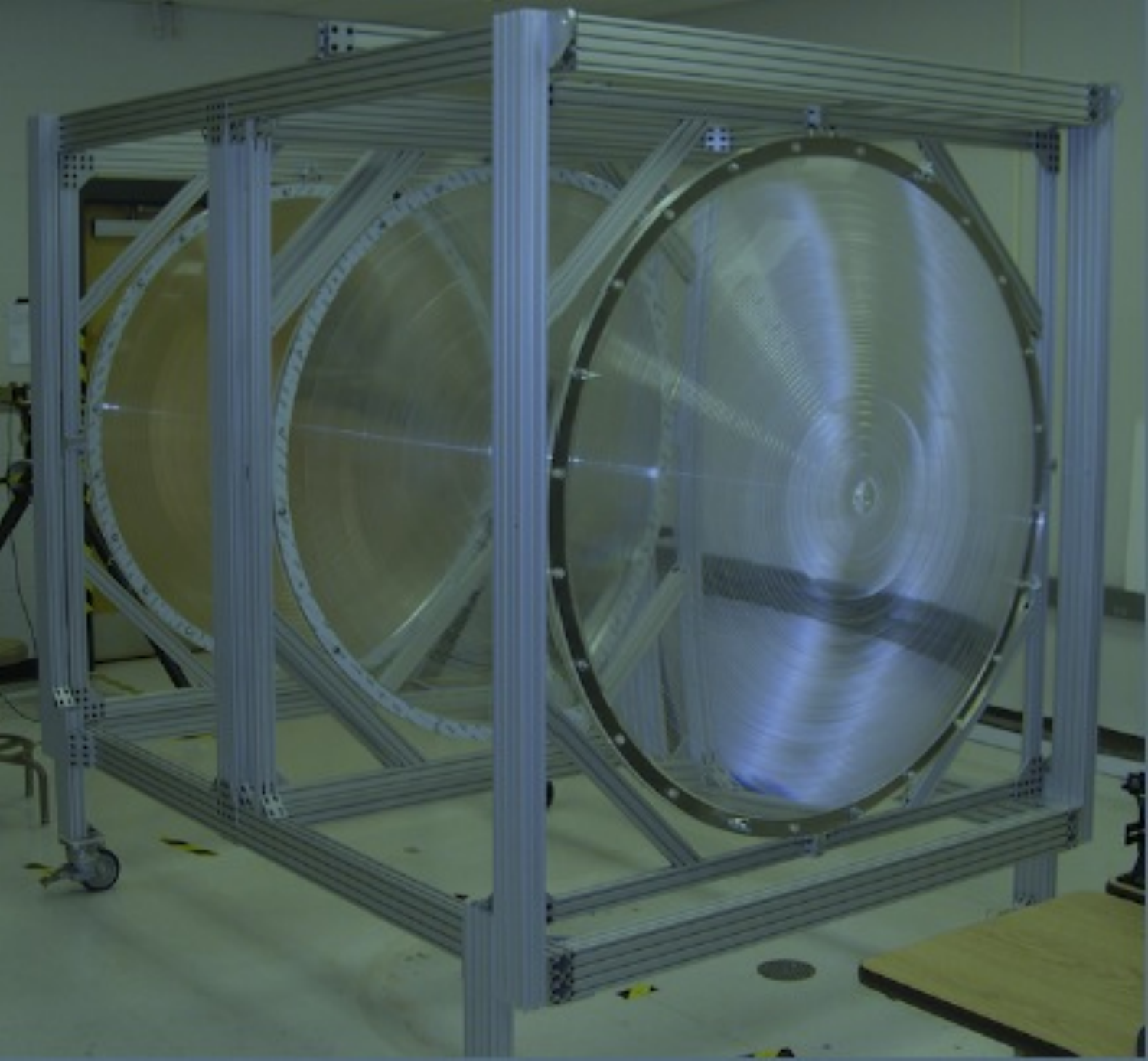}
  \caption{The Bread-board model of the JEM-EUSO optics tested at MSFC, NASA, Huntsville. The diameter is 1.5 m.}
  \label{figureoptics}
 \end{figure}

The current baseline optics consist of two curved, double sided, Fresnel lenses with 2.65 m external diameter, and of an intermediate curved precision Fresnel lens. The precision Fresnel lens, takes an advantage based on state of the art diffractive optics technology, and is used to reduce chromatic aberration. The combination of 3 Fresnel lenses allows a full angle FoV of 60 degrees with a resolution of 0.075 degrees, corresponding to a pixel of about 550 m on earth. PMMA, which has high UV transparency in the wavelength from 330nm to 430nm, and CYTOP are candidate materials.  Details of the optics are described in \cite{takizawa, Hachisu}. The focal surface (FS) is, in the current baseline, spherical with $\sim$2.5 m curvature radius, and 2.3 m diameter. A bread-board model has been already manufactured and tested at the MSFC of NASA in Huntsville (see figure \ref{figureoptics})

The Focal surface is organized in 137 Photodetector modules (PDMs), each one consisting of 9 Elementary cells (ECs). Each EC contains 4 multi-anode photomultiplier tubes (64-pixel MAPMT), with a quantum efficiency of about 40 \%. More than 5,000 MAPMTs are, therefore, integrated in the focal surface. The electronics record the signals generated by the UV photons of the EAS in the FS, providing a kinematic reproduction of each track. A new type of front-end ASIC has been developed, which has both functions of single photon counting and charge integration in a chip with 64 channels  \cite{Barrillon}.  Radiation tolerance of the electronic circuits in space environment is also required. 

\begin{figure}[!h]
  \centering
  \includegraphics[width=0.35\textwidth]{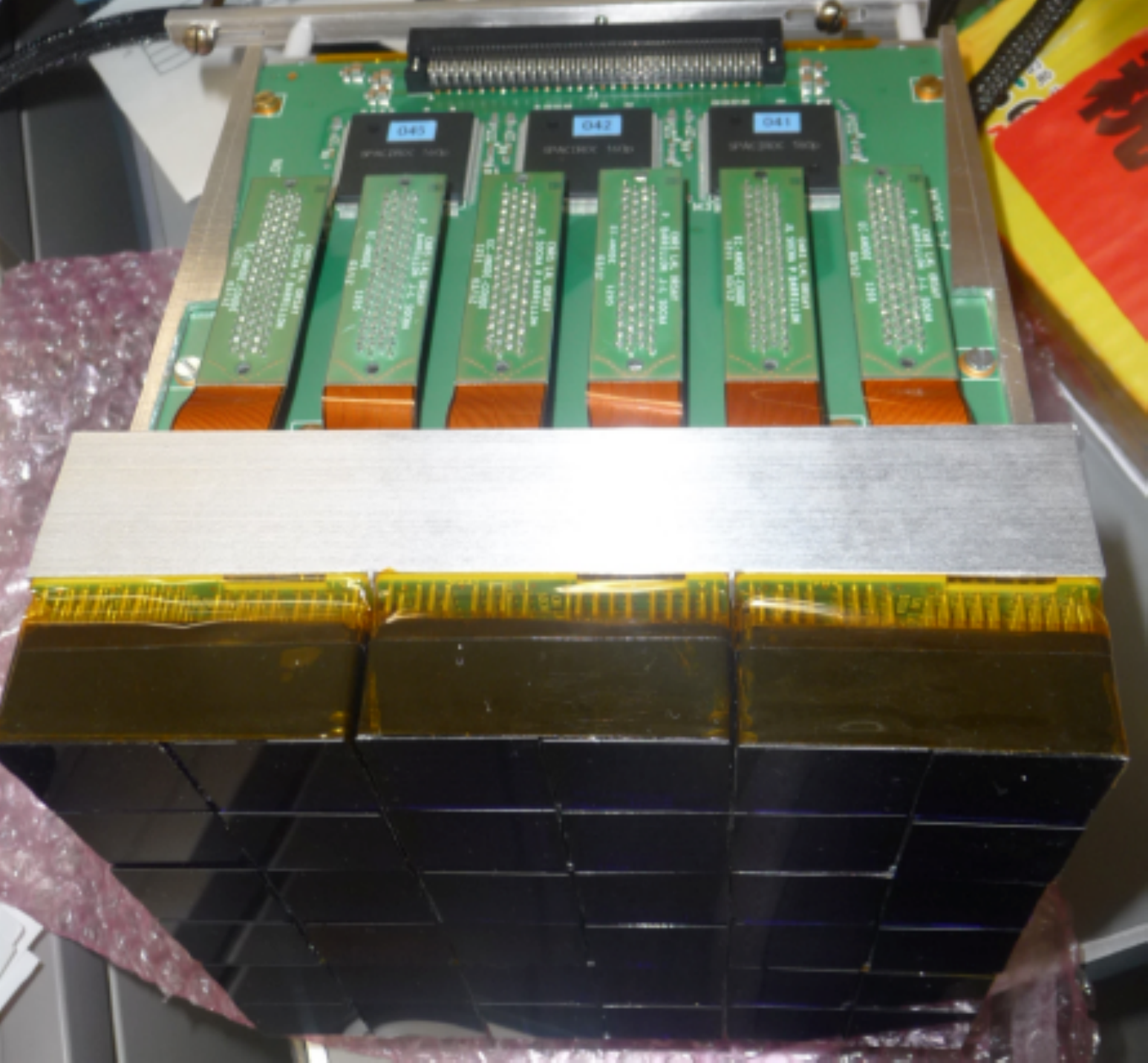}
  \caption{The first prototype of the JEM-EUSO PDM, already integrated in RIKEN. It will be tested as the focal surface of the TA-EUSO experiment, to be deployed at the TA site in UTAH in summer 2013.}
  \label{figurePDM}
 \end{figure}

The first prototype of the JEM-EUSO PDM, that will used for the TA-EUSO experiment, has been recently integrated in Wako (Japan) at RIKEN  \cite{Casolino2} (see figure \ref{figurePDM}).

The system is required to have high trigger efficiency and linearity over the $10^{19}-10^{21}$ eV range. A trigger logic based on two levels has been implemented. The logic seeks pattern features close to those of signal tracks we would expect from a moving EAS.  When a trigger is issued, the time frame of 128 GTU (gate time units, 2.5 $\mu$s) is saved to disc or transferred by the telemetry.

The AMS monitors the atmospheric scene of the FoV of the UV telescope \cite{Maria}. It consists of an IR camera, a Lidar, and the slow data of the main telescope to measure the cloud-top height with an accuracy better than 500 m. The calibration system monitors the efficiencies of the optics, the focal surface detector, and the data acquisition electronics. 

Calibration of the instrument will be possible by built-in LEDs and additional xenon flashers from ground.

\section{The Mission}
\begin{figure}[!h]
  \centering
  \includegraphics[width=0.35\textwidth]{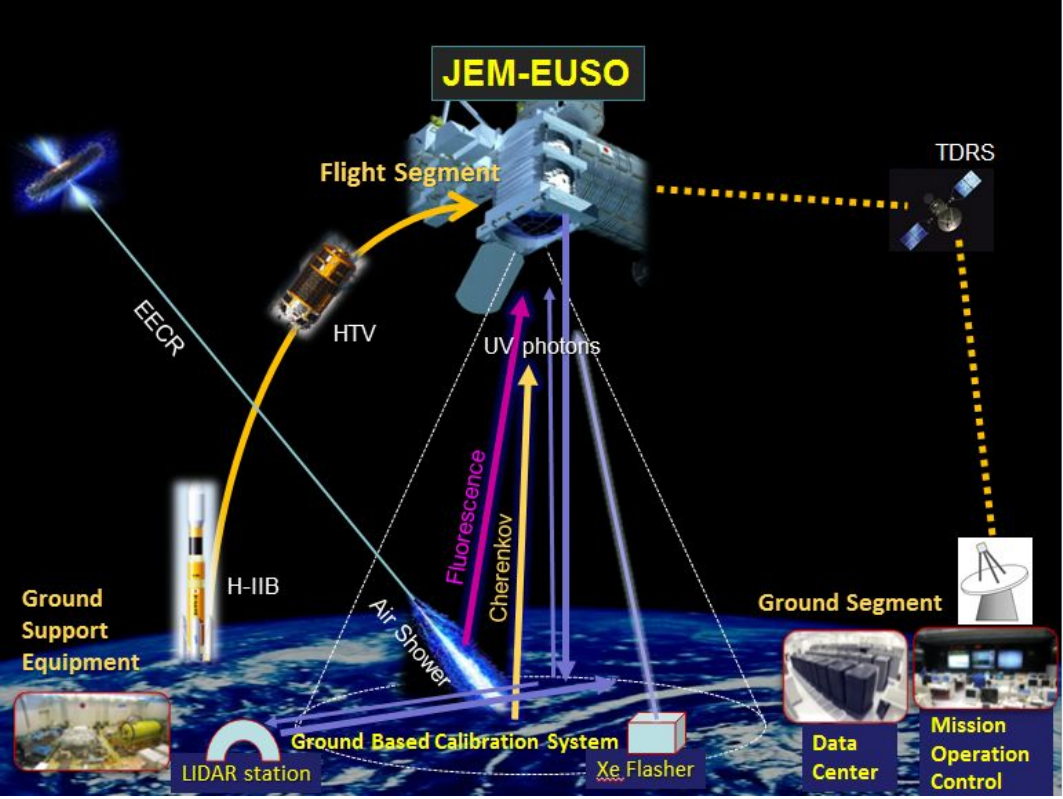}
  \caption{The main elements of the mission are summarized in the figure.}
  \label{figuremission}
 \end{figure}

The main elements of the mission are shown in \ref{figuremission}. 

\begin{figure}[!h]
  \centering
  \includegraphics[width=0.35\textwidth]{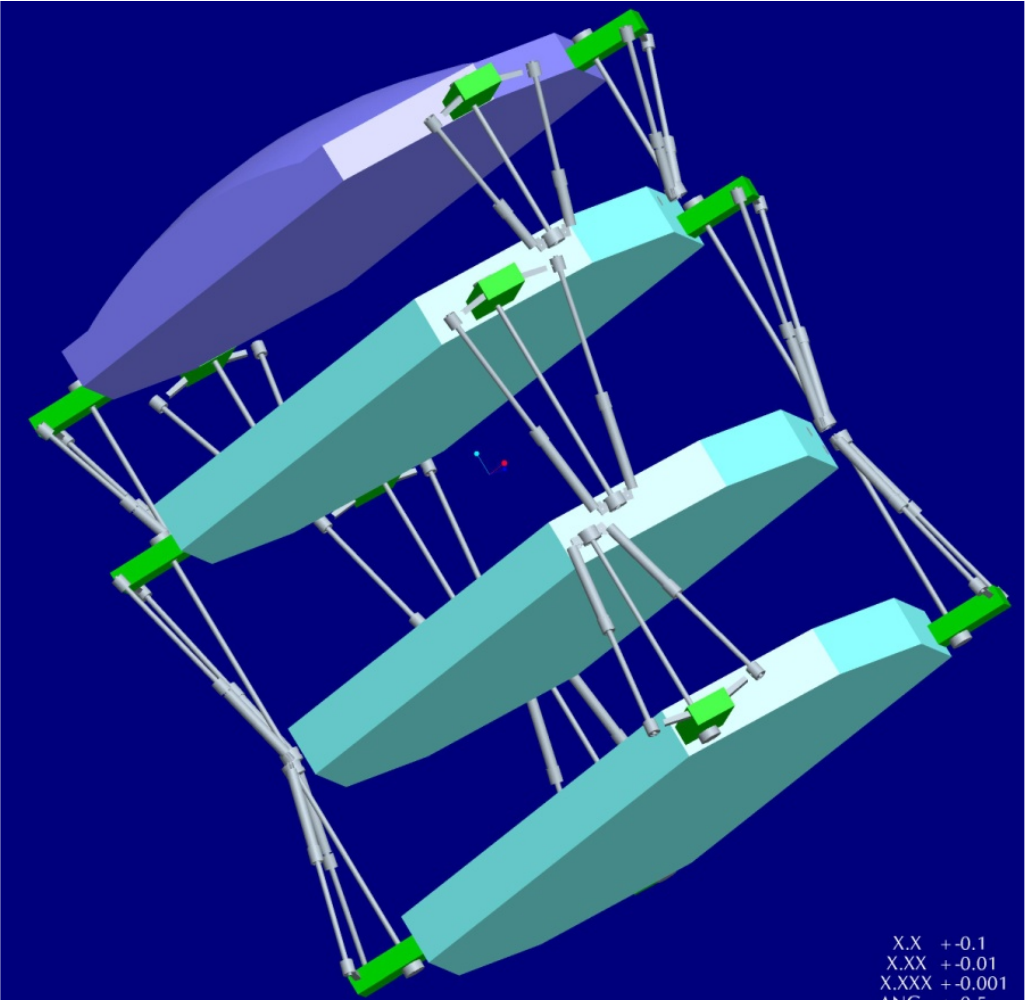}
  \caption{The Pyramid variant of the extension mechanism. Displacement screws move the lenses. The pyramidal structures provides the necessary strength and stiffness.}
  \label{figuredeployment}
 \end{figure}

According to the current baseline, JEM-EUSO shall be launched by an H2B rocket and will be conveyed to the ISS by the the unmanned resupply spacecraft H-II Transfer Vehicle (HTV).  It will then be attached, using the Canadian robotic arm, to one of the ports (baseline is port 9) for non-standard payloads of the Exposure Facility (EF) of the JEM. Such a scenario was successfully studied during the phase A study of the mission performed under the leadership of JAXA.
During launch and transportation, the instrument will be stored in a folded configuration, and will be deployed after the attachment procedure is successfully completed. The telescope structure and the deployment system are currently studied at the Skobeltsyn Institute of Nuclear Physics in Moscow. Three mechanical concepts for the extension mechanism have been developed. The so- called "Pyramid" variant is shown in figure  \ref{figuredeployment}.

 In alternative to the H2B-HTV scenario, the possibility of using the SpaceX Dragon spacecraft is under
consideration as an option for the transfer vehicle. The accommodation of JEM-EUSO in the trunk section of the SpaceX Dragon Spacecraft will require an optimization of the instrument baseline and slight modification will be necessary \cite{AdamsICRC}. In particular, through a careful study of the system, we are confidently reducing the weight of the instrument towards the 1.1-1.2 ton goal (including margins). These modifications will not impact the science performance of the mission. SpaceX began regular missions to deliver cargo to the International Space Station (ISS) in October 2012.

Data will be transmitted to the Mission Operation Center hosted by JAXA in the Tsukuba Space Center and managed by RIKEN with the support of the whole collaboration, via TDRS. We plan to establish several data centers in all major participating countries. 

According to the current plans, JEM-EUSO will be operated for three years in Nadir configuration (Nadir mode) to maximize statistics of events at the lowest energies in order to cross calibrate with the current generation of ground-based detectors. The instrument will then be tilted (about 35 degrees) with respect to Nadir, to maximize the statistics of events at the highest energies. 

During flight, JEM-EUSO will be calibrated, in addition to the on-board calibration system, using a ground-based Global Light System (GLS). The GLS is a worldwide network that combines ground-based Xenon flash lamps and steered UV lasers, which will generate benchmark optical signatures in the atmosphere with similar characteristics to the optical signals of cosmic ray EAS and with known energy, time, and direction (lasers). There will be 12 ground based units strategically placed at sites around the world. Six locations will have flashers and a steerable and remotely operated laser (GLS-XL), and 6 will only have flashers (GLS-X). Sites will be chosen for their low background light and altitude (above the planetary boundary layer) \cite{Wincke}.
 
\section{Mission Status}
The payload of the mission is currently being studied by an international collaboration, which includes more than 70 scientific institutions from 13 different countries and is led by RIKEN. Given the complexity of the mission, the participation of the major agencies involved with the ISS is essential. 

In 2010, JEM-EUSO has been included as a study in the ELIPS program of ESA. NASA is supporting the activities of the US team in the framework of APRA funds. The mission has been approved by the Tsiinimash Institute in Russia and has been submitted to ROSCOSMOS for implementation. Major funding agencies in Europe, Korea, and Mexico have been active in supporting the R\&D and the development of prototypes.    

Once completed, the payload will be delivered to JAXA. In the present scheme, JAXA is responsible, in coordination with the agencies playing a major role in the ISS, NASA, ESA and ROSCOSMOS, of the key aspects of the mission. 

\section{Status of Pathfinders}

While the studies for the main mission are actively continuing, the JEM-EUSO collaboration is developing and implementing two pathfinder experiments: the EUSO-TA and the EUSO-Balloon. The aim of the EUSO-TA project is to install a reduced prototype of the UV telescope in theTelescope
Array (TA) site in Black Rock Mesa, Utah, US. EUSO-TA will perform observations of ultraviolet light generated by cosmic ray showers and artificial sources. The detector consists of one PDM and the telescope is housed in a shed located in front of one of the fluorescence detectors of the Telescope Array collaboration, pointing in the direction of the ELF (Electron Light Source) and CLF (Central Laser Facility). EUSO-TA will be installed and start operations in summer 2013. Details can be found in \cite{Casolino2}. 

The EUSO-Balloon mission is a pathfinder mission of JEM-EUSO and consists of a series of stratospheric
balloon flights starting in 2014, and performed by the French Space Agency CNES. The payload of the EUSOBALLOON consists of a scaled version of the telescope and is developed by the JEM-EUSO consortium as a demonstrator for the technologies and methods featured in the forthcoming space instrument. With its Fresnel Optics and PDM, the instrument monitors a $12^{\circ} \times12^{\circ}$ field of view in a wavelength range between
290 and 430 nm, at a rate of 400,000 frames/sec. Details can be seen in \cite{VonBallmoos}.

\section{Conclusions}
JEM-EUSO is an ISS space-mission designed to to explore the extreme energies of our universe and its fundamental physics through the detection of UHECRs with high statistics. It is the first observatory with full-sky coverage and can achieve, depending on the mission lifetime, an exposure comparable close to $10^6$km$^2$ sr year. JEM-EUSO is currently designed to meet a launch date in 2017. A pathfinder for future missions, JEM-EUSO will pave the way to even larger space-based observatories, that will definitely explore the extremes of the Universe \cite{Santangelo}.

\section{Acknowledgements}
This work was supported by the Basic Science Interdisciplinary Research Projects of RIKEN and JSPS KAKENHI Grant (22340063, 23340081, and 24244042); by the Italian Ministry of Foreign Affairs, General Direction for the Cultural Promotion and Cooperation, by the Deutsches Zentrum fuer Luft- und Raumfahrt (DLR), and by Slovak Academy of Sciences MVTS JEM-EUSO as well as VEGA grant agency project 2/0081/10. The Spanish Consortium involved in the JEM-EUSO Space Mission is funded by MICINN under projects AYA2009-06037-E/ESP, AYA-ESP 2010-19082, AYA2011-29489-C03-01, AYA2012-39115-C03-01, CSD2009-00064 (Consolider MULTIDARK) and by Comunidad deMadrid (CAM) under project S2009/ESP- 1496.
The work was also supported in the framework of the ESA'S "JEM-EUSO" topical team activities and by the Helmholtz Alliance for Astroparticle Physics, HAP, funded by the Initiative and Networking Fund of the Helmholtz Association, Germany.

\clearpage


%% file: icrc2013-1128.tex



\title{The JEM-EUSO Instruments}

\shorttitle{JEM EUSO ICRC 2013}

\authors{
F. KAJINO$^{1,2}$,
M. CASOLINO$^{2,3}$,
T. EBISUZAKI$^{2}$,
J. ADAMS$^{4}$,
P. BALLMOOS$^{5}$,
M. BERTAINA$^{6}$,
M. CHRISTL$^{7}$,
S. DAGORET$^{8}$,
C. DE LA TAILLE$^{9}$,
M. FUKUSHIMA$^{10}$,
P. GORODETZKY$^{11}$,
A. HAUNGS$^{12}$,
N. INOUE$^{13}$,
Y. KAWASAKI$^{2}$,
K. KUDELA$^{14}$,
B. KHRENOV$^{15}$,
G. MEDINA-TANCO$^{16}$,
A. NERONOV$^{17}$,
H. OHMORI$^{2}$,
A. OLINTO$^{18}$,
G. OSTERIA$^{19}$,
M. PANASYUK$^{15}$,
E. PARIZOT$^{11}$,
I. PARK$^{20}$,
P. PICOZZA$^{2,3}$,
M. RICCI$^{21}$,
M. RODRIGUEZ FRIAS,$^{22}$,
H. SAGAWA$^{10}$,
A. SANTANGELO$^{23}$,
J. SZABELSKI$^{24}$,
Y. TAKIZAWA$^{2}$,
K. TSUNO$^{2}$,
G. VANKOVA-KIRILOVA$^{25}$,
S. WADA$^{2}$,
for the JEM-EUSO Collaboration.

}
\afiliations{
$^1$  Department of Physics, Konan University, Okamoto 8-9-1, Higashinada, Kobe 658-8501, Japan\\
$^2$  RIKEN, 2-1 Hirosawa, Wako351-0198, Japan \\
$^3$  Department of Physics, University of Rome Tor Vergata, Via della Ricerca Scientifica 1,  00133  Rome, Italy \\
$^4$  Department of Physics, University of Alabama, Huntsville, AL 35899, USA \\
$^5$  Institut de Recherche en Astrophysique et Planetologie, France \\
$^6$  Department of Physics, University of Torino, Via Giuria 1 10125 Torino, Italy \\
$^7$  NASA Marshall Space Flight Center, Huntsville, AL 35812, USA \\
$^8$  Laboratoire de Acc\'el\'erateur Lin\'eaire, University Paris-Sud 11, CNRS/IN2P3, France \\
$^9$  Institut Omega, LLR Aile 4, 91128 Palaiseau Cedex, France \\
$^{10}$  ICRR, University of Tokyo, 5-1-5 Kashiwa-no-Ha, Kashiwa, Chiba 277-8582, Japan \\
$^{11}$  APC, Univ. of Paris Diderot, CNRS/IN2P3, 10, rue A. Domon et L. Duquet, 75205 Paris Cedex 13, France \\
$^{12}$  Karlsruhe Institute of Technology, Germany \\
$^{13}$  Graduate School of Science and Engineering, 255 Shimo-Okubo, Sakura-ku, Saitama 338-8570, Japan \\
$^{14}$  Institute of Experimental Physics SAS, Watsonova 47, 040 01 Kosice, Slovakia \\
$^{15}$  SINP, Lomonosov Moscow State Univ., Leninskie Gory 1 str. 2, Moscow, 119991, Russia \\
$^{16}$  Inst. de Ciencias Nucleares, UNAM, AP 70-543 / CP 04510, Mexico D.F. \\
$^{17}$  University of Geneva, Sauverny, Switzerland \\
$^{18}$  Department of Astronomy and Astrophysics and EFI, University of Chicago, USA \\
$^{19}$  INFN Sezione di Napoli, I-80126 Napoli, Italy \\
$^{20}$  Department of Physics, Ewha Womans University, Seoul 120-750, Korea \\
$^{21}$  INFN Laboratori Nazionali Frascati, Via E. Fermi, 40 00044 Frascati Rome, Italy \\
$^{22}$  Space and Astroparticle Group, University of Alcala Ctra. Madrid-Barcelona, km. 33.6, E-28871, Alcala de Henares, Madrid, Spain \\
$^{23}$  Institut f\"ur Astronomie und Astrophysik, Universitat T\"ubingen, Sand 1, 72076 T\"ubingen, Germany \\
$^{24}$  Soltan Institute for Nuclear Studies, 90-950 Lodz, Box 447, Poland \\ 
$^{25}$  St. Kliment Ohridski University of Sofia, 5, James Bourchier Boul.,  SOFIA 1164, Bulgaria \\
}

\email{kajino@konan-u.ac.jp, casolino.marco@gmail.com, ebisu@postman.riken.jp} 

\abstract{
The Extreme Universe Space Observatory on the Japanese Experiment Module (JEM-EUSO) 
with a large and wide-angle telescope to be mounted on the International 
Space Station will open up "particle astronomy" from space. 
It will characterize Ultra High-Energy Cosmic Rays (UHECR) by detecting 
fluorescent and Cherenkov photons generated by air showers in the earth's atmosphere.  
The JEM-EUSO telescope consists of 3 light-weight 
optical Fresnel lenses with a diameter of about 2.5~m, 300~k channels of MAPMTs, frontend 
readout electronics, trigger electronics, and system electronics.  An infrared camera and  
a LIDAR system on-board and a global light system on the ground 
will also be used to monitor the earth's atmosphere and to calibrate the telescope instruments.
}


\maketitle

\section{Introduction}

JEM-EUSO on board the International Space Station (ISS) is a new type of observatory 
that uses the whole Earth as a giant detector to observe transient luminous phenomena 
in the earth's atmosphere caused by particles and waves coming from space.
JEM-EUSO telescope is designed to detect Extreme-Energy Cosmic Rays (EECR) that come into the atmosphere. 
They collide with atmospheric nuclei and produce extensive air showers (EAS). 
Charged particles in EAS excite nitrogen molecules and emit near ultra-violet (UV) photons. 
They also produce Cherenkov photons in a narrow cone of roughly $1 ^{\circ}$ along 
a trajectory of EAS. 
The telescope observes these photons from the ISS orbital altitude of about 400 km. 
Reflected Cherenkov photons at the ground are observed as a strong Cherenkov mark.
Viewing from the ISS orbit, the Field-of-View (FoV) of the telescope ($\pm 30 ^{\circ}$) corresponds 
to an observational area on the ground larger than  1.9 $\times 10^5$ km$^2$. 

The threshold energy to detect EECRs is as low as several $\times 10^{19}$ eV.
As EECRs with such energies will not bend much in the magnetic field of our galaxy and outer galaxy, 
we will be able to open up "charged particle astronomy"  to study origins and acceleration mechanism 
of EECRs. 

\begin{figure}[h]
  \centering
  \includegraphics[width=0.46 \textwidth]{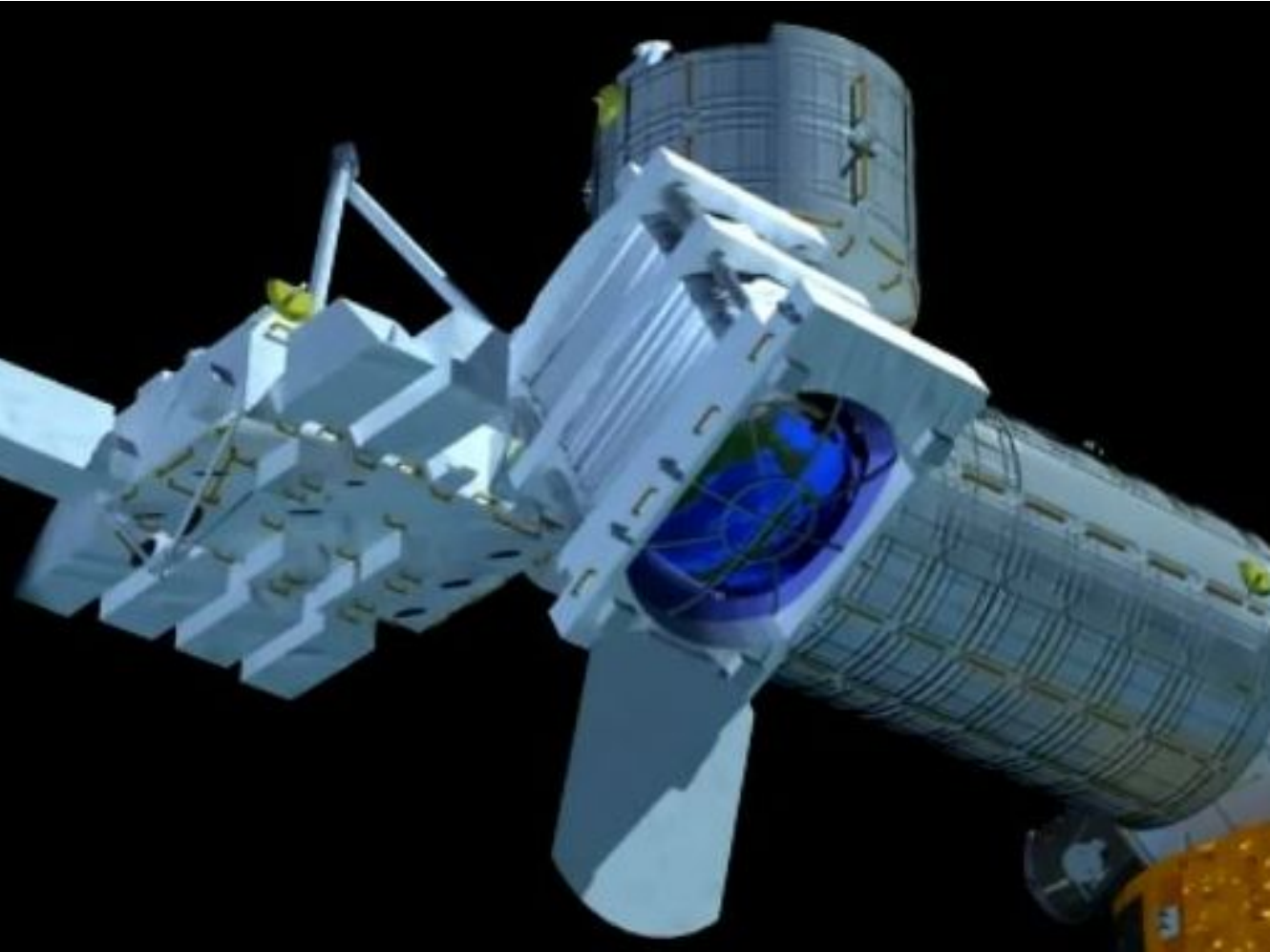}
  \caption{Illustrated view of the tilted mode of the JEM-EUSO telescope mounted on 
  the ISS}
  \label{jem-euso-iss-fig}
 \end{figure}

An increase in exposure is achieved by inclining the telescope from nadir to tilted mode
  (Figure \ref{jem-euso-iss-fig}) .
The first half of the mission lifetime will be devoted to observe lower energy cosmic rays 
with the nadir mode and the second half to observe higher energies using the tilted mode. 
JEM-EUSO will be launched by H2B rocket and conveyed by H-II Transfer Vehicle (HTV) to the ISS.
The SpaceX Dragon spacecraft is under consideration as an option for the transfer 
vehicle instead of the HTV \cite{bib:Adam}.
JEM-EUSO will be attached to the Exposure Facility (EF) of the Japanese Experiment Module (JEM). 

Details of JEM-EUSO mission\cite{bib:Taka,bib:Ebis,bib:Pico}, 
its science objectives\cite{bib:Medi}, 
its requirements and expected performances
\cite{mbib:EUSOperf,bib:Shin,bib:Supa,bib:Guzm,bib:Weil,bib:Bert,bib:Men2}
are reported elsewhere.

\section{JEM-EUSO System}
\begin{figure}[b]
  \centering
  \includegraphics[width=0.47 \textwidth]{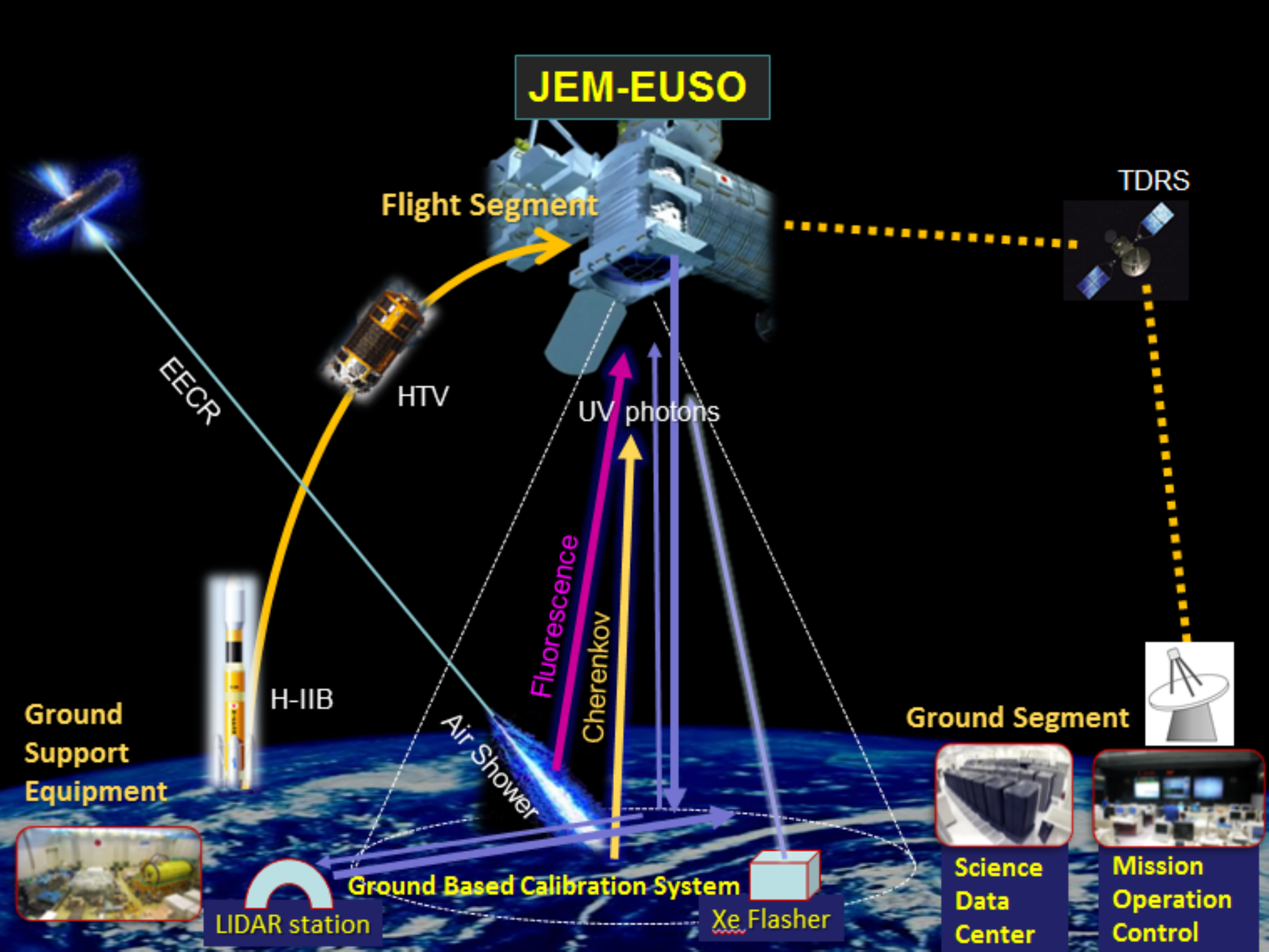}
  \caption{Conceptual view of the whole JEM-EUSO system}
  \label{jem-euso-sys1-fig}
 \end{figure}

\begin{figure}[h]
  \centering
  \includegraphics[width=0.47 \textwidth]{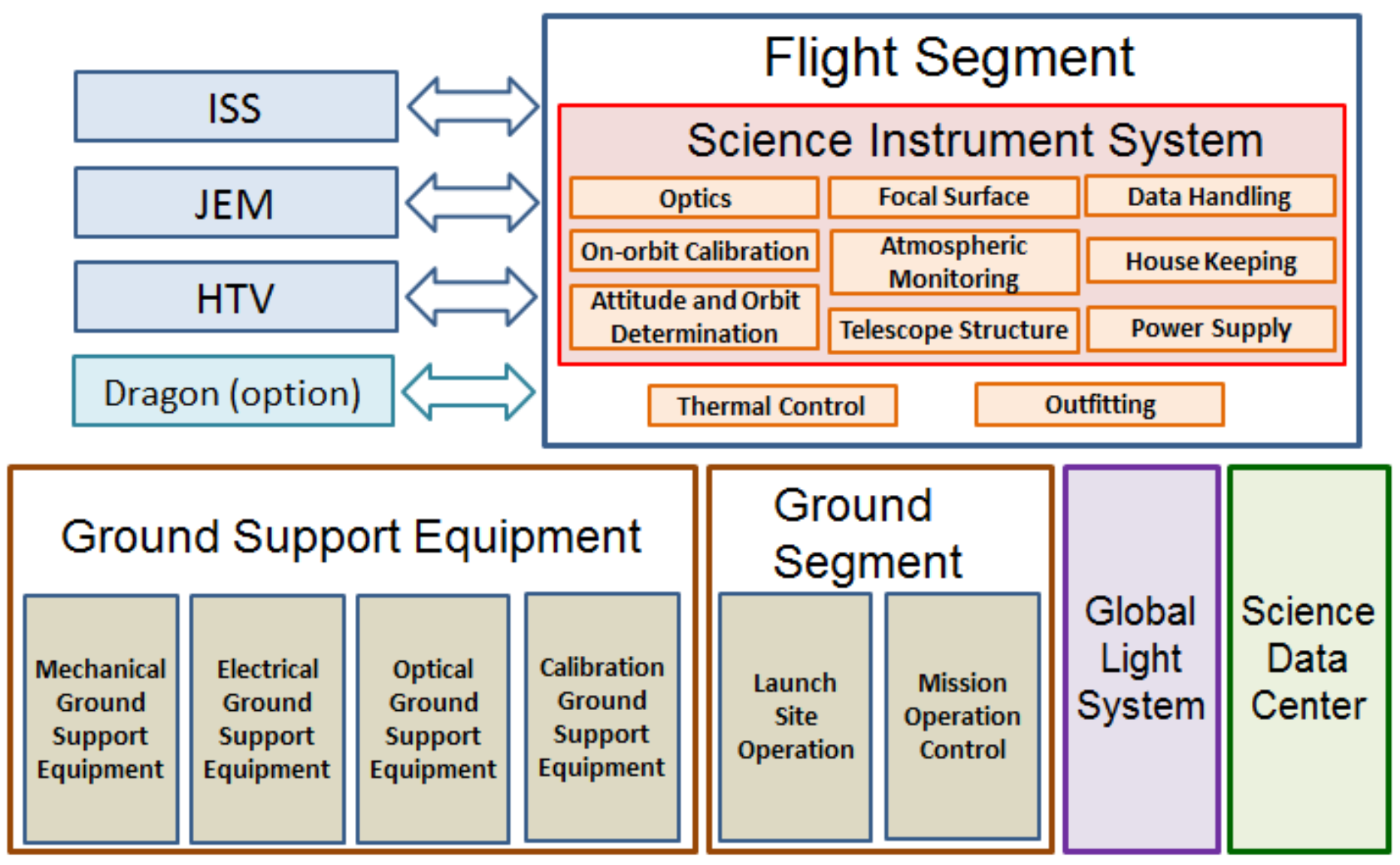}
  \caption{Overall JEM-EUSO system}
  \label{jem-euso-sys2-fig}
 \end{figure}

Conceptual view of the whole JEM-EUSO system is shown in Figure \ref{jem-euso-sys1-fig}. 
The JEM-EUSO system consists of a Flight Segment, Ground Support Equipment (GSE), 
Ground Segments (GS), a Global Light System (GLS) and a science data center which are shown 
in Figure \ref{jem-euso-sys2-fig}. 

The Flight Segment mainly consists of a Science Instrument System which basically 
consists of the following elements:
1) The JEM-EUSO telescope which is a large diameter telescope to observe EECR, ~
2)  Atmospheric Monitoring System, ~
3)  Calibration System.
Details of these systems are described in the following sections. 

GSE consists of mechanical, electrical, optical and calibration GSE. 
GSE supports the project during the manufacturing of the Flight Segment.  

The GS consists of Launch Site Operation and Mission Operation Control 
and supports launching and mission operation.
The GLS is used to calibrate the instruments while the mission is in operation by using a dozen of 
Xenon flashers installed on the ground.
This is done about once a day at each station, when JEM-EUSO passes overhead.

Ultraviolet lasers from the ground LIDAR stations are also used as a part of the GLS.
Data taken by the Science Instrument System on the ISS are sent 
to a Mission Operation Control (MOC) on the ground though a Tracking and Data Relay Satellite (TDRS), 
and then to a science data center.

\subsection{The JEM-EUSO telescope}
\begin{figure}[h]
  \centering
  \includegraphics[width=0.45 \textwidth]{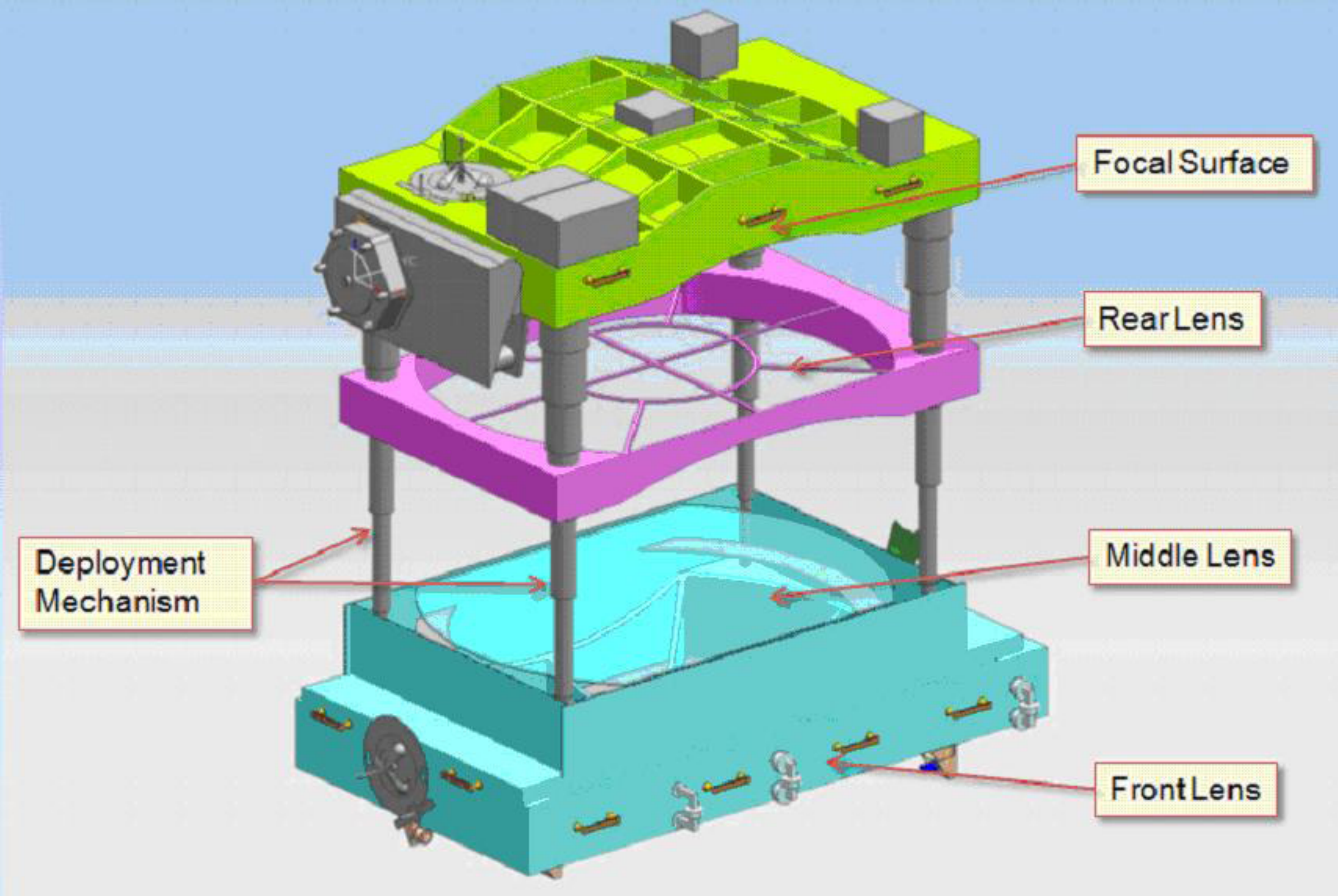}
  \caption{Conceptual view of the JEM-EUSO telescope}
  \label{jem-euso-tel-fig}
 \end{figure}

The Flight Segment of the JEM-EUSO mission forms a large aperture telescope.
This JEM-EUSO telescope is an extremely-fast, highly-pixelized, large-aperture and large-FoV 
digital camera, working in near-UV wavelength range (300 - 430~nm) with single photon counting capability. 
The telescope consists of four main parts: collecting optics with 3 lenses, a focal surface detector, 
electronics and a structure.  (Figure \ref{jem-euso-tel-fig})

The optics focuses the incident UV photons onto the focal surface with an angular resolution 
of $0.1^{\circ}$ . 
The focal surface detector converts the incident photons to electric pulses.  
The electronics counts the number of the pulses in a period less than 2.5 ~$\mu$s 
and records it as a brightness data. 
When a signal pattern of an EAS is found, a trigger is issued. 
This starts a sequence to send the brightness data of the triggered and surrounding 
pixels to the MOC. 

The structure encloses all the parts of the instruments and protects them from the outer 
harmful environment in space. 
It also preserves the optical lenses and the focal surface detector in the preset place. 
The telescope is stowed when it is launched and deployed in observation mode. 

Main parameters of the JEM-EUSO telescope are summarized in Table 1.


\begin{table}[h]
\begin{center}
\begin{tabular}{|l|l|}
\hline 
Field of View         & $ \pm 30 ^\circ $ \\ 
Observational area    & $>$ 1.4  $\times 10^5 $  ~km$^2$ \\ 
Optical bandwidth     & 300 - 430 ~nm  \\ 
Focal Surface area    & 4.5 ~m$^2$  \\ 
Number of pixels      & $ 3.2 \times 10^5 $   \\ 
Pixel size            &  2.9 ~mm   \\ 
Pixel size at ground & $ \sim$ 550 ~m   \\ 
Spatial resolution    & $ 0.07^\circ $  \\ 
Event time sampling   & $ 2.5 ~\mu$s  \\ 
Observational duty circle  & $\sim 20 ~\% $  \\ 
\hline 
\end{tabular}
\caption{Parameters of JEM-EUSO telescope}
\label{table_single}
\end{center}
\end{table}


\subsection{Optics}
Two curved double sided Fresnel lenses with 2.65~m external diameter, a precision middle Fresnel 
lens and a pupil constitute the optics of the JEM-EUSO telescope. 
The Fresnel lenses can provide a large-aperture, wide FoV optics with low mass and high 
UV light transmittance. 
The combination of 3 Fresnel lenses ahieves a full angle FoV of $60^{\circ}$ 
and an angular resolution of $0.07^{\circ}$. 
This resolution corresponds approximately to 550~m on the earth.
The material of the lens is CYTOP and UV transmitting PMMA which has high UV transparency 
in the wavelength from 330~nm to 400~nm. 
Precision Fresnel optics adopting a diffractive 
optic technology is used to suppress the color aberration.  
Details of the optics are described in \cite{bib:Tak1,bib:Tak2}.

\subsection{Focal Surface Detector}
The focal surface (FS) of JEM-EUSO has a spherical surface of about 2.3 m in diameter 
with about 2.5 m curvature radius, and it is covered with about 5,000 multi-anode 
photomultiplier tubes (MAPMTs) \cite{bib:Prie}. 
The FS detector consists of Photo-Detector Modules (PDMs), 
each of which consists of 9 Elementary Cells (ECs). 
The EC contains 4 MAPMT units \cite{bib:Blak}. 
137 PDMs are arranged in the FS \cite{bib:Kawa}.

A Cockcroft-Walton type high-voltage supply will be used to reduce power consumption 
including a circuit to protect MAPMTs from an instantaneous large light dose, such as 
a lightning flash \cite{bib:Karc}.

The MAPMTs developed for the JEM-EUSO mission are going to be tested 
in the TUS detector on a Russian space mission.

\subsection{Focal Surface Electronics}
The FS electronics system records the signals of UV photons generated by EECRs successively 
in time. 
A new type of frontend ASIC has been developed for this mission that functions
both as a single photon counter and as a charge integrator 
in a chip with 64 channels \cite{bib:Ahmm,bib:Bari}.  
The system is required to keep high trigger efficiency with a flexible trigger algorithm 
as well as a reasonable linearity over $10^{19}-10^{21} $~eV range.  
The requirements of very 
low power consumption must be fulfilled to manage $3.2 \times10^5$ signal channels.  
Radiation tolerance of the electronic circuits in the space environment is also required.

The FS electronics is configured in three levels corresponding to the hierarchy of the FS 
detector system: frontend electronics at an EC level, PDM electronics common to 9 EC units, 
and FS electronics to control 137 units of PDM electronics. 
Anode signals of the MAPMT are digitized and recorded in ring memories 
for each Gate Time Unit ($=~2.5~\mu$s) to wait for a trigger assertion,  
after which the data are read and are sent to control boards. 
JEM-EUSO uses a hierarchical trigger method to reduce the huge original data rate of about 10~GB/s.
Cluster Control Boards (CCB) are used at the last stage of the read-out structure and mainly 
perform further management and reduction of the data 
to 297~kbps for transmission of data from the ISS to the ground operation center \cite{bib:Baye}.

\subsection{Data Handling and Housekeeping Electronics}
The data acquisition and handling system is designed to maximize detector observation 
capabilities to meet the various scientific goals, monitor system status, autonomously
taking all actions to maintain optimal acquisition capabilities 
and handle off-nominal situations \cite{bib:Cas3}.

The data handling electronics includes Mission Data Processor (MDP), Telemetry Command Unit (TCU), 
Data Acquisition Interface (IDAQ), Clock \& Time Synchronization Board  \cite{bib:Scot}.

Main MDP tasks are: 
1) power on/off of all subsystems, ~
2) perform periodic calibrations, ~
3) acquire observation data from the FS detector and atmospheric monitor, ~
4) define trigger mode acquisition, ~
5) read Housekeeping (HK) data related to the mission system, ~
6) take care of real time contingency planning, ~
7) perform periodic Download/Downlink, ~
8) handle slow control 1553 commands.

The purpose of HK is to monitor and to relay control commands to the several ~
subsystems that constitute the JEM-EUSO instrument \cite{bib:Med2}.  

HK tasks include:  ~
(a) sensor monitoring of different subsystems, ~
(b) generation of alarms for the MDP,  ~
(c) distribution of telecommands to subsystems,  ~
(d) telemetry acquisition from subsystems,  ~
(e) monitoring of the status of subsystems.

\subsection{Atmospheric Monitoring System}
The Atmospheric Monitoring system (AM) provides information on the distribution 
and optical properties of the cloud and aerosol layers within the telescope FoV
\cite{bib:Anza,bib:Nero,bib:Tosc,bib:Mora}. 
The intensity of the fluorescent and Cherenkov light emitted from EAS at JEM-EUSO 
depends on the transparency of the atmosphere, the cloud coverage and the height of 
cloud top, etc..  These must be determined by the AM. 

In case of events above $10^{20}$ ~eV, the existence of clouds can be directly detected by 
the signals from the EAS. 
However, the monitoring of the cloud coverage by the AM is 
important to estimate the effective observing time with high accuracy and to increase 
the confidence level of the EECR flux.
The AM consists of an infrared camera (IR) and a LIght Detection And Ranging (LIDAR) device. 
Slow data of FS detector is also useful for monitoring the atmosphere. 

\subsection{Calibration System}
The calibration system measures the efficiencies of the optics, the focal surface detector 
and the data acquisition electronics with a precision necessary to determine energy 
and arrival direction of EECR. 
The calibration system consists of the following categories: 
1) pre-flight calibration, ~
2) on-board calibration, ~
3) calibration in flight with on-ground instruments, ~
4) atmospheric monitor calibration.

The pre-flight calibration of the detector will be done by measuring detection efficiency, 
uniformity, gain etc. with UV LED's \cite{bib:Goro}. 
To measure efficiencies of FS detector, several 
diffuse LED light sources with different wavelengths in the near UV region are placed 
on the support of the rear lens in front of the FS. 
To measure efficiencies of the lenses a similar light source is placed at the center of the FS. 
Reflected light at the inner surface of the lid is observed with the FS. 
In this way, the gain and the detection efficiency of the detector 
will be calibrated on board \cite{bib:Karu}.

The system can be calibrated with a dozen ground light sources when JEM-EUSO passes 
over them \cite{bib:Wien}. 
The amount of UV absorption in the atmosphere is measured with Xe flasher lamps. 
The systematic error in energy and direction determination will be empirically estimated, 
by observing emulated EAS images with a UV laser by the JEM-EUSO telescope. 
The transmittance of the atmosphere as a function of height will be also obtained.

Absolute in-flight calibration of the JEM-EUSO telescope with Moon light is also studied \cite{bib:Saka}.

The IR camera monitors the FoV by periodically taking pictures during observations.   
The IR data will help to estimate the effective area.

Studies of the UV night background estimation using simulations are anticipated \cite{bib:Mer3}.
The absolute fluorescence spectrum and yield need to be studied in order 
to determine the energies of EECR events seen by JEM-EUSO \cite{bib:Monn}.

\section{Pathfinder Experiments : TA-EUSO and EUSO-Balloon}
Two pathfinder experiments, TA-EUSO and EUSO-Balloon, are currently 
being developed that will contribute to the likely success of the JEM-EUSO mission.

TA-EUSO uses a ground-based telescope formed by one PDM and two Fresnel lenses 
to demonstrate and bring to maturity the technologies used for the JEM-EUSO telescope.
The TA-EUSO telescope has been set up and has measured the UV light at the Telescope Array (TA)
 site in Utah, USA in early 2013\cite{bib:Tak1,bib:Tak2,bib:Caso,bib:Cas2,bib:Haun}.

EUSO-Balloon will serve as a demonstrator for technologies and methods featured 
in the space instrument.
This balloon-borne instrument points toward the nadir from a float altitude of about 40 km. 
With its Fresnel optics and PDM, the instrument monitors a $12 \times 12$ degree 
wide field of view. The instrument is presently built. 
A first flight is scheduled in 2014 \cite{bib:Ball,bib:Men1,bib:Tak1,bib:Tak2,bib:Bal2,bib:Haun,bib:More}.

\section{Conclusion}
Phase A study (feasibility study and conceptual design) of the JEM-EUSO mission is 
in progress with an international collaboration of 13 countries at present.
Many new technological items have been developed and 
pathfinder experiments are being performed to realize the JEM-EUSO mission.

\vspace*{0.5cm}
{\footnotesize{{\bf Acknowledgment:}{This work was partially supported by Basic Science Interdisciplinary 
Research Projects of RIKEN and JSPS KAKENHI Grant (22340063, 23340081, and 
24244042), by the Italian Ministry of Foreign Affairs, General Direction 
for the Cultural Promotion and Cooperation, by the 'Helmholtz Alliance 
for Astroparticle Physics HAP' funded by the Initiative and Networking Fund 
of the Helmholtz Association, Germany, and by Slovak Academy  
of Sciences MVTS JEM-EUSO as well as VEGA grant agency project 2/0040/13.
The Spanish Consortium involved in the JEM-EUSO Space
Mission is funded by MICINN under projects AYA2009-
06037-E/ESP, AYA-ESP 2010-19082, AYA2011-29489-C03-
01, AYA2012-39115-C03-01, CSD2009-00064 (Consolider MULTIDARK)
and by Comunidad de Madrid (CAM) under project S2009/ESP-1496.
}}

}
\clearpage

%% file: icrc2013-0937.tex



\title{JEM-EUSO Science capabilities}

\shorttitle{JEM EUSO ICRC 2013}

\authors{
G. Medina-Tanco$^{1}$,
L. Anchordoqui$^{2}$,
A. Olinto$^{3}$
E. Parizot$^{4}$,
T. Weiler$^{5}$,
for the JEM-EUSO Collaboration.
}

\afiliations{
$^1$ Instituto de Ciencias Nucleares, UNAM, Mexico \\
$^2$ Department of Physics, University of Wisconsin-Milwaukee, Milwaukee, USA \\
$^3$ Dep. of Astron. \& Astrophys., Enrico Fermi Institute,
and Kavli Inst. for Cosmological Phys., The University of Chicago, USA \\
$^4$Laboratoire Astroparticule et Cosmologie, Universit\'e Paris 7 / CNRS, 10 rue A. Domon et L. Duquet, 75205 Paris Cedex 13, France \\
$^5$ Department of Physics and Astronomy, Vanderbilt University, Nashville, TN 37235, USA \\
}
 
\email{gmtanco@nucleares.unam.mx} 

\abstract{JEM-EUSO is a space telescope to be installed at the International Space Station to observe extensive air showers (EAS) in the EarthÕs atmosphere produced by cosmic rays of energies above $50$ EeV. JEM-EUSO will reach the unprecedented annual exposure at the highest energies of more than $6 \times 10^{4}$ km$^{2}$ sr yr with very nearly uniform dependence on declination over the Celestial Sphere. These capabilities go far beyond what can be practically achieved by ground-based observatories and enable an all sky study of anisotropies above 60 EeV where hints of anisotropies have been reported. The decrease in attenuation length of UHECRs with increasing energy implies that the extreme energy cosmic ray sky must be dominated by very few, relatively nearby sources. The full sky analysis of JEM-EUSO anisotropy patterns should unveil the closest of these extreme sources of the highest energy particles ever observed. These anisotropy patterns as a function of energy can also set constraints on the particle charge and  the effects of Galactic and extragalactic magnetic fields. The higher statistics measurement of the spectrum at extreme energies (above 100 EeV) can test if the maximum energy of these extreme accelerators reaches well beyond the GZK feature or if it coincides with the GZK effect further constraining the source characteristics.
JEM-EUSO will also study transient light events in the atmosphere related to meteors and atmospheric phenomena. In addition, JEM-EUSO will set limits on Lorenz Invariance violation and will search for nucleates and extreme energy photons and neutrinos that could lead to ground-breaking discoveries in fundamental physics.
}


\maketitle

\section{Introduction}

Although the study of ultrahigh energy cosmic rays (UHECRs), from 1 to 100 EeV (1 EeV = $10^{18}$ eV), has progressed considerably over the last decade, {\it not a single source} of these extreme events has been identified thus far. Current data indicates that only a significant increase in the exposure at the highest energies (above about 60 EeV) will allow a clear source identification \cite{Reviews}. Increasing the statistics of events at the highest energies to discover the first sources of UHECRs is the main goal of the Extreme Universe Space Observatory (EUSO) to be deployed on the Japanese Experiment Module (JEM) of the International Space Station (ISS) \cite{Santangelo_ICRC13}. 

JEM-EUSO will observe the ultraviolet fluorescence light emitted by atmospheric nitrogen excited by extensive air showers (EAS) through the use of an innovative wide field of view Fresnel optics telescope with a highly sensitive focal surface, complemented by an extensive real time atmospheric monitoring system\cite{Kajino_ICRC13}. The mission will reach the unprecedented annual exposure at the highest energies of more than $6 \times 10^{4}$ km$^{2}$ sr yr \cite{Shinozaki_ICRC13}, which is about 9 times the annual exposure of the largest observatory ever built, the Pierre Auger Observatory \cite{AugerExposure}. This order of magnitude increase in annual exposure is reached in the nadir configuration leaving open the possibility of a further increase in exposure at the highest energies in the tilted configuration.
The unprecedented exposure is also nearly uniform over the Celestial Sphere \cite{Shinozaki_ICRC13} enabling 
a full sky survey of possible sources.

\begin{figure}[t]
  \centering
  \includegraphics[width=0.38\textwidth]{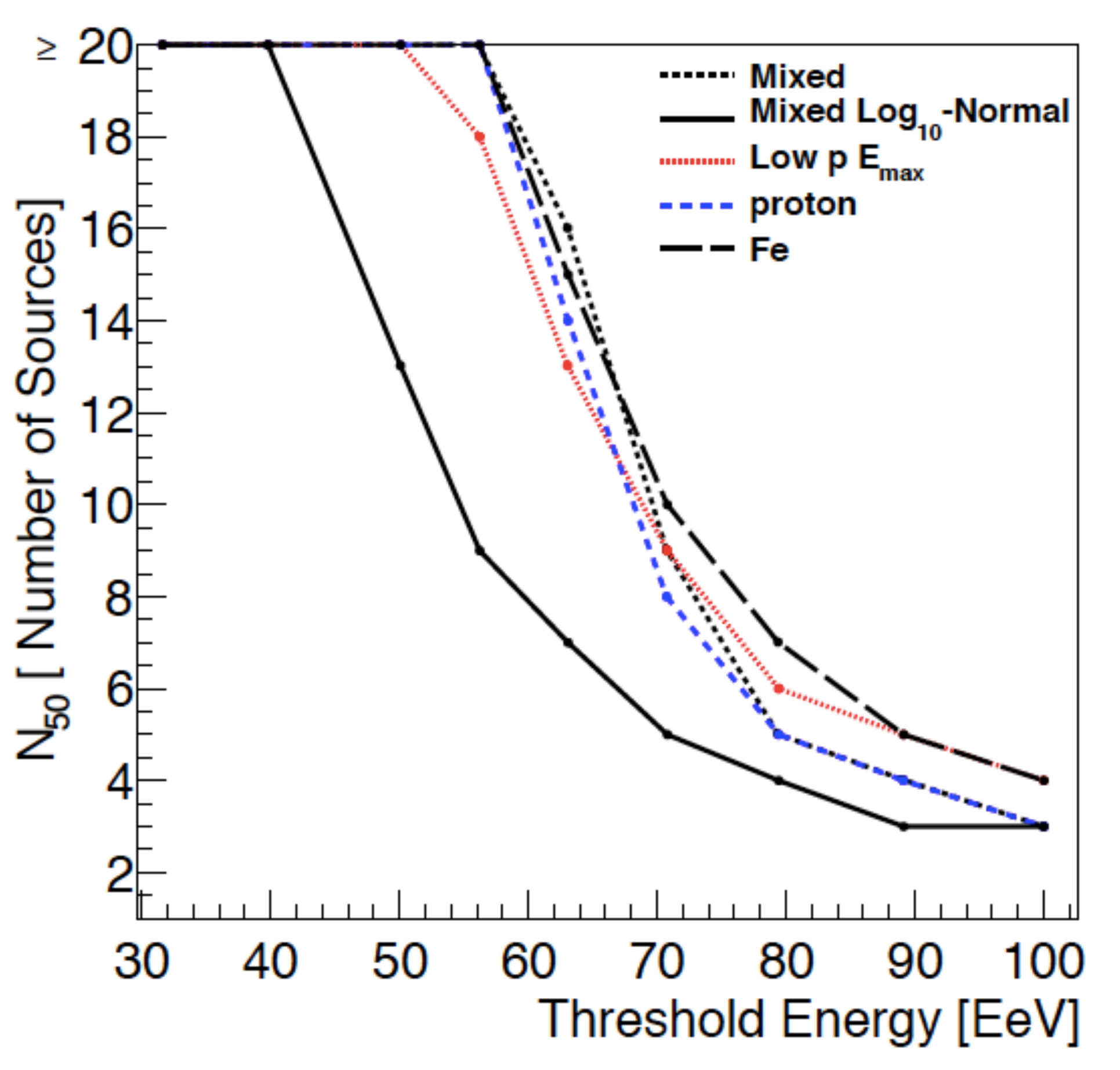}\\
 \vspace{-14pt}
\includegraphics[width=0.38\textwidth]{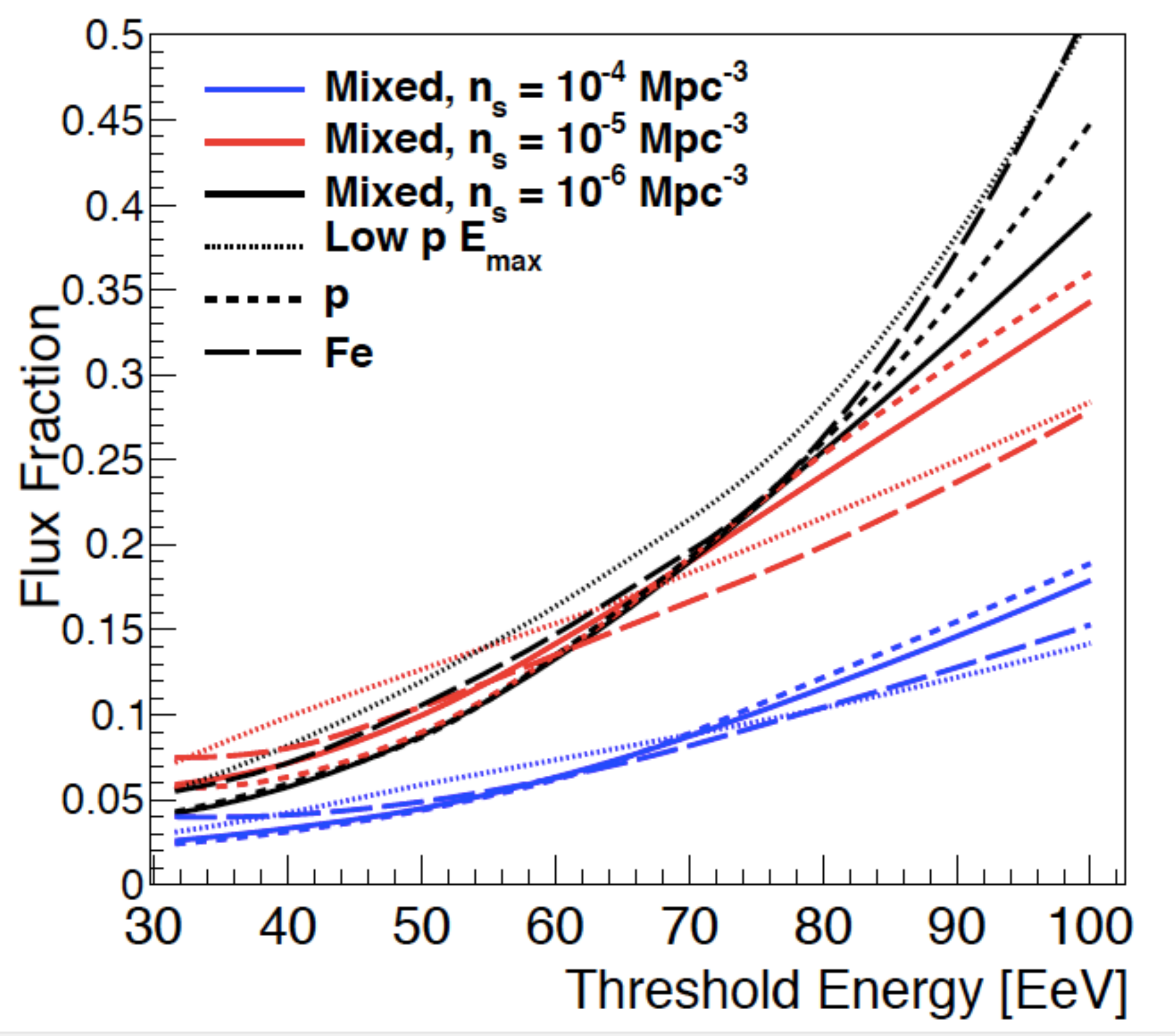}
  \caption{Dominance of the UHECR sky by a few individual sources due to horizon effects. Top: Plotted are the number of sources contributing 50\% of the flux as function of energy. Bottom: fraction of the flux contributed by the 3 brightest sources as a function of threshold energy. Several models and source densities are shown as in \cite{aaDOrfeuil_ICRC13}. }
\label{Fig:NumberSources}
\end{figure}

Recent progress in UHECR science is due to observations by giant ground arrays culminating with the 3,000 km$^2$ Auger Observatory in Mendoza, Argentina \cite{AugerInstrument}, the largest observatory worldwide, and the 700 km$^2$ Telescope Array (TA) in Utah, USA \cite{TAInstrument}, the largest in the northern hemisphere. These two leading observatories have made precise measurements of the spectrum  over a wide range of energy, each in their own hemispheres. Both observatories report spectra which are consistent in normalization and shape after an absolute energy scaling of about 20\% is applied (which is within the quoted systematic uncertainties). The reference spectrum where Auger and TA energy scales are averaged can be described by a triple power law fit where below the  {\it ankle} at about 4.8 EeV, the spectrum is $E^{-\gamma}$ with $\gamma = 3.3$ followed by a hardening with $\gamma = 2.7$ from the ankle up to a suppression at about 38 EeV when the spectrum softens to $\gamma = 4.2$ \cite{WGspectrum}. The ankle may be due to the transition from Galactic to extragalactic cosmic rays or possibly due to losses of cosmic ray protons producing electron-positron pairs in the cosmic microwave background (CMB). The suppression is consistent with the Greisen-Zatsepin-Kuzmin effect \cite{aaGZK} which is due to photo-pion production for protons interacting with the CMB or photo-dissociation of heavier nuclei on cosmic backgrounds (from microwave to ultraviolet). These energy losses limit the volume from which UHECRs can originate to be observed at Earth. 
The horizon for 60 EeV protons and iron are similar at $\sim 100$ Mpc. The attenuation length for intermediate nuclei between proton and iron is shorter. Therefore, the volume of  universe sampled by UHECRs, regardless of their composition, is local in cosmological terms and encompasses a region where the large scale matter distribution is inhomogeneous. The suppression may also be explained by the maximum energy of the accelerator, $E_{\rm max}$. 

\begin{figure}[t]
\centering
\includegraphics[width=\linewidth]{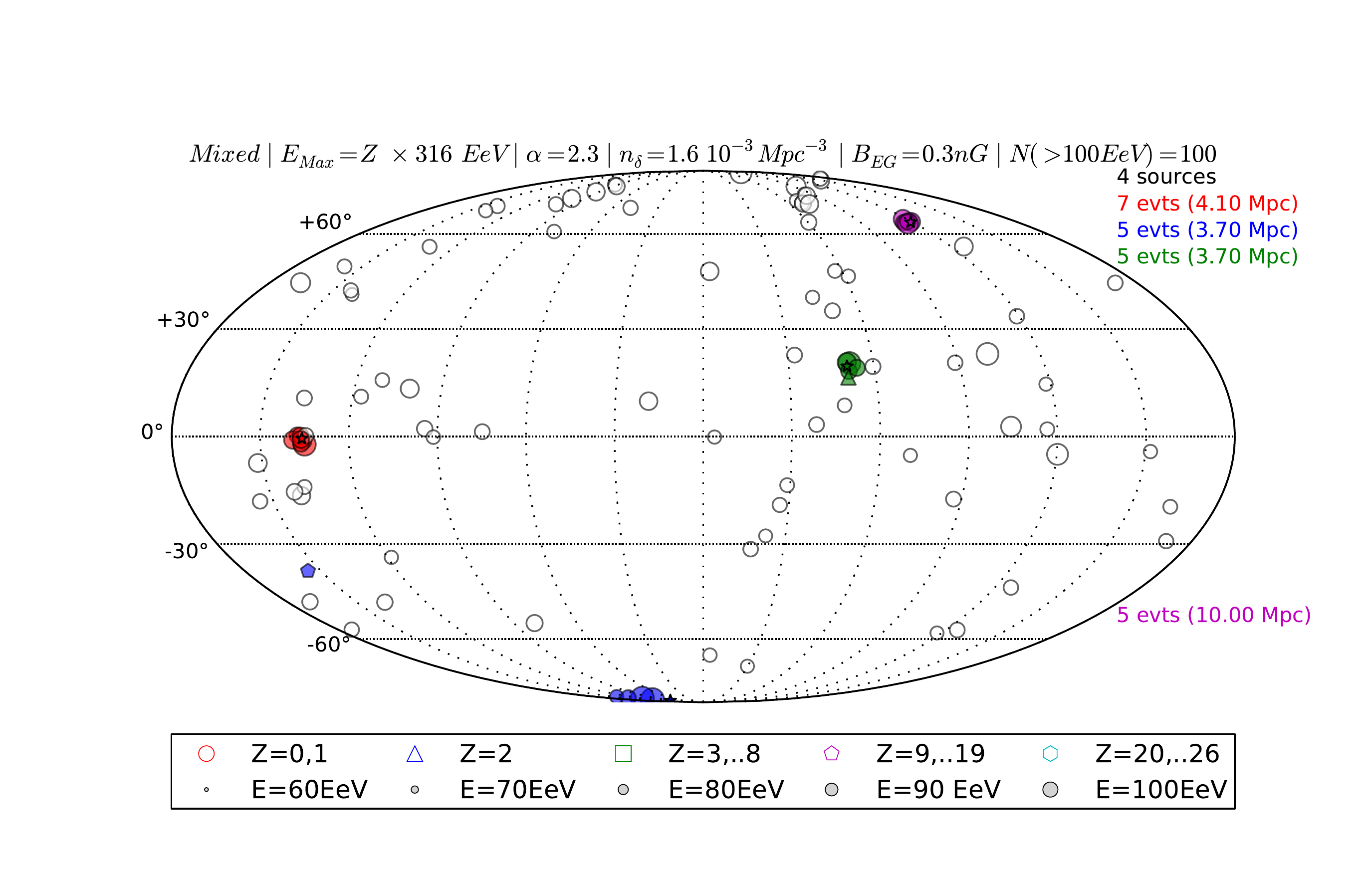}\\
\vspace{-14pt}
\caption{Example of a sky map of 100 UHECR events with energies above 100~EeV (using the Auger energy scale) obtained for total exposure of $300,000\,\mathrm{km}^{2}\,\mathrm{sr}\,\mathrm{yr}$  (5 years of JEM-EUSO). The proton dominated composition model  assumes a maximum energy $E_{Z,\max} = Z\times 316$ EeV,  an injection spectrum of $\alpha = 2.3$ and a source density of $10^{-3}\,\mathrm{Mpc}^{-3}$. The 4 sources contributing the largest fraction of events are color coded \cite{aaDOrfeuil_ICRC13}.}
\label{fig:protonCase}
\end{figure}

As theoretical models attempt a fit to the spectrum together with composition measurements, the GZK effect  is the main cause of the observed suppression for proton dominated models. For mixed composition models that fit the Auger composition,  $E_{26;\rm max}$ of iron (or $E_{p;\rm max}$ of protons divided by 26) is chosen to coincide with the suppression energy while the GZK effect still affects all components up to iron. In this sense, the maximum energy is the main driver for the suppression, although the GZK effect is still present \cite{Reviews}. 

Auger observes a trend toward heavier nuclei above about 5 EeV (or a change in hadronic interactions) while TA reports a proton dominated spectrum throughout their sensitivity range \cite{WGComp}. (It is possible that the change in shower properties observed by Auger is not due to a change in composition but instead to a change in the properties of particle interactions.) Finding the first sources with JEM-EUSO can help determine the composition with studies of source shape distortions at the highest energies while a high statistics measurement of the spectrum at 100 EeV can help select between a GZK effect and $E_{\rm max}$ cause of the suppression. In particular, the spectrum will display a significant recovery if the suppression is produced by the GZK effect and the sources have $E_{\rm max} \gg$ 100 EeV. 

The observed sky distribution of arrival directions show tantalizing hints of anisotropies above about 60 EeV. Both Auger and TA observe partial sky distributions (from the South by Auger and the North by TA) that show hints of the nearby large scale structure distribution above about 60 EeV, but the departure from isotropy continues to be at the low significance level. This is where JEM-EUSO can make significant contributions by increasing by an order of magnitude the exposure to cosmic rays with energies between 60 and 100 EeV, sometimes called extreme energy cosmic rays (EECRs).

JEM-EUSO will pioneer UHECR observations from space with a far greater exposure than any experiment on the ground. The 60 degree field of view will instantly monitor several 10$^6$ km$^3$ volume of the atmosphere, an order of magnitude larger than any current observatory. The corresponding quantitative jump in statistics will clarify the origin  of the UHECRs and probe particle interactions at energies well beyond those achievable by man-made accelerators. Furthermore, the JEM-EUSO mission will make important contributions to atmospheric phenomena including meteors by monitoring the  Earth's atmosphere in the ultraviolet with the main telescope and in the infrared with the telescope's atmospheric monitoring system. Among the exploratory objectives of JEM-EUSO are the search for high energy gamma rays and neutrinos that would be ground-breaking if detected. In addition, JEM-EUSO will set limits on the violation of Lorentz Invariance at relativistic factors up to $10^{11}$ and search for exotic events that may be caused by nucleorites and monopoles traversing the atmosphere.

\section{Main science objectives}

The main objective of JEM-EUSO is to begin the new field of particle astronomy and astrophysics by identifying the first sources of UHECRs. To reach that goal a 5 year mission will achieve an exposure of  $3 \times10^5$ km$^{2}$ sr yr at 100 EeV \cite{Shinozaki_ICRC13}. Each additional year of operation will add the equivalent of nine years of operation of a ground detector as large as Auger at extreme energies. Such exposure makes possible unprecedented anisotropy studies including the possible identification of individual nearby sources by high-statistics arrival direction analysis. It will also allow a higher statistics measurement of the energy spectrum at 100 EeV over the whole sky and the study of atmospheric and meteor phenomena.  A number of additional exploratory goals will be discussed in the next section.

The mysterious sources of UHECRs most certainly involve extreme physical processes in extreme extragalactic environments as very few known astrophysical objects can reach the requirements imposed by the observed spectrum, composition, and lack of strong anisotropies \cite{Reviews}. In particular, the lack of anisotropies towards the Galactic plane implies an extragalactic origin for protons above $\sim$ 1 EeV and above $\sim Z$ EeV for nuclei with charge $Z$, as discussed  by \cite{Giacinti} based on Auger limits on the dipole amplitude and reasonable models of Galactic magnetic fields.

As they traverse cosmological distances, UHECRs lose energy through interactions with cosmic photon backgrounds limiting the observable horizon to about 100 Mpc for energies above 60 EeV. The horizon effect limits the number of sources contributing to the observed flux for proposed source models as shown in Figure \ref{Fig:NumberSources}. This decrease in source number translates into an increase in anisotropies at the highest energies making source identification easier above energies of about 80 EeV. Thus, JEM-EUSO can discover the closest sources by a significant increase in statistics at extreme energies. 

The expected sky map of events that will be observed by JEM-EUSO depends strongly on the primary composition and the number density of sources in addition to other model parameters such as the injected spectrum and $E_{\max}$. An extensive study of predicted sky maps with JEM-EUSO statistics that are consistent with current data on spectrum, composition, and lack of strong anisotropies is found in \cite{aaDOrfeuil_ICRC13}. Figure \ref{fig:protonCase}  shows the sky map of a proton dominated case where 100 UHECR events are shown with energies above 100 EeV (using the Auger energy scale) obtained for an exposure of $300,000\,\mathrm{km}^{2}\,\mathrm{sr}\,\mathrm{yr}$  (i.e., 5 years of JEM-EUSO in Nadir mode). This model  assumes a maximum energy $E_{Z,\max} = Z\times 316$ EeV,  where $Z$ is the charge of nuclei between proton and iron. With such a large maximum energy, the spectrum is dominated by protons. The other model parameters are an injection spectral index of $\alpha = 2.3$ and a source density of $10^{-3}\,\mathrm{Mpc}^{-3}$ to be consistent with the lack of strong anisotropies in current observations. The 4 sources contributing the largest fraction of events are color coded and the clustering of events around the sources is clear. This proton case can be easily identified by JEM-EUSO. In this case, the change in observed shower properties reported by Auger should be interpreted as due to changes in particle interactions at the highest observed energies instead of due to composition changes. Also in this case, the observed suppression of the spectrum is due to the GZK cutoff, not $E_{\max}$, and JEM-EUSO may observe the recovery of the spectrum if $E_{\max}$ extends beyond 300 EeV.

\begin{figure*}[t!]
\centering
\includegraphics[width=0.75\linewidth]{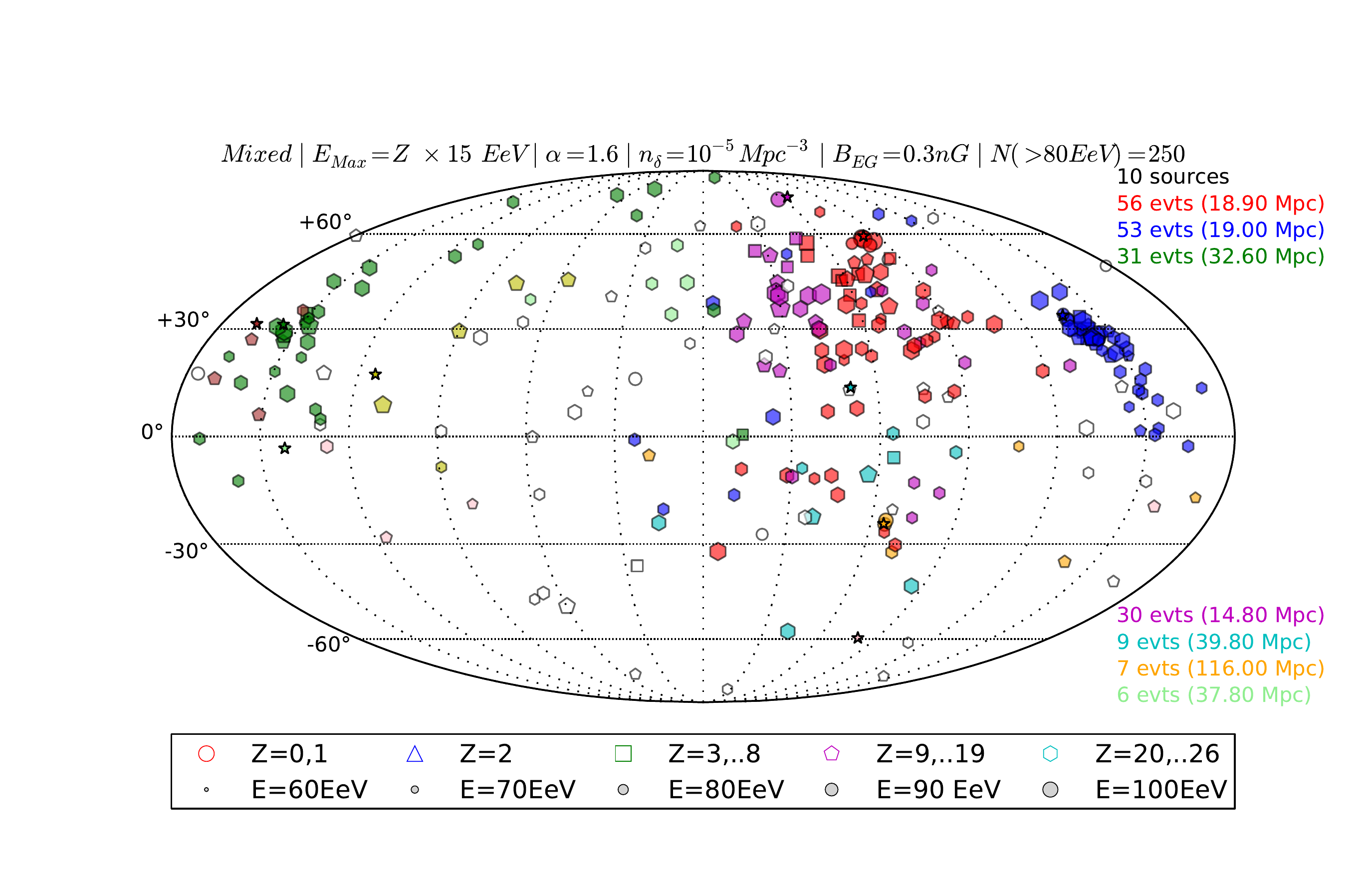}
\vspace{-14pt}
\caption{Example of a sky map of 250 UHECR events with energies above 80~EeV (using the Auger energy scale) obtained for total exposure of $300,000\,\mathrm{km}^{2}\,\mathrm{sr}\,\mathrm{yr}$  (i.e., 5 years of JEM-EUSO. This mixed composition model  assumes a maximum energy $E_{Z,\max} = Z\times 15$ EeV,  an injection spectrum of $\alpha = 1.6$, and a source density of $10^{-5}\,\mathrm{Mpc}^{-3}$. The 10 sources contributing the largest fraction of events are color coded \cite{aaDOrfeuil_ICRC13}.}
\label{fig:MCMid-5-80}
\end{figure*}

To fit the composition trend observed by Auger together with the spectrum and lack of strong anisotropies, mixed composition models are needed in which $E_{\max}$ for iron is close to the observed highest energy events. Figure \ref{fig:MCMid-5-80} shows such a mixed composition model in which the maximum energy  $E_{Z,\max} = Z\times 15$ EeV,  the injection spectral index is $\alpha = 1.6$, and the source density is $10^{-5}\,\mathrm{Mpc}^{-3}$  \cite{aaDOrfeuil_ICRC13}. In this case, 4 sources contribute $\sim 70$\% of events and generate significant anisotropies as shown by the distribution of 250 events above 80 EeV displayed in the figure. Identifying the sources in such a scenario will only be possible with an increase of statistics such as planned for JEM-EUSO. In this case, the observed suppression of the spectrum is due to the maximum energy reached by the accelerators, $E_{\max}$, instead of the GZK effect and the spectrum of these sources should not display a recovery. 

In addition to the significant increased exposure, an advantage of an orbiting observatory, such as JEM-EUSO, with respect to a ground observatory is the full sky coverage. An all-sky survey offers access to large scale multipoles such as dipoles and quadrupoles which are challenging for observations with partial sky coverages \cite{Weiler_ICRC13}. For example, a partial sky map may be unable to distinguish a dipole from a quadrupole depending on the orientation, while  a full sky survey can distinguish the two cases with the same statistics.  An example of the power of the combined statistics and full sky coverage of JEM-EUSO in extracting a high significance dipole is discussed in \cite{KICPsumm} where a 5$\sigma$ dipole detection is within reach of JEM-EUSO assuming the Auger anisotropy hint towards Centaurus A generates a dipole.

As shown by current experiments, there are a number of possible tests for anisotropies that can be applied when significant anisotropies are present. One possible such measurement is the variation due to cosmic variance of the spectrum at the highest energies. Large exposure over the full sky will allow the measurement of the spectrum variation as the sky is partitioned into different regions. As the energy increases, data from different hemispheres are dominated by different sources causing a detectable 
ensemble fluctuations.  The superior sensitivity of JEM-EUSO as compared to current ground observatories to ensemble fluctuations based on various assumptions about the CR source properties and distributions is discussed in \cite{ICRC239}. 
The size and fine details of the variance are sensitive to and therefore yield information about the density of sources, the proximity to the nearest source or source populations, and the composition of the highest energy CRs.  

In addition to searching for the mysterious sources of UHECRs, JEM-EUSO will monitor the Earth's dark atmosphere to observe atmospheric transient light events and meteor events. For example, meteor observations by JEM-EUSO will help derive the inventory and physical characterization of the population of small solar system bodies orbiting in the vicinity of the Earth. After decades of ground-based activities, JEM-EUSO mission may become the first space-based platform to the observe meteor events 
which are eminently ÓslowÓ events when compared to UHECR showers.

\section{Exploratory objectives}

In addition to studying the highest energy cosmic rays, JEM-EUSO is also capable of observing extreme energy cosmic photons and neutrinos \cite{ICRC:PhotonsNeutrinos}. EECR propagation through the cosmic background radiation produces  extreme energy gamma-rays (EEGRs) and neutrinos (EE$\nu$s) as a natural consequence of $\pi^{0}$ and charged  $\pi$ production respectively (usually called cosmogenic photons and neutrinos).  The attenuation length for EEGRs is very short depending on the cosmic radio background. The expected flux of EEGRs on Earth is small and highly model dependent, (e.g., nuclei primaries produce much fewer gamma-rays than proton primaries). JEM-EUSO will search for EEGRs events and place stronger constraints on their flux \cite{ICRC:PhotonsNeutrinos}. A detection of a higher than expected flux can be due to a new production mechanism such as top-down decay or annihilation \cite{Bhattacharjee:1998qc} or the breaking of Lorentz Invariance.  

Similarly to EEGRs, the detection of EE$\nu$s is another exploratory objective of the JEM-EUSO mission. The flux of cosmogenic neutrinos around 100 EeV is highly dependent on  $E_{\max}$ of cosmic rays. For high enough $E_{\max}$, a flux of cosmogenic neutrinos is within reach of the  JEM-EUSO mission \cite{KICPsumm}. A neutrino flux from extremely energetic sources may also be observed by JEM-EUSO.  The acceptance for EE$\nu$ events is well above current ground detectors. In addition, 
an order of magnitude larger acceptance results for Earth-skimming events transiting ocean compared to transiting land is discussed in \cite{PalomaresRuiz:2005}. Since ground-based observatories cannot observe ocean events, only space-based missions can realize the advantage of this possible enhancement of the acceptance over the ocean.

The observing strategy developed for JEM-EUSO to detect atmospheric and meteor events will also be sensitive to other hypothetical slow velocity events such as nuclearites or massive strangelets (quark nuggets with a fraction of strange quarks similar to up and down quarks).  JEM-EUSO is sensitive to nuclearites with mass $m > 10^{22}$ GeV/c$^2$  .  A null observation of these events will set strong limits on their flux, reaching one order of magnitude more stringent limits than current ones in only one day of observations \cite{Bertaina_ICRC13}. This search is a great example of the multi-disciplinary capabilities of the JEM-EUSO mission.

\clearpage


%% file: icrc2013-1171.tex



\def\labelitemi{--}

\title{EUSO-BALLOON : a pathfinder for observing UHECR's from space}

\shorttitle{The EUSO-BALLOON pathfinder}

\authors{
P. von Ballmoos$^{1}$, A. Santangelo$^{5}$,  J.H. Adams$^{19}$, P. Barrillon$^{2}$, J. Bayer$^{5}$, M. Bertaina$^{12}$, S. Blin-Bondil$^{2}$, F. Cafagna$^{7}$, M. Casolino$^{13,10,11}$, S. Dagoret-Campagne$^{2}$, P. Danto$^{4}$, A. Ebersoldt$^{6}$, T. Ebisuzaki$^{13}$, J. Evrard$^{4}$, Ph. Gorodetzky$^{3}$, A. Haungs$^{6,}$ A. Jung$^{14}$, Y. Kawasaki$^{13}$, H. Lim$^{14}$, 
G. Medina-Tanco$^{15}$, T. Omori$^{13}$, G. Osteria$^{9}$, E. Parizot$^{3}$, I.H. Park$^{14}$, P. Picozza$^{13,10,11}$, G. Prévôt$^{3}$, H. Prieto$^{13,17}$, M. Ricci$^{8}$, M.D. Rodríguez Frías$^{17}$, J. Szabelski$^{16}$, Y. Takizawa$^{13}$, K. Tsuno$^{13}$ 
for the JEM-EUSO Collaboration$^{20}$.}

\afiliations{
$^{1}$ Institut de Recherche en Astrophysique et Planétologie, Toulouse, France\\
$^{2}$ Laboratoire de l'Accélérateur Linéaire, Univ Paris Sud-11, CNRS/IN2P3, Orsay, France\\
$^{3}$  AstroParticule et Cosmologie, Univ Paris Diderot, CNRS/IN2P3, Paris, France\\
$^{4}$  Centre National d'Etudes Spatiales, Centre Spatial de Toulouse, France\\
$^{5}$  Institute for Astronomy and Astrophysics, Kepler Center, University of Tübingen, Germany\\
$^{6}$  Karlsruhe Institute of Technology (KIT), Germany\\
$^{7}$  Istituto Nazionale di Fisica Nucleare - Sezione di Bari, Italy\\
$^{8}$  Istituto Nazionale di Fisica Nucleare - Laboratori Nazionali di Frascati, Italy\\
$^{9}$  Istituto Nazionale di Fisica Nucleare - Sezione di Napoli, Italy\\
$^{10}$ Istituto Nazionale di Fisica Nucleare - Sezione di Roma Tor Vergata, Italy\\
$^{11}$ Universita’ di Roma Tor Vergata - Dipartimento di Fisica, Roma, Italy\\
$^{12}$ Dipartimento di Fisica dell’ Università di Torino and INFN Torino, Torino, Italy\\
$^{13}$ RIKEN Advanced Science Institute, Wako, Japan\\
$^{14}$ Sungkyunkwan University, Suwon-si, Kyung-gi-do, Republic of Korea\\
$^{15}$ Universidad Nacional Autónoma de México (UNAM), Mexico\\
$^{16}$ National Centre for Nuclear Research, Lodz, Poland\\
$^{17}$ Universidad de Alcalá (UAH), Madrid, Spain\\
$^{18}$ University of Alabama in Huntsville, Huntsville, USA\\
$^{19}$ http://jemeuso.riken.jp\\
}

\email{pvb@irap.omp.eu} 

\abstract{EUSO-BALLOON is a pathfinder mission for JEM-EUSO (Extreme Universe Space Observatory on-board the Japanese Experiment Module of the International Space Station). Through a series of stratospheric balloon flights starting in 2014, performed by the French Space Agency CNES, the JEM-EUSO consortium will demonstrate the key technologies and methods featured in its future space mission. As JEM-EUSO is designed to observe Ultra-High Energy Cosmic Rays (UHECR)-induced Extensive Air Showers by detecting their ultraviolet (UV) light tracks, EUSO-BALLOON is an imaging UV telescope too. The balloon-borne pathfinder points towards the nadir from a float altitude of about 40 km. With its Fresnel Optics and Photo-Detector Module, the instrument monitors a 12x12° wide field of view in a wavelength range between 290 and 430 nm, at a rate of 400'000 frames/sec. The objectives of EUSO-BALLOON are to perform a full end-to-end test of a JEM-EUSO prototype consisting of all the main subsystems of the space experiment, and to demonstrate the global detection chain while improving our knowledge of the atmospheric and terrestrial UV background. The balloon pathfinder also has the potential to detect for the first time, from above, UV-light generated by atmospheric air-showers, marking a milestone in the development of UHECR science, and paving the way for any future large scale, space-based UHECR observatory.}


\maketitle

\section{The Context for EUSO-BALLOON}

EUSO-BALLOON is a prototype of JEM-EUSO, the Extreme Universe Space Observatory to be hosted on-board the Japanese Experiment Module of the International Space Station (ISS). JEM-EUSO is designed to observe ultra high-energy cosmic rays (UHECRs) by looking downward to the Earth's atmosphere from the ISS, observing the UV fluorescence light of UHECR-induced Extensive Air Showers (EAS). These proceedings contain a number of detailed articles on JEM-EUSO, notably its status \cite{bib:Picozza_ICRC2013}, the science case \cite{bib:Medina-Tanco_ICRC2013}, and an overview on the instruments \cite{bib:Kajino_ICRC2013}. EUSO-BALLOON is developed by the JEM-EUSO consortium as a demonstrator for the technologies and methods featured in the forthcoming space instrument. Since JEM-EUSO's observation of UHECR-induced EAS is based on the detection of an UV light track (fluorescence emission of Nitrogen molecules excited by collisions with shower particles), EUSO-BALLOON is an imaging UV telescope as well. The balloon-borne instrument points towards the nadir from a float altitude of about 40 km. With its Fresnel Optics and Photo-Detector Module, the instrument monitors a 12x12° wide field of view in a wavelength range between 290 and 430 nm, at a rate of 400'000 frames/sec. The EUSO-BALLOON mission has been proposed by a collaboration of three French laboratories (APC, IRAP and LAL) involved in the international JEM-EUSO consortium. Balloon flights will be performed by the balloon division of the French Space Agency CNES, a first flight is scheduled for 2014.

\section{Objectives of the balloon flights}

EUSO-BALLOON will serve as a test-bench for the JEM-EUSO mission as well as any future mission dedicated to the observation of extensive air showers from space. The following objectives shall be attained in a series of balloon flights :

{\flushleft{\bf{A) technology demonstrator}}}\\
EUSO-BALLOON is a full scale end-to-end test of all the key technologies and instrumentation of JEM-EUSO. Crucial issues that will benefit from the balloon flights include the HV power supplies, the HV switches (HV relays commuting the HV in case a bright atmospheric event comes into the field of view and on a pixel), the Front-End Electronics (including the ASICs and FPGA), the on-board hardware and software algorithms involved in the triggering and recognition of cosmic-ray initiated air showers. 

{\flushleft{\bf{B) data acquisition and background study}}}\\
Although the physics and the detection technique of EAS through ultraviolet light (UV) emission is well established and used daily in ground based detectors, their observation from space has never been performed. Since JEM-EUSO uses the Earth’s atmosphere to observe UV (300-400 nm) fluorescence tracks and Cherenkov reflections from EAS, the observations will be sensitive to the variation of the background sources in the UV range. Whereas a number of background measurements have been performed by previous missions, even from space, no focusing instruments have been employed so far and, most importantly, spatial resolutions were extremely low, i.e. the “pixel size” was much larger. Important localized background signals could have been washed out by the integration over a large surface and, likewise, possible temporal variations on small scales were not observable, and thus went unconstrained. 
Measuring a representative background for JEM-EUSO has been the principal driver for determining the pixel size, and hence the global Field of View of EUSO-BALLOON. The EUSO Simulation and Analysis Framework (ESAF) has been adapted to simulate the response of the instrument (see e.g. \cite{bib:Mernik2_ICRC2013}). The configuration used for JEM-EUSO has been modified, scaling for the altitude of the instrument, changing the surface parameterization, introducing the new optical system and field of view (see Table 1).

Observing EAS from space will confirm the feasibility of the technique and provide valuable data for JEM-EUSO, and all future space-borne UHECR experiments. 

The B) objectives are thus:
\begin{itemize}
\item 	experimental confirmation of the effective background below 40 km observed with a pixel size on ground representative for JEM-EUSO (175 m x 175 m in a $\pm$ 6° field of view),
\item 	acquisition of UV signal and background in a format similar to JEM-EUSO,
\item 	testing of observational modes and switching algorithms,
\item 	testing/optimizing trigger algorithms with real observations, i.e. different ground-covers and time-variable background,
\item 	testing of the acquisition capability of the infrared-camera. 
\end{itemize}

\begin{table}[h]
\begin{center}
\begin{tabular}{|l|c|c|}
\hline & JEM-EUSO & BALLOON \\ \hline
Number of PDMs  &	14 & 31 \\ 
Flight Altitude [km]	& 420 & 40\\ 
Diameter of Optics [km]	& 2.5 &	1\\ 
Field of View / PDM  &	3.8°	 &12° \\ 
PDM@ground  [km] &	28.2 &  8.4\\ 
Field of View / pixel	  &  0.08° &0.25°\\ 
Pixel@ground  [km]	 &0.580	 & 0.175\\ 
Signal w/r  JEM-EUSO &1&17.6\\ 
BG w/r  JEM-EUSO &	1 &	0.9-1.8\\ 
S/$\sqrt{\rm{N}}$ w/r to JEM-EUSO	 & 1	& 20-10\\ 
Threshold Energy [eV]	& 3$\cdot10^{19}$ & 1.5-3$\cdot10^{18}$ \\ \hline
\end{tabular}
\caption{Comparison of the principle characteristics between JEM-EUSO and EUSO-BALLOON. The field of view of EUSO-BALLOON - and hence its pixel size - has been dimensioned to measure a background level comparable to the one expected for JEM-EUSO.}
\label{aatable_single}
\end{center}
\end{table}

{\flushleft{\bf{C) pioneering mission for JEM-EUSO }}}\\
A "bonus objective" for EUSO-BALLOON is the actual detection of one or several EAS by looking downward from the edge of space. Since detecting these obviously rare events is unlikely during a first short balloon flight (threshold $\simeq 10^{18}$ eV, see the paragraph on performance below), xenon-flashes and LASER-induced events will provide a proof of principle and a way to calibrate the threshold / sensitivity.

\section{Payload Overview}

\begin{figure}[t]
  \centering
  \includegraphics[width=0.5\textwidth]{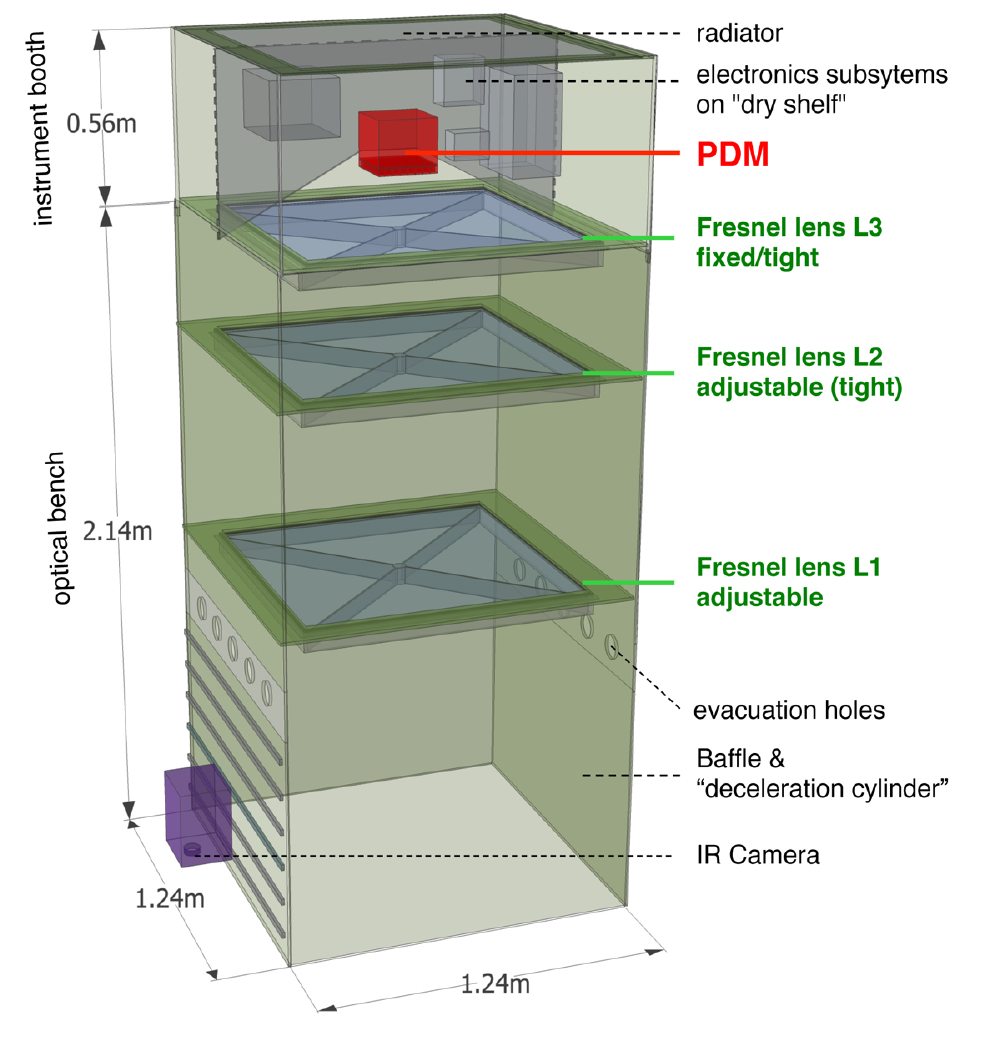}
  \caption{Schematic of the instrument, composed of an optical bench and a watertight instrument-booth.}
  \label{schematic1}
 \end{figure}

The general layout of EUSO-BALLOON is shown in Figure \ref{schematic1}, its main components are the optical bench and the instrument booth. An electronic block diagram of the entire instrument is shown in Figure 2. The development of all components and sub-assemblies\cite{Moretto_ICRC2013} is based on similar JEM-EUSO components and sub-assemblies. The total mass of the payload is about 320 kg; the battery packs will maintain constant power of 225 W during 24 hours of flight (which is more than enough for a first flight that is to last only one night).
The optical bench contains three Fresnel lenses made from 8 mm thick PMMA (UV transmitting polymethyl-methacrylate) with a front surface of 100 cm x 100 cm each. The EUSO-BALLOON optics has been designed to resemble the JEM-EUSO optics (i.e. three Fresnel lenses) : it is dimensioned to produce an RMS spot size smaller than the pixel size of the detector (i.e. 2.85  mm) and keep the background rate per pixel comparable to the one anticipated for JEM-EUSO (i.e. roughly 2 $\pm$1 photoelectrons per pixel in a 2.5$\mu$sec frame). Whereas L1 and L3 are aspherical Fresnel Lenses with focal lengths of 258.56 cm and 60.02 cm, respectively,  L2 is a diffractive lens with focal of 385.69 cm (focal lengths are reference values only, single lenses are not producing stigmatic images). Within the optical bench, the position of L1 and L2 can be adjusted along the optical (z-)axis. Together with the 15 cm x 15 cm focal plane detector (PDM, see below) the optics provides a field of view of $\pm$ 6°. A detailed description of the design and manufacturing of the balloon optics is given in \cite{bib:Takky_ICRC2013} and \cite{bib:Takky2_ICRC2013}.

Since EUSO-BALLOON will eventually observe above open water,  the payload has deliberately been designed to protect all sensitive equipment in the event of a water-landing. The instrument booth is made as a watertight capsule using the third Fresnel lens (L3) as a porthole. Besides the focal plane detector (PDM) and associated electronics (DP) which are described below, the instrument booth houses the telemetry system (SIREN), CNES specific instrumentation (ICDV, Hub), and two battery-packs.\\

{\flushleft{\bf{The Photo-Detector Module (PDM)}}}\\
UV light collected by the telescope is focused onto - and detected by - the PDM, which is composed of 36 MAPMT (Hamamatsu M64 multi-anode photomultipliers) containing 64 anodes each. Testing and sorting of the photomultipliers is detailed in \cite{bib:Blaksley_ICRC2013}. The PDM is organized in 3x3 Elementary Cells (EC) which in turn are composed of 2x2 MAPMTs. A UV color glass filter is bonded to the window of the MAPMT with optical glue. The filter (SCHOTT BG3 with anti-reflection coating) transmits UV light in a band between 290 and 430 nm. The EC unit supplies the voltages produced by the High Voltage Power Supply to each of its MAPMTs, collects the signals from their anodes and transmits them to the ASIC for processing. 

Each of the 2304 pixels (anodes) in the PDM is sensitive to single photons, and features a dynamic range of 6 orders of magnitude thanks to an adaptive gain. The dynodes are driven by Cockroft Walton High-Voltage generators. In order to protect the photodetectors against highly luminous events (lightnings, etc.), custom made High-Voltage switches are capable of reducing the gain in a few microseconds. 
The analogue signal from the MAPMTs anodes is continuously digitized and processed by the Front-End Electronics based on the "SPACIROC" ASIC (Spatial Photomultiplier Array Counting and Integrating Readout Chip, (see \cite{bib:Ahmad_ICRC2013}). The ASIC features a single photo-electron mode (SPE) as well a charge integration mode (KI - i.e. charge to time conversion permitting to measure the intensity of the photon flux). Data acquisition and readout are performed within a defined time slot called Gate Time Unit (GTU=2.5$\mu$s). This is fast enough to observe the speed-of-light phenomena in EAS. 
The output signals from the four ASICs of an EC unit are transmitted to the PDM board which can handle all 9 EC units. The hardware of the PDM board electronics includes an FPGA (Field Programmable Gate Array, the present baseline is the Virtex6 XC6VLX240T), which performs a first-level trigger algorithm (persistency track trigger). A shower candidate is triggered if there is an excess of signal above expected background fluctuations in a box of 3x3 pixels for few consecutive GTUs. The parameters will be adapted in flight as a function of the average background level.

{\flushleft{\bf{The Digital Processor (DP)}}}\\
The different sub-assemblies of the DP collect the PDM data, process them (trigger, time- and position-tagging), handle their on-board storage, and send a subset to the telemetry system. The DP also includes the housekeeping system. The CCB (Control Cluster Board) is developed around a Xilinx Virtex-4 FX-60, it collects the data from the PDM board, processes and classifies the received data, and performs a second level trigger filtering \cite{bib:Bayer_ICRC2013}. The DP then tags the events with their arrival time (UTC) and payload position (GPS). It also manages the Mass Memory for data storage, measures the operating- and dead-time of the instrument, provides signals for time synchronization of the event, performs housekeeping monitoring, and handles the interface with the telecommand/telemetry system.

An event selected by the two trigger levels represents roughly 330 kB of data. Since only a limited data rate can be transmitted to the ground through CNES' new NOSICA telemetry system, all data will be systematically stored on board. The mass storage is composed of two Solid-State Drives (SSD), each one with 1 TB capacity operating in fault-tolerant mode RAID-1 disks (Redundant Array of Independent Disks). The on-line and off-line data analysis is described in \cite{bib:Piotrowski_ICRC2013}. 

\begin{figure*}[!t]
\centering
\includegraphics[width=0.7\textwidth]{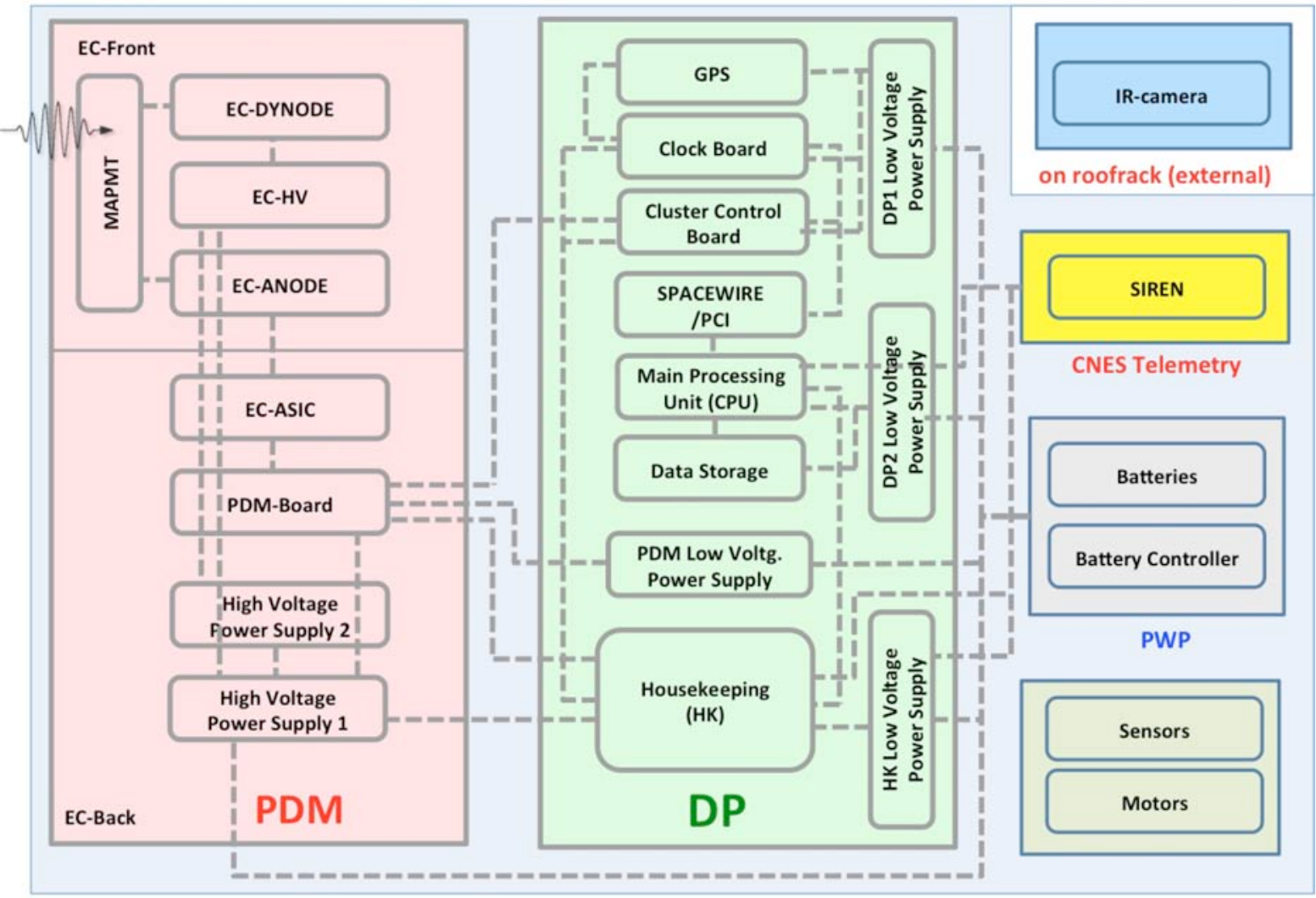}
 \caption{Functional Block Diagram of the EUSO-BALLOON Instrument.}
 \label{wide_fig8}
\end{figure*}

{\flushleft{\bf{Balloon operation}}}\\
During a first flight the payload will operate in nadir pointing mode, the spin rate will be determined by the natural azimuthal oscillations of the flight train. For later flights, the inclination of the pointing axis will be controlled between 0° and 30° with respect to the nadir and an azimuth motor will provide the possibility to perform revolutions with a spin rate of up to 3 rpm. Performing azimuthal revolutions will simulate a groundspeed comparable to the $\sim$7 km/s of the space-station, permitting a full scale test of the HV-switches : i.e. switching MAPMT voltages on/off within a few microseconds, as artificial and other light sources cross the field of view of the instrument.
As the first balloon flight shall take place from a new CNES launch base in Timmins, Canada (lat 48.5° N) a number of different groundcovers will be overflown, including various types of soil and vegetation, water, urban and industrial areas, and - very likely - clouds. EUSO-BALLOON should therefore be able to measure a representative variety of background conditions.

{\flushleft{\bf{Performance}}}\\
While the detection of Extensive Air Showers was not amongst the initial objectives for EUSO-BALLOON, the simulation showed that the instrument was able to detect and image Extensive Air Showers with energies above $10^{18}$eV. This threshold energy arises from the background estimate reported in \cite{bib:Sakaki_2008}.  A first analysis indicates that 0.2-0.3 event (E $> 2\cdot10^{18}$ eV) are expected to be observable during a night-flight of 10 hours. The uncertainty in the estimation assumes also the presence of a moderate cloud fraction. A clear detection will require long duration balloon flights - this is foreseen for subsequent launches and has become a further objective (C-level) for EUSO-BALLOON while the objective of the first flight is to focus on the A- and B-level objectives (see section 2).

In order to monitor the actual cloud covers, a co-aligned IR camera will observe the field of view of the main instrument (similar to the one used on JEM-EUSO, see \cite{bib:Frias_ICRC2013}, \cite{bib:Morales_ICRC2013}). 

\section{Project Organization and Status} 

EUSO-BALLOON is a mission of the French Space Agency CNES, led under the responsibility of the French team, which acts in coordination with the JEM-EUSO management. The instrument is designed and built entirely within the JEM-EUSO collaboration. As its pathfinder, EUSO-BALLOON is identical (PDM, triggers etc.) or similar (optics) to the main mission. All relevant institutions and international partners within the JEM-EUSO collaboration contribute to the instrument  according to their corresponding tasks and responsibilities within JEM-EUSO. 
A ground based prototype, very similar to EUSO-BALLON, has recently been integrated at RIKEN, Japan and installed on the Black Rock Mesa site of Telescope Array (TA), Utah. It is designed to cross calibrate the instrument with the TA Fluorescence Detectors through noise background comparison and during Lidar or electron beam shots. 
As this article is submitted, the Critical Design Review at CNES has been held, a qualification model of the entire electronics chain has shown to operate, and the Fresnel optics is under fabrication : the EUSO-BALLOON project is on track for its first balloon flight in 2014 !

\vspace*{0.5cm}
{\footnotesize{{\bf Acknowledgment :} {The authors acknowledge strong support from the French Space Agency CNES. The work was partially supported by Basic Science Interdisciplinary 
Research Projects of RIKEN and JSPS KAKENHI Grant (22340063, 23340081, and 
24244042), by the Italian Ministry of Foreign Affairs, General Direction 
for the Cultural Promotion and Cooperation, by the Helmholtz Alliance 
for Astroparticle Physics HAP' funded by the Initiative and Networking Fund 
of the Helmholtz Association, Germany.
The Spanish Consortium involved in the JEM-EUSO Space
Mission is funded by MICINN under projects AYA2009-
06037-E/ESP, AYA-ESP 2010-19082, AYA2011-29489-C03-
01, AYA2012-39115-C03-01, CSD2009-00064 (Consolider MULTIDARK)
and by Comunidad de Madrid (CAM) under project S2009/ESP-1496.
}}

}
\clearpage

%% file: icrc2013-1213.tex



\title{Calibration and testing of a prototype of the JEM-EUSO telescope on Telescope Array site}

\shorttitle{EUSO-TA prototype detector}

\authors{
M. Casolino$^{1,2}$, J. Adams$^{15}$, P. Barillon$^{3}$, J. Bayer$^{4}$, J. Belz$^5$,  M. Bertaina$^{13}$, F.Borotto$^{13}$, M.J.  Christl$^{16}$, G.  Distratis$^4$, A. Ebersoldt$^6$, T. Ebisuzaki$^{1}$, T. Fujii$^7$, M. Fukushima$^8$, G. Giraudo$^{13}$, D. Gottschall$^4$, D. Ikeda$^8$, A. Jung$^9$, F. Kajino$^{14}$, Y. Kawasaki$^{1}$, M. Marengo$^{13}$,  J. N. Matthews$^5$, T. Nonaka$^8$,  S. Ogio$^7$, G. Osteria$^{10,2}$, A. Pesoli$^{11}$, P. Picozza$^{1,2}$, L.~W. Piotrowski$^{1}$, H. Sagawa$^8$, V. Scotti$^{10,2}$, T. Shibata$^8$, K. Shinozaki$^{1,4}$, N. De Simone$^{2,11}$, P. Sokolsky$^5$, M. Takeda$^8$, Y. Takizawa$^{1}$, Y. Tameda$^8$, C. Tenzer$^4$  G. B. Thomson$^5$,  H. Tokuno$^{12}$, T. Tomida$^1$, Y. Tsunesada$^{12}$, and the JEM-EUSO collaboration.   
}

\afiliations{
$^1$ RIKEN, Wako, Japan \\
$^2$ Istituto Nazionale di Fisica Nucleare, Italy\\
$^3$ LAL, University of Paris-Sud, CNRS/IN2P3, Orsay, France\\
$^4$ University of Tubingen, Germany \\
$^5$ Institute for High Energy Astrophysics and Department of Physics, University of Utah, Salt Lake City, UT 84112-0830, USA\\
$^6$ Karlsruhe Institute of Technology (KIT), Germany\\
$^7$ Osaka City University, Faculty Asministration Department, Sugimoto-ku, Osaka, 558-8585\\
$^8$ Institute for Cosmic Ray Research, University of Tokyo, Kashiwa, Chiba 277-8582, Japan\\
$^9$ Ehwa Womans University, Seoul, South Korea \\
$^{10}$ University of Napoli, Italy \\
$^{11}$ University of Rome Tor Vergata \\
$^{12}$ Graduate School of Science and Engineering, Tokyo Institute of Technology, Meguro, Tokyo 152-
8551, Japan
$^{13}$  Istituto Nazionale di Fisica Nucleare, Sezione di Torino, Italy 
$^{14}$ Department of Physics, Konan University, Okamoto 8-9-1, Higashinada, Kobe 658-8501, Japan
$^{15}$ University of Alabama in Huntsville, Huntsville, USA\\
$^{16}$ NASA - Marshall Space Flight Center, USA\\
}

\email{casolino@riken.jp}

\abstract{The aim of the EUSO-TA project is  to install a prototype of the JEM-EUSO telescope in the
Telescope Array (TA) site in Black Rock Mesa, Utah, USA and perform observations of ultraviolet light generated by cosmic-ray showers and artificial sources.
The detector consists of one Photo Detector Module (PDM), identical to the 137 that will be present on the JEM-EUSO focal surface.
The PDM is composed of 36 Hamamatsu multi-anode photomultipliers (64 channels per
tube), for a total of 2304 channels. Front-End readout is performed by 36 ASICS, with
trigger and readout tasks performed by two FPGA boards that send the data to a CPU
and storage system. Two, $1$ meter side square Fresnel lenses provide a field of view of $8^{\circ}\times 8^{\circ}$. The telescope is housed in a shed  located in front of one of the
fluorescence detectors of the Telescope Array collaboration, pointing in the direction
of the ELF (Electron Light Source) and CLF (Central Laser Facility). The aim of the
project is to calibrate the response function of the EUSO telescope with the TA
fluorescence detector in presence of a shower of known intensity and distribution. An
initial run of about one year starting from summer 2013 is foreseen, during which we
expect to observe, triggered by TA electronics, a few cosmic ray events which
will be used to further refine the  calibration of the EUSO-TA with TA. Medium term plans include the
increase of  the number of PDM and therefore the field of view.}


\maketitle

\section{Introduction}
The Extreme Universe Space Observatory on the Japanese Experiment Module (JEM-EUSO) of the International Space Station (ISS) is the first mission that will study   Ultra High-Energy Cosmic Rays (UHECR) from space\cite{astra0,2009NJPh...11f5009T}. JEM-EUSO will observe  Extensive Air Showers (EAS) produced by UHECRs traversing the Earth's atmosphere from above. For each event, the detector will make accurate measurements of the energy, arrival direction and nature of the primary particle using a target volume far greater than what is achievable from ground. The corresponding increase in statistics\cite{mario} will help to clarify the origin and sources of UHECRs as well as the environment traversed during production and propagation. Possibly, this  will bring new light onto particle physics mechanisms operating at energies well beyond those achievable by man-made accelerators.
The spectrum of scientific goals of the JEM-EUSO mission includes the detection of high-energy gamma rays and neutrinos, the study of cosmic magnetic fields, and tests of relativity and quantum gravity effects at extreme energies. In parallel JEM-EUSO will systematically perform  observations of the  surface of the Earth in  the infra-red and ultra-violet ranges, studying  atmospheric phenomena (Transient Luminous Effects).
The apparatus is a 2 ton detector using Fresnel-based optics to focus the ultraviolet (UV) light from EAS on a focal surface  composed of about 6,000 multianode photomultipliers for a total of
  $\simeq 3\cdot 10^5$ channels.
In the framework of the EUSO project, a number of prototype detectors are being realized to calibrate the detector response, test its performance in air and  space, raise the Technological Readiness Level of some of the  components and improve our knowledge of the various detectors.
These projects include a series of stratospheric balloon flights (EUSO-BALLOON) and a ground calibration with the Telescope Array collaboration (EUSO-TA).

\begin{figure}
\begin{center}
\resizebox{0.9\columnwidth}{!}{
  \includegraphics{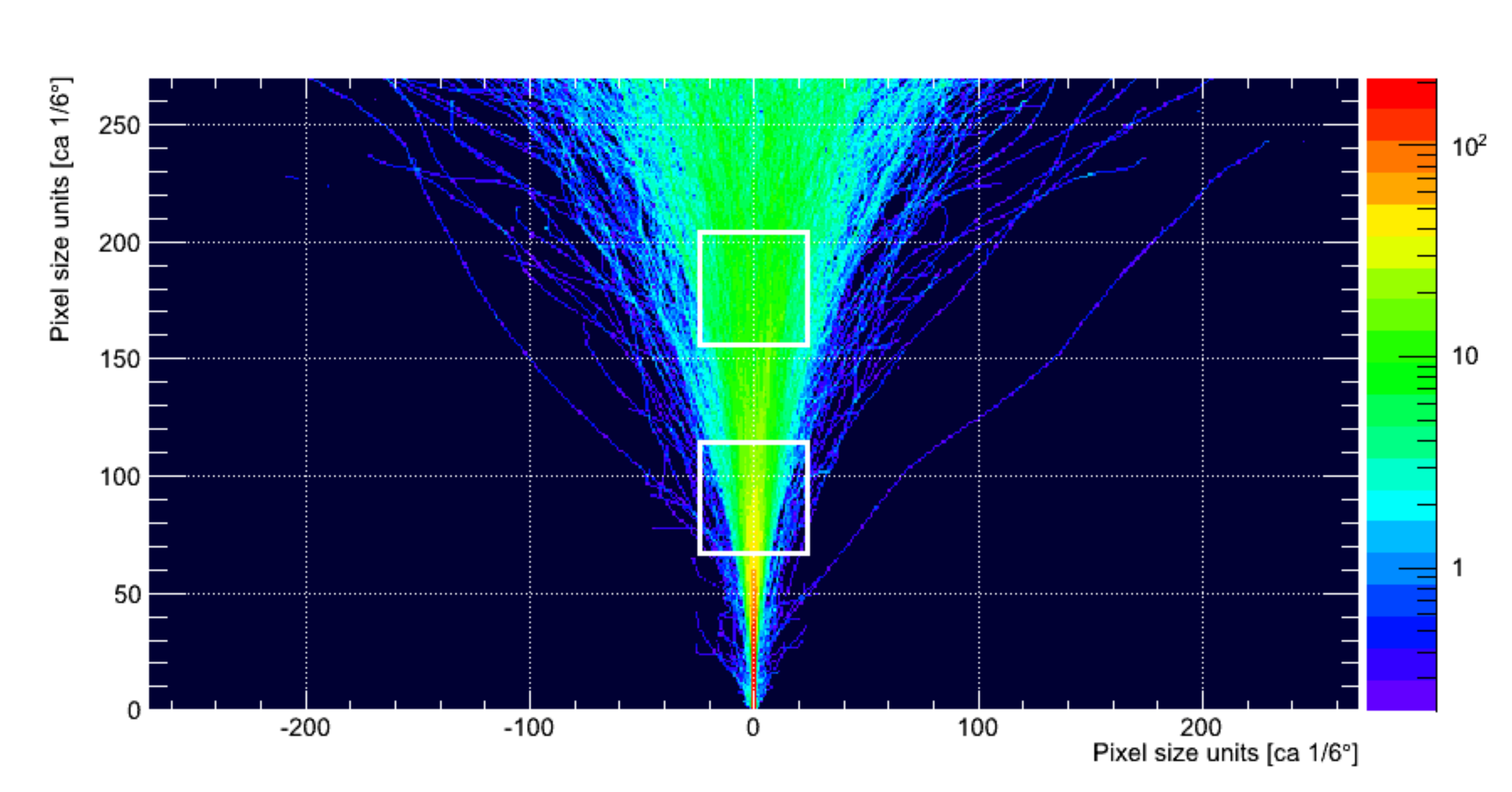} }
\caption{Simulation of the UV photons generated by the TA ELS as seen from the EUSO-TA detector. The white square shows the field-of-view of the EUSO-TA telescope. }
\label{fig:1}       
\end{center}
\end{figure}

\section{EUSO-TA}
The EUSO-TA project aims to install a fully functional prototype of JEM-EUSO in Black Rock Mesa, the site of one of the fluorescence light detectors of the Telescope Array collaboration. From there it will observe artificial light (laser and electron-generated UV) and events coming from cosmic rays.

The aim of the project is to calibrate this Ground-EUSO detector prototype using the signals of UV light coming from  known sources. They include artificial light coming from fluorescence emitted by electrons accelerated at the Electron Light Source (ELS, a compact electron linear accelerator) and from a  laser at the CLF (Central Laser Facility). The response to artificial light will be correlated with that of the TA fluorescence light telescope in order to calibrate the  response with that of TA and reduce the systematic errors of the measurement.

Also, UV light coming from cosmic ray events will be detected by Ground-EUSO with an external trigger coming from TA. In this case studies of the transversal profile of the shower will be performed. Note that in the first stage of the project, the use of one PDM will only allow to see part of the shower (Figure \ref{fig:1}), albeit with a higher spatial resolution than TA Fluorescence detector. In subsequent stages the addition of other PDM detectors (with the same optics) will be considered, to enlarge the field of view.

\begin{figure}
\begin{center}
\resizebox{0.9\columnwidth}{!}{
  \includegraphics{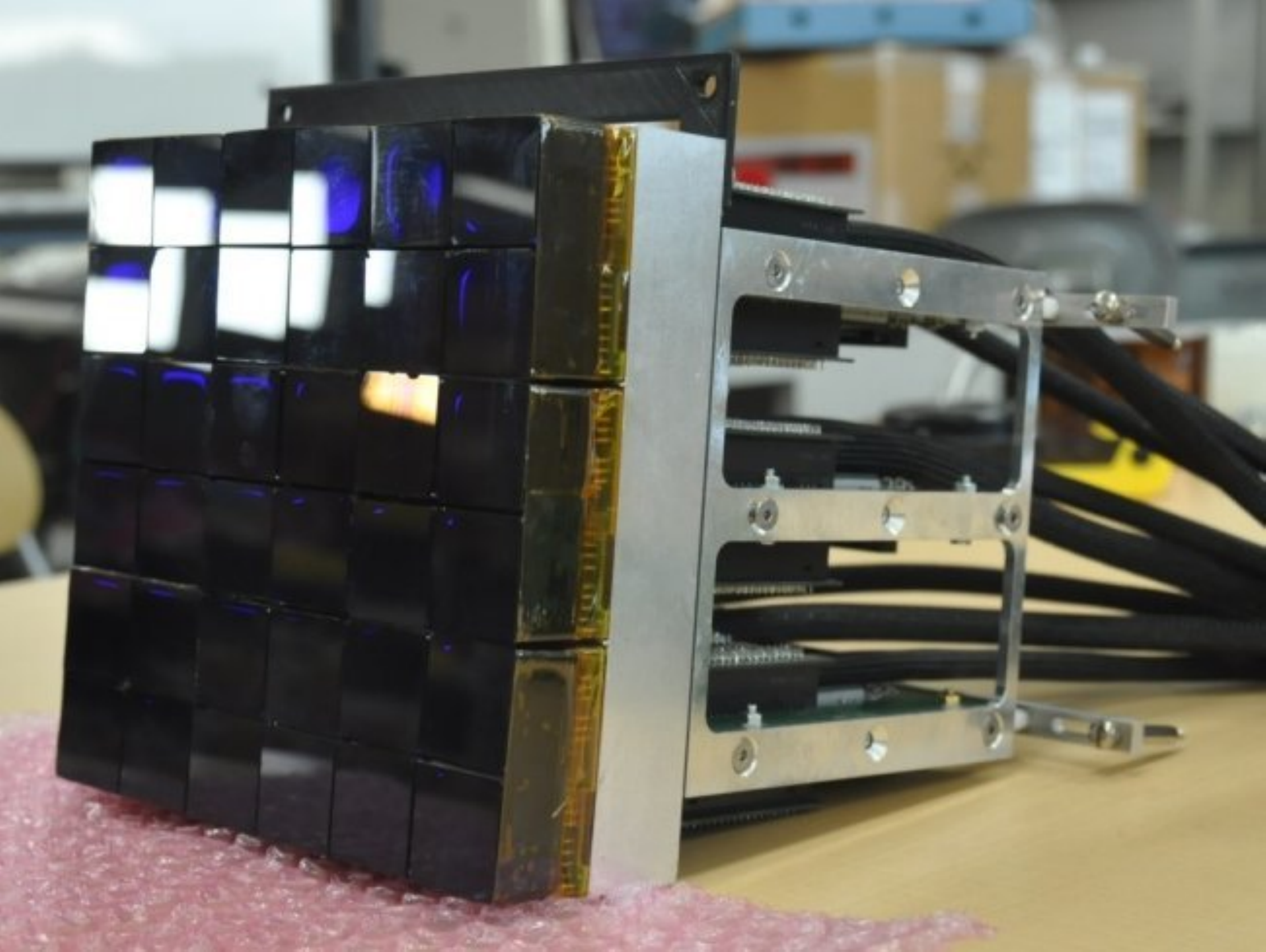}  }
 \caption{The full focal surface of EUSO-TA consisting of one PDM.} \label{full_ta_pdm}
\label{fig:3}       
\end{center}
\end{figure}

The Ground-EUSO telescope is housed in a container about 20 m in front of the TA fluorescence detector of Black Rock Mesa. As mentioned, both the ELS and the CLF lie in the telescope field of view, so that light from these sources is seen by the two double sided Fresnel lenses optical system. The optical signal is focused on and  detected by  the PDM (Figure \ref{fig:3}).
The PDM is attached to the telescope with alignment accuracy better than 0.1$^{\circ}$.

\begin{figure}
\begin{center}
\resizebox{0.9\columnwidth}{!}{
  \includegraphics{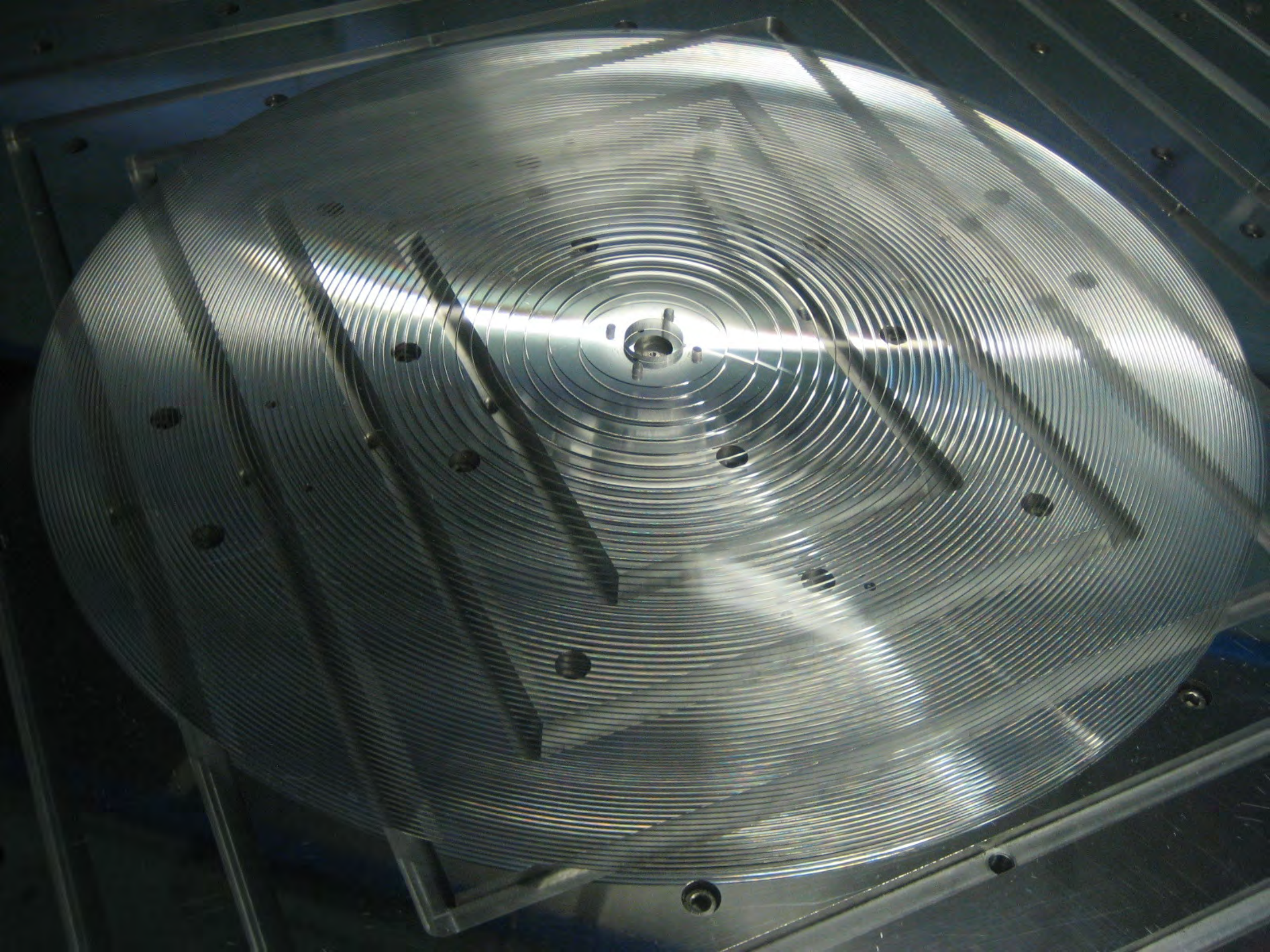}  }
 \caption{Picture of the rear Fresnel lens during manufacturing phase. } \label{zzfig7}
\end{center}
\end{figure}

The optical system consists of two square Fresnel lenses, 1 meter side (Figure \ref{zzfig7}), focusing the light in a $\pm 4^{\circ}$ field of view on one PDM (Photo-Detector-Module) of 2304 pixels.
Each PDM has a focal surface of $13.6 \times 13.6$ cm and is composed of 36 Hamamatsu multi-anode photomultipliers (64 channels per
tube), for a total of 2304 channels. They are arranged in a $6\times 6$ element array, with front-end readout  performed by 36 ASICs (SPACIROC, Spatial Photomultiplier Array Counting and Integrating Readout Chip).
Readout tasks are performed by an FPGA board that stores the data in a 100 GTU (Gate Time Unit. 1 GTU = 2.5$\mathrm{\mu s}$) round buffer.

In case of an event (trigger, pedestal or calibration)  data are sent to a second FPGA for further processing and interfacing with the CPU.
Note that the electronic system and the 2.5$\mathrm{\mu s}$ sampling rate are designed for observation of UHECR showers from the altitude of 400 km of the ISS. Therefore the relative proximity of cosmic ray events and the limited field of view seen from ground is such that the shower is visible only in one - two time frames (GTUs), compared to dozens of GTUs for JEM-EUSO. Therefore, a dedicated trigger system for the ground detector will be realized. However, given the limited number of GTUs available, this solution is expected to be of low efficiency -- therefore an external trigger, generated from   TA trigger electronics, will   be used for the main cosmic ray acquisition.

\begin{figure}
\begin{center}
\resizebox{0.9\columnwidth}{!}{
  \includegraphics{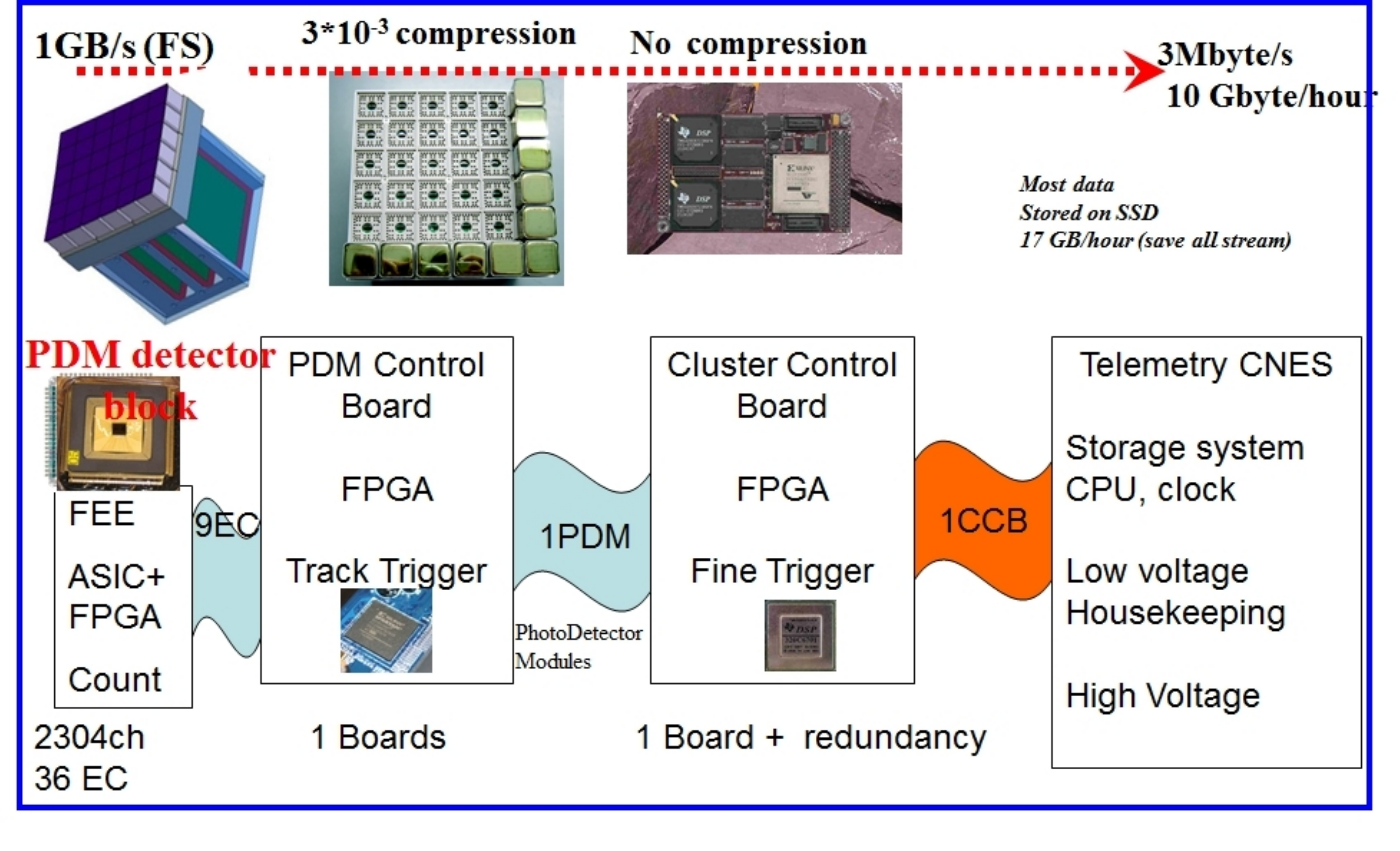}  }
 \caption{Data Reduction scheme. Each of the $\simeq $ 36  Multi-Anode Photo Multipliers (MAPMT) of the focal surface is read by an ASIC digitizing the photoelectron signal.  A 6*6 array of MAPMT is present and read by  an FPGA which performs first level triggering and rejects noise by three orders of magnitude. The general acquisition and data storage is performed by   the main CPU (right). } \label{zzfig6}
\end{center}
\end{figure}

\subsection{Data Acquisition and Reduction}
The data acquisition system  (Figure \ref{zzfig6}) is an
architecture capable of reducing data at each level
through a series of triggers controlling an increasingly growing
area of the focal surface \cite{Casolino2010516,astra}. On ground it is necessary to reduce the 1  Gbyte/s background output
of the Focal Surface (FS) to 10-100 Gbyte/day  which can be
stored for off-line analysis.  Each board and data exchange protocol is therefore
capable of handling the data and sending them to the higher level of processing if
they satisfy the trigger conditions.
This structure is similar to that expected on board  the International Space Station, where most of the triggers will be due to noise.

An ASIC chip performs  photo-electron signal readout and conversion for the 64 channels of the MAPMT.
It has two main purposes: counting the number of photons reaching each pixel of the MAPMTs and measuring the intensity of photon flux by performing charge to time (Q-to-T) conversion.  The first version of the ASIC is the result of the collaboration between OMEGA/LAL-Orsay, France, RIKEN, ISAS/JAXA and Konan University, Japan. It has 64 channels preamplifier with independent gain (8-bit) adjustment in order to correct for the non-uniformity of the 64 MAPMT anodes; photon counting for each channel with a system managing $100\%$ trigger efficiency for a charge greater than 50 fC (~1/3 p.e for a MAPMT gain of $10^6$) and a double pulse resolution as close as possible to 15 ns.

The Q-to-T converter  has an input charge range of  2  - 200 pC (12.5 - 1250 p.e.). The last dynode signal, produced by the MAPMT, as well as 8 internal channels corresponding to the sum of up to 8 channels signals, have to be processed.  The chip has a low power consumption of about 1 mW/channel.

An FPGA (Xilinx XC6VLX240T) board handles   first level trigger   data on a PDM level (reading 36 MAPMTs). Background  events are reduced by a factor $10^3$. Second level triggering algorithms are implemented by the  CCB (Cluster
Control Board), DSPs with about 1Gflop computing capability which
further process the data.  At this level background is rejected by another factor $10^3$.
The CPU has a
relatively low processing power (100 MHz) since it  is in charge  of
the general handling of the experiment. The CPU is part of the Storage and Control Unit System (SCU), the evolution of a similar system used for PAMELA  and   composed of a number of
boards devoted to different tasks: CPU main board,  mass
Memory (8 Gbyte),  internal and external housekeeping interfaces (CAN bus), Hard Disk storage.
Data acquisition and status of the apparatus can be monitored remotely from a PC in the counting room of TA building. All hardware and software resets, as well as power cycling, can be performed remotely in order to avoid access to the EUSO-TA container during data acquisition in case of malfunction.
Only high level data, coming from artificial light sources and cosmic ray events will be transferred via network, whereas the calibration and pedestal raw data will be physically transferred to a higher (Grid-based) link location.

\subsection{Slow control and Housekeeping}

\begin{figure}
\begin{center}
\resizebox{0.9\columnwidth}{!}{
  \includegraphics{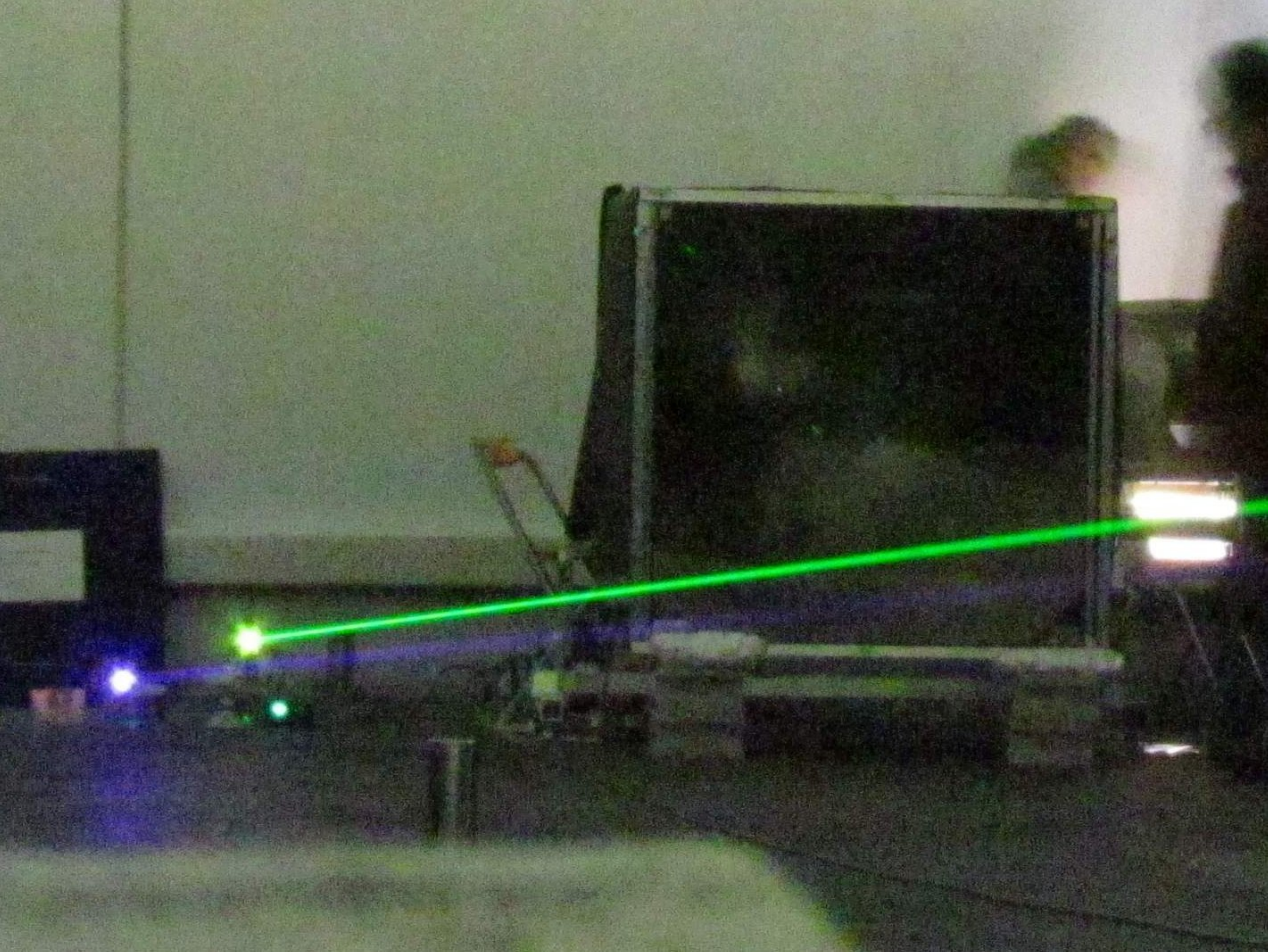}  }
 \caption{Testing of the integrated EUSO-TA on the roof of one of the RIKEN buildings, pointing at the UV and green laser spots on a wall.} \label{roof_laser}
\end{center}
\end{figure}

\begin{figure}
\begin{center}
\resizebox{0.99\columnwidth}{!}{  \includegraphics{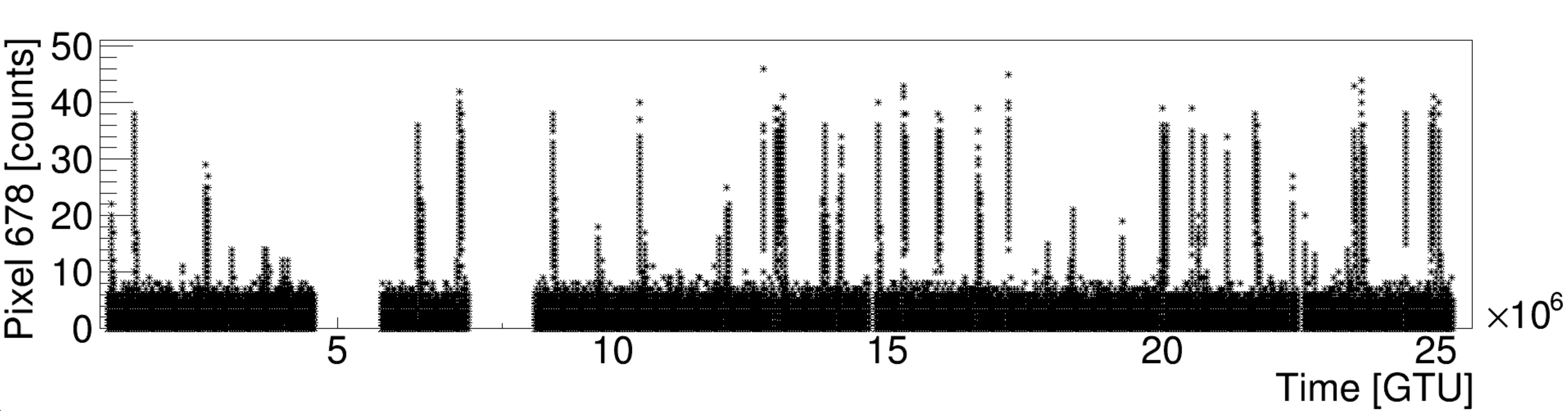}  }
 \caption{Number of photon counts per GTU (1 GTU = 2.5 $\mathrm{\mu s}$) detected by the EUSO-TA telescope during tests on the roof in RIKEN. The higher counts visible on the plot are caused by a moving UV laser light spot displayed on the wall, coming into the field of view of the apparatus. The empty areas are due to pauses in the acquisition introduced in the debugging phase. From the plot it is possible to estimate the UV background in cloudy conditions in the Wako area to be about 8 photons/GTU.} \label{moving_laser_time}
\end{center}
\end{figure}

\begin{figure}
\begin{center}
\resizebox{0.9\columnwidth}{!}{
  \includegraphics{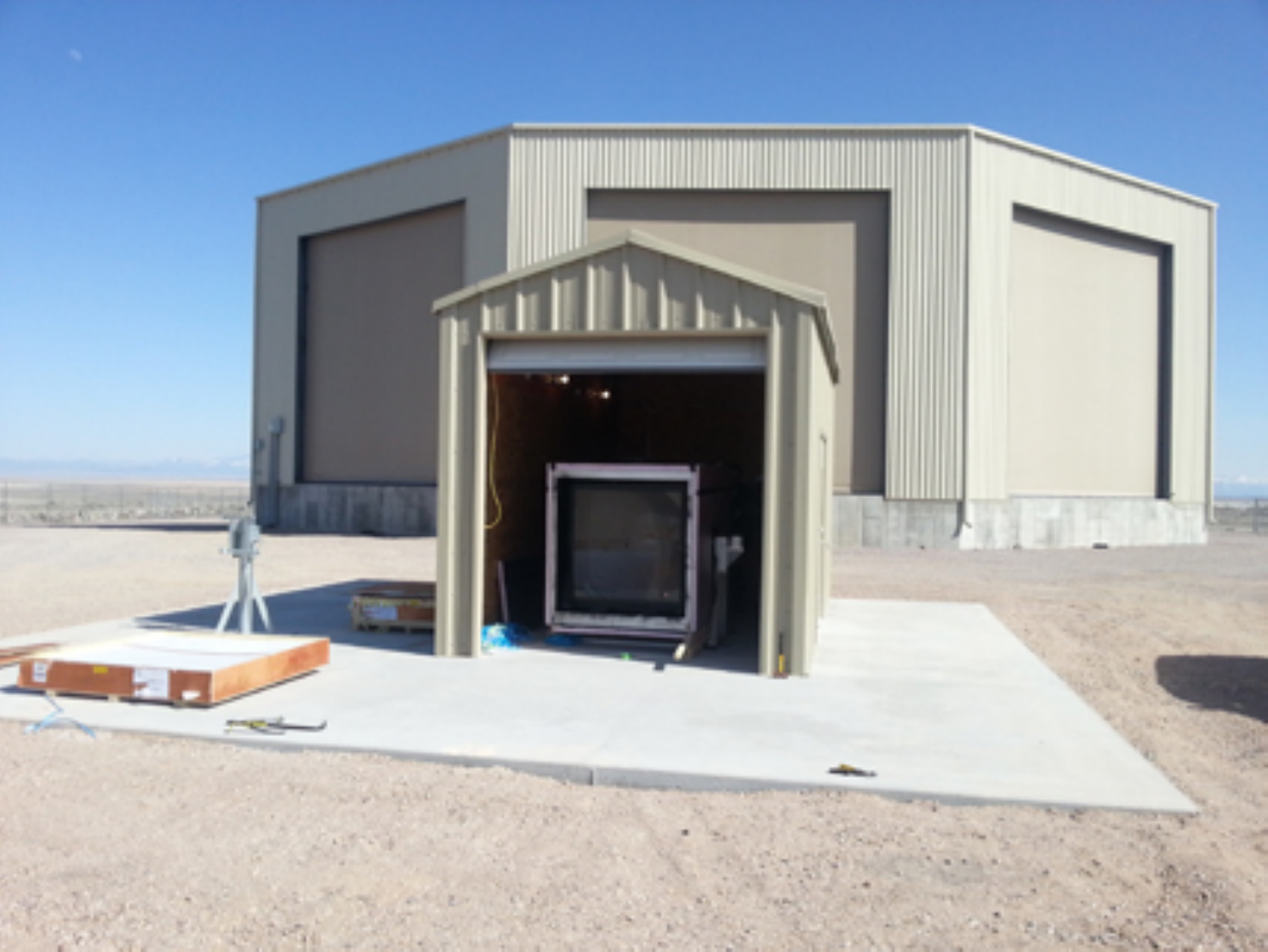}  }
 \caption{The EUSO-TA telescope in the final stage of assembling in the Black Rock Mesa Telescope Array site. In the background, the TA fluorescence telescope.} \label{utah}
\end{center}
\end{figure}

The housekeeping module is connected to the CPU
with the task to distribute commands to the various detectors  and to collect
telemetry for them in order to monitor in real-time the status of the experiment and optimize its
observational parameters.
The module is capable of handling, single, periodic and time-tagged instruction according to the CPU   commands. For
instance all relays to toggle secondary power supplies and
subsystems are controlled by high level signals. This approach has
the advantage of a great degree of flexibility keeping at the same
time a strong robustness and reliability.

\section{Initial integration and tests of the EUSO-TA detector}

In January 2013 a first integration of the whole detector, from optics to read-out systems  was performed in RIKEN, Japan. The data processing elements, produced by different institutions in different countries, were connected and configured to work with each other. This completed  the chain of data acquisition, starting with the request for data  sent from CPU, through CCB, PDM and EC-ASIC to MAPMT, detecting single photons with the MAPMT and sending back the photon counts through all the electronics to the CPU which then stored them in an appropriate file. The task required adjusting the hardware interfaces between the elements but also developing proper protocol for the exchange of commands and data between the hardware elements. For the purpose of these tests only a single MAPMT was used.

Upon successful completion of tests of the electronics with MAPMTs illuminated by an artificial light source in light-sealed conditions, we performed tests in an environment close to the final experiment site conditions. The whole telescope was assembled, including the focal surface consisting of a single MAPMT and the two Fresnel lenses. The mechanical frame of the telescope allows changes in the vertical pointing, which was crucial for measuring different types of signal, as well as for properly focusing the image on the focal surface.

First, we  acquired single expositions of the night sky in order to measure the ultraviolet background on-site. The second task was a continuous exposure to a variable light source, resembling the conditions of the EAS observation. For this purpose we  used a spot of the fast moving laser projected on the wall surrounding the roof of the building (Figure \ref{roof_laser}). The whole data acquisition process was successful and the signal was stored by the CPU for further analysis. The detector registered the changes in the illuminance due to the spot of the laser coming in the field of view. Apart from the laser signal, we could see the constant UV background registered by the MAPMT pixel (Figure \ref{moving_laser_time}). From these measurements it was possible to estimate the UV background in cloudy conditions in the Wako area to be about 8 photons/GTU, with signal from the laser reaching up to 35 photons/GTU. A more precise measurement of the background will have to be performed in Utah.

\section{Installation and tests of EUSO-TA in the TA site}

Between February and  March 2013  the telescope housing, mechanical structure of the TA-EUSO telescope and its  optical system  were installed in Black Rock Mesa, on the site of the   TA fluorescence telescope. The two Fresnel lenses (Figure \ref{utah}) were installed and aligned. Preliminary tests with a single MAPMT connected to a test system were performed to test  the installation. The second stage of the installation, scheduled for summer 2013, will be dedicated to installing the full focal surface and data processing hardware. This will allow for the target, automatic data gathering and analysis, with initial tests performed remotely from the TA control room.

\section{Conclusion}

JEM-EUSO aims to perform a high-statistic UHECR measurement from space for the first time. Due to the innovative character of the experiment, a number of tests  with prototype detectors have been performed.  EUSO-TA is an on-ground, smaller version of JEM-EUSO, built to observe EAS from the Telescope Array site in Utah, USA. In RIKEN we have successfully integrated the full chain of the data acquisition data processing system with optics. The tests performed on the roof of the building in RIKEN were successful, showing the ability to register variable UV light source and store the data for offline analysis. The optics of the telescope have been installed at the destination site of TA and night background data was acquired with a test readout. Currently we   plan to install the electronics and start systematic data taking in summer 2013.

{\footnotesize{{\bf Acknowledgment:}{This work was partially supported by 1) Basic Science Interdisciplinary
Research Projects of RIKEN and JSPS KAKENHI Grant (22340063, 23340081, and
24244042). 2) Italian Ministry of Foreign Affairs grant. Part of the Montecarlo simulations  were performed  using the RIKEN Integrated Cluster of Clusters (RICC) facility. }}

}
\clearpage

%% file: icrc2013-1072.tex



\title{Atmospheric Monitoring system of the JEM-EUSO telescope}

\shorttitle{LIDAR simulation for JEM-EUSO}

\authors{
A. Neronov$^{1}$,
M. D. Rodriguez Fr\'ias$^{2}$,
S. Toscano$^{1}$,
S. Wada$^{3}$
for the JEM-EUSO Collaboration.
}

\afiliations{
$^1$ ISDC Data Centre for Astrophysics, Versoix, Switzerland \\
$^2$ SPace \& AStroparticle (SPAS) Group, UAH, Madrid, Spain \\
$^3$ RIKEN Advanced Science Institute, Japan\\
}

\email{Andrii.Neronov@unige.ch}

\abstract{The JEM-EUSO fluorescence telescope will observe UV emission from Ultra High Energy Cosmic Ray (UHECR)
 induced Extensive Air Showers (EAS) from space. Observation with a space-based telescope has an advantage compared to the ground-based observations, because the EAS signal from the upper atmosphere above 10 km altitude (above the top of the Troposphere) is never obscured by optically thick clouds for such a telescope. Nevertheless, proper interpretation of the UV signal from the lower parts of some 60-70\% EAS detected by JEM-EUSO, including the reconstruction of the energy, direction and identity of the UHECR particle, requires a detailed knowledge of the influence of clouds and aerosols on the detected UV signal. The Atmospheric Monitoring system of JEM-EUSO will use the LIDAR, operating in the UV band, an infrared camera, the UV images of the night sky obtained by the JEM-EUSO telescope itself, as well as real time global meteorological data and models to deduce the distribution and properties of clouds and aerosol layers in the atmospheric volumes around the location of each triggered EAS event. In this contribution we describe the set-up of JEM-EUSO Atmospheric Monitoring System and characterise its performance. In addition, we show that the reconstruction of UHECR events will be possible also for events occurring in cloudy sky conditions if the data of the Atmospheric Monitoring are taken into account.}


\maketitle

\section{Introduction}
JEM-EUSO (the Extreme Universe Space Observatory on-board the Japanese Experiment Module) on the International Space Station (ISS) is a new space mission which aims to discover the origin of the Ultra High Energy Cosmic Rays (UHECRs) with energy above $10^{19}$ eV \cite{ExposurePaper, tPicozza_ICRC2013}. It is a refractive telescope with the aperture $\simeq$ 2.5 m which it will detect fluorescence UV emission from Extensive Air Showers (EAS) produced by UHECRs penetrating in the atmosphere.
The properties of the primary UHECR particles (energy, type, arrival direction) will be derived from the imaging and timing properties of the UV emission from the EAS track in the atmosphere. The amount of both fluorescence and Cherenkov signals reaching JEM-EUSO depends on the extinction and scattering properties of the atmosphere. A correct reconstruction of the UHECR energy and of the type of the primary cosmic ray particle requires, therefore, information about absorption and scattering of the UV light. \\
Also, the presence of clouds and aerosols layers will alter the physical properties of the atmosphere. Uncertainties on the knowledge of extinction and scattering coefficients related to the variable meteorological conditions introduce distortions of the UV signal from the EAS leading to systematic errors in the determination of the properties of UHECR from the UV light profiles \cite{Lupe_ICRC2011}. \\
In particular, presence of optically thin cloud layers between the EAS and JEM-EUSO telescope reduces the overall intensity of UV light leading to an under-estimate of the UHECR energy. EAS penetration into an optically thick cloud produces strong enhancement of the scattered Cherenkov light emission from the shower, which can be misinterpreted as Cherenkov light reflection from the ground/sea. This again leads to a wrong estimate of the depth of the EAS maximum in the atmosphere. \\
Since the ISS is moving with an orbital velocity of $\sim 7$ km/sec, JEM-EUSO will experience all possible weather conditions. The AM system will continuously monitor the variable atmospheric conditions in JEM-EUSO Field of View (FoV) during the entire UHECRs data taking period, providing information on cloud cover and optical properties of cloud/aerosol layers at the time and location of the EAS. \\
In this contribution we describe the set up of the AM system of JEM-EUSO and its expected performance.

\section{Atmospheric Monitoring system}
\begin{figure}[t]
  \centering
  \includegraphics[width=0.5\textwidth]{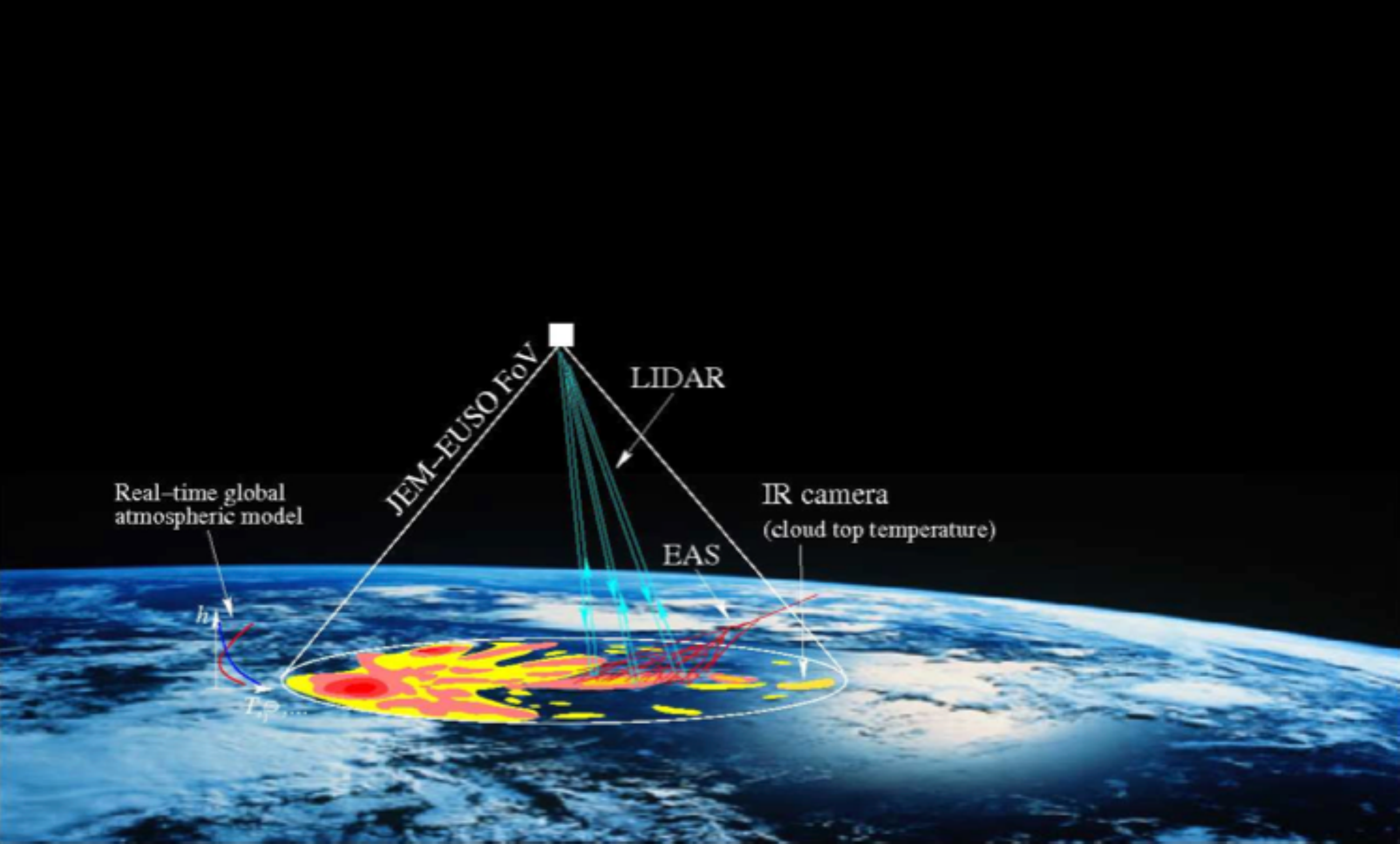}
  \caption{Sketch of the concept of Atmospheric Monitoring in JEM-EUSO.}
  \label{fig:AMS}
 \end{figure}
The goal of the AM system is to obtain information on the distribution and optical properties of clouds and aerosol layers inside the JEM-EUSO FoV \cite{Maria-CERN}.\\
The basic requirements on the precision of the measurements of the clouds and aerosol layers characteristics are obtained from the general requirements on the precision of measurements of the UHECRs properties: \emph{(A1)} measurement of UHECR energy with precision better than 30\%; \emph{(A2)} measurement of the depth of the shower maximum with precision better than 120 g/ cm$^2$.  \\
Since the energy of the UHECR is proportional to the overall intensity of the UV fluorescence emission from the EAS, the uncertainty of the energy measurement (A1) will depend on the uncertainty in the determination of the extinction properties in the atmosphere. Adopting a maximum 15\% uncertainty as a reference value (so that the error introduced by this uncertainty is sub-dominant), one can conclude that the measurement of the optical depth profile of the atmosphere around the EAS location has to be $\Delta \tau \geq$~0.15.\\
In addition, the depth of the shower maximum will be affected by the uncertainty in the determination of the location and physical properties of clouds and aerosol layers, in such a way that uncertainties in the determination of both extinction and scattering properties of these features will directly affect the precision of $X_{max}$ measurement (A2). Imposing a maximum of 60~g/cm2 as a possible contribution to the uncertainty of $X_{max}$ measurement, lead to the conclusion that a measurement of the cloud top with an accuracy $\Delta H \leq 500$~m is required. \\

The AM system will include \cite{Andrii_ICRC2011}: 
\begin{enumerate}
\item
an Infrared (IR) camera;
\item
a LIght Detection And Ranging (LIDAR) device;
\item 
global atmospheric models generated from the analysis of all available meteorological data by global weather services such as the National Centers for Environmental Predictions (NCEP), the Global Modeling and Assimilation Office (GMAO) and the European Centre for Medium-Range Weather Forecasts (ECMWF).
\end{enumerate}
\noindent
The principle of the AM system in JEM-EUSO is illustrated in Fig~\ref{fig:AMS}. The JEM-EUSO telescope will observe the EAS development only during nighttime. The IR camera will monitor the entire FoV  to detect the presence of clouds and to obtain the cloud cover and cloud top altitude during the observation period of the JEM-EUSO main instrument. The LIDAR will be shot in several directions around the location of each triggered EAS event and it will measure the optical depth profiles of the atmosphere in these selected direction, with the ranging accuracy of $375/cos(\theta_z)$ m, where $\theta_z$ is the angle between the direction of the laser beam and the nadir. The power of the laser will be adjusted in such a way that cloud/aerosol layers with optical depth $\tau \geq 0.15$ at $355$ nm wavelength will be detectable.\\
The LIDAR measurements are complementary to the measurements taken by the infrared camera. From one hand the IR camera will provide an overall picture of the optically thick cloud coverage in the JEM-EUSO FoV, which is not possible to measure with the LIDAR, since this device can retrieve the optical properties of the atmosphere only for a certain direction. On the other hand the LIDAR will provide information (altitude and optical depth) of optically thin clouds and aerosol layers, which cannot be identified by the IR camera.  \\
Finally real-time atmospheric profiles from global models will be used as an input for the off-line analysis of the LIDAR data, calibration of the IR camera and modelling of the EAS development in the atmosphere and its reconstruction.  

\section{Infrared Camera}

\begin{figure}[t]
  \centering
  \includegraphics[width=0.5\textwidth]{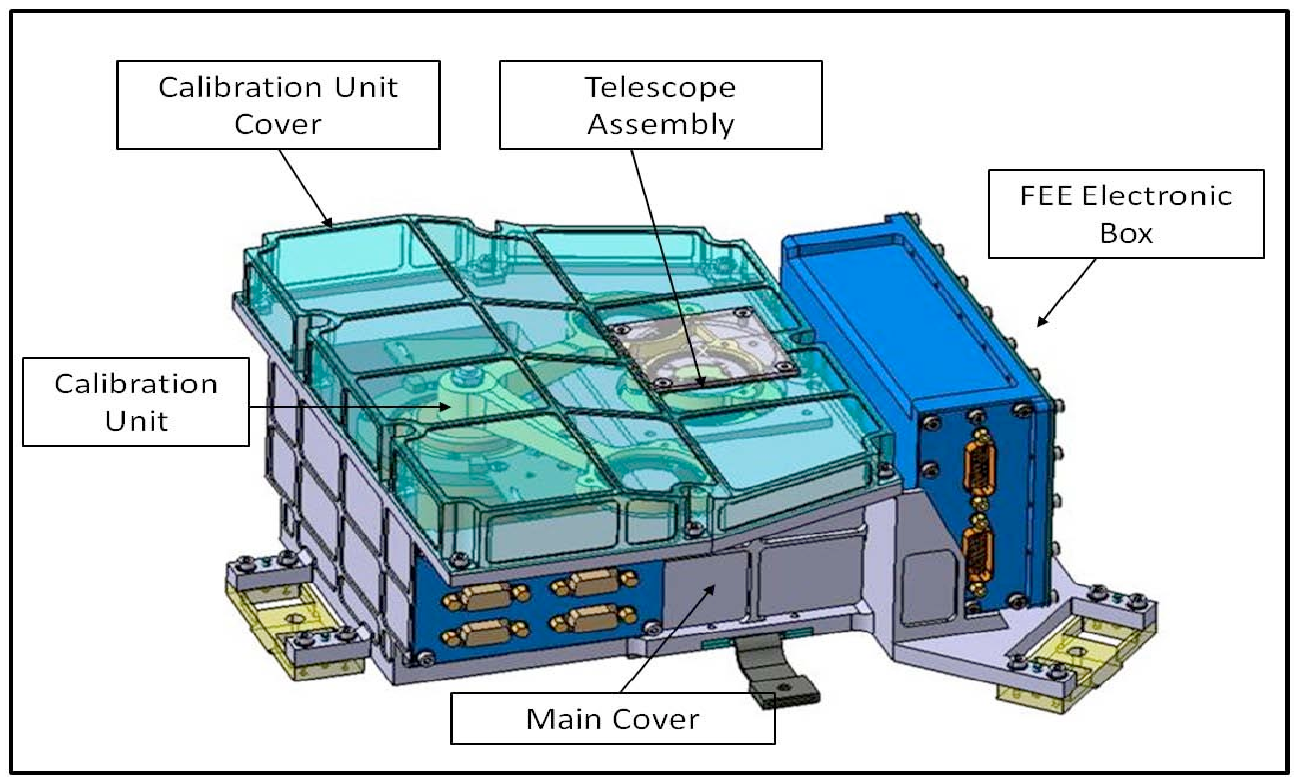}
  \caption{Illustrative picture of the IR Camera design}
  \label{fig:IRCamera}
 \end{figure}
The IR camera of JEM-EUSO is an infrared imaging system aimed to detect the presence of clouds in the FoV of the JEM-EUSO main telescope and to obtain the cloud cover and cloud top altitude during the observation period of the JEM-EUSO main instrument. It will consist of a refractive optics made of germanium and an uncooled $\mu$bolometer array detector \cite{jose_ICRC2011}. Interferometer filters will limit the wavelength band to 10-12.5 $\mu$m. In the current configuration, two $\delta\lambda$ = 1 $\mu$m wide filters (centred at 10.8 $\mu$m and 12 $\mu$m ) will be used to increase the precision of the radiative temperature measurement. The FoV of the IR-Camera will totally match the FoV of the main JEM-EUSO telescope. The angular resolution, which corresponds to one pixel, is about 0.1$^\circ$. A temperature-controlled shutter in the camera and mirrors are used to calibrate background noise and gains of the detector to achieve an absolute temperature accuracy of 3 K. \\
To accomplish the mission and scientific requirements for the Infrared camera, a System Preliminary Design (SPD) of a prototype of the IR camera is under development by the Spanish consortium involved in JEM-EUSO \cite{IRCamera_ICRC2013}. A schematic illustration of the camera hardware can be seen in Fig.~\ref{fig:IRCamera}.

\section{LIDAR}

The LIDAR is composed of a transmission and receiving system. The transmission system comprises a Nd:YAG laser and a pointing mechanism to steer the laser beam in the direction of triggered EAS events. \\
The specifications of the laser unit of the LIDAR transmission system are similar to the laser ranging devices on board of several satellites for atmospheric sounding purposes such as the NASA's satellite CALIPSO \cite{CALIPSO}. The main difference from the previous space-based lasers is that the operational wavelength ($\lambda$ = 355 nm) will be the third harmonic of the Nd:YAG laser, rather than the first, at 1064 nm, or the second, at  532 nm, harmonics, as in existing systems. LIDAR measurements should probe the atmosphere at the location of each triggered EAS event. To get this information, the LIDAR will have a re-pointing capability. The laser beam will be repointed in the direction of EAS candidate events following each EAS trigger of the JEM-EUSO telescope. Re-pointing of the laser beam will be done with the help of a steering mirror with two angular degrees of freedom and maximal tilting angle $\pm$15$^\circ$,  needed to point the laser beam anywhere within the JEM-EUSO FoV. The laser backscattered signal will be received by the main JEM-EUSO telescope which is well suited for detection of the 355 nm wavelength. Any Multi-Anode Photo-Multiplier Tube (MAPMT) in the focal surface of JEM-EUSO telescope could temporarily serve as the LIDAR signal detector; a special LIDAR trigger is foreseen in the Focal surface electronics of JEM-EUSO detector. 

A summary of the specifications needed for the entire system is reported in table~\ref{tab:LIDAR}.
\begin{table}[h]
\begin{center}
\begin{tabular}{|l|l|}
\hline 
Parameter			&		 Specification	\\ \hline
Wavelength 			& 		$355$ nm				 	\\ \hline
Repetition Rate	 	& 		$1$ Hz					 \\ \hline
Pulse width   	   		& 		$15$ ns		            		\\ \hline
Pulse energy 	  		& 		$20$ mJ/pulse		     		\\ \hline
Beam divergence		&		 $0.2$ mrad 				\\ \hline
Receiver        		 	& 		JEM-EUSO telescope	 	\\ \hline
Detector                 	 	& 		MAPMT (JEM-EUSO)	  	\\ \hline
Range resolution (nadir)  	&		 $375$ m 				  	\\ \hline
Steering of output beam 	&		 $\pm 30^\circ$ from vertical 	\\ \hline
Mass   				&		 $14$ kg				     	\\ \hline
Dimension  			&		 $450\times350\times250$ mm\\ \hline
Power				&		$< 20$ W					\\ \hline

\end{tabular}
\caption{Specification for the JEM-EUSO LIDAR.}
\label{tab:LIDAR}
\end{center}
\end{table}

\begin{figure}[t]
  \centering
  \includegraphics[width=0.5\textwidth]{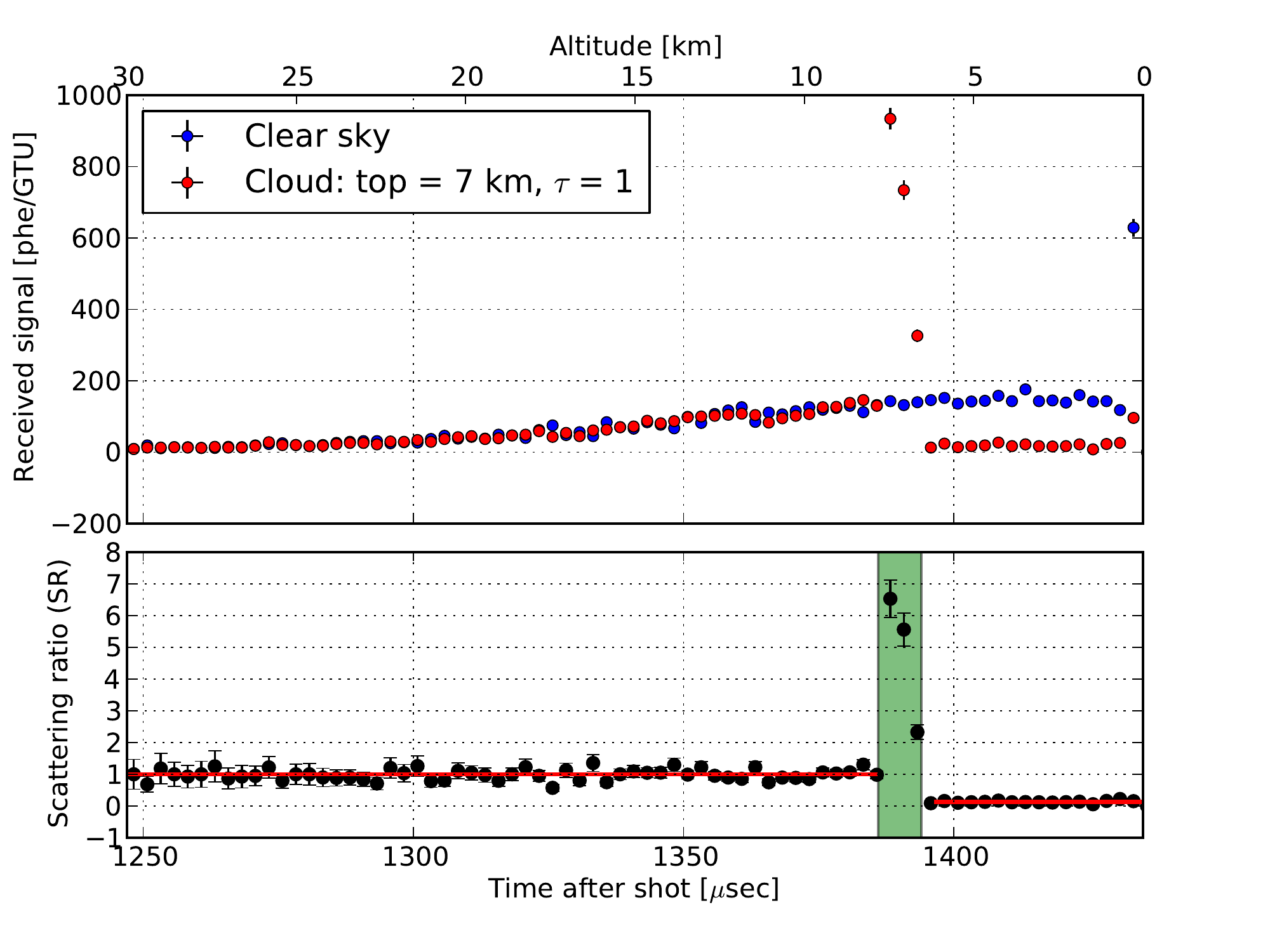}
  \caption{\emph{Top}: LIDAR backscattered signal in clear sky (blue) and in the presence of a cloud (red) as a function of time. \emph{Bottom} Scattering ratio (SR) for the case of LIDAR shooting an optically thick ($\tau = 1$) cloud located at an altitude of 7 km. The cloud mark region is highlighted with a green box. A fit of the SR is shown as a red line. The cloud optical depth is retrieved by this fit procedure.}
  \label{fig:LIDARRatio}
 \end{figure}

\begin{figure}[t]
  \centering
  \includegraphics[width=0.5\textwidth]{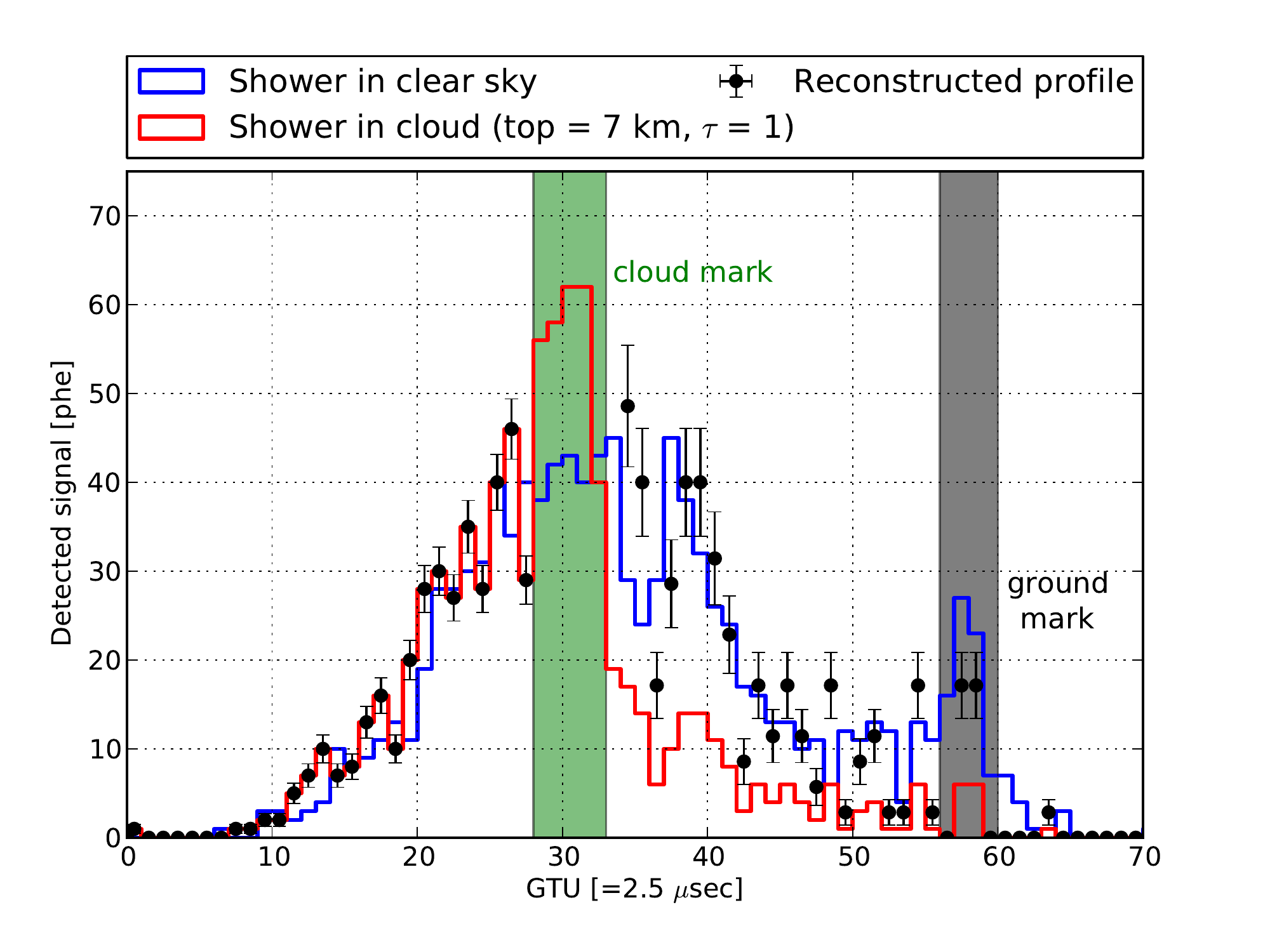}
  \caption{Reconstructed time profile (black points) of  10$^{20}$ eV EAS together with the clear sky (blue) and cloud affected (red) profiles for an EAS of E = $10^{20}$ eV and $\theta = 60^\circ$. Error bars are statistical only. The loss of information inside the cloud does not allow for a good reconstruction of the profile in that region. The ground mark (gray band) is almost entirely recovered by this procedure.}
  \label{fig:RecoShowerProfile}
 \end{figure}

Measurements of the laser backscattered signal with time resolution of 2.5 $\mu$s (representing the duration of the time unit, named Gate Time Unit or GTU, of the pixel-level digital trigger) will provide a range resolution of 375 m in nadir direction. The energy of the laser pulse will be adjusted in such a way that the backscattered signal will have enough statistics for the detection and measurement of the optical depth of optically thin clouds with $\tau \leq$0.15 at large off- axis angles. \\
In order to study the capability of the system in retrieving the physical properties of atmospheric features such as cloud and aerosol layers a simulation of the LIDAR has been implemented inside the ESAF Simulation Framework used for the JEM-EUSO mission \cite{Toscano_ICRC2013}. This simulation chain has been used to study the features of the LIDAR backscattered signal and to reproduce a real-case observation in which the EAS profile shows observable deviations from the ``clear sky case'' and needs to be corrected.\\ 
In fact, once an EAS is detected, LIDAR is used to monitor whether the shower developed in clear sky or not. If the presence of a cloud is detected the backscattered signal from the laser is used to measure the cloud optical depth.     
Examples of the simulated laser backscattered signal as it would appear in the JEM-EUSO detector are shown in Fig.~\ref{fig:LIDARRatio}. The top panel shows the comparison of signal in case of clear sky (blue points) and in presence of the cloud (red points) as a function of the time after shooting the laser and the altitude. The presence of a cloud at $\sim$7 km is clearly detected by the LIDAR as an increasing of backscattered signal coming from that region. The bottom panel shows the so-called LIDAR Scattering Ratio (SR), the ratio between the backscattered signal detected in the real condition and a reference profile represented by the backscattered signal in clear sky. 
 Fitting the SR in the region below the cloud allows to measure the optical depth ($\tau$) of the cloud, simply using  the formula: $SR = -log(2 \tau)$.\\
 Once the cloud is detected and its optical depth measured the EAS profile can be corrected. The result of this correction is reported in Fig~\ref{fig:RecoShowerProfile}. 
 The profile of the detected photo-electron signal is shown as a function of GTU for a shower generated by a UHE proton with E = $10^{20}$ eV and $\theta = 60^\circ$. The blue points represent the shower time profile in clear sky conditions and it is characterized by the presence of a feature at $\sim$ 60 GTU, the ``ground mark", due to Cherenkov photons hitting the ground and reflected back to the JEM-EUSO focal surface. The second profile, in red, represents the shower crossing an optically thick cloud ($\tau = 1$) located at an altitude of 7~km. The cloud mark (green band), generated by photons reflected by the cloud top layers, is visible this time, while the Cherenkov mark from the ground is strongly suppressed. The reconstructed profile is shown by the black point. Statistical errors are calculated propagating the error on the optical depth measurement. Lack of information on the optical depth profile inside the cloud does not allow for a correct reconstruction of the shower profile in that region. It is worth to notice that the ground mark is almost entirely recovered by this analysis procedure. 
 
\section{Global Atmospheric Models}
\label{sec:GAM}
Analysis of both the IR camera and LIDAR data can be improved if initial values of the physical parameters of the atmosphere (temperature and pressure profiles, humidity, wind speed, etc) in the monitored region are known \cite{Keilhauer2012}. Weather forecasting services across the world, such as ECWMF \cite{ECWMF} in Europe or GMAO \cite{GMAO} and NCEP \cite{NCEP} in US systematically collect all the available meteorological data (from weather stations, meteorological balloons, satellite and aircraft measurements) to use them as input data for the Global Atmospheric Models (GAM), which are computer generated models of atmospheric conditions at the entire Earth. The product of the model is an estimation of the state of the atmosphere, or state variables at any given point on a latitude-longitude grid and at different times. This calculation takes into account the real-time conditions of the atmosphere as boundary condition for the global model. As result, data products, e.g. temperature, pressure and humidity profiles, are available. \\
Currently some efforts have been made by the collaboration to investigate the possibility of using data from the Global Data Assimilation System (GDAS) of NCEP \cite{GDAS}. GDAS data have been successfully incorporated in JEM-EUSO simulation of the air showers, and the final goal is to incorporate them in the EAS reconstruction analysis and in the analysis of the data of the LIDAR and IR camera in JEM-EUSO.

\section{Conclusions} 
JEM-EUSO is a next-generation fluorescence telescope which will detect UHECRs induced EAS from space. To correctly reconstruct the EAS profile, knowledge of the atmospheric condition at the location of the shower is needed. The AM system of JEM-EUSO, which includes the IR camera, the LIDAR and global atmospheric model data, will provide sufficient information on the state of the atmosphere around the location of EAS events. This information will be used to correct the profiles of cloud-affected EAS events for the effect of clouds and aerosol layers, so that most of them could be retained for the UHECR data analysis.


\clearpage

%% file: icrc2013-1250.tex




\title{Estimated exposure of UHECR observation by the JEM-EUSO mission}

\shorttitle{Overview of UHECR observation performance by JEM-EUSO mission}

\hyphenpenalty=10000\relax


\authors{
K. Shinozaki$^{1}$, M.E.~Bertaina $^{2}$, S.~Biktemerova$^{3}$, P.~Bobik$^{4}$,
F.~Fenu$^{1,5}$, A.~G\'uzman$^{1}$, F.~Guarino$^{6}$, G.~Medina~Tanco$^{7}$, 
T.~Mernik$^{1}$, J.A.~Morales~de~los~Rios~Pappa$^{8,5}$, D.~Naumov$^{3}$, 
G.~Sa\'ez C\'ano$^{8}$, N.~Sakaki$^{9}$, A.~Santangelo$^{1}$, S.~Toscano$^{10}$
and L.~Valore$^{6}$ for the JEM-EUSO Collaboration}

\afiliations{
  $^1$Institut f\"ur Astronomie und Astrophysik, Eberhard Karls Universit\"at 
  T\"ubingen, Sand 1, 72076 T\"ubingen, Germany\\
  $^2$Department of General Physics, University of Torino, Via P. Giuria 1, 
  10125 Turin, Italy\\
  $^3$Joint Institute for Nuclear Research, Joliot-Curie 6, 141980 Dubna, 
  Russia\\
  $^4$Insteitute of Experimental Physics, Slovak Academy of Science, Watsonova
  47, 040 01, Kosice, Slovakia\\
  $^5$Computational Astrophysics Laboratory, RIKEN, 2-1 Hirosawa, Wako 351-0198, Japan\\ 
  $^{6}$Istituto Nazionale di Fisica Nucleare - Sezione di Napoli, 
  via Cintia, 80126 Naples, Italy\\
  $^7$Instituto de Ciencias Nucleares, Universitat Natcional Autonoma Mexico, 
  Apartado Postal 70-543, Mexico D.F. 04510, Mexico\\
  $^8$Space and Astroparticles Group, University of Alcala, 28807 Alcala de 
  Henares, Spain\\
  $^9$Institut f\"ur Kernphysik, Kalsruhe Institut f\'ur Technologie, 
  Hermann-vol-Hlmholtz Plaz 1, 76344 Eggenstein-Lepopldshafen, Germany\\
  $^{10}$ISDC Data Centre for Astrophysics, Chemin d'Ecogia 16, 1290 
  Versoix, Switzerland\\}
\email{kenjikry@astro.uni-tuebingen.de} 

\abstract{The nature of the ultra-high energy cosmic rays (UHECRs) remains
  unsolved mystery mainly due to severely low fluxes for ground-based 
  observatories. The JEM-EUSO (Extreme Universe Space Observatory on-board the
  Japanese Experiment~Module) mission operates huge-aperture UHECR observation 
  via extensive air shower (EAS) observation from the International Space 
  Station. To evaluate the performance in exposure, a large number of EAS 
  simulations are generated taking into account EAS properties, background 
  noise, role of the cloud and the configuration of the JEM-EUSO telescope.
  The results show that observation to the nadir direction reaches about 9 
  times annual exposures compared to that is achieved by the largest existing 
  detector. The enhancement of exposure by tilting the telescope is also 
  demonstrated. Operating on the orbit allows the full coverage of the 
  Celestial at high degree of uniformity for the analysis of the UHECR arrival
  direction distribution.}


\maketitle

\section{Introduction}

The nature of the ultra-high energy cosmic rays (UHECRs) remains a long lasting
mystery in astrophysics \cite{rev}. So far, extremely low fluxes of UHECRs have
constrained effective observation and no origin has been identified. In 
ground-based observatories, the  observable region and efficiency 
in surveying the Celestial Sphere depend on geographical location. Thus, 
dramatic increases of effective areas with all-sky coverage capability are 
highly desired in forthcoming era of UHECR physics. 

The JEM-EUSO (Extreme Universe Space Observatory on-board the Japanese 
Experiment Module) mission \cite{aTakahashi} is a novel approach by fluorescence
technique in space to investigate science objectives for UHECRs \cite{gmt}. In
the present work, we aim at estimating the fundamental performance, mainly 
focusing on the expected exposure of the mission in various operational 
conditions. The exposure is a basic measure to evaluate the statistics of observed UHECR 
events. We discuss the key pretties and factors that determine the exposure 
such as EAS development, condition within the field of view (FOV), detector 
responses and observation time. 

\section{Apparatus}

The JEM-EUSO observatory is the ensemble of the UV telescope, referred to as 
`main telescope,' the atmospheric monitoring (AM) system~\cite{ams2} and other
sub-system instruments. It is designed to operate on the JEM {\it Kibo} module
of the International Space Station (ISS)~\cite{ISS}. Orbiting at a nominal 
altitude $H_0 \sim 400$~km from the Earth's surface (hereafter, meant for an 
assumed ellipsoid), it revolves every $\sim 90$~min at sub-satellite speed of 
$\sim~7$~km~s$^{-1}$. According to inclination, the ISS operation extends
latitudes within $\pm 51.6^\circ$. During the operation, the JEM-EUSO 
instrument may be pointed to the nadir, referred to as `nadir
mode' or tilted astern, `tilt mode.' 

The main telescope is designed to detect moving tracks of the ultra-violet (UV)
photons produced in extensive air showers (EASs). It consists of a 4.5-m$^2$ 
Fresnel optics \cite{opt}
 viewed by the focal surface (FS) detector \cite{kaj}.
It is formed by 137 photo-detector modules (PDMs). 
Each PDM is a set of 36 multi-anode photomultiplier tubes (MAPMTs) with 
64~pixels. The effective FOV $\omega_{\rm FOV}$ is 
~$\sim~0.85$~sr with a spatial resolution of $0.075^\circ$ 
equivalent to $\sim 0.5$~km on the surface. The time resolution 
is 2.5~$\upmu$s called gate time unit~(GTU). Due to a limited 
telemetry budget, two levels of trigger algorithms~\cite{catalano} 
are operated to search every PDM for stationary and transient excesses of EAS 
signals over prevailing background in the nighttime atmosphere~\cite{slavo}. Threshold levels for
trigger criteria are dynamically set to fit permissible fake trigger rates at an order of $\sim 0.1$ Hz.

\begin{figure}[t]
  \begin{center}
    \includegraphics*[width=0.46\textwidth]{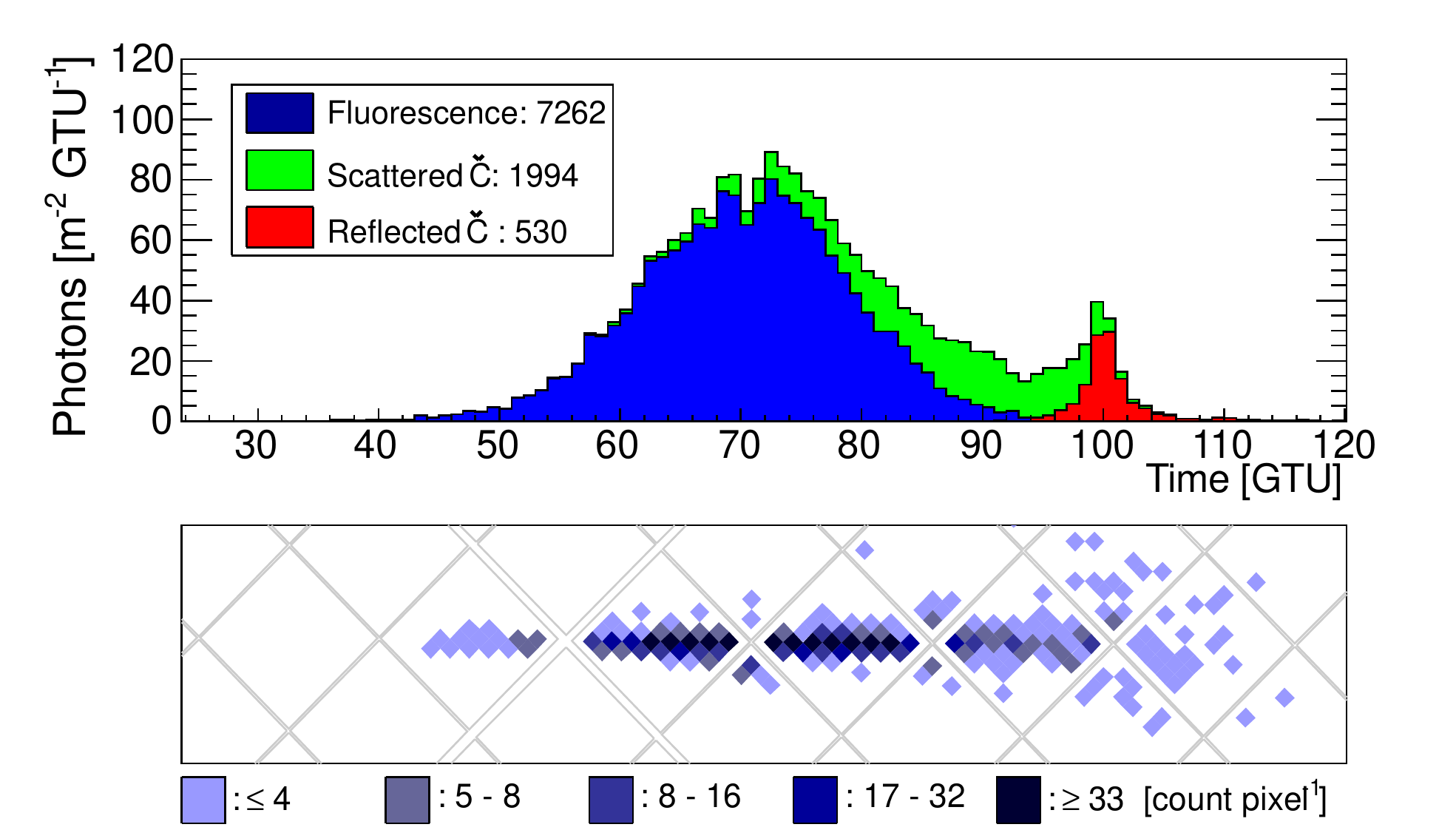} 
  \end{center}
  \caption{Arrival time distribution of photons (top panel) from a typical EAS
    of $E_0~=~10^{20}$~eV from $\Theta~=~60^\circ$ and time-integrated 
    signals on the FS detector (bottom). In the top panel, components of 
    fluorescence photons, and scattered and reflected Cherenkov photons are 
    shown by different histograms. In the bottom panel, signal counts per pixel
    are indicated by filled squared. The gray lines denote the MAPMT 
    boundaries. The horizontal position along the EAS axis corresponds to the 
    arrival time on the top panel.}
  \label{fig:prof2}
\end{figure}

\section{Observed properties of EAS}

In the present work to simulate EAS and detector response, we employ the ESAF 
(EUSO Simulation and Analysis Framework) package~\cite{esf} adapted into the 
JEM-EUSO baseline configuration (see Refs.~\cite{kaj} for details). 
The altitude of JEM-EUSO is set to be 400~km for the nadir mode and 
the case of 350~km is also tested. The tilt mode cases are also investigated 
for relevant interests. The clear atmosphere condition is assumed in 
the simulation. The role of clouds and the influence on the exposure are 
separately taken into account\cite{adams}. As a nominal assumption, the background level from night 
glow in the dark night $I_{\rm BG}$ is set to 500 photons 
m$^{-2}$~sr$^{-1}$~ns$^{-1}$ ~\cite{slavo}.

In Figure~\ref{fig:prof2}, the top panel shows the arrival time distribution
of photons from a typical EAS. Fluorescence and 
scattered and reflected Cherenkov light components are shown by the different
histograms. The sample is the case for an EAS of energy $E~=~10^{20}$ eV from zenith
angle $\Theta~=60^\circ$. The horizontal axis denotes the absolute time and
is set 100 GTUs at the time that shower particles on the axis reach the surface
level. The bottom panel displays the time-integrated image of signals on the
FS detector. Signal counts per pixel are indicated by the filled squares with 
MAPMT boundaries shown by gray lines. The horizontal position corresponds to 
the arrival time on the top panel.

In UHECR observation from space, fluorescence light is the dominant component 
of signals and its luminosity is almost proportional to the energy deposited by
the EAS particles. A part of Cherenkov light that is scattered in the 
atmosphere is also observed. In addition, the space-based observation also 
detects the reflection of Cherenkov photons from land or water as well as
cloud. Those reflection signals, referred to as `Cherenkov footprint,' provide
a piece of information on the position and timing of the EAS reaching such 
boundaries. 

For EASs of a given energy, intrinsic observable properties are dominantly 
determined by the zenith angle of EASs. For larger zenith angles, EASs 
result in signals more intense with longer apparent track and duration. These 
effects are all in favor to the trigger algorithms as well as 
subsequent event reconstruction. Within FOV of the main telescope, the 
vignetting of the optics depends on the direction of EASs with respect to the 
optical axis. Also the EASs in the direction of FOV edge is more distant and 
the overall effect reduces the signal intensity \cite{adams}. 

\begin{figure}[t!]
  \begin{center}
    \includegraphics[width=0.46\textwidth]{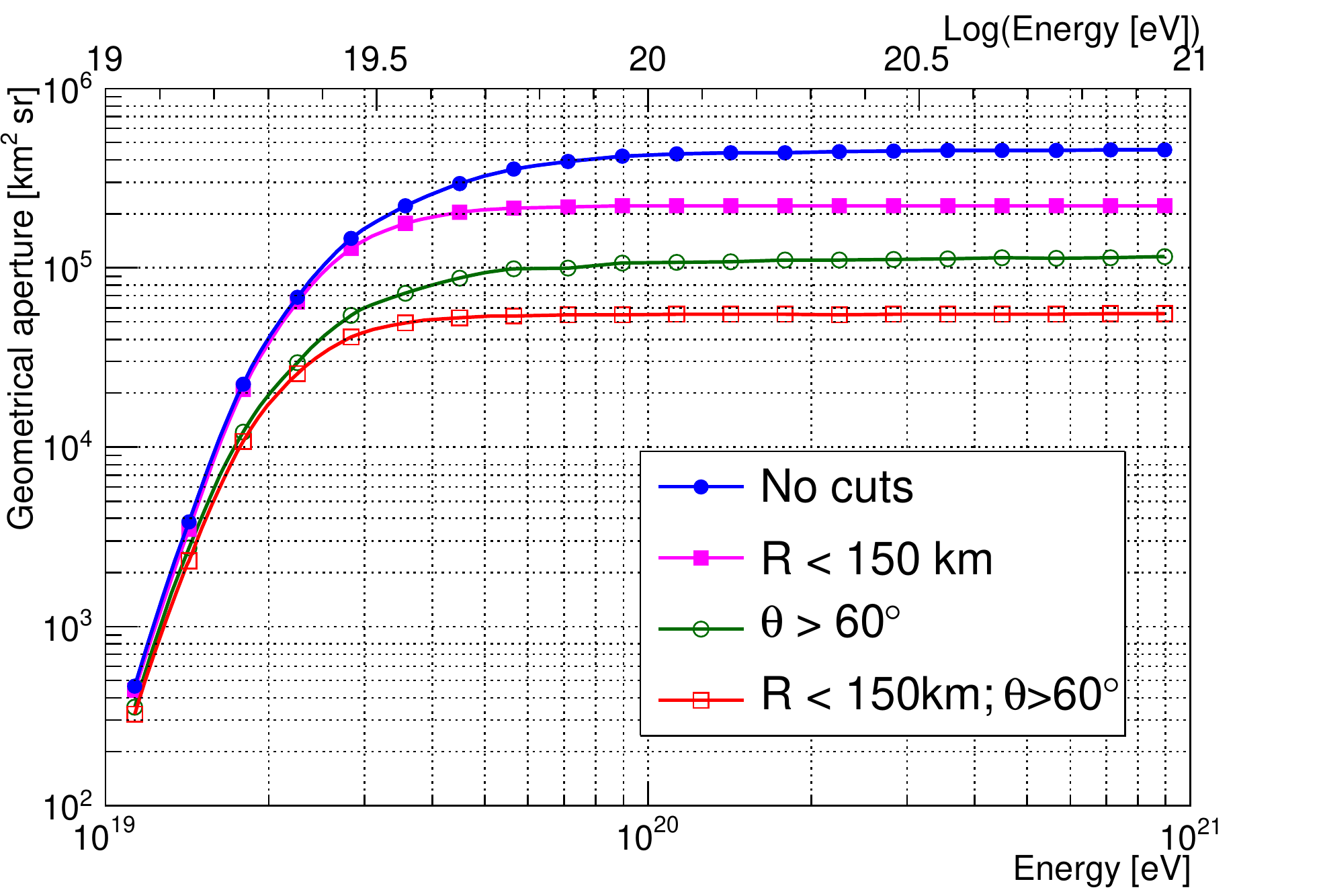}
  \end{center}
  \caption{Geometrical aperture as a function of energy. Closed circles 
    show the case without cuts. Closed squares and open circles indicate for 
    the cuts of $R<150$~km and $\Theta~>~60^\circ$. Open squares are for 
    the combined case.}
  \label{aafig2}
 \end{figure}

\section{Geometrical aperture}

To estimate the geometrical aperture, a large number of EASs are simulated 
over a far larger area than that seen by optics, 
namely $\omega_{\rm FOV}~\cdot~{H_0}^2 \sim~1.4\times~10^5$~km$^2$ for 
$H_0~=~400$~km. The geometrical aperture $A(E)$ is 
wwritten as 
a function of 
energy $E$ by the product of simulated area, solid angle acceptance and probability of
trigger obtained by simulations. Simple geometrical cuts  may be applied on 
displacement $R$ of the impact location of the EAS from the projected center
of the FOV on surface and on the lower limit of zenith angles 
$\Theta_{\rm cut}$. The subsection of the aperture for such cuts is
given in the same way for an area of $\pi R^2$ and a solid angle 
acceptance of $\pi \cdot \cos^2 \Theta_{\rm cut}$ [sr]. The latter is a result 
of an integral of solid angle element weighted by projected area from EAS
point of view. These cuts are used to select the events with larger signals.
Unlike ground-based observatories, the spaced-based ones are more 
sensitive to larger zenith angle EASs.

Figure~\ref{aafig2} shows the geometrical aperture as a function of energy for 
$H_0~=~400$~km without geometrical cuts (closed circles) 
a cut of $R~<~150$~km (closed squares), of $\theta~<~60^\circ$-cut (open 
circles) and their combination (open squares).

The geometrical aperture without cut reaches the plateau around 
$(6-7)\times~10^{19}$~eV. From the FOV of the optics, the saturated 
aperture is $\sim~4.3\times 10^5$~km$^2$~sr. Slight increase seen at the 
highest energies is due to a little contribution by the EASs part of which 
cross the FOV. Applying the geometrical cuts helps lower the energy where the 
aperture saturates. With both $R<150$~km and $\Theta~>~60^\circ$ cuts, though 
it reduces the saturated aperture to be about an eighth of that without cuts, 
a constant aperture is achieved at $\sim~3\times~10^{19}$~eV. Extension of plateau region towards lower energies allows a cross-check of the flux
measured by the full sample of events in the specific range of energies. 
Consequently, the overlapping energy range between JEM-EUSO and ground-based 
observatories is enlarged.

Results shown above are for the case of $H_0~=~400$~km. Among the orbital 
elements of the ISS, the altitude $H_0$ varies throughout the time and is 
not predictable for the era of the mission. As the observation area in the 
nadir mode is scaled by ${H_0}^2$, operation at lower altitudes correspondingly
lessens the saturated aperture. In the tilt mode, however, tilting the 
telescope increases the projected area of FOV on the Earth's surface. With 
tilting angle $\xi~\lesssim~30^\circ,$ it grows proportional to $\sim
(\cos\xi)^{-3}$. This allows recovery of observation areas in lower altitude 
operation. In even larger tilting angles, the effect of the Earth's curvature 
further amplifies this factor. At $H_0~=~400$~km, the projected observation 
area for $\xi \sim 40^\circ$ where the edge of FOV does not see the sky above 
the local horizon reaches $\sim 6$ times of that of the nadir mode. 

\begin{figure}
  \begin{center}
    \includegraphics[width=0.46\textwidth]{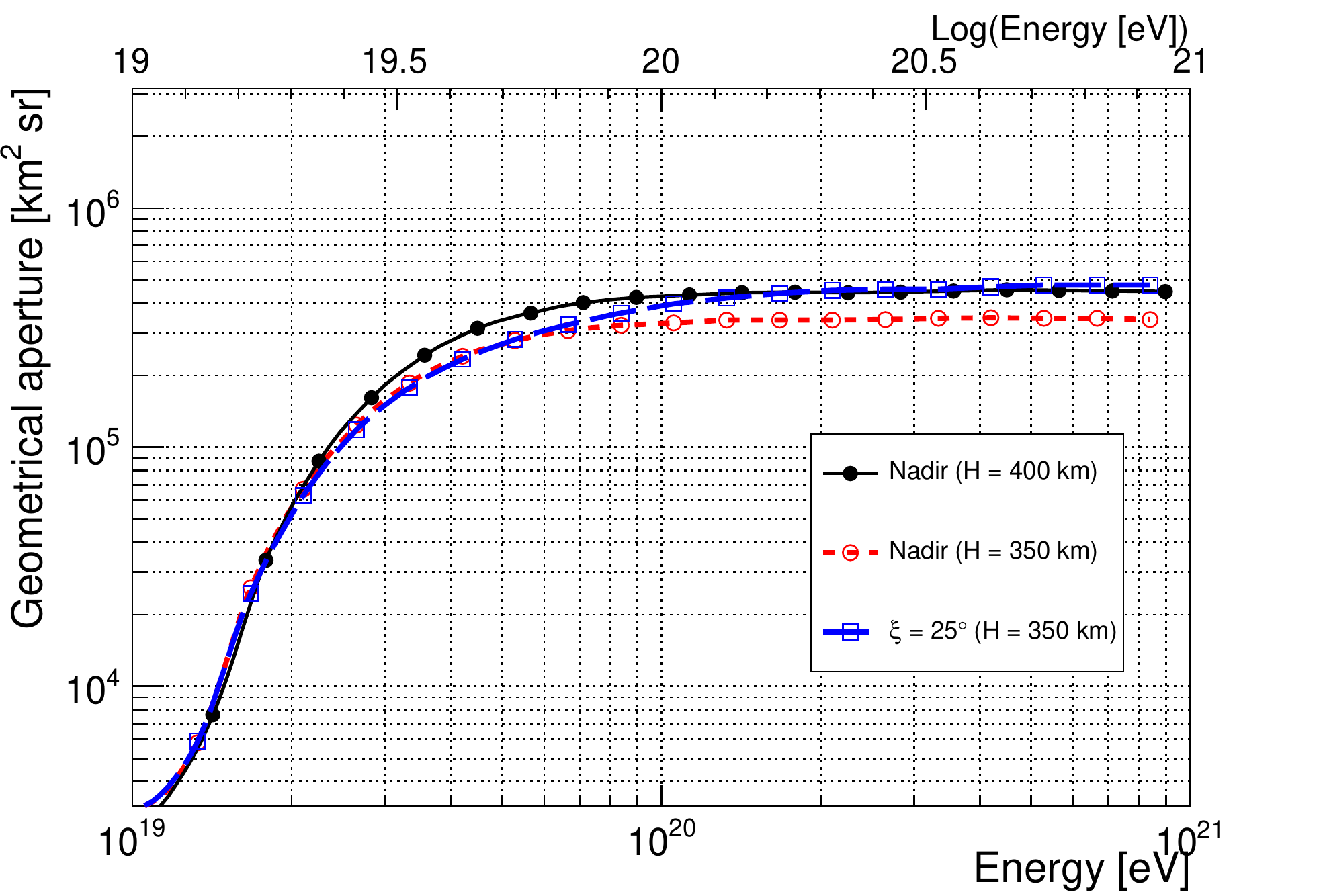}
  \end{center}
  \caption{Aperture as a function of energy for the nadir mode at 
    $H_0~=~400$~km (closed circles) and 350~km (open circles) as well as one 
    for a quasi-nadir mode with $\xi~=~25^\circ$ at $H_0~=~350$~km (open 
    squares). }
  \label{fig:app350}
\end{figure}

In Figure~\ref{fig:app350}, the apertures as a function of energy are shown for
nadir mode at $H_0~=~400$~km (closed circles) and 350~km (open circles). For the
latter, the case for $\xi~=~25^\circ$ is also indicated by the open squares.

In this example of $H_0~=~350$~km, the saturated aperture decreases by 
$\sim 30$~\% in comparison to the case of $H_0~=~400$~km. Note that in this 
case the threshold in energy lowers by the same factor as the distance to 
EAS is closer. Also by tilting the telescope by $\sim 25^\circ,$ referred to as 
`quasi-nadir mode,' the saturated aperture is similar to that of nadir 
mode at $H_0~=~400$~km. 

\section{Exposure}
\label{exp-exp}

\begin{figure}[t]
  \begin{center}
    \includegraphics[width=0.46\textwidth]{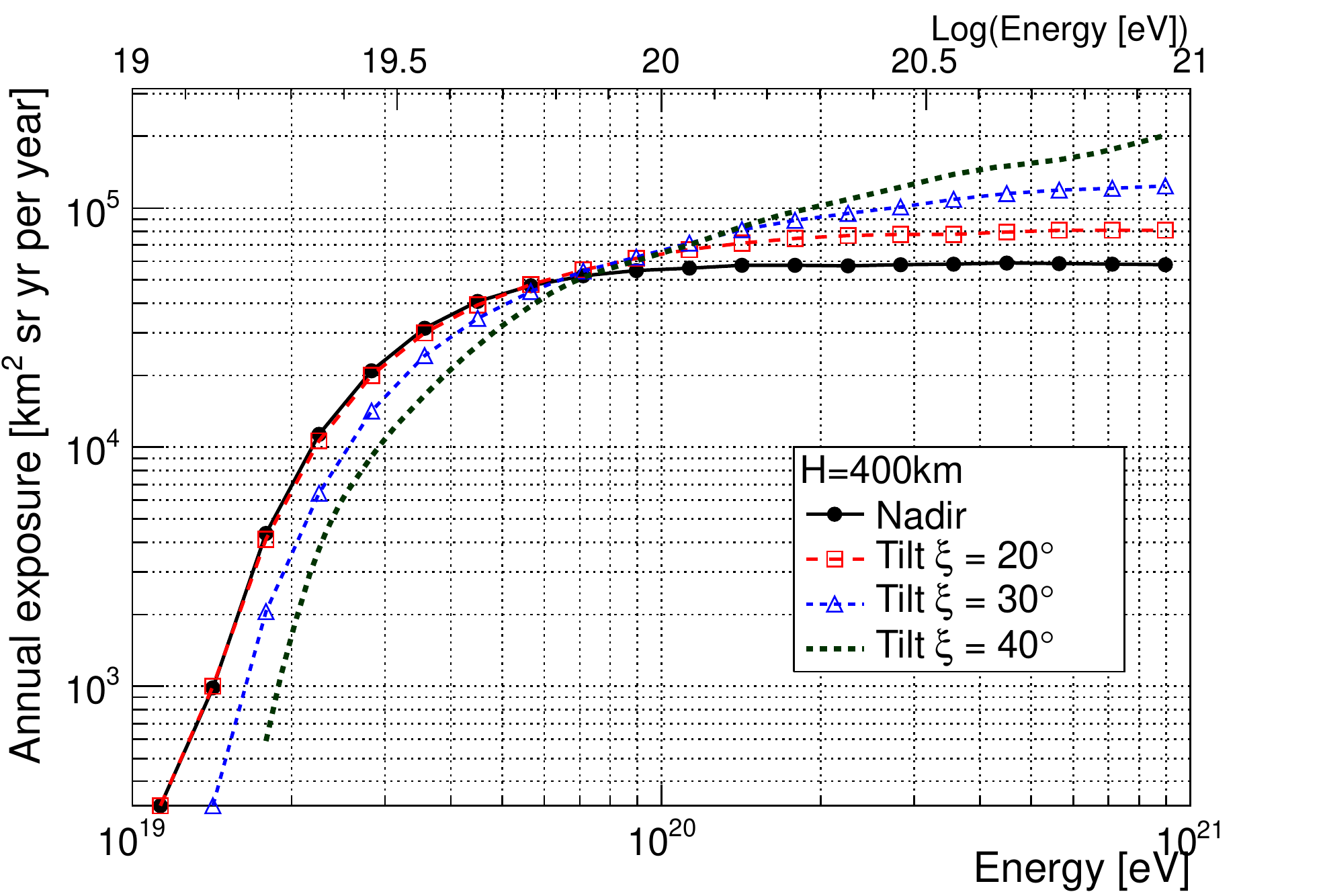}
  \end{center}
  \caption{Annual exposure as a function of energy for nadir mode (closed 
    circles) and tilt modes for $\xi~=~20^\circ$ (open squares) and $30^\circ$
    (open triangle). The case for $\xi~=~40^\circ$ is shown as a reference. 
    $H_0 = 400$~km is assumed. }
  \label{fig:app400}
\end{figure}

In order to estimate the number of events observed by the JEM-EUSO mission,
the exposure growth per unit time such as one-year operation is an essential 
measure. Apart from the nominal geometrical aperture, reduction factors
in exposure and observation on-time should be taken in account. In the present
work, the exposure per year of operation for events that
trigger JEM-EUSO, defined as the `annual exposure' 
is evaluated as follows:
\begin{equation}
\mbox{(Annual exposure)}\equiv A(E) \cdot \kappa_{\rm C} \cdot \eta_0 
\cdot (1 - f_{\rm loc})\cdot \mbox{(1~yr)},
\label{eqn:annexp}
\end{equation}
where $\kappa_{\rm C}~\approx~0.72$ is a cloud efficiency, $\eta_0~\approx~0.2$
is an observational duty cycle and $f_{\rm loc}\approx~0.1$ is the fraction of
locally light-polluted areas. The details of those factors have been 
intensively investigated in Ref.~\cite{adams}. The cloud efficiency is 
defined as the ratio in trigger aperture between the cases with and without 
cloud coverage taken into account. In this estimation, the visibility of the 
EAS maximum above or through the cloud is required. The observational duty 
cycle is the fraction of time in which the background level $I_{\rm BG}$ is below
a given threshold level. Here $I_{\rm BG}~<$~1500~{photons}~m$^{-2}$~sr$^{-1}$~ns$^{-1}$ is 
assumed to suppress high back-scattered moonlight. Note that it does not 
limit the operation of the main telescope and is conservative for EASs at 
highest energies. The factor for the local light coverages applies to man-made 
light such as cities. The reduction by the occurrence of aurorae is also 
included.

The exposure for tilt mode is also evaluated. As discussed in the previous 
section, the observation area increases in the tilt mode, while the observable
time may be reduced since both the ISS and the region within FOV should be in 
Earth's umbra. In the present work, we deliver as first results based on the 
direct application of Eq.~(\ref{eqn:annexp}).

Figure~\ref{fig:app400} shows the annual exposures at $H_0 = 400$~km as a
function of energy for the nadir mode (closed circles) and tilt modes for 
$\xi~=~20^\circ$ (open squares) and $30^\circ$ (open triangle). The case for 
$\xi~=~40^\circ$ is shown as a reference. 

According to Eq.~(\ref{eqn:annexp}), the conversion factor between exposure and
geometrical apertures $\sim~0.13$~yr. The operational inefficiencies due to the events such as rocket docking, lid operation, detector maintenance or aging, 
etc. as well as quality cuts on reconstruction are not yet taken into account. 
The latest results on the reconstruction are addressed in Ref.~\cite{rec}. 
Therefore, the present results constitute an upper limit on the effective 
exposure of the instrument for the assumed conditions.

In the case of the nadir mode, the annual exposure  at saturated level is 
expected to be $\sim 9$~times greater than that of the Pierre Auger 
Observatory with the corresponding value of about 7000~km$^2$~sr yr \cite{pao}.
Because of the steeply rising aperture at lower energies, the subsets of
data with reduced but flat exposure are used down to $\sim (2-3)\times 
10^{19}$~eV to cross-check 
with measurements by other ground-based experiments. 
It is important to underline that the most stringent cuts shown in 
Figure~\ref{aafig2} correspond to an annual exposure comparable to that of Auger 
even at $\sim 3\times 10^{19}$~eV. This means that statistically similar data 
samples are obtained. In this way, it allows to have a comparison of UHECR 
fluxes for one entire energy-decade above this energy.

The quasi-nadir mode such as $\xi = 20^\circ$ case allows to slightly 
increase the exposure above $10^{20}$~eV. This is an interesting option to 
recover the exposure from unexpected operational inefficiencies. In the tilt 
mode at $\xi = 30^\circ$, the increase of exposure is a factor of 
$\sim 2$ at $\sim 3\times 10^{20}$~eV. 

So far, a constant background level 
has been assumed, while it is variable with time. By increasing the maximum 
acceptable background level, the observational time increases as well. The 
trigger system is capable of dynamically adjusting the thresholds to cope with
variable background intensity~\cite{catalano}. The exposure function shifts in
energy proportional to $\sim\sqrt{I_{\rm BG}}$ as it depends on Poissonian 
fluctuations of the average background level. Along with possible tilt mode 
operation, this is particularly useful to explore the extreme energy ranges 
with elongated exposures.

\begin{figure}[t]
  \begin{center}
    \includegraphics[width=0.46\textwidth]{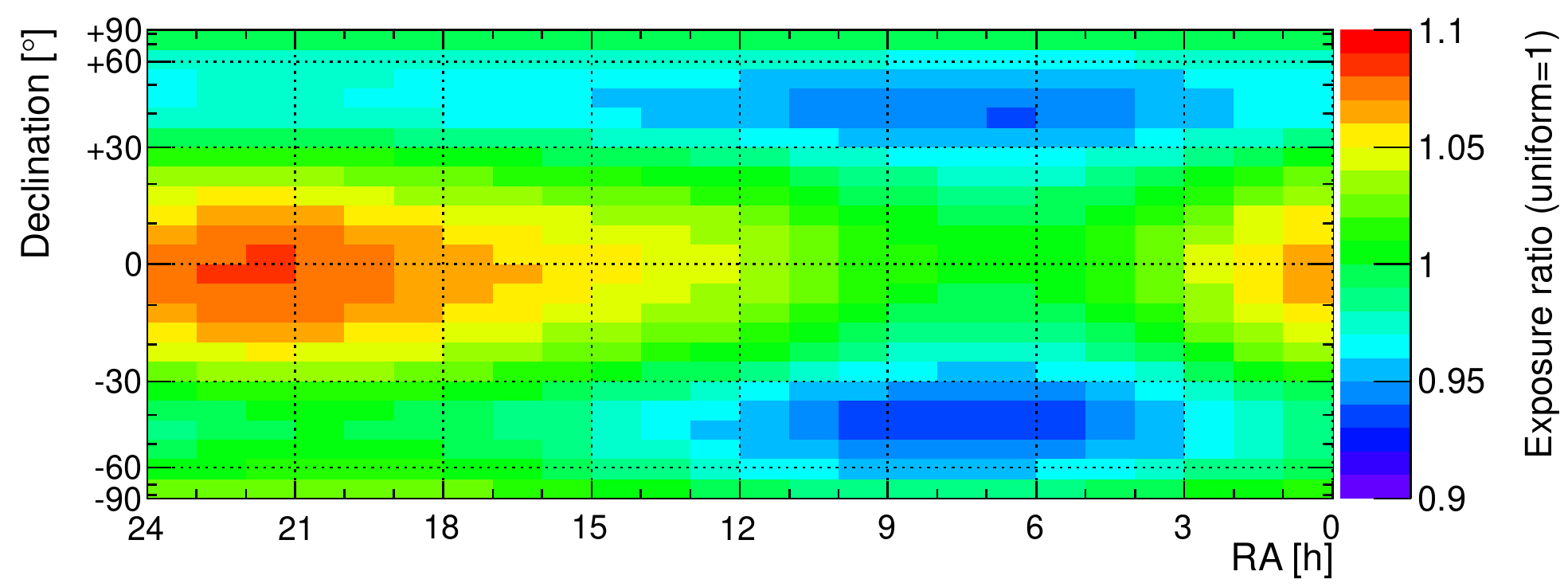}
  \end{center}
  \caption{Relative exposure as a function of declination and right ascension 
    in color scale. The impact of cloud coverage is investigated from TOVS 
    database \cite{TOVS}. The nadir mode is assumed for this analysis.}
  \label{aafig3}
\end{figure}

In the end, it is also interesting to mention that unlike ground-based 
observatories, the ISS orbit and sensitivities to large 
zenith angle EASs allow the full coverage of the entire Celestial Sphere. 
Knowing nighttime duration along with locational dependence of the cloud
coverage, the expected exposure distribution may be computed.

In Figure~\ref{aafig3} the exposure distribution is shown on the Celestial Sphere
for highest energy EASs. The unity corresponds to uniformity. The impact of the
cloud is investigated from the TOVS database~\cite{TOVS}. In this analysis, we
assume the nadir mode and the cases with cloud-top altitudes lower than 3.2~km
as fiducial regions. 

The distribution of exposure is primarily determined by astronomical 
factors. In each latitude, the observable sky regions in declination 
are limited. The  
distribution in declination is a result of integration over latitudes within 
$\pm 51.6^\circ$ taking into account that the resident time of the ISS.
The distribution in right ascension (RA) intrinsically 
has non-uniformity due to longer twilight around the time of solstices 
although the effect is not dramatic as seen in ground-based observatories. This
results in deficits around 6~h and 18~h in RA. 

Another factor is due to global cloud distribution and its seasonal variation.
In the present work, we assume the clear atmosphere and clouds at low altitudes
that allows visibility of the EAS maximum for $\Theta~>~30^\circ$. Favorable 
scenes are found in dry land and ocean in part. On the contrary, there are less
efficient areas near Equator like rainforest. The figure shows the summary of 
overall effects. In first one-year operation, the entire Celestial Sphere
is completely observed with a $\pm 10\%$ uniformity.

\section{Summary and outlook}

In the present work, we discuss an overview of the JEM-EUSO performance 
focusing on the expected exposure. Generating a large number of EAS events by 
the ESAF package, one-year operation for $H_0 = 400$~km in the nadir mode 
results in about 9-year exposures by Auger at highest energies. With subsection
of the aperture, annual exposure similar to larger than that of the Auger is 
achieved at energies 
$\sim(3-4)\times~10^{19}$~eV. This allows cross-check with those data. The 
performance in quasi-nadir and tilt modes are also considered. The former
effectively allows a recovery of the observation area in case of lower 
altitude operation and the latter enhances exposures at highest energies. Thanks to the
ISS orbit, the all-sky survey with the JEM-EUSO telescope alone is achievable 
at high degree of uniformity. This is a distinct feature compared with 
ground-based observatories and significantly reduces uncertainties in source 
search efforts.

The scope of the present work focuses on the trigger, while the performance in 
reconstruction and the role of the AM system are discussed 
in Refs~\cite{ams2,rec}. The aperture and exposure herein are 
derived with specific assumptions on the detector properties, background level,
EAS development etc. Recent introduction of the JEM-EUSO configuration into the Offline code~\cite{offline} will systematically allow cross-checks in
simulation processes. \\

\section*{Acknowledment}

The present work has been partly supported by the ESA Topical Team Activities
Fund and by the DLR, Deutsches Zentrum f\"ur Luft- und Raumfahrt, Germany, by 
the Italian Ministry of Foreign Affairs General Direction for the Cultural 
Promotion and Cooperation 
and by Slovak Academy of Sciences MVTS 
JEM-EUSO as well as VEGA grant agency project 2/0081/10. The computing facility
of RICC (RIKEN Integrated Cluster of Clusters) has been used for the simulation study.

\clearpage

%% file: icrc2013-0461.tex



\hyphenation{
con-sists
sho-wers
cha-rac-te-ris-tics
e-ner-ge-tic
ac-count
ac-counts
re-gions
ge-ne-ra-ted
Bo-ttom
do-mi-na-ted
ta-ken
co-rres-pon-ding
im-prove-ment
com-pa-ring
e-ner-gies
e-ner-gy
du-ring
pi-ons
pre-dict
rea-sons
fluc-tu-at-ions
}

\title{Identification of extreme energy photons with JEM-EUSO}

\shorttitle{Identification of extreme energy photons with JEM-EUSO}

\authors{
A. D. Supanitsky$^{1,2}$,
G. Medina-Tanco$^{2}$
for the JEM-EUSO Collaboration.
}

\afiliations{
$^1$ Instituto de Astronom\'ia y F\'isica del Espacio (IAFE), CONICET-UBA, Argentina.\\
$^2$ Departamento de F\'isica de Altas Energ\'ias, Instituto de Ciencias Nucleares, Universidad
Nacional Aut\'onoma de M\'exico, A. P. 70-543, 04510, M\'exico, D. F., M\'exico. \\
}

\email{supanitsky@iafe.uba.ar}

\abstract{
Extensive searches for ultra-high-energy photons have been performed by past and current cosmic-ray observatories. 
Nevertheless, at present no firm candidates have been found. All candidate events are compatible with proton primaries, 
which are the principal source of backgrounds for their identification. As a result, several upper limits on their integral 
photon flux have been obtained. Besides other theoretically possible sources, at least a flux of ultra-high-energy photons 
is expected as a result of the interactions suffered by cosmic rays during propagation through intergalactic medium. However, 
current upper limits do not reach the flux expected by the corresponding astrophysical models. Extreme Universe Space Observatory 
on board Japanese Experimental Module (JEM-EUSO), is an orbital fluorescence telescope intended to observe the most energetic 
component of cosmic rays ($E \gtrsim 10^{19.7}$ eV), planned to be installed on the International Space Station. By design, 
the instrument is also sensitive to photons and neutrinos. In this work we study the potential of JEM-EUSO for photon searches. 
We obtain the upper limits on photon fractions in a total of expected events (under the assumption that there are no photons in 
the samples) for different combinations of observation time in the Nadir and Tilted modes of the telescope. For the calculation 
of the upper limits we use a statistical method based on the parameter $X_{max}$, the atmospheric depth for which the maximum 
development of the shower of a primary particle is attained.
}


\maketitle

\section{Introduction}

Extreme energy photons can originate in different astrophysical contexts. They can be produced as a consequence 
of the interactions suffered by cosmic rays during their propagation through intergalactic medium, on route to 
the Earth (see for instance Ref. \cite{Gelmini:08}). These energetic photons are generated by the decay of neutral 
pions produced by the interactions of cosmic rays with the low energy photons of the radiation field that fills 
the Universe. Extreme energy photons can also be produced by the interactions of cosmic rays in their acceleration 
sites. In this case the neutral pions can be produced by the interaction of cosmic rays with intense radiation 
fields and also with ambient protons present in the acceleration regions \cite{Murase:09}. Another possibility 
is the production of extreme energy photons in the decay of super heavy relic particles or topological defects (see for 
instance \cite{Aloisio:04}). However, these type of top-down scenarios are disfavored by present data \cite{augerSD}.   
It is worth noting that at present there is no ultra high energy photon unambiguously identified.

High energy photons can generate extensive air showers when they interact with the molecules of the atmosphere. At the 
highest energies the characteristics of such air showers are dominated by the Landau and Pomeranchuk 
\cite{aLandau:53a,aLandau:53b}, Migdal \cite{aMigdal:56} (LPM) effect and pre-showering (i.e., photon splitting) in the 
Earth's magnetic field (see Ref. \cite{Risse:07} for a review). In this work we briefly discuss the characteristics of 
the longitudinal profiles of the extensive air showers generated by these energetic photons. We also calculate the 
expected upper limits on the photon fraction in the integral cosmic ray flux, assuming that there is no photon in the 
samples, expected for the JEM-EUSO mission \cite{Takahashi:09}. We use the atmospheric depth of the maximum development 
of the showers, $X_{max}$, which can be reconstructed from future JEM-EUSO data, as the parameter to discriminate 
between showers initiated by protons and photons. We use an extension of the method proposed in Refs. 
\cite{aSupanitsky:11,SupanitskyICRC11} to calculate the expected upper limits.

\section{Characteristics of photon showers}

One of the most sensitive parameters to the nature of primary cosmic rays is the atmospheric depth of the points at 
which the showers reach the maximum development. It can be reconstructed from data taken by the fluorescence 
telescopes like JEM-EUSO.    

A shower library of protons and photons is generated by using the program CONEX \cite{aconex} (v2r2.3). It consists of 
$1.1 \times 10^{5}$ proton showers following a power law energy spectrum of spectral index $\gamma = -1$ in the interval of
[$10^{19.7}, 10^{21}$] eV. The arrival directions of the showers are distributed uniformly. Also $1.5 \times 10^{5}$ photon 
showers are generated under the same conditions but in this case the impact points of the showers are uniformly distributed 
over the Earth's surface in order to properly take into account the pre-showering effect in the geomagnetic field. The 
hadronic interaction model used to generate the showers is QGSJET-II \cite{aQGSJETII}.

\begin{figure}[!h]
\centering
\includegraphics[width=0.45\textwidth]{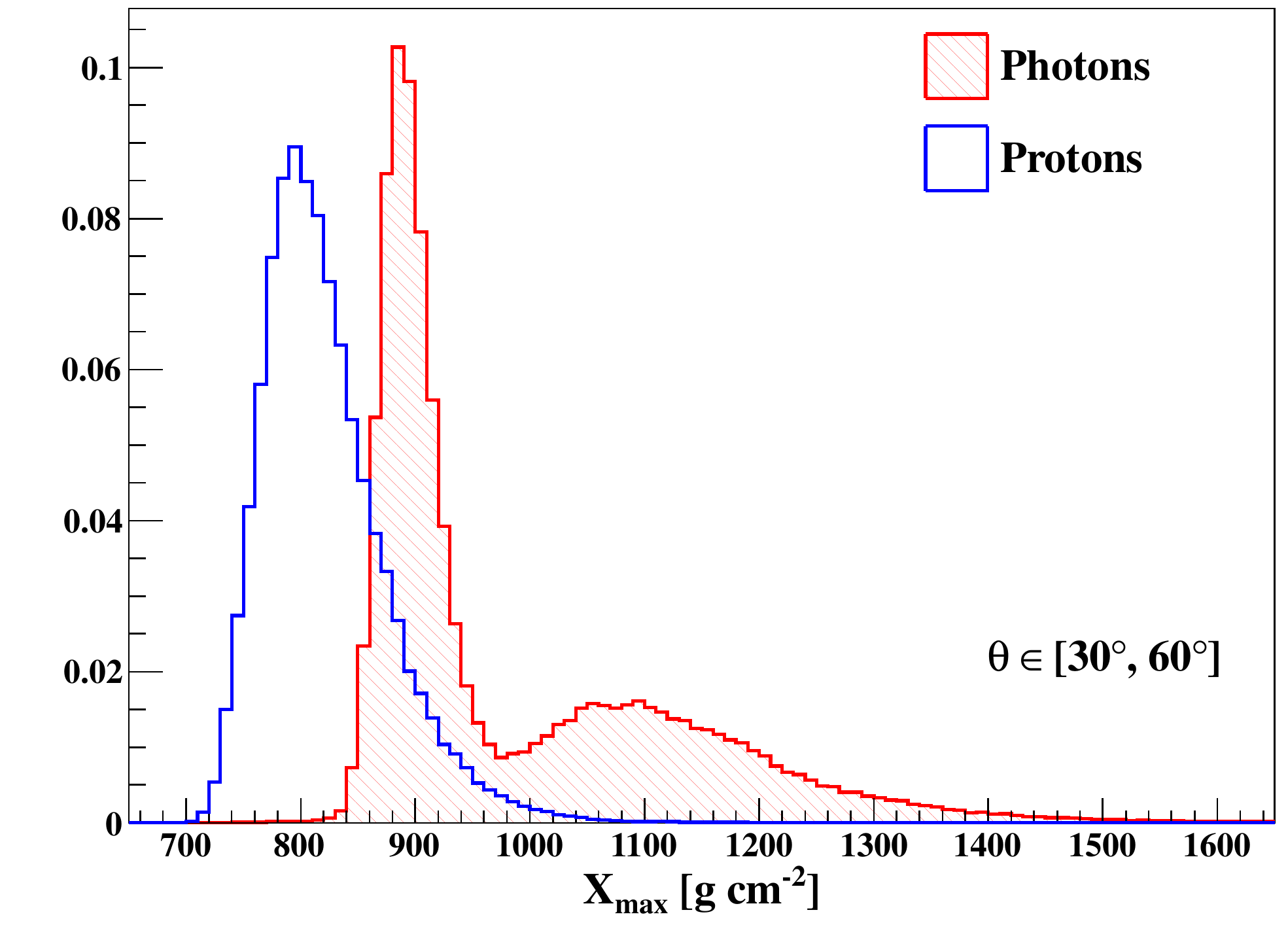}
\includegraphics[width=0.45\textwidth]{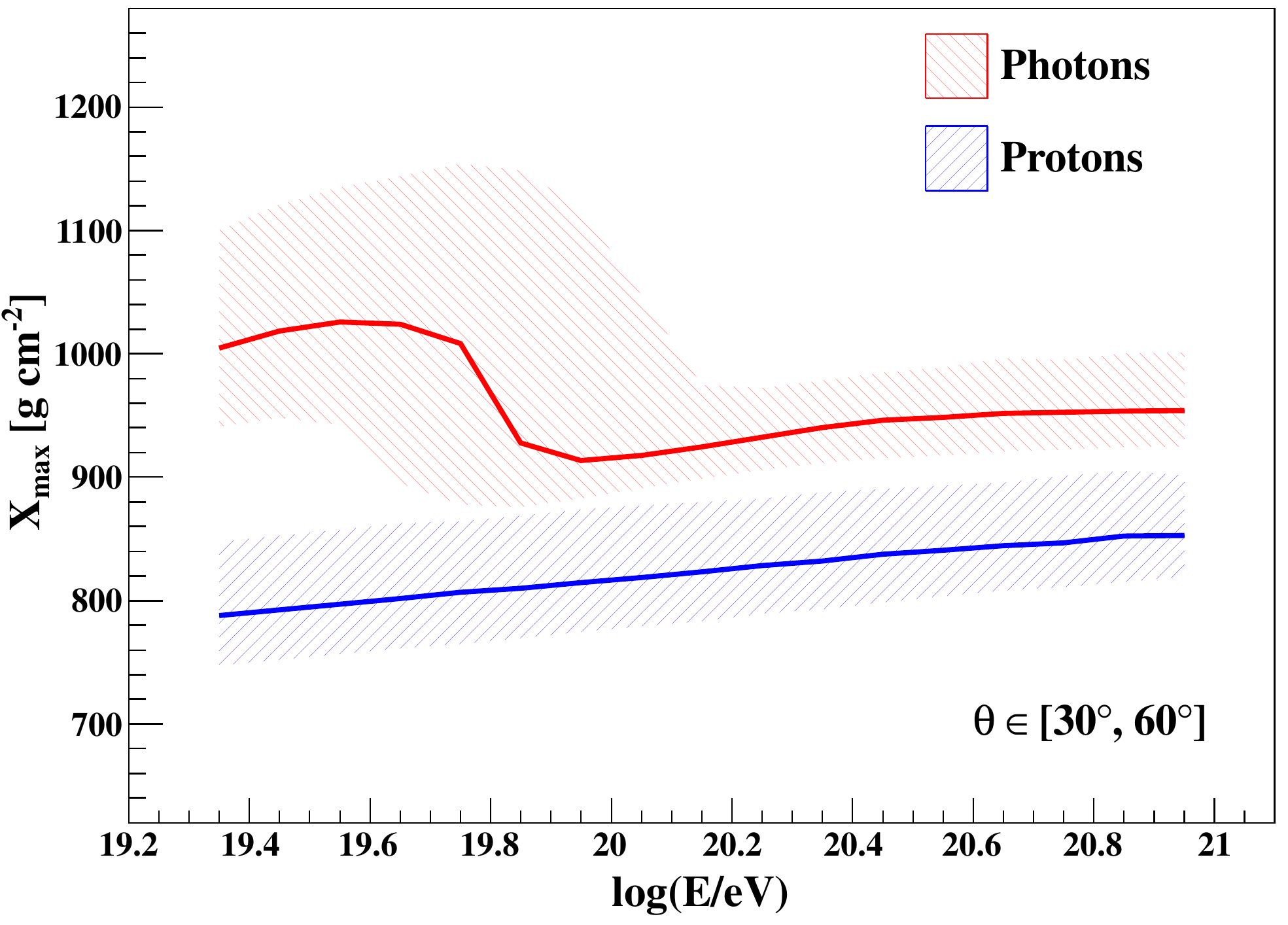}
\caption{Top panel: $X_{max}$ distributions for proton and photon primaries of $E \in [10^{19.8}, 10^{20}]$ eV and 
$\theta \in [30^{\circ}, 60^{\circ}]$. Bottom panel: Median and region of $68\%$ probability of the $X_{max}$ distributions
as a function of the logarithm of the primary energy for protons and photons with $\theta \in [30^{\circ}, 60^{\circ}]$.}
\label{XmaxDist}
\end{figure}

The top panel of figure \ref{XmaxDist} shows the distributions of $X_{max}$ for proton and photon showers with 
$E \in [10^{19.8}, 10^{20}]$ eV and zenith angle of the shower $\theta \in [30^{\circ}, 60^{\circ}]$. It can be seen that 
the distribution of photons has two components; one corresponds to photons that suffered from photon splitting and the other 
one corresponds to the photons that do not suffered from photon splitting. Note that the values of $X_{max}$ for the photons 
that are converted in the geomagnetic field are smaller and present smaller fluctuations.   

The bottom panel of figure \ref{XmaxDist} shows the median and the region of the central $68\%$ probability containment of the 
$X_{max}$ distribution as a function of the logarithm of primary energy for protons and photons with 
$\theta \in [30^{\circ}, 60^{\circ}]$. It can be seen that for energies below $\sim 10^{19.5}$ eV the LPM effect dominantly affects 
the $X_max$ distribution, i.e. the photon splitting is negligible. From $\sim 10^{19.5}$ eV to $\sim 10^{20.1}$ eV the $X_{max}$ 
distribution is composed by the two populations of photons, and for energies above $\sim 10^{20.1}$ eV all photons are converted in 
the geomagnetic field. Note that the discrimination power between protons and photons by the parameter $X_{max}$ increases with primary 
energy in the region where all photons are converted in the geomagnetic field.  

\section{Expected upper limits}

\begin{figure*}[t]
\centering
\includegraphics[width=0.8\textwidth]{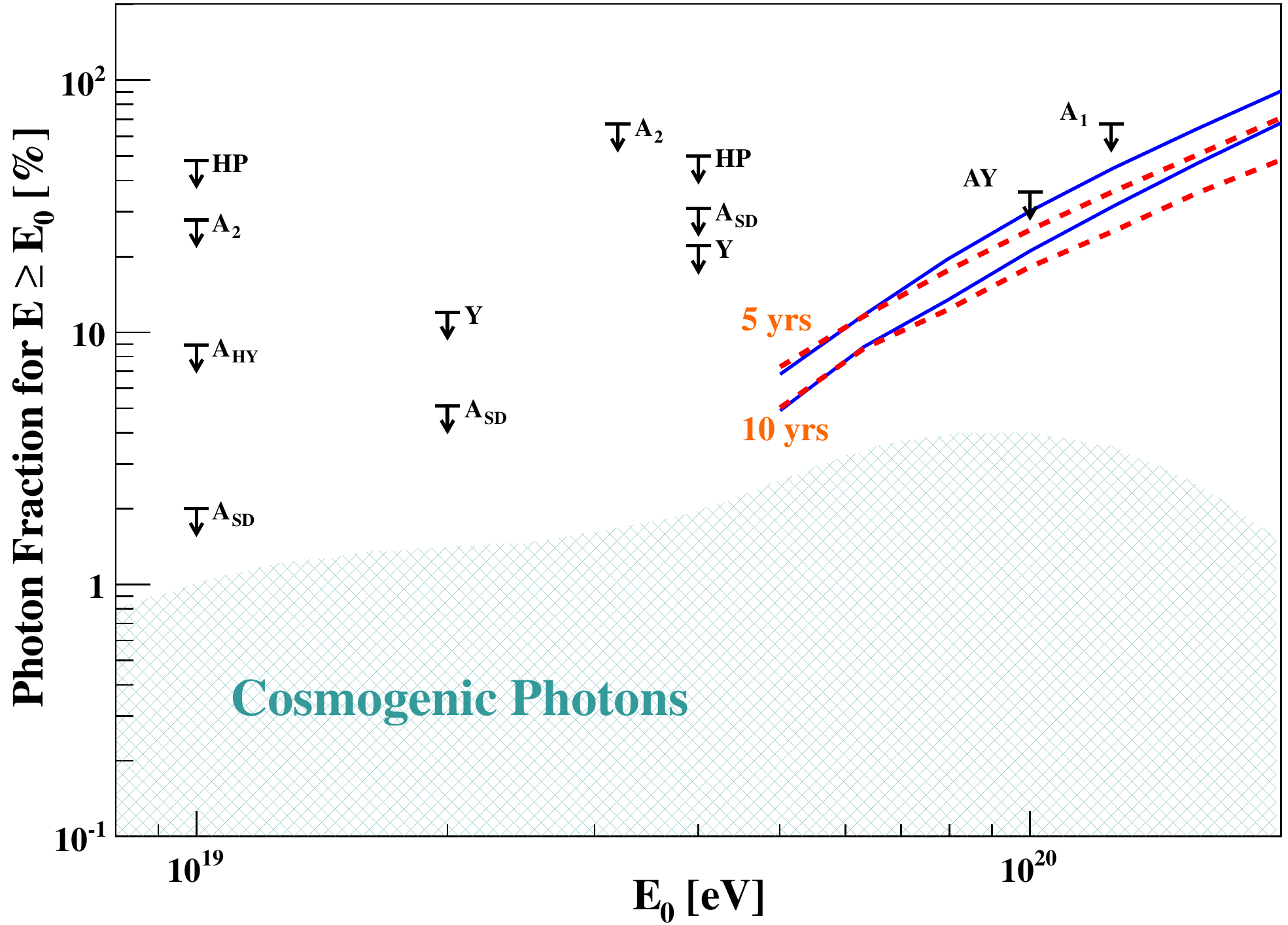}
\caption{Upper limits on the fraction of photons in the integral cosmic ray flux at 95\% confidence level as a function
of primary energy. Solid lines correspond to the expected upper limits, $\mathcal{F}_{\gamma}^\xi(E_0)$, obtained by using the $\xi_{X_{max}}$
method for JEM-EUSO in the Nadir mode and dot-dash-dot lines correspond to a combination between the Nadir and Tilted modes (see text). A
Gaussian uncertainty on the determination of the $X_{max}$ parameter of 100 g cm$^{-2}$ is assumed for calculations. The shadow region is the
prediction for the cosmogenic photons \cite{Gelmini:08}. Black arrows are experimental limits, HP: Haverah Park \cite{Ave};
A$_1$, A$_2$: AGASA \cite{Risse:05,Shinozaki:02}; A$_{\textrm{HY}}$, A$_{\textrm{SD}}$: Auger \cite{augerHY,augerSD};
AY: AGASA-Yakutsk \cite{Rubtsov:06}; Y: Yakutsk \cite{Glushkov:07}.}
\label{UpperL}
\end{figure*}

Assuming that the JEM-EUSO exposure is the same as the one for protons \cite{JemEusoExp:13} it is possible to calculate the expected 
number of photon events above a given energy threshold. Table \ref{Nph} shows the expected number of photons with energies above 
$10^{19.6}$ eV, calculated considering the most optimistic photon flux taken from Ref.~\cite{Gelmini:08} (it corresponds to the curve 
on the top of the shadowed region of Fig.~\ref{UpperL}). The calculation is done for four cases: The observation of 5 and 10 years in 
the Nadir mode, 1 year in the Nadir mode and 4 years in the Tilted mode, and 1 year in the Nadir mode and 9 years in the Tilted mode. 
The tilted angle used for the calculation is $40^\circ$. Note that the number of events for the cases including observation in the 
Tilted mode the expected number of events is slightly smaller than the ones corresponding to observation in the Nadir mode. This is 
due to the fact that the exposure for $40^\circ$ of the Tilted angle is smaller than the one corresponding to the Nadir mode for 
energies higher than $\sim 10^{19.85}$ eV. For higher threshold energies this tendency is inverted.  
\begin{table}[h]
\begin{center}
\begin{tabular}{|c|c|c|c|}
\hline 
5 yrs N & 10 yrs N & 1 yr N+4 yrs T & 1 yr N+9 yrs T \\ \hline
88      & 177      & 85               & 168               \\ \hline
\end{tabular}
\caption{Expected number of photon events for energies larger than $10^{19.6}$ eV. N corresponds to the Nadir 
mode of observation and T to the Tilted one. The cosmogenic photon flux corresponds to the most optimistic 
case taken from Ref.~\cite{Gelmini:08}. The exposure used for the calculation is the one calculated for proton 
primaries.}
\label{Nph}
\end{center}
\end{table}
    
Nevertheless, there are astrophysical models that predict a much smaller flux of cosmogenic photons, specially the ones that includes 
heavier nuclei in the composition injected by the sources (see for instance \cite{Hooper:11}). Also, fluctuations of a relatively large 
flux can still produce a null detection even for a non-null statistical expectation. For these reasons and also to compare with the 
existing upper limits on photon fractions obtained by different experiments the case in which there are no photons in the samples is 
studied below.  

In an ideal condition where it is known that there are no photons in a given sample of $N$ events the upper limit to the photon 
fraction can be easily calculated and it is given by \cite{Risse:07},
\begin{equation}
\mathcal{F}_{\gamma} = 1-(1-\alpha)^{1/N}
\label{Fuplideal}
\end{equation}
where $\alpha$ is the confidence level of rejection. However, in practice, the probability of the existence of photons must be 
realistically estimated through some observational technique which involves the determination of experimental parameters, like 
$X_{max}$, which leads unavoidably to less restrictive upper limits than in the ideal case.

The method used to calculate the upper limits of the photon fraction by using the $X_{max}$ parameter is based on the abundance 
estimator first introduced in \cite{Supanitsky:09},
\begin{equation}
\xi_{X_{max}} = \frac{1}{N} \sum_{i=1}^{N} \frac{f_{\gamma}(X_{max}^i)}{f_{\gamma}(X_{max}^i)+f_{pr}(X_{max}^i)}
\label{xidef}
\end{equation}
where $f_\gamma(X_{max})$ and $f_{pr}(X_{max})$ are the distributions of $X_{max}$ for photons and protons, respectively, $X_{max}^{i}$ 
are experimental values of $X_{max}$ of the $i$th event, and $N$ is the sample size.

For samples of a large size it is possible to calculate the upper limit to the photon fraction for the case in which there is no photon
in the sample by using the $\xi_{X_{max}}$ parameter analytically \cite{SupanitskyICRC11}. However in this work the Monte Carlo technique 
is used for the calculations which is valid for samples of any size. 

Given a sample of the $X_{max}$ parameter of size $N$, the upper limit on the photon fraction, $\mathcal{F}_{\gamma}^\xi$, is obtained 
as the solution of the following equation,
\begin{equation}
\int_{\textrm{med}(\xi_{X_{max}}^{pr}(N))}^{1} d\xi\ P(\xi|\mathcal{F}_{\gamma}^\xi, N) = \alpha
\label{ULxi}
\end{equation}
where $\textrm{med}(\xi_{X_{max}}^{pr}(N))$ is the median of $\xi_{X_{max}}$ assuming that there are only protons in the sample,
$P(\xi|\mathcal{F}_{\gamma}^\xi, N)$ is the distribution function of $\xi_{X_{max}}$ for a photon abundance 
$c_\gamma = \mathcal{F}_{\gamma}^\xi$, and $\alpha$ is the rejection probability.

The distribution function $P(\xi|c_{\gamma}, N)$ is obtained by means of a Monte Carlo simulation. Given the sample size
$N$ and the photon abundance $c_\gamma$ a large number of $X_{max}$ samples is used to estimate the distribution function. The number 
of photons in a sample, $N_\gamma$, is obtained by sampling a binomial distribution of probability $c_\gamma$ and total number of 
events $N$. Then, $N_\gamma$ values of $X_{max}$ are taken at random from the $X_{max}$ distribution of photons and $N_{pr}=N-N_\gamma$ 
values of $X_{max}$ are also taken at random from that of protons. The value of $\xi_{X_{max}}$ in a given sample is obtained by using 
the Eq. (\ref{xidef}). The distribution functions needed to calculate $\xi_{X_{max}}$ are obtained from the simulated data by using the 
non-parametric method of kernel superposition with adaptive bandwidth \cite{Silvermann:86,Supanitsky:09}. The number of events expected 
above a given energy threshold are calculated by using the broken-power law fit of the Auger energy spectrum \cite{AugerSpe:11} and the 
exposure of JEM-EUSO \cite{JemEusoExp:13}.

Figure \ref{UpperL} shows the upper limits on the photon fraction in the integral flux at 95\% confidence level. The zenith angle of the 
showers is in the interval $[45^\circ, 90^\circ]$ and a Gaussian uncertainty on the determination of $X_{max}$ of 100 g cm$^{-2}$ is assumed 
for the calculation. Note that the uncertainty on the determination of $X_{max}$ considered is a conservative value (perhaps overestimated) 
for the resolution expected for the JEM-EUSO mission. The solid lines show the expected upper limits for the cases in which the observation 
is done in the Nadir mode during 5 and 10 years. The dashed lines correspond to the observation of one year in the Nadir mode and 4 and 9 
years in the Tilted mode for 5 and 10 years of the total observation time, respectively. The tilted angle used for the calculation is 
$40^\circ$. The arrows correspond to the upper limits obtained for several experiments and the shadowed region corresponds to the expectation 
for the cosmogenic photons taken from Ref. \cite{Gelmini:08}. The expected photon fraction obtained in Ref. \cite{Gelmini:08} is calculated 
assuming a power law energy spectrum of nucleons at injection, a uniform distribution of sources in the universe and no evolution of the 
sources with redshift. The normalization of the spectrum is obtained by fitting the Hires data (see Ref. \cite{Gelmini:08} for details).     

The expected upper limits are more restrictive for increasing values of the number of events. The number of events collected by observing 
in the Tilted mode increases at the highest energies and decreases at lower energies comparing with the observation in the Nadir mode. For 
$40^\circ$ of the tilted angle the number of events detected in one year of observation in the Nadir mode is compatible with that in the 
Tilted mode for an energy of $\sim 10^{19.85}$ eV. For energies higher than that the difference in the number of events increases. This 
difference in the number of observed events is responsible for the improvement on the expected upper limits obtained when the observation 
in the Tilted mode is considered.

In order to compare the expected upper limits of the photon fraction in the integral cosmic ray flux obtained by using the $\xi_{X_{max}}$ 
method with the ideal case in which it is known that there is no photon in the samples, the following parameter is defined,
\begin{equation}
R(E_0) = \frac{\mathcal{F}_{\gamma}^\xi(E_0)}{\mathcal{F}_{\gamma}(E_0)},
\label{aaRatio}   
\end{equation}
where $\mathcal{F}_{\gamma}(E_0)$ is given by Eq. (\ref{Fuplideal}).

Figure \ref{aaRatio} shows $R$ as a function of the logarithm of primary energy for the four cases considered. Note that all curves decrease 
with primary energy because the discrimination power between protons and photons of the parameter $X_{max}$ increases with primary 
energy (see bottom panel of figure \ref{XmaxDist}). It can also be seen that for a given energy, $R$ is larger for samples with larger number 
of events. This is due to the fact that the upper limits obtained for the ideal case decrease faster as a function of the number of events 
than the ones corresponding to the realistic case which is caused by the limited discrimination power of $X_{max}$ to separate protons from 
photons.   
\begin{figure}[!h]
\centering
\includegraphics[width=0.45\textwidth]{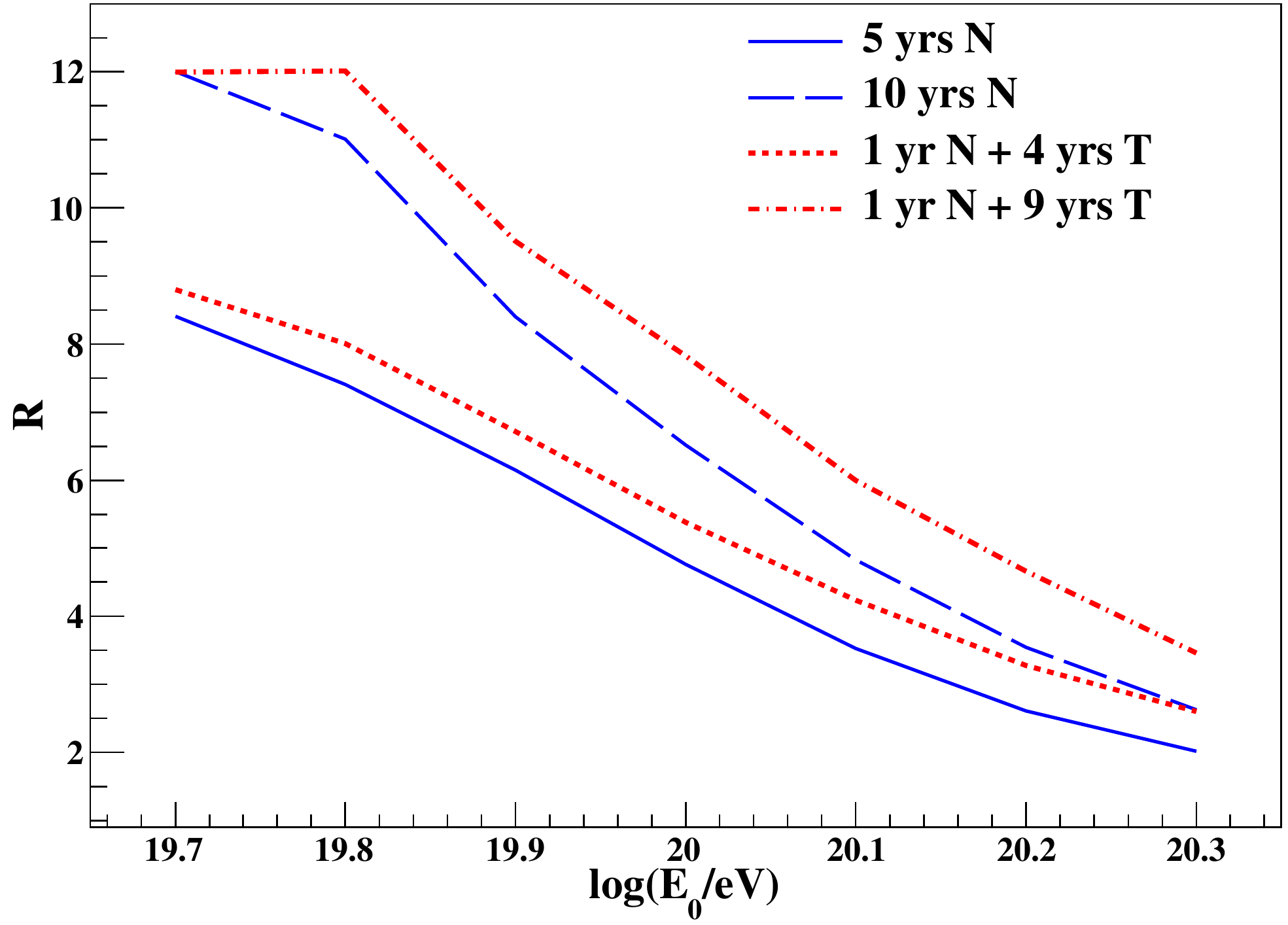}
\caption{Ratio between the expected upper limits of the photon fractions considering experimental configurations and that in the ideal 
case (see text).}
\label{abRatio}
\end{figure}

Note that improving the methods to discriminate between protons and photons more stringent upper limits can be obtained. More sophisticated 
techniques are under development at present to make progress in this direction.

\section{Conclusions}

In this work we have studied the characteristics of the photon showers in the energy range relevant to the JEM-EUSO mission,
which is important for the development of methods for photon identification. We have also presented the attainable upper limits 
on the photon fraction in the integral cosmic ray flux by using an extension of a method developed earlier. We have shown that
for 5 and 10 years of observation it is possible to obtain more stringent upper limits than the existing ones at present.    
Comparing with the ideal case in which it is known that there is no photon in the samples we have shown that there is still room
for improvement which is at present work in progress.  


\clearpage


%% file: icrc2013-0533.tex


\title{A study on JEM-EUSO's trigger probability for neutrino-initiated EAS}

\shorttitle{JEM-EUSO's $\nu$ trigger probability}

\authors{
Alejandro Guzm\'an $^{1}$,
Alberto Supanitsky $^{2,3}$,
Elias Iwotschkin $^{1}$,
Thomas Mernik $^{1}$, 
Francesco Fenu $^{1}$,
Gustavo Medina-Tanco,$^{3}$,
Andrea Santangelo $^{1}$,
for the JEM-EUSO Collaboration.
}

\afiliations{
$^1$ Insitute f\"ur Astronomie und Astrophysik, Univerist\"at T\"ubingen, Germany \\
$^2$ Instituto de Astronom\'ia y F\'isica del Espacio, CONICET-UBA, Argentina \\
$^3$ Insituto de Ciencias Nucleares, UNAM, M\'exico \\
}

\email{guzman@astro.uni-tuebingen.de, dalesupa@gmail.com}

\abstract{Neutrinos at ultra high energies (UHE) are expected as by-products of the interaction UHE Cosmic Rays (UHECRs). Whether 
these interactions happen in astrophysical sources or whilst their propagation, UHE neutrinos retain valuable information about the 
origin and propagation of UHECRs and about their sources. Those elusive particles can be detected by very large exposure observatories 
that are currently being operated or, as is the case of JEM-EUSO, designed. Currently designed to be hosted on-board the International 
Space Station, the JEM-EUSO mission will pioneer the observation of the Extensive Air Shower’s (EAS) fluorescent light from space. Hence, 
specific studies tailored to address all the peculiarities of such mission are necessary to assess JEM-EUSO’s capabilities. In this paper 
we perform a simulations study of the trigger probability for neutrino-initiated EAS. The simulations are carried out within the EUSO’s 
Simulation and Analysis Framework (ESAF), which is a software tool specifically developed bearing space borne missions in mind and in 
particular EUSO-like observatories. The shower longitudinal profiles are produced by a combination of the PYTHIA interaction code and 
CONEX shower simulator. The resulting EAS are then integrated to ESAF’s simulation chain and subsequent triggering conditions analyzed.
}


\maketitle


\section{Introduction}
 Neutrinos can be generated as by-products of the
interaction of  cosmic rays during their propagation through the 
intergalactic medium or by interactions in the acceleration
sites \cite{Ostapchenko:08}. The showers initiated by ultra high energy(UHE) neutrinos 
can be observed by orbital detectors like the Extreme Universe Space Observatory,  
on-board the Japanese Experimental Module (JEM-EUSO)\cite{icrcstatus}. The 
identification of the events initiated by UHE neutrinos is based on 
the different characteristics of the longitudinal profiles of the Extensive Air Shower(EAS) generated by these primaries. In this work we present the response of the  
JEM-EUSO telescope to these type of showers and discuss the trigger probability 
for horizontal neutrino showers.

\subsection{JEM-EUSO}

JEM-EUSO is a space based UV telescope devoted to the observation of ultra high energy cosmic
ray induced air showers in the Earth's atmosphere. It will be mounted on board the Japanese 
Module of the International Space Station (ISS), orbiting the Earth at an altitude of $\sim 400$ km. 
JEM-EUSO will study the energy region around $10^{20}$ eV \cite{PerfPaper}, 
allowing it to study the sources and their spectra with high precision \cite{icrcscience}. The duration 
of the mission is scheduled to be a minimum of 3 years. During this time, JEM-EUSO will observe 
several hundreds of events with energies $>5\times10^{19}$ eV  \cite{PerfPaper}. 

The JEM-EUSO  instrument consists of a refractive optics of three Fresnel lenses focusing the 
UV photons onto the focal surface (FS) detector. The focal surface detector is made of 137 individual 
photo-detector modules (PDMs) . Each PDM is formed by 36 multi-anode photomultiplier tubes (MAPMT). 
Each MAPMT has $(8\times 8=)$~64 pixels.
Two levels of trigger algorithms are operated to search each PDM for stationary and transient 
excesses over background. The telescope is equipped with an atmospheric monitoring system using 
LIDAR and IR-camera data to record the state of the atmosphere and infer the altitude of possible 
clouds inside the field of view (FOV) \cite{icrcAMS}. More details on the specifications of the 
detector can be found on \cite{icrcinstruments} .


\section{Neutrino shower simulations}

High energy neutrinos that propagate in the Earth atmosphere can interact
with protons and neutrons of the air molecules. There are two possible channels
for this interaction, charged current (CC) and 
neutral current (NC),
\begin{eqnarray}
\textrm{CC:} \ \ \ \nu_\ell+N &\rightarrow& \ell+X  \\ 
\textrm{NC:} \ \ \ \nu_\ell+N &\rightarrow& \nu_\ell+X, 
\end{eqnarray}
Here $N$ is a nucleon (proton or neutron), $\nu_\ell$ is a neutrino of the
family $\ell$, $\ell$ is the corresponding lepton and $X$ is the hadronic part
of the processes. For the scope of this work we shall concentrate only in the 
electron neutrino $\nu_e$.

For this work, the simulation of the neutrino nucleon interaction is performed by 
using the PYTHIA code \cite{pythia} linked with the library LHAPDF \cite{lhapdf}
to be able to use different extrapolations of the parton distribution functions. 
PYTHIA is an event generator, intended for high-energy processes with particular 
emphasis on the detailed simulation of quantum chromodynamics (QCD) parton showers 
and the fragmentation process. The set of PDFs used for these simulations is CTEQ66 
\cite{cteq6}.

The energy fraction taken by the leading particle after a CC or NC interaction
depends on the primary energy of the incident neutrino. For example the energy 
fraction taken by the  electron   increases with the energy of the incident 
electron neutrino reaching values close to $0.82$ at 
 $E_{\nu} = 10^{20}$ eV
\cite{Supanitsky:11}.

\begin{figure}[!h]
\centering
\includegraphics[width=.4\textwidth]{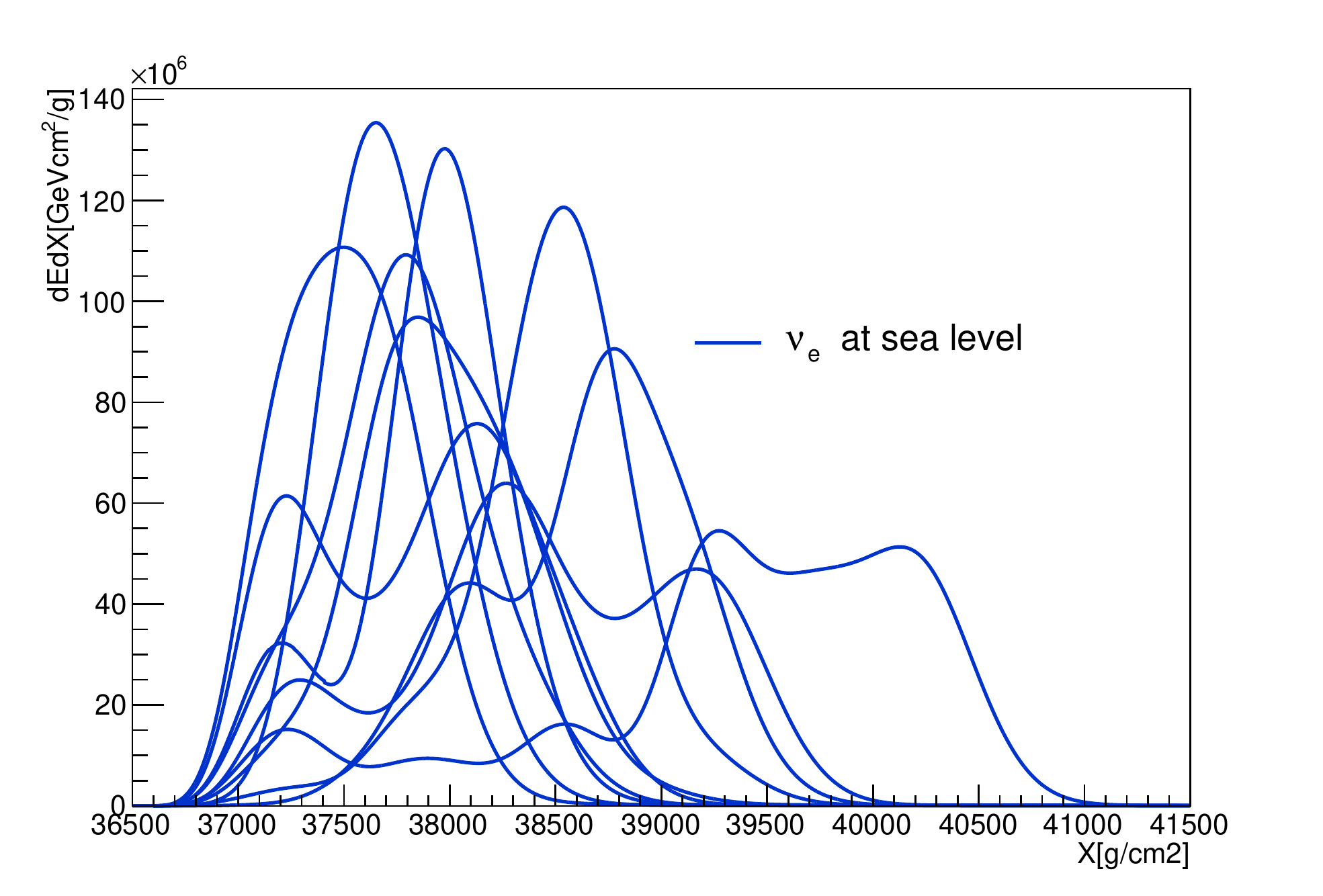}
\includegraphics[width=.4\textwidth]{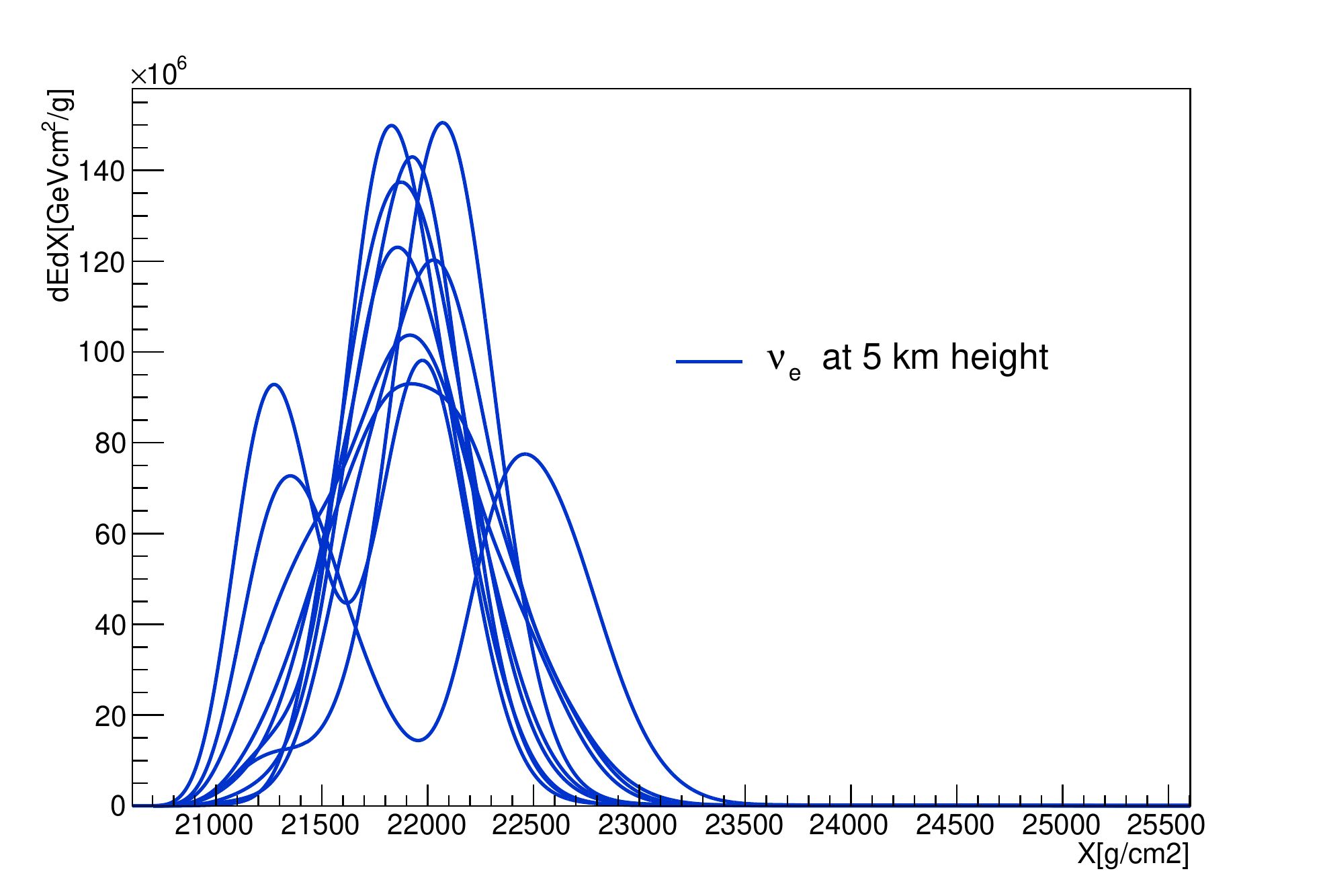}
\includegraphics[width=.4\textwidth]{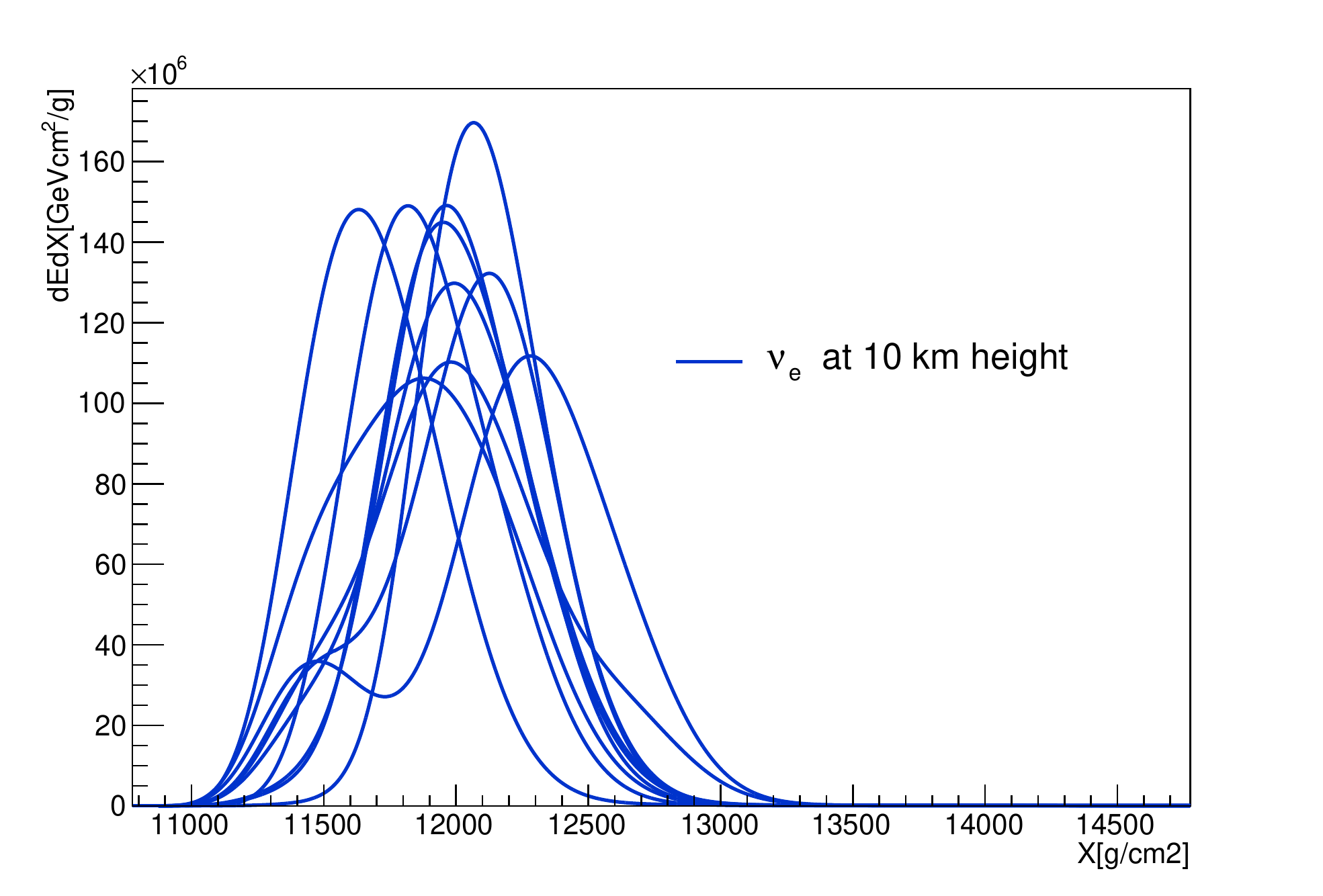}
\caption{Longitudinal profiles for horizontal electron neutrino showers with $E_\nu = 10^{20}$ eV. The neutrino injection point is contained on the vertical axis of JEM-EUSO in the Nadir mode. At sea level (top), 
at an altitude of 5km (middle) and 10 km (bottom).}
\label{longiprofiles}
\end{figure}

Neutrinos can initiate atmospheric air showers when they interact with the nucleons of
the air molecules. The CC interactions are the most important for the space observations
because in the NC interactions most of the energy is taken by a secondary neutrino that 
could produce an observable air shower just in case it suffers a subsequent CC interaction. 
In this work the showers initiated by CC interactions are just considered. Note that the 
probability that a neutrino interacts in the atmosphere increases with the zenith angle 
because of the increase of the number of target nucleons.

Electron neutrino showers are simulated following Ref. \cite{Supanitsky:11}. The
secondary particles produced in the interaction are used as input in the program CONEX
\cite{conex} (v2r2.3), in order to simulate the shower development. The high energy
hadronic interaction model used for the shower simulations is QGSJET-II \cite{QGSJETII}. 

Because the mean free path of neutrinos propagating in the atmosphere is very large,
they can interact very deeply, after traversing a large amount of matter. An orbital
detector like JEM-EUSO can also detect
horizontal showers that do not hit the ground. In particular, horizontal neutrinos can
interact at higher altitudes producing a shower observable by the detector. In fig.
\ref{longiprofiles} we show the energy deposit $\frac{dE}{dX}$ as a
function of $X$, where $X$ is the atmospheric depth  in grams per square centimeter,
for horizontal electron neutrino showers of $E_\nu=10^{20}$ eV.

The profiles corresponding to low altitudes are very broad which can present several 
peaks and large fluctuations. This behavior is due to the Landau and Pomeranchuk
\cite{Landau:53a,Landau:53b}, Migdal \cite{Migdal:56} (LPM) effect, which is very important 
inside dense regions of the atmosphere and for electromagnetic particles, electrons in 
this case, which take about 82\% of the parent neutrino energy. This effect can be 
appreciated in the longitudinal profiles shown in fig. \ref{longiprofiles}. As the altitude 
increases the fluctuations are reduced and, on average, the profiles become thinner. This 
is due to the fact that the LPM effect become progressively less important with decreasing  
atmospheric density.

\begin{figure}[!h]
\centering
\begin{subfigure}[b]{0.35\textwidth}
\centering
\includegraphics[width=1\textwidth]{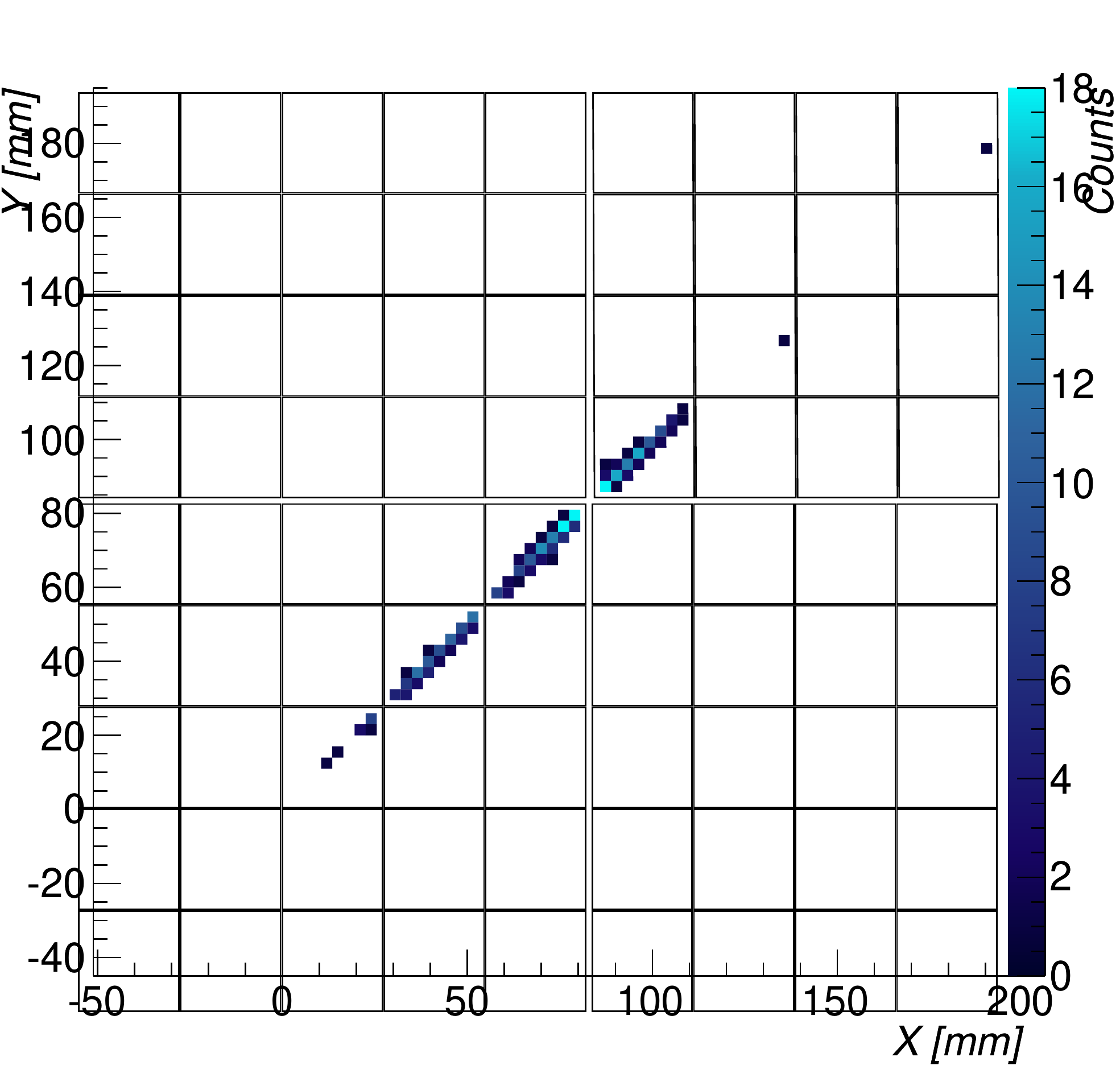}
\caption{Neutrino initiated EAS's fluorescence signal track  on the focal surface of the
detector. Height is 5 km}
\end{subfigure}
\begin{subfigure}[b]{0.5\textwidth}
\centering
\includegraphics[width=.6\textwidth]{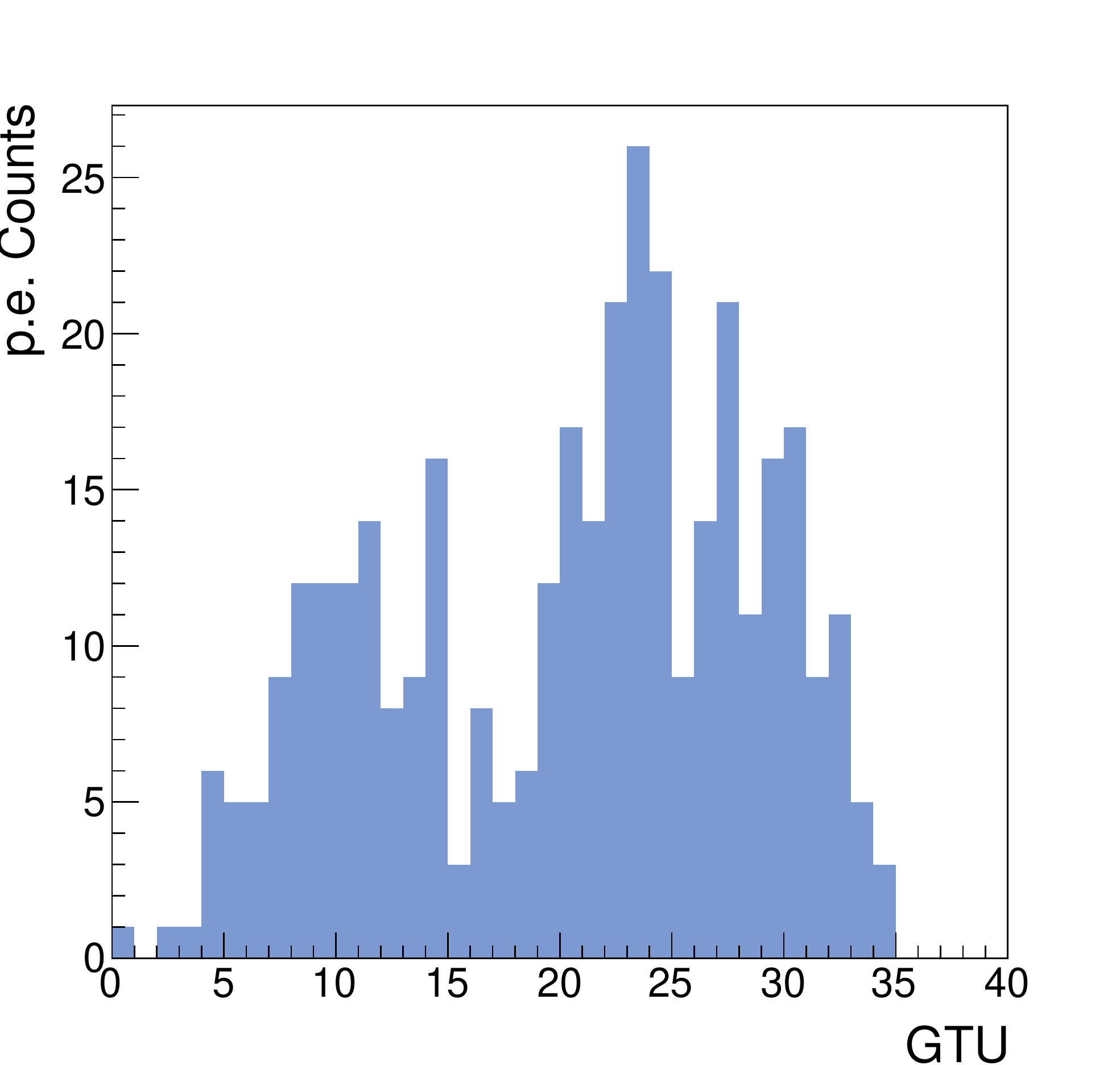}
\caption{Detected signal  function of time for the same (above) shower. Time measured in GTU's (see text).}
\end{subfigure}
\caption{ESAF simulated neutrino EAS with a strong LPM effect as  seen by the JEM-EUSO
 detector. The energy of the neutrino is $ 10^{20}$eV and the injection height is  5 km.}
\label{strongLPM}       
\end{figure}


\section{Detector simulations}
 
   The EUSO Simulation \&  Analysis Framework (ESAF) \cite{berat} is a 
   modular  software built upon the ROOT
   framework,  designed to  simulate UHECR
   detectors.  The simulations  account for  the physical processes 
   that take place during the
   development of an EAS. Each simulation covers the whole chain, the
   longitudinal development of the EAS itself, the fluorescence and
   Cerenkov light produced at the shower site, the atmospheric
   propagation of  photons, as well as the processes  within the
   detector,  i.e. the propagation of photons through the optics, the 
   response  of the electronics, and the  triggering algorithms. On a
   second stage, ESAF also provides the tools  for reconstructing the 
   simulated events  based on the recorded information of the detector's 
   response. The detector's response in time is fixed to a value of 2.5 $\mu$s,
   called Gate Time Unit (GTU).

   In order to bring into context the main scope of the present work, we would like to explain
   a  bit  in detail the JEM-EUSO's trigger. The trigger of the detector is implemented in ESAF
   and consist of two levels. The first, the 
\emph{Persistent Track Trigger (PTT)}, searches for groups of pixels (3 $\times$ 3) in which a 
signal appears longer than average background counts and the total count rate is higher than a 
preset threshold.
The second trigger level, the \emph{Linear Tracking Trigger} (LTT), looks for patterns 
that could be signal tracks by moving an integration box along a set of pre-defined directions. 
The LTT issues a triggering flag if the maximum integral along those tracks is above a preset 
threshold.

\begin{figure}[!h]
\centering
\begin{subfigure}[b]{0.35\textwidth}
\centering
\includegraphics[width=1\textwidth]{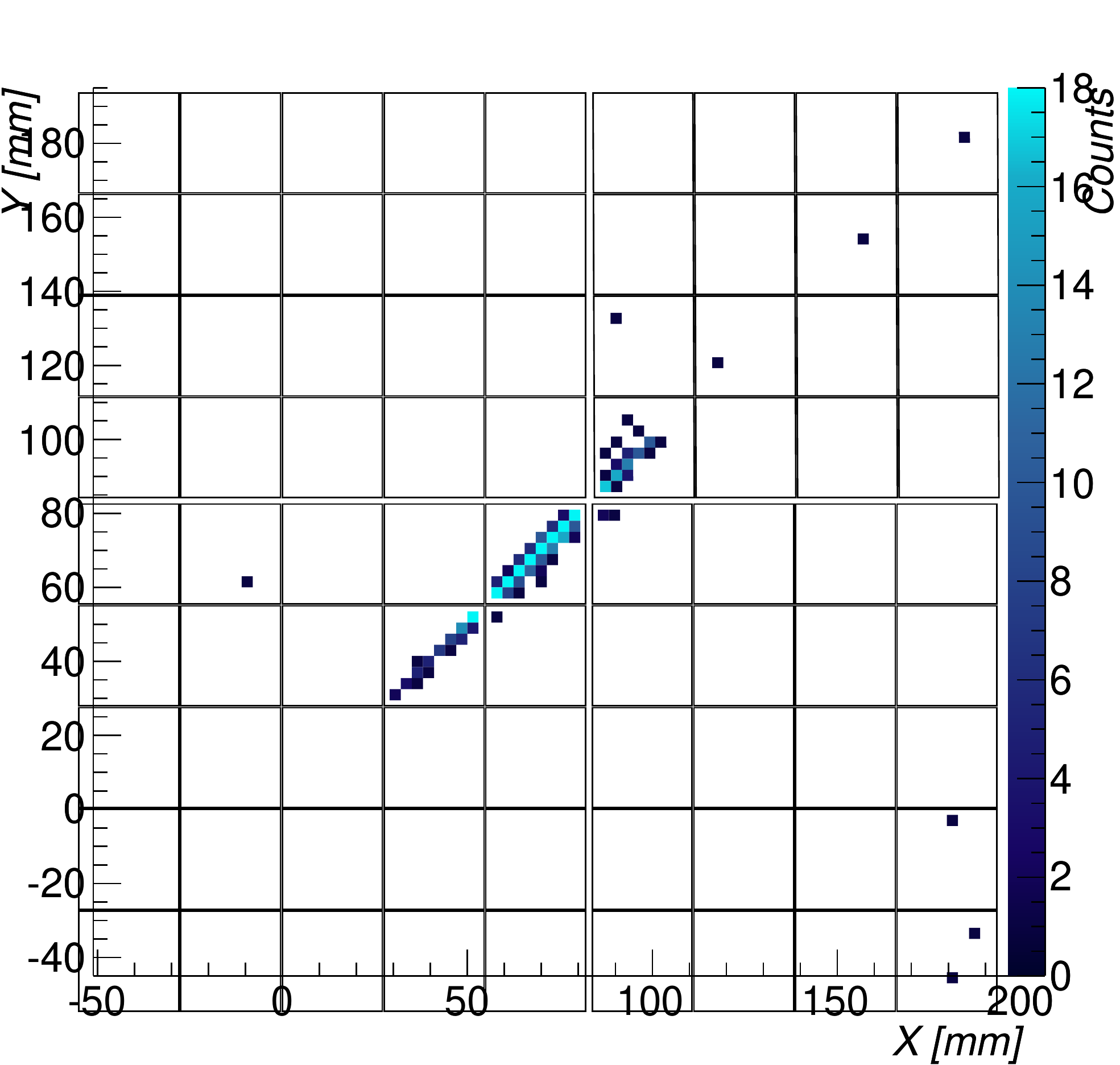}
\caption{Neutrino initiated EAS's fluorescence signal track  on the focal surface of the
detector. Height is 5 km}
\end{subfigure}
\begin{subfigure}[b]{0.5\textwidth}
\centering
\includegraphics[width=.6\textwidth]{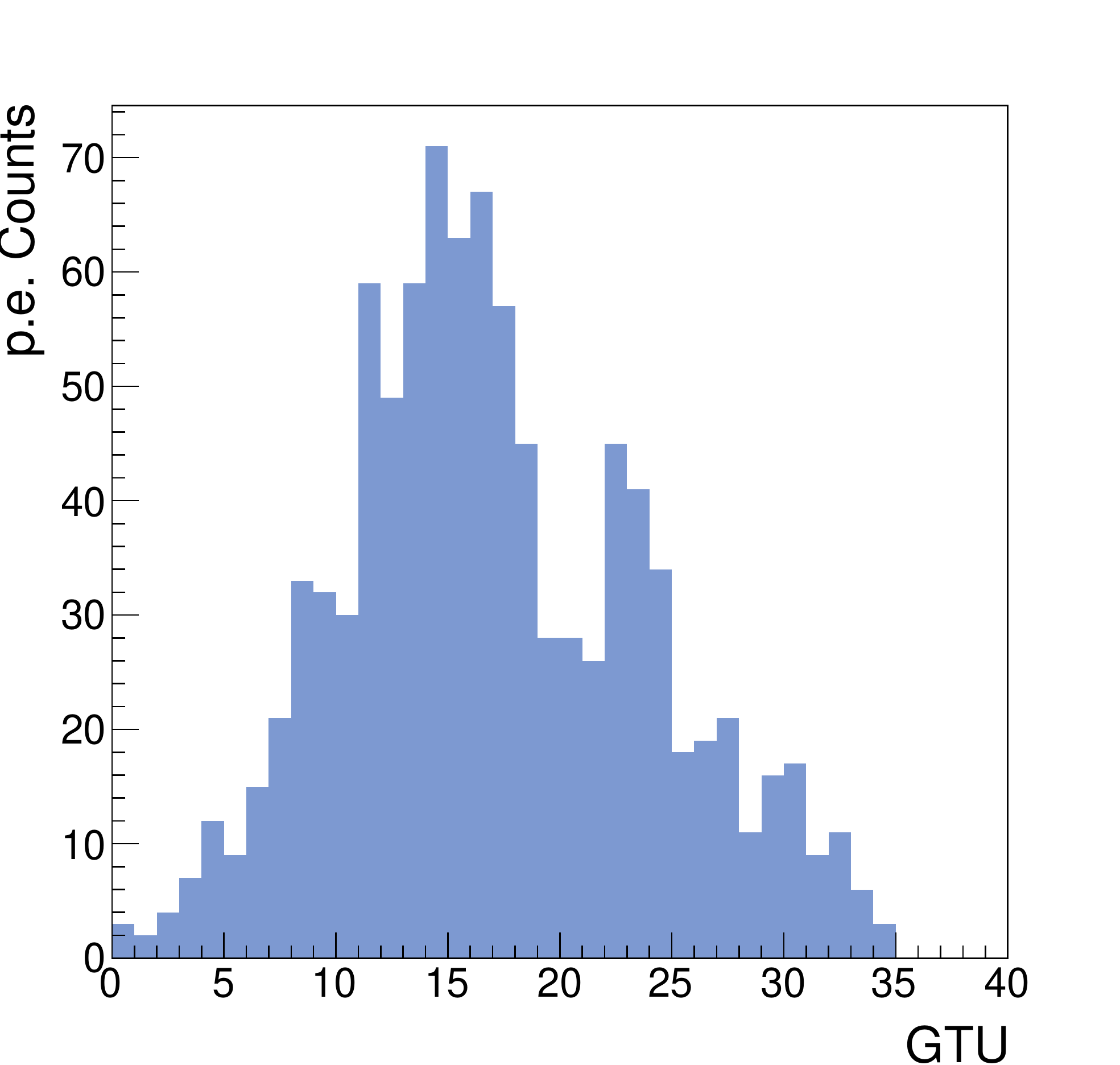}
\caption{Detected signal as a function of time for the  signal track above.}
\end{subfigure}
\caption{ESAF simulated neutrino EAS with a small LPM effect as  seen by  JEM-EUSO. The energy 
of the neutrino is $ 10^{20}$ eV and the injection height is 5 km.}
\label{smallLPM}       
\end{figure}

\begin{figure}
\centering
\includegraphics[width=.42\textwidth]{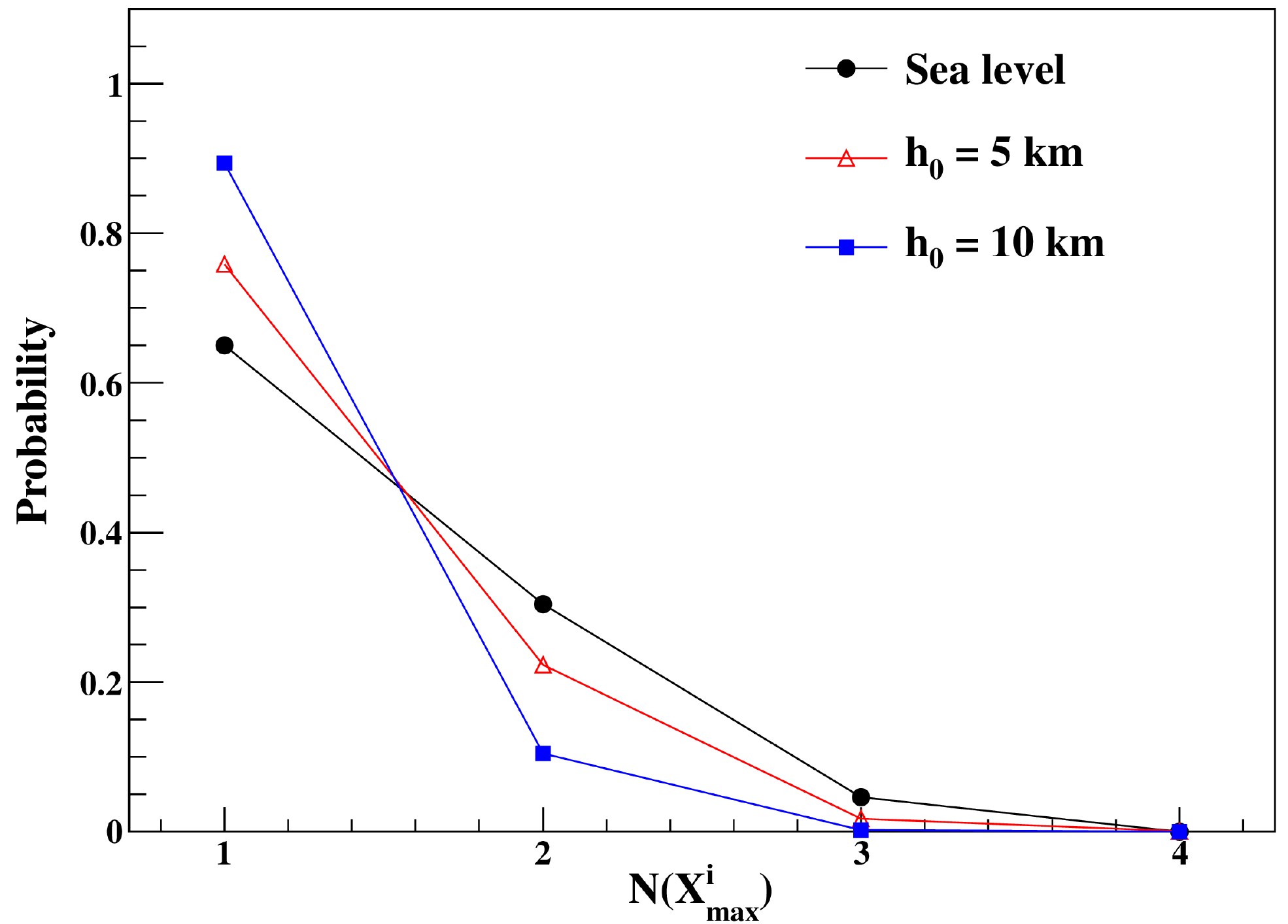}
\caption{Probability to find profiles with $N(X_{max}^i)$ peaks for $\nu_e$ showers occurring at sea level, 
5 km and 10 km height.}
\label{npeaks}
\end{figure}

In fig. \ref{strongLPM} we show the expected signal track  without the background, for a 
neutrino shower of $10^{20}eV$. This shower was  injected at 5 km above sea level, but still,
in this case the LPM effect diminishes the maximum of the photon counts. This is more evident
if we compare fig. \ref{strongLPM} with fig. \ref{smallLPM}. In the latter the LPM effect 
is not so drastic, therefore the track seems a bit shorter but brighter. In both cases the
gaps between PMTs introduce artificial peaks in the light curves, whose origin  should not be 
mistaken as multiple peaks coming from the LPM effect. We also  calculated the  probability of 
a neutrino shower having N maxima. This is shown in fig. \ref{npeaks}. As expected from 
fig. \ref{longiprofiles}, at lower altitudes the probability of having a shower with multiple 
peaks increases.

We simulated a set of $\nu_e $ initiated showers at energies between
$10^{19.25}$eV and $10^{20.5}$eV. All showers were horizontal ($\Theta = 90^\circ
$) and  their first interaction point was either at  sea level, 3 km, 5 km or 10 km above sea level. This
last statement, translates to grammages for the first interaction point of $\approx 36500 $ g cm$^{-2}$ 
at sea level, $\approx 26139$ g cm$^{-2}$ for 3 km of altitude, $\approx 20600$ g cm$^{-2}$ for 5 km of altitude, and $\approx 10776$ g cm$^{-2}$ for 
10 km of altitude. 

\begin{figure}[!h]
\centering
\includegraphics[width=.51\textwidth, height=.45\textwidth]{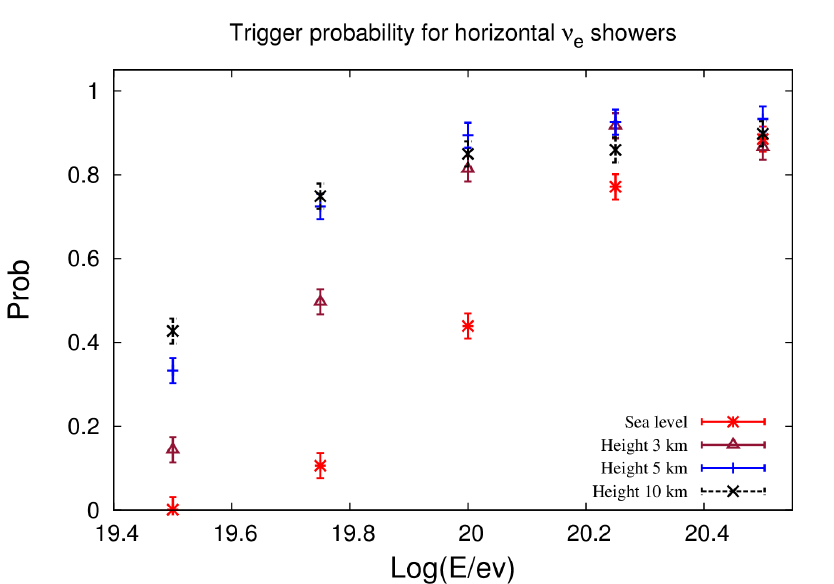}
\caption{Trigger probabilities as a function of energy for horizontal electron neutrino showers.  }
\label{trigprob}
\end{figure}

We proceeded to set the first point of interaction  randomly within  JEM-EUSO's Field
of View (FoV). With the aid of ESAF we can  make a statistical test on how many of the
injected showers actually activated the trigger  mechanisms of  JEM-EUSO. The trigger
probability is defined as   product of the ratio of the injected  neutrino-EAS to  the triggered ones , times the ratio of the observed area  to the injection area.
The trigger probability  is  shown in fig. \ref{trigprob}, for  injection altitudes of zero, five and ten kilometers.

The reduction in the  probability  for showers  injected at sea level is a consequence of a
diversity of factors. The  atmospheric attenuation will be  significantly enhanced since  
most of the atmospheric mass is concentrated at lower altitudes. Secondly the shower is developing 
farther away from the detector. And last but not least, the broadening  of the longitudinal profile 
due to the LPM creates a longer signal track, but as a trade-off suffers from worse signal to noise 
ratio (see fig. \ref{longiprofiles}).

\section{Conclusions}

In the current work  we prepared  neutrino simulations and injected them into the
JEM-EUSO  simulation framework. This allowed us to study  JEM-EUSO's response  to EAS
initiated by UHE neutrinos. For this purpose we combined  well established simulating tools.
For the   EAS simulator we used CONEX, for the neutrino interaction we used PYTHIA and for the 
rest of the chain ESAF. With these tools at hand we studied the expected trigger probability 
for deeply interacting neutrino showers at sea level, 3 km, 5 km and 10 km altitudes, as case studies.  
Though not definitive, this types of test  are  part of the first  full end-to-end simulation studies carried out to calculate  the expected number of UHE neutrino events under the current JEM-EUSO instrument's  configuration.

\section{Acknowledgments}

The present work has been conducted under support of Deutsches Zentrum f\" Luft und Raumfahrt
(DLR). Moreover it has partly been funded by the European Space Agency (ESA) Topical Team
Activities Fond.
We wish to thank RICC, the RIKEN Integrated Cluster of Cluster in Tokyo, Japan for the
allocation of computing resources and service.

\clearpage
\newpage


%% file: icrc2013-0631.tex




\title{Sensitivity of orbiting JEM-EUSO  to large-scale cosmic-ray anisotropies}

\shorttitle{JEM-EUSO anisotropy study}

\authors{
Thomas J. Weiler$^{1}$,
Peter B. Denton$^{1}$,
Luis A. Anchordoqui$^{2}$,
Andreas A. Berlind$^{1}$,
and 
Matthew Richardson$^{1}$,
for the JEM-EUSO Collaboration.
}

\afiliations{
$^1$Department of Physics \& Astronomy, Vanderbilt University, Nashville, TN, 37235 USA \\
$^2$ Physics Department, University of Wisconsin-Milwaukee, Milwaukee,  WI, 53201 USA 
}

\email{tom.weiler@vanderbilt.edu}

\abstract{
The two main advantages of space-based observation of extreme-energy ($\gsim 10^{19}$~eV) 
cosmic-rays (EECRs) over ground-based observatories 
are the increased field of view, and the all-sky coverage with nearly uniform systematics 
of an orbiting observatory.  The former guarantees increased statistics,
whereas the latter enables a partitioning of the sky into spherical harmonics.  
We have begun an investigation, using the spherical harmonic technique, 
of the reach of \J\  into potential anisotropies in the extreme-energy cosmic-ray sky-map.  
The technique is explained here, and first results are presented.
The discovery of anisotropies would help to identify the long-sought origin of EECRs.
}
%


\newcommand{\J}{JEM-EUSO}
\newcommand{\A}{PAO}

\newcommand{\beq}[1]{\begin{equation}\label{#1}}
\newcommand{\eeq}{\end{equation}}
\newcommand{\bea}[1]{\begin{eqnarray} \label{#1}}
\newcommand{\eea}{\end{eqnarray}}
\newcommand{\ba}{\begin{array}}
\newcommand{\ea}{\end{array}}
\newcommand{\nn}{\nonumber}
\newcommand{\rf}[1]{(\ref{#1})}
\newcommand{\rarr}{\rightarrow}
\newcommand{\be}{\begin{equation}}
\newcommand{\ee}{\end{equation}}

\newcommand{\lsim}{\stackrel{<}{\sim}}
\newcommand{\gsim}{\stackrel{>}{\sim}}

\maketitle

\section{Introduction}

The Extreme Universe Space Observatory (EUSO) is a consortium of 250 Ph.D. researchers from 25 institutions, 
spanning 13 countries.
It is a down-looking telescope optimized for near-ultraviolet fluorescence produced by extended air showers in the atmosphere of the Earth.
EUSO is proposed to occupy the Japanese Experiment Module (JEM) on the International Space Station (ISS),
and collect up to 1000 cosmic ray (CR) events at and above 55~EeV ($1\,{\rm EeV}=10^{18}$~eV) 
over a $~5$~year lifetime, far surpassing the reach of any ground-based project.  

\J\ brings two new, major advantages to the search for the origins of EECRs.
One advantage is the large field of view (FOV), attainable only with a space-based observatory.
With a $60^\circ$ opening angle for the telescope, the down-pointing (``nadir'') FOV is 
\vspace*{-0.3cm}
\beq{nadirFOV}
\pi(h_{\rm ISS}\tan(30^\circ))^2\approx h_{\rm ISS}^2\approx 150,000\,{\rm km}^2\,.
\eeq
%
Tilting the telescope turns the circular FOV given in Eq.~\rf{nadirFOV}
into a larger elliptical FOV.  The price paid for ``tilt mode'' is an increase in the threshold energy of the experiment.

The second advantage is the coverage of the full sky ($4\pi$~steriadians)
with nearly constant systematic errors on the energy and angle resolution, 
again attainable only with a space-based observatory.
(Combined data from ground-based observatories in the Northern and Southern hemispheres may offer full-sky coverage,
but not uniformity of systematics.) 
This poster pursues all-sky studies of possible spatial anisotropies.
The reach benefits from the $4\pi$~sky coverage, 
but also from the increased statistics resulting from the greater FOV.
A longer study will soon be completed and published~\cite{DABRW}.

In addition to the two advantages of space-based observation just listed, 
a third feature provided by a space-based mission may turn out to be significant.  
It is the increased acceptance for Earth-skimming neutrinos when the skimming 
chord 
transits ocean rather than land.
On this latter topic, just one study has been published~\cite{PalomaresRuiz:2005xw}.
The study concludes that an order of magnitude larger acceptance
results for Earth-skimming events transiting ocean compared to transiting land.
Ground-based observatories will not realize this benefit, 
since they  cannot view ocean chords.

\section{All-sky coverage and anisotropy}

As emphasized by Sommers over a dozen years ago~\cite{Sommers:2000us}, 
an all-sky survey offers a rigorous expansion in spherical harmonics,
of the normalized spatial event distribution
$I(\Omega)$, where $\Omega$ denotes the pair of latitude ($\theta$) and longitude ($\phi$) angles:
\vspace*{-0.3cm}
\beq{SphHarm}
I(\Omega)\equiv \frac{N(\Omega)}{\int d\Omega\, N(\Omega)} =\sum_{\ell=0}^\infty \, \sum_{|m| \le l} a_{\ell m}\,Y_{\ell m}(\Omega)\,,
\eeq
i.e., the set $\{ Y_{\ell m} \}$ is complete.~\footnote
{Averaging the $a^2_{\ell m}$ over the $(2\ell+1)$ values of $m$ defines the rotationally-invariant ``power spectrum'' in the 
single variable $\ell$, 
%
$C(\ell)= \frac{1}{2\ell+1}\sum_{|m|\le\ell}\,a^2_{\ell m}$.
%
We do not use this formalism here.
}
We are interested in the real valued $Y_{\ell m}$'s, defined as 
$P^\ell_m(x) (\sqrt{2}\cos(m\phi))$ for positive $m$, 
$P^\ell_{|m|}(x) (\sqrt{2}\sin(|m|\phi))$ for negative $m$, 
and $P_\ell(x)$ for $m=0$.
Here, $P^\ell_m$ is the associated Legendre polynomial, 
$P_\ell=P^\ell_{m=0}$ is the regular Legendre polynomial,
and $x\equiv \cos\theta$.

%
%
The lowest multipole is the $\ell=0$ monopole, 
equal to the all-sky flux.
The higher multipoles ($\ell\ge 1$) and their amplitudes $a_{\ell m}$ correspond to anisotropies.
Guaranteed by the orthogonality of the $Y_{\ell m}$'s,
the higher multipoles when integrated over the whole sky equate to zero.

\begin{figure}[t]
  \centering
  \includegraphics[height=0.04\textheight,width=0.15\textwidth]{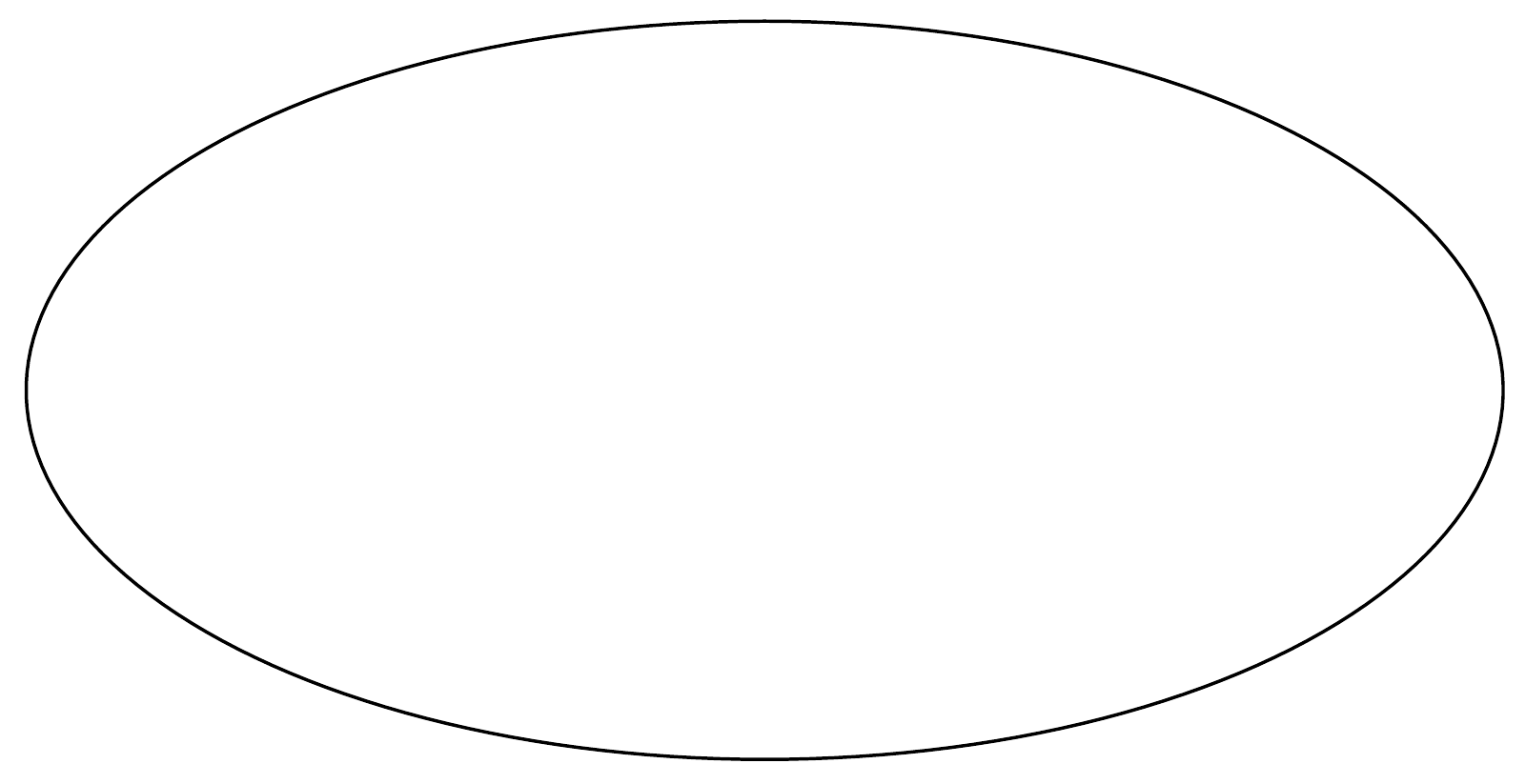}
  \includegraphics[height=0.04\textheight,width=0.15\textwidth]{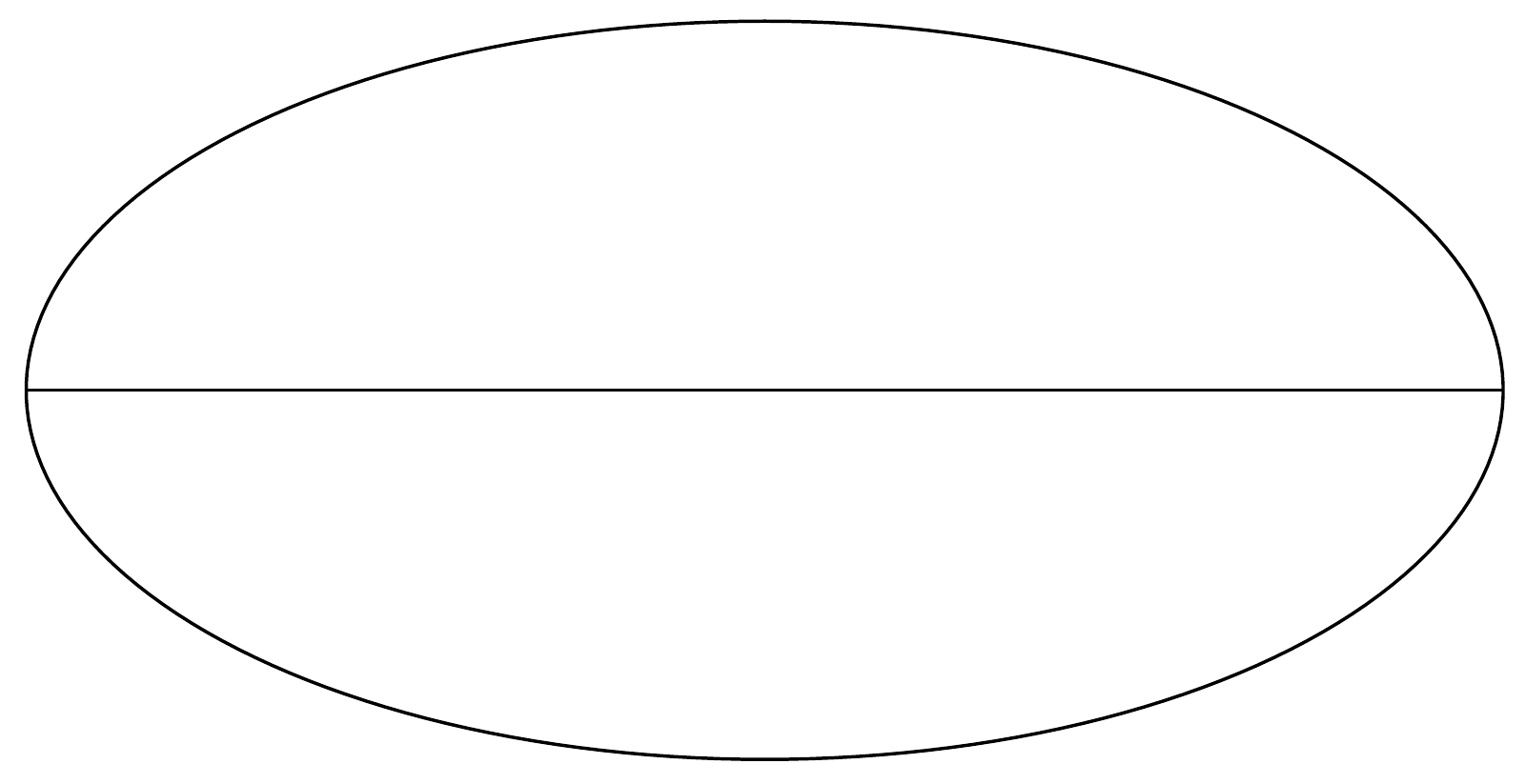}
  \includegraphics[height=0.04\textheight,width=0.15\textwidth]{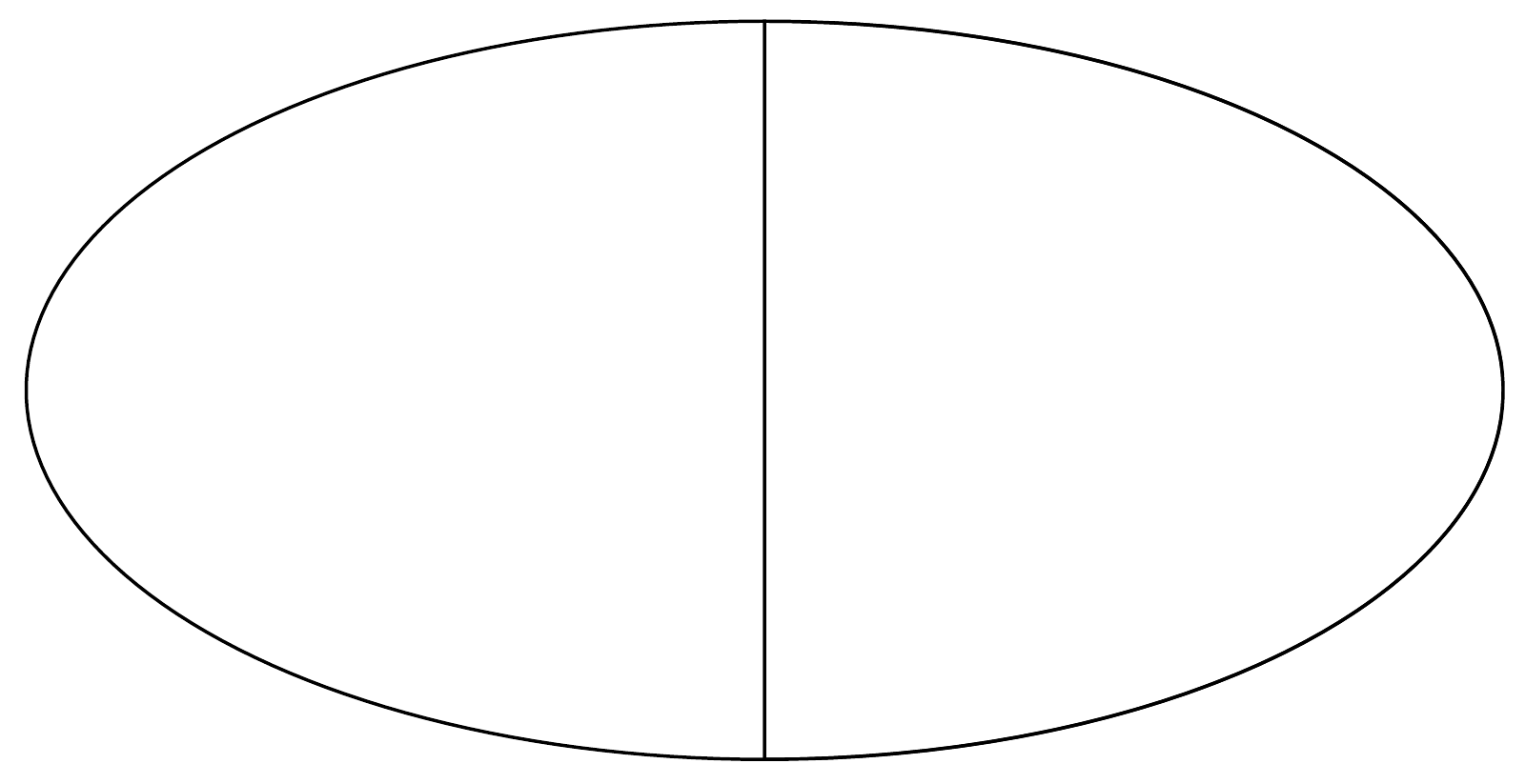}
  \caption{Nodal lines separating surplus and deficit regions of sky,
  for (left) $\ell=0,\ m=0$ monopole, and $\ell=1,\ m=0$ (middle) and $m=1$ (right) dipoles. }
  \label{fig1}
 \end{figure}

\begin{figure}[t]
  \centering
  \includegraphics[height=0.04\textheight,width=0.15\textwidth]{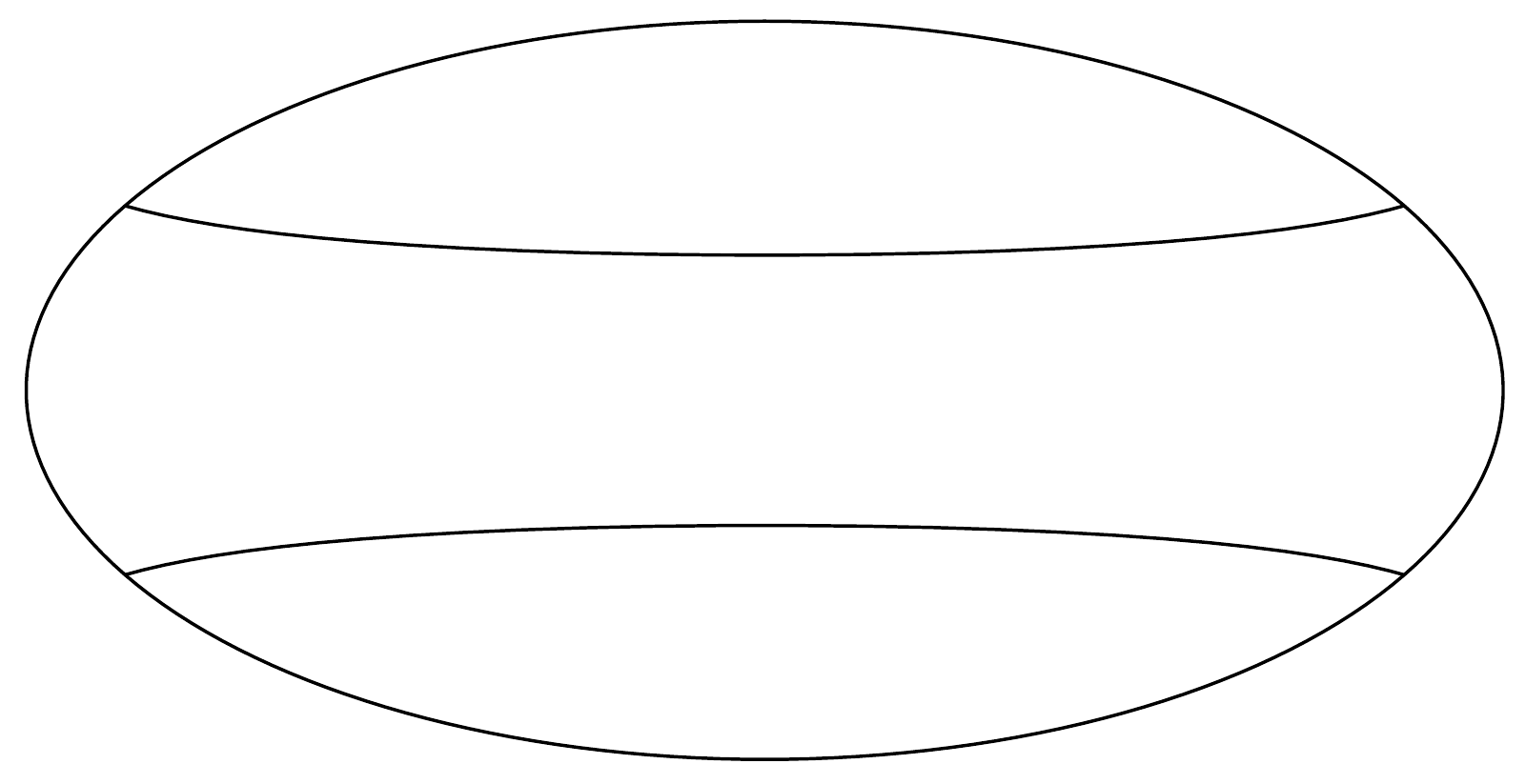}
    \includegraphics[height=0.04\textheight,width=0.15\textwidth]{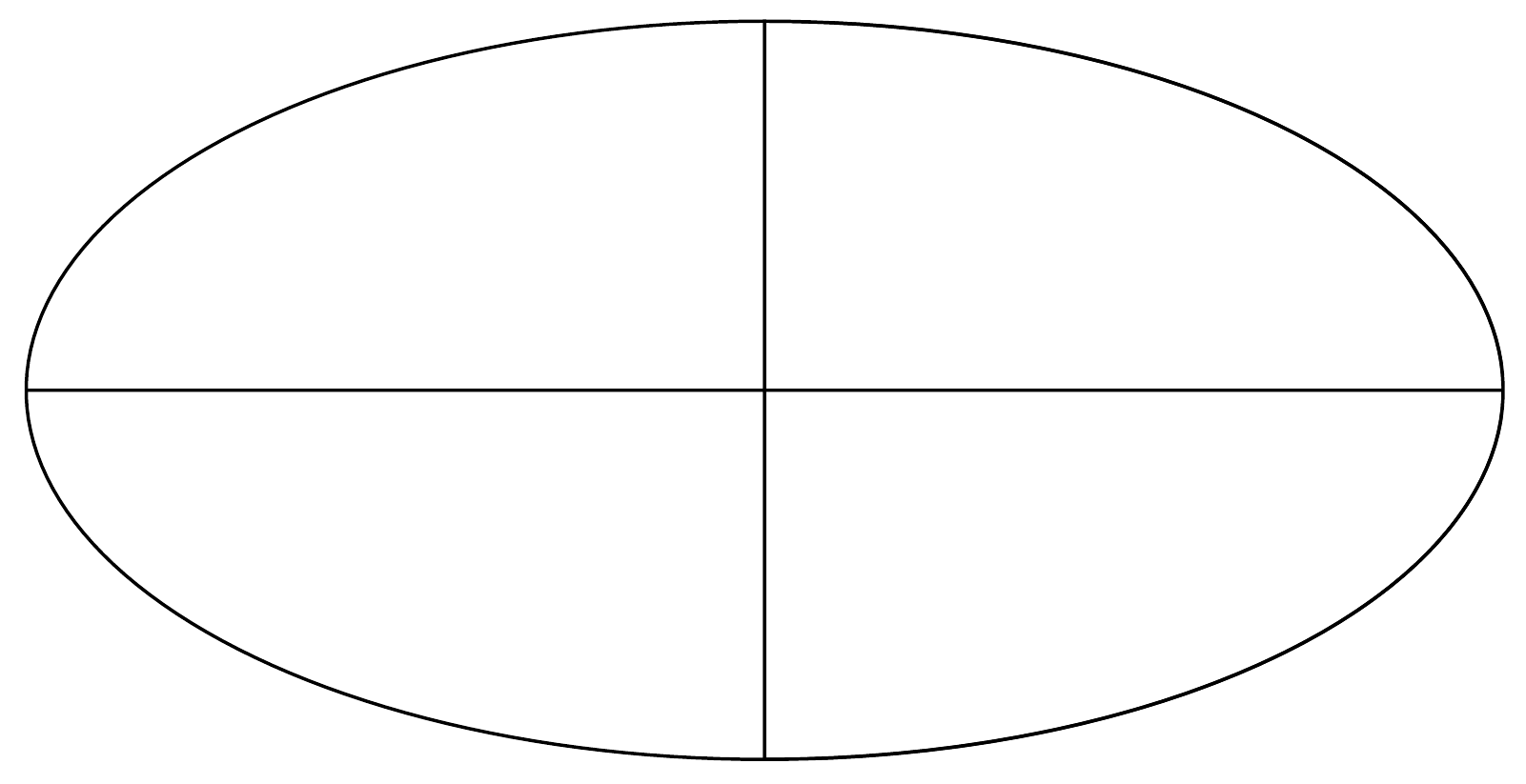}
      \includegraphics[height=0.04\textheight,width=0.15\textwidth]{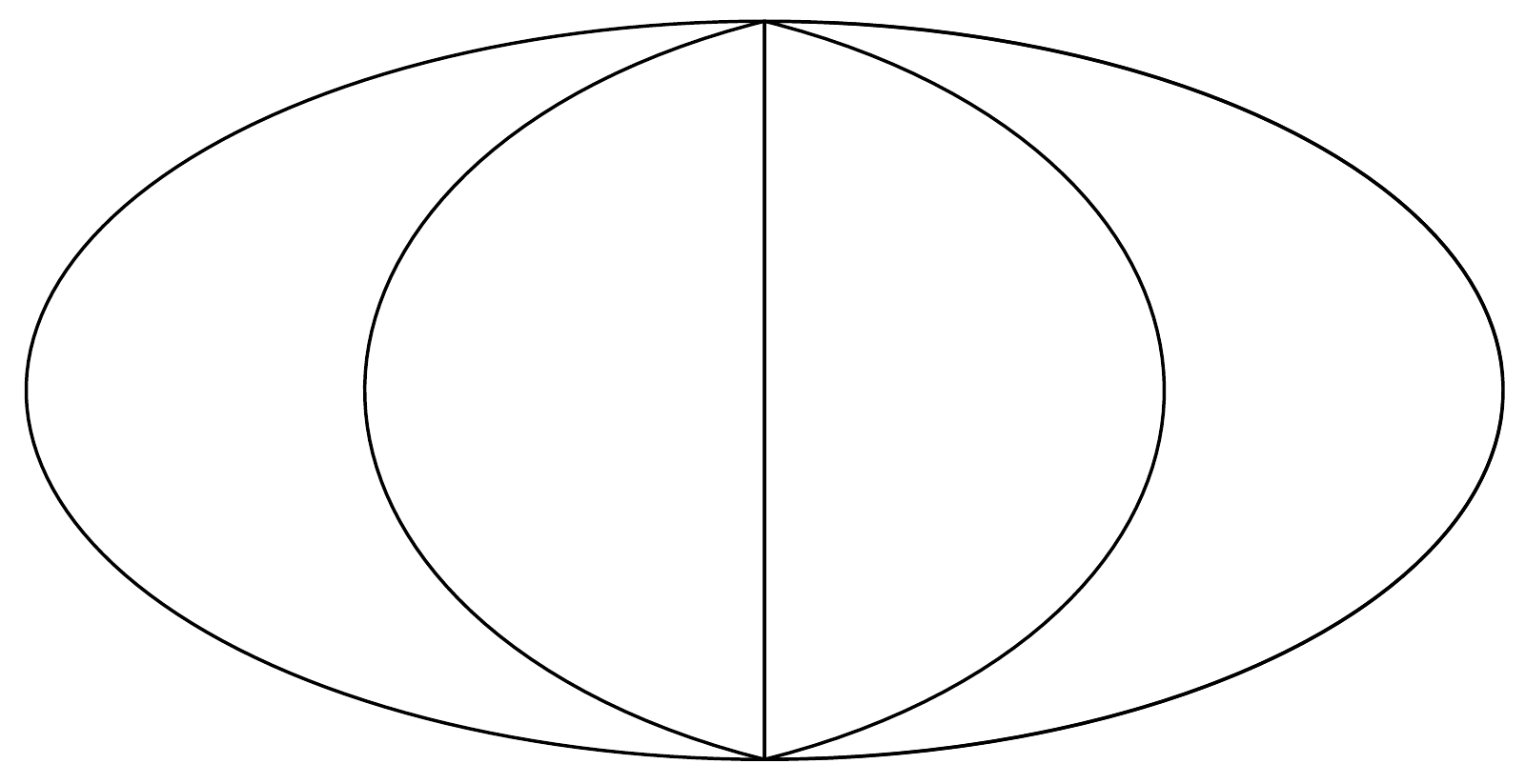}
  \caption{Nodal lines separating surplus/deficit regions of sky, 
  for $\ell=2,\ m=0,\ 1,\ 2$ quadrupoles, respectively. }
  \label{fig2}
 \end{figure}

\begin{figure}[t]
  \centering
  \includegraphics[width=0.11\textwidth]{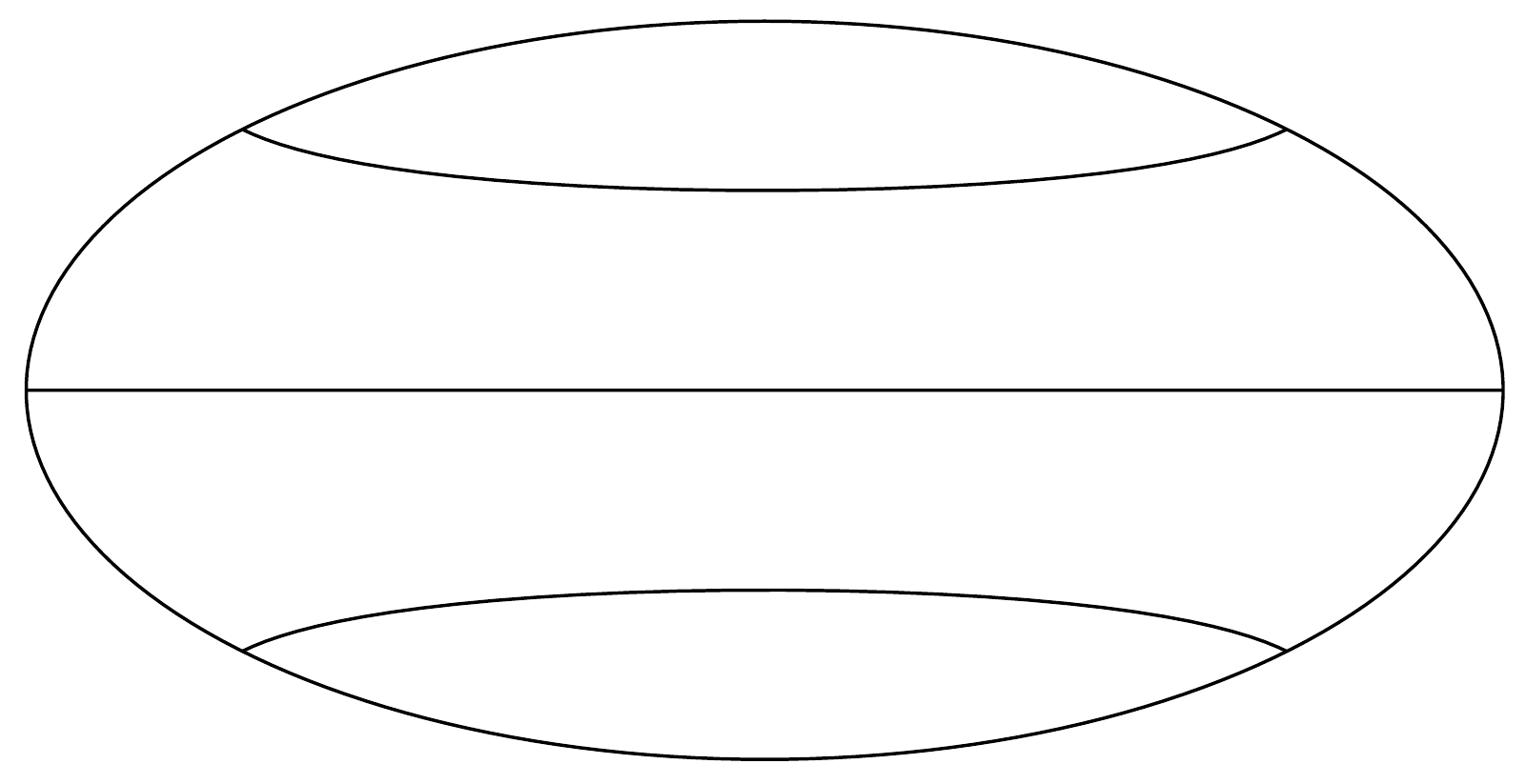}
    \includegraphics[width=0.11\textwidth]{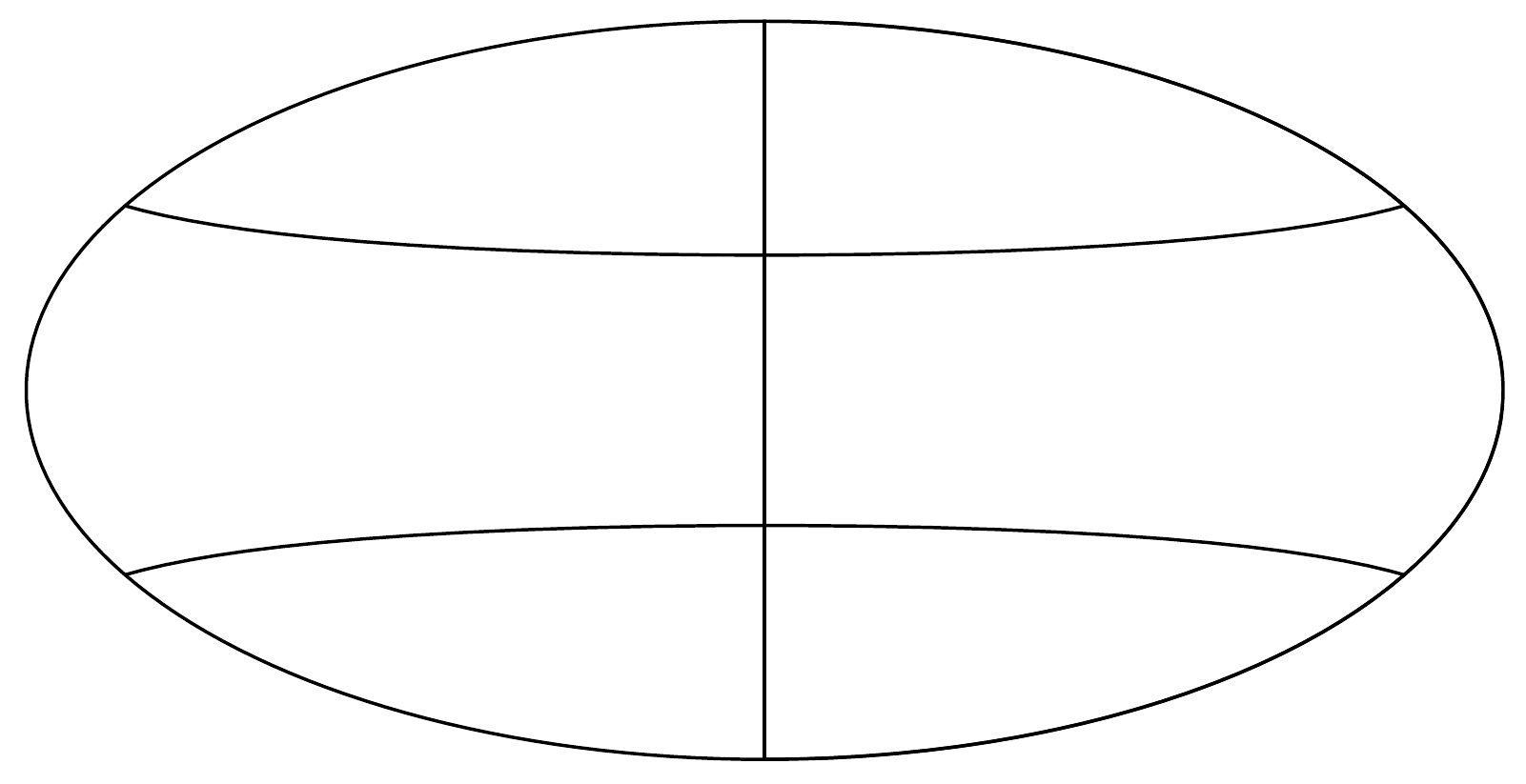} 
     \includegraphics[width=0.11\textwidth]{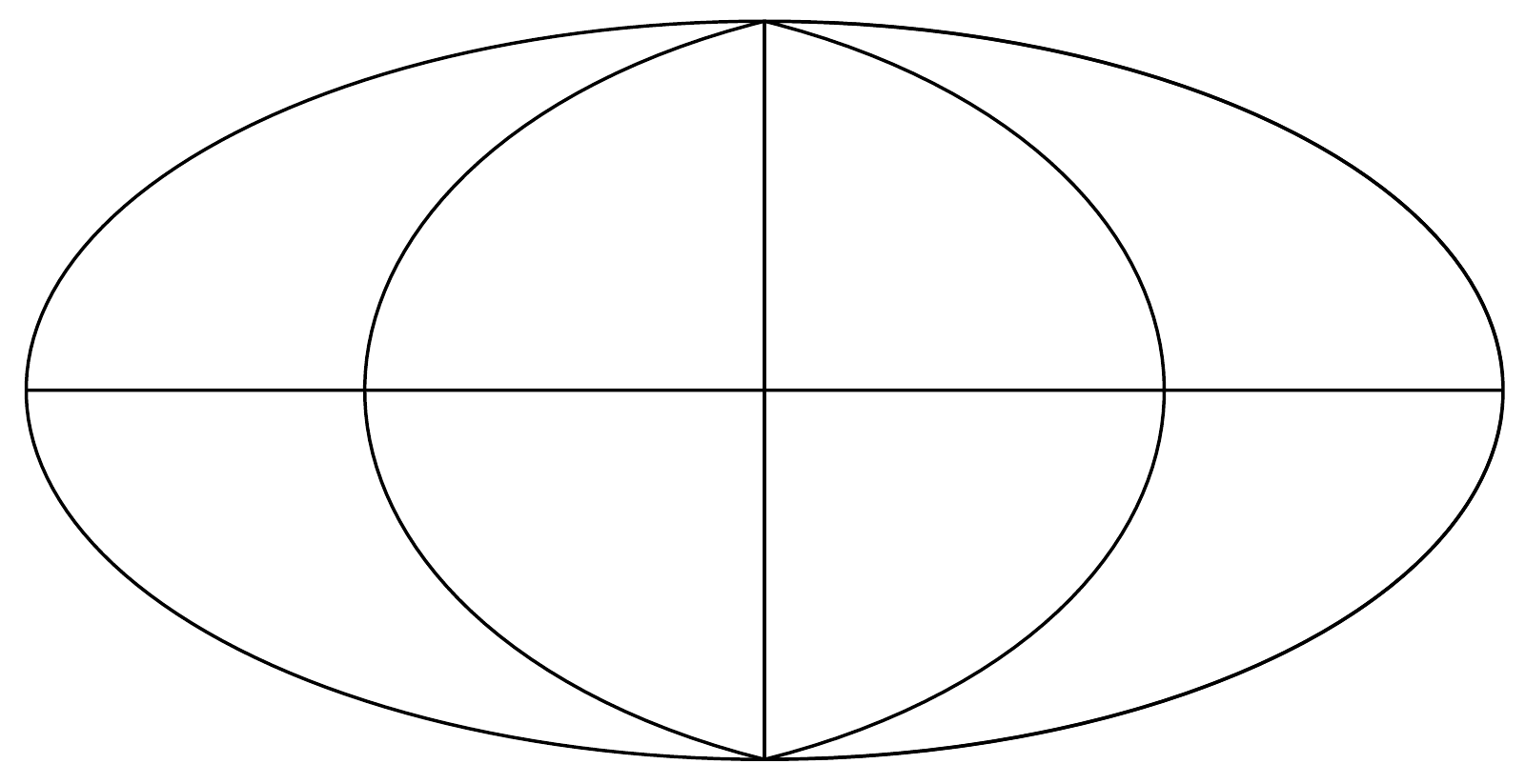}
       \includegraphics[width=0.11\textwidth]{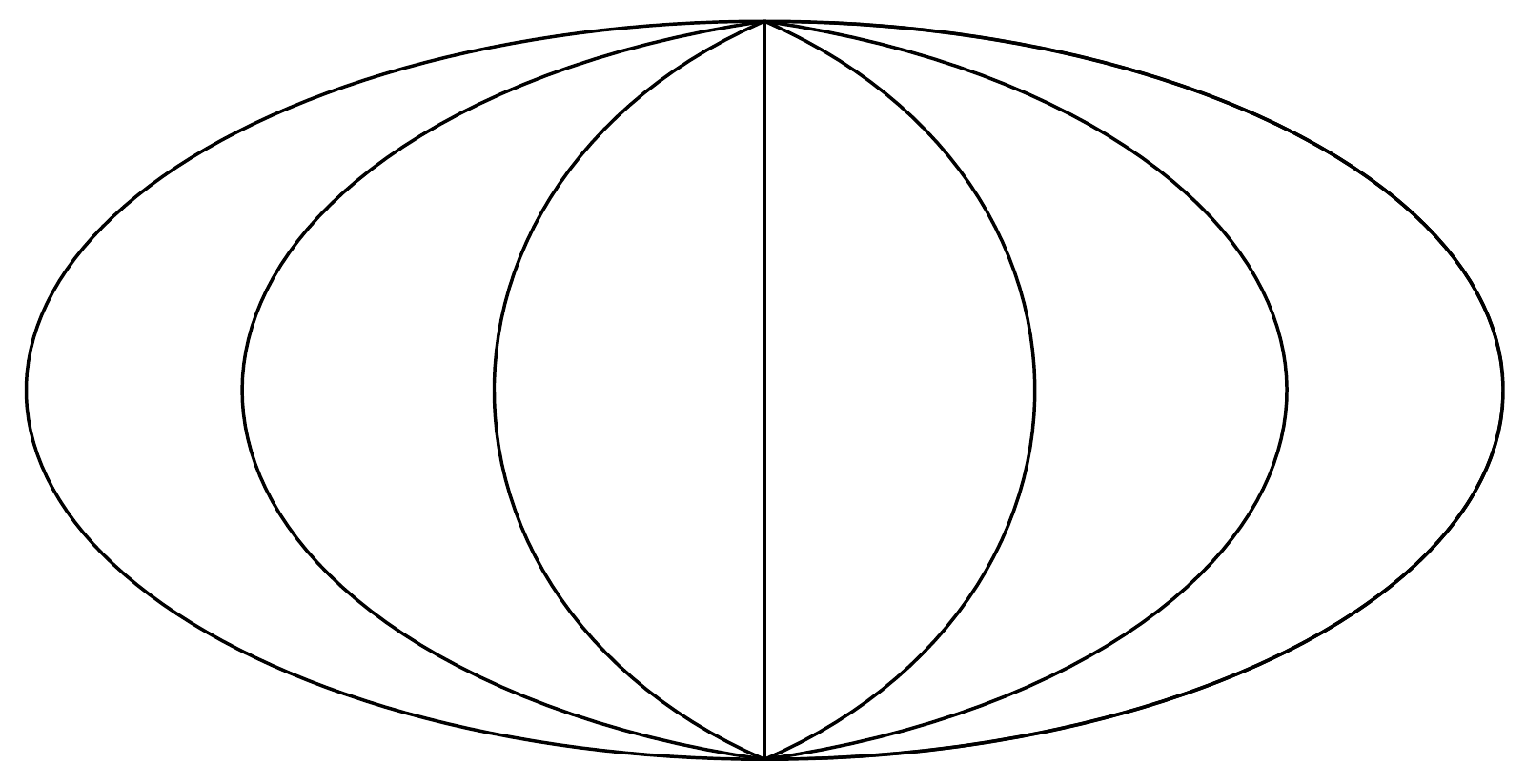}
  \caption{Nodal lines separating surplus/deficit regions of sky, 
  for $\ell=3,\ m=0,\ 1,\ 2,\ 3$, respectively.}
  \label{fig3}
 \end{figure}

A nonzero $m$ corresponds to $2\,|m|$ longitudinal ``slices'' ($|m|$ nodal meridians).
A nonzero $\ell$ and $m$ corresponds to $\ell+1-|m|$ latitudinal ``zones'' ($\ell-|m|$ nodal latitudes).
In Figs.~(\ref{fig1}-\ref{fig3})
we show the partitioning described by some low-multipole moments.
The configurations with $(\ell,-|m|)$ are related to those with $(\ell,+|m|)$ by a longitudinal phase advance
$\phi\rarr\phi+\frac{\pi}{2}$, or $\cos\phi\rarr\sin\phi$.

 
\section{Previous anisotropy searches}
The first full-sky large anisotropy search was  based on the combined Northern and Southern data from the SUGAR and AGASA experiments taken during a 10~yr period.
 Nearly uniform exposure to the entire sky resulted.  
 No significant deviation from isotropy was seen, even at energies beyond 
 $4 \times 10^{19}$~eV~\cite{Anchordoqui:2003bx}. 
 More recently, the Pierre Auger Collaboration carried out various searches for large scale anisotropies in the distribution of arrival directions of cosmic rays above $10^{18}$ eV~\cite{Abreu:2011ve,Auger:2012an}. 
 The latest study was performed as a function of both declination and right ascension in several energy ranges 
 above $10^{18}$ eV, and reported in terms of dipole and quadrupole amplitudes. 
 Again no significant deviation from isotropy was revealed. 
 Assuming that any cosmic ray anisotropy is dominated by dipole and quadrupole moments in this energy range, the Pierre Auger Collaboration derived upper limits on their amplitudes. 
 Such upper limits challenge an origin of cosmic rays above $10^{18}$ eV from non-transient galactic sources densely distributed in the galactic disk~\cite{Abreu:2012lva}.  
At the energies exceeding $6 \times 10^{19}$~eV, however, hints for a dipole anisotropy 
may be emerging~\cite{Anchordoqui:2011ks}.

It must be emphasized that because previous data were so sparse at energies which will be accessible to \J, 
upper limits on anisotropy were necessarily restricted to energies below the threshold of \J.
 \J\ expects many more events at $\sim 10^{20}$~eV, allowing an enhanced anisotropy reach.
In addition, \J\ events will have a higher rigidity ${\cal R}=E/Z$, and so will be less bent by magnetic fields;
this may be helpful in identifying particular sources on the sky.


 \begin{figure*}[t]
  \centering
   \includegraphics[width=0.26\textwidth]{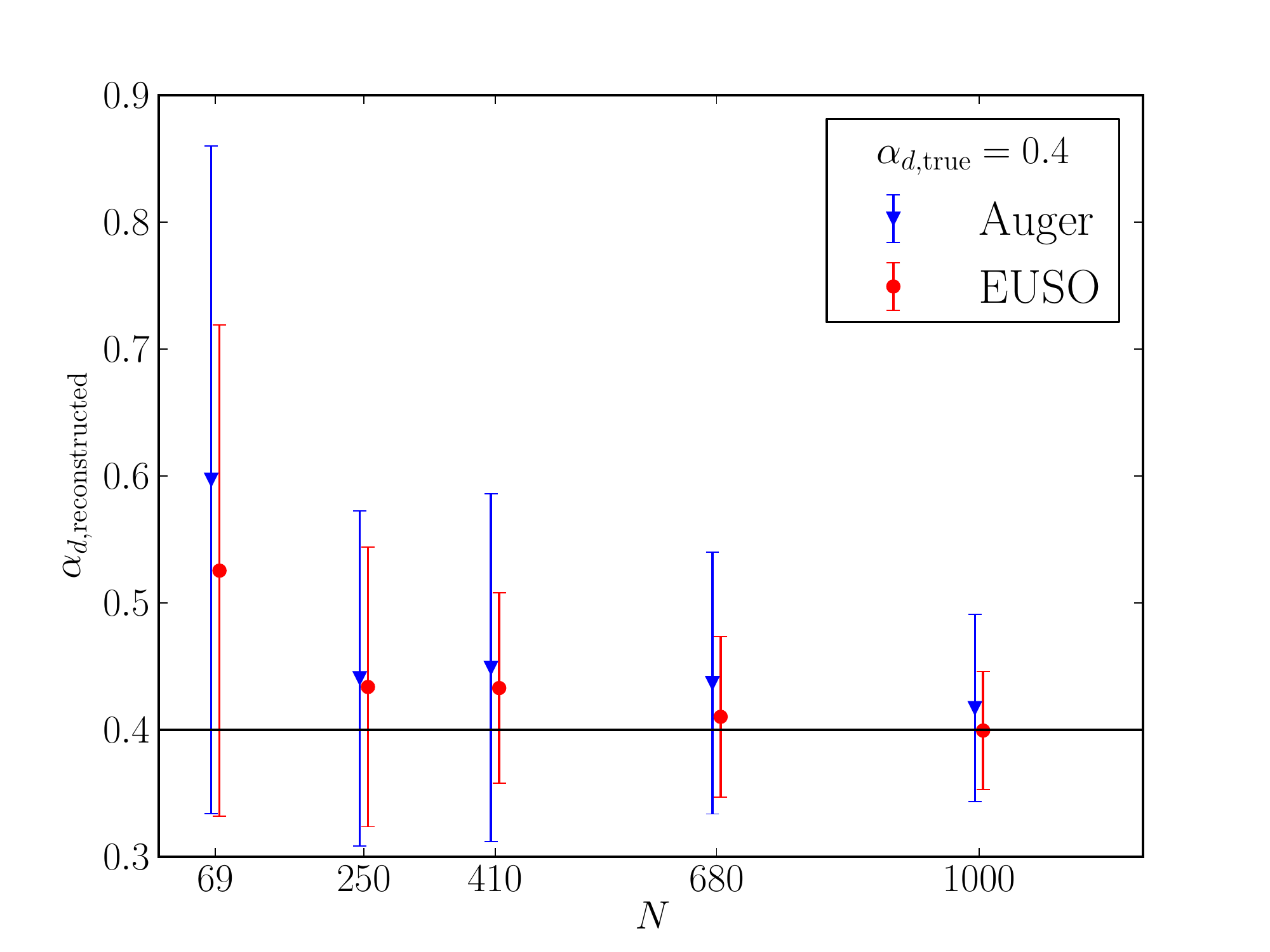}
  \includegraphics[width=0.23\textwidth]{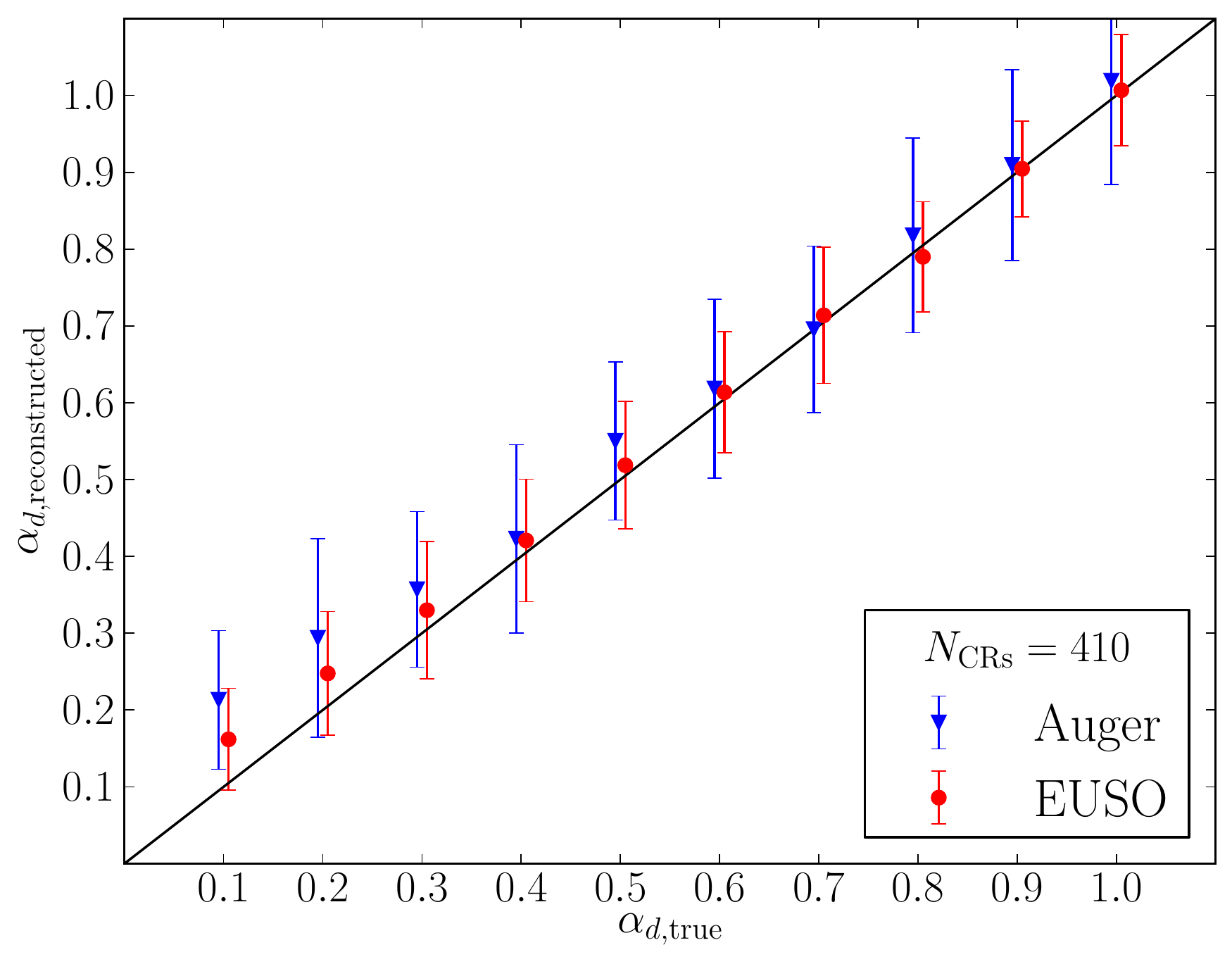}
    \includegraphics[width=0.23\textwidth]{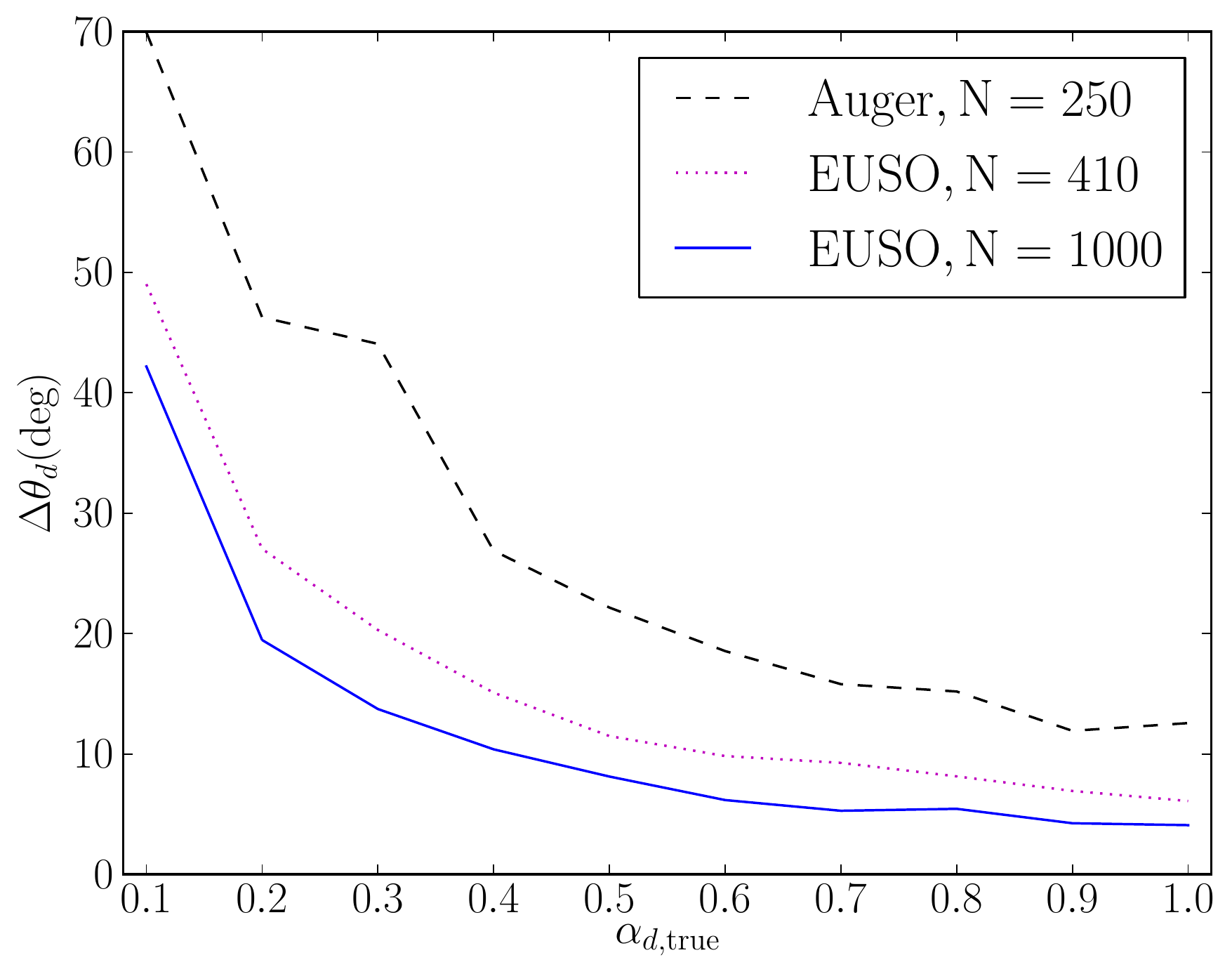}
   \includegraphics[width=0.26\textwidth]{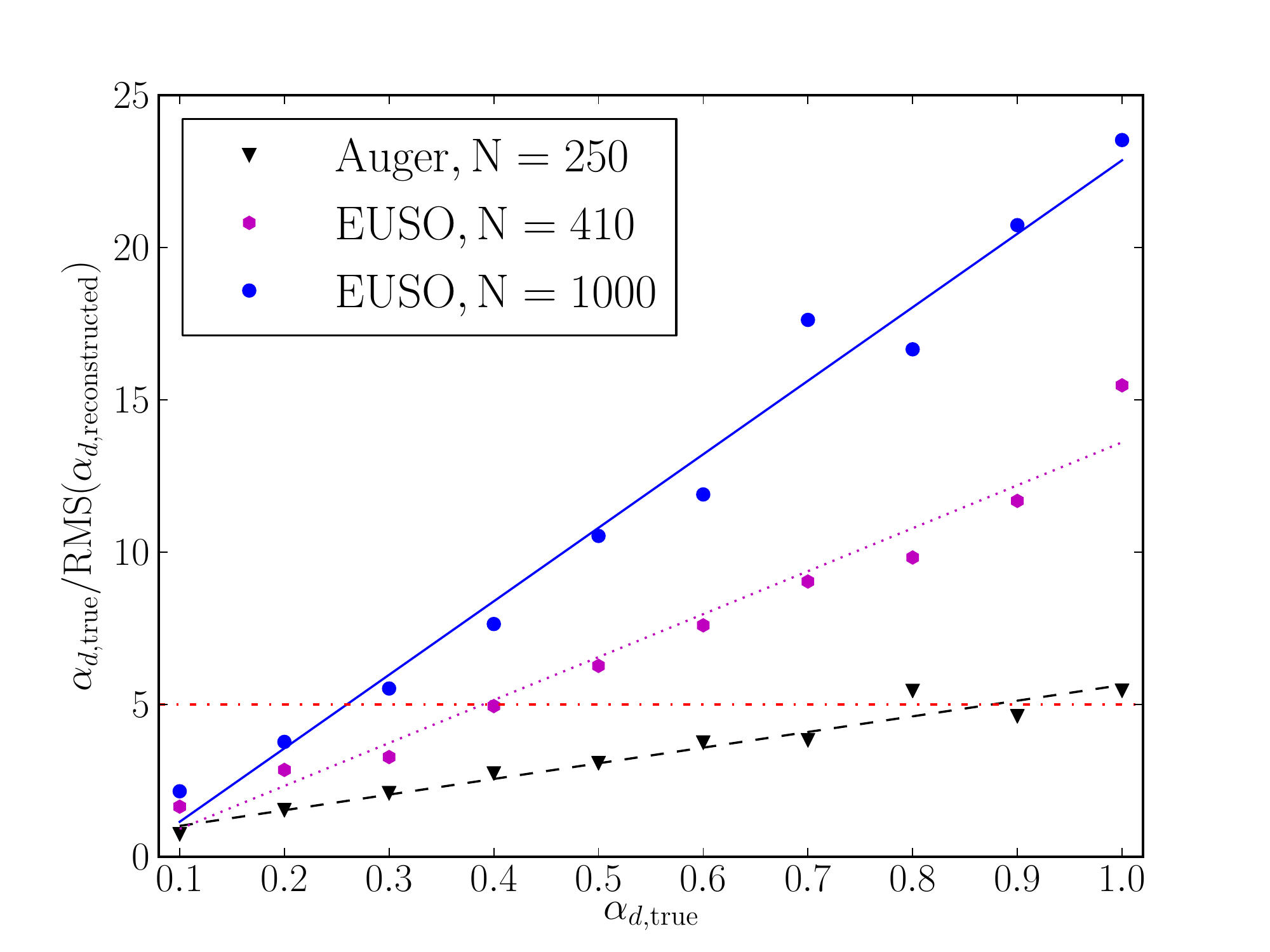}
  \caption{Reconstruction of  the dipole amplitude (left panels) and angle (third panel), for \A\ and \J.
  Discovery reach (right panel) of \J\ and \A, with 5-$\sigma$ horizontal line.} 
  \label{fig4}
 \end{figure*}
 
%
%

\begin{figure*}[t]
  \centering
   \includegraphics[width=0.26\textwidth]{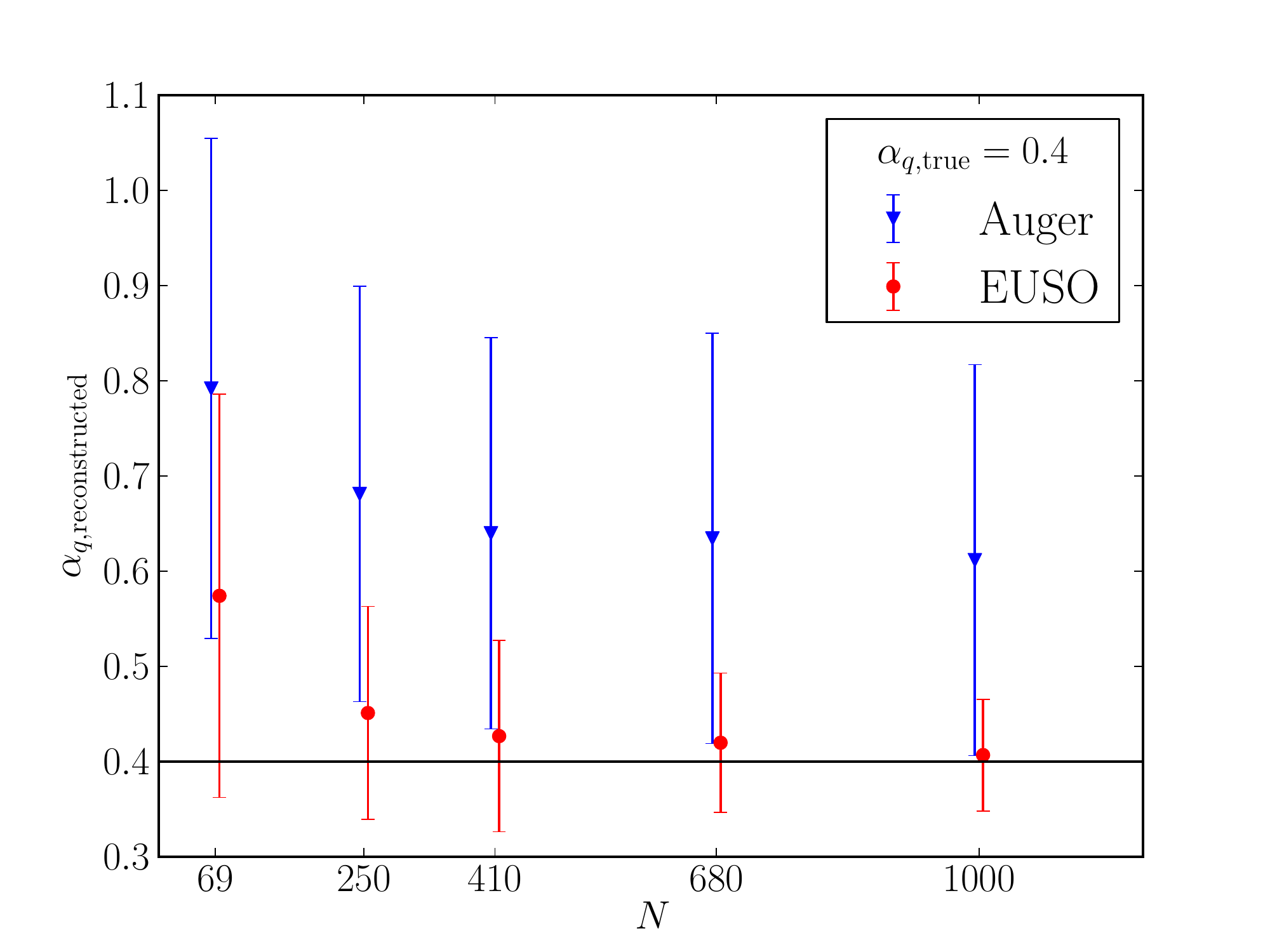}
  \includegraphics[width=0.23\textwidth]{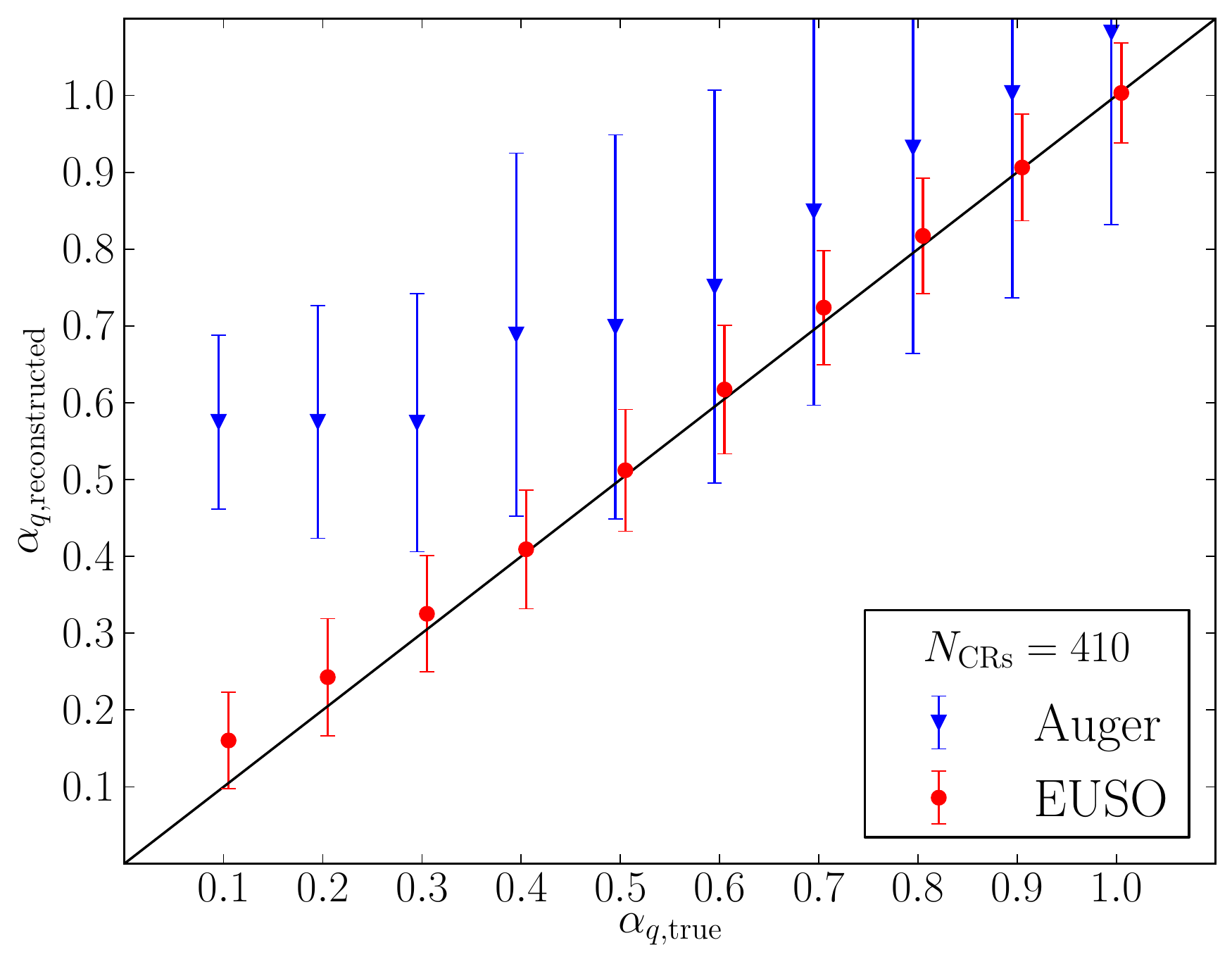}
      \includegraphics[width=0.23\textwidth]{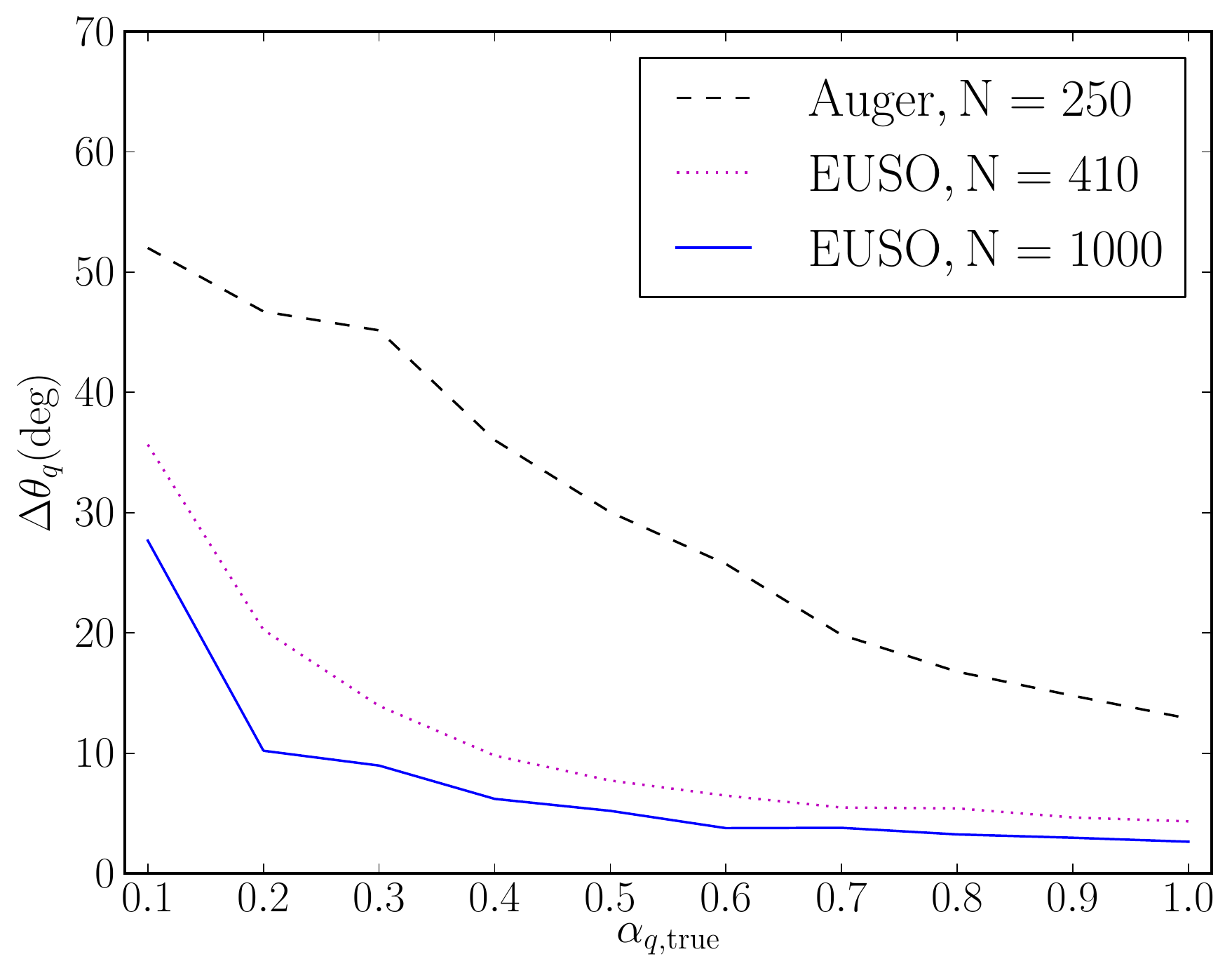}
   \includegraphics[width=0.26\textwidth]{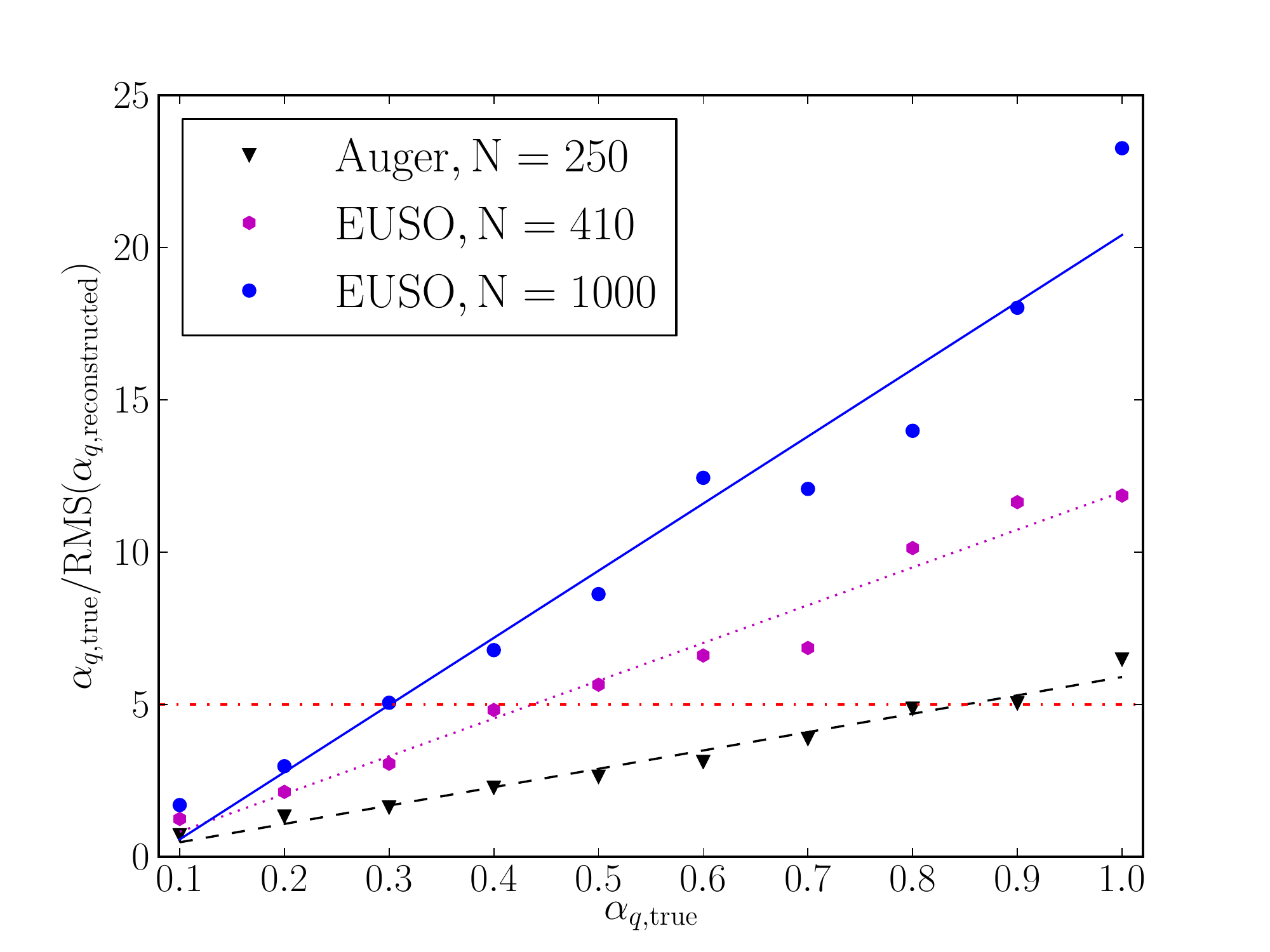}
  \caption{Reconstruction of  the $Y_{20}$ quadrupole amplitude (left panels) and angle (third panel), for \A\ and \J.
  Discovery reach (right panel) of \J\ and \A, with 5-$\sigma$ horizontal line.}
  \label{fig5}
 \end{figure*}
 
%
 %

\section{Comparison of all-sky \J\ to half-sky \A}

The ground-based  the Pierre Auger Observatory (\A) is an excellent, ground-breaking experiment.
However, in the natural progression of science, 
\A\ will be superseded by space-based observatories.
\J\ is designed to be first of its class, building on the successes of \A.

The two main advantages of \J\ over \A\ 
are the (i) greater FOV leading to a greater exposure at EE,
and the (ii) all-sky nature of the orbiting, space-based observatory.  
We briefly explore the advantage of the enhanced exposure first.
We consider data samples of 69, 250, 410, 680, and 1000 events.
The 69 events sample is that presently published by \A\ for events accumulated over three years at and above 55EeV.
The annual rate of such events at \A\ is $\sim 69/3=23$.
Thus, the 250 event sample is what \A\ could attain in ten years of running.

\A\ has a FOV of 3,000 km$^2$.
The \J\ FOV, given in Eq.~(\ref{nadirFOV}), is 50 times larger for instantaneous measurements (e.g., for observing transient sources).
Multiplying the \J\ event rate by an expected 18\% duty cycle, we arrive at a time-averaged nine-fold increase in acceptance  for \J\ 
compared to \A, at energies where the \J\ efficiency has peaked (at and above $~\sim 100$~EeV).  
Including the \J\ efficiency down to 55~EeV reduces the factor of 9 to a factor ~6 at and above 55~EeV. 
We arrive at the 410 event sample as the \J\ expectation at and above 55~EeV after three years running in nadir mode
(or, as is under discussion, in tilt mode with a reduced aperture/PDM count).
A 680 event sample is then expected for five years of \J\ in a combination of nadir and tilt mode.
Finally, the event rate at an energy measured by High Resolution Fly's Eye (HiRes) is known to exceed that of \A\ by~50\%.
This leads to a five-year event rate at \J\ of about 1000 events.  

Now we turn to the $4\pi$~advantage.
Commonly, a major component of the anisotropy is defined via a max/min directional asymmetry,
 %
 $\alpha\equiv \frac{I_{\max} - I_{\min}}{I_{\max} + I_{\min}}$.
 %
 For a monopole plus dipole distribution $1+\alpha_D\cos\theta$, one
 readily finds that $\alpha_D=\alpha$.
For a monopole plus quadrupole distribution  $1-C\cos^2\theta$ (no dipole),
 one finds 
 that  $\alpha=\frac{C}{2-C}$, and 
 $C=\frac{2\alpha}{1+\alpha}$.
 %

In Fig.~(\ref{fig4}) we compare the capability of \J\ and \A\ to reconstruct a dipole anisotropy.
In this comparison, both advantages of \J,
namely the increased FOV and the $4\pi$~sky coverage, are evident.
Dipole (plus monopole) data are constructed in the following way.

Reconstruction plots are composed in the following way.
First, we choose a dipole amplitude $\alpha_{\rm true}$ (relative to the monopole amplitude),  and a dipole direction.
The latter is randomly oriented, and defines the axis for the polar angle $\theta$. 
Then, a given number of events, 69, 250, 410, 680, or 1000, are randomly distributed within the weighting
$1+\alpha_{\rm true}\cos\theta$.
Next, the fitting algorithm determines, as best it can, reconstructed values for $\alpha$ and for the dipole direction.
This process is repeated 100 times, each time with a different randomly oriented dipole direction.
Results are averaged, and presented in the figures.
Our observation point, the Earth, is located at the center of the dipole distribution.
A dipole distribution might be indicative of a single, dominant cosmic-ray source.

In the leftmost panel of Fig.~(\ref{fig4}) are shown the error bars that result from a reconstruction of the dipole amplitude
(we have chosen $\alpha_{\rm true}=0.4$ for illustration), for the various $N$-event samples.
The errors in $\Delta\alpha$ are seen to scale as $1/\sqrt{N}$.
More significantly, the reconstruction errors in \A\ for the dipole amplitude are almost twice those of \J, 
due to the limited sky-coverage of \A\ (and even worse for the quadrupole, to be analyzed next)~\cite{ParizotMethod}.  
Moreover, if the dipole were aligned with the zenith angle of \A, 
then \A\ could easily confuse a quadrupole with a half-dipole.
There is no such ambiguity with the $4\pi$ coverage of \J.

The second panel of Fig.~(\ref{fig4}) shows the error in reconstruction of the dipole amplitude,
for a fixed event number $N=410$~events.  
Of course, \A\ will not attain an event sample of 410, but the figure correctly displays the 
loss of quality when $4\pi$ acceptance is reduced to that of \A.
The third panel of Fig.~(\ref{fig4}) shows the reconstruction errors on the dipole direction,
 versus the dipole amplitude, for three event samples, \A\ with 250 events,
 and \J\ with 410 and 1000 events.  
 Not surprisingly, the ability to reconstruct the dipole direction improves dramatically with an increase in the dipole amplitude.
 For $\alpha$ exceeding 0.2, the 410 event \J\ sample has less that half the error of the 
 250 event \A\ sample,
 and the 1000 event sample has less than a third the error of the \A\ sample.

The final panel in Fig.~(\ref{fig4}) gives the number of $\sigma$ for reconstruction of the dipole amplitude,
as a function of the true dipole amplitude.
The discriminatory power of $4\pi$ \J\ is obvious.
A discovery claim (5$\sigma$) is evident to \A\ only for dipole amplitudes above 0.80.
However, 410-event \J\ (3yrs) can claim discovery for an amplitude down to 0.40,
and 1000-event \J\ can claim discovery all the way down to 0.28.
The latter sample can reveal 3-$\sigma$ ``evidence'' for an 
amplitude down to 0.20.
  
   \begin{figure*}[t]
  \centering
  \includegraphics[height=0.12\textheight,width=0.38\textwidth]{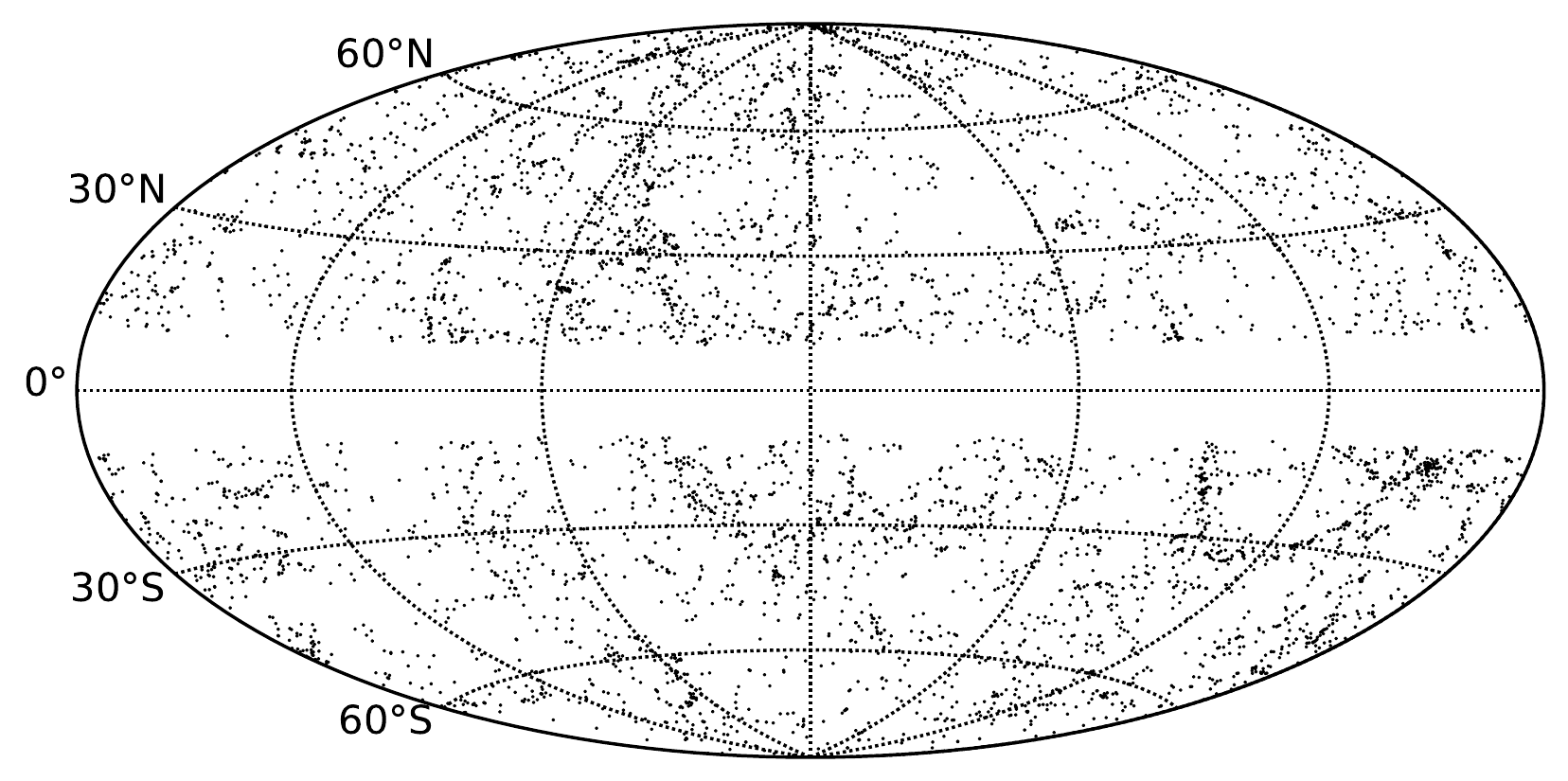}
  \includegraphics[height=0.12\textheight,width=0.28\textwidth]{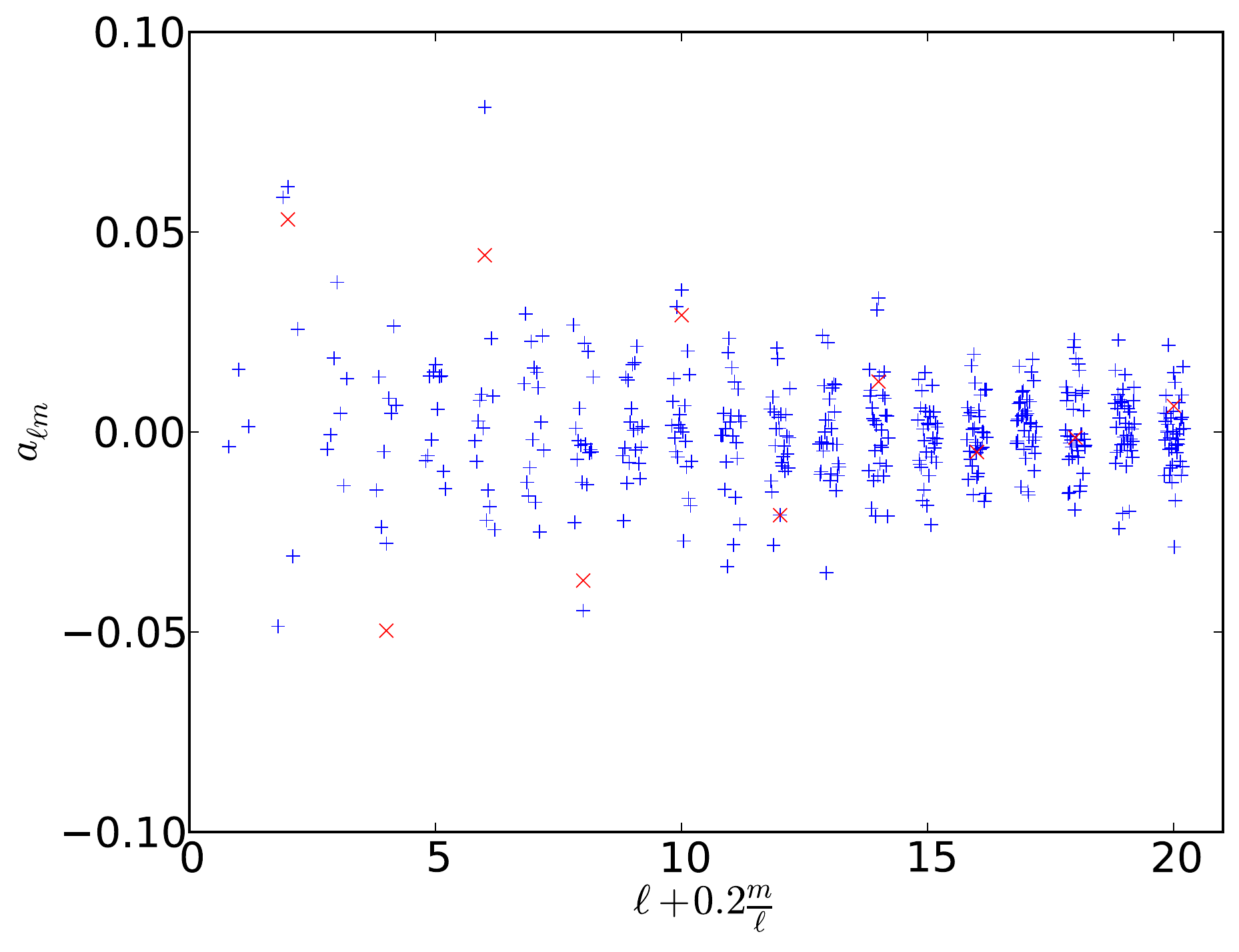}
   \includegraphics[height=0.12\textheight,width=0.28\textwidth]{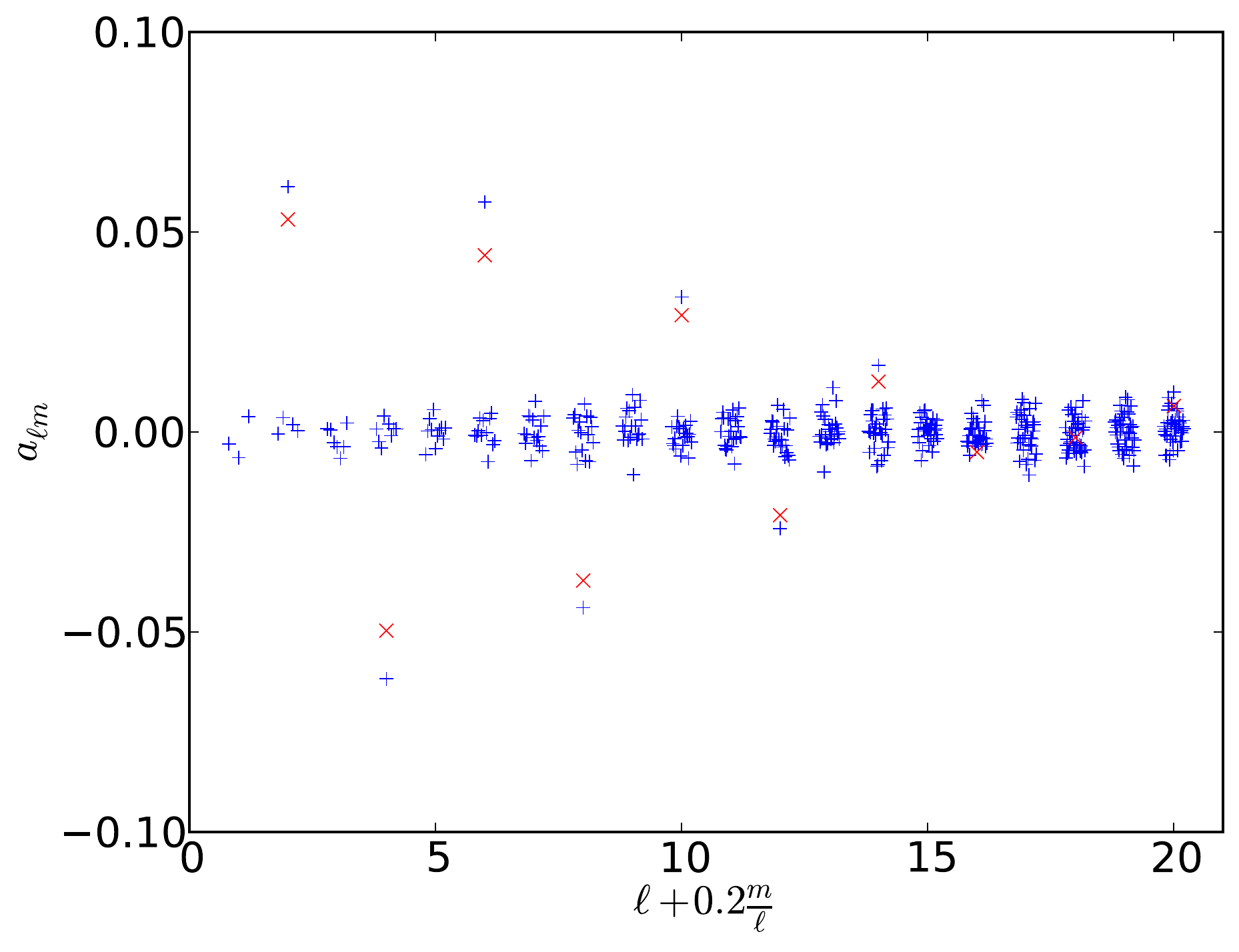}
  \caption{
  Sky-map of 
  (left panel) the ``5310 Galaxy-source'' distribution, taken from the 2MRS catalog of galaxies~\cite{2MRS} 
  out to z=0.03 (115~Mpc),
  excluding the Galactic plane at $|b| \le 10^\circ$; and the fitted $a_{\ell m}$ values (center panel).
  (right panel) $a_{\ell m}$'s from randomly distributed 5310-event ``isotropic'' data, excluding the same strip of Galactic plane.
  The $\ell/m$-dependency of the abscissa in the $a_{\ell m}$ plots is chosen~\cite{Sommers:2000us} to increase the visibility
  of  the $(2\ell+1)\ m$ values at fixed $\ell$. 
In both $a_{\ell m}$~panels, the red x's denote analytic $a_{2n\,0}$'s calculated 
for an isotropic distribution minus the Galactic plane at $|b| \le 10^\circ$. 
  }
  \label{fig6}
 \end{figure*}
 

 In Fig.~(\ref{fig5}) we repeat the reconstruction comparison of \J\ and \A,
 but this time for a quadrupole distribution.
 Our results are an average over 100 trials of a random but weighted distribution of a 
 purely isotropic monopole and anisotropic quadrupole, the latter having a randomly chosen orientation,
 again with the Earth located at the center of the distribution.
 There is no dipole in the input data set.
 A quadrupole amplitude might be indicative of a dominant Galactic distribution of sources,
 or even of a distribution of dominant sources in the Supergalactic Plane.
 A general quadrupole ($\ell =2$) has five allowed $m$ values.
Here we consider a pure $m=0$ distribution of events, 
i.e., a quadrupole with azimuthal symmetry about the quadrupole axis, 
as shown in the first sky-map of Fig.~(\ref{fig2}).  
 We have again chosen $\alpha=0.4$ for illustration.
The total distribution of events mimics an oblate spheroid with
$1-\frac{4}{7}\cos^2\theta$.

It is evident in the leftmost panel of Fig.~(\ref{fig5}) it is seen how poorly a partial-sky observatory will reconstruct any quadrupole.
For $\ell=2$, $m=0$, the sky is partitioned into three latitudinal segments (refer to Fig.~(\ref{fig2})),
one of which is barely seen if at all by a ground-based observatory.
On the other hand, space-based \J\ does very well at quadruple reconstruction,
even with a relatively small event sample.
In the second panel of Fig.~(\ref{fig5}) we display the error in the reconstruction of the quadrupole amplitude versus the 
true amplitude; the number of events is fixed at 410. 
 \J\ reproduces the quadrupole all the way down to small amplitude,
 while a ground-based instrument has roughly three times the error bar 
 of \J\ above an amplitude of 0.4, and struggles mightily to reconstruct amplitudes below 0.4.
  With N=410 events, the \J\ error in quadrupole amplitude reconstruction is about 0.07 
  for any amplitude value from 0.1 to one.

 The third panel of Fig.~(\ref{fig5}) displays the directional reconstruction of the quadrupole main axis.
 The same three event numbers that were considered in the analogous panel of Fig~(\ref{fig4}) are reconsidered here.
Again, limited-sky \A\ does poorly and all-sky \J\ does well. 
 With the 410 event sample, the angular error in directional reconstruction 
 with all-sky \J\ falls from $20^\circ$ at  $\alpha_Q=0.2$,
 to just $4^\circ$ at large amplitude.  
 For the 1000~event sample, the corresponding numbers are $10^\circ$ and $3^\circ$.
 
The rightmost panel of Fig.~(\ref{fig5}) shows that \J\ can claim 5-$\sigma$ discovery of a quadrupole amplitude 
as low as 0.4 with 410 events, and as low as 0.3 with 1000 events.
Also with 1000 events, \J\ is sensitive to a 3-$\sigma$ indication down to an amplitude of 0.2.
On the other hand,  \A\ is incapable of claiming a quadrupole discovery 
unless the quadrupole amplitude is maximum, an unlikely value.

\section{Future studies}

 In the near future, we will include some additional complicating, real world aspects of the 
 spherical harmonic search for anisotropy.
 One relates to the energy resolution for individual events.
 With a spectrum falling steeply in energy, 
a spill-over of a lower-energy bin with isotropic events into a higher-energy bin 
with potentially anisotropic events will dilute the signal.
While more study of this issue is warranted, this seems not to be a serious concern.
\A\ quotes an energy resolution $\Delta E/E$ of 22\%, with 12\% statistical, and 
about 20\% systematic.  
Simulations in \J\  to determine the energy resolution at EE are ongoing,
with the present upper limit $\sim 30\%$.
Also, the data sample of \J\ used in anisotropy studies is not bound to the 55~EeV threshold that 
\A\ chose due to its limited statistics.  With more statistics expected, \J\ can choose a higher-energy threshold.
Of course, any anisotropy will turn on gradually in energy, and simulations must include this fact.

Other probable non-concerns are the systematic error in the angular resolution of \J\ ($\lsim 3^\circ$),
and the bending of proton trajectories at EE, 
given firstly by the random walk equation through extragalactic magnetic domains of strength 
$B_{\rm nG}$ in units of nanoGuass, and coherence size $\lambda$,
\vspace*{-0.3cm}
\beq{magbending}
\delta\theta^\circ = 0.8\,Z\, \left( \frac{B_{\rm nG}}{E_{20}}\right)\, \sqrt{\frac{D\,\lambda}{10\,{\rm Mpc}^2}}\,;
\eeq
Here, Z and $E_{20}$ are the CR charge and energy in units of 100~EeV, respectively, 
and $D$ is the distance traveled by the CR.
One sees that for a proton ($Z=1$) at 100~EeV, the natural unit of bending is a degree in the extragalactic magnetic field.
On the other hand, heavy nuclei trajectories may be so severely bent as to eliminate 
even large-scale event anisotropies.  
So our hope hangs on protons being dominant at EE.
Complicating the issue is that the bending may be more or less if  filamentary structure or voids are encountered enroute.
Subsequent to the transit of extragalactic space, the CR encounters the Galactic magnetic field,
about which more is known.

Our near-future studies will incorporate 
energy resolution effects,
realistic estimates of Galactic and extragalactic magnetic fields, and ``GZK'' energy losses on cosmic radiation fields.
To incorporate these effects, we will add energy and direction assignments
for the individual simulated events in the context of two models, which we call 
the ``Galaxy source model'' and the ``Single source model''.
For the Galaxy source model, the simulated source data 
is weighted to the 2MRS all-sky catalog of galaxies out to $z=0.03$ (about 115~Mpc)~\cite{2MRS}.  
For the Single source model, the simulated source data is fixed to a single source on the sky.
For each data set, we will propagate the CRs to Earth and  
perform a multipole analysis of the resulting sky map.

The flavor  of our work in progress can be gleaned from Fig.~(\ref{fig6}) .
The first panel in Fig.~(\ref{fig6}) presents the sky-map of the 5310 galaxies 
present in the 2MRS survey, which reaches out to z=0.03, about 115~Mpc.
In the middle panel 
are shown the $a_{\ell m}$'s that result from the Galaxy-source model.
As a control, the right panel displays the $a_{\ell m}$'s that result from an isotropic distribution of 5310 events.  
In both analyses, the Galactic Plane at declination below $|b| \le 10^\circ$ is omitted.
An eyeball comparison of the two panels shows the power of the $a_{\ell m}$'s to reveal anisotropy.
(Of course, \J\ will accumulate 1000 events or less above 55~EeV, so the errors on the $a_{\ell m}$'s 
will be larger by $\sim 2-2.5$.)
%
 
%
%

\section{Conclusions}
The two main advantages of space-based observation of EECRs over ground-based observatories 
are increased FOV and $4\pi$ sky coverage with uniform systematics.
The former guarantees increased statistics,
whereas the latter enables a partitioning of the sky into spherical harmonics.  
We have begun an investigation, using the spherical harmonic technique, 
of the reach of \J\  into potential anisotropies in the EECR sky-map.  
The discovery of anisotropies would help to identify the long-sought source(s) of EECRs.

\vspace*{0.1cm}
{
{\footnotesize{{\bf Acknowledgment:}{This work is supported in part by a Vanderbilt Discovery Grant (TJW and PBD), 
NSF CAREER PHY-1053663 and NASA 11-APRA11-0058 Awards (LAA),
Alfred P. Sloan Foundation (AAB), and NSF GAANN fellowship (MR).}

}}
\clearpage

%% file: icrc2013-0272.tex


\title{Nuclearite observations with JEM-EUSO}


\shorttitle{Nuclearite observations with JEM-EUSO}

\authors{
M.~Bertaina$^{1,2}$,
G.~Bruno$^{1,3}$,
M.~Casonato$^{1,3}$,
A.~Cellino$^{2,3}$,
F.~Ronga$^{4}$,
for the JEM-EUSO Collaboration $^{5}$
}

\afiliations{
$^1$ University of Torino, Italy\\   
$^2$ INFN Sez. Torino, Italy\\   
$^3$ INAF-OATO, Italy\\
$^4$ INFN-LNF Frascati, Italy\\
$^5$ http://jemeuso.riken.jp\\

}

\email{bertaina@to.infn.it} 


\abstract{
JEM-EUSO is expected to produce results of utmost importance for a wide and
heterogeneous scientific community which includes theoretical and
experimental physicists, high-energy astrophysicists, solar system
specialists and experts of atmospheric phenomena.
The main objective of the mission is to detect extremely high energy
cosmic rays, gamma rays, and neutrinos. However, the detector is sensitive also
to much-slower-velocity events such as `nuclearites' or other massive 
quark-nuggets particles with interaction similar to nuclearites, which consist
of neutral matter including a strange quark among its constituents.
We focus in this paper on nuclearites because they are an example of 
particles already studied and searched for by other experiments.
In this contribution we show that JEM-EUSO is sensitive to `nuclearites'
with mass $m$ $>$ 10$^{22}$ GeV/c$^2$ and that a null observation of 
those class
of events in just one full day of data taking will allow to set
limits on their flux one order of magnitude more stringent than what
has been obtained so far by other experiments.
This search can be done at  practically no extra cost and is a great example of 
the multi-disciplinary capabilities of the JEM-EUSO mission.}

\maketitle

\section{Introduction}
During the last decade a very large experimental and theoretical
effort has been devoted to understand the problem of dark matter
(DM). Recently, composite objects consisting of light quarks in a
color super-conducting phase have been suggested.  In addition,
super-heavy DM anti-quark nuggets could exist and could perhaps
solve the matter-antimatter asymmetry \cite{Gorham12}; the
detection of such anti-quark nuggets by cosmic ray experiments is
discussed in \cite{Lawson11}. 
Recently the possibility to have meteor-like
compact ultradense quark-nuggets objects dressed by normal matter  has been 
suggested \cite{Labun:2011wn}.
The energy loss predicted for
super-heavy DM particles varies in different models, but it is
likely that such particles could be confused with meteors, since the
velocity, $270$$~\rm{km~s^{-1}}$, is higher, but of the same order of magnitude
of the fastest meteors.

\begin{table*}
\caption{Experimental techniques, locations, representative
experiments, sensitive area and nuclearite mass thresholds computed
for $v=270$$~\rm{km~s^{-1}}$.}
\begin{center}
\label{nuexp}
\begin{tabular}{ccccc}
\hline
 Technique  & location & Experiment & S($m^2$) & m$_{th}$ (g) \\
 \hline\hline
thermo-acoustic     &   sea level    &\cite{explorer}   &  $\sim1 $    & 
$10^{-13}$\\
damage      &  mountain 5230 m a.s.l.     &\cite{SLIM}          & 427     & $5\cdot 10^{-14}$ \\
light in oil& underground 3700 hg $cm^{-2}$ &\cite{MACRO}         & $\sim$700
   & $2\cdot 10^{-10}$ \\
light in water & underwater 2500 hg $cm^{-2}$ & \cite{ANTARES} & $\sim$$10^5$
& $2\cdot10^{-10}$\\
earth or moon-quakes & earth/moon inner &\cite{Herrin06} & $\sim 10^{11}$ & $\sim$ $10^{4}$ \\
\hline
\end{tabular}
\end{center}
   \vspace{-0.5cm}
\end{table*}
Here, we will focus our attention only on the kind of very
massive particle called `nuclearite'. This consists of neutral matter
including a strange quark among its constituents. We make this
choice because nuclearites are an example of particles already
searched for by other experiments, and for which we can be able to
compute some expected performance improvements which should be
possible using JEM-EUSO as a possible detector.

Nuggets of Strange Quark Matter (SQM), composed of approximately the
same numbers of up, down and strange quarks could be the true ground
state of quantum chromodynamics \cite{Witten84,Alcock:1988re}.

According to \cite{derujula} nuclearites are considered to be large
strange quark nuggets, with overall neutrality ensured by an
electron cloud which surrounds the nuclearite core, forming a sort
of atom. Nuclearites with galactic velocities are protected by their
surrounding electrons against direct interactions with the atoms
they might hit.

As a consequence, the principal energy-loss mechanism for a
nuclearite passing through matter is atomic collision. For a massive
nuclearite the energy-loss rate is:

\begin{equation}
\label{eq:rel}
\frac{dE}{dx}= -A \rho v^{2}
\end{equation}
where $\rho$ is the density of the traversed medium, $v$ the
nuclearite velocity and $A$ is its effective cross-sectional area.
The effective area can be obtained by the nuclearite density
$\rho_{N}$. For a small nuclearite of mass less than $1.5$ ${\rm ng}$, the
cross-section area $A$ is controlled by its surrounding cloud of electrons 
which is never smaller than $10^{-8}$ cm:

\begin{equation}
\label{eq:sigma}
A =\left\{ \begin{array}{ll}
\pi \cdot {10}^{-16} \,\textrm{cm$^{2}$}    &  for ~ m < 1.5 ~{\rm ng}
  \\
\pi {\left(\displaystyle{\frac{3 m}{4 \pi {\rho}_{N}} }\right)}^{2/3} & for ~ m > 1.5 ~{\rm ng}
\end{array} \right.
\end{equation}
where \mbox{$\rho_{N}= 3.6\cdot10^{14}$~ {\rm g cm}$^{-3}$} is the nuclearite
density and $m$ its mass.

According to Eq. \ref{eq:rel}, nuclearites having galactic velocity
and mass heavier than $10^{-14}$$~{\rm g}$ penetrate the atmosphere, while
those heavier than 0.1$~{\rm g}$ pass freely though an Earth diameter. Eq.
\ref{eq:rel} has been used by \cite{derujula} to compute the amount
of visible light emitted in the atmosphere, assuming that the light
is emitted as a black-body radiation from an expanding cylindrical
thermal shock wave and to compute therefore the apparent magnitude
as defined for meteors.

The efficiency of the light emission due to the black body radiation is inversely proportional
to the medium density: this cancels the density dependence of the energy-loss and therefore
in most of the nuclearite path in the atmosphere the light emission is constant with  height.
According to \cite{derujula} the upper limit to
the altitude ($h_{max}$) at which nuclearites effectively generate
light is described by the following relation:
\begin{equation}
h_{max} = 2.7 \ln ( m / 1.2 \times 10^{-5} ~g ) ~~{\rm km}
   \label{eqn:height}
\end{equation}
and for altitudes less then $h_{max}$ the light emitted is constant.
For the range of masses 0.1-100 ${\rm g}$,  $h_{max}$ is expected
to be located between 24 km and 60 km.

It turns out  therefore that there are three important
differences that can help to discriminate between nuclearites and
meteors. The first one is that the amount of light emitted by
nuclearites  is constant at  $h\le  h_{max} $,
the second
difference is that a nuclearite of mass bigger than 0.1 $g$ can move
upward and this is extremely unlikely for a meteor; the third difference  is that 
the absolute value of the velocity is higher, with a maximum value of  
$\sim570~\rm{km~s^{-1}}$, while
meteors are limited to $\sim72~\rm{km~s^{-1}}$.

Nuclearites and similar particles, as for example neutral
Q-ball\cite{Kusenko98}, have been searched for using different
approaches. The experiments can be characterized by the detection
area (S) and by the minimum nuclearite mass that can be detected
(m$_{th}$), usually computed for a speed of $270$$~\rm{km~s^{-1}}$. Many
techniques, summarized in Table~\ref{nuexp} have been used to detect
nuclearites: acoustic emission due to the thermal shock in  aluminum
gravitational wave cylindrical detectors, damages in plastic
materials like CR39, Makrofol or Lexan, light emission in oil or sea
water, seismic waves induced by big nuclearites. Due to the
uncertainties in the energy losses it is important to have different
techniques to detect such exotic particles. Table~\ref{nuexp} lists
the different techniques and a representative experiment of each
technique. It is not aimed at being a full list of  the experiments
done so far to search nuclearites, but it is a reasonable summary of
the state of the art in this field.

\section{General description of JEM-EUSO payload and focal plane assembly}
A general description of the JEM-EUSO telescope \cite{Kajino2}
has already been given elsewhere in this volume. We recall here only the
essential points related to the meteor and nuclearite detection.
The role of the JEM-EUSO telescope \cite{Kajino} is to act as an
extremely-fast ($\sim \mu$s) and highly-pixelized ($\sim3\times10^5$
pixels) digital camera with a large aperture (a diameter of about
2.5m) and a wide field of view (FoV) of $60^\circ$. It works in
near-UV wavelengths ($290$--$430$ nm).

The optics focuses the incident UV photons onto the focal surface.
The focal surface detector converts incident photons into electric
pulses. The electronics counts the number of pulses in time
intervals of $2.5$ $\mu$s (Gate Time Unit - GTU) and records it. When a
signal pattern is
found, a trigger is issued. This starts a sequence which eventually
transmits to the ground operation center the signal data recorded
within (and surrounding) a selected pixel region.

The combination of 3 Fresnel lenses has an angular resolution of
0.07$^\circ$. This resolution corresponds approximately to a linear
size of $550$ m on the ground beneath the ISS located at an altitude
above ground of about 400 km.

The Focal Surface (FS) of JEM-EUSO has a spherical shape of about
$2.3$ m in diameter with about $2.5$ m curvature radius, and it is
covered by $\sim 5,000$ multi-anode photomultiplier tubes (MAPMTs).
The FS detector consists of Photo-Detector Modules (PDMs), each of
which consists of $9$ Elementary Cells (ECs). Each EC contains $4$
units of MAPMT (Hamamatsu R11265-03-M64, 2 inches in size, with
$8\times 8$ pixels). A total of $137$ PDMs are arranged on the FS. A
Cockcroft-Walton-type high-voltage supply is used to suppress power
consumption, including a circuit to protect the photomultipliers
from instantaneous bursts of light, like in the case of lightning or
bright fireball phenomena.

The FS electronics system records the signals of UV photons
generated by cosmic rays successively in time. A new type of front-end
ASIC has been developed for this mission, which has both functions
of single photon counting and charge integration in a chip with $64$
channels. The FS electronics is configured in three levels
corresponding to the hierarchy of the FS detector system: front-end
electronics at EC level, PDM electronics common to $9$ EC units, and
FS electronics to control $137$ units of PDM electronics. Anode
signals of the MAPMT are digitized and recorded in ring memories for
each GTU to wait for a trigger
assertion, then, the data are read and sent to control boards.
JEM-EUSO uses a hierarchical trigger method to reduce the huge
original data rate of $\sim$10 GB/s down to $297$ kbps, needed to
transmit data from the ISS to the ground operation center.

\begin{table*}
  \caption{For different absolute magnitudes (M) of
meteors in visible light, the corresponding flux in the $U$-band are
shown (according to the Flux Density Converter of the Spitzer
Science Center; details can be found at the web site
http://ssc.spitzer.caltech.edu/warmmission/propkit/pet/magtojy/index.html).
The corresponding number of photons per second,
the number of photo-electrons per GTU, the typical mass of the
meteor, and the number of events expected to be observed by JEM-EUSO
(the latter is computed assuming a duty cycle of $0.2$) are also
shown.}
  \label{mag}
  \begin{center}
  \begin{tabular}{cccccc}
    \hline
     magnitude & U-band flux & photons & photo-electrons &
mass & collisions in \\
 (M)    & (erg/s/cm$^2$/A) & (s$^{-1}$) & (GTU=2.5$\mu$s)$^{-1}$ &
(g) & JEM-EUSO FoV \\
    \hline
    7 & 6.7$\cdot$10$^{-12}$ &  4.3$\cdot$10$^{7}$ & 4 & 2$\cdot$10$^{-3}$ & 1/s   \\
    5 & 4.2$\cdot$10$^{-11}$ &  2.7$\cdot$10$^{8}$ & 23 & 10$^{-2}$ & 6/min   \\
    0 & 4.2$\cdot$10$^{-9}$ &  2.7$\cdot$10$^{10}$ & 2300 & 1 & 0.27/orbit   \\
   -5 & 4.2$\cdot$10$^{-7}$ &  2.7$\cdot$10$^{12}$ & 2.3$\cdot$10$^5$ & 100 & 6.3/year   \\
    \hline
  \end{tabular}
  \end{center}
   \vspace{-0.5cm}
\end{table*}

\section{Simulations}
One of the exploratory objectives of the JEM-EUSO mission is the 
observation
of atmospheric phenomena such as meteors. For this reason
a very simple model of meteor phenomena has been preliminarily
developed in order to make it possible to
carry out a campaign of numerical simulations aimed at analyzing the
kind of signals which may be produced on the JEM-EUSO focal plane in
a variety of possible observing scenarios.
This simulator is quite useful to estimate the sensitivity of 
JEM-EUSO in observing nuclearites. 
 The results presented in the following are in fact
rescaled from the simulations conducted for meteors \cite{Cellino}.  

Currently, the response of the detector, including optics and
focusing, and the response of the photomultipliers in the focal
surface, is parameterized. An overall throughput efficiency of 10\%
is assumed. An optical point spread function (PSF) of $\sim$2.5$~\rm{mm}$
is assumed. Cross-talk, pixel-to-pixel non-uniformity response in
gain of the order of 10\%, as well as poissonian fluctuations of the
night glow background are introduced in the simulations. The FS is
considered to be a uniform layer of MAPMTs.

Table~\ref{mag} summarizes the relation between meteor
absolute magnitude, photon flux, number of photo-electrons at the
maximum of the development, mass and expected number of events in
the FoV of JEM-EUSO in the nadir mode.

\begin{figure}
  \begin{center}
  \includegraphics[width=0.40\textwidth]{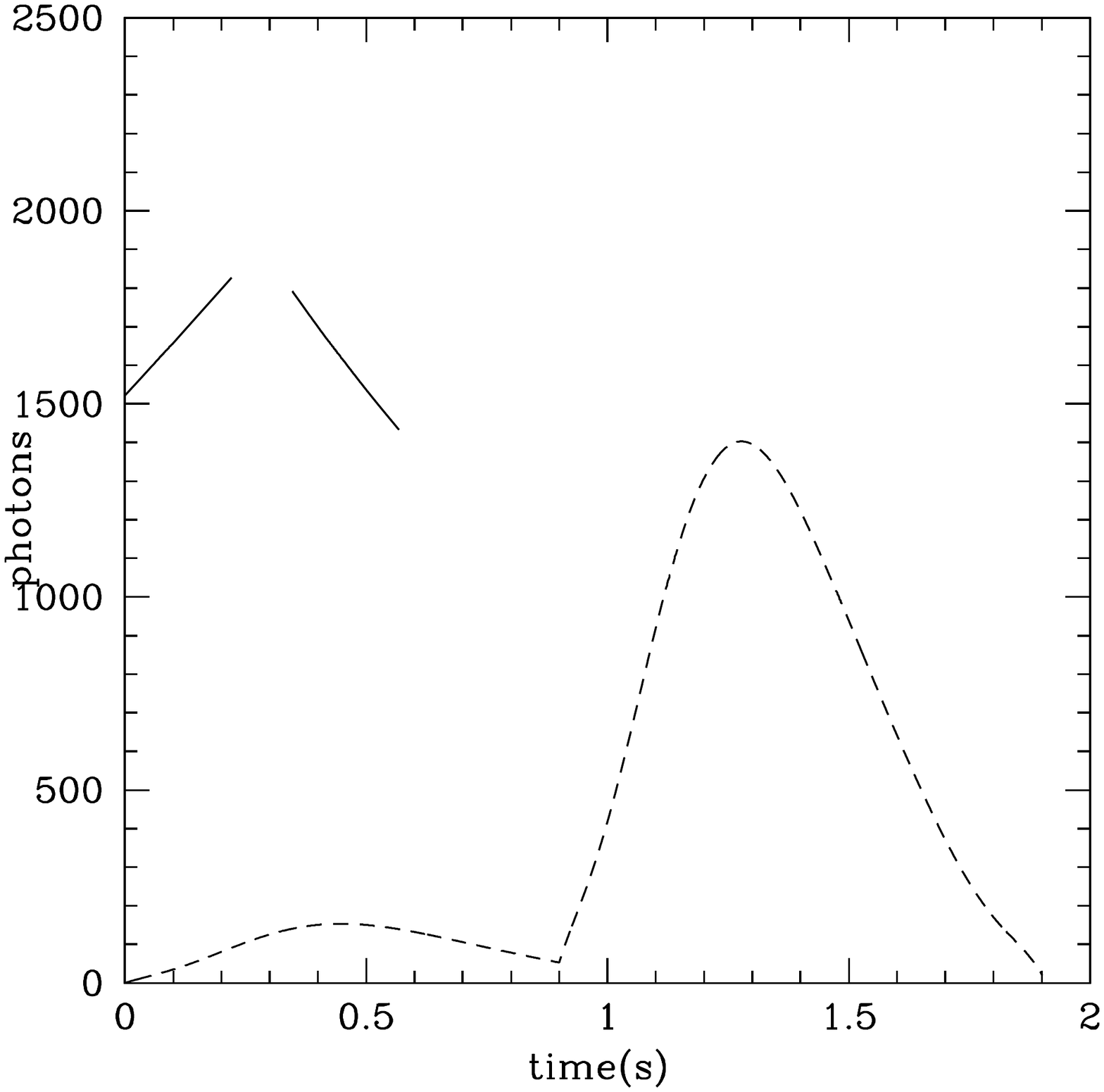}
  \end{center}
  \caption{Comparison between the light profile of two nuclearites (thick
lines) and that of a meteor (dashed line). 
The nuclearite has a mass of $m$ = 20 $\rm g$ and velocity of 250 
$\rm{km~s^{-1}}$ and it is simulated up-ward 
going (left curve) and down-ward going (right curve) with $\theta$ = 
45$^\circ$ inclination from the vertical. The magnitude of the meteor is 
M = -1,
velocity 70 $\rm{km~s^{-1}}$, and $\theta$ = 45$^\circ$ inclination as 
well.}
  \label{fig:nucl-mas}
\end{figure}
The simulation work carried out so far suggests that
JEM-EUSO could be able to detect meteors down to absolute visual magnitudes
of the order of $6-7$, a limit which in absolute terms is not better
than the performances of the best ground-based facilities, but which
becomes very interesting for meteor science by considering the large
FoV and high duty cycle of the JEM-EUSO detector.

\section{JEM-EUSO sensitivity to nuclearites}
A dedicated simulation of the
signals produced by nuclearites moving through the atmosphere is currently
under way for a detailed assessment of the expected performances of
JEM-EUSO in detecting and recording these events. However, the results
already obtained for meteors can be used to draw some general and
preliminary conclusions.

First of all, the detection sensitivity to nuclearites can be
extrapolated from previous results obtained for the meteors. In
particular, the absolute visual magnitude (M) of an atmospheric 
nuclearite can be computed according
to \cite{derujula} as:
\begin{equation}
   M = 15.8 - 1.67 \cdot log_{10}(m/1\mu g).
   \label{eqn:magnitude}
\end{equation}
We recall that the absolute magnitude of a meteor corresponds to the apparent
magnitude measured on the ground if the meteor is seen at the zenith and at
an height of 100 km. 
By inverting equation~\ref{eqn:magnitude} it is 
possible to estimate the minimum mass of the nuclearite detectable by 
JEM-EUSO in terms of absolute magnitude, which is independent of the 
distance $h$. This is reasonable at a first level of approximation, because
the maximum difference of apparent magnitudes of the same event in different
locations of the field of view is $\Delta M_{app}$ $<$ 1.
Results indicate that JEM-EUSO is sensitive to objects having mass
$m$ $>$ 0.1 g when working in single photon-counting mode, and to $m$
$>$ 3 -- 30 g when working in charge integration mode. There is
of course a dependence upon the sky background luminosity (mainly
due to Moon phase).

For what concerns the triggering strategy to be adopted for these
events, the same algorithms already developed for meteors can be
used, simply varying the total sampling time in order to take into
account the shorter duration of the phenomenon (and correspondingly
shorter track length). Assuming to be in most unfavorable
conditions, namely a nuclearite starting to emit at an height of
$60$ km, and moving along a trajectory having a zenith angle such
that the track crosses the entire PDM along its diagonal ($\sim$42
km) before landing at ground, and taking into account a velocity of
$250$$~\rm{km~s^{-1}}$, it follows that the total duration of the phenomenon is
only $\sim 0.3$ s. Therefore, in charge integration mode (KI mode)
the optimized condition would be to record the signal during 1024
GTUs, sampled at a rate of one every 128 GTUs, whereas in single
photon counting mode, one should record 128 GTU, with a sampling
rate of one every 1024 GTUs. In both cases the total integrated time
is $\sim 0.33$ s (one GTU being equal to $2.5$ $\mu$s). This is a very
conservative estimation. In fact, by integrating the signal accumulated
in 128 GTU, instead of samping it, a much better performance
of the instrument would be obtained.  

The most important criterion 
to distinguish nuclearites from meteors is based
on their velocity. Meteors have much slower speeds (in general
below 72$~\rm{km~s^{-1}}$ ). As it is shown in \cite{Cellino}, already
at trigger level it is possible to estimate the projected velocity
of the signal on the FS with reasonable uncertainty. A subsequent
data analysis of the recorded signals 
will certainly provide much more accurate results. Although
it is not possible to derive directly from the data the 3D velocity
vector of the source, a limit can be set to the recorded projected
velocity ($v_{proj}$). By requiring that $v_{proj} > v_{proj}^{min}$
and assuming that the velocity of the nuclearite is $v = 250$$~\rm{km~s^{-1}}$,
choosing a value for $v_{proj}^{min}$ automatically sets a limit on
the zenith angle of the track $\theta_{min} =
arcsin(v_{proj}^{min}/v)$ and relative acceptance $R_{acc}= (1+\cos(2
\cdot \theta_{min}))/2$. Table~\ref{tab:nucl-acc} shows the relative
acceptance as a function of different possible choices of
$v_{proj}^{min}$.
\begin{table}
  \caption{{Impact on the relative acceptance
($R_{acc}$) (see text) as a function of different possible choices
of $v_{proj}^{min}$.},  assuming a nuclearite velocity $v=250$$~\rm{km~s^{-1}}$. }
  \label{tab:nucl-acc}
  \begin{center}
  \begin{tabular}{ccc}
    \hline
    \hline
     $v_{proj}^{min}$ & $\theta_{min}$ & $R_{acc}$ \\
      ($~\rm{km~s^{-1}}$) & (deg.)  & (\%) \\
    \hline
    100 & 23.6 & 84 \\
    130 & 31.3 & 73 \\
    160 & 39.8 & 59 \\
    190 & 49.5 & 42 \\
    220 & 61.6 & 23 \\
    \hline
  \end{tabular}
  \end{center}
\end{table}
It turns out that even a very tight cut on $v_{proj}^{min} > 160$
$\rm{km~s^{-1}}$ makes the acceptance to decrease only by about a factor of 2.

Another important fact to be taken into account is that nuclearites
tend to develop at lower heights in the atmosphere compared to
meteors. Moreover, any possible evidence of tracks moving upwards
would be a clear sign of a nuclearite. The light profile looks also quite
different (see Figure~\ref{fig:nucl-mas}).
\begin{figure}[!h]
  \begin{center}
  \includegraphics[width=0.42\textwidth]{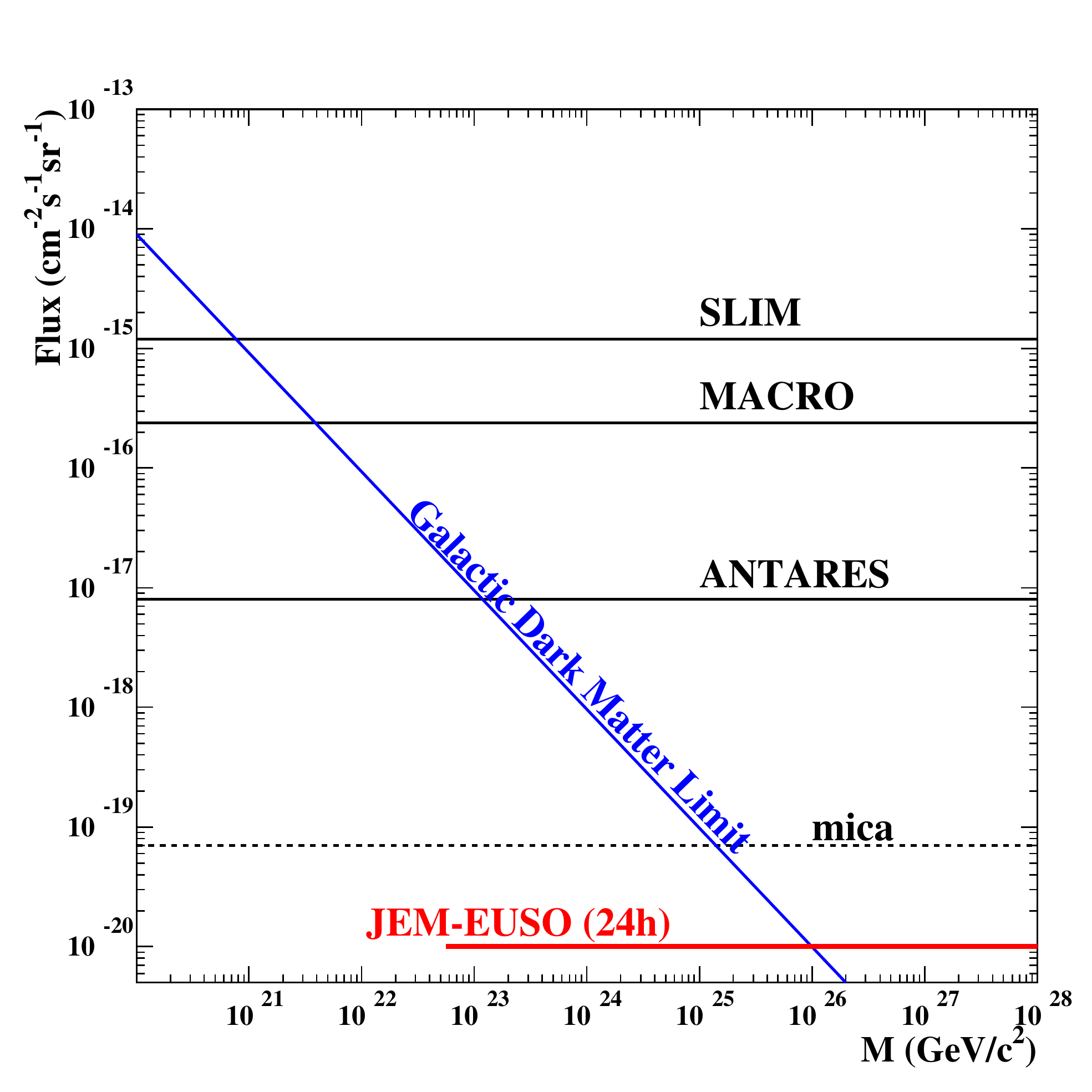}
  \end{center}
  \caption{The JEM-EUSO $90$\% confidence level upper limit on the flux of
nuclearites resulting from null detection over $24$ hours of JEM-EUSO
operations. The limits of other experiments \cite{SLIM}, \cite{MACRO},
\cite{ANTARES}, \cite{Price:1988ge} are also shown for a comparison.
The old mica limits \cite{Price:1988ge} are dependent from several additional
assumptions, respect to the other experiments.}
  \label{fig:nucl-sens}
\end{figure}

We can expect therefore that JEM-EUSO will be able to set very
stringent limits on the flux of nuclearites, even after short
acquisition times, due to the tremendous instantaneous exposure
(A $\sim 5 \times 10^{20}$ cm$^2$ s sr) of the instrument. Even
adopting a very severe rejection criterion, such as $v_{proj}^{min}$
$>$ 190$~\rm{km~s^{-1}}$, from Table~\ref{tab:nucl-acc} we can infer that for a
$24$ h accumulation time, a null detection would set a limit in flux
at the 90\% confidence level of the order of 10$^{-20}$ cm$^{-2}$
s$^{-1}$ sr$^{-1}$ (see Figure~\ref{fig:nucl-sens}).

\section{Conclusions}
Our preliminary analysis concerning the possible detection of
nuclearites indicate that JEM-EUSO will be sensitive to nuclearites
with mass higher than a few 10$^{22}$ GeV/c$^2$ and will be able, after
a run time of only $24$ h, to provide limits on nuclearite flux
lower by one order of magnitude with respect to the limits of the
experiments carried out so far, and lower than the dark matter limit.

 \vspace*{0.5cm}
{
{\footnotesize{\bf Acknowledgment:}{
This work was partially supported by the Italian Ministry of Foreign
Affairs, General Direction for the Cultural Promotion and
Cooperation. 
}

}}
\clearpage

%% file: icrc2013-0875.tex


\title{ESAF-Simulation of the EUSO-Balloon}

\shorttitle{EUSO-Balloon}

\authors{
T. Mernik$^{1,2}$,
A. Guzman$^{1,2}$,
A. Santangelo$^{1,2}$,
K. Shinozaki$^{1,2}$,
N. Sakaki$^{3}$,
C. Moretto$^{4}$,
D. Monnier-Ragaigne$^{4}$,
H. Miyamoto$^{4}$,
S. Dagoret-Campagne$^{4}$,
C. Catalano$^{5}$,
P. von Ballmoos$^{5}$,
for the JEM-EUSO Collaboration.
}

\afiliations{
$^1$ Institute for Astronomy and Astrophysics (IAAT), Kepler Center, Universit\"at T\"ubingen, Germany \\
$^2$ RIKEN Advanced Science Institute, Wako, Japan \\
$^3$ Karlsruhe Institute of Technology (KIT), Karlsruhe, Germany \\
$^4$ Laboratoire de l'Acc\'el\'erateur Lin\'eaire, Univ Paris Sud-11, CNRS/IN2P3, France \\ 
$^5$ Institut de Recherche en Astrophysique et Plan\'etologie (IRAP), France \\
}

\email{mernik@astro.uni-tuebingen.de}

\abstract{The EUSO-Balloon is a balloon borne ultraviolet (UV) telescope, which is being developed as a pathfinder of the JEM-EUSO mission the \emph{Extreme 
Universe Space Observatory onboard the Japanese Experiment Module} on the International Space Station (ISS).
Designed as a scaled version of JEM-EUSO, the EUSO-Balloon will serve as a technology demonstrator. From 2014 on, it is planned to conduct a number of missions, between a few and several tens of hours at an altitude of approximately 40 km.
Besides proving the robustness of the JEM-EUSO technology, it will perform UV background studies under many different ground conditions
and potentially observe extensive air showers (EAS) induced by ultra-high-energy cosmic rays (UHECR) with energies of the order of $10^{18}$ eV. 
The detector design consists of a system of Fresnel lenses focusing the incoming  300 - 400 nm UV fluorescence photons onto an array of multi-anode photomultipliers. Generated 
photoelectrons are then readout by the front end electronics, converted into digital data and saved to disc if a trigger is issued.
The ESAF (EUSO Simulation and Analysis Framework) software package is designed to simulate space based observation of EAS, taking into account every physical process from EAS 
generation, propagation of light in atmosphere, detector response and eventually reconstruction. EUSO-Balloon specifications such as the optics  and dedicated electronics components have been implemented in the code to study the expected instrument behavior and its ability to resolve the UHECR arrival direction. 

In this article we describe ESAF simulations of the EUSO-Balloon. Furthermore, we present a first estimate of the expected spatial resolution performance of the instrument.
}


\maketitle

\section{Introduction}
JEM-EUSO is a space based UV telescope developed for the detection of ultra-high-energy cosmic rays (UHECR) \cite{bbtakahashi,bbcasolino}.
It will be attached on board the Japanese Experiment Module
at the ISS. JEM-EUSO will monitor from space the earth's atmosphere to search for ultra high energy cosmic ray (UHECR) induced extensive air showers (EAS). 
The EUSO-Balloon is a pathfinder mission for the JEM-EUSO instrument \cite{eusoballoon}.
It is a scaled version of the space detector using the same optical and electronics components. It will prove the feasibility of the JEM-EUSO mission 
by demonstrating the robustness of technological key elements under quasi-space conditions. During a number of envisaged campaigns above different ground conditions it will
deliver background data and test the trigger implemented. Moreover, the balloon detector might detect a few UHECR events of the order of $10^{18}$ eV. These events would 
be the first of their kind ever observed from space.  

\section{EUSO-Balloon}
\begin{figure}[h!]
  \centering
  \includegraphics[width=0.4\textwidth]{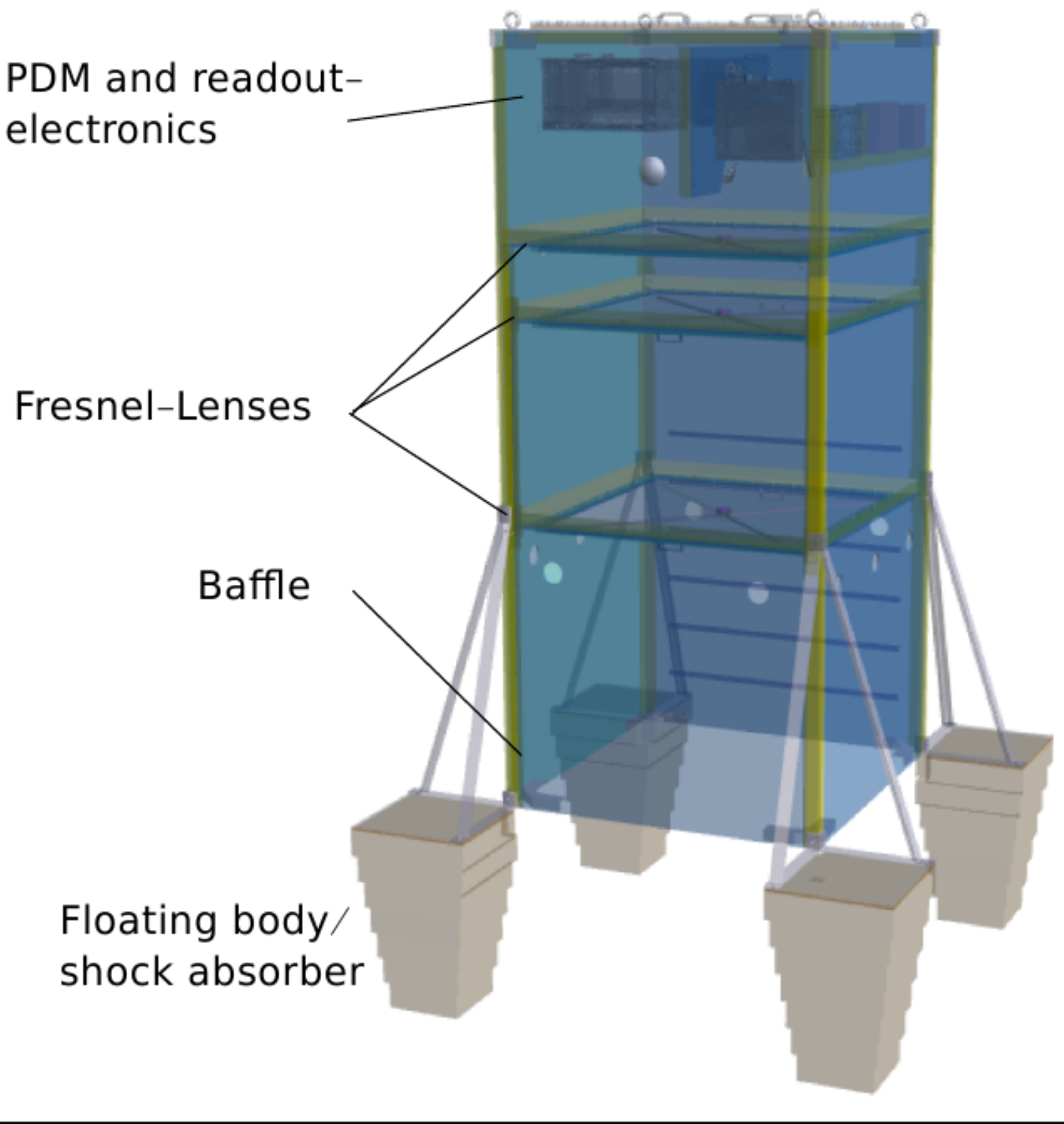}
  \caption{The EUSO-Balloon gondola. Three Fresnel lenses focus the incoming UV shower photons on a PDM. All components used are identical or 
scaled versions to the JEM-EUSO components.}
  \label{gondola}
 \end{figure}

Like the JEM-EUSO detector, the EUSO-Balloon is an UV telescope using a refractive optics of Fresnel lenses to focus the incoming photons
in the wavelength range of 300 to 430 nm onto a \emph{photo detector module} (PDM) consisting of an array of 36 \emph{multi anode photomultiplier tubes} (MAPMT). 
Each MAPMT has 8 $\times$ 8 = 64 pixels. The instrument will have an exposure of 7$ \cdot 10^{7}$ km$^2$ sr  s.  For the general setup see Fig. \ref{gondola}.
The trigger logic implemented on the cluster control board continuously seeks for pattern characteristics meeting those of signal tracks we would
expect from a moving EAS.  When a trigger is issued, the time frame of 128 GTU (gate time units, 2.5 $\mu$s) is saved to disc or 
transferred by the telemetry for analysis. In collaboration with the French space agency (CNES), multiple balloon campaigns are planned 
from 2014 on. Altitudes of approximately 40 km will be reached. 
The main scientific objectives are first of all to prove the reliability of the proposed JEM-EUSO components in quasi-space conditions. This includes
the optics as well as the readout electronics and the atmospheric monitoring system. Moreover, the EUSO-Balloon will perform a background measurement
in the near UV in various conditions.  Possible ground scenarios include snow, forest and ocean. 
With the help of a laser device we will create artificial light tracks in the atmosphere comparable to those released by EAS,
to test the instrument's capability of triggering and reconstructing air showers.

\section{ESAF}
The EUSO Simulation and Analysis Framework is a software package for the simulation of UHECR space detectors \cite{bbesaf}. It is an object-oriented C++ code, based on ROOT \cite{bbroot}.
Its development at the time of the EUSO  mission (\cite{bbeuso}) was stopped in 2004. 
During the JEM-EUSO study, we have reactivated the code and improved parts of the software when necessary.
New algorithms have been implemented as well as new hardware components. Moreover, we have implemented the EUSO-Balloon specifications for this pathfinder mission. 
ESAF is now capable of simulating the entire chain of processes 
to be taken into account during the measurement of UHECR air showers by the EUSO-Balloon.
Primary particle/atmosphere interaction,  development of the EAS, creation of UV - fluorescence and Cherenkov photons and their propagation to the detector are
simulated by the dedicated modules. Second part of the simulation accounts for the processes inside the detector. Photons passing through the optics and being
focused on photon detection module (PDM) is accounted for by a ray tracing module. Following that the photon/ photomultiplier tube (PMT) interaction and electronics 
response is simulated.

\section{The Reconstruction Framework}
Cosmic ray induced EAS emit fluorescence light isotropically in all directions plus a beamed Cherenkov component. 
Parts of that light go directly to the telescope. Other components are reflected diffusely from ground or scattered towards the detector.
The UV photons reaching the entrance pupil of the instrument propagate through the optics and 
activate the photomultiplier tubes arranged on the focal surface. When the readout electronics recognizes certain patterns a trigger is issued. The signal is then processed and transmitted to earth for analysis and reconstruction.

In ESAF different modules are dedicated to the single stages during the evaluation of the signal. First of all, the signal has to be disentangled from noise. Following that direction and energy reconstruction algorithms can be applied.

\subsection{Pattern Recognition}
The fluorescence signal will appear as a faint moving spot of the focal surface of the telescope embedded in the background generated by night 
glow, city light, weather phenomena and other sources. The extraction of the signal track and the 
determination of its spatio-temporal behavior remains crucial for any further analysis aiming at reconstructing the arrival direction or 
energy of the primary. There are two possible algorithms for the pattern recognition:  
\begin{itemize}

\item \emph{PWISE}, an algorithm that analyses every pixel individually for significant deviation from background fluctuations.

\item \emph{LTT-PreClustering}, a technique that searches for accumulations of counts that are arranged along a line. 

\end{itemize}
Both have been implemented in ESAF and can be used either alone or in combination.

\subsubsection*{PWISE}
The Peak and Window Searching Technique (PWISE) selects 
photon-counts coming from the EAS, and at the same time it filters out 
multiple-scattered photons which results in a ``fuzzy'' image of the track.  
This effect appears as a consequence of their shifted arrival time due to the
multiple scattering.
\begin{description}
\item[\bf{Step 1}] For each pixel, PWISE  only considers pixels whose highest
photon-count (peak) is above a certain  threshold (\emph{peak-threshold}).
\item[\bf{Step 2}] Next  PWISE searches for the time window with the highest
 signal-to-noise ratio (SNR).
\item[ \bf{Step 3}] We check if the maximum SNR is above a given 
\emph{SNR-threshold}. Only if the SNR is above the threshold we select 
the photon-counts within the time window that maximizes SNR. The selected 
photon-counts are then passed on to the next reconstruction module.
 \end{description}

\subsubsection*{LTT-PreClustering}
The Linear Tracking Trigger (LTT) Pre-Clustering technique can improve the performance of the angular reconstruction when 
applied in combination with the actual pattern recognition. It is a refined version of
the logic implemented in the 2$^{\rm nd}$ level trigger.
It selects the pixels on the focal surface containing the highest number of 
counts. Then it searches for the track that maximizes counts by moving an 
integration box along a predefined set of directions intersecting this point.
Pixels outside this track are ignored by the following pattern recognition.
For details see \cite{bbbertaina}.

\subsection{Direction Reconstruction}
\begin{figure}[!h]
  \centering
  \includegraphics[width=3.in]{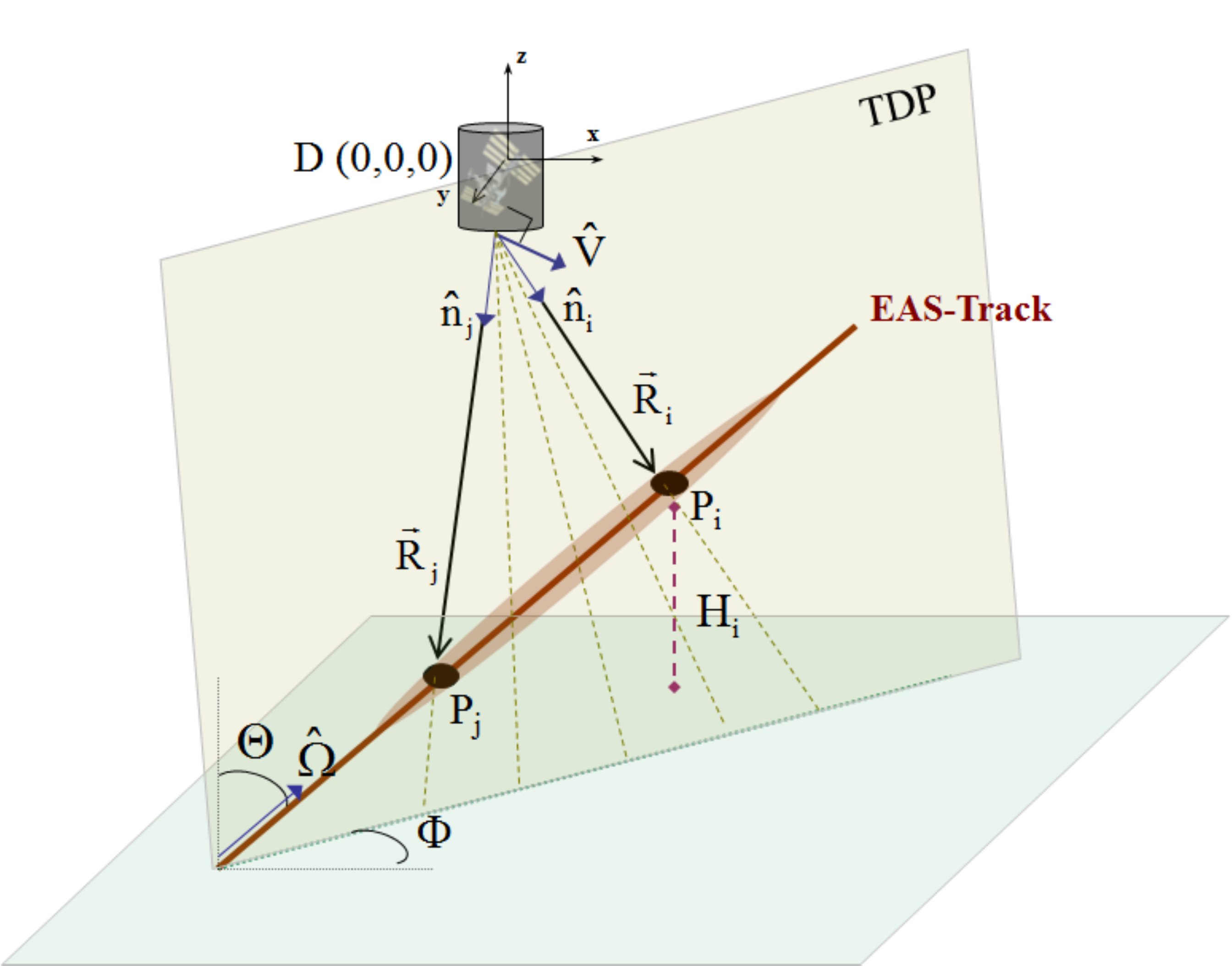}
  \caption{EAS observed with JEM-EUSO: Within the track-detector-plane (TDP), photons emitted at different times t$_j > t_i$ reach the 
detector from certain directions $\hat n_i$, $\hat n_j$ after traversing R$_i$, R$_j$ in atmosphere. From the timing information and arrival angle of the shower photons, the direction of the primary $\hat \Omega(\Theta,\Phi)$ can be determined.}
  \label{qqrecosyst}
 \end{figure}

From the geometrical properties of the signal track on the focal surface the arrival direction of the primary can be computed by a variety 
of methods implemented in ESAF as described in more detail in \cite{bbbertaina} and \cite{angurecoperformance}. Fig. \ref{qqrecosyst} shows the system of the EAS and the detector.
In the current configuration there are 5 different algorithms implemented in ESAF. Their performances depend on conditions such as 
energy and zenith angle of the primary UHECR but also on atmospheric conditions:
\begin{itemize}

\item \emph{Analytical Approximate 1}: The angular velocities of the signal track in the x-t and y-t planes are linearly fitted. 
The arrival angle of the primary is derived by geometrical estimations. 

\item \emph{Analytical Approximate 2}: The angular velocity of the signal track on the z-t plane is linearly fitted. 
The arrival angle of the primary is derived by geometrical estimations. 

\item \emph{Numerical Exact 1}: a $\chi^2$ minimization is performed between the activation times of pixels induced by the actual 
signal to those induced by a signal track theoretically computed. 

\item \emph{Numerical Exact 2}: a $\chi^2$ minimization is performed between arrival angles of photons coming from the actual 
signal to those induced by a signal track theoretically computed.  

\item \emph{Analytically Exact 1}: without prior knowledge of the TDP, this method reconstructs the direction of the primary by using 
 using the exact relations between pixel directions in the FOV and photon's arrival times. 

\end{itemize}

\section{Balloon Simulations}
\begin{figure}[!h]
  \centering
  \includegraphics[width=0.4\textwidth]{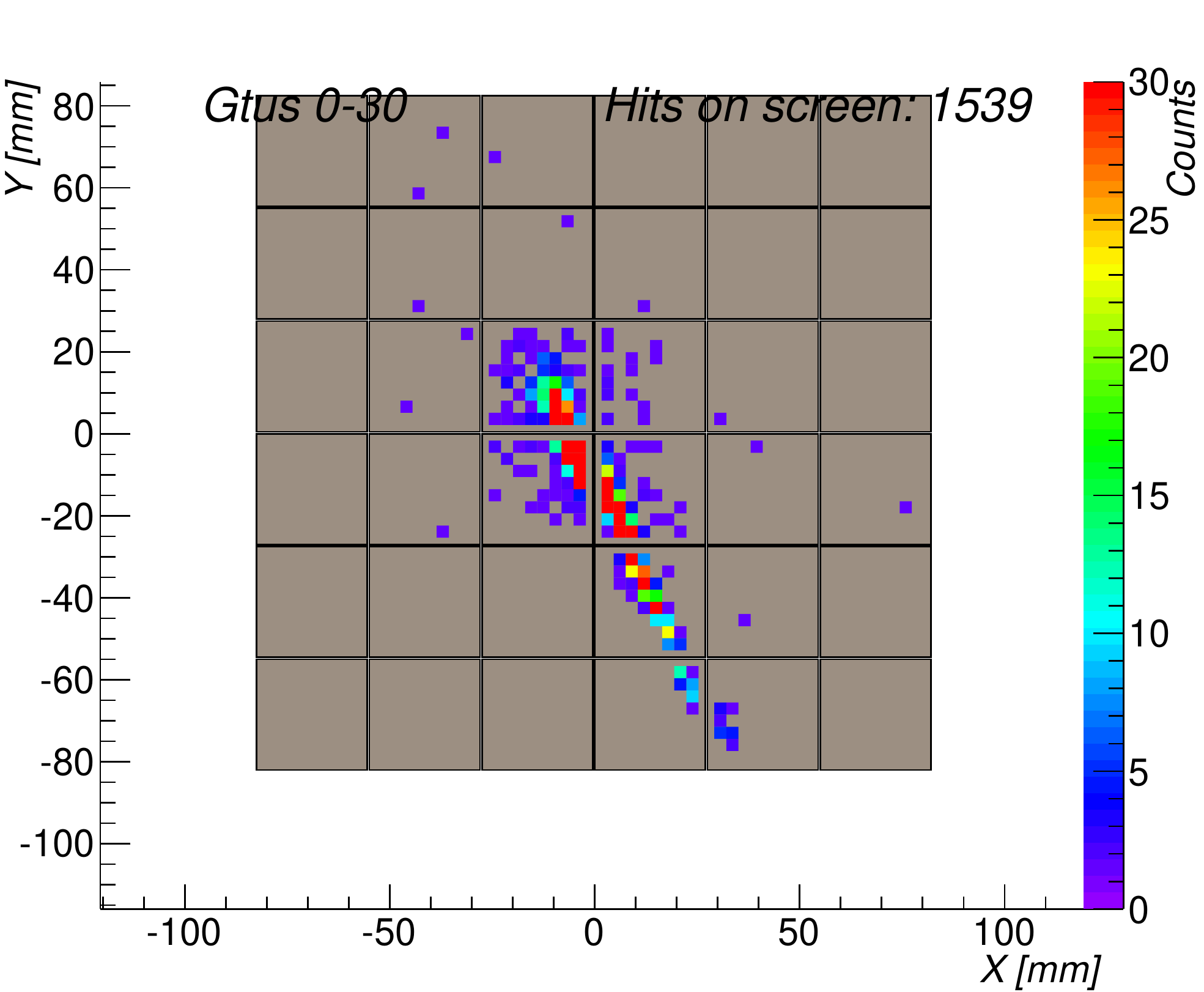}
  \caption{Example of EAS signal track for a proton shower with an energy of $10^{19}$ eV and zenith angle $\Theta=30^{\circ}$ entirely within the FOV.}
  \label{eastrack}
 \end{figure}

Basic parameters for the simulation of the EUSO-Balloon are
\begin{itemize}

\item altitude= 40 km

\item background: 500 photons m$^{-2}$ ns$^{-1}$ sr$^{-1}$, uniformly distributed

\item  field of view 12$^{\circ}$ $\times$ 12$^{\circ}$

\end{itemize}
The background has been chosen in accordance to the data of the BABY balloon mission \cite{baby}. It is simulated only at the electronics level
in order to save computing time. The area in which the events have been simulated was greater than the projection of the balloon FOV on the earth's surface to check for trigger of stray light photons and to analyze the behavior of signal tracks that traverse the FOV only partially. A typical shower event seen by the EUSO-Ballon can be seen in Fig. \ref{eastrack}.
A high statistics of UHE proton events with energies from $10^{18}$ eV to  $10^{19}$ eV have been simulated.
We have chosen zenith angles between 10$^{\circ}$ and 60$^{\circ}$. All events were distributed randomly having their 
impact point within an area of 10 $\times $ 10 km. The FOV projected on ground corresponds to an area of 8.4 $\times $ 8.4 km.

 \section{Angular Resolution Estimates}
Even though in reality the probability to measure UHECR generated EAS is relatively low, we have simulated a large number of showers
to make a statistical study of the expected angular resolution capabilities of the instrument.
Out of the simulated  12301 events with uniformly distributed energies, inclinations and impact points, 2623 have been triggered. 
The reason for the low number of triggering events is that only about 25\% of the events have significant parts of the shower track within the
FOV of the telescope. Out of the triggered events 2480 can be reconstructed.
We regard events as successfully reconstructed if the pattern recognition module is able to identify enough counts as signal
and the following fit of the track direction module converges. Of course, even in these cases the value of the reconstructed 
directions might have a relatively large error.

We measure the angular resolution by $\Delta \Theta= \Theta_{reconstructed} - \Theta_{simulated}$ (zenith angle)
and $\Delta \Phi= \Phi_{reconstructed} - \Phi_{simulated}$ (azimuth). 
 \begin{figure*}[t]
  \centering
  \includegraphics[width=\textwidth]{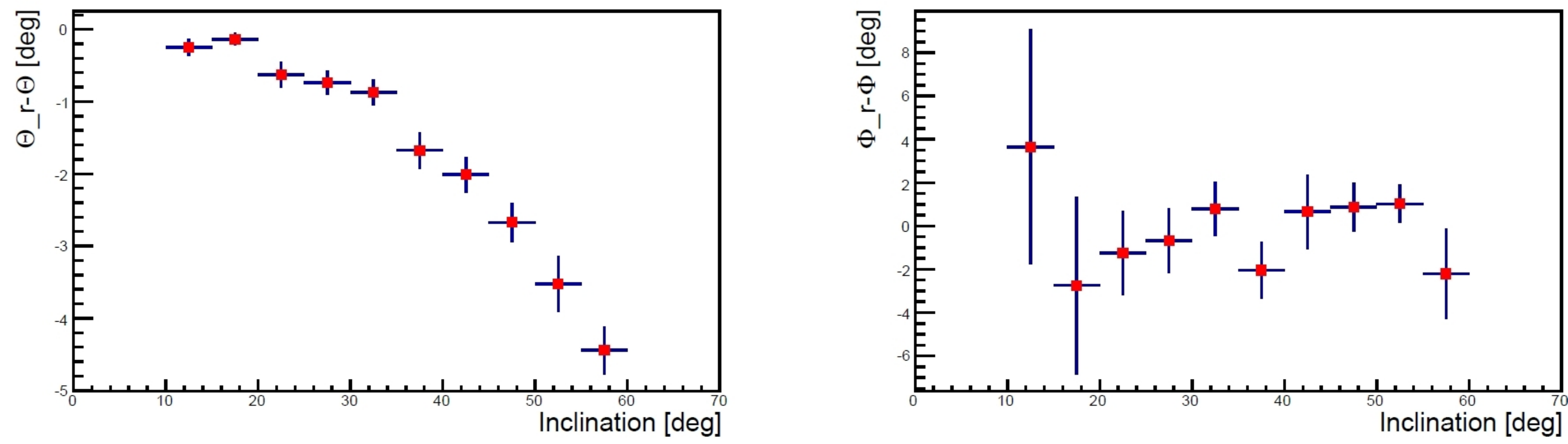}
  \caption{Mean value and standard deviation of $\Delta \Theta= \Theta_{reconstructed} - \Theta_{simulated}$ (left)
and $\Delta \Phi= \Phi_{reconstructed} - \Phi_{simulated}$ (right) plotted against the true zenith angle (inclination).}
  \label{thetaphires}
 \end{figure*} 
In Fig. \ref{thetaphires},w e can clearly see that the direction of UHECR can be resolved sufficiently when the zenith angle is between a little larger than 10$^{\circ}$ up to approx. 50$^{\circ}$. The lower limit is due to the fact, that the visible track on the FS is too short to make a meaningful fit which is the base for angular reconstruction. For zenith angles exceeding about 50$^{\circ}$, the shower track does not fit entirely on the PDM, therefore we lose information.

\begin{figure}[!h]
  \centering
  \includegraphics[width=0.45\textwidth]{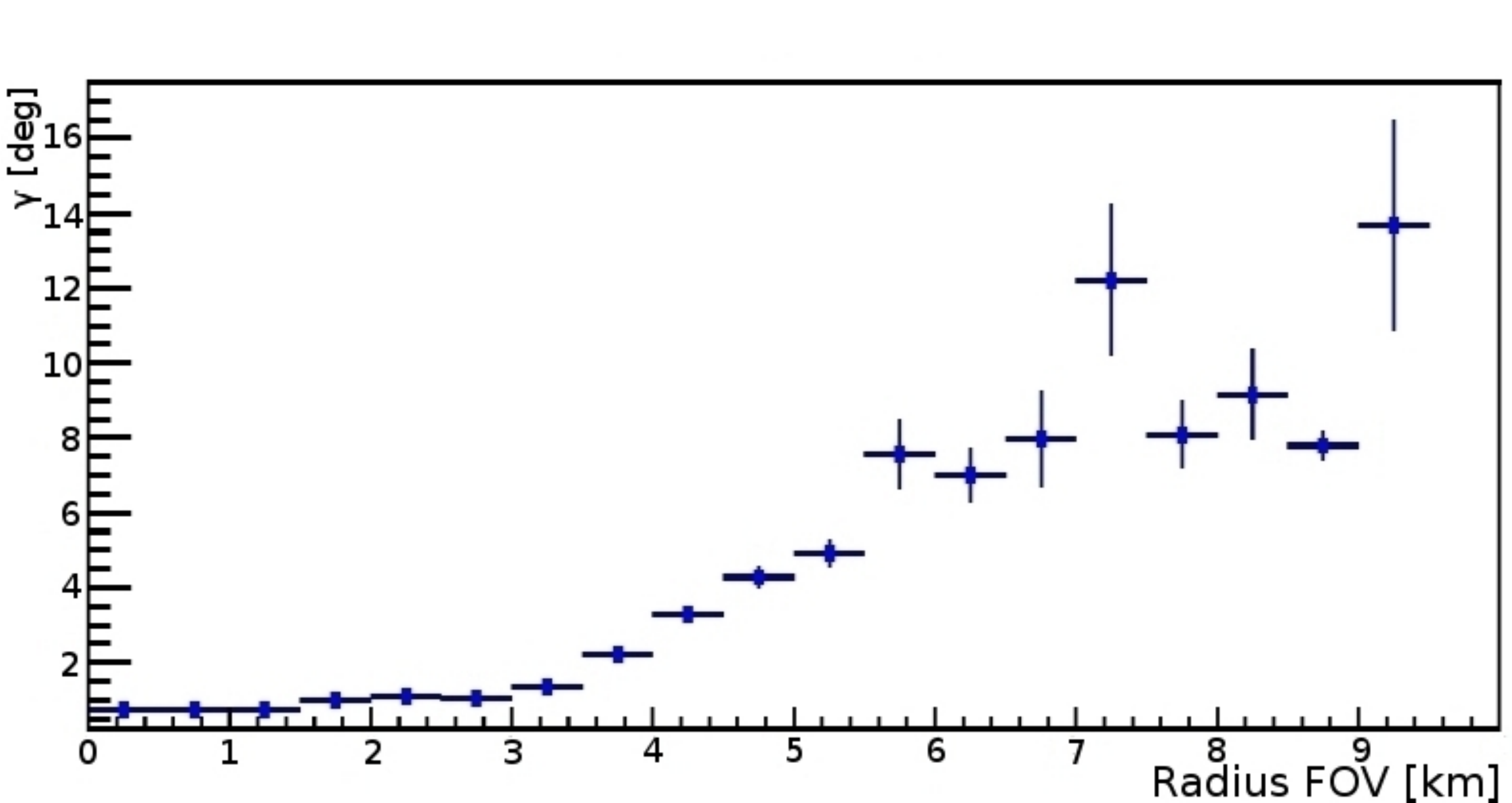}
  \caption{Expected angular resolution expressed by $\gamma$ vs. radius of FOV projected on ground. Data points indicate the mean value, error bars represent the standard deviation.} 
  \label{gammarad}
 \end{figure}

To evaluate this effect, we plot $\gamma$ (the angle between the true shower direction and the reconstructed in three-dimensional vector space) against the radius of the FOV projected on ground. See Fig. \ref{gammarad}. Obviously, the probability that parts of the signal are lost, increases at the edge of the FOV. Therefore the uncertainty in direction reconstruction increases.

\section{Conclusions}
The EUSO-Balloon specifications have been successfully implemented in the ESAF software package. We have carried out a large number of test to ensure
proper treatment of the single components within the simulations. To evaluate the expected angular resolution of the detector we have conducted a 
study with UHE protons as primary particles. We have shown how typical air showers would appear on the instruments focal surface. 
These test demonstrate the instrument's ability to detect EAS. With this study we also constrained conditions for which angular reconstruction
of the UHECR is possible. 

As long as the shower track remains inside $\pm$5 $\times$ $\pm$ 5 km, the direction can be resolved within small errors. 
In an area larger we can still give reasonable estimates. It is important to point out that the results
of this study are rather conservative. In reality only a small amount of data will be available. Therefore a more careful analysis 
of each event will allow to estimate the direction of the UHECR more precisely.

We can conclude that in case of a triggering UHECR event we will be able with a high probability to reconstruct its arrival direction within reasonable 
error boundaries.

\section{Acknowledgements}
We wish to thank RICC the RIKEN Integrated Cluster of Clusters for an allocation of computing resources.

This work has partly been funded by the ESA topical team activities fond, the DLR - Deutsches Zentrum f\"ur Luft- und Raumfahrt and by the
Helmholtz Alliance for Astroparticle Physics, Germany.

\clearpage

%% file: icrc2013-0777.tex


\title{Simulating the JEM-EUSO Mission: Expected Reconstruction Performance}

\shorttitle{JEM-EUSO}

\authors{
T. Mernik$^{1,2}$,
A. Guzman$^{1,2}$,
F. Fenu$^{1,2}$,
K. Shinozaki$^{1,2}$,
A. Santangelo$^{1,2}$,
M.E. Bertaina$^{3,4}$,

for the JEM-EUSO Collaboration.
}

\afiliations{
$^1$ Institute for Astronomy and Astrophysics, Kepler Center, University of T\"ubingen, Germany \\
$^2$ RIKEN Advanced Science Institute, Wako, Japan \\
$^3$ Dipartimento di Fisica, Universit\'a di Torino, Italy \\
$^4$ INFN, Istituto Nazionale di Fisica Nucleare - Sezione di Torino, Italy
}

\email{mernik@astro.uni-tuebingen.de}

\abstract{JEM-EUSO (Extreme Universe Space Observatory onboard the Japanese Experiment Module) is a space borne UV-telescope which will be mounted on the ISS 
(International Space Station) in 2017.
It is designed for the observation of extensive air showers (EAS) induced by ultra-high-energy cosmic rays (UHECR) above an energy of a few $10^{19}$ eV by using the earth's
atmosphere as a large detector. Due to the amount of monitored target volume it gains an effective aperture of approximately $4 \cdot 10^5$ km$^2$ sr (in nadir mode).
Thus, during the time of the mission JEM-EUSO will measure several hundred of events $>$ $5 \cdot 10^{19}$ eV and improve the statistics in this 
part of the UHECR spectrum significantly.

The EUSO Simulation and Analysis Framework (ESAF) is a software for the simulation of space-based UHECR detectors.
Each of its modules is devoted to a specific aspect of EAS generation and detection: Interaction of the primary in atmosphere,
air shower development, light transport to the telescope, propagation of photons within the instrument and detector response.
From the recorded data the properties of the primary (energy, arrival direction and species) can be reconstructed.

In this article we describe the simulation of the JEM-EUSO mission and illustrate reconstruction strategies used in ESAF.
Furthermore we present the expected instrument performance in terms of resolution of the atmospheric depth of the shower maximum (Xmax), energy and angular resolution.
}


\maketitle

\section{Introduction}
JEM-EUSO is a space based UV telescope 
for the detection of ultra-high-energy cosmic rays \cite{wwtakahashi,wwcasolino}. 
Mounted at the Japanese Experiment Module on the ISS, JEM-EUSO will monitor the earth's atmosphere for extensive air showers (EAS)
that are created when UHECR interact with the nitrogen molecules of the atmosphere.
By measuring the fluorescence and Cherenkov photons of the shower, analysis allows to reconstruct the shower properties and therefore
energy, arrival direction and kind of the primary UHECR particle.
Using the well established UV detection technique on a much larger volume of air than any of the earth based experiments are capable of,
JEM-EUSO will reach an instantaneous aperture of about $4 \cdot 10^{5} km^{2} sr$ (nadir mode) \cite{AdamsJr201376}.
Hence, during the lifetime of the mission, JEM-EUSO will observe several hundreds of UHECR events with energies exceeding 
$5 \cdot 10^{19}$ eV.

\section{JEM-EUSO}
The JEM-EUSO detector is a UV telescope using a refractive optics of Fresnel lenses to focus the incoming photons
in the wavelength range of 300 to 430 nm onto a \emph{photo detector module} (PDM) consisting of an array of \emph{ multi anode 
photomultiplier tubes} (MAPMT).
The trigger logic implemented at the cluster control board continuously seeks for pattern characteristics meeting those of signal tracks we would
expect from a moving EAS.  When a trigger is issued, the time frame of 128 GTU (gate time units, 2.5 $\mu$s) is saved to disc or transferred by the telemetry.
Due to the wide field of view (FOV) of 2 $\times$ 30$^{\circ}$ (circular shape with cuts at the sides) and the altitude of approximately 400 km, a large target volume
can be monitored from space. Permanent surveillance of atmospheric conditions such as clouds is done by the infrared camera and LIDAR system. 
Calibration of the instrument will be possible by built in LEDs and additional xenon flashers from ground.

\section{Simulations}
The EUSO Simulation and Analysis Framework is a software package for the simulation of UHECR space detectors\cite{esaf}. It is an object-oriented C++ code, based
on ROOT \cite{root}.
Its development at the time of the EUSO  mission (\cite{euso}) was stopped in 2004. 
During the JEM-EUSO study, we have reactivated the code and improved parts of the software when necessary.
New algorithms have been implemented for pattern recognition, angular, X$_{max}$ and energy reconstruction. 
The framework can perform entire end-to-end simulations, starting from
primary particle/atmosphere interaction,  development of the EAS, creation of UV - fluorescence and Cherenkov photons and their propagation to the detector.
 
Second part of the simulation accounts for the processes inside the detector. The photons trajectories through the optics is simulated by a ray tracing module. Following that, the MAPMT and electronics response are simulated.

Important facts for the simulation of the JEM-EUSO mission are:
\begin{enumerate}

\item Altitude= 400 km

\item Background: uniform, 500 photons  m$^{-2}$ sr$^{-1}$ ns$^{-1}$ 

\item Field of View: circular, 30$^{\circ}$ radius with side cut optics

\end{enumerate}
The background  is simulated only at the electronics level
in order to save computing time. The area in which the impact points of the showers are distributed is greater than the corresponding projection of the telescope's FOV on the earth surface. This is done for the purpose of checking
for triggers of stray light photons and analysis of the behavior of signal tracks traversing the FOV only to some extent. 

We have simulated proton showers with discrete energies: $5 \cdot 10^{19}$ eV, $7 \cdot 10^{19}$ eV, $10^{20}$ eV, $3 \cdot 10^{20}$ eV and $10^{21}$ eV.
The zenith angles are 30$^{\circ}$, 45$^{\circ}$, 60$^{\circ}$ and 75$^{\circ}$.

For any event reconstruction, a pattern recognition module is applied first of all, to separate 
the signal tracks from background. Then, the pixel counts are forwarded to the track direction reconstruction module.
There is a choice of different angular reconstruction algorithms implemented in ESAF. For details see \cite{bertaina}.
For a successful event reconstruction, we need to have enough information on the shower track in terms of 
photons. I.e., when the pattern recognition module is not able to find a sufficient number of photon counts
belonging to the signal, the reconstruction is not possible.
 
\section{Reconstruction}
The traversing air shower appears on the focal surface as a faint spot moving very quickly. The signal is embedded in background coming from the atmospheric
night glow, but also weather phenomena and bypassing cities contribute. Moreover, the scattered EAS photons contribute significantly to the background around the signal track. 

For a successful event reconstruction, the signal has to be disentangled from background, first of all. Afterwards, direction and energy reconstruction 
algorithms can be applied.

\subsection{PWISE}
\begin{figure}[!h]
  \centering
  \includegraphics[width=0.4\textwidth]{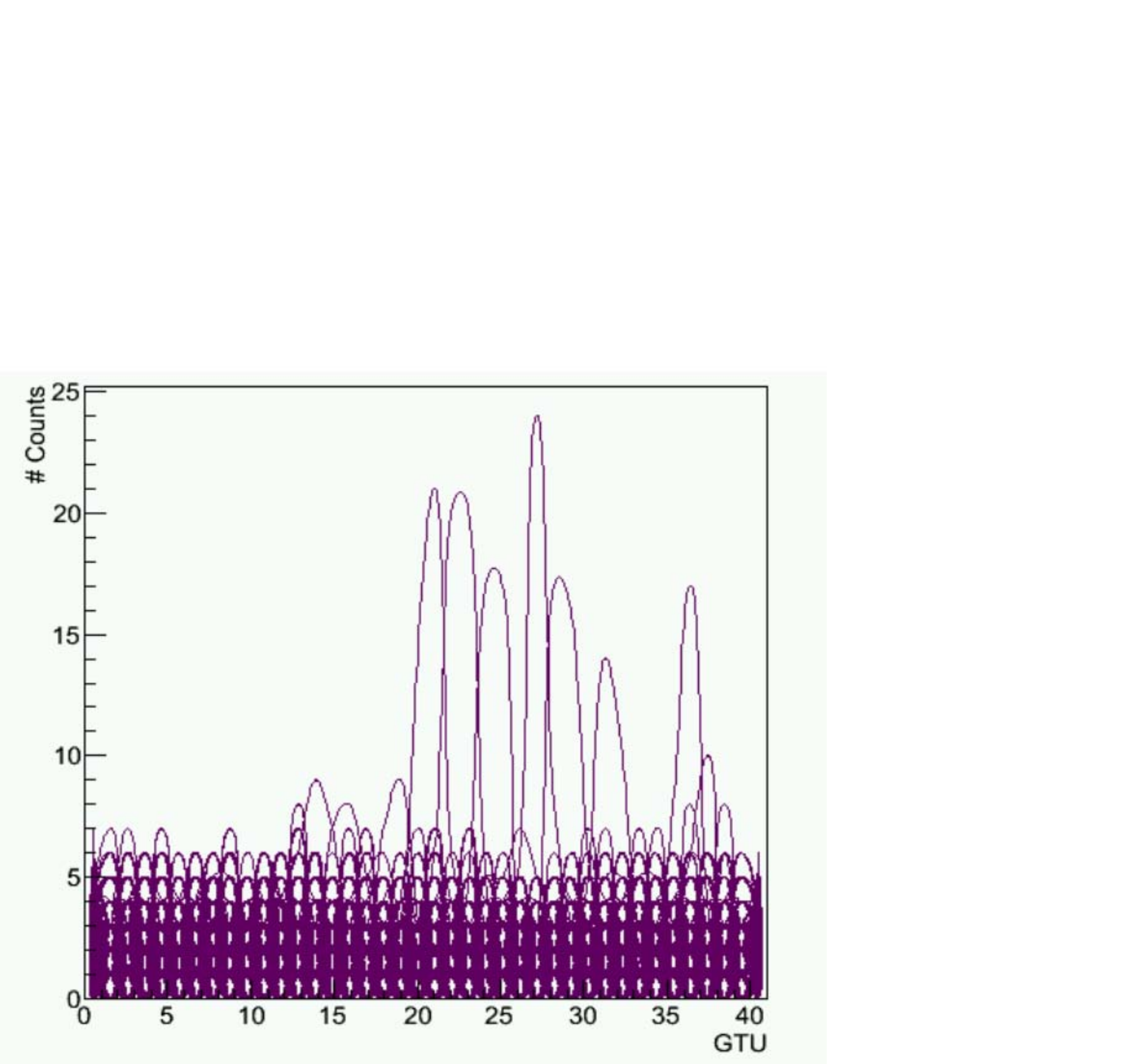}
  \caption{PWISE algorithm: For each pixel the count distribution in time is analyzed. Here background counts up to 5 can be seen intermixed with signal counts which
are significantly higher and happen during a certain time interval. }
  \label{pwise}
 \end{figure}

The \emph{Peak and WIndow SEarching Technique} (PWISE) selects 
photon-counts coming from the EAS, and at the same time it filters out 
multiple-scattered photons which results in a ``fuzzy'' image of the track.  
This effect appears as a consequence of their shifted arrival time due to the
multiple scattering.
PWISE is an algorithm that analyses each pixel individually for possible signal traces (See Fig. \ref{pwise})
\begin{description}
\item[\bf{Step 1}] For each pixel, PWISE  only considers pixels whose highest
photon-count (peak) is above a certain  threshold (\emph{peak-threshold}).
\item[\bf{Step 2}] Next  PWISE searches for the time window with the highest
 signal-to-noise ratio (SNR).
\item[ \bf{Step 3}] We check if the maximum SNR is above a given 
\emph{SNR-threshold}. Only if the SNR is above the threshold we select 
the photon-counts within the time window that maximizes SNR. The selected 
photon-counts are then passed on to the next reconstruction module.
 \end{description}

\subsection{Angular Reconstruction}
\begin{figure}[!h]
  \centering
  \includegraphics[width=3.in]{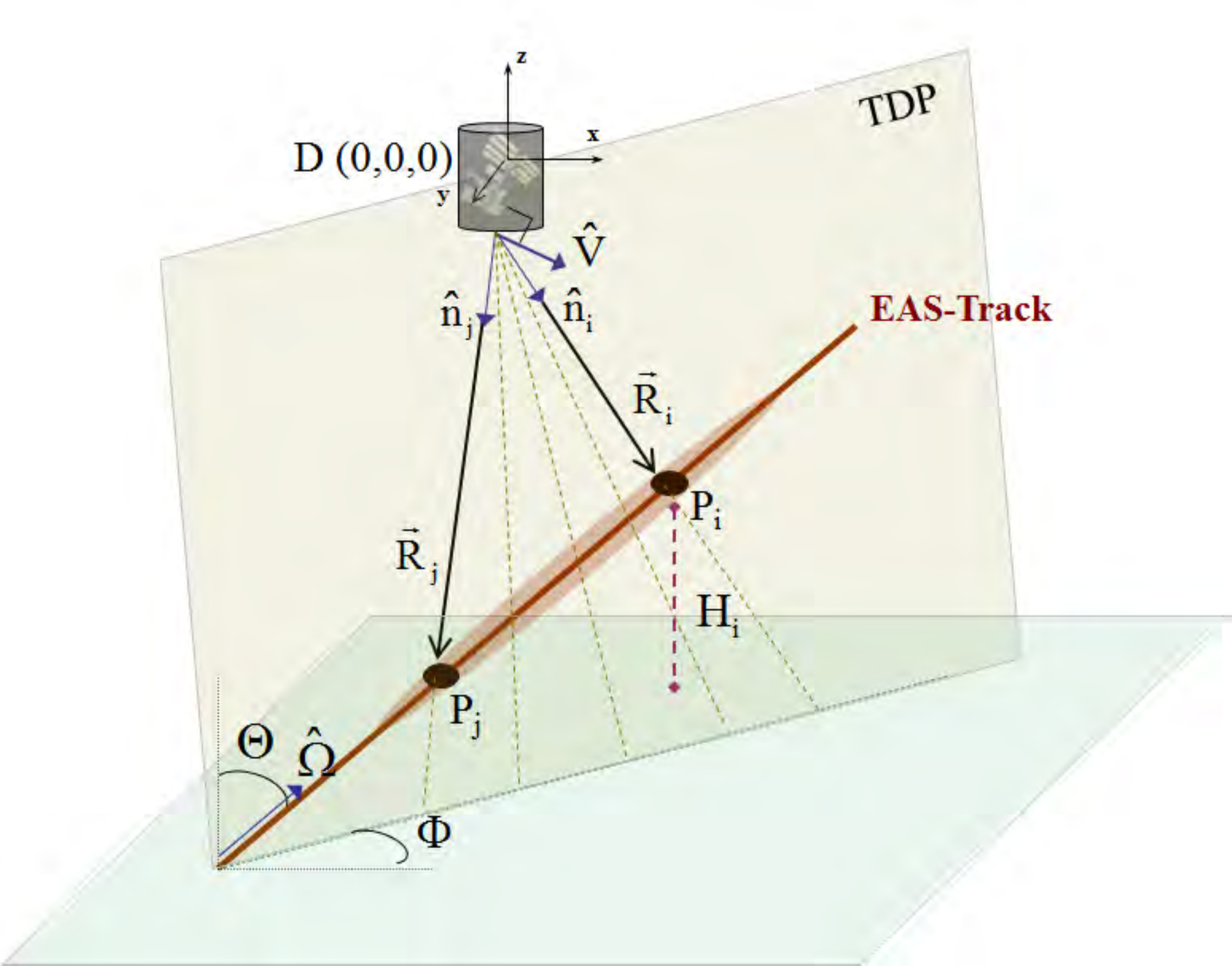}
  \caption{Sketch of the system: Air shower - detector, to reconstruct the arrival direction of the UHECR: Within the track-detector-plane (TDP) defined by the unit vector $\hat V$, photons emitted at different times t$_j > t_i$ reach the 
detector from certain directions $\hat n_i$, $\hat n_j$ after traversing R$_i$, R$_j$ in atmosphere. From the timing information and arrival angle of the 
shower photons, the direction of the primary $\hat \Omega(\Theta,\Phi)$ can be determined.}
  \label{recosyst}
 \end{figure}

From the geometrical properties of the signal track on the focal surface the arrival direction of the primary can be computed by a variety 
of methods implemented in ESAF as described in more detail in \cite{bertaina,wwangurecoperformance,esaf}. Fig. \ref{recosyst} shows the coordinate systems of the EAS and the detector.
In the current configuration there are 5 different algorithms implemented in ESAF. Their performances depend on conditions such as 
energy and inclination of the primary UHECR but also on atmospheric conditions:
\begin{itemize}

\item \emph{Analytical Approximate 1}: The angular velocities of the signal track in the x-t and y-t planes are linearly fitted. 
The arrival angle of the primary is derived by geometrical estimations. 

\item \emph{Analytical Approximate 2}: The angular velocity of the signal track on the z-t plane is linearly fitted. 
The arrival angle of the primary is derived by geometrical estimations. 

\item \emph{Numerical Exact 1}: a $\chi^2$ minimization is performed between the activation times of pixel induced by the actual 
signal to those induced by a signal track theoretically computed. 

\item \emph{Numerical Exact 2}: a $\chi^2$ minimization is performed between arrival angles of photons coming from the actual 
signal to those induced by a signal track theoretically computed.  

\item \emph{Analytically Exact 1}: a $\chi^2$ without prior knowledge of the TDP, this method reconstructs the direction of the primary by using 
 using the exact relations between pixel directions in the FOV and photon's arrival times.  

\end{itemize}

\subsection{Energy and Xmax Reconstruction}
The \emph{PmtToShowerReco} is the energy and $X_{max}$ reconstruction algorithm developed in the framework of the JEM-EUSO mission.
It is structured in several subsequent steps each of which performs a sub-task. Such an algorithm can however be seen just as a
initial guess on the shower parameters. After the trigger signal has been issued, the time profile of detected counts must be reconstructed also taking into 
account the losses occurring due to gaps between the single PDMs. The background contamination must be estimated and subtracted.
Following that, the light curves at the focal surface and at the pupil must be estimated after a detailed modeling of the detector response.
The shower geometry in atmosphere must be reconstructed according to the timing of the shower track on the focal surface. 

\begin{figure}[!h]
  \centering
  \includegraphics[width=0.4\textwidth]{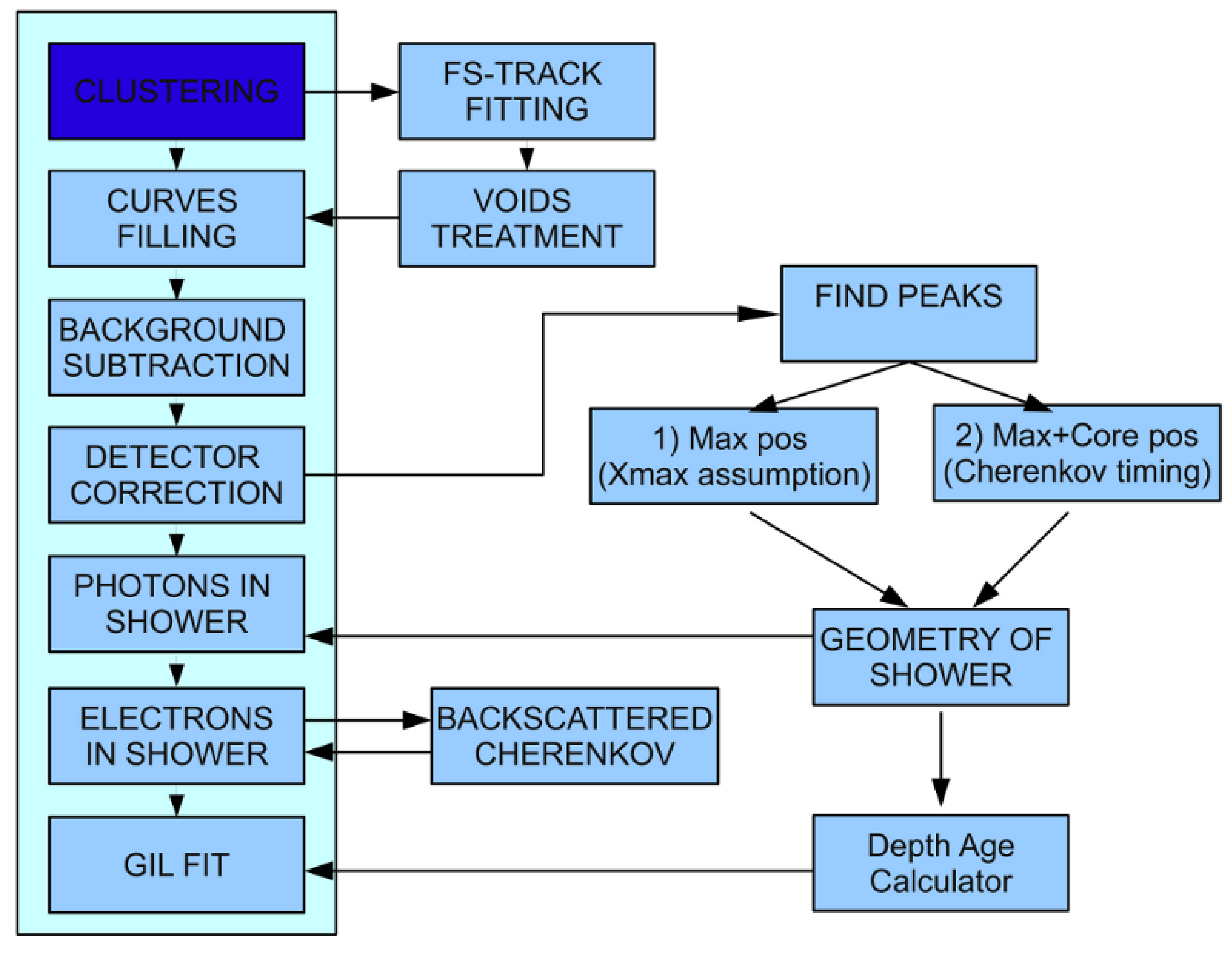}
  \caption{Scheme of energy and X$_{max}$ reconstruction} 
  \label{erecoscheme}
 \end{figure}

Several methods have been developed for this purpose. Essentially, the first method uses the time separation between the fluorescence maximum and the Cherenkov peak to give an estimate of the maximum altitude. The second, assumes a parametrization of the maximum slant depth and the reconstructed shower direction to give an assessment of the altitude of the maximum. This parameter allows on its own the reconstruction of the shower geometry for each time of its development. Therefore, at this point a parametrization of the energy distribution of the secondary particles is possible given the standard cosmic ray theory. Eventually, the light spectrum is computed together with the fluorescence and Cherenkov yield. See Fig. \ref{erecoscheme}

Given the reconstructed spectrum and the geometry of the shower it is possible to calculate the number of photons at the shower position.
Knowing fluorescence and Cherenkov yield allows the calculation of the number of electrons in the shower. Such a curve can then be fitted with any standard 
shower parametrization to obtain the energy and X$_{max}$ parameters. Nevertheless, such a method is affected by the statistical quality of the counts 
curve and therefore the backward iteration of the entire chain is needed in order to find the best fitting parameters and a confidence interval.

\section{Expected Reconstruction Performance}

 \subsection{Angular Resolution}
\begin{figure}[t]
  \centering
  \includegraphics[width=0.4\textwidth]{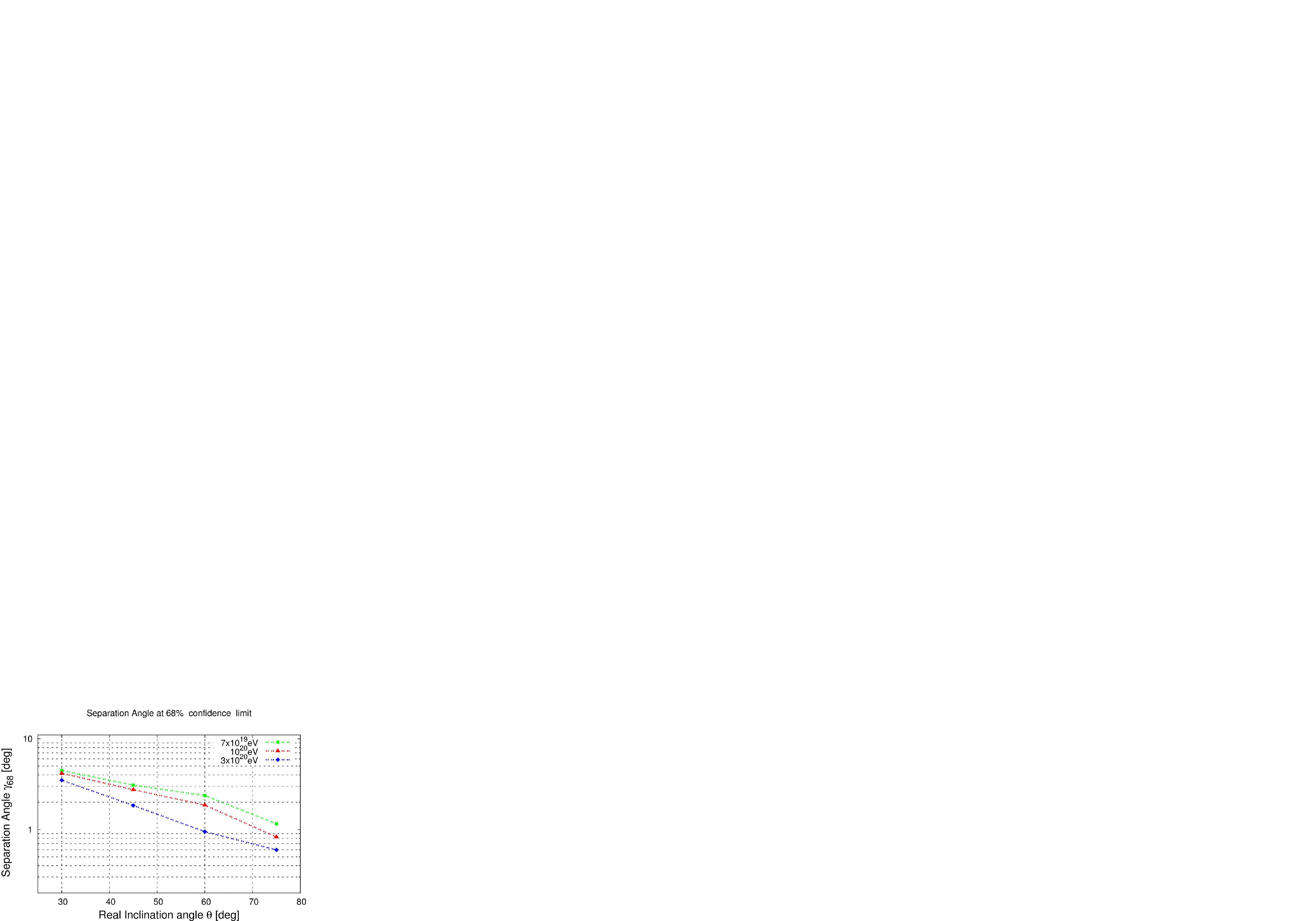}
  \caption{The estimated angular resolution performance in terms of$\gamma_{68}$ plotted against the true inclination for three different energies. \cite{bertaina} }
  \label{gamma68}
 \end{figure}

We have simulated a large number of showers to make a statistical study of the expected angular resolution capabilities of the instrument.
For each combination of energy and inclination we have used a statistics of about 700 events to minimize statistical fluctuations.
As a measure for the expected angular resolution we use $\gamma_{68}$. The separation angle $\gamma$ is the angle between the true and the reconstructed direction 
in three-dimensional vector space. It is positive by definition.  $\gamma_{68}$ means that 68 \% of the events reconstructed have a separation angle less than this value. It depends on the true
inclination of the shower and on its energy. Both affect the brightness of the track and therefore the amount of information we have to reconstruct it.
See Fig. \ref{gamma68}.

\begin{figure}[!h]
  \centering
  \includegraphics[width=0.4\textwidth]{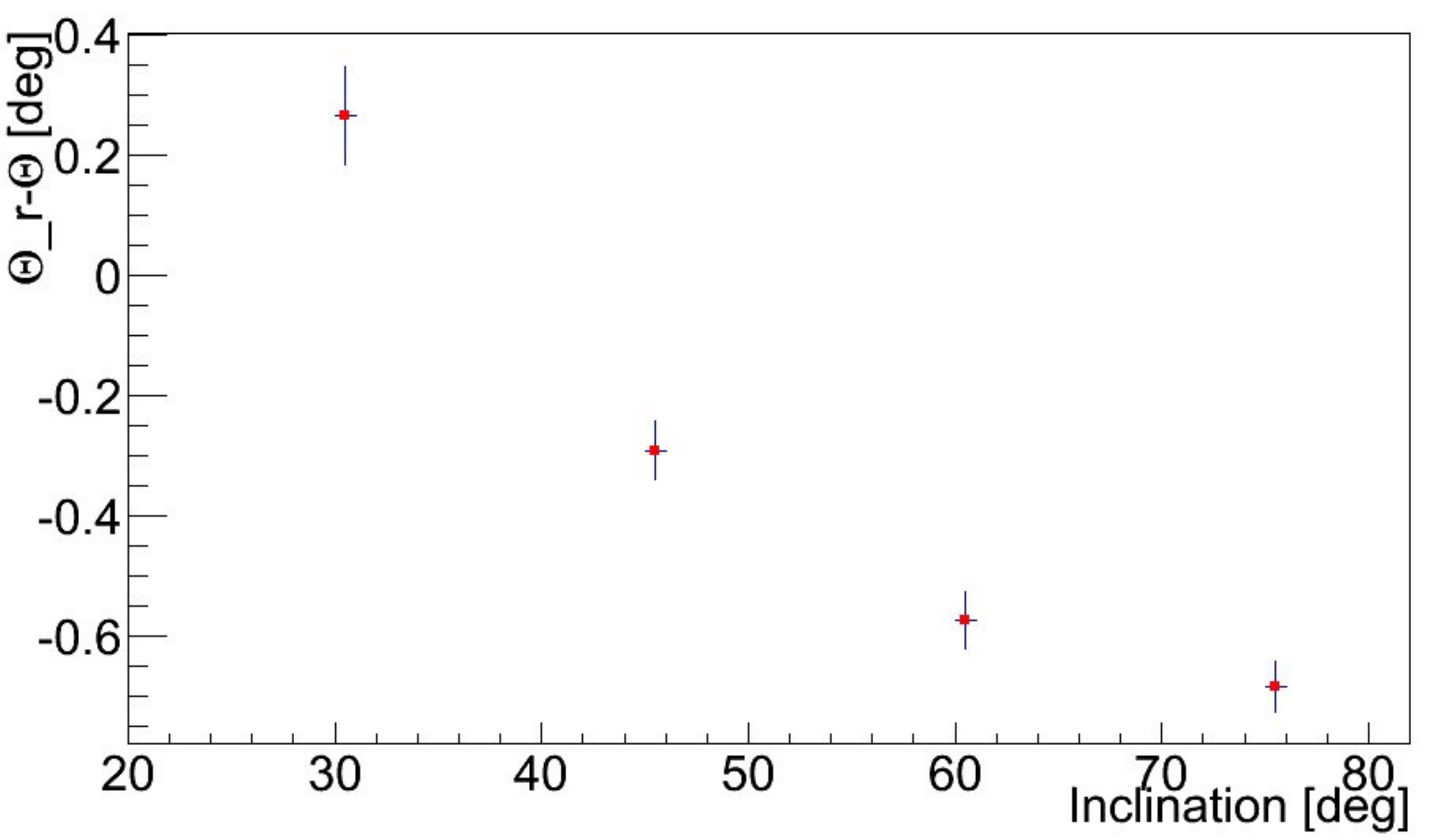}
  \caption{Expected $\Theta$ resolution expressed by mean value of $\Delta \Theta= \Theta_{reconstructed} - \Theta_{simulated}$ and standard deviation as function of the true zenith angle. } 
  \label{thetasyst}
 \end{figure}

However, considering the $\gamma_{68}$ value does not reveal information about the contributions of the different components ($\Theta$ and $\Phi$).
Thus, we also look at the determination error in $\Theta$ expressed by $\Delta \Theta= \Theta_{reconstructed} - \Theta_{simulated}$ as shown
in Fig. \ref{thetasyst}.

\subsection{Energy and Xmax Resolution Estimates} 
For the  energy and X$_{max}$ resolution study we have used another set of events. 
Here 8000 events have been simulated with an energy of $10^{20}$ eV and several inclination angles. 
The impact point of these events on ground have been selected to be in an area of  $\pm$ 20km both in X and Y (the projection onto the earth's surface).

\begin{figure}[!h]
  \centering
  \includegraphics[width=0.4\textwidth]{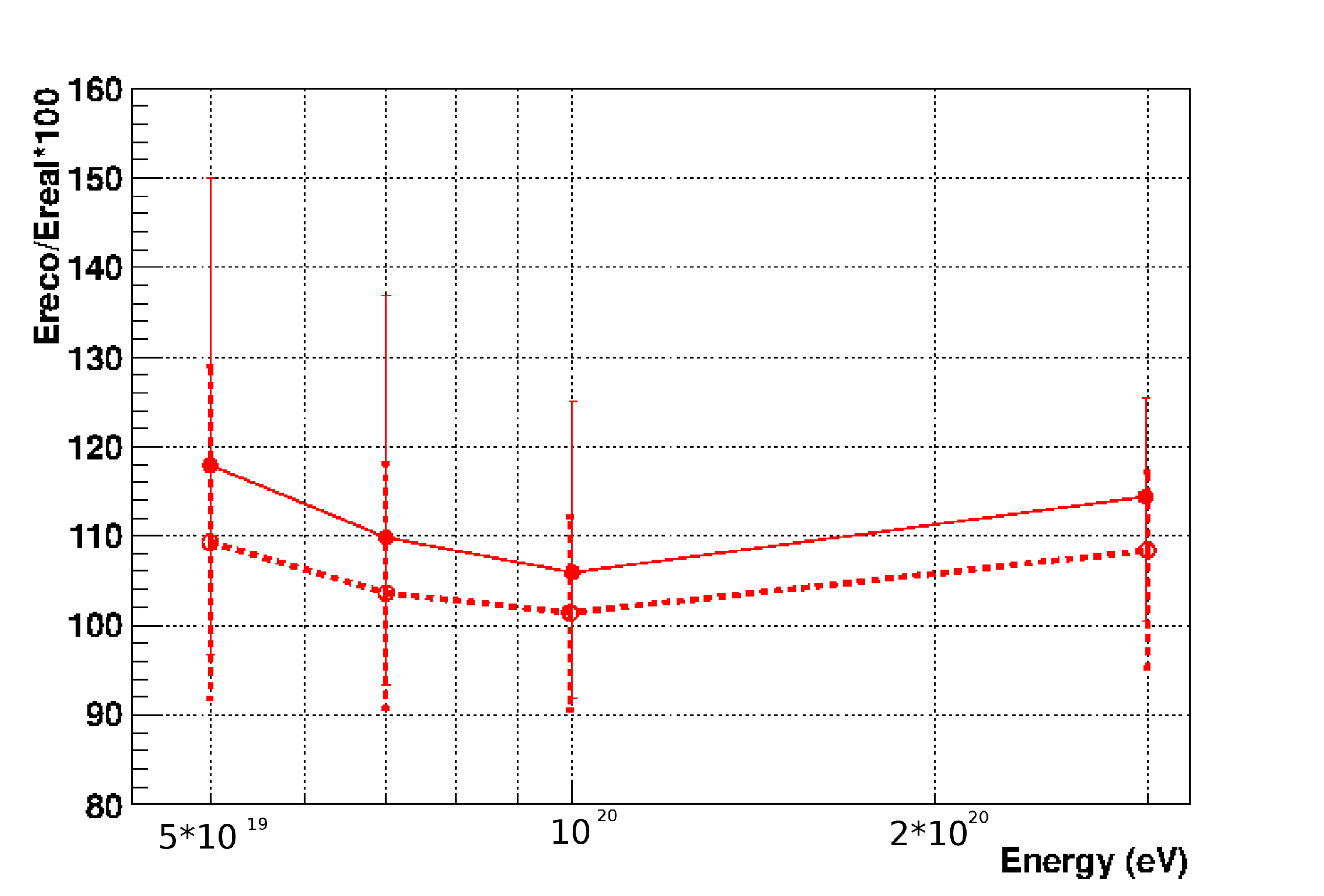}
  \caption{A preliminary estimate of the expected energy resolution of JEM-EUSO with Cherenkov method for 45$^{\circ}$ inclined events. 
  Data points show the median, the error bars show 68\% confidence limit. solid line: all FOV, dashed line: core position of events inside 100 km radius from centre.}
  \label{eres}
 \end{figure}

Due to the very peculiar condition under which the events have been simulated, this small study cannot be regarded as an estimate of JEM-EUSO's energy resolution. In fact, these examples must be seen as a proof that the algorithms are operational.
For illustrations for energy reconstruction see Fig. \ref{eres}.

\begin{figure}[!h]
  \centering
  \includegraphics[width=0.4\textwidth]{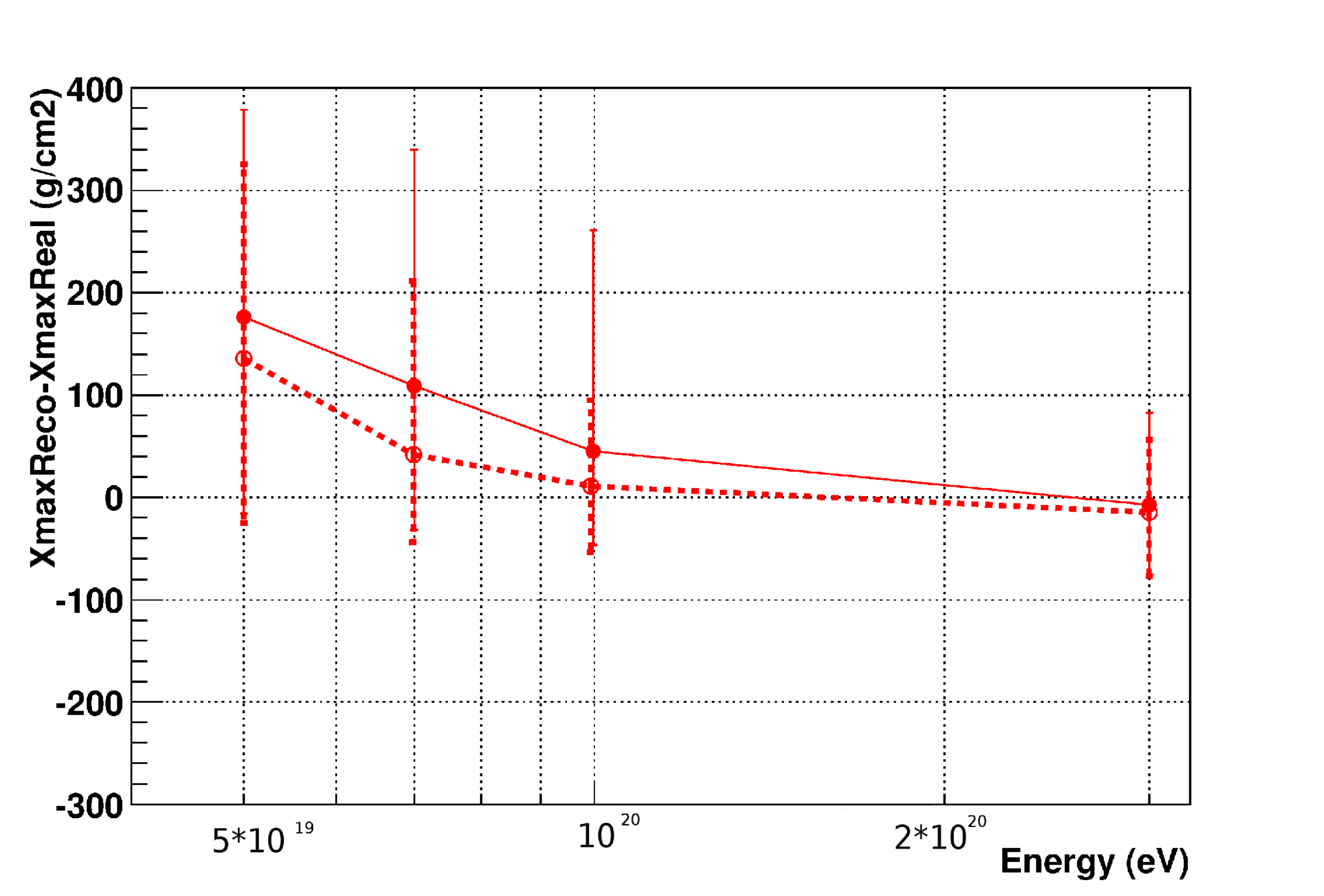}
  \caption{A preliminary estimate of the expected X$_{max}$ resolution of JEM-EUSO with Cherenkov method for 45$^{\circ}$ inclined events. 
Data points show the median, the error bars show 68\% confidence limit. solid line: all FOV, dashed line: core position of events inside 100 km radius from centre.}
  \label{xres}
 \end{figure}

The X$_{max}$ study uses the same statistics of events as in the energy study described above, since the  X$_{max}$ algorithm is part of the same reconstruction module. Fig. \ref{xres} shows a preliminary estimate of the expected X$_{max}$ resolution in the central region of the FOV.

\section{Conclusions}
We have updated and improved the ESAF software package in order to investigate the JEM-EUSO technical specifications and scientific requirements. 
Now, ESAF provides an independent and parallel assessment of the JEM-EUSO performance.
We have made a complete end-to-end simulation and analysis of a large statistics of events for the JEM-EUSO detector. Hence, we have demonstrated the robustness and 
precision of the code. 

Such methods are now under assessment to provide their performances for a much wider class of events, such as neutrinos, iron nuclei or $\gamma$ rays. 
The results presented in this paper are compliant with the scientific requirements of the mission.
However, the reconstruction performance estimates given in this study must be understood as rather conservative. 
Once, real data will be available, it will be investigated on an event by event basis. Finetuning of the parameters of the algorithms will allow for a more precise estimation of direction, energy and X$_{max}$ of the UHECR.
Moreover, at the moment further effort is made to implement more refined pattern recognition modules and angular reconstruction algorithms. 
Thus, we believe that our result will even improve for certain conditions.

\vspace*{0.3cm}

{\bf Acknowledgements:}
This work has partly been funded by the ESA topical team activities fond and
the DLR, Deutsches Zentrum f\"ur Luft- und Raumfahrt. We wish to thank RICC the RIKEN integrated cluster of 
clusters for an allocation of computing resources.
Moreover, FF would like to express his gratitude for being supported by the IPA Program of RIKEN.

\clearpage


%% file: icrc2013-1283.tex


\title{Simulations and the analysis of fake trigger events background in JEM-EUSO experiment}

\shorttitle{Fake trigger background study}

\authors{
Svetlana Biktemerova$^{1}$,
Blahoslav Pastircak$^{2}$,
Mario Bertaina$^{3}$,
Pavol Bobik$^{2}$,
Francesco Fenu$^{4,5}$,
Karel Kudela$^{2}$,
Marian Putis$^{2}$,
Miroslav Staron$^{2}$,
Kenji Shinozaki$^{5,4}$,
for the JEM-EUSO Collaboration.
}

\afiliations{
$^1$ Joint Institute for Nuclear Research, Joliot-Curie 6, 141980 Dubna, Russia \\
$^2$ Institute of Experimental Physics Slovak Academy of Sciences, Watsonova 47, 04001 Kosice, Slovakia  \\
$^3$ Department of General Physics, University of Torino, Torino, Italy  \\
$^4$ Institute for Astronomy and Astrophysics, Kepler Center, University of Tubingen, Sand 1, 72076 Tubingen, Germany \\
$^5$ RIKEN, 2-1 Hirosawa, 351-0198 Wako, Japan \\
}

\email{sveta.biktemerova@gmail.com}

\abstract
{
The goal of the trigger system is to detect the occurrence of scientifically valuable signal  
among very huge background noise detected by JEM-EUSO telescope. The UV background registered 
by JEM-EUSO is randomly distributed. We study if these random processes produce fake pattern, 
which could be mistakenly interpreted as extreme energy cosmic rays events. For this purpose 
very huge amount of measurements on one photo detection module with only detector noise were 
simulated. To distinguish between such simulated fake events and real extreme energy cosmic 
rays events we have applied Hough transform pattern recognition method. The presented results 
provide reasonable estimation, that background cannot produce a patterns whose can be mistaken with real event.
}


\maketitle

\section{Introduction}
\label{sec:introduction}

The JEM-EUSO \cite{bib:jem-euso} is an Extreme Energy Cosmic Rays (EECR) experiment whose main 
purpose is the study of the End of Cosmic Rays spectrum above the GZK cut-off. The detector is 
basically a large field of view UV camera, pointing toward the earth atmosphere, to detect and 
measure the fluorescence light imprint produced by development at speed of light of Extensive 
Air Showers (EAS). Typically, for a $10^{20} eV$ EAS, a few thousands photons are expected on 
the JEM-EUSO detector focal surface (FS). However, the background photons are much more than 
those of signal. Therefore the background reduction is essential for such space observatory of 
EECRs. It is the aim of the trigger to try to extract the signal from the background sea. The 
electronics will have to reject as much counts as possible without rejecting the signal itself. 
Fortunately the signal has some peculiar characteristics that can be used to distinguish it. 
The shower generate a spot moving on the focal surface. On the other hand, the background is 
distributed randomly. Despite of it is necessary to assess, if the random processes do not 
produce fake patterns, which could be mistakenly interpreted as EECR events. For this purpose a 
huge amount of measurements with only background events have to be simulated. The obtained 
results  would be consequently analysed by several pattern recognition algorithms to verify the 
probability of registration a fake trigger events in several trigger conditions. 

\section{Trigger}
\label{sec:trigger}

The role of the trigger is to select EAS events rejecting the random background. The random 
hits come from the fluorescence photons having undergone Mie and Rayleigh scattering in the 
atmosphere induced by the night glow, the air glow, the moon light and light cities and the 
reflected  stars light. This background needs to be greatly reduced. To reject it, JEM-EUSO 
electronics operate with several trigger levels. The trigger scheme relies on the partitioning 
of the FS in subsections.

The FS is covered by a large numbers of photo-detector tubes mechanically structured in series 
of similar pieces, the one embedded in the others. The largest piece is a photodetector module 
(PDM). The whole FS is made of 137 such PDM's. Each PDM structure is itself a squared matrix of 
3x3 smaller elements called elementary cells (EC). Each EC is a squared matrix of $2 \times 2$ 
multianode photomultipliers. An EC is a $12 \times 12$ pixel matrix, corresponding to 144 
channels. PDM is a $36 \times 36$ pixel matrix corresponding to 1296 channels. Each PDM probe 
a squared pad of $27km  \times 27km$, which is large enough to contain a substantial part 
of the imaged trace under investigation (this depends on the zenith and eenergy of the EAS).  
The FS has in total 177600 channels. The simulations and following analysis have been performed 
for above described particular configuration (M36 configuration) of the detector, where the 
photomultipliers had 36 channels instead of 64 channels (M64 configuration) as it is the 
prefered option now.

The Table \ref{tab:trig} gives a possible reduction of the trigger rates that could be 
achieved at various trigger levels \cite{bib:trig0}, \cite{bib:trig1} for M36 configuration. 

\begin{table}[h]
\begin{center}
\begin{tabular}{|l|c|c|}
\hline \bf Level & \bf Triggers rate & \bf Triggers rate \\ 
   & \bf  PDM level [Hz]  & \bf  FS level [Hz] \\ \hline
Photon trig (channel)   & $\sim 5.2 \times 10^{8}$ & $\sim 7.8 \times 10^{10}$ \\ \hline
Counting trig (EC)  & $\sim 7.1 \times 10^{5}$ & $\sim 1.1 \times 10^{8}$ \\ \hline
$1^{st}$  ~(PDM) &   &  \\ 
Persistency trigger& $\sim 7 $& $\sim 10^{3}$ \\ \hline
$2^{nd}$  ~(PDM cluster) &   &  \\ 
Linear track trigger & $ \sim 6.7 \times 10^{-4}$ & $\sim 10^{-1} $ \\ \hline
EECR expected rate  & $ \sim 6.7 \times 10^{-6}$ & $ \sim 10^{-3}$\\ \hline
\end{tabular}
\caption{ The trigger rate reduction on different trigger levels
}
\label{tab:trig}
\end{center}
\end{table}

General JEM-EUSO trigger philosophy asks for a system trigger organized into two main trigger 
levels. The system trigger works on the statistical properties of the incoming photon flux in 
order to detect the interesting events hindered in the background, basing on their position and 
time correlation.

The $1^{st}$ trigger level mainly operates to remove most of the background fluctuations by 
requiring a locally persistent signal above over a few GTU's duration. GTU is the gate time 
unit of the value $2.5 \mu s$, which is the temporal time resolution of detector electronics. 
In the $1^{st}$ level trigger named also PTT (Persistency Track Trigger) are the pixels grouped 
in boxes of $3 \times 3$. A trigger is issued if for 5 consecutive GTU's there is at least one 
pixel in the box with an activity higher than a preset threshold and the total number of 
detected photoelectrons in the box is higher than a preset value. These two values are set as a 
function of the average noise level in order to keep the rate of triggers on fake events at a 
few Hz per PDM.

The role of the $2^{nd}$ trigger level - Linear Track Trigger (LTT) is to find some tracks 
segments in three dimensions from the list of pixels provided by the first level, for each GTU 
time bin. The track speed has to be compatible with a point travelling at speed of light in 
whatever direction it propagates. So it follows the movement of the EAS spot inside the PDM 
over some predefined time, to distinguish this unique pattern of an EAS from the background. 
From a PTT trigger, the PDM electronics will send a starting point, which contains the pixel 
coordinates and the GTU which generated the trigger. The LTT algorithm will then define a small 
box around it, move the box from GTU to GTU and integrate the photon counting values. When the 
excess of integrated value above the background exceeds the threshold, an  LTT trigger will be 
issued. Currently it is foreseen to have a total of 67 starting points for the integration, 
which are distributed equally over time and position around this box. Each integration will be 
performed over $\pm 7$ GTU's for a predefined set of directions. The background-dependent 
threshold on the total number of counts inside the track is defined to reduce the level of fake 
events to a rate of 0.1 Hz per FS. These two trigger levels combined together reduce therefore 
the rate of signals on the level of $ 10^{9}$ at PDM level.

\section{Simulations}
\label{sec:simulations}

As already pointed, a crucial aspect of each simulation is the background.  In presence of a 
background a certain number of Fake Trigger Rates (FTR) is expected. Aim of a trigger algorithm 
is to reduce this rate without affecting too much the real events rate. The PTT and LTT trigger 
algorithms were implemented in ESAF - general simulation and analysis framework of  JEM-EUSO 
experiment \cite{bib:esaf}. These algorithms are optimized using stand alone Monte Carlo 
simulations to minimize the fake trigger rate against average background level. The standalone 
simlations are much faster and coud be performed in parallel in comparison with ESAF.

The massive simulation results of FTR obtained by fast and standalone simulation code, which 
contains the trigger algorithm together with background generation input were performed and the 
results from obtained data will be presented. In the code one PDM was simulated. The PTT and 
LTT trigger algorithms were implemented. The background source is the Poisson distribution of 
average $500~ photons~  m^{-2}  s^{-1}  sr^{-1}) = 2.1~ photons/pixel/GTU$.
Code is fast, but since to produce very huge statistics, it has to be run in parallel. Minimal 
needed statistics obtained by a year of continuos computing on nearly full PC cluster (over 200 
CPU cores), optimally several years (not possible to run continuosly).

\begin{figure}[!t]
  \centering
  \includegraphics[width=0.4\textwidth]{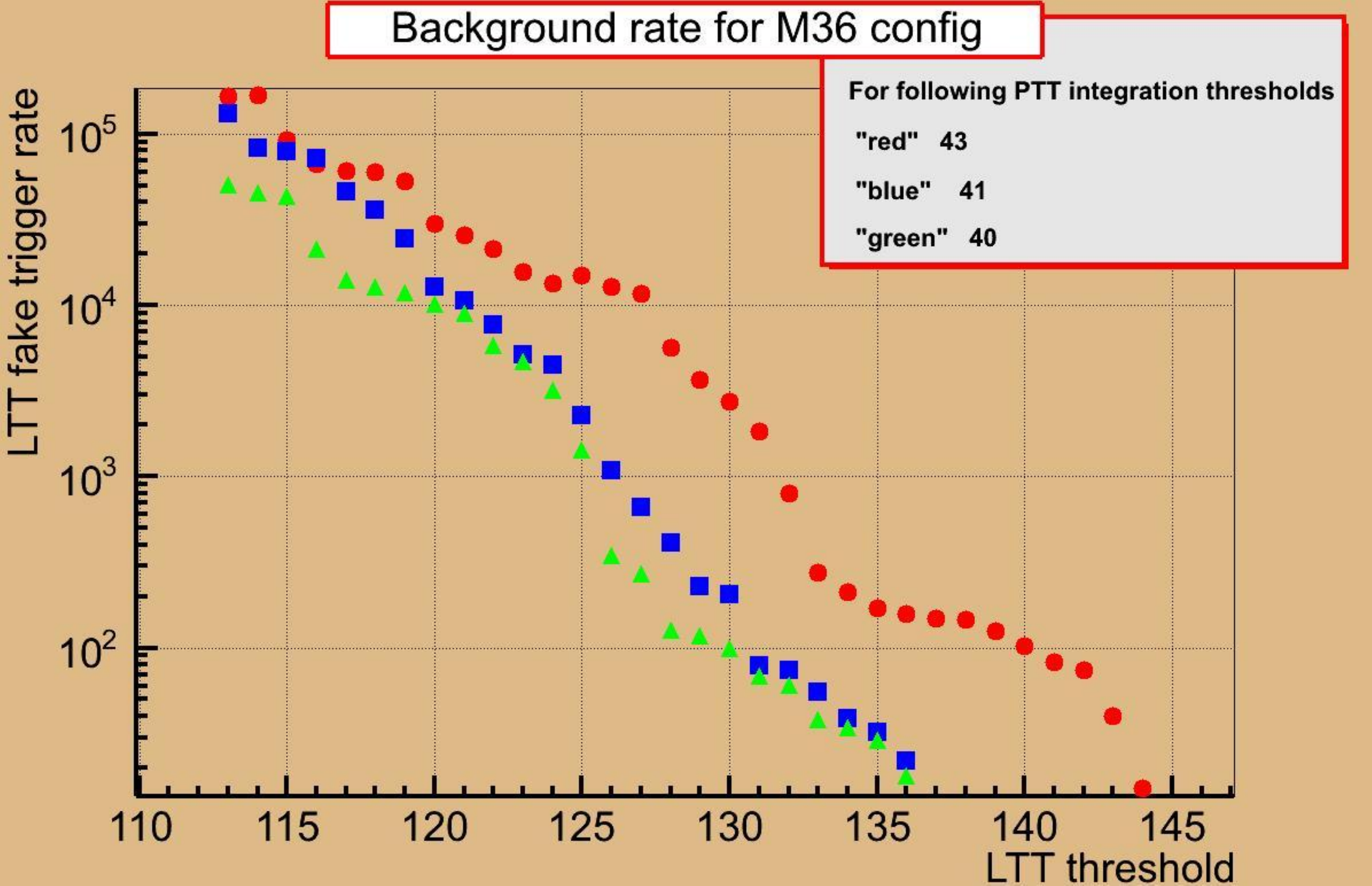}
  \caption{Thresholds for M36 configuration}
  \label{fig:M36}
 \end{figure}

\begin{figure}[!t]
  \centering
  \includegraphics[width=0.4\textwidth]{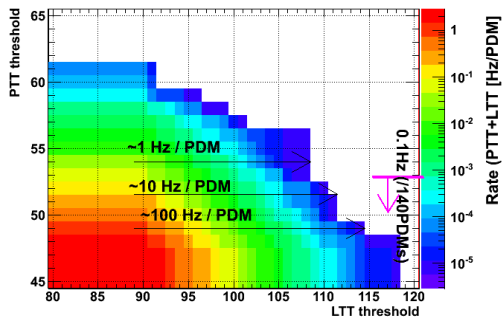}
  \caption{Thresholds for M64 configuration}
  \label{fig:M64_2d}
 \end{figure}

Firstly, the threshold levels for triggers has to be adjusted to fit within the permissible fake
trigger rates by a large amount of background simulations.  This was done for two configurations
of PDM mentioned in \ref{sec:trigger}, for M36 configuration ($36 \times 36$ pixels) and M64 
configuration ($64  \times 64$ pixels). It has to be noticed here, that we have consequently 
performed simulations for both configurations, but the analysis only for M36 configuration 
is presented in this  paper. In Figures \ref{fig:M36} and \ref{fig:M64_2d} these results for 
the FTR depending on threshold values for PTT and LTT trigger are shown.

The PTT and LTT threshold values of $PTT_{integr} = 43$, $LTT_{integr} = 145$ for M36 
configuration and $PTT_{integr} = 52$, $LTT_{integr} = 115$ for M64 configuration have been 
setup and used in massive simulations. The accumulated amount of data for M36 configuration is 
$10^{12}$ GTU's actually and among them 12000 LTT triggers and 750000 PTT triggers have been 
obtained. The statistics for M64 configuration is $5 \times 10^{10}$ upto now. 

Stored are the events filtered on PTT and LTT levels. Corresponding two files with an 
information on pixel positions, time and number of counts are written, when the thresholds are 
reached. Average size of the LTT output used in the following analysis is 250 MB per $10^{9}$ 
GTU's.

\section {Analysis}

\subsection{Pattern recognition}

\begin{figure}[t]
  \centering
  \includegraphics[width=0.4\textwidth]{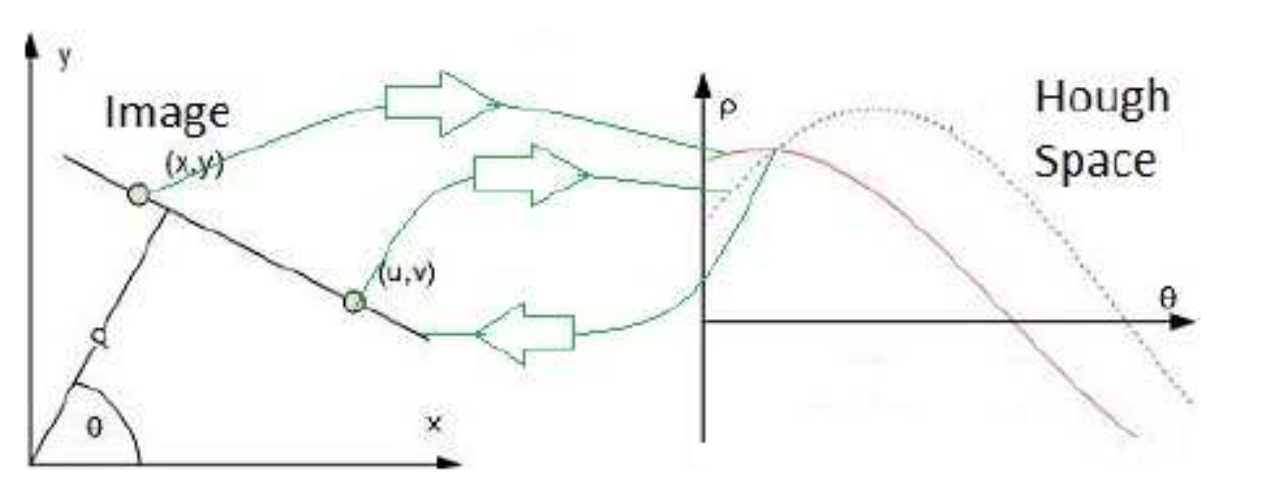}
  \caption{Hough transform.}
  \label{fig:hough}
 \end{figure}

To verify whether the data obtained by simulation of random background could not contain random 
fake patterns whose can be mistaken as real events, we have applied pattern recognition methods 
for signal tracks. The signal track on the FS contains information about the observed air  
shower and consequently about the primary EECR particle itself. It is a distribution of counts 
in space and time. There are possible several algorithms for the pattern recognition. The  
presented analysis have been performed by using the  Hough Transform (HT), developed to  
identify prefixed shapes within noise by transforming the relevant parameters to Hough space 
and back. The HT is an algorithm for the discrimination of certain shapes (even incomplete 
ones) from others \cite{bib:hough}. The longest pattern can be abstracted as a straight line. 
For each data point the HT assumes a number of lines passing through it. These lines can be 
parametrized by their distance from the origin of the coordinate system $\varrho$ and the 
angle $\Theta$ between its normal and the x-axis (Fig. \ref{fig:hough}, left). Transformed 
into the Hough space, a two dimensional parameter space spanned by $\varrho$ and $\Theta$  
each data point represents a sinusoidal curve (Fig. \ref{fig:hough}, right). The intersection 
points of the many sinusoidals are summed up in an accumulator. The intersection point that 
drews in most of the counts is then transformed back into the image space, where it corresponds 
to a straight line passing through as many data points as possible. 

\subsection{Results}

The simulation results described in section~\ref{sec:simulations} have been analyzed by HT
and consequently by modified HT.

Firstly, we have developed and checked the method on purely uniformly distributed random 
values. A large number of matrices $8 \times 8$ (like PMT) were generated. Two pattern 
characteristics are of interest:

\begin{itemize}
\item $pattern~ length = No.~ of~ pixels~ over~ threshold$
\item $avg~ pattern ~value = \sum ~ pixel~ values / ~pattern~ length$
\end{itemize}

\begin{figure}[t!]
  \centering
  \includegraphics[width=0.4\textwidth]{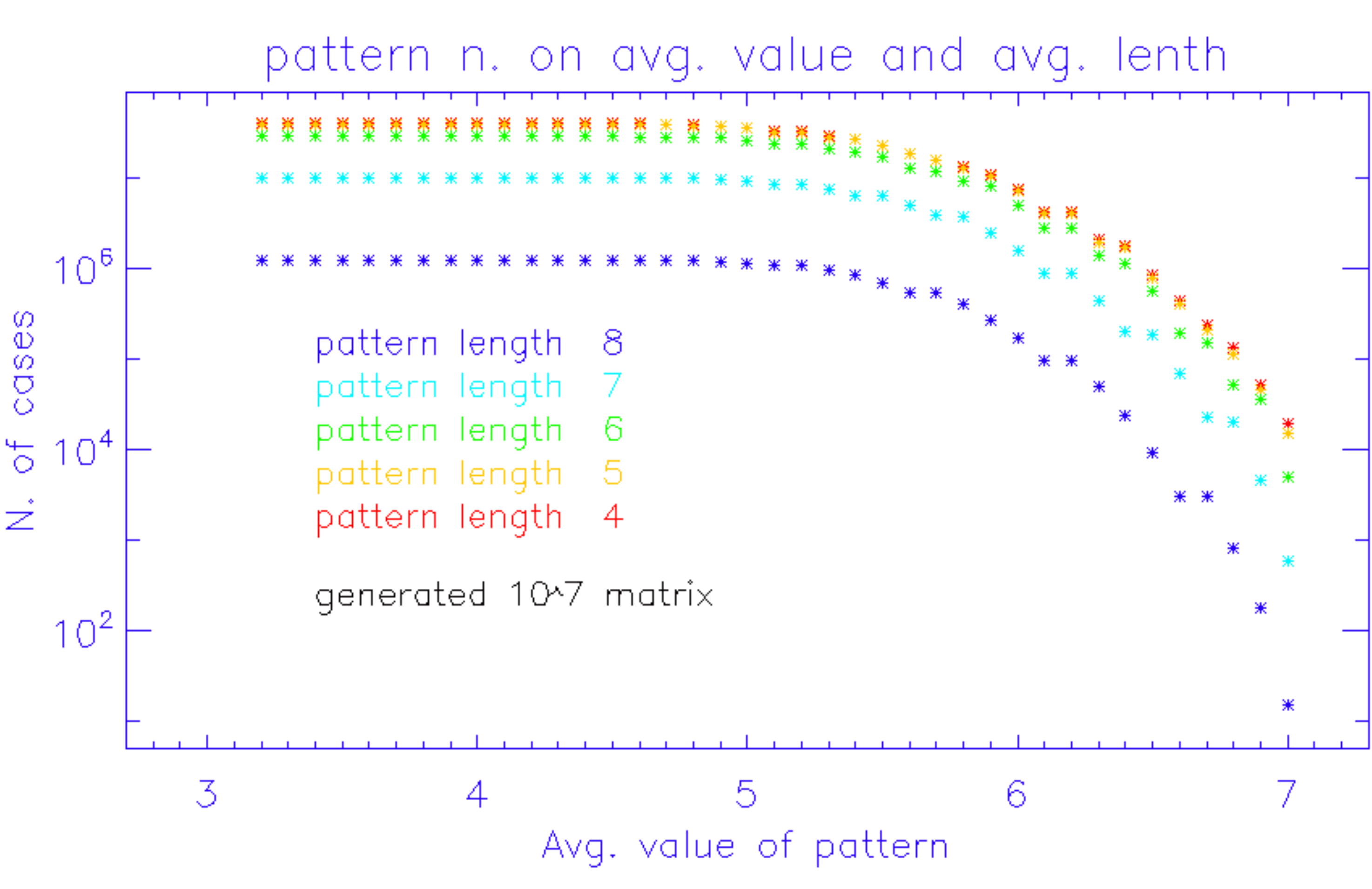}
  \caption{The numbers of recognized patterns depending on their lenghts and average values}
  \label{fig:r1}
 \end{figure}

Method was firstly tested by putting by hand small amount of patterns to huge amount of 
generated background. The method reliably detected artificial patterns. In Figure \ref{fig:r1} 
it is shown the number of detected patterns dependence over selected average pattern value 
for several pattern lengths (4 - 8). It can be seen, that for $10^{7}$ generated  $8 \times 8$ 
matrices, around 20 matrices with fake pattern with the length of 8 pixels with average pixel 
value (all pixels at maximum) is found. It could be simply verified. The probabillity that 
matrix pixel has some value is 1/8. Any 8 pixel configuration, so lineal pattern 8 pixel long, 
too appears with a probability $(1/8)^{8} = 5.96 \times 10^{-8}$. Such lineal patterns are 32, 
then the result is 19.07, compatible with the simulation result 20.
 
However, classic HT cannot distinguish between continuos and disconnected patterns. Thus the 
number of recognized patterns is overestimated. It was needed to improve the algorithm for the
JEM-EUSO purpose to be able to differ between such patterns. It was done on the basis of pixel 
distance. 

In the next step we have tested the modified HT algorithm on the LTT triggers obtained in 
section \ref{sec:simulations}. For each LTT trigger we have 31 matrices of $36 \times 36$ - 
the actual snapshot and for 15 anterior and 15 posterior in GTU. 

The real shower appears as a light speed moving point. On the basis of this we have developed a 
strategy of folding for above  mentioned matrices to recognize the pattern created by  moving 
point. We have divided atmosphere to the cells equivalent to pixel projection of JEM-EUSO PMT 
pixels on Earth surface (i.e. $0.75 \times 0.75~ km$ in nadir mode of detector). We evaluate a 
projection of moving light point created by shower on Earth surface and time when pixel is 
observed in GTU unit for a set of zenith and axial angles of incoming particle.

Every direction of incoming EECR particle is equivalent to a set of projections in consequetive 
GTUs. Then for one incoming direction we can take only columns where moving light point is
visible. We combine a new matrix from the stored 31 matrices from selected columns.

Pattern recognition method is then applied to this new matrix. We build over the 31 stored 
matrices a set of new matrices for selected incoming angles of primary cosmic rays. Such an 
analysis is applied to all simulated sets of 31 matrices passed LTT trigger. The method 
validity was verified by artificial patterns with known incoming direction added to tested data 
set. All artificial patterns were found by the method.

\begin{figure}[t!]
  \centering
  \includegraphics[width=0.4\textwidth]{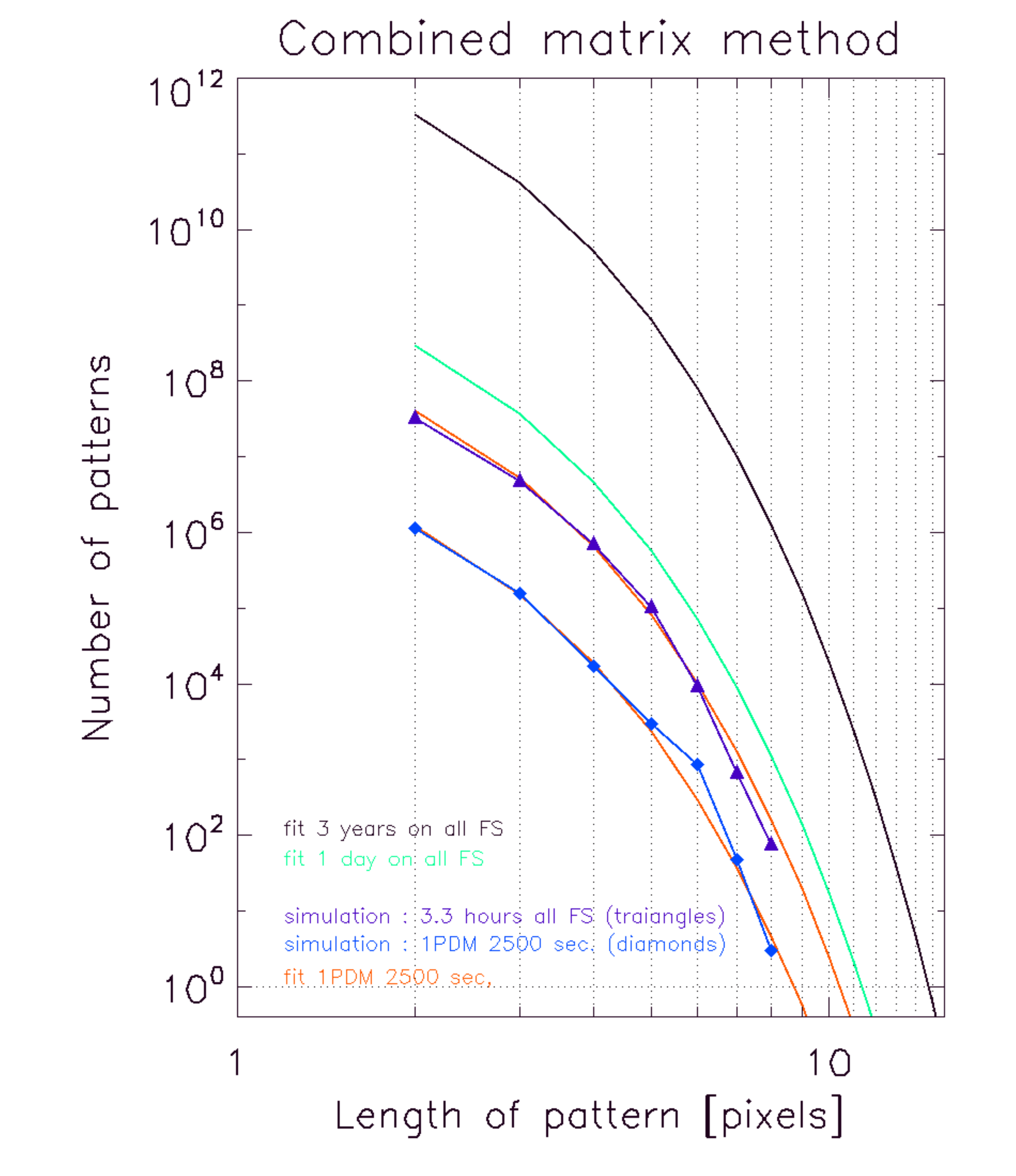}
  \caption{Approximation to longer time periods of measurements for dependences the numbers of recognized patterns on their lengths}
  \label{fig:r2}
 \end{figure}

Finally we go through $10^{12}$ simulated GTU's on one PDM. This is equivalent to 3.3 hours 
measurement of all 137 PDMs of JEM-EUSO detector. The result from full analysis of these 3.3 
hours measurements of all detector is presented by magenta line with triangles. The number of 
recognized patterns as a function of pattern length is presented on the Figure \ref{fig:r2}.
Example of analysis result for $10^{9}$ GTU's run equivalent to 2500 second measurements at one 
PDM of detector is presented by blue line with diamonds.  We fit both of them by statistically 
motivated function:

\begin{equation}
N_{p}~(L_{p})  \sim  (1/N_{pix})^{L_{p}},
\end{equation}

where $N_{p}$ is number of recognized patterns, $L_{p}$ is the patern length and $N_{pix}$ is 
number of possible pixel values. We set a number of possible pixel values to 8 following a 
histogram of pixel values. This approximation conservatively estimate number of patterns for 
longer patterns recognized in analysed data set. If we scale approximation to one day 
measurement of all detector (green line on Figure \ref{fig:r2}.), we can find the few patterns 
with the length of 11 and maybe one with the length of 12 pixels. Further approximation 
scalling to full planned 3 years of JEM-EUSO operation, we will find only one pattern with the 
length of 15 pixels. This 15 pixels on ground means 7.65 km long projection of shower. Showers 
created by more inclined and higher energetic particles are more easy to recognize and 
reconstruct. Let's assume the worst case when we will have particle with energy 
$5 \times 10^{19} eV$ and with maximum zenith angle. Particle with such energy can create 
first pixel visible by detector at altitude 13 km. If fake pattern will be 7.65 km long with 
first visible point at altitude 13 km, then zenith angle of primary particle is 30.5 degrees. 
Thus the fake pattern during 3 years of measurement can be mistaken by particle with zenith 
angle maximaly 30.5 degrees. 

\begin{figure}[t!]
  \centering
  \includegraphics[width=0.4\textwidth]{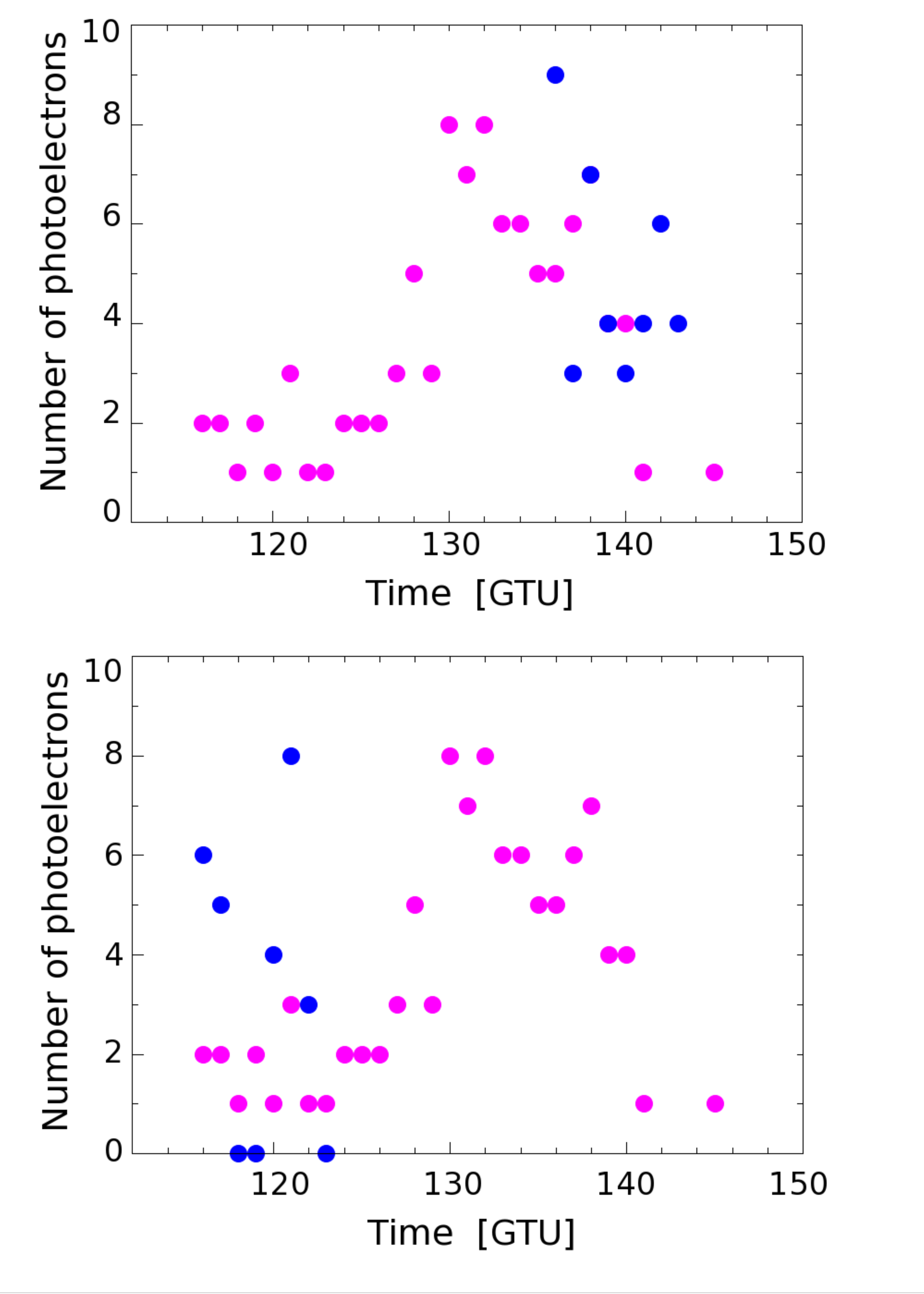}
  \caption{Time flow of two selected showers. Simulated shower is denoted by purple circles, longest recognized patterns by blue circles.}
  \label{fig:r3}
 \end{figure}

We evaluate a sum of intensity around moving pixel in longest recognized patterns. The sum 
of 4 pixels - one from pattern and 3 from surrounding pixels (2x2 matrix) - in a triggered 
matrix was evaluated for every GTU when pattern was found. The results are presented at Figure 
\ref{fig:r3}, where pattern intensity time evolution is compared with a simulated showers 
created by primary particle with energy $5 \times 10^{19} eV$. While simulated shower 
evolution has a typical time profile with increasing intensity  to maximum point ($H_{max}$) 
and then decreasing, the recognized patterns has different noisy shape and shorter length.

We can conclude that during 3 years of measurement we cannot mistake fake trigger to real event.
 
\section{Conclusions}

We present analysis of simulated UV background of JEM-EUSO detector. Data set of triggered UV 
BG was analysed by Hough pattern recognition method. The results show dependence of recognized 
patterns lengths on time of measurement. Longest patterns from backgroud are comparable with 
events with zenit angle smaller than 30 degrees. Patterns from backgroud can not be mistaken 
with real event time profile because they have not regular shape with typical maximum but noisy 
shape. In conclusion during planed 3 years of measurement the background cannot produce a 
patterns whose can be mistaken with real event. 

\vspace*{0.5cm}
{
\footnotesize{{\bf Acknowledgment:}
{
This work was supported by Slovak Academy of Sciences MVTS JEM-EUSO as well as VEGA grant 
agency project 2/0076/13.
} }

}

\clearpage

%% file: icrc2013-1282.tex



\def\va{\vec{n}}
\def\vn{\vec{V}}
\def\vz{\vec{z}}
\def\vR{\vec{R}}
\def\vw{\vec{\Omega}}
\title{Pattern recognition and direction reconstruction for JEM-EUSO experiment}

\shorttitle{JEM EUSO ICRC 2013}

\authors{
S. Biktemerova$^{1}$,
M. Gonchar$^{1}$,
S. Sharakin$^{2}$
for the JEM-EUSO Collaboration.
}

\afiliations{
$^1$ Joint Institute for Nuclear Research\\
$^2$ Skobeltsyn Institute of Nuclear Physics, Lomonosov Moscow State University
}

\email{$^1$ sveta.biktemerova@gmail.com, maxim.mg.gonchar@gmail.com} 
\abstract{JEM-EUSO is a space based observatory, that will detect light produced by an extensive air shower (EAS) 
after interaction of a cosmic ray particle with atmosphere. The fluorescent and Cherenkov light produced in EAS 
is focused on a focal surface by a system of Fresnel lenses. The focal surface is covered by a set of multi-pixel 
photomultipliers. For the experiment preparation and the data analysis a dedicated software ESAF is used. 
ESAF is a robust simulation code which includes libraries for the EAS simulation, fluorescent and Cerenkov light production,
its propagation in the atmosphere and the detector response. In order to reconstruct the direction of the primary cosmic
rays we need a pattern recognition algorithm able to find EAS on the 'image' on the focal surface. In this work we develop
algorithms of pattern recognition and angular reconstruction. The results are presented in this proceedings. }


\maketitle

\section{Introduction}
JEM-EUSO is a space based observatory, that is aimed at observation of ultra high energy cosmic ray (UHECR). The telescope will be 
mounted to the International Space Station (ISS) at altitude of approximately 400~km. It will collect fluorescent 
and Cerenkov light produced by extensive air shower (EAS) after interaction of a primary cosmic ray particle with atmosphere  \cite{jemeuso}. 
To collect the light from EAS a system of three Fresnel lenses and a focal surface is used. The focal surface 
is covered by a set of multi-pixel photomultipliers. The description of the detector in more detail can be found in \cite{casolino} and \cite{dagoret}.

One of the main scientific goals of the JEM-EUSO mission is the identification of individual cosmic ray sources using the arrival direction of the primary particle
and study acceleration mechanisms with the observed events \cite{science}. JEM-EUSO is able to study UHECR with energies above  $5\times10^{19}$~eV \cite{performance}.
The cosmic rays of such a high energy are weakly deflected by the galactic magnetic field and they
can be traced back to their origin by their measured arrival direction with accuracy better than a few degrees. 
Therefore precise estimation of the arrival direction of the UHECRs is among of the major scientific objectives of the mission.

One of the difficulties of arrival direction reconstruction is to distinguish the signal from the background: the atmosphere night glow and city light. In order to discriminate the 
signal in ESAF (Euso Simulation and Analysis Framework) \cite{naumov} several algorithms of the pattern recognition are developed.
In this article we focus only one of them: the ``Track finding method''.

\section{Pattern recognition}
\subsection{Track finding method}
The track finding method is an additional algorithm in ESAF that makes it possible to find a shower track on the focal plane.
A track on the focal plane is a sequence of pixels ordered in time and lying along some direction. 
A signal track is a track corresponding to the EAS signal. 

This method uses the photon-count distribution on the focal plane at each time step. 
The time steps in  which this information is kept are called gate time units (GTU).  
The GTU length is fixed by JEM-EUSO's electronic response, and  its nominal value is $2.5 \mu s$.
Thus we have a ``snapshot'' of the focal plane with
photon-count distribution for each GTU. The task of the algorithm is to find a point that moves uniformly along a straight line 
on the focal plane using a sequence of snapshots.
The algorithm creates a set of all possible track candidates, of which the best one is chosen.
To build each track the algorithm uses the principles of Kalman filter~\cite{kalman}. 

Let us consider the technique of the algorithm in more detail.
The algorithm operates sequentially with all snapshot pairs.
For each snapshot the pixels with large number of counts are selected.
As soon as we have a set of selected pixels the algorithm attempts to connect all possible pairs 
of pixels between two snapshots into track segments. Thus it tries to connect all pairs of points, which 
satisfy criteria of distance, duration and deviation from track line. If a point satisfies all the criteria it is added to the track.
When the track contains at least two pixels it is fitted with a line on each step. The line is used in ``deviation from track line'' criterion.

\begin{figure*}[!t]
 \centering
\begin{subfigure}[b]{0.45\textwidth}
\centering
\includegraphics[width=\textwidth]{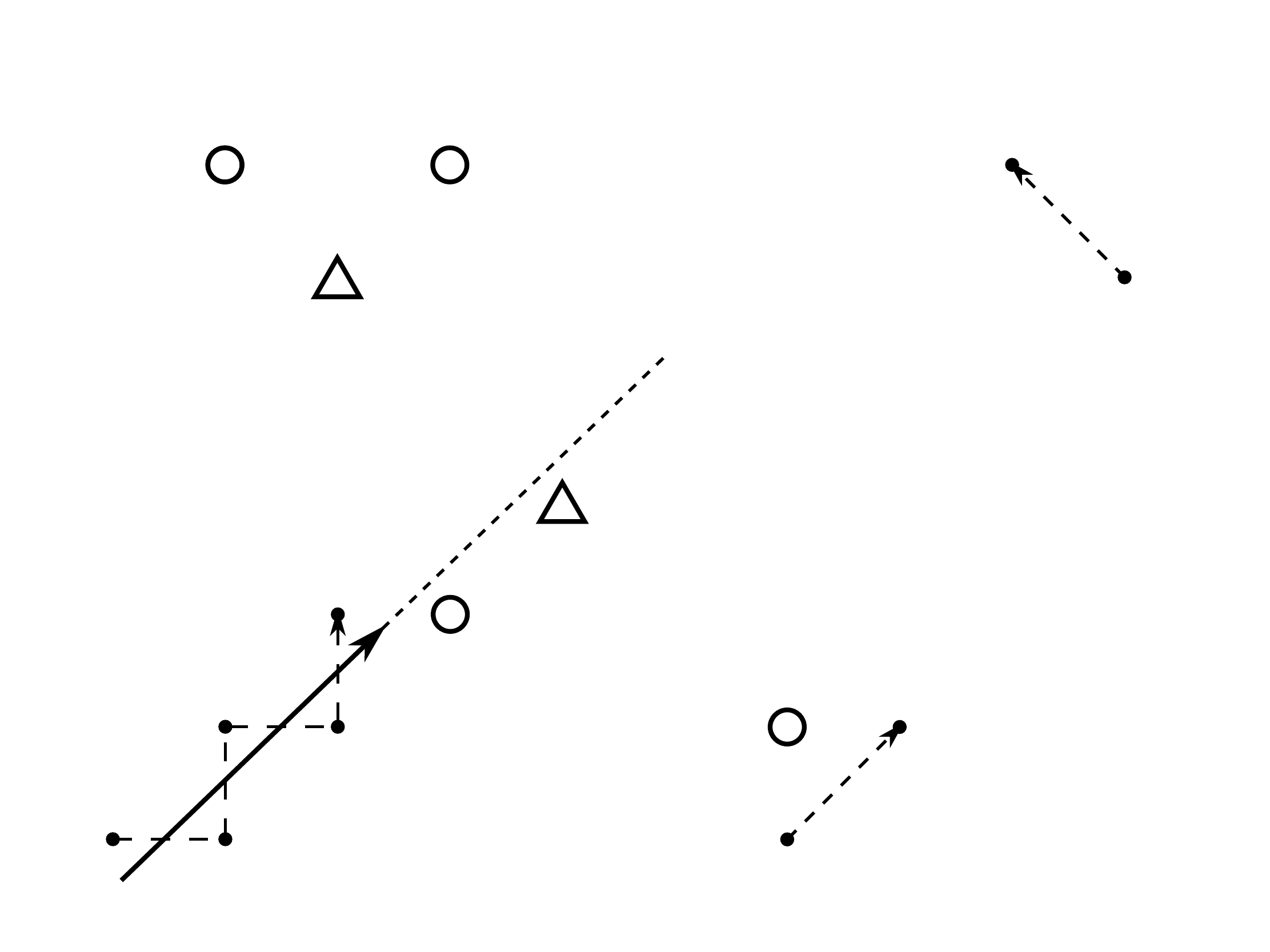}
\caption{Before processing next snapshots.}
\label{f:sel1}
\end{subfigure}%
\hspace{0.3cm}
\begin{subfigure}[b]{0.45\textwidth}
\centering
\includegraphics[width=\textwidth]{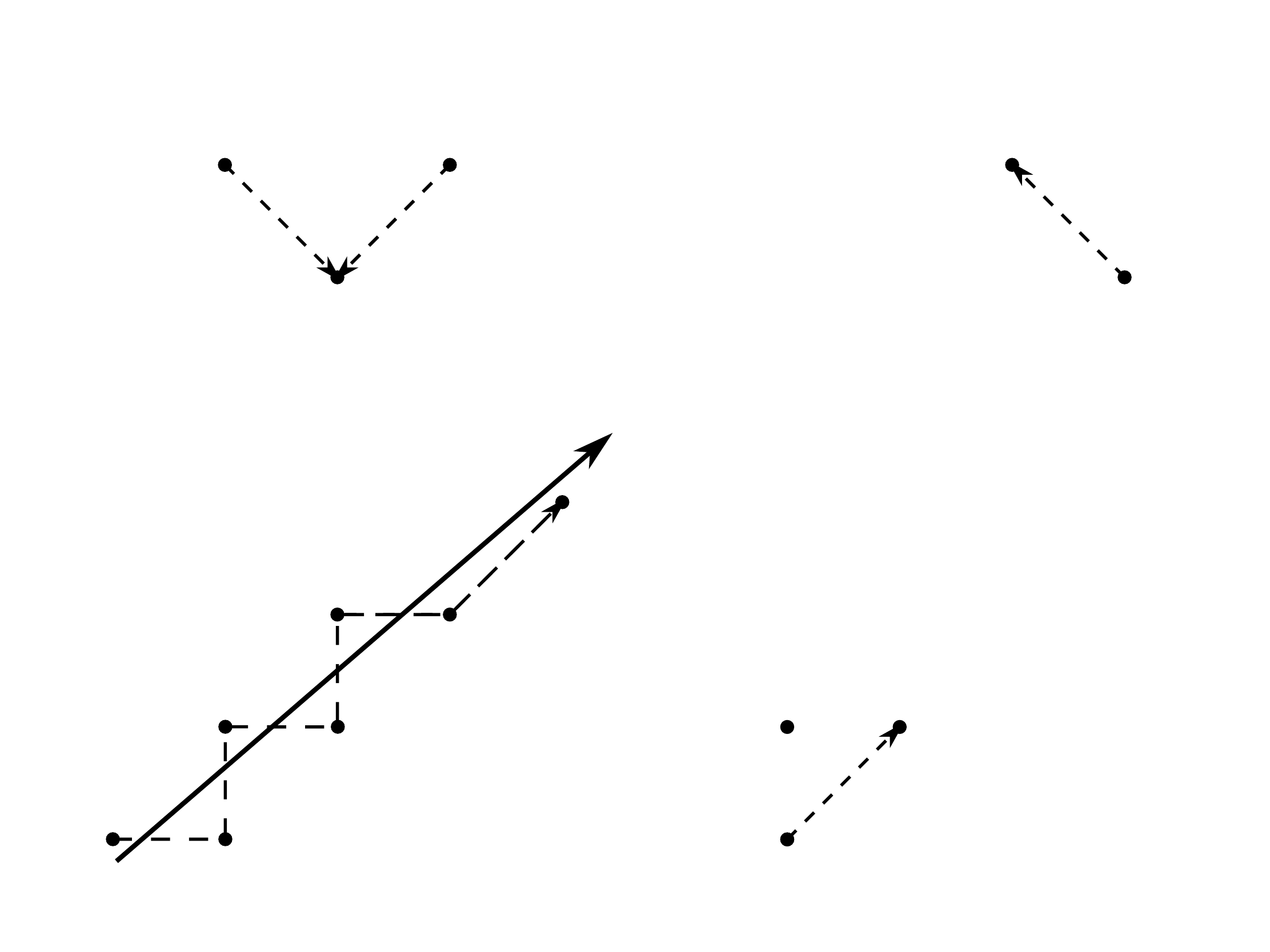}
\caption{After processing next snapshots.}
\label{f:sel2}
\end{subfigure}
 \caption{The scheme of track finding method. Figure~\ref{f:sel1} represents three already found tracks (dashed lines) with their 
pixels (black dots) and a fitted line for the track containing more than 2 pixels. Selected pixels which will be added to the tracks
on next iterations are drawn with circles (+1 GTU) and triangles (+2 GTU). Figure~\ref{f:sel2} represents the same set of data, but 
after the addition of new pixels: two more short tracks are found, one track is extended and one pixel is ignored since it's not matched 
to any track.}
 \label{sel}
\end{figure*}

In the Fig.~\ref{sel} a rough scheme of the algorithm is shown. 
In the Fig.~\ref{f:sel1} three obtained tracks and selected pixels that can be
added to the tracks are shown. On the next step the selection criteria are used to add new pixels to existing tracks and create new ones~\ref{f:sel2}.

Track candidates are selected not only from two consecutive snapshots: the algorithm is able to look back for 5 GTU in order to find track segments.

The algorithm does not distinguish between the signal and background pixels. However, the background pixels are distributed randomly and
the probability of these pixels to be connected into a single track decreases vastly with the track length.   
In addition, occasionally a background pixel can be added to the signal track, thereby spoiling it. In this case the problem is solved by
copying the track before point addition. Thus we have two tracks: one of them does not have a bad point and another has. This provides a way 
of continuing track reconstruction even after the addition of a improperly aligned pixel. 

\begin{figure*}[!t]
\centering
\begin{subfigure}[b]{0.35\textwidth}
\centering
\includegraphics[trim=120 63 172 80,width=\textwidth,clip]{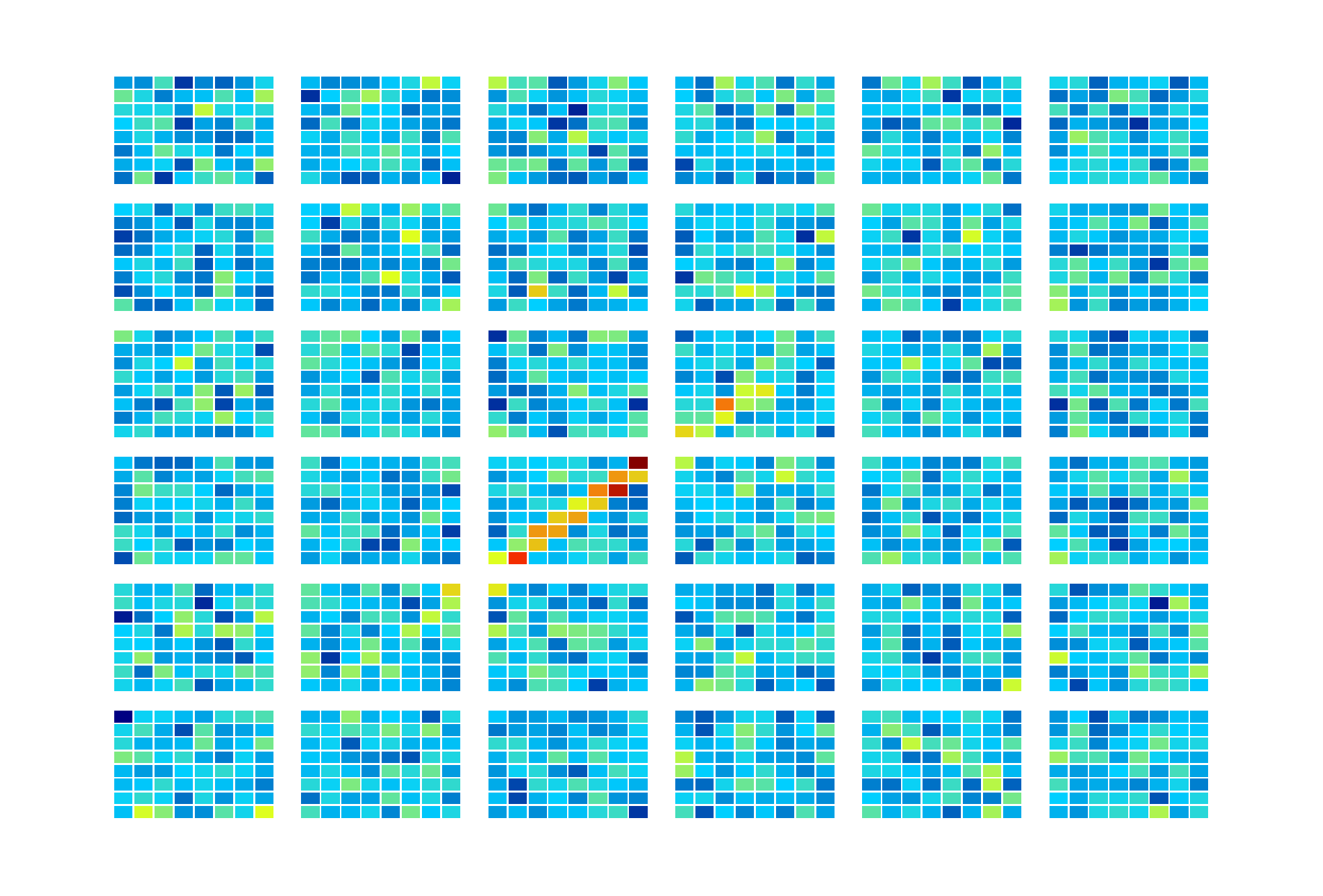}
\label{pat1}
\end{subfigure}%
\hspace{0.3cm}
\begin{subfigure}[b]{0.35\textwidth}
\centering
\includegraphics[trim=120 63 172 80,width=\textwidth,clip]{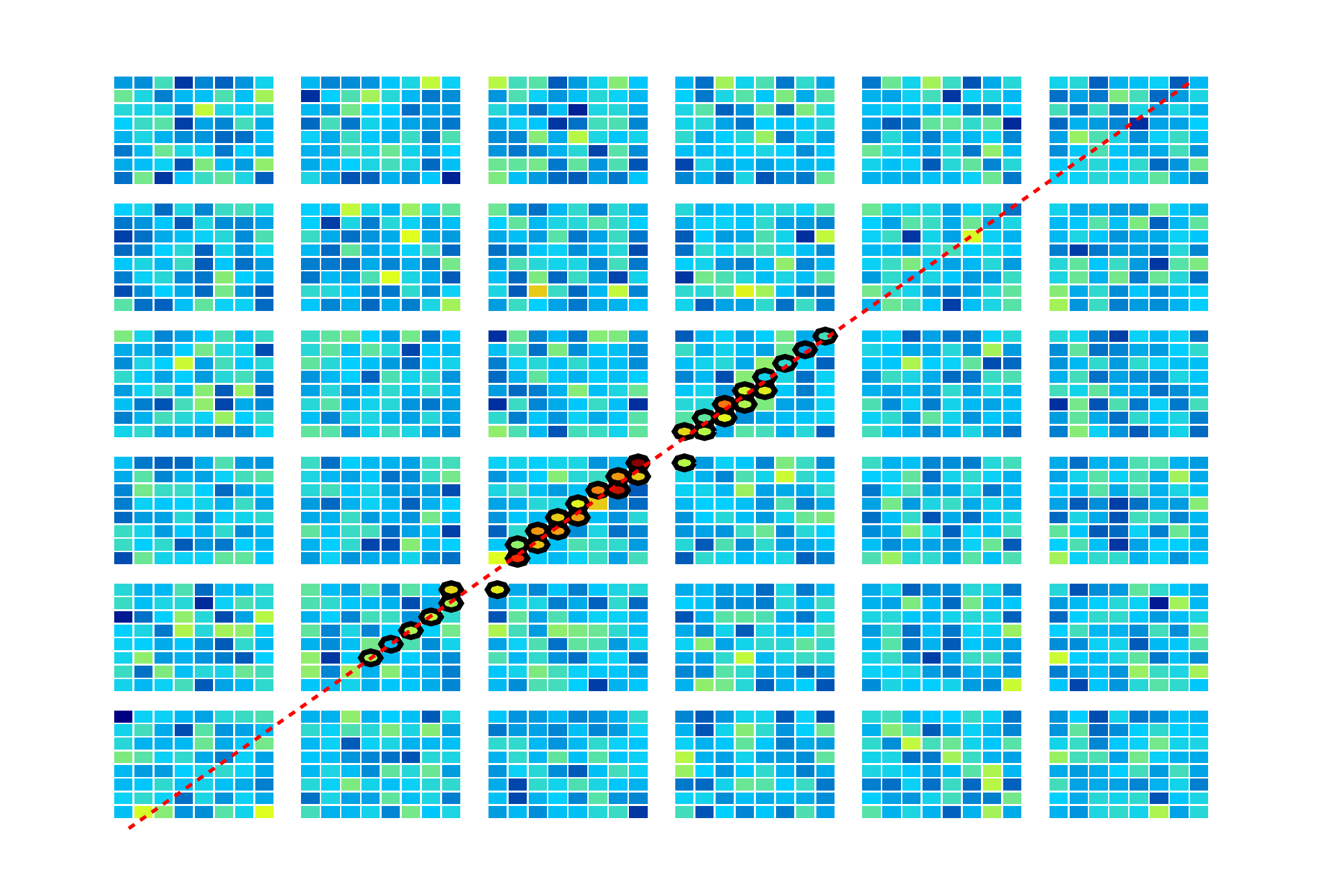}
\label{pat2}
\end{subfigure}
\begin{subfigure}[b]{0.1\textwidth}
\centering
\includegraphics[trim=520 0 0 0,scale=0.45,clip]{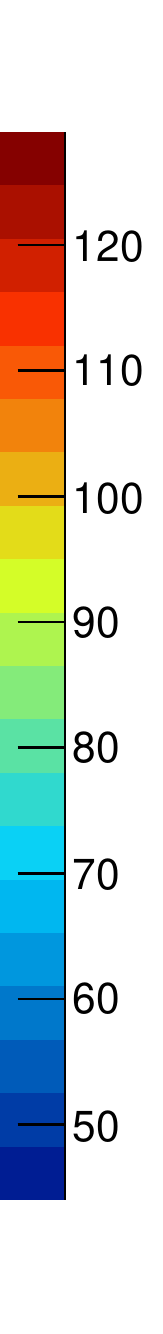}
\label{sel2}
\end{subfigure}
 \caption{The integrated signal on the focal surface from EAS with $E = 1\cdot10^{20}$~eV and $\theta = 60^{\circ}$.  
The pixels selected by the algorithm are marked with black circles. The dotted red line represent the obtained track line. Pixels color corresponds to the p. e. count.}
 \label{pat}
\end{figure*}

In the end of the procedure we have a large set of tracks. Nearly the entire set is composed of short tracks, which are occasioned by the
accidental coincidence of background pixels as well as the fragments of the signal track, that are ``spoiled'' by addition of
background pixels. The signal track is selected as a track with maximal summary number of counts: it corresponds to the longest found 
straight track with highest signal and containing no time leaps. In the Fig.~\ref{pat} a result of algorithm application to a MC event is shown.

Further, one can define the selection criteria that are used in the algorithm in more detail:
\begin{description}
\item[\bf{Pixel selection}] The number of selected pixels on each step is an adaptive quantity: the number of selected
pixels with same number of counts on each snapshot should be less than 32. In the Fig.~\ref{fig:pedist} the average distribution of p.~e. 
counts for signal and background pixels for events with energies $7\cdot10^{19}$ and $3\cdot10^{20}$ eV and incident angles $30^{\circ}$ and $75^{\circ}$ are shown.
One can see that the chosen cut on number of counts selects a big portion of background pixels in addition to signal ones: 
the main purpose of this cut is to limit the number of track candidates in memory.
\begin{figure}[ht]
 \centering\includegraphics[width=0.5\textwidth]{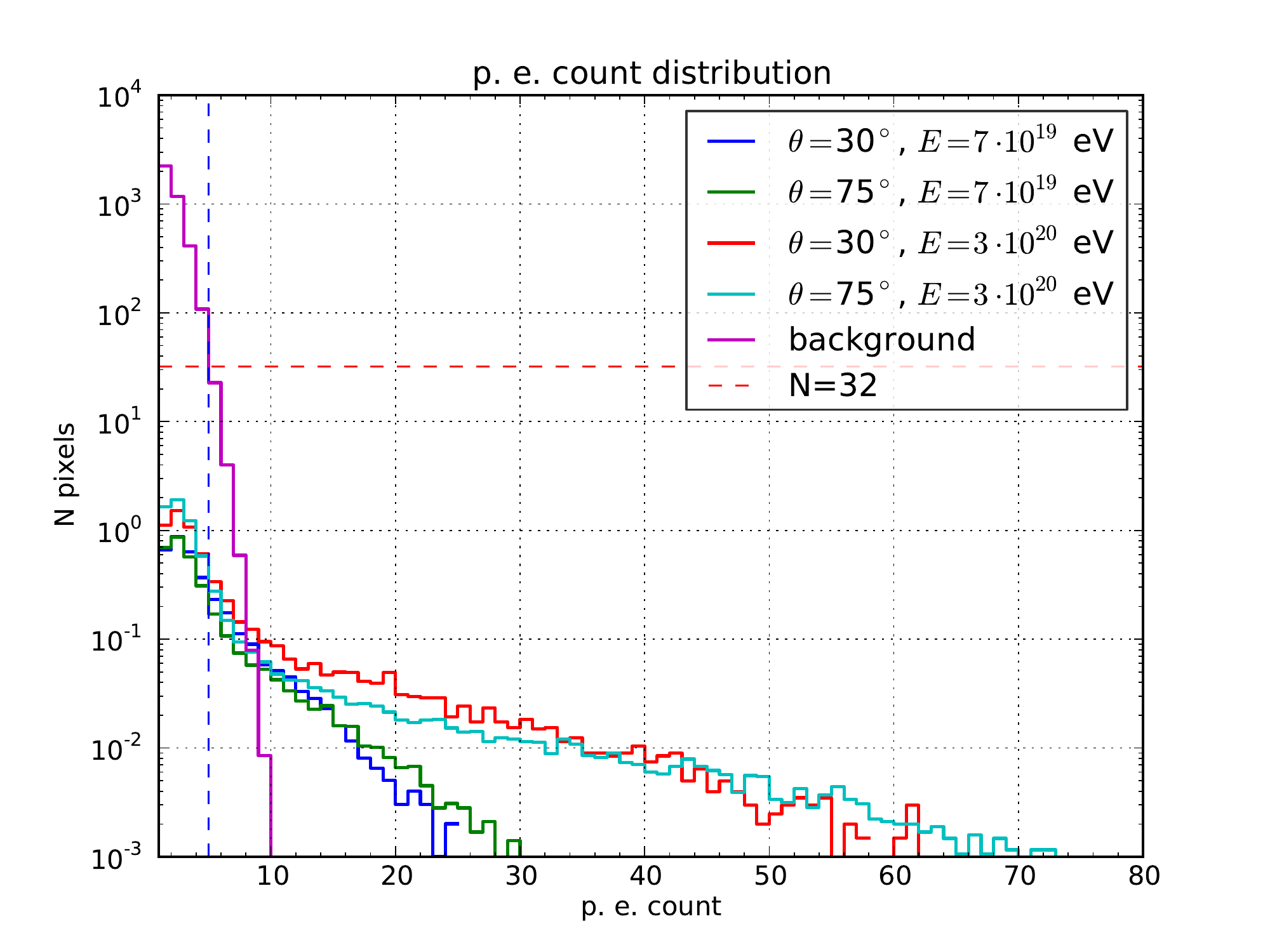}
 \caption{The average distribution of p.~e. counts for signal and background pixels for $\sim2800$ events with energies $7\cdot10^{19}$ and $3\cdot10^{20}$~eV 
 and incident angles $30^{\circ}$ and $75^{\circ}$. Red dashed line represents the chosen cut on number of counts.}
 \label{fig:pedist}
\end{figure}

\item[\bf{Distance}]  
In the beginning of the procedure the maximal distance between two connected pixels is equal to 2 pixel diagonals.
If track average velocity exceeds one pixel diagonal per GTU the additional cut on distance is applied: the distance to the new pixel divided by delta GTU
should be less than doubled track velocity.

\item[\bf{Duration}] The duration between two connected pixels should be less than 5 GTU.
This number is based on the geometry of the focal plane and velocity of the track: the gap between photomultipliers is 
not large enough to produce a delay in signal of more than 5 GTU. 
\item[\bf{Deviation from the track line}] A distance between the pixel and the fitted line should be less than 2 pixels in size.
\end{description}

The constants for this algorithms are chosen based on geometrical estimations and in the future can be tuned based on simulation results.
Currently the algorithm was tested on the MC simulation and reconstructs proper tracks for all the triggered events.

\section{ Angular  Reconstruction.}

After the signal discrimination basic information about the track on the focal
plane is available.  For each pixel on the focal surface that is determined as
`signal' the number of produced photo-electrons $N_{i}^{\text{p.e.}}$, their timing
information $t_i$ and photons arrival direction $\va_i$ are known.

The first step of the angular reconstruction is the estimation of the Track Detector Plane (TDP).
It is the plane that contains the shower track and the detector itself. 
\subsection{TDP determination algorithm}
Based on $\va_i$ the unit vector pointing on the shower maximum $\va_{\text{max}}$
can be obtained and the TDP can be computed in the following way.
The TDP is determined by it's normal $\vn$. It can be made up of two unit
vectors $\va_{\text{max}}$ (pointing to the shower maximum) and $\va_i$ (see Fig. \ref{fig:tdp}):
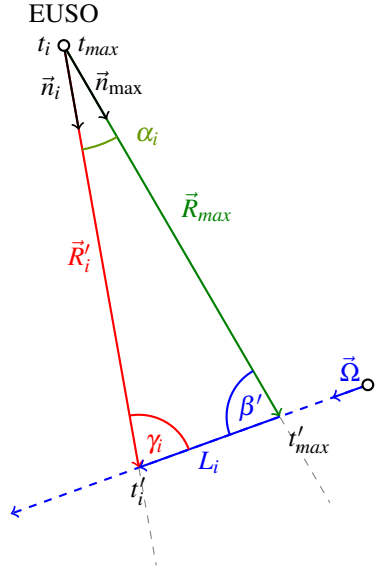
\begin{figure}[t]
 \centering
\begin{tikzpicture} 
  \tikzset{mypoints/.style={fill=white,draw=black,thick}} 
  \def\ptsize{2.0pt}
  \def\a{3} \def\b{1.75}
  \def\xp{-5.0} \def\yp{4.5} 
  \def\xf{-1.0} \def\yf{0} \def\th{110}
  \def\lenA{5}
  \pgfmathsetmacro{\xe}{\xf-\lenA*sin(\th)}
  \pgfmathsetmacro{\ye}{\yf+\lenA*cos(\th)}
  \pgfmathsetmacro{\xeu}{\xf-0.5*sin(\th)}
  \pgfmathsetmacro{\yeu}{\yf+0.5*cos(\th)}
  \def\len{7} \def\dirA{-80} \def\dirB{-60}

  \coordinate[label={[label distance=5pt]above:EUSO},
  label={[label distance=2pt]right:$t_{max}$},
  label={[label distance=1pt]left:$t_i$}] (P) at (\xp,\yp);

  \coordinate (F) at (\xf,\yf);
  \coordinate (Fu) at (\xeu,\yeu);
  \coordinate (E) at (\xe,\ye);
  \draw[name path=lineFE,blue,dashed,thick,->] (F)--(E);
  \draw[name path=lineFE,blue,thick,->] (F)--(Fu) node[midway,above] {$\vw$};



  \coordinate (A2) at ({\xp+\len*cos(\dirB)},{\yp+\len*sin(\dirB)});
  \coordinate (A3) at ({\xp+\len*cos(\dirA)},{\yp+\len*sin(\dirA)});
  \coordinate (D) at (\xp,{\yp-\len});
  \draw[name path=linePA2,gray,dashed] (P)--(A2);
  \draw[name path=linePA3,gray,dashed] (P)--(A3);

  %
  %
  \coordinate[label=below right:$t_{max}'$] (T0) at (intersection of F--E and P--A2);
  \coordinate[label=below:$t_i'$] (Ti) at (intersection of F--E and P--A3);

  %
  %
  \coordinate (TD) at (intersection of F--E and P--D);
  \draw[name path=linePT0,thick,->,color=green!50!black] 
      (P)--(T0) node[midway,auto] {$\vR_{max}$};
  \draw[name path=linePTi,thick,red,->] (P)--(Ti) node[midway,left] {$\vR_i'$};
  \draw[name path=linePTiv,thick,black,->] (P)--($ (P)!.2!(Ti) $) node[midway,left] {$\va_i$};
  \draw[name path=linePTiv,thick,black,->] (P)--($ (P)!.2!(T0) $) node[midway,right] {$\va_{\text{max}}$};
  \draw[name path=lineT0i,thick,blue,->] (T0)--(Ti) node[midway,below] {$L_i$};

  %
  %
  \def\arclen{1.4}
  \draw[thick,color=green!60!red] ({\xp+\arclen*cos(\dirB)},{\yp+\arclen*sin(\dirB)}) arc (\dirB:\dirA:\arclen);
  \draw[color=green!60!red] ({\xp+\arclen*cos(\dirB)+0.4},{\yp+\arclen*sin(\dirB)}) node {$\alpha_i$};

  %
  %
  \def\arclen{0.7cm}
  \draw[blue,thick] 
  let
  \p1 = (T0)
  in
  ({\x1-\arclen*cos(\dirB)},{\y1-\arclen*sin(\dirB)}) arc (180+\dirB:90+\th:\arclen);
  \draw[blue] 
  let
  \p1 = (T0)
  in
  (\x1-0.38cm,\y1+0.12cm) node {$\beta'$};

  %
  %
  \draw[red,thick] 
  let
  \p1 = (Ti)
  in
  ({\x1-\arclen*cos(90+\th)},{\y1-\arclen*sin(90+\th)}) arc (\th-90:180+\dirA:\arclen);
  \draw[red] 
  let
  \p1 = (Ti)
  in
  (\x1+0.21cm,\y1+0.31cm) node {$\gamma_i$};

  \foreach \p in {F,P}
  \fill[mypoints] (\p) circle (\ptsize);
\end{tikzpicture} 
\caption{The scheme of angular reconstruction algorithm.}
\label{fig:tdp}
\end{figure}

\begin{equation}
 \vn_i = \frac{\va_i\times \va_{\text{max}}}{\sin(\alpha_i)}
\end{equation}
where $\alpha_i$ is angle between $\va_{\text{max}}$ and $\va_i$. 
The normal $\vn(\theta_{\vn}, \varphi_{\vn})$ describing TDP is found by maximizing the
sum of scalar products of $\vn$ and $\vn_i$: $C = \sum\limits_i ( \vn \vn_i )$.
All $\vn_i$ are chosen to point in the same half-sphere, so all scalars have the same sign.
This can be done analytically by requiring first derivatives of $C$ by
$\varphi_{\vn}$ and $\theta_{\vn}$ to be zero. Thus the TDB is given by the following
equations:
\begin{eqnarray}
    &\varphi_{\vn} = \arctan \left( \frac{\sum\limits_i\frac{n^i_y}{\sin\alpha_i}}{\sum\limits_i\frac{n^i_x}{\sin\alpha_i}}\right) , \theta_{\vn} = \arctan\left(\frac{\sum\limits_i\frac{n^i_{\perp}}{\sin\alpha_i}}
    {\sum\limits_i\frac{n^i_z}{\sin\alpha_i}}\right)
\end{eqnarray}

Once TDP is found the task of finding 3-dimensional shower direction vector
$\vw$ is reduced to the 2-dimensional case with a single parameter $\beta'$. As one can see from Fig.~\ref{fig:tdp} the
$\beta'$ is the plane angle between vector that points to shower maximum $\vR_{max}$ and shower direction $\vw$.
The shower direction vector $\vw = -( \sin\theta\cos\varphi, \sin\theta \sin\varphi, \cos\theta) $ can be found by rotating $\va_{max}$ around the calculated
$\vn(\theta_{\vn}, \varphi_{\vn})$ on the angle $\beta'-\pi$.

\subsection{Direction reconstruction algorighm}
As soon as the TDP is found the expected value of $\va_i' =
\frac{\vR_i'}{|\vR_i'|}$ can be obtained:

\begin{equation}
\vR_i' = \vR_{\text{max}} + \vw\cdot L_i
\label{eq:ri}
\end{equation}
where the $L_i$ is the distance which shower passes during time $\Delta t = t_i - t_{\text{max}}$. 
$L_i$ is given by eq.:

\begin{equation}
L_i = c\Delta t + R_{\text{max}} - R_i'
\end{equation}

The length of the expected vector $\vR_i'$ then can be found by taking the
square of eq.~\ref{eq:ri}. Thus $R_i'$ is given by the following equation:

\begin{equation}
R_i' = \frac{\left(\vR_{\text{max}} + \vw\left(c\Delta t+R_{\text{max}}\right)\right)^2}{2\left(\vR_{\text{max}}\cdot\vw+c\Delta t+R_{\text{max}}\right)}
\end{equation}

The distance $R_{\text{max}}$ between the detector and shower maximum can be obtained
using equation \ref{rmax} in which the altitude of the EAS maximum $H_{\text{max}}$
is computed using relation between the time width of the signal on the focal plane $\sigma$ and air density 
$\rho(H_{\text{max}}^{\text{fluo}})$ in the atmosphere at which EAS develops~\cite{naumov}. 
R$_{earth}$ and $H_{ISS}$ denotes the earth radius and altitude of the ISS, respectively. The angle $\theta_{max}$ corresponds to the
angle between two unit vectors $\va_{max}$ and vector pointing from ISS to the center of Earth.    

\begin{multline}
R_{max}= \left( R_{earth}+H_{ISS} \right) \cdot \cos \theta_{max} - \\
 -\sqrt{\left( R_{earth} + H_{max} \right)^2-\left( \left( R_{earth} + H_{ISS} \right) \cdot \sin \theta_{max} \right)^2}
\label{rmax} 
\end{multline}

Equations \ref{r1} and \ref{r2} were obtained using the GIL parametrization for the longitudinal
development of the number of charged particles.  $\xi_{\text{max}}$ is a dimensionless
parameter, $E$ is the energy of the primary particle, $A$ --- its atomic
number, $X_0 = 37.15$~$g/cm^{-2}$ --- air radiation length, $E_c =
81$~MeV (critical energy), $a = 1.7$, $b = 0.76$. These values are chosen
based on CORSIKA-QGSJET-II results \cite{Corsika}.

\begin{eqnarray}
\label{r1} 
&\sigma = \sqrt{2\xi_{\text{max}}}\frac{X_0\left(1+\va_{\text{max}}\cdot\vw\right)}{\rho(H_{\text{max}}^{\text{fluo}})c} \\
&\xi_{\text{max}} = a + b\ln\left(E/E_c - \ln A\right)
\label{r2} 
\end{eqnarray}

Considering that parameter $\xi_{\text{max}}$ depends on UHECR energy logarithmically,
it can be taken into account in an iterative procedure or the energy can be set
to a mean expected value. The $\sigma$ can be estimated from the information
about the signal or can be assumed as a minimization parameter.
  
As soon as we calculate $\rho(H_{\text{max}}^{\text{fluo}})$ the altitude of the EAS maximum
becomes known. This method of the altitude shower maximum reconstruction is
correct for any kind of particle.

Thus we have two minimization parameters $\beta'$ which along with TDP
determines $\vw$ and $\rho(H_{\text{max}}^{\text{fluo}})$ which determines $H_{\text{max}}$.

Since the expected value of the photons arrival direction $\va_i'(\theta^{\text{FOV}}_{\text{expected}}, \varphi^{\text{FOV}}_{\text{expected}})$ is computed, we can
minimize the $\chi$ function, that is defined as:

\begin{equation}
\chi = \sum\limits_{i=1}^{n_{gtu}}\frac{\left(\va_i - \va_i'\right)^2 N_{i}^{\text{p.e.}}}{(\sigma_{\Delta t}^2 + \sigma_{\text{pix}}^2)_i},
\end{equation}
where $\sigma_{\Delta t} = |\va_{i+1} - \va_{i}| = \sqrt{2(1-cos\alpha)} $ is calculated as variation of $\va_i$ within time of 1 GTU, where $\alpha$ is angle between 
$\va_{i+1}$ and $\va_{i}$. $\sigma_{\text{pix}} = \sqrt{2(1-cos\gamma)} $ is calculated as variation  of $\va_i$ inside a single pixel field of view $\Omega^{pix}_{FOV}$, where $\gamma$ is the cone angle
that one can calculate using following equation $\gamma\approx\sqrt{4\Omega/\pi}$. 
Both assumptions overestimate the real error and will be improved in future.
\begin{figure}[t]
 \centering\includegraphics[width=0.5\textwidth]{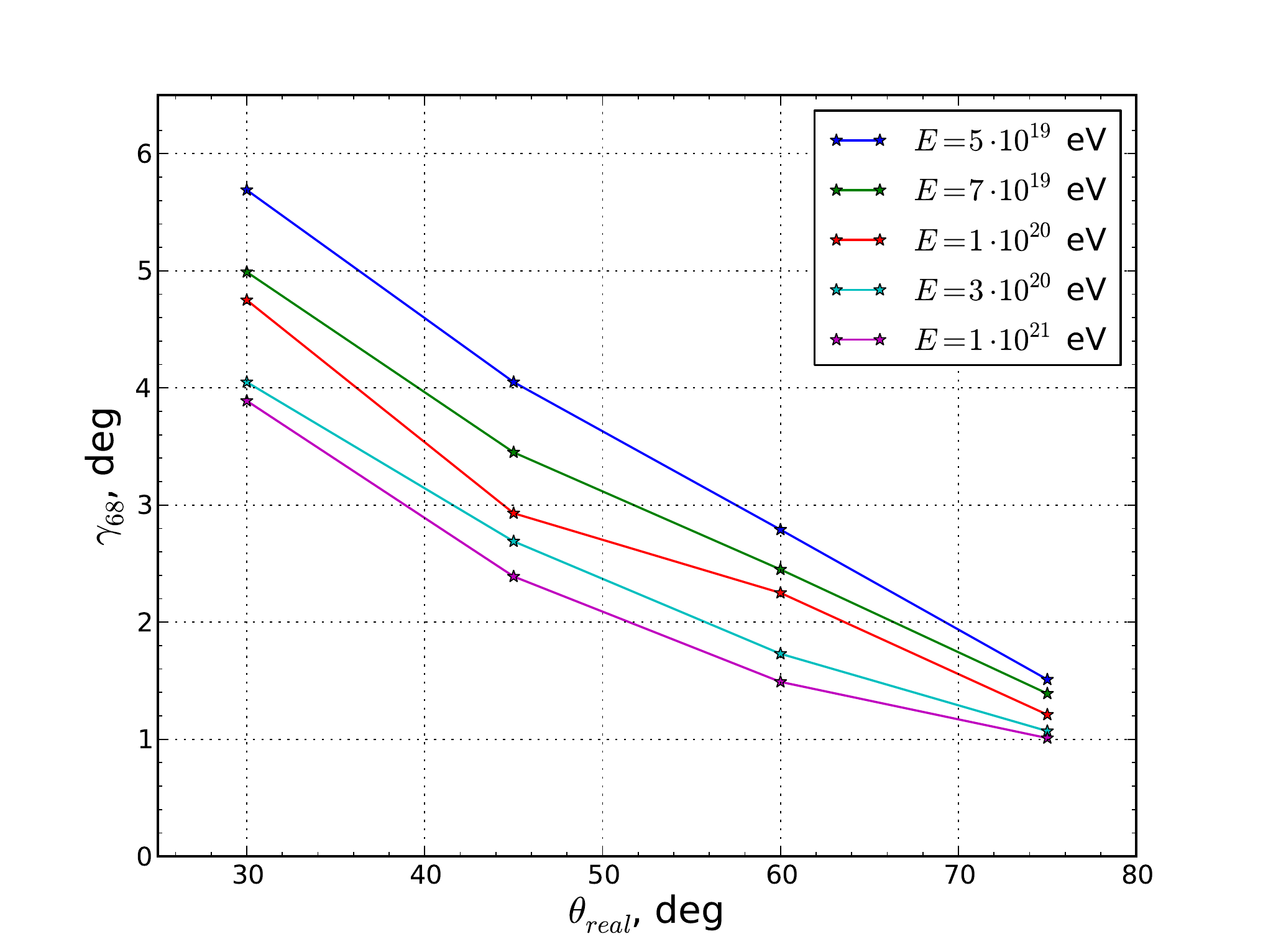}
 \caption{$\gamma_{68}$ for different energies and $\theta$ configurations.}
 \label{fig:gamma}
\end{figure}

To estimate the expected angular resolution of JEM-EUSO the angle $\gamma $
between the injected shower axis  and the reconstructed one is compared. 
We define  $\gamma _{68}$ as the value at which the cumulative distribution of $\gamma$ reaches 0.68. 
The systematic errors and statistical fluctuations are included within the
definition of $\gamma _{68}$. This parameter is used to see the overall performance of our 
reconstruction capabilities. The expected angular resolution without any selection cuts for different 
energies and EAS zenith angles $\theta$ is presented in the Fig. \ref{fig:gamma}.
The azimuth shower angle $\varphi$ is simulated randomly. 

\section{Conclusions}
Currently the pattern recognition algorithm was tested on the MC simulation and reconstructs proper tracks for all the triggered events.
The pattern recognition parameters will be fine-tuned in order to increase performance and minimize memory footprint.

The direction reconstruction accuracy satisfies the experiment 
requirements: $\gamma_{68} < 2.5^{\circ}$ for $\theta = 60^{\circ}$. The results are comparable with other direction reconstruction algorithms
used in ESAF. The errors used in $\chi^2$ calculation are to be updated.

\vspace*{0.5cm}
{
\footnotesize{{\bf Acknowledgment:}{This work was partially supported by Basic Science Interdisciplinary 
Research Projects of RIKEN and JSPS KAKENHI Grant (22340063, 23340081, and 
24244042), by the Italian Ministry of Foreign Affairs, General Direction 
for the Cultural Promotion and Cooperation, by the 'Helmholtz Alliance 
for Astroparticle Physics HAP' funded by the Initiative and Networking Fund 
of the Helmholtz Association, Germany, and by Slovak Academy  
of Sciences MVTS JEM-EUSO as well as VEGA grant agency project 2/0081/10.
The Spanish Consortium involved in the JEM-EUSO Space
Mission is funded by MICINN under projects AYA2009-
06037-E/ESP, AYA-ESP 2010-19082, AYA2011-29489-C03-
01, AYA2012-39115-C03-01, CSD2009-00064 (Consolider MULTIDARK)
and by Comunidad de Madrid (CAM) under project S2009/ESP-1496.
The work was partially supported by JINR grant No. 13-902-07.
}}

}

\clearpage

%% file: icrc2013-0713.tex



\title{On-line and off-line data analysis for the EUSO-TA and EUSO-BALLOON experiments}

\shorttitle{On-line and off-line data analysis for the EUSO-TA and EUSO-BALLOON experiments}

\authors{
Lech Wiktor Piotrowski$^{1}$ and Alessandro Pesoli$^{2}$\\
for the JEM-EUSO Collaboration.
}

\afiliations{
$^1$ RIKEN, Wako-shi, Saitama, Japan \\
$^2$ INFN and University Rome Tor Vergata, Rome, Italy
}

\email{lech.piotrowski@riken.jp} 

\abstract{JEM-EUSO, with EUSO-TA and BALLOON-EUSO prototypes, is an orbital detector of extremely energetic cosmic rays, which will be attached to a Japanese Experiment Module of International Space Station. It detects cosmic ray induced showers in the atmosphere by detecting the emitted fluorescence and Cherenkov UV radiation with 2.5 $\mathrm{\mu s}$ time resolution.

We show here the basics of communication protocol used to exchange the commands and data between the subsystems of the detector and the software used for on-line and off-line analysis. The data acquisition starts with calibration, in which we acquire data necessary to obtain the position of the single photoelectron peak. The calibration is an iterative procedure of changing the gain and threshold to obtain similar detection efficiency for each pixel. Then, in a standard acquisition mode, each exposure is quickly analyzed by a multi-level trigger and appropriate packets with all the hardware information are being stored into a file. The file is then transmitted to the Earth for off-line analysis. 

The off-line analysis and visualization involves two steps. First, the binary packets included in the file received from space have to be converted to a human readable format. Then for specific, most common tasks a designated software can be used. We chose ROOT TTree as the format for storing off-line data. The software for visualization and analysis serves the role of controlling the performance of the system. Its functionality includes among others: visualizing the photoelectron counts for the whole focal surface, analysis of photon counts over time (lightcurves), drawing pixel and exposition statistics, analysis of calibration S-curves. Additionally, the flow of commands exchanged internally in the system can be visualized, to allow tracing the possible errors and improving the system capabilities.}


\maketitle

\section{Introduction}

Cosmic rays -- high-energy particles of extra-terrestrial origin -- are dynamically developing area of science since their discovery by Wulf in 1909\cite{bib:wulf}, and following observations by Pacini in 1911\cite{bib:pacini} and Hess in 1912\cite{bib:hess}. Further research has extended our knowledge about this phenomenon significantly, revealing most of the energy spectrum, composition and sources of this radiation. However, the extremely high energy tail of the energy spectrum remains mysterious. Among others, we are still not sure if there is and what is the cut-off energy and what in the Universe can accelerate particles to such huge velocities.

The main reason for those questions remaining unanswered is the flux of cosmic rays reaching Earth, which drops significantly with their energy. If one considers protons, nuclei and electrons, the rate for $10^9$ eV particles is $\frac{10000}{\mathrm{m^2\cdot s}}$, but above $10^{20}$ eV it drops to $\frac{1}{\mathrm{km^2\cdot 100 yr}}$, which makes the study in the highest energies regime extremely difficult. To obtain a statistically significant result in an average experiment lifespan, one has to monitor at least thousands square kilometers of ground. Experiments dedicated for this task, such as Pierre Auger Observatory in Argentina \cite{bib:auger} and Telescope Array in United Stated \cite{bib:ta} have been constructed and have already presented important results.

Mentioned detectors observe atmospheric particle showers induced by the primary cosmic ray, with a hybrid method. They directly detect some of the secondary particles hitting on-ground scintillation detectors and indirectly observe the cascade development with wide field of view UV telescopes, registering the produced fluorescence light. The ground surface coverage of these facilities is limited by the possible extension of detectors network. For the UV telescopes the volume of observed atmosphere is limited by their field of view and the proximity of atmospheric surface, as well as several other factors. The coverage of the telescope could be significantly increased with increase of the distance to the atmospheric surface. Introducing observations of shower-induced UV light from the orbit makes this idea, standing behind JEM-EUSO experiment, possible.

The key part of the JEM-EUSO experiment is an UV telescope\cite{bib:EUSOperf}, consisting of curved focal surface assembled from about 6000 Multi Anode Photomultiplier Tubes (MAPMT), 64 pixels each. This, in combination with Fresnel lenses based optics, gives $\pm 30^{\circ}$ field of view and $4.5'$ angular resolution. The temporal resolution is 2.5 $\mathrm{\mu s}$. The apparatus will be attached to the Japanese Module on the International Space Station. Orbiting on the altitude of $\sim 400$ km, the observational aperture is a circle of radius of about 230 km on the ground, which is larger by a factor of $\sim 56$ then the Pierre Auger Observatory's. Changing the inclination of the telescope can further increase the aperture. These observational parameters give JEM-EUSO a chance for significantly extending the scientific knowledge in the area of cosmic radiation.

The JEM-EUSO launch will be preceded by prototype experiments, EUSO-TA and BALLOON-EUSO, which will be used for testing the subsystems and estimating the performance of the final experiment. The EUSO-TA on-ground experiment is being installed in Utah, United States in the site of Telescope Array, which allows for comparing results from limited focal surface and smaller lenses with this experiment. The BALLOON-EUSO will be an attempt to perform dedicated observations of cosmic ray induced atmospheric showers from above for the first time. The apparatus will consist of limited focal surface and smaller lenses.

\section{Communication protocol for commands and data}

The UV light reaching the JEM-EUSO is focused on the focal surface and registered by the MAPMT. Simplifying, the signal from the number of MAPMTs increasing with each stage of processing, goes through EC-ASIC (Elementary Cell Application Specific Integrated Circuit) \cite{bib:ASIC} unit to a Photodetector Module (PDM) and then to Control Cluster Board (CCB) \cite{bib:CCB} and Central Processing Unit (CPU) \cite{bib:instrumental}. The CPU prepares the acquired data for sending to Earth and proceeding off-line analysis. A reverse flow of information is also required, where higher-level modules send requests for specific tasks to lower-level modules and retrieve feedback.

\begin{figure}[t]
 \centering
 \includegraphics[width=0.5\textwidth]{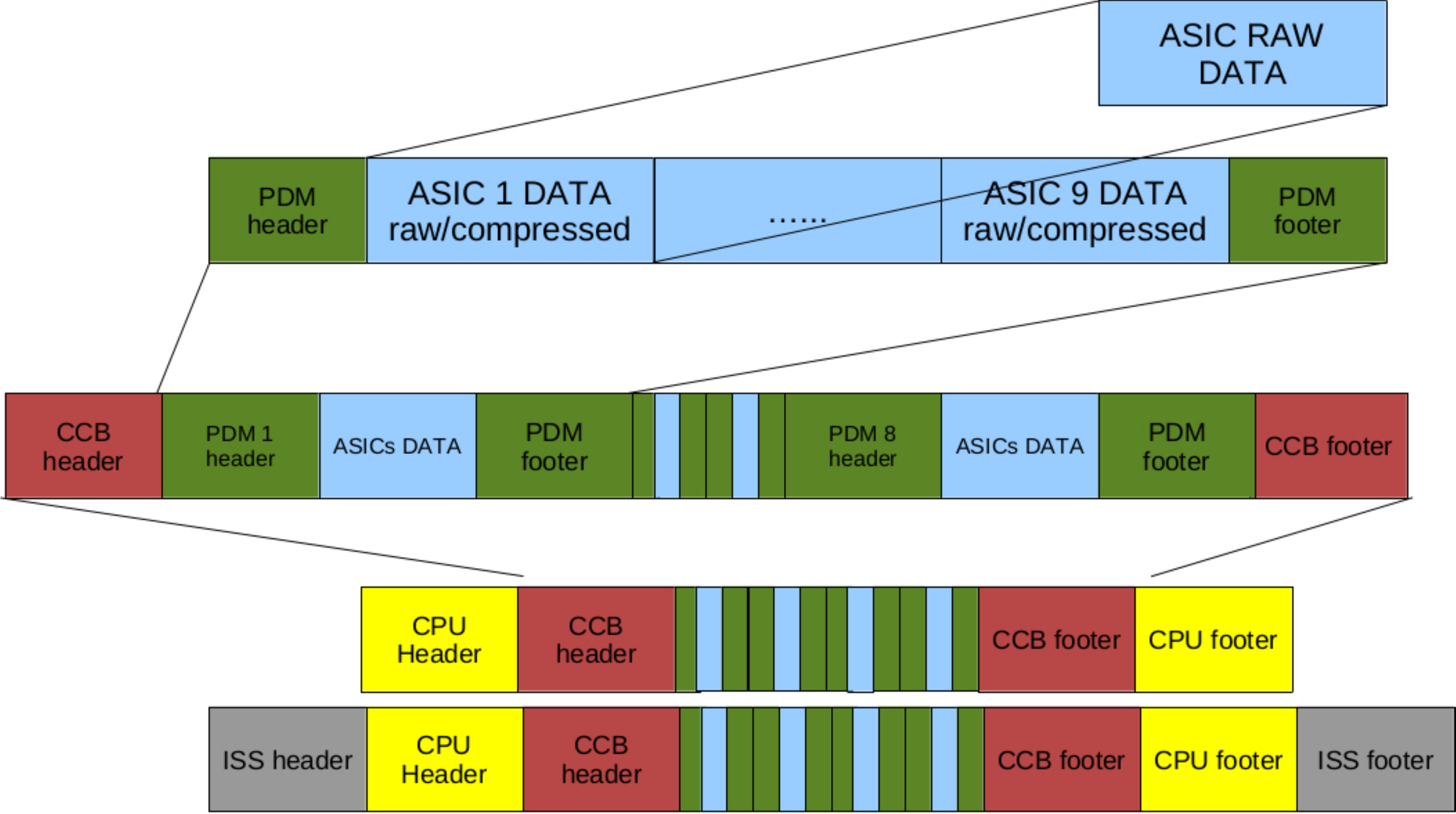}
 \caption{A scheme of ``matroska'' style packet organization in JEM-EUSO experiments. Data packets from lower level subsystem are joined and encapsulated into higher level subsystem packet.}
 \label{fig:matroska_packets}
\end{figure}

The exchange of information is made with encapsulated packets, organized in a ``matroska'' style. The amount of information contained in a packet increases with the level of processing (fig. \ref{fig:matroska_packets}). Each processing unit has a C language structure for storing the information it provides. Pointers to the structures of the lower level units are stored in the higher level unit structure, which adds its own data. For conveying commands, packets containing appropriate C language structures are sent from higher level units down, each lower level unit ensuring that the order gets to all the requested subunits. 

\section{Data acquisition process}

The JEM-EUSO apparatus can work in two data acquisition modes. The default is reacting to the internal trigger based on a multi-level algorithm constantly analyzing incoming data. The Level 1 trigger is the fast trigger issued by the PDM board, the more sophisticated Level 2 trigger being issued by the CCB board. The second mode is a reaction to an external trigger, which can be used for working in coincidence with other experiments. It is important for a prototype, EUSO-TA telescope which has very limited self triggering capabilities due to small ($\pm 4^{\circ}$) field of view. In current configuration the external trigger is received by a time synchronization subsystem called clockboard and is directly passed to CCB \cite{bib:CCB}. Additionally, CPU may requests snapshots of data or issue triggers to lower level subsystems on its own.

Obtaining proper measurements in data acquisition mode requires prior calibration of the apparatus, to properly count the amount of photons hitting each pixel. For this tasks CPU requests and analyses specific data for each pixel, which will be explained in more detail in sec.\ref{sec:online_analysis}. For the BALLOON experiment calibration will be performed offline, prior to launch. For the JEM-EUSO and EUSO-TA experiments, the frequency with which the calibration will be performed has not yet been decided and will depend on the behaviour of subsystems.

Except acquiring data on trigger and on calibration request, data acquisition process includes checking the status of apparatus elements and reaction to alarms concerning them -- so called housekeeping. The final stage of data acquisition process is sending the chosen selection of data to Earth.

\section{Online analysis of data}
\label{sec:online_analysis}

Due to the in-space power consumption restrictions for JEM-EUSO and weight limits for BALLOON-EUSO, the CPU of EUSO experiments has very limited processing capabilities. Therefore the on-board software has to be characterized by low memory, processing power and other resources requirements. To fulfill this demand, the on-line data analysis and management software is written in C/C++ using only simple libraries. On the other hand, the goal for the software is to be fast and reliable to fulfill the scientific objectives of the mission.

\subsection{Calibration}

\begin{figure}[t]
 \centering
 \includegraphics[width=0.4\textwidth]{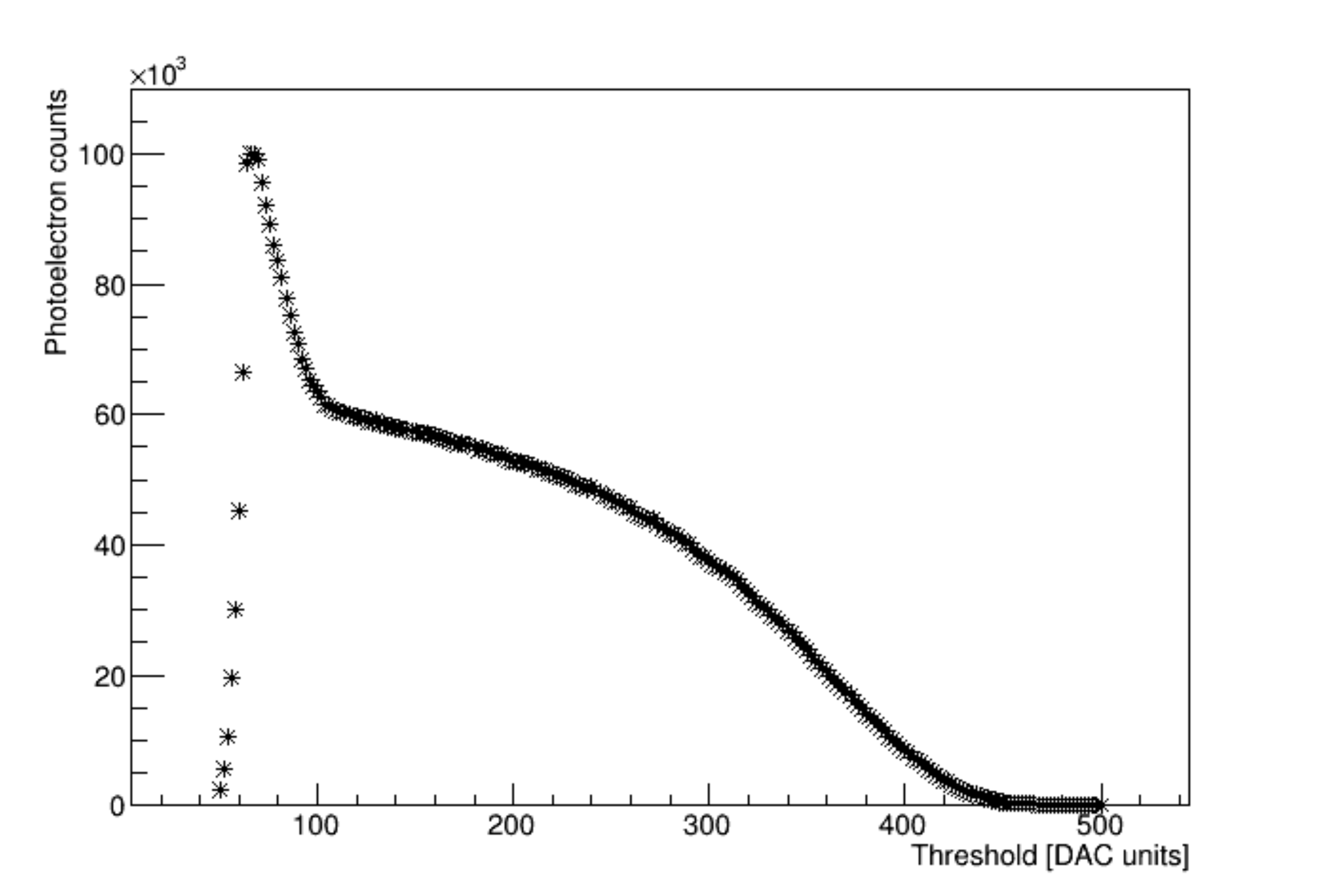}
 \caption{A sample S-curve -- sum of photon counts for specific number of measurements over the threshold for single photoelectron detection (in Digital to Analogue Converter (DAC) units).}
 \label{fig:scurve}
\end{figure}

\begin{figure}[t]
 \centering
 \includegraphics[width=0.4\textwidth]{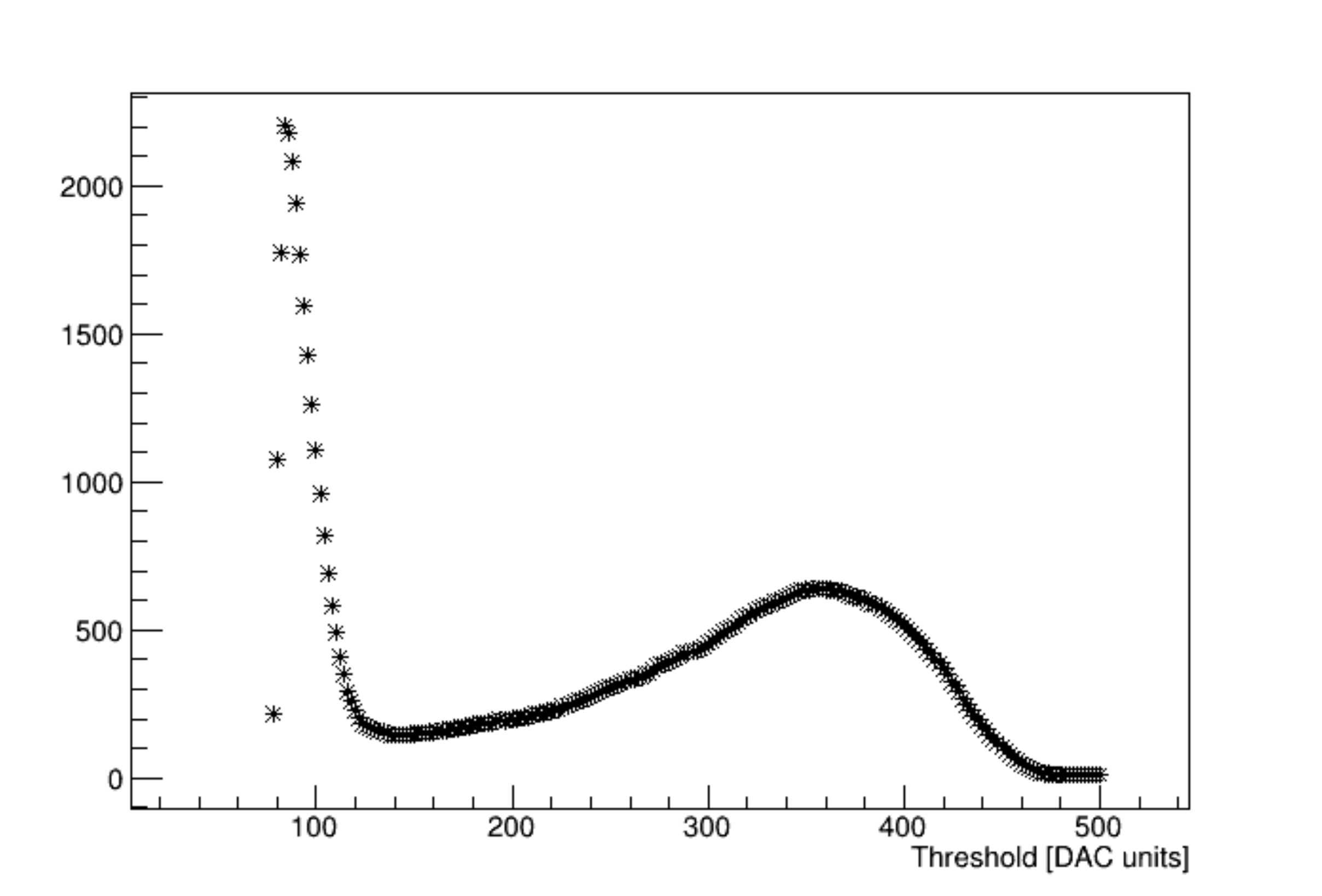}
 \caption{A sample single pixel ``single photoelectron spectrum'' measured in laboratory, for threshold levels represented in DAC units. The noise peak on the left and single photoelectron peak on the right are clearly visible. The distance between peaks is determined by the total pixel gain. The minimum between the peaks is the single photoelectron valley -- the optimal position of threshold for single photoelectron detection.}
 \label{fig:single_pe_spectrum}
\end{figure}

Most of the data getting past the level 1 and 2 triggers is intended for offline analysis. However, as mentioned above, standard data gathering modes require prior calibration. Therefore the main computational task for data analysis of CPU is performing the calibration, which is an iterative process of sending commands and acquiring data from lower level subsystems.

The goal of the calibration is to obtain a similar overall gain for the single photon detection for all apparatus' pixels. In general we want to reject the noise from PMT and EC-ASIC electronics and accept more intense signal coming from a single photon. This requires analysis of so called S-curves -- explained below -- obtained in the process.

An S-curve contains Digital to Analogue Converter (DAC) signal over threshold for single photoelectron detection. To obtain it CPU requests measurement of a high number of frames for each threshold setting, to minimize statistical fluctuations\footnote{The actual number of frames requested for each threshold level has yet to be determined, but initial measurements show that the single photoelectron signal should be around $0.1\%-1\%$ of the maximal signal, requiring order of 10000-100000 GTUs} (fig. \ref{fig:scurve}). A ``single photoelectron spectrum'', which reveals separated peaks for the electronics noise and single photoelectron, is the derivative of an S-curve (fig. \ref{fig:single_pe_spectrum}). The optimal threshold for single photon detection lies between the peaks.

However, in JEM-EUSO experiments a single threshold must be set for all PMT pixels. Therefore the calibration procedure must determine an optimal amplifier gain for all pixels to have the same total gain, which roughly corresponds to obtaining a similar ``single photoelectron spectrum'' shape. For this task the algorithm calculates the median position of the single photoelectron peak for all pixels and adjusts the gain of all pixels to move the peak to the median position. After the adjustment the measurement and median calculation is repeated to determine if the similarity of efficiency is satisfactory. If not, adjustment values are calculated and utilized again, until satisfactory performance is obtained.

Finding the single photoelectron peak requires calculating the derivative of data, which due to time constrains and thus low statistics may be noisy. Therefore more sophisticated, noise suppressing algorithms have to be used. Currently we have obtained the best results using a Super Lanczos Low-Noise Differentiator\cite{bib:hamming}, but better procedures may be implemented in the future. The procedure of peak finding is simply finding a highest value of the ``single photoelectron spectrum'' which lies after the single photoelectron valley (smallest value after the noise peak).

\subsection{Event data}

The calibration algorithm operates on data packets that contain the average of GTUs taken for each threshold setting for each pixel of the focal surface. In the case of real event data the packet format is slightly different -- it contains the count of photons detected by each pixel of the focal surface in each GTU. Additionally data from KI (charge-to-time converter, for higher signals) is also supplied.

The event data packets, each containing data for 128 consecutive GTUs are added sequentially to a file, which also contains calibration packets, commands exchanged between the subsystems and other information. On request, the file is transferred to Earth.

\section{Basic visualisation and offline analysis of data}

\begin{figure}[t]
 \centering
 \includegraphics[width=0.4\textwidth]{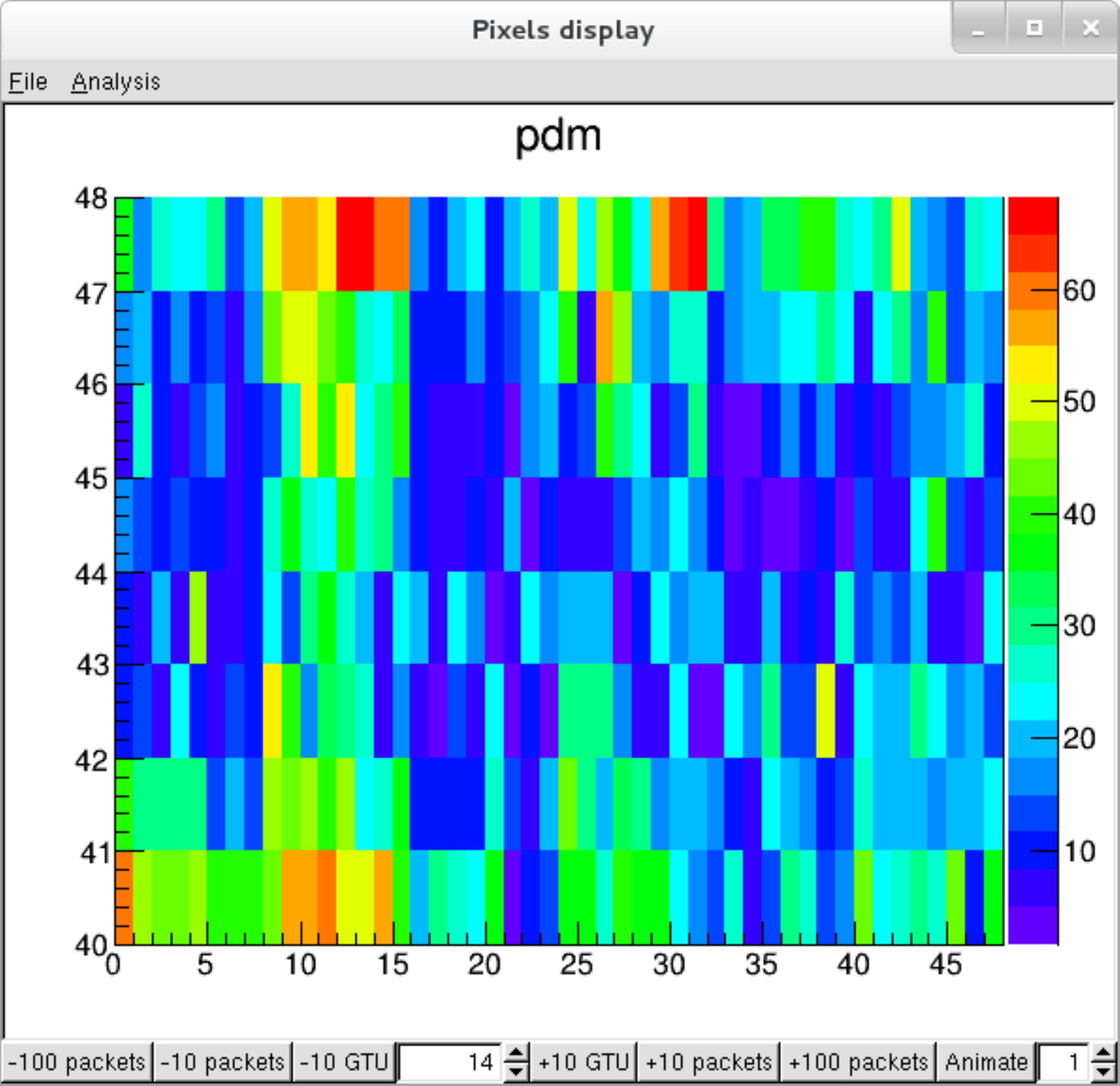}
 \caption{Screenshot of the main window of the ``etos'' program, showing photoelectron counts for pixels of 6 PMTs, for single GTU.}
 \label{fig:etos_main}
\end{figure}

\begin{figure}[t]
 \centering
 \includegraphics[width=0.4\textwidth]{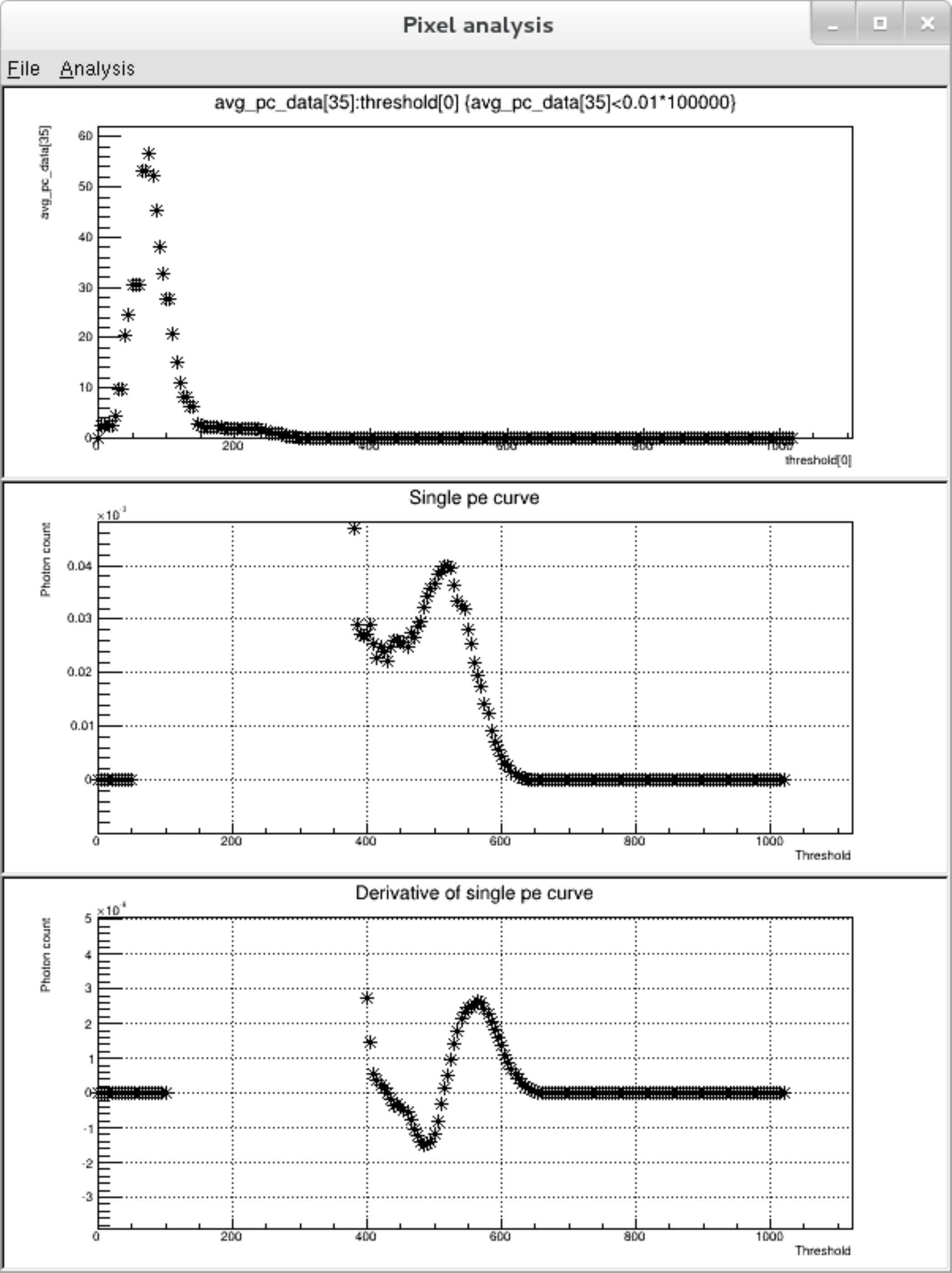}
 \caption{Screenshot of the ``Pixel analysis'' window of the ``etos'' program, displaying the analysis of the calibration for a single pixel. The top plot shows an S-curve -- the counts over threshold value plot, the middle is a ``single photoelectron spectrum'' -- a derivative of an S-curve, and the bottom is the second derivative of an S-curve. Single photoelectron peak is clearly visible in the middle of the ``single photoelectron spectrum''.}
 \label{fig:etos_calib_curves}
\end{figure}

The data is sent to Earth as packets which are simply C-structures in a binary format contained in a single file, not convenient for offline analysis and visualization. Therefore we have decided to convert the data on-ground to Cern ROOT\cite{bib:root} TTree format, which is used by most of JEM-EUSO offline analysis software. The TTree format is convenient both for use inside dedicated software as well as for quick look into the packets' contents. The program for the conversion is called ``etot'' (EUSO to TTree) and simply analyses all the packets in the file sequentially converting them and assigning to separate TTrees for calibration, event data, etc.

The resulting TTrees can be read by a basic visualization and analysis software ``etos'' (EUSO to Screen). The main tasks of the program are to provide quick look into the data, displaying the photon counts on the focal surface for specific GTU (fig. \ref{fig:etos_main}) or change of counts over GTUs for specific pixel. Similarly, in case of calibration, it can display average value for focal surface pixels for specific threshold or average over threshold for specific pixels (fig. \ref{fig:etos_calib_curves}.

Current features of ``etos'' include:
\begin{itemize}
	\item visualization of the focal surface\footnote{Currently for single PDM board containing 36 PMTs for EUSO-TA purposes, but this can be easily extended in the future} for event data
	\item visualization of the focal surface for calibration data
	\item browsing through GTUs/thresholds for focal surface visualiation
	\item display of lightcurve -- photoelectron count over time (GTU) -- for chosen pixel
	\item display of S-curve
	\item display of first derivative and second derivative of the S-curve
	\item single photoelectron peak and valley finding
	\item calculation of lightcurve periodogram
	\item analysis of pixel's single photoelectron peak behaviour with gain change
	\item display of focal surface pixels statistics
\end{itemize}

The features currently being under development include animating focal surface display over GTU's threshold, summing and averaging the photon count over specific time and setting threshold for pixel value display. The functionality can be and will be extended with future users' requests. ``etos'' is written in Python programming language with bindings to the ROOT framework -- PyROOT. In future we plan to include visualisation of data and commands flow to ``etos'' for system monitoring purposes. 
 
\section{Conclusions}

JEM-EUSO telescope is an ongoing development requiring dedicated tools for online and offline data analysis, which can be thoroughly tested with prototype EUSO-TA and BALLOON-EUSO experiments. The main computational task of the CPU's online analysis is performing calibration. The main part of the calibration algorithm, utilizing efficient low-noise derivative has been implemented and shows promising results. The tuning and integration with the full Data Processing system is ongoing.

The basic offline analysis and visualization software -- ``etot'' and ``etos'' programs are successfully functioning and already proved a valuable tool for analysis of the laboratory results from the EUSO-TA apparatus. Their simplicity allows for quickly extending their functionality in future, as needed.

\vspace*{0.5cm}
{
\footnotesize{{\bf Acknowledgment:}{This work was partially supported by Basic Science Interdisciplinary 
Research Projects of RIKEN and JSPS KAKENHI Grant (22340063, 23340081, and 
24244042), by the Italian Ministry of Foreign Affairs, General Direction 
for the Cultural Promotion and Cooperation, by the 'Helmholtz Alliance 
for Astroparticle Physics HAP' funded by the Initiative and Networking Fund 
of the Helmholtz Association, Germany, and by Slovak Academy  
of Sciences MVTS JEM-EUSO as well as VEGA grant agency project 2/0081/10.
The Spanish Consortium involved in the JEM-EUSO Space
Mission is funded by MICINN under projects AYA2009-
06037-E/ESP, AYA-ESP 2010-19082, AYA2011-29489-C03-
01, AYA2012-39115-C03-01, CSD2009-00064 (Consolider MULTIDARK)
and by Comunidad de Madrid (CAM) under project S2009/ESP-1496.

LWP acknowledges the support from JSPS Postdoctoral Fellowship.           
for Foreign Researchers.
}}

}

\clearpage

%% file: icrc2013-0858.tex



\title{Absolute calibrations of the Focal Surface of the Jem-Euso Telescope}

\shorttitle{calibration for JEM-EUSO }

\authors{
P. Gorodetzky$^{1}$, C. Blaksley$^{1}$, S. Dagoret-Campagne$^{2}$, M. Fukushima$^{6}$, A. Haungs$^{3}$, D. Ikeda$^{6}$, A. Insolia$^{4}$, M. Karus$^{3}$, Y. Kawasaki$^{5}$, H. Miyamoto$^{2}$, L. Piotrowski$^{5}$, H. Sagawa$^{6}$, N. Sakaki$^{4}$, A. Segreto$^{4}$, M. Takeda$^{6}$, Y. Takizawa$^{4}$, Y. Tsunesoda$^{7}$, T. Tynieniecva$^{8}$, L. Wiencke$^{9}$
for the JEM-EUSO Collaboration$^{10}$.
}

\afiliations{
$^1$ AstroParticule et Cosmologie, Univ Paris Diderot, CNRS/IN2P3, Paris, France\\
$^2$ Laboratoire de l'Acc\'el\'erateur Lin\'eaire, Universit\'e Paris Sud-11, CNRS/IN2P3, Orsay, France\\
$^3$ Karlsruhe Institute of Technology (KIT), Germany\\
$^4$ Dipartimento di Fisica e Astronomia - Universita di Catania, Italy, Istituto Nazionale di Fisica Nucleare - Sezione di Catania, Italy\\
$^{5}$ RIKEN Advanced Science Institute, Wako, Japan\\
$^{6}$ Institute for Cosmic Ray Research, University of Tokyo, Kashiwa, Japan\\
$^{7}$ Interactive Research Center of Science, Tokyo Institute of Technology, Tokyo, Japan\\
$^{8}$ National Centre for Nuclear Research, Lodz, Poland\\
$^{9}$ Colorado School of Mines, Golden, USA\\
$^{10}$ http://jemeuso.riken.jp
}

\email{philippe.gorodetzky@cern.ch} 

\abstract{Jem-Euso will detect light from Fluorescence induced by Cosmic Ray showers. How do we go back from the MAPMT signals to the shower energy ? This paper describes the real meaning of what is called "calibration" when it is only a measurement of an intrinsic property: that is here the PMT efficiency. 
}



\maketitle

\section{Introduction}

The project Jem-Euso is basically a large aperture UV telescope looking from the International Space Station at the earth atmosphere to detect the nitrogen fluorescence light induced by the myriads of charged particles created in an Ultra-High Energy Shower. It is composed of large (5m$^2$) plastic Fresnel lenses, focusing the light on a spherical focal surface (FS) composed of roughly 5000  64-anodes photomultipliers (PMT) for a total of 350000 independent pixels. The field of view is $\pm 30^o$, for a ground observed surface, when in nadir mode, of 200 000 km$^2$.
The light is produced by the shower through nitrogen fluorescence at altitudes between roughly 0$-$10~km. The shower max is at 3$-$4~km. The standard model gives the first constant in the relation between the shower energy $E$ to the number of charged particles $N$ around the shower maximum depth $X_{max}$. The second constant relates the light produced by a charged particle: the fluorescence yield $YF$~\cite{bib:fluo}. This light has to travel upwards through the atmosphere. Its characteristic transmission parameter $T_A$ will be given by the Atmospheric Monitoring system AMS~\cite{bib:ams}. The light has to go through the lenses with a transmission $T_L$. Finally, the light hits the PMT pixels and is transformed into measured charges. This is called pixel efficiency $\epsilon_{pix}$. It is expressed in Amperes / Watts (A/W) in the general case of light strong enough to induce pile-up, where the resulting charges is the product of the number of photoelectrons produced simultaneously by the photocathode, by the gain of the tube. In the case of weak light (we will see  soon this is valid for Jem-Euso), where we have no pile-up, the charge per photoelectron is only the gain. Then the efficiency can be expressed in photoelectrons per incident photons (a typical value for the Jem-Euso PMTs is 0.1 A / W = 0.3 pe / ph at 400 nm).  
 We will now describe how we measure $\epsilon_{pix}$ in an absolute and precise way.

\section{Jem-Euso mode of operation}
For the main physics of Jem-Euso: shower observation, the time occupancy by photoelectrons is small: one pulse is 2~ns at its base (1~ns at mid height). The maximum rate (given by CORSIKA simulations) for a $10^{20}$ eV shower is around 30 photoelectrons (pe) per Gate Time Unit (GTU = 2.5~$\mu$s), which is 1.2 pe per 100 ns. The usual background (light of the stars reflected at earth surface) and light glow is 20 times less. This is why Jem-Euso will be in single photo-electron (SPE) mode. In this SPE mode, the spectrum is composed of two peaks: the first one, narrow, is the pedestal and its width is due to the integration of the base line when there is a trigger (due for instance to light in other pixels but no light in the examined one). The second peak is due to the integration of the pulse arriving when there is light present on that pixel, corresponding to a SPE. fig.\ref{fig:Fig1} is an example of these spectra. This obeys the Poisson statistics and, from the number of SPE measured in a GTU, one can retrieve the true number of incident photons. The surface of the SPE peak is then the number of photoelectrons out of the photocathode and having reached the first dynode.

In the case of strong light (cities, lightnings, meteors...), the pile up can become important if we keep the SPE mode. The first consequence is damage to the PMT (too much current at the anodes). Second, impossibility to measure the phenomenon responsible for this strong light because of saturation. Jem-Euso is equipped with a system of fast switches which reduce the value of the voltage applied to the photocathode in less than 2~$\mu$s. The voltages applied to the other dynodes stay constant, hence, it is the collection efficiency which is reduced. If, for example, the cathode voltage is reduced by some 200~V to be equal to the first dynode voltage, the collection efficiency is diminished by a factor~100. But the tube having still the same voltages applied to its dynodes, the gain is unchanged. Only the number of photoelectrons reaching the first dynode is reduced hence the tube is still working in SPE mode.

The advantage of this method compared to the classical reduction of all dynodes voltages is its speed, because the capacity involved is small.

\section{How to produce pure SPE spectra in the laboratory?}

\begin{figure*}[!t]
  \centering
  \includegraphics[width=\textwidth]{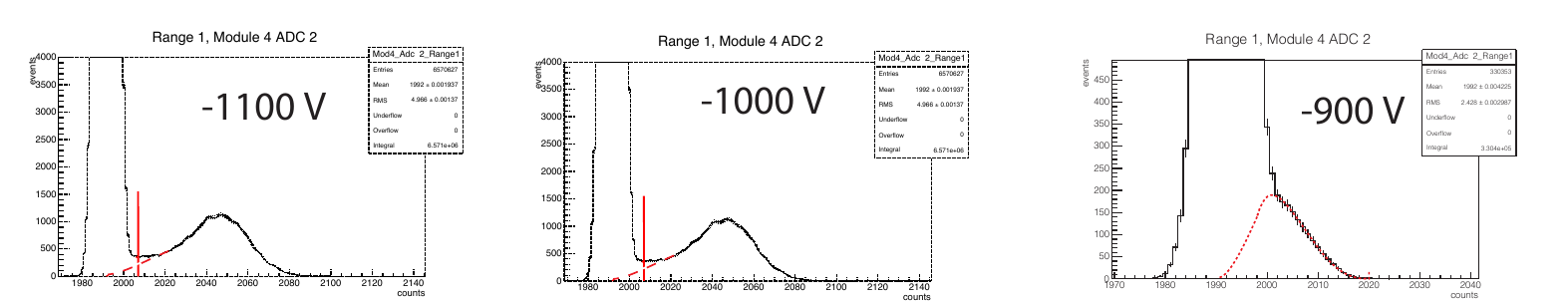}
  \caption{Single photoelectron (SPE) spectra of one pixel of a M64 PMT at 3 voltages. Threshold efficiency corresponds to the SPE peak surface above the discriminator threshold, while full efficiency corresponds to the surface of the SPE peak extrapolated to 0 (dotted curve). It is impossible to extract the efficiency at -900 V with precision.}
  \label{fig:Fig1}
 \end{figure*}

First, a little bit of philosophy. There are two ways to measure the absolute efficiency of a PMT pixel: a) send a known light, or b) compare to a known detector. It is extremely difficult to have a uniform precisely known beam on a small surface even if this non uniformity can be corrected offline, with some effort. For instance, consider the reference~\cite{bib:moon} explaining how to use the moon as an absolute calibration tool when in-flight. There, we cannot use b) (cannot replace the focal surface with a known calibrated detector) and this moon method is not easy. Method a) cannot calibrate to better than 5$-$10\%, so b) in the lab, which can reach 2\% is by far the best solution and will be described later.
LEDs give fast signals, and it is easy to produce light pulses in times as short as 5$-$10~ns. The secret is to pump more current in them (up to 1~A in 10~ns). If the amount of light produced per pulse is such that we have 1 photoelectron per 50 pulses, then the Poisson law tells that there will be less than 1 pulse of two photoelectrons together for 100 pulses corresponding to one photoelectron. If one records a spectrum by integrating around the light pulse, in a short gate (10$-$20~ns), then the pedestal will be $P_0$ events, the one photoelectron will be $P_1$, and its surface will be $P_0$ / 50 and the 2 photoelectron peak $P_2$ will be $P_1$ / 100. This is roughly what is shown in the left of fig.\ref{fig:Fig1}.
$P_0$ results from the integration of the electronic noise, which is small, so the $P_0$ peak is narrow. $P_1$ is the result of a multiplication of electrons in the PMT up to a few 10$^6$. Its width is governed by the multiplication factor of the first dynode which is around 5, so that its $\sigma(P_1)$ is about $1 / \sqrt{5}$ = 0.45. So $\sigma(P_1)/$~
position ($P_1$) is roughly 0.45. $P_2$ (which cannot be seen at the 1\% level) has little bit smaller width.
The gain is the distance expressed in charges between the $P_1$ and $P_0$ peaks. So, to arrive to an absolute measurement, it is important to measure what is the channel width in $fC$ of each QDC used (we have four 16-channel QDC to look at one PMT tube at a time). See paper~\cite{bib:pmt} in these proceedings.
The surface of the $P_1$ peak divided by the number of light pulses is close to the pixel efficiency $\epsilon_{pix}$. We do not speak of PMT efficiency, because all the pixels are independent. It is the product of the transmission ($>$ 99\%) of the 2 mm BG3 filter glued on the photocathode glass times the glass transmission ($>$ 99\%) times the photocathode, or quantum efficiency $\epsilon_Q$, times the probability for an electron escaping the photocathode to reach the first dynode, called collection efficiency $\epsilon_{coll}$ (about 70\%, but varies with voltage) times the probability to go from one dynode to the next (here 100\%). So, basically, $\epsilon_{pix} = \epsilon_{Q} \times \epsilon_{coll}$. The 4\% factor of the light loss between air (or vacuum) and the BG3 filter is not mentioned here, because it is inherent to this experiment, and it will not vary. The evaporated photocathode thickness can vary along the pixels, and the collection efficiency, due to electrostatic effects can also vary along the pixels.

Now, the surface of the $P_1$ peak can be taken in two ways: the exact one $S_{exact}$ with an extrapolation of $P_1$ to its origin (the $P_0$ position, see fig.\ref{fig:Fig1}), which is time consuming (Polya fit). Or, we can set a threshold (shown in fig.\ref{fig:Fig1} in red) in the valley between $P_0$ and $P_1$ at 1/3 of the distance, and say that everything above that threshold, $S_{thres}$, is representative of $P_1$ surface with a correction. This is what is done with Jem-Euso DAQ : the threshold mode. At the laboratory, as the physicist time is cheap, one determines what is the ratio of the surface above the threshold to the exact total surface. 
Another factor contributing to the PMT efficiency is the size of the cloud of electrons after multiplication. Inside the PMT, the electron "shower" reaching the anode is slightly larger than a pixel. When a pixel (2.88$\times$2.88 mm) is illuminated in its center (with a 0.1~mm precision) by a 0.3~mm light beam, we see that 4\% of the events lie on the pixels neighbors to the illuminated one. The true efficiency $\epsilon_{true}$ is then the efficiency of the pixel illuminated determined by $S_{exact}$ plus 4\%. $\epsilon_{true}$ will be then what we call the real pixel efficiency $\epsilon_{pix}$.
Finally, a very important point, generally not seen by many experiments: $\epsilon_{pix}$ varies with the high voltage applied to the dynodes. This is due mainly to the electrostatic effects of the field between the photocathode and the first dynode (responsible for $\epsilon_{coll}$, the collection efficiency). The tube being square and not cylindrical, this field is not uniform and the non-uniformity will vary with the high voltage. So, it is mandatory to measure the gains of the pixels.

\begin{figure}
  \centering
  \includegraphics[width=0.5\textwidth]{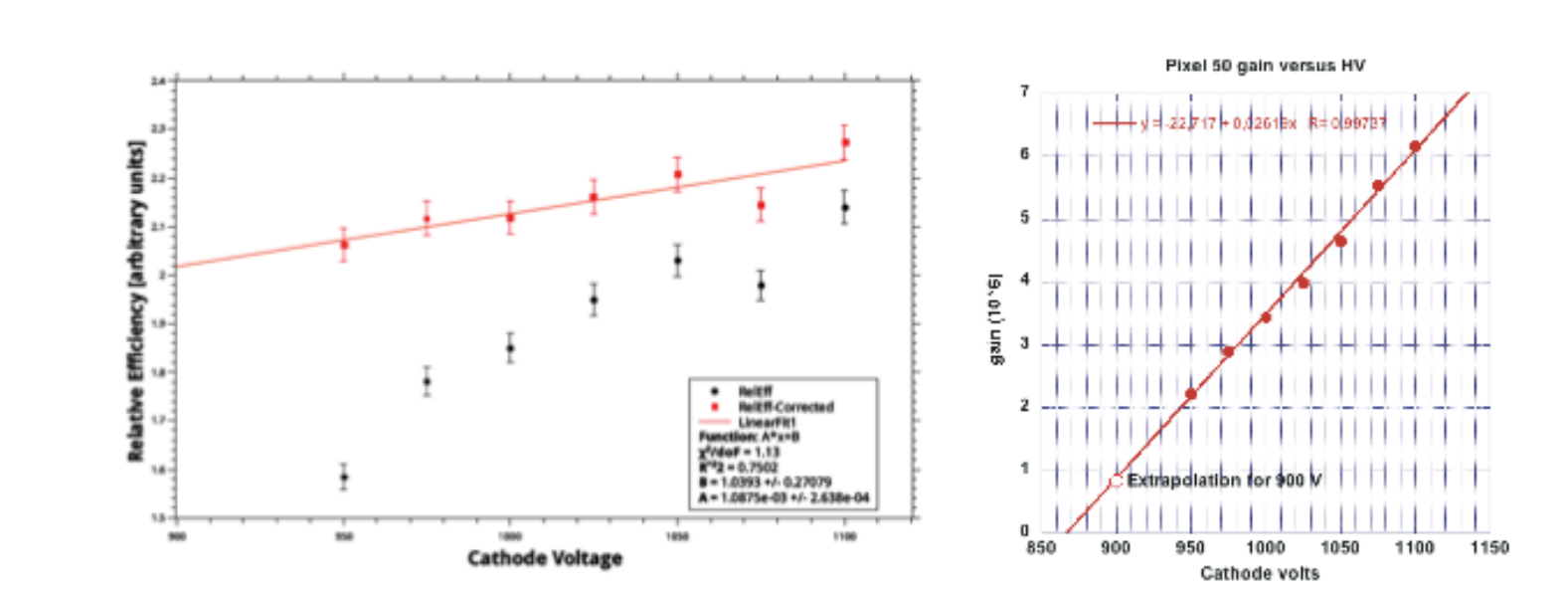}
  \caption{Left: red: $\epsilon_{pix}$ and black $E_{thresh}$ for different HV, (open circles); Right: gain versus HV with extrapolation to -900 V.}
  \label{fig:Fig2}
 \end{figure}

Fig.\ref{fig:Fig2} shows in red the true pixel efficiency (one pixel of a PMT) for different voltages and the correspondence between gain and voltage is given in the figure on the right. In black, we see how the efficiency measured above the threshold varies faster with the voltage. This arises from the fact that the threshold is constant, and as the gain is reduced, the pedestal does not move, and the SPE peak gets "eaten" by the threshold (as illustrated in the far right spectrum of fig.\ref{fig:Fig1}. But beware: the spectra were taken at the laboratory with CAEN~QDC~C1205 with a channel width of about~19~fC, so that we had to be between -1100~V and -1000~V to have enough gain. In Jem-Euso, the front-end electronics~\cite{bib:balloonelec} (home made ASICs~\cite{bib:asic}) are much more sensitive, and at -900~V, we have spectra similar to the one of fig.\ref{fig:Fig1} at -1000~V. So, in Jem-Euso operation, the PMTs being read by the ASICs, the black curve of the left figure has to be shifted towards the left by about 100~V. 
Finally, from fig.\ref{fig:Fig2}, right and left, one can conclude that between 1100~V and 900~V, the gain varies by a factor of~7 and $\epsilon_{pix}$ by~8\%.

\section{Experimental laboratory set-up}

\begin{figure}
  \centering
  \includegraphics[width=0.4\textwidth]{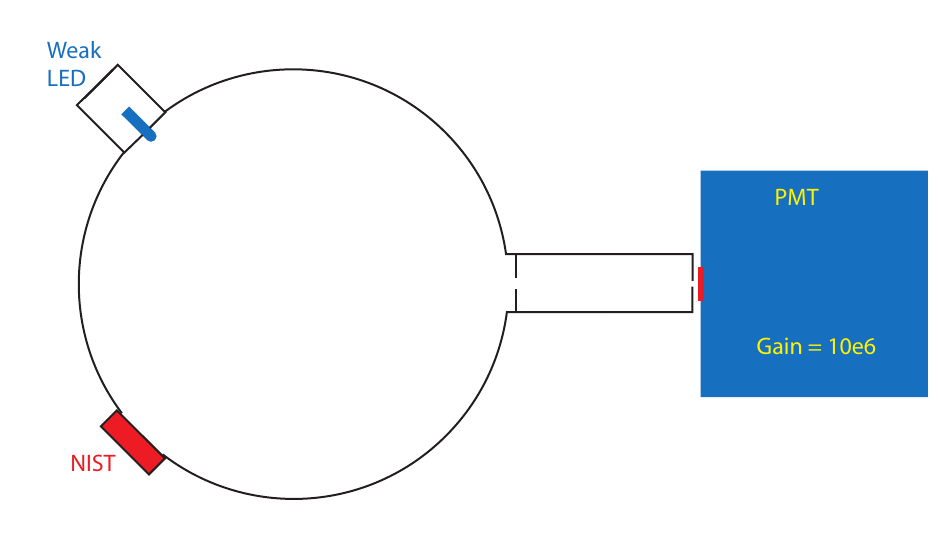}
  \caption{Integrating sphere arrangement with PMT.}
  \label{fig:Fig3}
 \end{figure}

The LEDs used to send the light are not totally stable. Their yield varies with temperature. It is mandatory to monitor the light intensity with a NIST photodiode. This device being non-polarized is extremely stable over 10~decades, ranging from 1~pW to 10~mW at 400~nm.
The light from the LED has to be splitted in a very stable way between the PMT and the NIST photodiode. The best (most stable and easy to set-up) splitter is an integrating sphere. The basic principle is that the amount of light going through a port is proportional to the port surface.
 We use a 10~cm diameter sphere with 3 ports, one for the LED whose light fills the sphere, one for the NIST and one to send the beam to the PMT pixel. This is sketched in fig.\ref{fig:Fig3}. Notice the collimator made of two holes, one of 1~mm diameter next to the sphere, and the other of 0.25~mm next to the PMT. They are separated by 20~mm. This way the light is reduced by a factor of about 10$^5$, to take into account the 10$^6$ gain of the PMT. The power read on the NIST, SPE mode, is then around 1~nW, when the noise is around a few pW.

\begin{figure}
  \centering
  \includegraphics[width=0.4\textwidth]{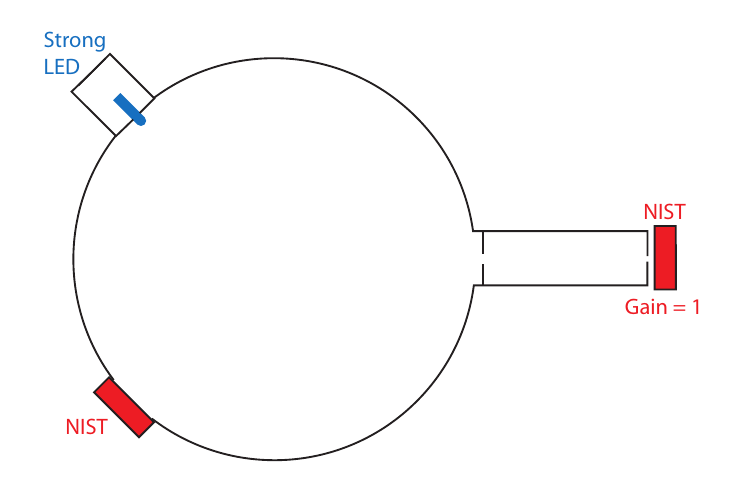}
  \caption{Integrating sphere arrangement with second NIST.}
  \label{fig:Fig4}
 \end{figure}

With this set-up the LED intensity is fixed to a level where we have roughly 100 pedestals $P_0$ for one $P_1$.
The ratio $R$ of the number of photons reaching the NIST to the number reaching the PMT can be calculated, but with an accuracy not better than 10\%. So, it is more accurate to measure $R$.
To do that, we use the same set-up with a second NIST photodiode replacing the PMT. See fig.\ref{fig:Fig4}.

Now the LED is 10$^6$ times stronger to compensate for the low gain of the NIST (about 0.5). The response of the 2 NIST photodiodes is given by two pico-ammeters, and it is best to take snapshots of these two instruments together, to get rid of fluctuations.
Having measured that ratio R between the two NISTs, we apply it to the measurement of fig.\ref{fig:Fig3}. We know how many pulses have been sent to the LED in a run. We know then what is the number of photons hitting the NIST on the sphere, hence, through the ratio,  we know how many photons hit the PMT. We determine the full $P_1$ spectrum surface which gives the number of created SPE and dividing this number by the number of incident photons, we get $\epsilon_{pix}$. We add 4\% (contribution of neighbors pixels) to this efficiency and we finally get the full absolute pixel efficiency. The accuracy on this efficiency is 2\% (the NIST on the sphere uncertainty cancels out in the ratio).

\section{Application to a given PMT }

\begin{figure}
  \centering
  \includegraphics[width=0.31\textwidth]{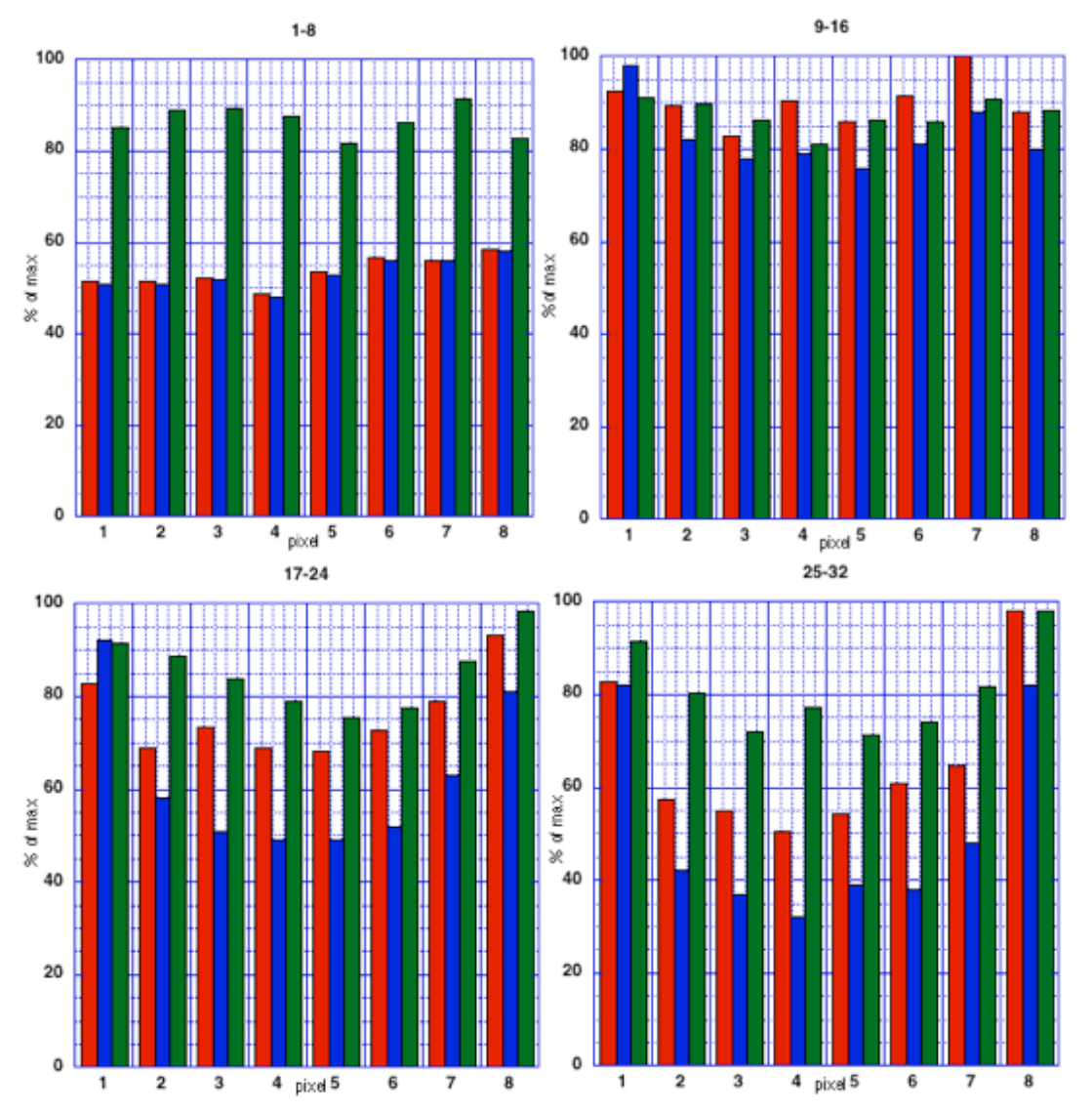}
\includegraphics[width=0.31\textwidth]{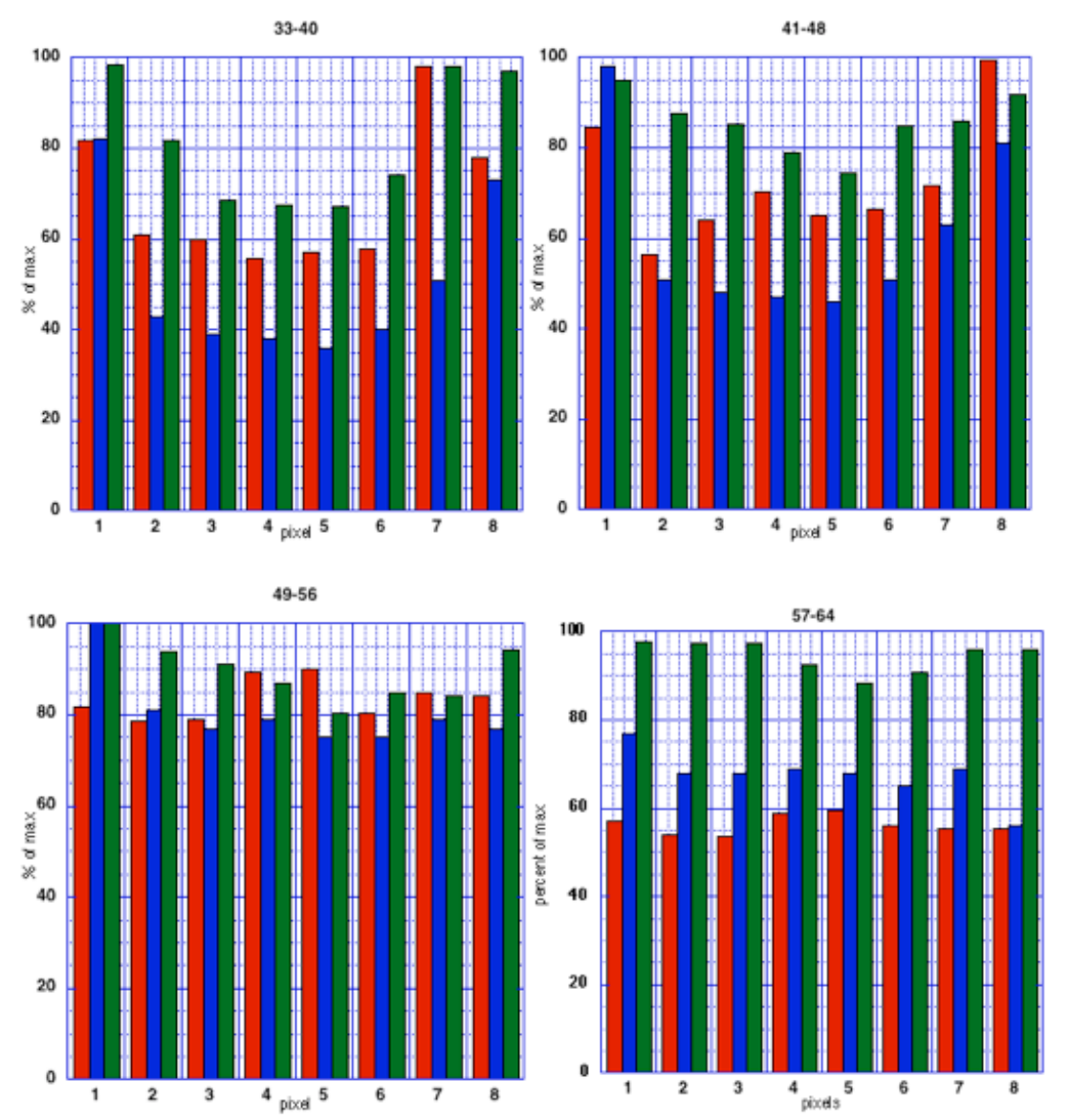}
  \caption{Gain in red (our data) and blue (Hamamatsu data) and efficiency in green of the 64 pixels of a tube. All data are normalised to their own maximum.}
  \label{fig:Fig5}
 \end{figure}

In blue, fig.\ref{fig:Fig5} shows Hamamatsu gain data taken with a strong light (not in SPE mode), at 1000~V. In red, our gain measurements also at 1000~V, and in green, the pixels efficiencies measured with the described set-up. Both gains and efficiencies exhibit strong variations.

\section{Calibration of Euso-Balloon focal surface}

In the Euso-Ballon instrument~\cite{bib:balloonmiss},\cite{bib:balloondet},\cite{bib:balloonelec}, the crucial absolute efficiencies of the pixels in a PMT are being measured (together with the gain) with an accuracy of 2\%. This is done for 2 values of high voltage: -1100~V and -1000~V when using the CAEN QDCs. They are measured at -900~V with the front-end ASIC. However, this last method, due to very slow laboratory DAQ (involving a USB connection between the test board reading the ASICs and a PC controlled by a dedicated LabView software for electronic tests), the acquisition time turns to be nearly two days to get with sufficient statistics the 64 pixels efficiencies. To overcome this limitation, we do that measurement on one pixel only, which we call a "NIST pixel". Then we move the PMT to 30~cm of the integrating sphere, so that the cos$^4\theta$ law valid for lambertian sources makes the illumination uniformity better than~99\%. This way we can measure the whole EC-unit at once, saving a lot of time. This has another advantage: the EC-unit being potted, the field at the photocathode level could be slightly distorted due to the dielectric rigidity of the potting material. The accuracy then will be around 3\%.
Finally, to check that the efficiency follows the bi-alkali curves, these efficiencies are measured with 3~LEDs of different wavelengths, between 300 and 400~nm.

\section{Conclusion}

The pixels having been "calibrated", that is their gain for a given HV, and $\epsilon_{pix}$ for that HV have been determined with a great accuracy in the laboratory. Naturally, the ASIC channel used for a given pixel has to be kept during the flight (keep the same connections).
Then, the instrument flies, and the measurements made on earth are still valid until something in the gain or in the efficiencies varies.

a)	{\it Gains:} to check during the flight if the gains have changed, the focal surface will be illuminated over its whole surface by a few 1~inch integrating spheres equipped with LEDs of different wavelengths (another port will be equipped with a NIST photodiode, but this is not used in the gain measurement). The uniformity of illumination has no importance as long as we are in SPE mode with negligible $P_2$. The spheres will be disposed at the periphery of the last lens before the focal surface, with the output light directed towards this FS. The spectrum of each pixel is realised through S-curves made by changing the ASIC discriminator threshold. The gain of each pixel is then measured in an absolute way, with a 1\% precision. The causes of gain changes maybe radiations and cosmic rays going through the dynodes.

b)	{\it Efficiencies:} The surface of the SPE peaks in an S-curve is equal to the number of counts given by the discriminator when the threshold is at 1/3 of the pedestal-SPE peak (in the valley of fig.\ref{fig:Fig1}). On earth, before launch, we will have made this experiment of illuminating the FS with the small spheres, and by comparison with the beforehand measured pixel efficiencies, we will have a correspondence table of these absolute efficiencies with the discriminators counts. In order to ensure that the LEDs variations are taken into account, the NIST diodes in the small spheres will be recorded. These NIST photodiodes are non-polarized, hence very robust to radiations and temperature changes. So, in flight, it will be easy to see if any pixel efficiency has changed significantly (more than 2\%) and the new efficiency will be known. The transparency of the lenses is part of the efficiency. This will be checked with another small integrating sphere positioned on the lid (the calibrations are made during the ISS day, lid closed), and sending its light to the FS through the lenses. If the gain has changed significantly, and the HV of one EC-unit had to be modified, then the measured efficiencies would reflect the change. If there is a big change in efficiency ($>$ 20\%), then we would like to make a new absolute efficiency measurement. This can be done with the moon, or with external sources, but with a precision not better than 20\%. The causes of big changes during five years of flight are mainly radiations, dust and oxygen.
The philosophy used on flight is explained in papers~\cite{bib:upcalib},\cite{bib:moon} and \cite{bib:lasercalib} in these proceedings.

\clearpage


%% file: icrc2013-0628.tex


\title{Photomultiplier Tube Sorting for JEM-EUSO and EUSO-Balloon}

\shorttitle{PMT Sorting for JEM-EUSO}

\authors{
c. Blaksley$^{1}$,
P. Gorodetzky$^{1}$,
for the JEM-EUSO Collaboration.
}

\afiliations{
$^1$ Laboratoire Astroparticule et Cosmologie (APC), Universit\'e Paris 7/CNRS, 10 rue A. Domon et L. Duquet, 75205 Paris Cedex 13, France \\
}

\email{blaksley@in2p3.fr}

\abstract{The detector portion of JEM-EUSO is a focal surface made of 137 photo-detection modules of 9 elementary cells (EC), each composed of 4 Hamamatsu R11265-M64 multi-anode photomultiplier tubes (PMTs).
JEM-EUSO's daughter experiment, EUSO-balloon, is a path-finder mission composed of a single JEM-EUSO photo-detection module with optics in a balloon-borne gondola.
Each EC is powered by a single Cockcroft-Walton type high voltage power supply, and the gain of the EC can be adjusted as a unit by changing the power supply output. 
The ASIC readout electronics include a preamplifier which allows the gain of each pixel within the PMT to be equalized. 
There is up to a factor of 4 variation in gain between PMTs, and around a 20\% variation in gain from pixel to pixel within a PMT.  
The gain and efficiency of each PMT is measured in single photon electron mode, and they are sorted so that each EC can be build from PMTs with a similar enough gain that all 256 pixels can be equalized using the dynamic range of the ASIC preamp.
Sorting the PMTs in this way also allows a rejection defective PMTs. 
For JEM-EUSO the sorting requires measuring the gain and quantum efficiency of 64 pixels for over 5,000 photomultiplier tubes. 
The sorting of 40 PMTs for EUSO-balloon, serving as model and test run for future sorting for JEM-EUSO, included the building and calibration of a data acquisition system,
the measurement of spectra in single photoelectron mode, and final analysis of the 64 resulting spectra for each of 40 PMTs.}


\maketitle

\section{Introduction}
The JEM-EUSO experiment is a ultra high energy cosmic ray (UHECR) observatory which will be placed on the International Space Station.
It will observe fluorescence photons created in extensive air showers (EAS) induced by UHECR with energies above $10^{20}$ eV.
The heart of the JEM-EUSO instrument \cite{ bib:PicozzaJEMEUSOmisson, bib:KajinoJEMEUSOinstrument}
is a focal surface composed of 137 photo-detection modules (PDM), the smallest self-triggering element. Each PDM is composed of
9 elementary cells (EC), the smallest flat surface, with each EC composed of 4 Hamamatsu M64 multi-anode photomultiplier tubes (PMTs). The readout of each PMT is 
through the dedicated \textit{Spatial Photomultiplier Array Counting and
Integrating Readout Chip} (SPACIROC) ASIC which has been developed for JEM-EUSO \cite{bib:AhmmadASICex, bib:AhmmadASICicrc}.

EUSO-Balloon is a JEM-EUSO path-finder mission
led by the French JEM-EUSO collaboration.
The EUSO-Balloon detector is composed of a single JEM-EUSO PDM with optics launched in a balloon-borne gondola \cite{bib:VonBallmoosEUSOballon}.
The philosophy in designing and building EUSO-Balloon is to follow as closely as possible the actual hardware design and requirements of JEM-EUSO.

\begin{figure}[t]
  \centering
  \includegraphics[width=0.475\textwidth]{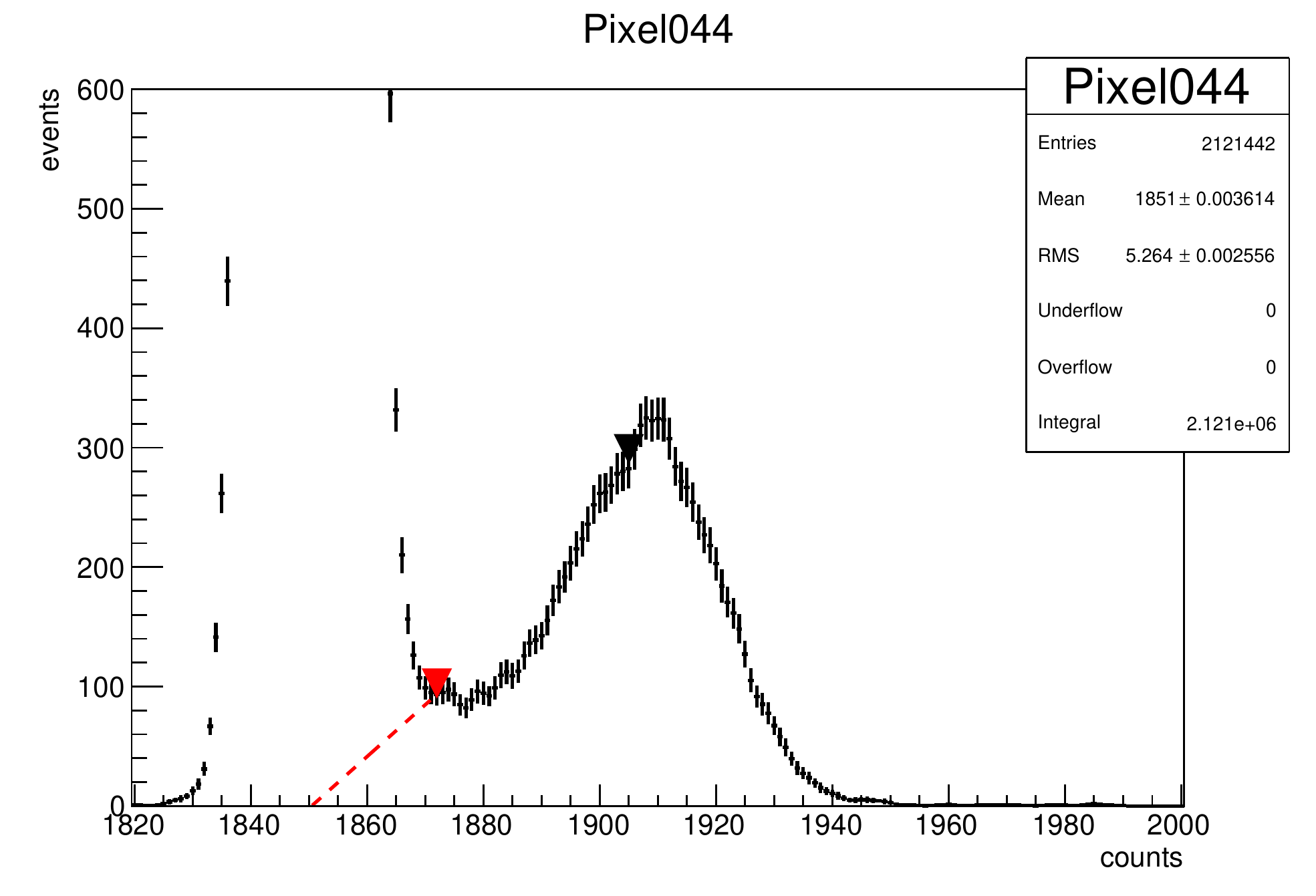}
  \caption{An example of a single photoelectron (pe) spectrum for one pixel of a M64 PMT, taken by the DAQ system discussed in section \ref{sec:DAQ}. The spectrum is shown as a histogram of the number of events (see section \ref{sec:SPEgainEff})
  with a given charge, in counts, returned by QDC. The first peak is the pedestal, corresponding to events in which no pe are collected at the anode of the pixel. The second peak, on the right, corresponds to
  events in which 1  pe is emitted. The 2 pe peak is not visible as a consequence of the Poisson statistics (see text). The gain of the pixel is the difference between the mean of the 0 pe and 1 pe peaks, while
 the efficiency of the pixel is proportional to the surface of the 1 pe peak. Here the mean of the 1 pe is shown by the black marker, with the valley between the two peaks taken at the location of the red marker. The red line shows a correction to the surface 
 of the 1 pe peak assuming a linear dropoff from the valley.}
  \label{fig:SpeExample}
 \end{figure}

\section{Sorting as Calibration}
Each EC of 4 PMTs is powered by its own Cockcroft-Walton type high voltage power supply (HVPS) which has been designed to meet the strict power consumption requirements of the JEM-EUSO mission \cite{bib:HPVSICRC2011}.
The gain of the 4 PMTs within each EC can be adjusted together by regulating the HVPS output.
In addition, the SPACIROC readout chip of each PMT includes a preamplifier with a range of a factor of 2, allowing the gain of each pixel to be modified individually.

Each PMT can be characterized by its measured gain and efficiency\footnote{Where the efficiency can be either relative, i.e. compared to the other PMTs in the set, or absolute.}. 
There is a factor of $\approx$ 4 variation in average gain between individual M64 PMTs.
These differences in gain come from small manufacturing variations in the multiplication stage of the PMT, where a relatively small
change in electrostatics between the dynodes can have a large impact on the overall gain of the PMT.
Within each PMT there is a further variation of $\approx$ 25\% in gain from pixel to pixel. This variation is due mainly to the change in electrostatics with location on the photocathode. 
These same electrostatic variations also affect the total efficiency of the PMT, as they
modify the efficiency of collecting converted photoelectrons (pe) from the photocathode into the multiplication stage of the PMT. Not only will the average efficiency vary from PMT to PMT, but the 
relative efficiency of each pixel within the PMT will vary as well.
  
The measured properties of the SPACIROC ASIC, particularly in terms of photon counting linearity, require that the working gain of the PMT
be close to $1\cdot10^{6}$. Within a PMT the gain variation from pixel to pixel can be compensated in most cases by using the dynamic range of the ASIC preamp, but the larger factor of 4 variation between PMTs cannot. The PMTs must
be sorted by gain so that each EC can be build from PMTs with gains (at same HV) which are similar enough that all 256 pixels can be brought to a working gain of $\simeq1\cdot10^{6}$ using the ASIC preamp.

In addition, we have found during laboratory tests that in a small number (less than 1 in 10) of M64 PMTs one of the first dynodes draws a large current regardless of the incident light.
This is due a to a low resistance between the dynode and the cathode, on the 
order of $\simeq10$ M$\Omega$ rather than several G$\Omega$.
These PMTs still function for light detection (as evidenced by the fact that they passed Hamamatsu testing), but the large current drawn by the dynode is a potential problem, and these PMTs can not be used.

Due to this, sorting a sufficiently large number, in principle every PMT to be used in the experiment, is necessary before any EC can be constructed. 
This sorting can be thought of as a calibration of the EC units. 
For JEM-EUSO, sorting the PMTs requires measuring the gain and efficiency of 64 pixels for over 5,000 photomultiplier tubes.
In EUSO-balloon the sorting of a total of 40 PMTs serves as model and test run for future sorting for JEM-EUSO. It includes: 
building and calibration of a data acquisition (DAQ) system,
taking spectra for each pixel of each PMT, and final analysis to determine the gain and (relative) efficiency.

\section{Single Photon Gain and Efficiency}
\label{sec:SPEgainEff}
In order to measure both the gain and efficiency separately we work in single photoelectron mode. 
Using the technique of Lefeuvre et al. \cite{Lefeuvre:2007jq, LefeuvreThesis} the gain can be measured with a total error of $\simeq 2 \%$, and the absolute efficiency can be measured to around $3 \%$ by using a comparison with 2 NIST photodiodes 
The PMTs to be sorted are received and measured with the BG3 filter glued on. 
The response of the PMT is measured as a charge spectrum using charge to digital conversion (QDC) electronics. 
A detailed explanation of the measurement technique can be found in Gorodetzky et al. (\cite{bib:JEMEUSOcalib}, this conference).

An example of a measured single photoelectron spectra is shown in figure \ref{fig:SpeExample}. The gain, the average number of electrons arriving at the anode for a collected photoelectron, is the difference in charge between the
0 pe (the pedestal) and 1 pe peak. The total efficiency is the product of the quantum efficiency, i.e. the efficiency of converting photons to electrons at the cathode, and the collection efficiency, that of collecting emitted pe onto the first dynode. 
The total efficiency is proportional to the surface of the 1 pe peak.

For the error to be at the $1 \%$ level the number of 2 pe events (gates during which 2 pe are created and collected) in the spectrum must be negligible. 
As the emission of a pe is a Poisson process, for the number of 2 pe to be negligible requires that the average rate of pe be such that 
\begin{eqnarray*}
 n_{1\textrm{pe}} \leq 0.01\times n_{0\textrm{pe}}.
\end{eqnarray*}
As a $1 \%$ statistical error requires $10^{4}$ signal events, and working in single photoelectron mode requires a signal to background ratio of $1 \%$, we 
require at least $10^{6}$ events per spectrum.
To sort a reasonable number of PMT per day, the time per PMT should be on the order of minutes, which then requires a DAQ rate in the range of kHz.

The fact that we must measure 64 spectra in parallel (one for each pixel), makes the use of amplifiers difficult and, more importantly, expensive. This means that the resolution of the QDCs is a key component of the measurement.
The M64 PMT has an typical gain of $\simeq 1\cdot10^{6}$ at a cathode voltage of 900 V. To be sure that the worst pixels of each PMT can be measured, we sort the PMTs at a cathode voltage of 1100 V, where the gain is factor $\approx 7$ higher. 
At a gain of $1\cdot10^{6}$ the separation between the 0 pe and 1 pe peak is 160 fC, and the resolution of the QDC must be high enough to divide this charge into a enough bins that the 0 pe and 1 pe peak can be reliably separated.      

\section{The Data Acquisition System}
\label{sec:DAQ}

The need for a data acquisition rate of several kHz can be easily satisfied by CAMAC or VME hardware. While VME is more modern and faster, we have found that the best available VME QDCs have a conversion resolution in the range of 100 fC per count.
The high transfer rate of digital VME modules, and the fact that they are not shielded, makes VME less suited to precision charge measurements than CAMAC. 
To meet the needs of PMT sorting we used the CAEN model C1205 CAMAC QDC . The C1205 is a Wilkinson-type QDC with 3 independent charge ranges.
The lowest of these is 0 to 80 pC with a 12 bit resolution, giving a theoretical conversion of 21 fC per QDC count. 
Each C1205 has 16 channels with inputs in Lemo format, meaning that 4 modules are needed per PMT. The Lemo format of the QDC inputs is an advantage in terms of signal quality and the ease with which a single pixel can manipulated.

In our DAQ the CAMAC crate is interfaced to VME using a CBD 8210 CAMAC branch driver, and the VME is readout by Motorola VME processor board. 
Using CAMAC has the advantage of its relative simplicity and the existence of a large library of CAMAC hardware in our laboratory.
Readout of the CAMAC through VME gives a high speed and the possibility to include VME hardware if needed.

\begin{figure}[t]
  \centering
  \includegraphics[width=0.4\textwidth]{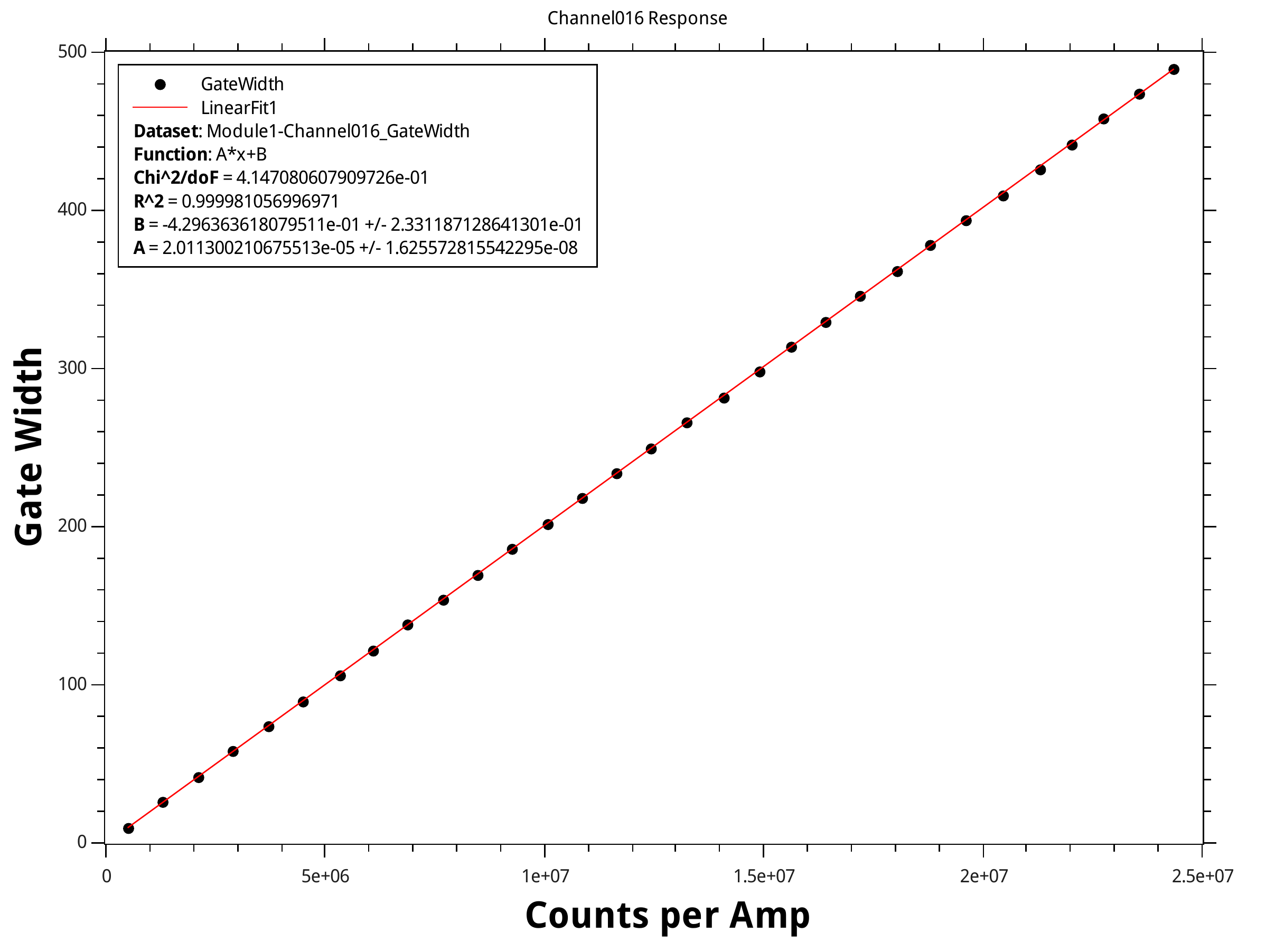}
  \caption{Example of the measured conversion curve for one input (here channel 16) of one QDC module. To have the required accuracy, and to account for the true integral nonlinearity of the QDC, the response is measured for $\geq30$ different input
charges across the full range of the QDC. The measured data is shown by the black dots. The value of the conversion slope is taken from a least-squares fit to the data, shown in red.}
  \label{fig:QDC_mod1-16_response}
 \end{figure}

\subsection{QDC Characterization}
To extract the absolute gain from the measured spectra the conversion from QDC counts to Coulombs must be known with an accuracy of $1 \%$ or better. 
The conversion quoted by the manufacturer does not include the linearity properties of the QDC, nor the uniformity of channels within the same module, which is $\pm~5 \%$. Any differential 
nonlinearity, the variation in width of each charge bin, in the QDC response can be reduced by using a built in sliding scale technique \cite{bib:SlidingScale}.
The integral nonlinearity, the total deviation from a linear response, must be measured across the full range of the QDC.

The response curve of each channel of each QDC was measured using a DC level with a resistor in series. The current through the input was measured with a picoammeter.
An integration gate was created using a digital pulse generator, and for each gate width two spectra were taken, one with a high level and one at $\simeq$ 0 mV to take the QDC pedestal. 
The resulting response curve for one channel of one module is shown in figure \ref{fig:QDC_mod1-16_response}, plotted with the gate width in ordinates and the ratio of the QDC counts returned to the measured current in abscissa.
The conversion slope was determined by performing a least squares fit of the measured curve.
In order to reach a $1 \%$ accuracy on the slope, at least 30 data points per curve where needed, making more than 3840 measurements to complete a full characterization of all 64 QDC channels.
To made this feasible, the readout of the picoammeter, control of the DC level, and the setting of the pulse generator where interfaced directly into the DAQ software, and the measurement of each response curve was scripted using the DAQ run control. 

\begin{figure}[ht]
  \centering
  \includegraphics[width=0.495\textwidth]{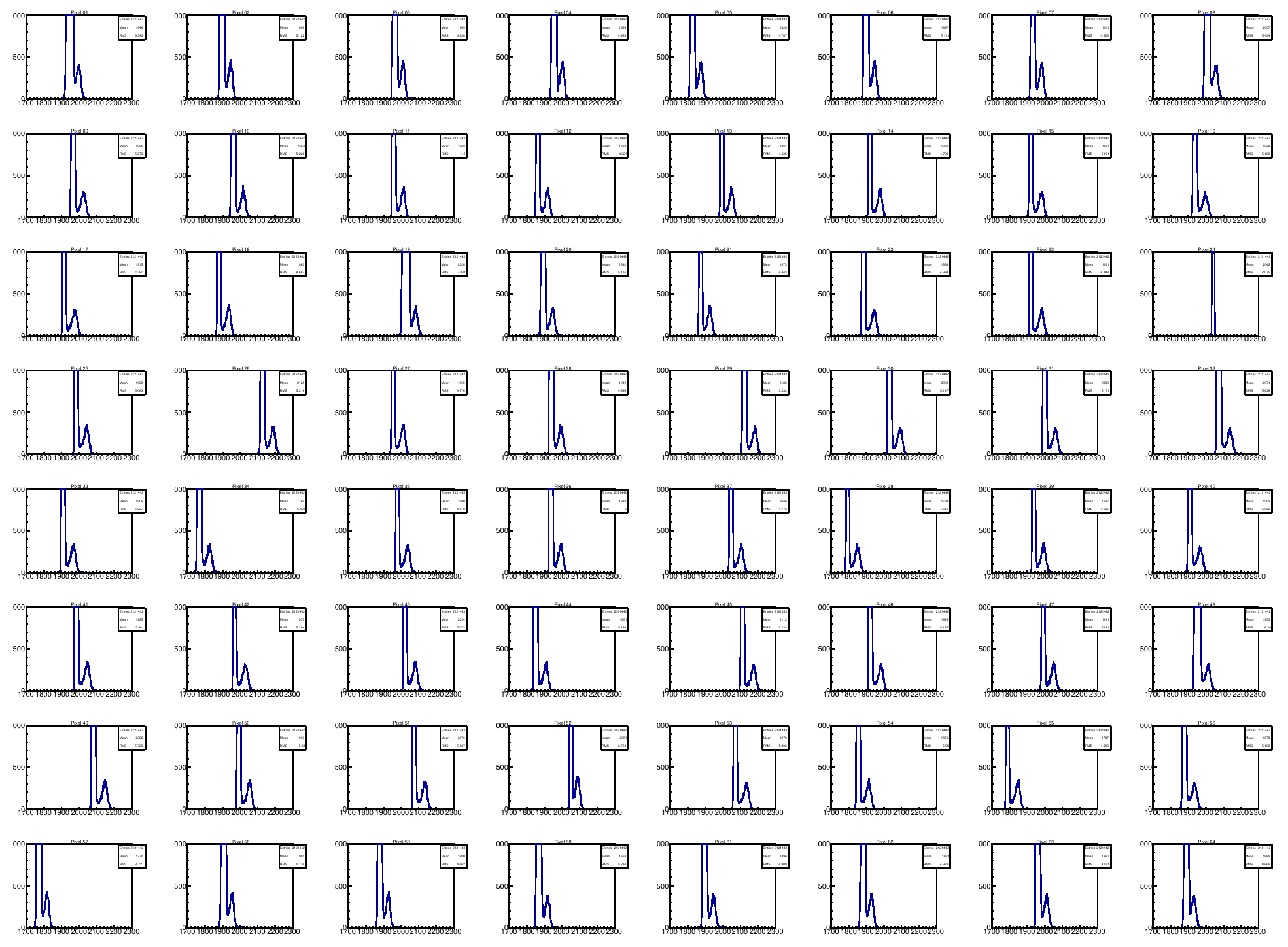}
  \caption{64 spectra taken for one M64 PMT in a single run. Each pixel shows a good spe spectrum with a clearly visible pedestal and 1 pe peak.}
  \label{fig:64Spectra}
 \end{figure}

\begin{figure}[ht!]
  \centering
  \includegraphics[width=0.495\textwidth]{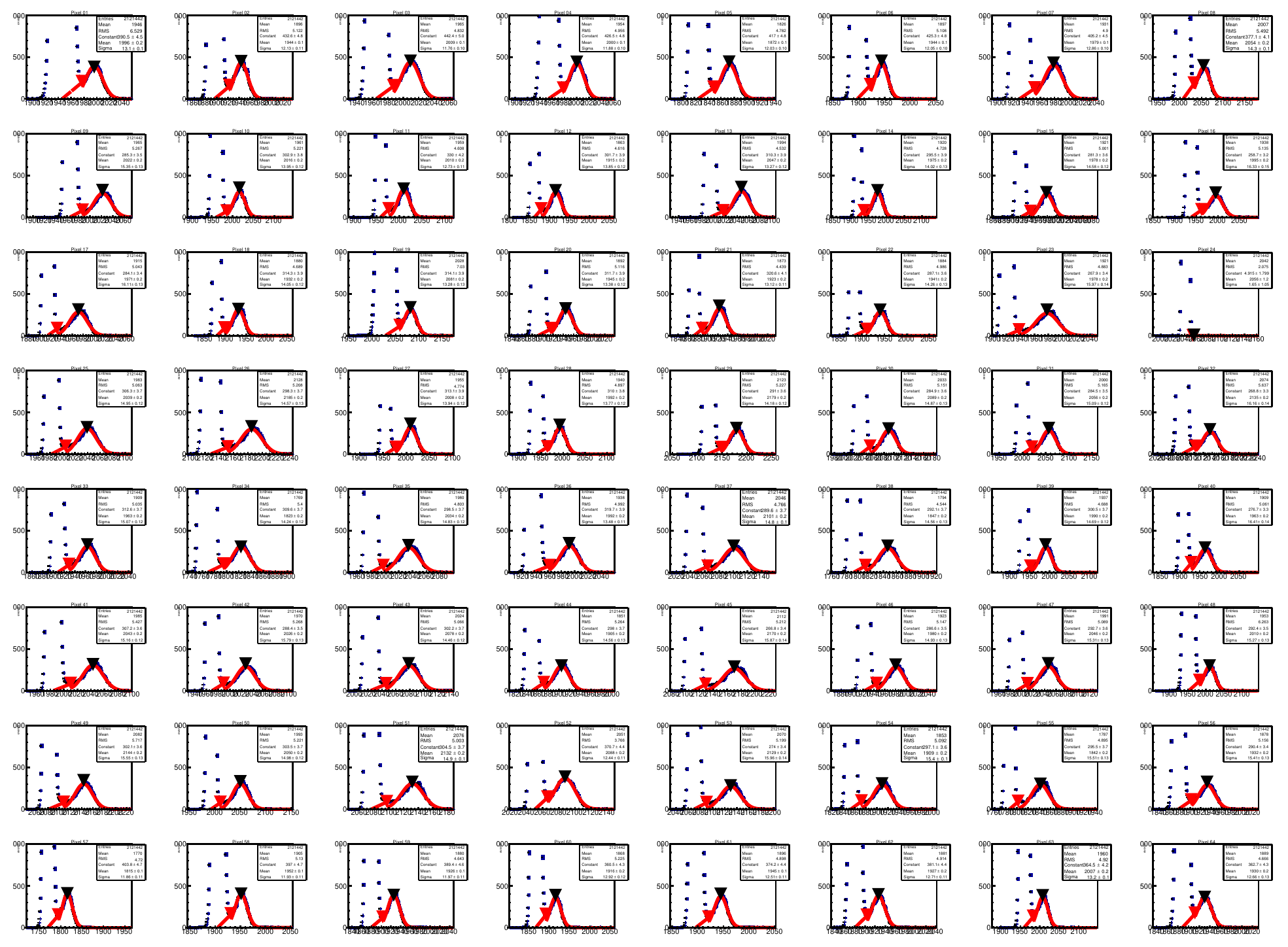}
  \caption{The 64 spectra of figure \ref{fig:64Spectra} after analysis by the routine discussed in section \ref{subsec:DAQsoftware}. For each pixel the location of the 1 pe peak mean and the valley have been found and marked. The red line 
shows a Gaussian fit to the 1 pe peak as a reference. One spectrum of these 64 can be seen in detail in figure \ref{fig:SpeExample}.}
  \label{fig:64SpectraAnalyzed}
 \end{figure}

\subsection{Data Acquisition Software}
\label{subsec:DAQsoftware}
The data acquisition software has been written especially for this setup in C/C$^{++}$ using the MIDAS data acquisition framework \cite{bib:MIDASref}.
The software is divided into front-end programs which collect data, and back-end programs which handle background processes, run control, data analysis, and storage. 
The front-end which controls the CAMAC crate runs directly on the VME processor board and connects through Ethernet to the desktop which histograms, analyzes, and stores the data.
Although in our case all the software has been developed on Linux, the entire system is portable and capable of running on any operating system. 

This setup allows the acquisition to proceed to a rate of 2 kHz for all 64 channels in parallel with $100 \%$ efficiency. In this configuration, the rate is limited by the CAMAC signal definitions. 
As the C1205 is compatible with the FAST-CAMAC standard, a upgrade of the CAMAC crate controller to a FAST-CAMAC compatible model would increase the readout rate by a factor of 2 - 10 ,\cite{bib:FASTCAMAC}.

Other hardware is also interfaced to the DAQ software, such as the periodic readout of the NIST photodiode and control of the XY movement on which the light source is mounted. A full feed-back loop between the run analysis and control 
is possible. This allows complex tasks such as automatically centering on a given pixel with micron precision, and scanning the full photocathode of a PMT pixel by pixel. 
 
An example of 64 spectra measured during one run are shown in figure \ref{fig:64Spectra}\footnote{There is no response in pixel 24 because the anode pin of the PMT was bent when it was inserted into the base}.
Analysis poses a particular problem, as analyzing 64 spectra by hand for 5000 PMTs would be a huge task. We therefore developed a simple and robust
software analysis using ROOT which searches within each measured spectrum for the valley between the 0 and 1 pe peak. The routine avoids using any fitting procedures 
to reduce the possibility of unphysical results which would require an external control by a physicist.

This analysis is performed automatically at the end of the run, and the routine has access to the NIST measurement results and the characteristics of the QDC so that it directly outputs a measured gain and efficiency for each pixel.
The 64 spectra in figure \ref{fig:64Spectra} can be seen after analysis in figure \ref{fig:64SpectraAnalyzed}. 
For each pixel the valley, shown by the red marker, has been found and the mean of the
1 pe peak has been determined, shown by the black marker. The red line shows a Gaussian fit to the 1 pe part, which is used only as a cross check.

\section{Conclusion}
The DAQ setup described works extremely well, and has already been used in the sorting of a number of PMTs. Results such as those shown in figures \ref{fig:SpeExample}, \ref{fig:64Spectra}, and \ref{fig:64SpectraAnalyzed}
are typical.  In addition to its use in sorting, this setup is a powerful tool 
to be used in all the photodetection test bench activities at APC. 

In the near future the same system will be used to test the completed EC units for EUSO-balloon. 
This test is necessary because the potting which surrounds the ECs changes the electrostatic properties of the PMTs in the EC. 
The DAQ system developed for the sorting will be used to measure the gain of each pixel in the EC and the absolute efficiency of several pixels within each PMT.
The relative efficiency of all pixels will then be measured using the ASIC readout electronics and this relative efficiency will be converted to an absolute one
using the absolute efficiencies of the pixels measured with the QDC.  
In the farther future the tools developed here will be leveraged towards the task of sorting 5000 PMTs for JEM-EUSO. Here the flexibility and power of the DAQ system will be extremely important, especially in scaling up
to simultaneously measure multiple PMTs.

\clearpage

%% file: icrc2013-0545.tex



\title{On-board calibration system of the JEM-EUSO mission} 

\shorttitle{M.\,Karus \etal JEM-EUSO on-board calibration system}

\authors{
M.\,Karus$^{1}$,
N.\,Sakaki$^{1}$,
A.\,Haungs$^{1}$,
P.\,Gorodetzky$^{2}$,
A.\,Ebersoldt$^{3}$,
H.\,Schieler$^{1}$,
H.\,Sagawa$^{4}$
for the JEM-EUSO collaboration.
}

\afiliations{
$^1$Institut f\"ur Kernphysik (IKP), Karlsruhe Institute of Technology (KIT), Karlsruhe, Germany\\
$^2$Laboratoire AstroParticule et Cosmologie (APC), Paris, France\\
$^3$Institut f\"ur Prozessdatenverarbeitung und Elektronik (IPE), Karlsruhe Institute of Technology (KIT), Karlsruhe, Germany\\
$^4$Institute for Cosmic Ray Research (ICRR), University of Tokyo, Tokyo, Japan\\
}

\email{Michael.Karus@kit.edu}

\abstract{In order to unveil the mystery of ultra-high energy cosmic rays (UHECRs), JEM-EUSO (Extreme Universe Space Observatory on-board Japanese Experiment Module) will observe extensive air showers induced by UHECRs from the International Space Station (ISS) orbit with a huge acceptance. Calibration of the JEM-EUSO instrument, which consists of Fresnel optics and a focal surface detector with 5,000 multi-anode photomultiplier tubes (MAPMTs), 300,000 channels in total, is very important to discuss the origin of UHECRs precisely with the observed results.
The performance of the detector should always be monitored on orbit. Since the on-board resource is very limited, on-board calibration is in principle a relative one. For that purpose, a few uniform light sources with UV-LEDs and integrating spheres will be settled along the edge of the lens facing the focal surface (FS). Very uniform light is available thanks to the integrating sphere and the light intensity will be monitored in real-time by a photo diode attached to each sphere.
The same light sources will be put along the edge of the FS and will illuminate the entrance pupil to monitor the transmission of the optics. The performance of the detector itself and the optics will be measured in the ISS days as required.
The present development status of the calibration device will be reported together with the expected performance.
}


\maketitle
{\setlist{noitemsep}

\section{Introduction}

The Extreme Space Observatory on-board the Japanese Experiment Module (JEM-EUSO) is an UV-fluorescence telescope that will be installed at the International Space Station (ISS) in 2017 \cite{ybib:EUSOpurp}. The JEM-EUSO telescope consists of three Fresnel lenses and a focal surface (FS) and has a field of view of $60^\circ$. From the ISS-orbit ($\approx400\,\mathrm{km}$ altitude) the JEM-EUSO telescope will be able to observe a surface area of around $1.4 \times 10^5\,\mathrm{km}^2$. The FS consists of roughly $5,000$\,Multi-anode photomultiplier tubes (MAPMTs) of which each has $8\times8$\,pixels and is glued with an UV-filter that transmits UV-light from $330-400\,\mathrm{nm}$. Four MAPMTs form one elementary cell (EC) and nine ECs form one photodetector module (PDM). 137 of these PDMs form the whole FS of the telescope.

The main function of the JEM-EUSO telescope is the observation of extensive air showers (EASs) induced by ultra-high energy cosmic rays (UHECRs) with energies above $5\times 10^{19}\,\mathrm{eV}$ \cite{ybib:EUSOperf}. The main component of EASs are electrons which excite Nitrogen molecules of the atmosphere and thus produce isotropic fluorescence light. The particles in EASs also travel faster than the speed of light in air and thus produce Cherenkov light directed towards the Earth. The ultraviolet fluorescence light as well as reflected and scattered Cherenkov light will be detected by the JEM-EUSO telescope.

To estimate the energy of the primary particle the fluorescence yield from electrons which has been measured formerly \cite{ybib:Ave2013
} will be used. Furthermore there are also several quantities related to the detector itself \cite{ybib:EUSOcal-abs}
: quantum efficiency and collection efficiency of the detector, probability for a photon to be contained in a pixel, transmission of the Fresnel lens system and of the optical filter, trigger efficiencies of the electronics, atmospheric transmission and the aperture of the telescope.

%

These quantities have to be measured very precisely before the mission start and have to be monitored throughout the whole mission to have a good understanding of the detector performance at all times. Therefore several systems will be used: pre-flight calibration \cite{ybib:EUSOcal-abs}, on-board calibration, in-flight calibration with external light sources \cite{ybib:EUSOgloballight} and an on-board atmospheric monitoring system~(AMS) \cite{ybib:EUSO-ams}. The AMS will be attached to the telescope and will consist of an IR-camera \cite{ybib:EUSO-IR} to measure the cloud coverage in the field of view and an UV-laser to measure the height of these clouds. The other three subsystems are included in the calibration system of JEM-EUSO.

\section{Calibration system}
The calibration system consists of three subsystems that are prepared by collaborators from Japan, Germany, France, United States of America, Italy and Mexico. Since an absolute calibration is very difficult to maintain for the whole mission time of JEM-EUSO, it is imperative to monitor changes in the detector. Therefore the on-board calibration system will be used to do a relative calibration of the detector with respect to the absolute pre-flight calibration \cite{ybib:EUSOcal-abs}. It is also planned to use external light sources like the Moon for further in-flight calibration \cite{ybib:EUSOmoon}. 
There will also be on-ground lasers to check the trigger efficiency and the error in the reconstruction of the arrival direction. Xe-flashers will be used in combination with the AMS devices to measure local atmospheric conditions, e.\,g. absorption of photons in the atmosphere.

\subsection{On-ground reference light source}
In order to measure changes in the detection efficiency of the JEM-EUSO detector a reference light source with known optical output is needed. Therefore a prototype on-ground reference light source was built.
It consists of a 3-port $13.5\,\mathrm{cm}~(5.3\,\mathrm{inch})$ diameter integrating sphere, with two $2.54\,\mathrm{cm}$ (one inch) exit-ports and a $6.35\,\mathrm{cm}~(2.5\,\mathrm{inch})$ entrance-port. An UV-LED-array is mounted light-tight to the entrance-port and a NIST-calibrated photo diode and a collimator are mounted light-tight to the exit-ports (Fig.\,\ref{fig:pre-flight.cal.ratio}).

The UV-light ($\approx375\,\mathrm{nm}$) from the LED-array is diffusely reflected inside the integrating sphere and distributed uniformly over the inner surface of the sphere. The sphere's inside is made of Spectralon \cite{ybib:LS-spectralon} which reflects 98\% of UV-light in the region of $300-430\,\mathrm{nm}$. 
The integrating sphere behaves as a beam splitter and a diffuser. The fraction of photons leaving the sphere from one port is proportional to the area of the port itself \cite{ybib:LS-sphere}. Therefore both exit-ports emit the same number of photons $\mathrm{N_{Sphere}}$. This is measured at one exit-port with a NIST-calibrated photo diode (Photo Diode 1).
The collimator at the second exit-port is there to reduce the photon flux from the exit-port. This is necessary because the light source will illuminate MAPMTs and their gain is around a factor of $10^6$ bigger than the gain of the photo diode.

The optical output of the light source is measured by a second NIST-calibrated photo diode (Photo Diode 2) as the number of photons N that are emitted by the light source. The ratio of both photon numbers gives the collimator factor R of about $10^{-6}$. Because of the low gain of the photo diode and the strong collimator reduction the whole LED-array is set to continuously emit light. Because the collimator factor R only depends on the collimator geometry and was measured very precisely with the second NIST-calibrated photo diode, the number of photons N emitted by the reference light source can be calculated via the measurements of the number of photons $\mathrm{N_{Sphere}}$ inside the sphere. With this the number of emitted photons N is known and the light source can be used to illuminate one or more MAPMTs. For high-gain sensors in front of the reference light source, only one LED of the LED-array will be used while being pulsed by a LED-driver. Then the number of detected photons by this sensor is $\mathrm{N_{PMT}}$.

\begin{figure}[t]
  \centering
  \includegraphics[width=0.45\textwidth]{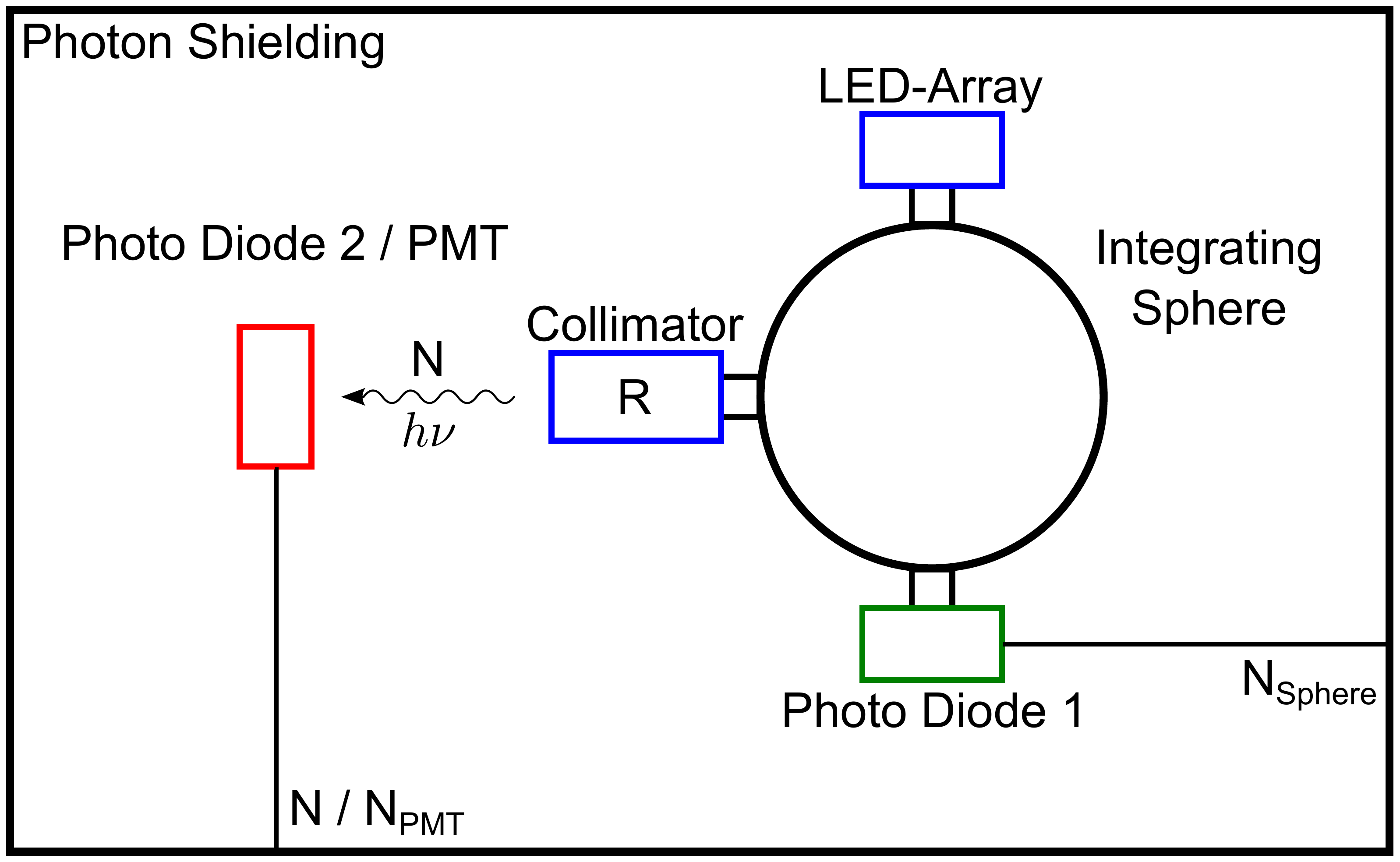}
  \caption{Sketch of the on-ground reference light source. The number of photons before and after the collimator is measured via two NIST-calibrated photo diodes. The emitted number of photons by the reference light source is calculated via the signal from the first NIST-calibrated photo diode.}
  \label{fig:pre-flight.cal.ratio}
\end{figure}

With this set-up the number of photons emitted by the reference light source can be calculated via the collimator factor and the signal from the first NIST-calibrated photo diode. The ratio of the number of photons detected by the pixels of the MAPMTs and the total number of photons leaving the reference light source at the collimator gives the detection efficiency of every pixel of the MAPMTs. The gain can be obtained by measuring the single photo-electron spectra of every MAPMT \cite{ybib:EUSOcal-abs}.

\subsection{On-board calibration system}
The on-board calibration system will be installed into the JEM-EUSO telescope to monitor changes in the detection efficiency of the detector and in the transmission of the optics. This calibration will be relative to the absolute calibration that was done pre-flight. The on-board system will consist of several small identical diffuse light sources that will be placed at different locations inside the telescope~(Fig.\,\ref{fig:on-board.system}).

The on-board light sources will be built with an integrating sphere with a diameter of $2.54\,\mathrm{cm}$ (one inch), one or more UV-LEDs with $300-430\,\mathrm{nm}$, a LED driver and a NIST-calibrated photo diode to monitor variations of the light intensity~(Fig.\,\ref{fig:on-board.lightsource}). The coating on the inside of the integrating sphere will be Spectralon
. The optical output from one source will be a Lambertian distribution with a maximum emitting angle from the optical axis of the source. This maximum angle is dependent on the shape and size of the pinhole that will be put on the exit-port of the sphere. The on-board light sources will be characterised pre-flight with the on-ground calibration system shown above.
\begin{figure}[t]
  \centering
  \includegraphics[width=0.40\textwidth]{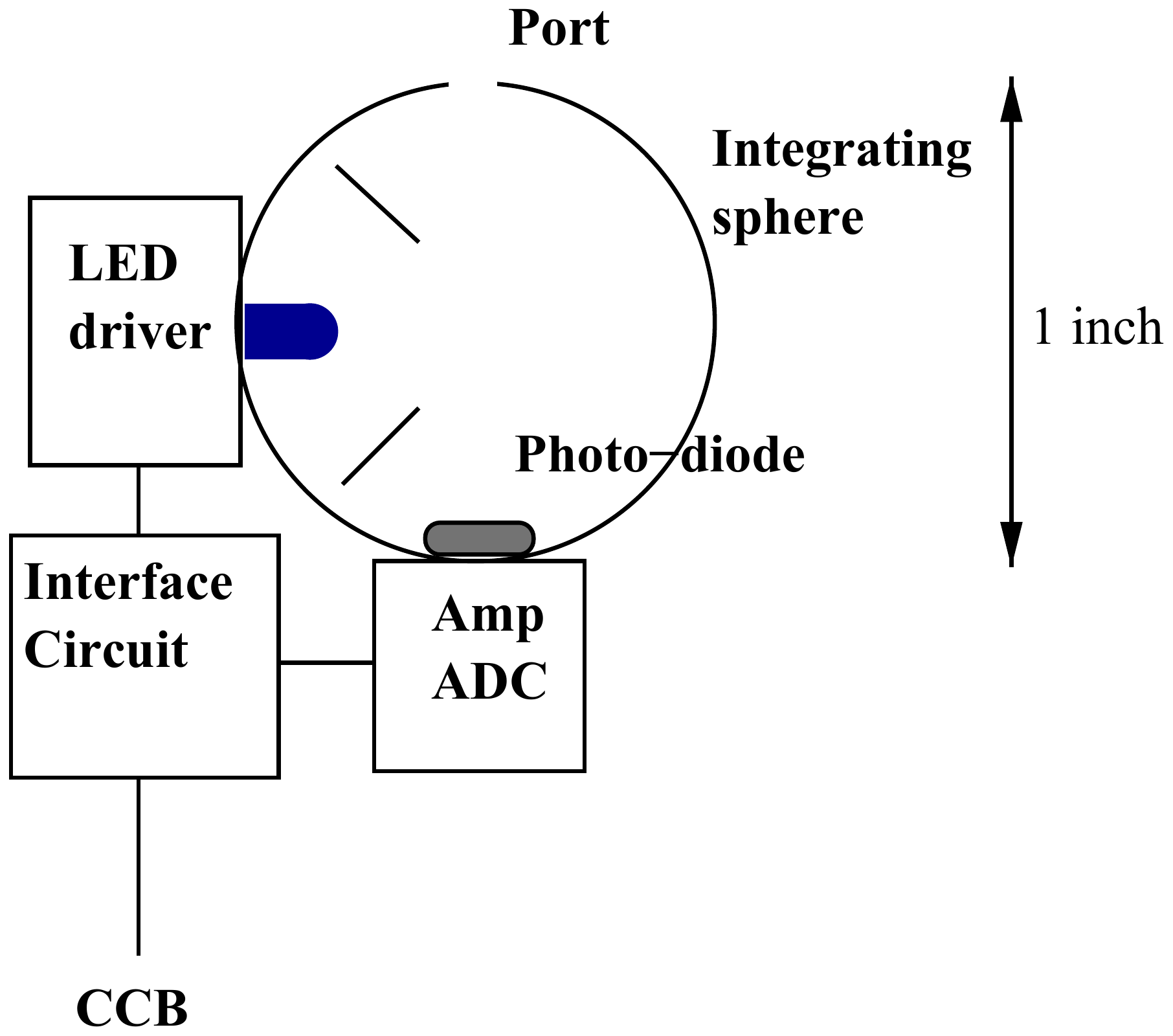}
  \caption{Schematic view of the on-board light source. The light source consists of one ore more UV-LEDs, a monitor photo diode, LED driver electronics, readout electronics and an interface circuit to a cluster control board (CCB).}
  \label{fig:on-board.lightsource}
\end{figure}

To measure the detector calibration change, several identical diffuse light sources will be placed at the edge of the third Fresnel lens to achieve a direct illumination of the FS~(Fig.\,\ref{fig:on-board.system}\,(a)). The intensity will be set to single photo-electron mode and the relative change of the detection efficiency will be measured while the gain of the MAPMTs will be measured absolutely. The threshold level for the counting will be adjusted if a large variation in gain is found.

The efficiency of the optics will be measured with identical light sources placed at the edge of the FS facing the rear side of the third lens~(Fig.\,\ref{fig:on-board.system}\,(b)). The UV-light from the light sources will pass the optics, be reflected at the diffuse lid (sand-blasted Aluminium) and pass the optics a second time. The MAPMTs at the FS will detect a fraction of the emitted photons. The time variation of the performance of the optics and the detector will be obtained at the same time in this measurement. Changes in the optical system can be obtained after subtracting the degradation of the detector.
\begin{figure*}[!t]
  \centering
  \includegraphics[width=0.65\textwidth]{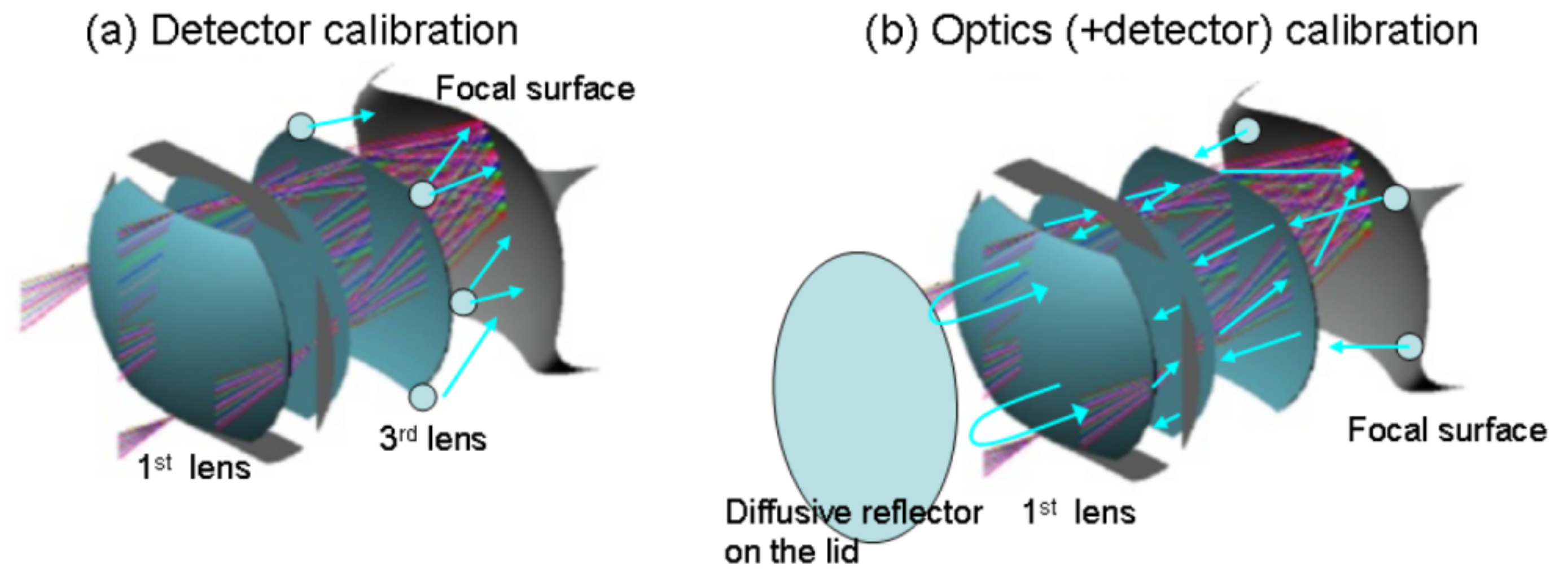}
  \caption{On-board calibration system. Diffuse UV-light sources made of integrating spheres will be set at the position shown in the panels (a) and (b), and the time variation of the efficiency of the optics and the detector will be monitored. (a) Several light sources will be set along the edge of the rear lens to illuminate the focal surface (FS) directly. The relative change of the detector efficiency will be taken. (b) Four light sources are placed along the edge of the FS to illuminate the rear lens. The light is reflected back at the diffusing surface on the lid and is detected by the focal surface detector. Here, convolution of the efficiency of the optics and that
of the detector will be obtained.}
  \label{fig:on-board.system}
\end{figure*}

\subsection{Expected performance}
In order to gain a better understanding of the requirements for the on-board calibration system, raytracing simulations where made for the direct illumination and the illumination through the optics. The goal was to achieve a very uniform illumination pattern on the FS. For both cases different set-ups were used. The required intensities for the sources were estimated to operate the on-board system in the single photo-electron regime.

\textbf{Direct illumination:} Four light sources were placed at the centres of the four edges of the rear lens. Each light source faced the FS and had a Lambertian optical output. The maximum emitting angle from the optical axis of the source was set to $60^\circ$. The inclination of the optical axis of the source from the optical axis of the telescope was set to $50^\circ$. The resulting illumination pattern is very uniform (Fig.\,\ref{fig:sim-illu-four}). Additional simulations where done with a single source failing, resulting in a non-uniform illumination pattern on the FS. 
However the resulting ratio of the intensity was about a factor of two and is still acceptable for the on-board calibration. With four light sources, the on-board calibration with direct illumination will be redundant.  

\begin{figure}[t]
  \centering
  \includegraphics[width=0.5\textwidth]{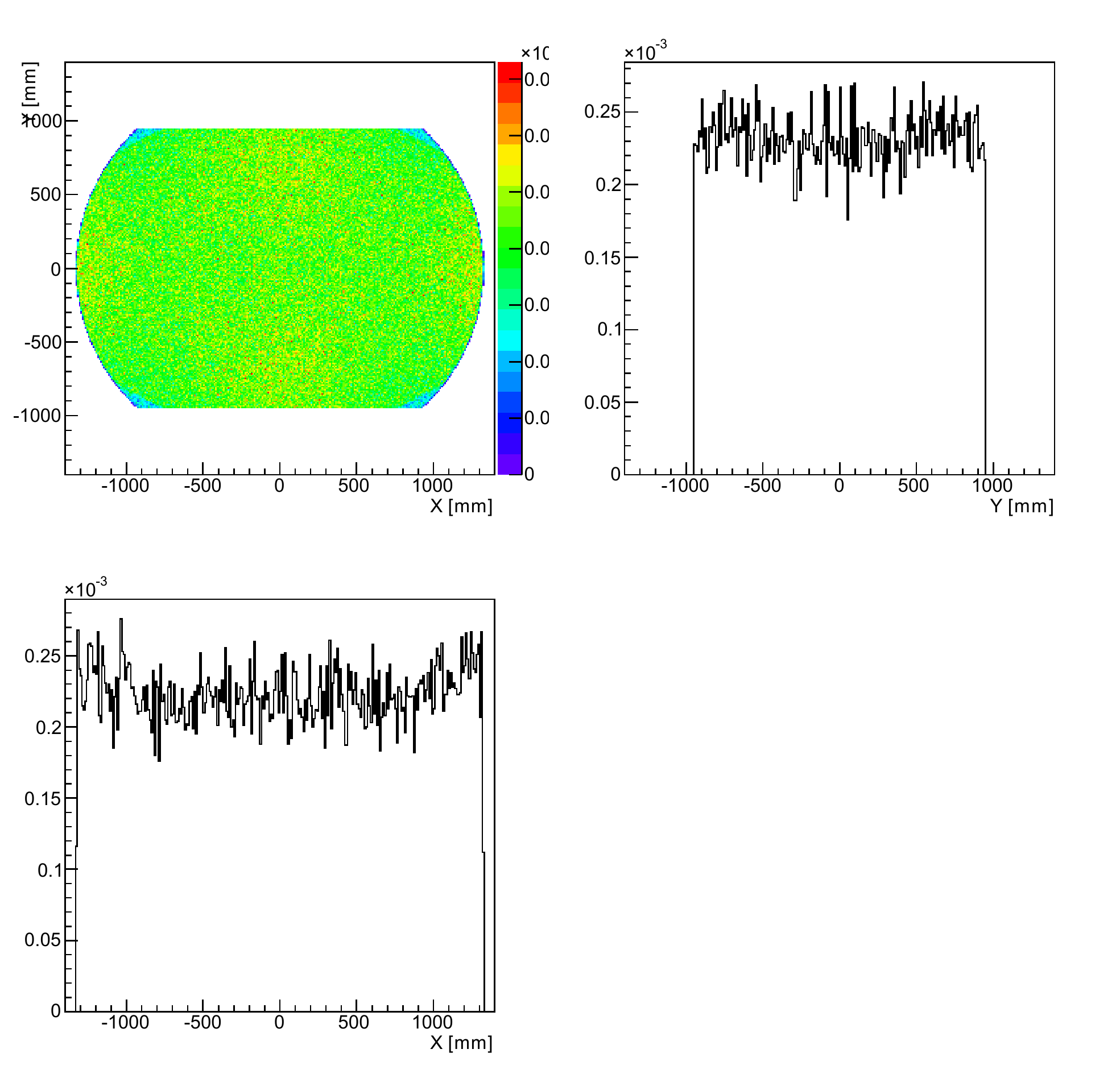}
  \caption{Light intensity distribution on the focal surface (FS) with four light sources at ($0\,\mathrm{mm}, \pm 945\,\mathrm{mm}$) and ($\pm 1320\,\mathrm{mm}, 0\,\mathrm{mm}$). The upper left panel shows the 2-dimensional distribution. The colour scale shows the photon detection probability in a $1\,\mathrm{cm}^2$ area, when the light sources emit one photon. The upper right panel shows the projected histogram of a vertical $5\,\mathrm{cm}$ wide stripe at x=0. The vertical axis shows the probability of a photon reaching a $5\times1\,\mathrm{cm}^2$ area. The lower left panel shows the same histogram for a horizontal $5\,\mathrm{cm}$ wide stripe at y=0.}
  \label{fig:sim-illu-four}
\end{figure}


\textbf{Illumination through optics:}
For this simulation one light source was placed at the centre of the bottom edge of the FS facing the rear lens. The optical output of this source was again a Lambertian distribution. The maximum emission angle was narrowed to $10^\circ$. Therefore a suitable pinhole will be designed. The inclination of both optical axes was set to be $10^\circ$. The material of the lenses is PMMA with the respective optical properties. The reflectivity of the diffusive lid was set to 50\%. The resulting illumination pattern on the FS is composed of three bunches at different arrival times
~(Fig.\,\ref{fig:sim-optics-four}). The first two bunches result from reflections at the rear and middle lenses. The last bunch is a superposition of reflections at the front lens and the lid. Raytracing showed that photons which were reflected only from the lid create a diffuse image on the FS. A discrimination of these photons is not possible, however one can choose photons that arrive near the center of the FS. These have a higher probability to have been reflected only by the lid.

\begin{figure}[t]
  \centering
  \includegraphics[width=0.5\textwidth]{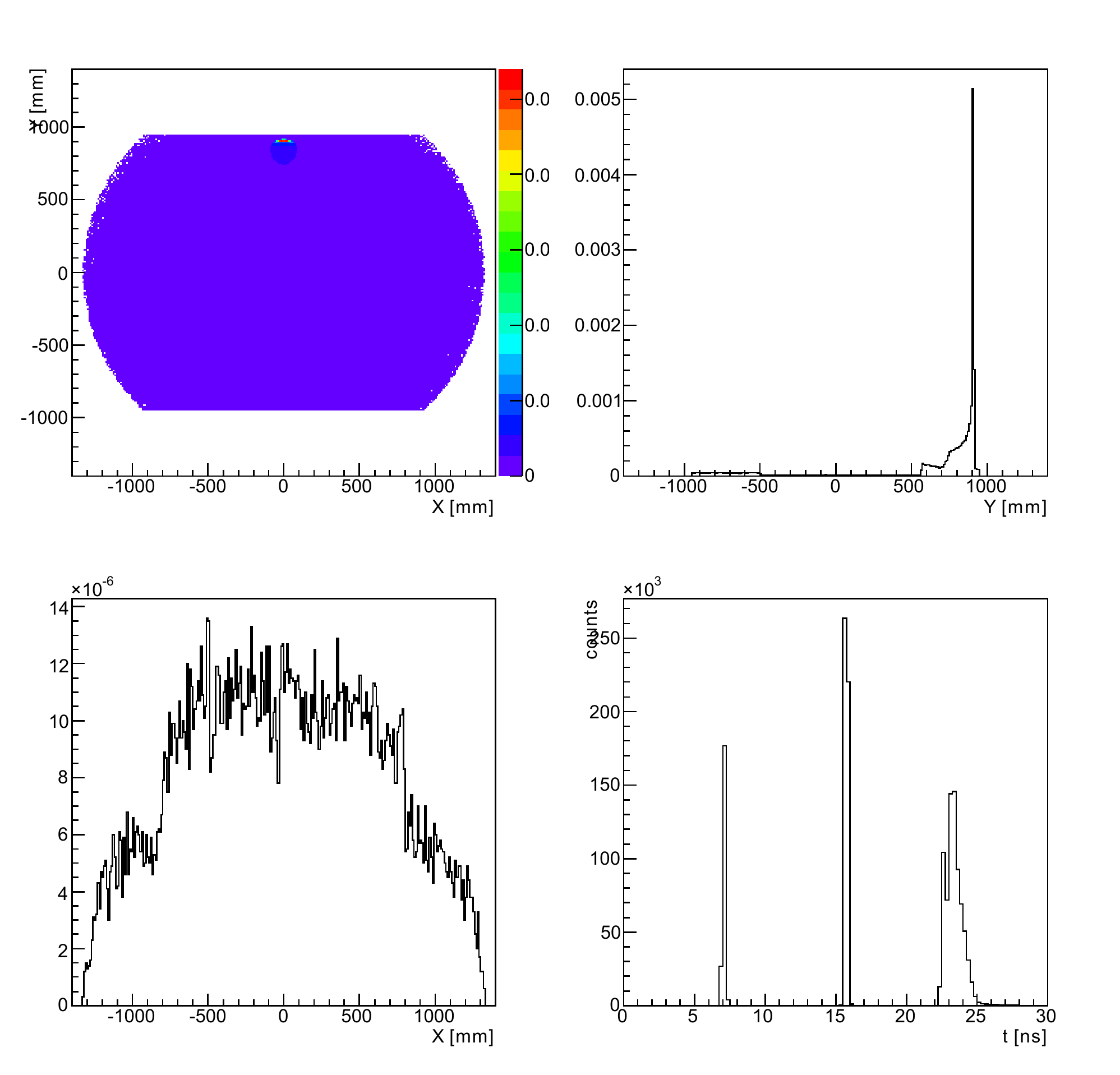}
  \caption{Light pattern on the focal surface (FS) for the optics transmittance measurement configuration. The upper left panel shows the 2-dimensional illumination pattern on the FS, the upper right projection along y-axis and the lower left projection along x-axis, as described before. The lower right panel shows the arrival time distribution of photons at the FS.}
  \label{fig:sim-optics-four}
\end{figure}

\textbf{Light source intensity:}
First the light intensity estimation for the direct measurement will follow. A typical UV-LED with light emission at $380\,\mathrm{nm}$ and an optical output power of $1\,\mathrm{mW}$ is attached to an integrating sphere with a diameter of $25\,\mathrm{mm}$ and Spectralon coating. The exit-port of the sphere is a pinhole of $1\,\mathrm{mm}$ diameter. The resulting light intensity follows from \cite{ybib:LS-sphere}:
\begin{eqnarray*}
	\Phi = \Phi_\mathrm{i} \cdot \frac{A_\mathrm{port}}{A_\mathrm{sphere}} \cdot \frac{\rho}{1-\rho \left( 1 - \frac{\Sigma A_\mathrm{port}}{A_\mathrm{sphere}} \right)} \cdot sin^2 \theta~~~~,
\end{eqnarray*}
with the input flux $\Phi_\mathrm{i}$, the exit-port area $A_\mathrm{port}$, the internal sphere area $A_\mathrm{sphere}$, the reflective index $\rho$ and the emission angle $\theta$ of the port. With the reflective index $\rho = 0.98$ for Spectralon and an emission angle of $60^\circ$ (Lambertian distribution), one source will emit about $10^{13}\,\mathrm{photons/s}$. The raytracing simulation results (Fig.\,\ref{fig:sim-illu-four}) gave approximately $3\times 10^{-4}\,\mathrm{photons/PMT}$ for the detection probability on the FS. Combining these two results gives for the number of photons per pixel (px) and gate time unit (GTU$=2.5\,\mathrm{\mu s}$):
\begin{eqnarray*}
	10^{13} \times 3\cdot 10^{-4}\,\frac{\mathrm{ph}}{64\,\mathrm{px}} \times 2.5\cdot 10^{-6} \frac{1}{\mathrm{GTU}} = 120 \frac{\mathrm{ph}}{\mathrm{GTU} \cdot \mathrm{px}}~~~.
\end{eqnarray*}
A conservative assumption of 20\% for the detector detection efficiency results in $24$\,p.e./GTU/px. In order to avoid overlapping of single photo-electron pulses we require $\approx1$\,p.e./GTU/px. This can be achieved by reducing the intensity of the LED via the LED-voltage or the reduction of the duty cycle via the LED driver. For higher intensities ($> 300$\,p.e./GUT/px) one LED with 1\,mW is not enough. Here a more powerful LED (15\,mW) or several LEDs are necessary.

The light intensity estimation for the optics transmittance is similar to the estimation above with a $1\,\mathrm{mW}$ LED. Here the emission angle was reduced to $10^\circ$ and the detection probability was reduced to about $10^{-7}$\,ph/GTU/px. This leads to $0.025$\,p.e./GTU/px that are detected. For this calculation, the photons which hit the wall were assumed to be absorbed completely. In the real case with a finite reflectivity of the walls, the discrimination of reflected photons from the lid becomes very difficult.  

\section{Summary}
The calibration system of JEM-EUSO with the focus on the on-board calibration was presented. The on-board calibration is very important to monitor changes in the detector throughout the whole mission time. It consists of several UV-light sources that are placed at different positions inside the telescope. Raytracing simulations have pointed out that the direct illumination of the focal surface produces a very uniform illumination with four light sources. The calibration can still be continued if one light source fails. The indirect illumination through the optics needs further investigation. Furthermore other configurations like a light source on the lid will be studied in detail. Intensity calculations showed that the desired photo-electron/GTU levels can be achieved with $15\,\mathrm{mW}$ LEDs. After completion of the light source prototype it will be tested with the on-ground calibration device.

\vspace*{0.3cm}
{
\footnotesize{{\bf Acknowledgment:}{This work was partially supported by Basic Science Interdisciplinary Research Projects of RIKEN and JSPS KAKENHI Grant (22340063, 23340081, and 24244042), by the Italian Ministry of Foreign Affairs, General Direction for the Cultural Promotion and Cooperation, by the 'Helmholtz Alliance for Astroparticle Physics HAP' funded by the Initiative and Networking Fund of the Helmholtz Association, Germany, and by Slovak Academy of Sciences MVTS JEM-EUSO as well as VEGA grant agency project 2/0081/10. The Spanish Consortium involved in the JEM-EUSO Space Mission is funded by MICINN under projects AYA2009-06037-E/ESP, AYA-ESP 2010-19082, AYA2011-29489-C03-
01, AYA2012-39115-C03-01, CSD2009-00064 (Consolider MULTIDARK) and by Comunidad de Madrid (CAM) under project S2009/ESP-1496.}}

}}

\clearpage


%% file: icrc2013-0546.tex



\title{Absolute In-flight Calibration of the JEM-EUSO Telescope with the
Moonlight}

\shorttitle{Absolute Calibration of JEM-EUSO with the Moonlight}

\authors{
N. Sakaki$^{1}$,
P. Gorodetzky$^{2}$,
T. Ebisuzaki$^{3}$,
A. Haungs$^{1}$
for the JEM-EUSO Collaboration.
}

\afiliations{
$^1$ Karlsruhe Institute of Technology (KIT), IKP, D-76021 Karlsruhe, Germany\\
$^2$ Laboratoire Astroparticule et Cosmologie (APC), Universit\'e Paris Diderot - Paris 7, 75205 Paris, France\\
$^3$ RIKEN, 2-1 Hirosawa, Wako, Saitama 351-0198, Japan
}

\email{naoto.sakaki@kit.edu}

\abstract{%
JEM-EUSO (Extreme Universe Space Observatory on-board Japanese
Experiment Module)
is a space observatory on the International Space Station (ISS) to observe
ultra high energy cosmic rays (UHECRs) in the future.
UHECRs induce cascade showers
(extensive air showers; EASs) in the atmosphere.
The main component of the shower particle is electron.
The electrons excite Nitrogen molecules to emit fluorescence mainly
in the range between 300 and 400 nm. The JEM-EUSO telescope sees
the fluorescence from the ISS orbit.
In this paper, absolute calibration of the JEM-EUSO telescope with the
moonlight will be discussed.
The moon is known to be a very stable and well studied natural light
source and has been used to calibrate on-orbit sensors so far.
The observation of UHECRs with JEM-EUSO requires dark nights in
principle,
therefore moonlight nights are available for the calibration purpose. The
expected number of photoelectrons was found to be several tens to hundreds
per 2.5~$\mu$s for the full moon. That number is within the dynamic range of
the JEM-EUSO electronics. The detail will be discussed in the paper.
}


\maketitle

\section{Introduction}
JEM-EUSO is a ultra-high energy cosmic ray observatory on
orbit whose launch is foreseen
in 2017\cite{bib:eusogen13,bib:eusosci13,bib:eusodet13}. 
The fluorescence and \v{C}erenkov light generated by extensive air
shower (EAS) particles induced by cosmic rays will be detected with a wide
field of view telescope on the International Space Station.
Calibration is important to reconstruct accurately 
cosmic ray energy and arrival
direction and so on to study the ultra-high energy universe in detail.
At present, $\sim10^{20}$eV particles are available only in the nature,
so the EAS property can be studied with cosmic ray observation or by
simulation work. If once how many particles and their energy
distribution are known, we can estimate the produced fluorescence and
\v{C}erenkov light because the fluorescence yield has been measured
precisely enough in recent works\cite{bib:fwicrc13,bib:airfly13}.

 The JEM-EUSO telescope consists of three Fresnel lenses, a focal
 surface (FS) detector 
with 5,000 multi anode photomultiplier tubes (MAPMTs) of 0.3M pixels in total
and data acquisition electronics\cite{bib:eusodet13}. In order to discuss
the cosmic ray origins in detail, detector calibration is mandatory.
For the calibration of the JEM-EUSO telescope, onboard UV-LED light sources
\cite{bib:euso-onboard13} and on-ground light sources (Xe flashers and
355nm LIDARs)\cite{bib:gls2013} will be deployed.
One of the advantages of onboard
calibration system (light source) is the availability whenever it is
necessary and the controllability, but the concern about the degradation
 with time of the onboard calibration system
 is always accompanied as a light source for
the absolute calibration. For the case of on-ground light sources,
attenuation in the atmosphere should be evaluated in addition to the
calibration of the light sources themselves.
Natural light sources may be a good candidate
for the absolute calibration. For JEM-EUSO, the moonlight may be a good
light source\cite{bib:ebisu-moon}. The moonlight has been found to be
stable enough\cite{bib:moon-stability} to be used for the calibration
purpose of instruments on orbit in the atmospheric science.
In fact, several sensors on satellites have been calibrated with the
moonlight
(e.g. Refs.\cite{bib:GOMElunar,bib:planetB,bib:MODISmoon03,bib:MODISmoon07,bib:SeaWiFScalib12}).

\section{Absolute Calibration with the Moonlight}
\subsection{Introduction}
We consider the conditions which are necessary to observe the moonlight
with JEM-EUSO.
In order to take calibration data with the moonlight, the following
procedure is assumed.
\begin{enumerate}
\item Detect cloud candidates with the slow mode
\item Take Infrared (IR) images with the onboard IR camera\cite{bib:euso-ams}
 and choose a cloud at high altitude
\item Shoot onboard laser\cite{bib:euso-ams} at the cloud
      to determine the height and the reflectance
\item Record the MAPMT signals of the reflected moonlight on the cloud
 and evaluate the overall photon detection efficiency of the JEM-EUSO
 telescope
\end{enumerate}
In principle, JEM-EUSO is designed to be triggered to take data
by cosmic ray air showers which will last for 100 $\mu$s typically
and continuous background light intensity will not be
triggered. Therefore a dedicated mode is necessary. The slow mode,
which is the monitoring of the pixel signal rate every 3.5 seconds 
for the observation of transient luminous events (TLEs), may be also
applicable for the cloud monitor.

The cloud should be chosen at high altitude $>10$km
to reduce the uncertainty in the atmospheric transmission around the
ground. Above 10km, the scattering is usually well described only by the
Rayleigh scattering and the attenuation of light will be smaller.
This selection may be done in offline analysis. The IR images
are also important to know the area covered by the cloud.

The sensors usually used in the atmospheric science can see the moon
because they use 
steerable mirrors to observe targets. However, since JEM-EUSO can see at a
fixed direction towards the earth, the earth albedo could be a
major source of error in the absolute calibration. 

\subsection{Moon Irradiance}
\subsubsection{Introduction}
The brightness of the moon is considered to be very stable.
The photometric stability of the lunar surface was studied by
H.H. Kieffer\cite{bib:moon-stability}. It was found that events that would
change the moon surface brightness (integration in a visible band)
by 1\% would occur once per 1.4 Gyr. Therefore the apparent brightness
changes according to the geometrical configuration of the sun, the
moon and JEM-EUSO.
Because of its stability, H.H. Kieffer and T.C. Stone developed a
disk-averaged moon reflectance model (ROLO) from 350~nm to 2450~nm
based on the 6-year observation on ground for the calibration of
on-orbit sensors of the Earth science\cite{bib:moon-irrad}.
With this model, we have estimated the moon irradiance for the JEM-EUSO
observation.

\subsubsection{Irradiance calculation}
The irradiance was calculated as the product of
the solar irradiance at the moon and the moon disk averaged reflectance.
C.A. Gueymard compiled 24 years of irradiance
measurements into a composite spectrum up to 1~mm, 
whose resolution is 0.5~nm in $280-400$~nm and 1~nm in
$400-1000$~nm\cite{bib:solar-spect}.

\begin{figure}[htb]
\begin{center}
 \includegraphics[width=.5\textwidth]{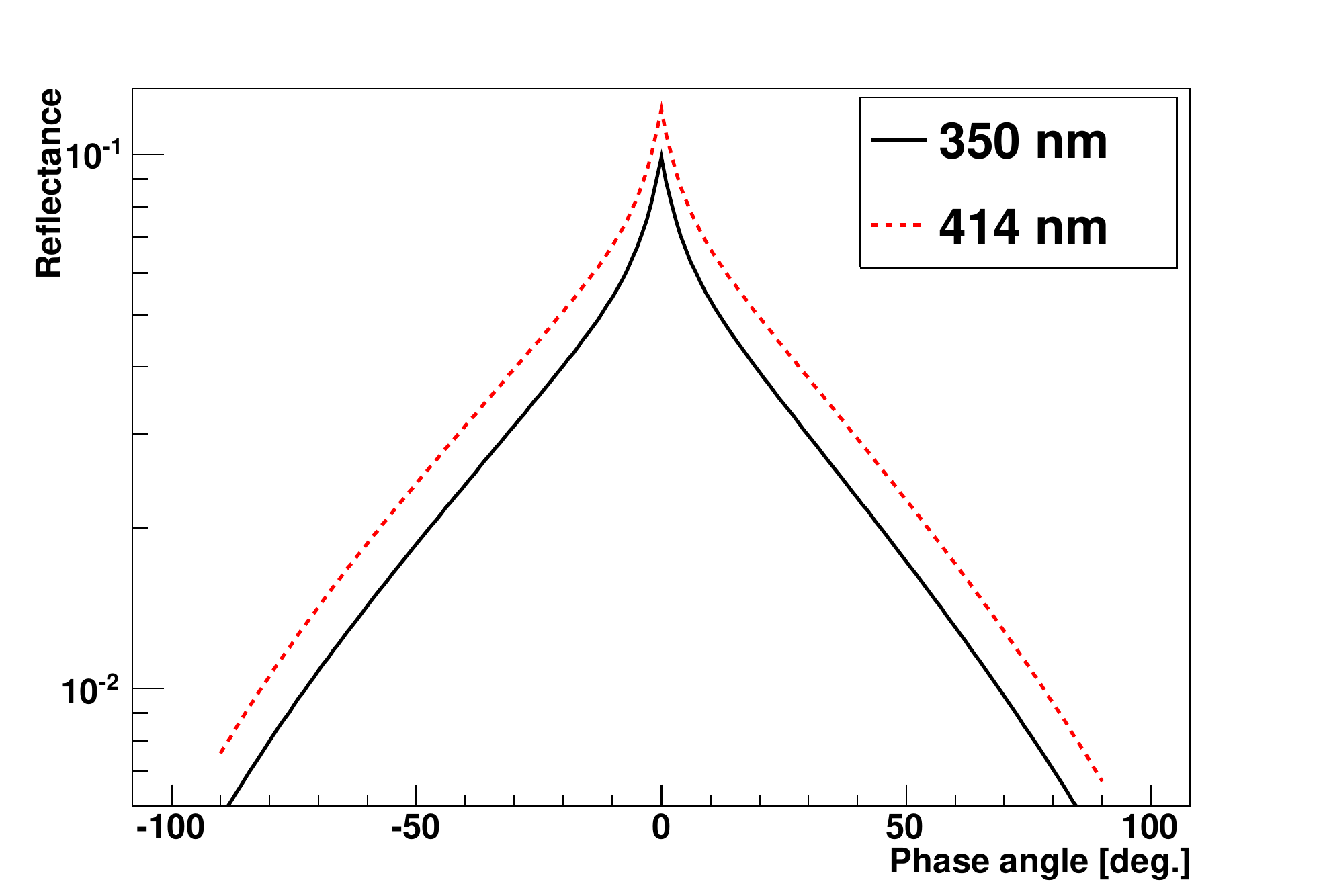}
\end{center}
\caption{Lunar disk-averaged reflectance as a function of phase angle
 for 355~nm (solid line) and 414~nm (dotted line)
 calculated with the ROLO model.%
\cite{bib:moon-irrad}}
\label{fig:lunarrefphase}
\end{figure}

\begin{figure}[htb]
\begin{center}
 \includegraphics[width=.5\textwidth]{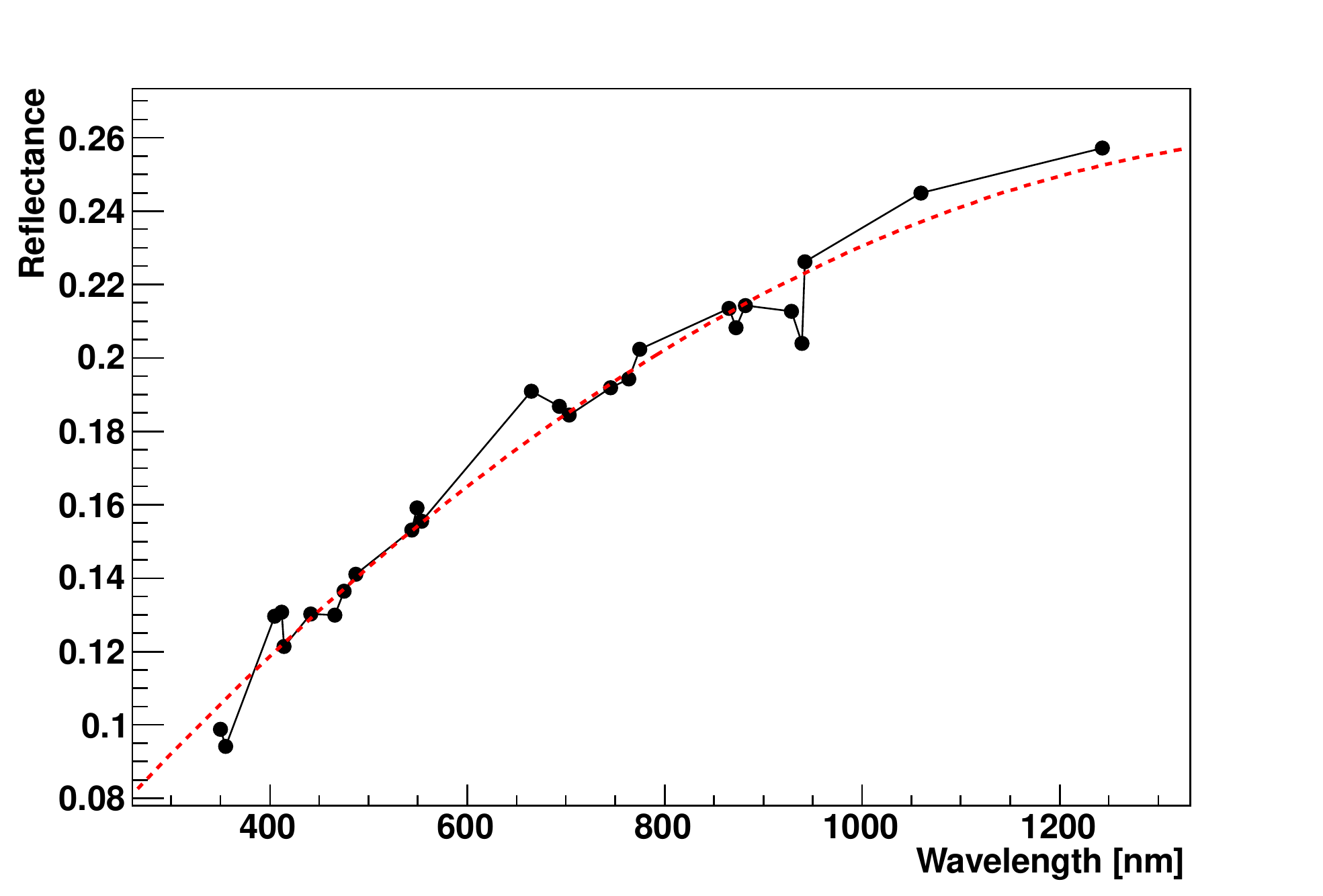}
\end{center}
\caption{Lunar reflectance spectrum at phase angle 0 calculated with the
 ROLO model\cite{bib:moon-irrad}. The red dotted curve shows the best fitted
 quadratic function.}
\label{fig:lunarrefspect}
\end{figure}

In the ROLO model, disk averaged spectral reflectance ($\rho_k$) of the
moon at wavelength $\lambda_k$ was modeled empirically as follows.

\begin{eqnarray}
\ln \rho_k &=& \sum^3_{i=0} a_{ik}g^i + \sum^3_{j=1} b_{jk}\Phi^{2j-1} +
 c_1\theta +c_2\phi + c_3\Phi\theta \nonumber\\ 
&& +c_4\Phi\phi + d_{1k}e^{-g/p_1} + d_{2k}e^{-g/p_2} \nonumber\\
&& + d_{3k}\cos[(g-p_3)/p_4]\ , \label{eq:reflect}
\end{eqnarray}
where $g$ is the phase angle, $\theta$ and
$\phi$ are the selenographic latitude and longitude of the observer,
and $\Phi$ is the selenographic longitude of the Sun. $c_i$ and $p_j$
 are wavelength independent constants, and $a_{ik}$, $b_{jk}$ and
 $d_{lk}$ are wavelength dependent constants. These constants are
 given in the paper\cite{bib:moon-irrad}. In this paper, for
 simplicity $\theta=0$ and $\phi=0$ were assumed and $\Phi$ was
 approximated as $-g$. Figs.\ref{fig:lunarrefphase} and
 \ref{fig:lunarrefspect} show phase angle dependence of the lunar
 reflectance at 350~nm and 414~nm and the reflectance
 spectrum at phase angle 0 degree. The reflectance increases with
 increasing wavelength. The phase angle dependencies are
 similar to each other at
 350~nm and 414~nm. Since the spectrum calculated with the model
 showed irregularity, the data points were fitted with a quadratic
 function shown by the red dotted curve in Fig.\ref{fig:lunarrefspect}.

\begin{figure}[htb]
\begin{center}
 \includegraphics[width=.5\textwidth]{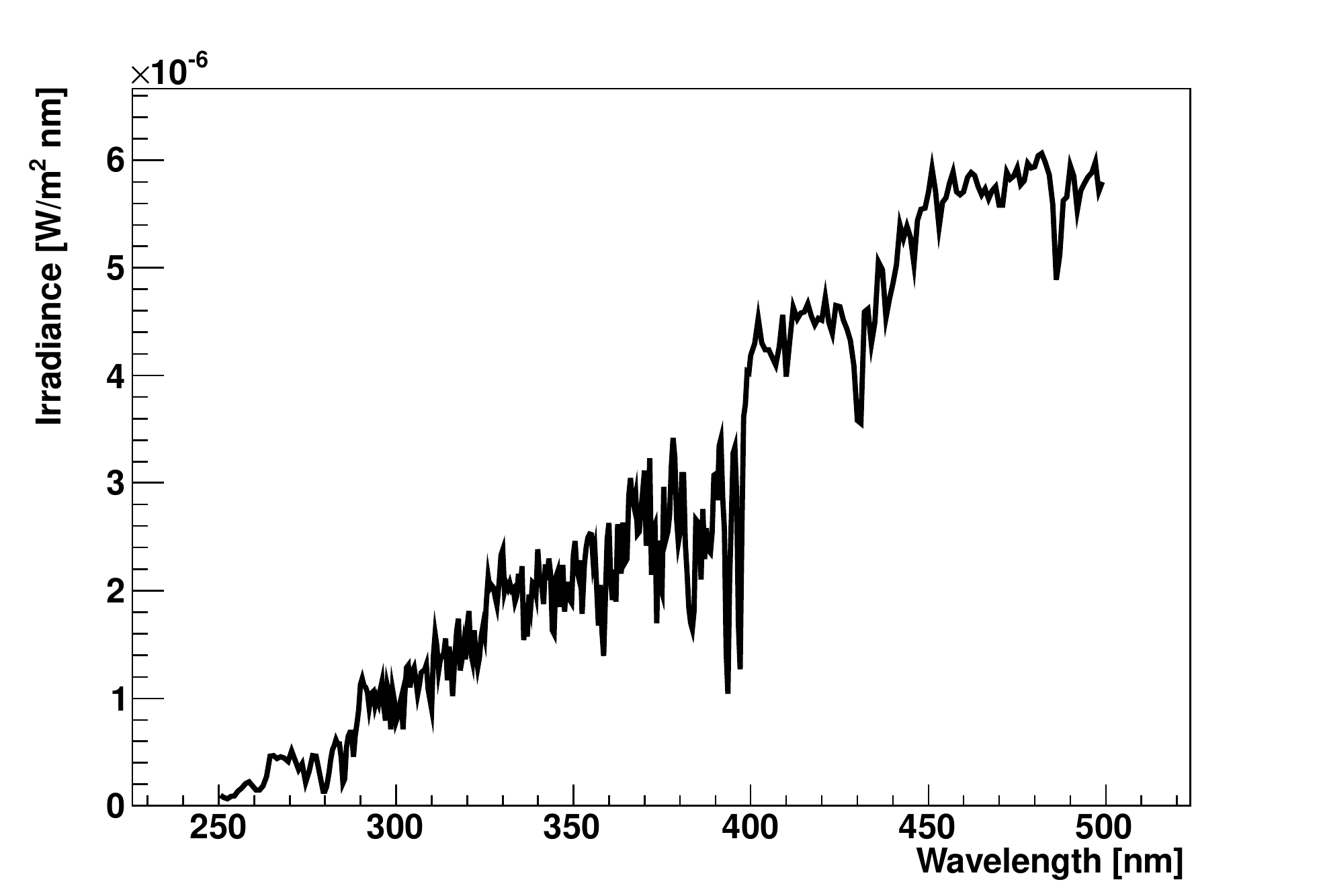}
\end{center}
\caption{Spectral irradiance of the full moon at the top of the
 atmosphere of the earth.}
\label{fig:lunarspectTOA}
\end{figure}

The irradiance ($I_k$) at the top of the atmosphere at wavelength band $k$ 
can be calculated with the following equation.
\begin{equation}
I_k=\frac{1}{\pi} \rho_k\Omega_\mathrm{M}E_k
 \left(\frac{1\mathrm{AU}}{D_\mathrm{S-M}}\right)^2\left(\frac{384,400\mathrm{km}}{D_\mathrm{E-M}}\right)^2\ ,
\label{eq:irradiance}
\end{equation}
where $\Omega_\mathrm{M}$ is the solid angle of the Moon
($=6.4177\times10^{-5}$sr) and $E_k$ is the solar spectral irradiance.
$D_\mathrm{S-M}$ and $D_\mathrm{E-M}$ are the distances of Sun-Moon
and Earth-Moon, respectively. In this paper, $D_\mathrm{S-M}$ and
$D_\mathrm{E-M}$ are  assumed to be 1~AU and 384,400~km, respectively.
The resultant spectral irradiance of the full moon
at the top of the atmosphere is shown in Fig.\ref{fig:lunarspectTOA}.
The irradiance between 300~nm and 400~nm is
$3.72\times10^5$ photons/ns~m$^2$.
In Ref.\cite{bib:EUSO-duty04}, it was estimated as $2.437\times10^5$~
photons/ns~m$^2$ at full moon.
They used the same model, ROLO model, but without
the opposition effect that the moon brightness increases rapidly
as the phase angle reaches zero.
They reported that the brightness would increase by $\sim35$\% if the
opposition effect was taken into account. Another major reason for the
difference is the wavelength dependence of the reflectance.
In our calculation, the slope is smaller. It was estimated as $\sim$10\% effect.

\subsection{Cloud Reflection}
If the optical thickness of the cloud is larger than unity ($\tau > 1$),
the reflected intensity does not depend very much on the incident angle
or the emergent angle\cite{bib:chandrasekhar,bib:liou1973}. Here the emergent
angle is assumed up to $\sim$30 deg. which is the half angle of the
JEM-EUSO field of view. However, the reflectance varies very much from
a few \% to $\sim$100\% depending on the optical thickness\cite{bib:cloud-thick}.
This variation may introduce a large error directly in the calibration.
The reflectance should be determined as precisely as possible
together with the cloud top height.
Hereafter we assume the cloud reflection as of Lambertian type, which is a
good approximation for thick clouds, and
the reflectance as $\rho_\mathrm{cloud}$.
In order to avoid the variability of the atmosphere
transmittance, only thick clouds above $\sim$10~km were assumed as already
mentioned.

The appearance probability of clouds with optical depth larger than one
was studied in the JEM-EUSO performance paper\cite{bib:jeuso-performance}.
Based on the database by TOVS (TIROS-N (Television Infra-Red Observation
Satellite) Operational Vertical Sounder) \cite{bib:TOVS}, it was
estimated as about 7\%
along the ISS track at night\cite{bib:jeuso-performance}. There must be
quite a good chances in a year to observe the reflected moonlight on
thick clouds.
\subsection{Expected Signal for the Calibration}
\begin{figure}[htb]
\begin{center}
 \includegraphics[width=.5\textwidth]{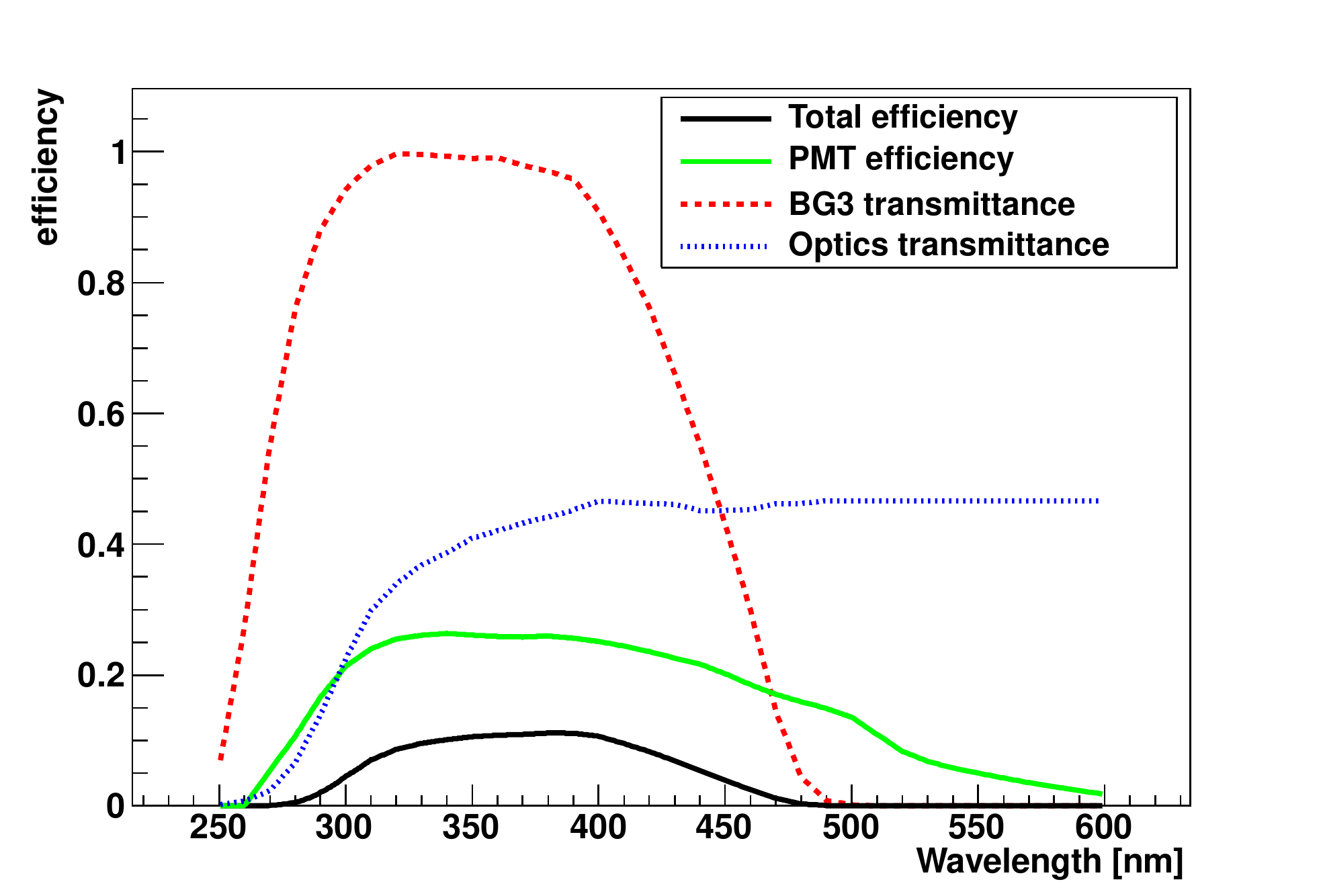}
\end{center}
\caption{The JEM-EUSO spectral efficiency assumed in the calculation.
 The optics efficiency for the light parallel to the optical axis is
 plotted with a blue dotted line, the filter (BG3) transmittance
 with a red dotted line, PMT efficiency (QE$\times$CE)
 with a green solid line and the
 total efficiency with a black solid line.}
\label{fig:efficiency}
\end{figure}

\begin{figure}[htb]
\begin{center}
 \includegraphics[width=.5\textwidth]{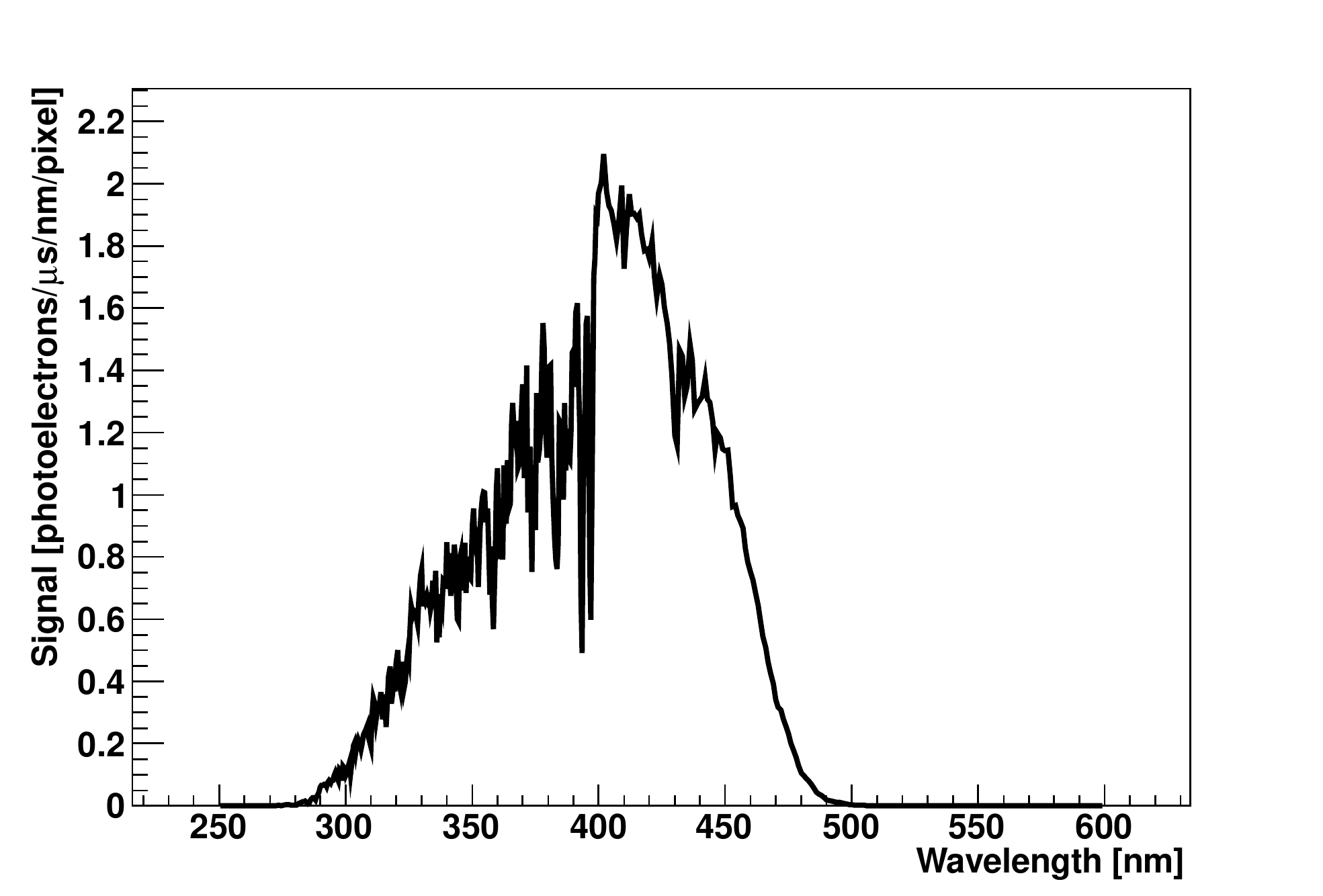}
\end{center}
\caption{The expected full-moon spectrum observed with JEM-EUSO at the
 zenith and with the cloud reflectance 1.}
\label{fig:moon-euso}
\end{figure}

The expected photoelectron rate in a pixel, $S_k$, can be derived as follows:
\begin{equation}
S_k=I_k \cos\theta_0\cdot \rho_\mathrm{cloud}/\pi \cdot T_\mathrm{atm}^2
 \cdot A_\mathrm{pixFOV} \cdot
 \Omega_\mathrm{EUSO}\cdot \epsilon_\mathrm{EUSO}\ ,
\label{eq:signal}
\end{equation}
where $I_k$ is the irradiance of the full moon shown in
 Eq.(\ref{eq:irradiance}), $\theta_0$ is the zenith angle of the moon,
 $T_\mathrm{atm}$ is the trasmittance in the atmosphere above the cloud,
 $A_\mathrm{pixFOV}$ is the area on the cloud
which one pixel (FoV: $0^\mathrm{o}.073 \times
 0^\mathrm{o}.073$ for the center pixel in the JEM-EUSO FS) sees,
$\Omega_\mathrm{EUSO}$ is the solid angle of the JEM-EUSO at
400~km seen from the cloud height,
 $\epsilon_\mathrm{EUSO}$ is the efficiency of the
JEM-EUSO telescope, which is the product of the optics
efficiency ($T_\mathrm{Opt}$), the filter transmittance ($T_\mathrm{BG3}$),
and the MAPMT efficiency ($\mathrm{QE}\times\mathrm{CE}$). The
efficiencies are plotted in Fig.\ref{fig:efficiency}. Here, the cloud at
 the nadir was considered as a typical case 
and then the optics efficiency was taken for the light parallel to the optical axis.
The resultant full-moon spectrum observed by JEM-EUSO is shown in
Fig.\ref{fig:moon-euso}. The expected signal rate for the full moon is
$449\rho_\mathrm{cloud}\cos\theta_0$ [photoelectrons (p.e.)/GTU/pixel],
where GTU stands for
the gate time unit (2.5$\mu$s) which JEM-EUSO counts
 the signal for. For the case of the half moon, the signal decreases to
$24.5\rho_\mathrm{cloud}\cos\theta_0$ [p.e./GTU/pixel].
Since the dynamic range of the JEM-EUSO electronics is from 1 p.e. to
$>300$ p.e.\cite{bib:eusodet13},
the signal intensity of the moon is almost within the
dynamic range. In order to accumulate 10,000 p.e., it takes 25~GTU for
the full moon at zenith and 10,000 GTU for the half moon at zenith angle of
60~deg. and with the cloud reflectance of 10\%.
The displacement of the ISS in
10,000~GTU will be shorter than 200~meters, which is smaller than the
projected size of 1 pixel on ground. As a consequence, the moon could be
a good calibration light source.
%
\section{Discussion}
If the data is accumulated for the consecutive 100 GTU, the total number
of detected photoelectrons is about 10,000 p.e. for the full moon at the
zenith.
Therefore the statistical error will be 1\% or less.
With the help of the ROLO model, the degradation of the Sea-viewing Wide
Field-of-view Sensor (SeaWiFS) on orbit has been determined at 0.1\% level
successfully\cite{bib:seawifs-err}. However, there are still large discrepancies
($\sim\pm10$\% in visible and near infrared band) in the absolute irradiance 
between the model prediction and the observed
data\cite{bib:moon-irrad}. S.D. Miller and R.E. Turner have developed another
lunar reflectance model independently and reported $15$\%
smaller and 3\% larger at 350~nm
at phase angle 2.9 and 34.9 degrees calculated with their model
than with ROLO\cite{bib:Miller2009}. At present, the systematic uncertainty
of the moon intensity seems at $10-15$\% level. In order to utilize the
moon as a standard absolute candle and to ensure the traceability, a new
observation with better accuracy has been proposed\cite{bib:LUSI}.
In the future, the spectral irradiance model will be expected to be improved.

Another source of the systematic error would be the reflectance of the
cloud. The reflected signal of a shot of the onboard 20~mJ 355~nm LASER would be 
$28,000\rho_\mathrm{cloud}$~p.e., which is much larger than the dynamic
range and it is difficult to determine the reflectance of the cloud.
If a satellite for the atmospheric science observes the same location as
JEM-EUSO and the data is available later, it could be used.

Another possible reflector would be ocean or desert under clear sky. The
``reflectance'' of clear sky can be measured with the onboard LASER,
which is about 30\%. In this case, Lambertian reflection of the surface
and wavelength independent reflection may not be applicable, so that the
observation site should be chosen carefully. Soil, water, vegetation
have surface albedo about 5\% in near UV range. The sensitivity of the
top of the atmosphere (TOA) reflectance to surface albedo in
Ref.\cite{bib:sciamachy}. The coefficient is $0.1-0.2$, that means 
100\% error in the surface albedo leads to $10-20$\% error in the
reflectance. Therefore clear sky above surface with small albedo may be
a possible condition for the calibration.
If the TOA reflectance can be
determined with systematic error of 15\%, the total systematics will be
about 20\% since the errors of the other factors will be negligibly
small because they are geometrical ones in principal.

\section{Summary}
For the space based experiments to observe EAS on orbit
like JEM-EUSO, it is difficult to obtain
reliable and stable light source for absolute calibration. 
In this paper, possibility of the moon as a calibration light
source has been studied. In the full moon period, EAS observation by
JEM-EUSO is very difficult because of the large background level.
The observed intensity of the moonlight was estimated as 
$449\rho_\mathrm{cloud}\cos\theta_0$ [p.e./GTU/pixel] for the full moon
light reflected on the cloud at the nadir, where
$\rho_\mathrm{cloud}$ is the reflectance of the cloud and $\theta_0$ is
the zenith angle of the moon. This intensity is almost within the dynamic range
of the JEM-EUSO electronics. Since the moonlight is very stable, the
error of the relative calibration (i.e. the degradation in time) will depend
on the error of the earth reflectance. For the absolute calibration,
systematic error in the moonlight remains $10-15$\% at
present. Therefore if the reflectance could be determined with $<15$\%
error, the total systematic error in the absolute calibration with the
moonlight would be about 20\%. There are a few activities to improve the
moon reflectance model for the calibration purpose of on-orbit
instruments, the systematic error in the moonlight may be smaller at the
time of observation by JEM-EUSO.

\vspace*{0.5cm}
{
\footnotesize{{\bf Acknowledgment:}{%
This work was partially supported by Basic Science Interdisciplinary 
Research Projects of RIKEN and JSPS KAKENHI Grant (22340063, 23340081,
and 
24244042), by the Italian Ministry of Foreign Affairs, General Direction 
for the Cultural Promotion and Cooperation, by the 'Helmholtz Alliance 
for Astroparticle Physics HAP' funded by the Initiative and Networking
Fund of the Helmholtz Association, Germany, and by Slovak Academy  
of Sciences MVTS JEM-EUSO as well as VEGA grant agency project
2/0081/10. The Spanish Consortium involved in the JEM-EUSO Space
Mission is funded by MICINN under projects AYA2009-06037-E/ESP,
AYA-ESP 2010-19082, AYA2011-29489-C03-01, AYA2012-39115-C03-01,
CSD2009-00064 (Consolider MULTIDARK)
and by Comunidad de Madrid (CAM) under project S2009/ESP-1496.
}}

}
\clearpage

%% file: icrc2013-0818.tex



\title{The JEM-EUSO Global Light System}

\shorttitle{JEM-EUSO GLS}

\authors{
L.~Wiencke$^{1}$,
J.~H.~Adams$^{2}$,
M.~Christl$^{3}$,
S.~Csorna$^{4}$,
F.~Sarazin$^{1}$
for the JEM-EUSO Collaboration with
J.~Bogulski$^{1}$, T.~Horvath$^{1}$, R.~Larsen$^{1}$, 
W.~Naslund$^{1}$, Z.~Norman$^{1}$, G.~Pasqualino$^{1}$.
}

\afiliations{
$^1$ Colorado School of Mines \\
$^2$ University Alabama in Huntsville \\
$^3$ NASA/Marshal Space Flight Center\\
$^4$ Vanderbilt University
}

\email{lwiencke@mines.edu}

\abstract{
The Global Light System (GLS) is a network of ground-based Xenon flashlamps and steered UV lasers to validate the key functions of the JEM-EUSO cosmic ray fluorescence detector that is planned for the international space station. These functions include triggering efficiency, the accuracy of intrinsic luminosity measurements, and the reconstructed pointing direction accuracy. GLS units will generate benchmark optical signatures in the atmosphere with similar characteristics to the optical signals of cosmic ray EASs. The lasers will generate tracks and the flashers will generate point flashes. But unlike air showers, the number of laser and flasher pulses, their energy, precise time, direction (lasers) can be specified. JEM-EUSO will reconstruct the pointing directions of the lasers and the energy of the lasers and flashlamps to monitor the detector triggers, and accuracy of energy and direction reconstruction. 12 GLS units will be deployed at selected sites around the globe. The JEM-EUSO footprint will pass over a GLS unit on average once per (near) moonless night under clear conditions for appropriately selected sites. The 12 units will be supplemented by campaign style measurements with an airborne unit that will be flown over the open ocean at selected altitudes under JEM-EUSO. A GLS prototype in an airplane will support a high-altitude balloon flight in 2014 of a prototype JEM-EUSO telescope. We describe the concept and system design and report on the status of prototyping and the selection process for candidates 
sites. }


\maketitle

\section{Introduction}
JEM-EUSO is a pioneering air fluorescence experiment planned for the international space station. It will map the full sky by measuring cosmic ray extensive air showers (EAS) above $5\times10^{19}$ eV with a single instrument of unprecedented detection aperture. The goal is to identify the highest energy cosmic accelerators. Orbiting the globe every 90 minutes, JEM-EUSO will look down on the atmosphere from an altitude of about 350 km. During dark periods JEM-EUSO will record the optical signatures of EASs and other UV optical atmospheric transients that occur within its moving footprint of some 150,000 km$^{2}$.  
 \begin{figure}[t]
  \centering
  \includegraphics[width=0.4\textwidth]{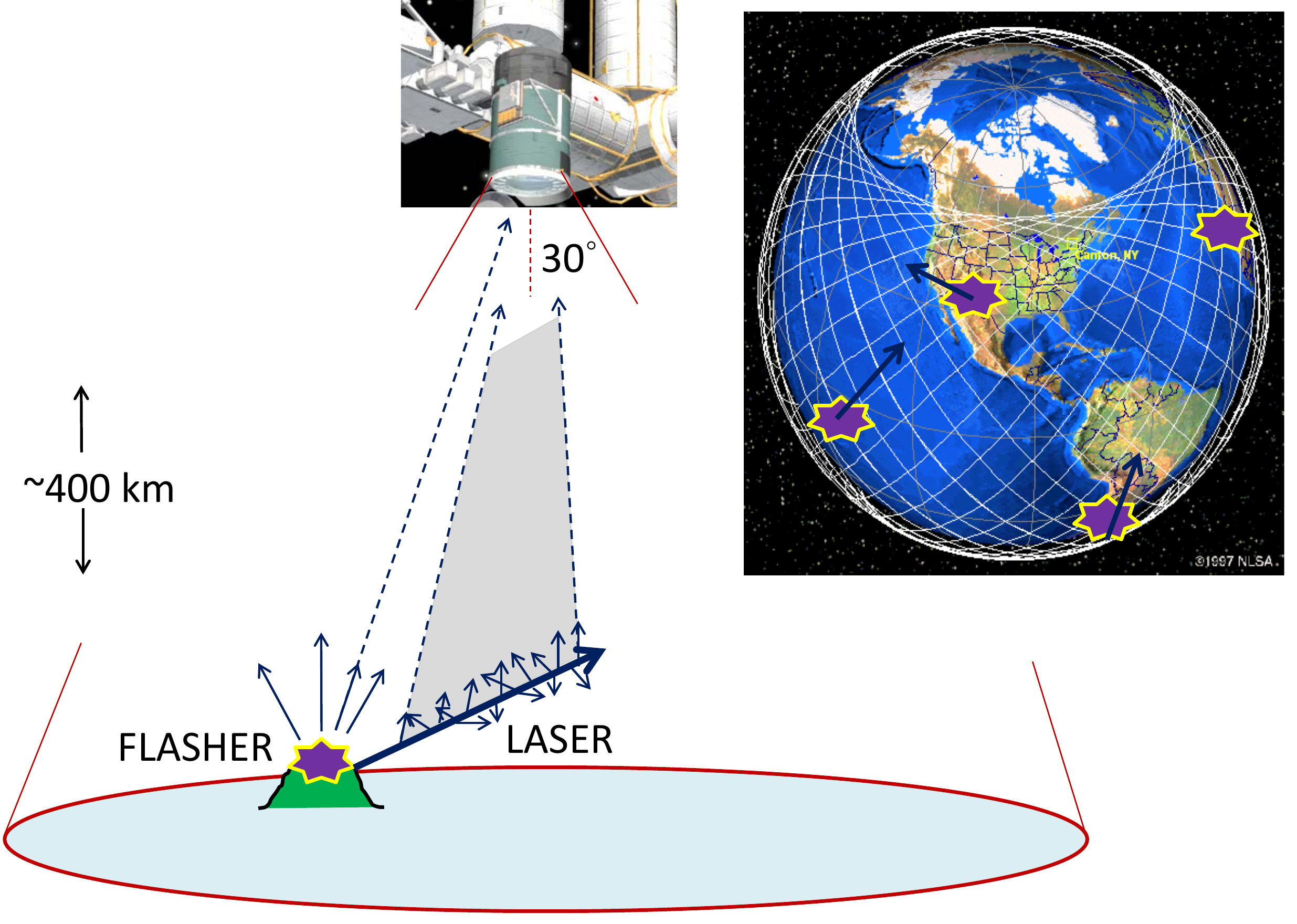}
  \caption{The Global Light System of ground-based calibrated xenon flashlamps and lasers will be measured by JEM-EUSO to monitor its performance during the mission.}
  \label{fig:GLScartoon}
 \end{figure}
\section{The Global Light System}
JEM-EUSO will also record optical signatures generated by a global network of calibrated UV light sources called the Global Light System (GLS). UV light from GLS xenon flashlamps will appear as optical point sources dominated by direct transmission. Light scattered out of the beams of UV lasers aimed across the JEM-EUSO field of view will appear as tracks. The technique (Fig. \ref{fig:GLScartoon}), draws on the experience of the ground based fluorescence detectors of Fly's Eye\cite{Baltrusaitis:1985mx}, HiRes\cite{HiRes:NIM200}, and Pierre Auger\cite{Abraham:2009pm}. These experiments used flashlamps\cite{Wiencke:1999} and lasers\cite{FICK:2005yn} in various configurations as part of their science programs.  Their data demonstrated \cite{Bird:1994,Cannon:2003wf, 2010zzl,2013atm} that lasers observed from the side as "test beams" produce a luminosity that is comparable to the EASs that are expected to be above the JEM-EUSO energy threshold.  

Unlike cosmic EAS events which are essentially random, the properties of GLS flashes and laser shots can be programmed in advance and measured independently at their source. The properties include the absolute time, energy, wavelength, and direction (lasers). These independent measurements can then be compared, event by event, to the space measurements obtained by reconstructing the JEM-EUSO GLS events. In this way the GLS will be used to monitor and validate key parameters of the detector and the data analysis chain. These parameters include:
\begin{itemize}
\item Triggering efficiency
\item Accuracy of EAS intrinsic luminosity measurements
\item Pointing accuracy of EAS arrival directions (depends on the absolute timing, pointing and focus of the instrument)
\end{itemize}
\begin{figure}[t]
  \includegraphics[width=0.40\textwidth]{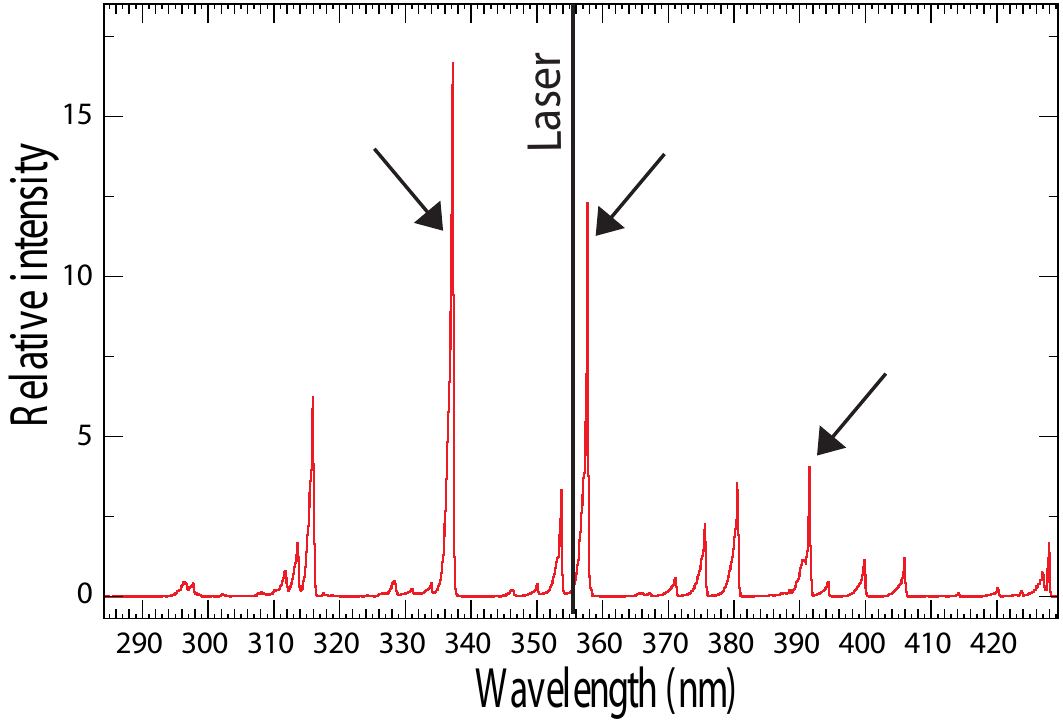}
  \caption{The filtered wavelengths (337, 357, 391 nm) of the GLS xenon flashlamps and laser (355 nm) are indicated on the fluorescence spectrum of electrons in air. (Spectrum shown is from reference \cite{Ave2007}.)}
  \label{fig:GLSwavelength}
 \end{figure}
\subsection{Configuration and Applications}
The GLS will include 12 ground based stations around the globe. All sites will include calibrated Xenon flashlamps. Six of the 12 sites will include a steerable laser system. In addition, a portable system with a laser and flashlamps will be deployed occasionally by aircraft. The wavelengths of the sources will overlap major lines in the fluorescence spectrum of electrons in air (Fig. \ref{fig:GLSwavelength}). The ground based systems will be operated and programmed remotely. 

\begin{figure}
  \includegraphics[width=0.43\textwidth]{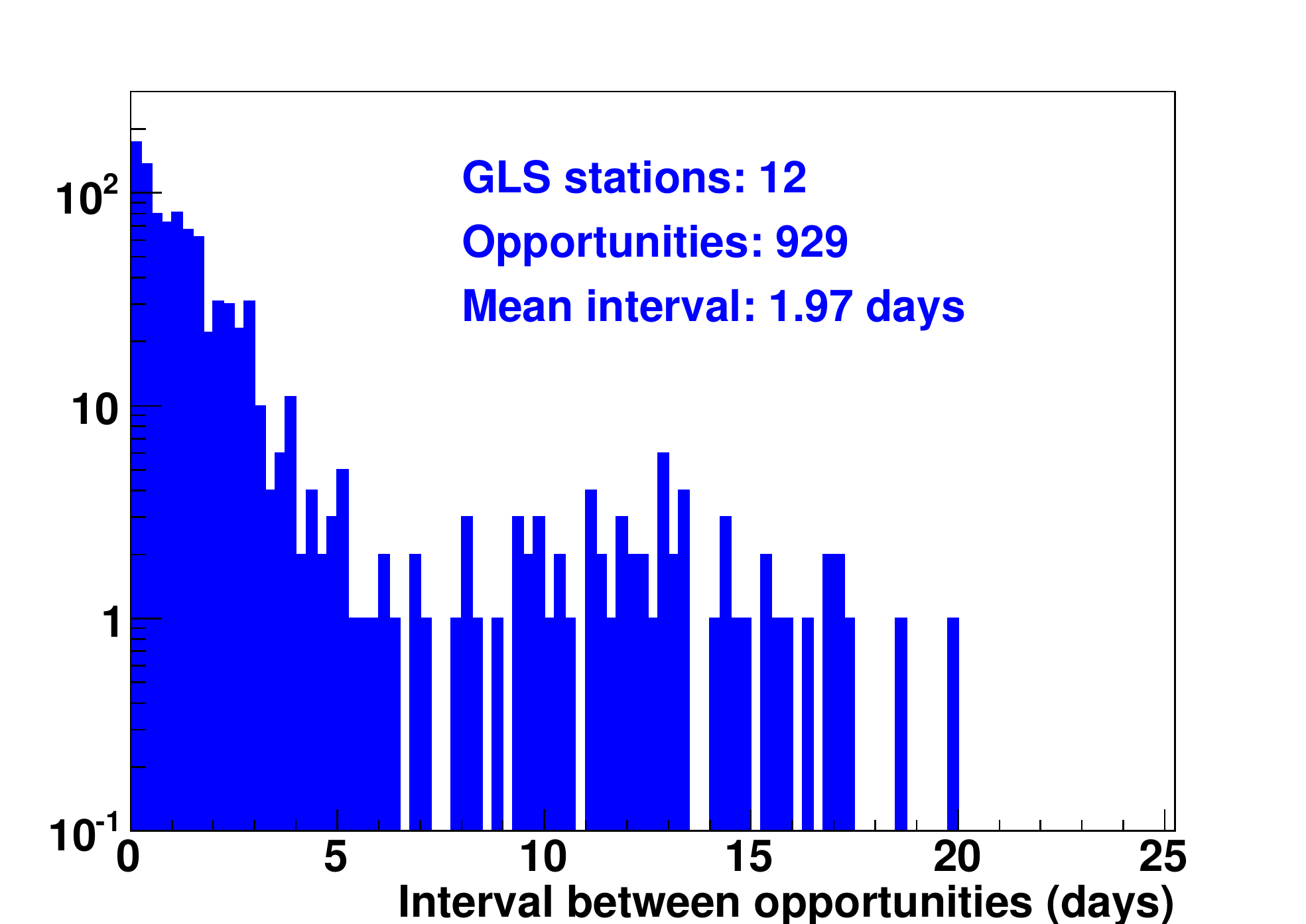}
  \caption{The estimated time between successful measurement opportunities (see text) for 12 GLS stations over 5 years of operations. The first measurement after the end of bright moon periods (which introduce a gap of about 6 days in observations) populate the tail of the distribution.}
  \label{fig:times}
 \end{figure}
Over its mission JEM-EUSO will make about the same number of passes over GLS stations under good conditions as the number of EASs that it will record above $5\times10^{19}eV$.
JEM-EUSO will pass over a ground station during dark clear viewing conditions about every 48 hours on average. This estimate of times between measurement opportunities (Fig. \ref{fig:times}) was obtained by a model that required the sun be at least 18 deg below the horizon, the illumination of the moon be less than 50\% and assumed a chance of clear viewing conditions of 33\%. The average crossing time of the JEM-EUSO footprint over a GLS station is about 60 seconds. The combined stations will alternate laser and flasher pulses, for a total rate of 20~Hz to provide nearly a
continuous set of measurements across the JEM-EUSO field of view. For each trigger, the on board atmospheric monitoring system will be activated automatically and acquire an IR camera image and a LIDAR shot aimed at the location of the GLS site~\cite{Adams:2003cu}. 

There will occasionally be very clear conditions when the measured total optical depth is not significantly greater than the molecular optical depth. The latter can be determined
accurately~\cite{Abreu:2012zg} from the global data assimilation
system (GDAS)~\cite{NOAA}. In these cases, the intrinsic luminosity
resolution can be measured using track-like signatures by comparing
the laser energy as reconstructed by JEM-EUSO and as measured at the laser.

GLS lasers will also be programmed to generate an artificial full sky map of potential cosmic accelerators. This will be done by firing shots in the direction of astronomical objects of interest that will include, for example, Cen-A, Virgo, and the galactic center.  A sky map of laser track directions as reconstructed by JEM-EUSO will be accumulated over the mission. Clusters of points and their spread about the directions of the programmed targets will provide a simple but comprehensive validation of the absolute EAS pointing accuracy reconstruction of the JEM-EUSO instrument, including the correct generation and transfer of absolute time stamps through the data acquisition and analysis chains.

\subsection{Sites}
\begin{figure}
  \includegraphics[width=0.48\textwidth]{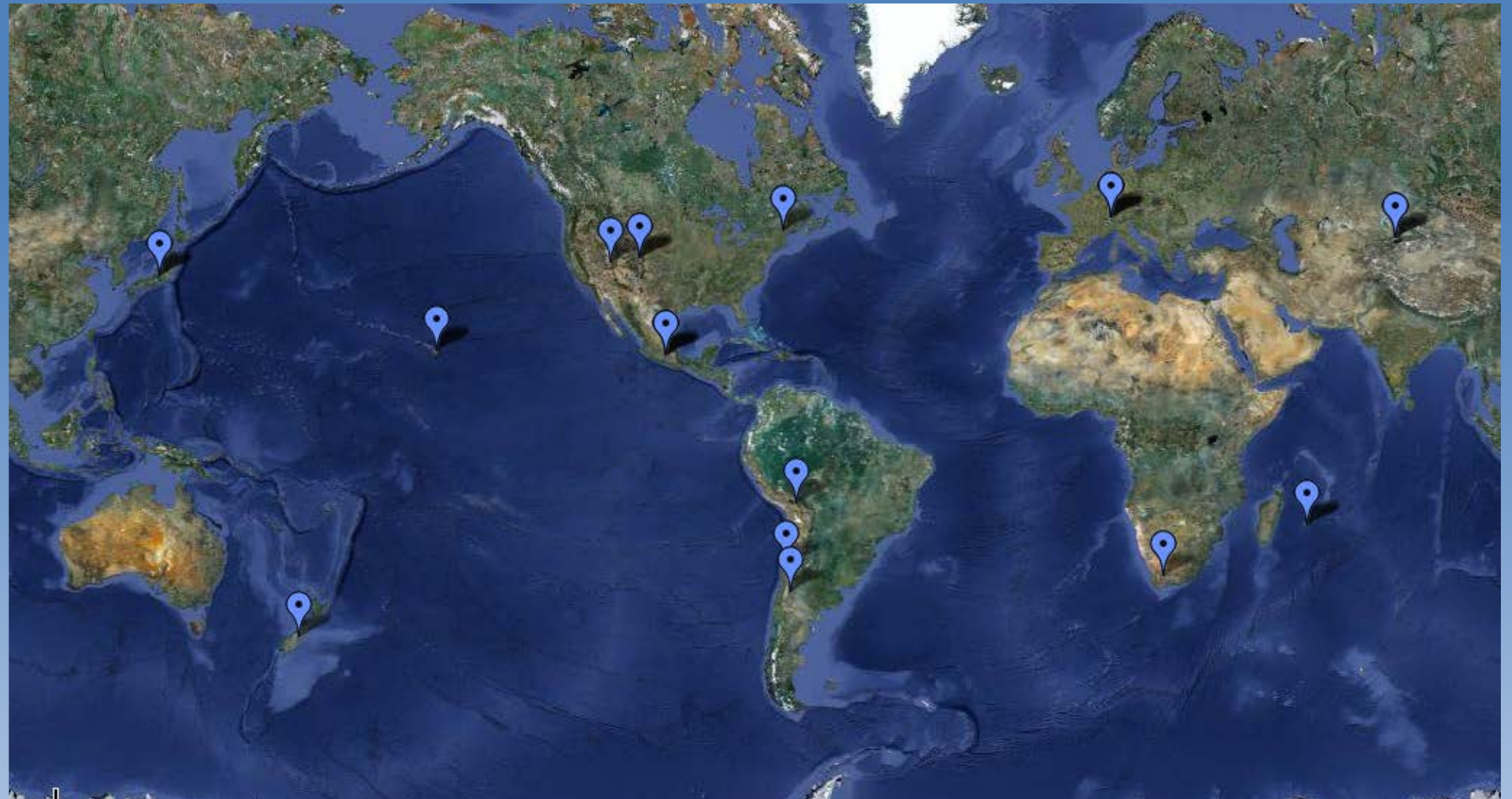}
  \caption{Some of the candidate GLS station locations. }
  \label{fig:GLSmap}
 \end{figure}

GLS sites will be selected to represent the variety of terrestrial backgrounds over which JEM-EUSO is expected to measure EASs. Selection criteria for sites include low light backgrounds, an altitude higher than the typical planetary aerosol boundary layer for that site, physical and legal access, a communications link, and some maintenance support. Shipping logistics and door to door costs will also be considered. Oceans represent the bulk of the dark sky regions. Consequently sites that satisfy the selection criteria and are located on isolated mountainous islands are especially desirable. Sites with existing scientific installations, including atmospheric monitoring are also quite desirable. The map in figure \ref{fig:GLSmap} shows some of the possible candidate sites locations which are also listed in table \ref{tab:locations}.
\begin{table}[h]
\begin{center}
\begin{tabular}{|l|c|c|}
\hline Site & Latitude & Elev. (km) \\ \hline
Jungfraujoch (Switzerland)   & 47$^\circ$N   & 3.9 \\ \hline
Alma-Ata (Kazakhstan) & 44$^\circ$N   & 3.0 \\ \hline
Mt Evans (CO, USA) &  39$^\circ$N   & 3.0 \\ \hline
Mt Norikura (Japan) &  30$^\circ$N   & 2.9 \\ \hline
Mauna Kea (HI, USA)  &  20$^\circ$N   & 3.0 \\ \hline
Nevado de Toluca (Mexico) &  19$^\circ$N   & 3.4 \\ \hline
Chacaltaya (Bolivia) & 16$^\circ$S   & 5.3 \\ \hline
La R\'{e}union (France) &21$^\circ$S   & 1.0 \\ \hline
Cerro Tololo (Chile) &30$^\circ$S   & 2.2 \\ \hline
Sutherland (South Africa) &32$^\circ$S   & 1.8 \\ \hline
Pampa Amarilla (Argentina) &35$^\circ$S   & 1.4 \\ \hline
South Island (New Zealand) &43$^\circ$S   & 1.0 \\ \hline
\end{tabular}
\caption{Some of the possible candidate GLS site locations.}
\label{tab:locations}
\end{center}
\end{table}
\subsection {Xenon Flashlamp}
Since JEM-EUSO will look down on the atmosphere, intrinsic luminosity can be monitored directly with flashlamps \cite{calib_2011}. 
All 12 GLS stations will include 4 individual flashlamps (Hamamatsu L6604). The L6604 model features a highly stable output with $<3\%$ shot-to-shot stability, a stable lifetime of more than 10$^{7}$ pulses and $<3\%$ degradation over the lifetime of the mission~\cite{Hamamatsu}. The light pattern from each flash is smoothly distributed over a wide field of view. These key performance parameters have been verified in laboratory tests. Three flashlamps will be filtered to match the primary lines indicated in figure \ref{fig:GLSwavelength}, and the fourth will use a broad band (Shott BG3) transmission filter identical to the filter planned for the JEM-EUSO detector.
\begin{figure}
  \centering
  \includegraphics[width=0.46\textwidth]{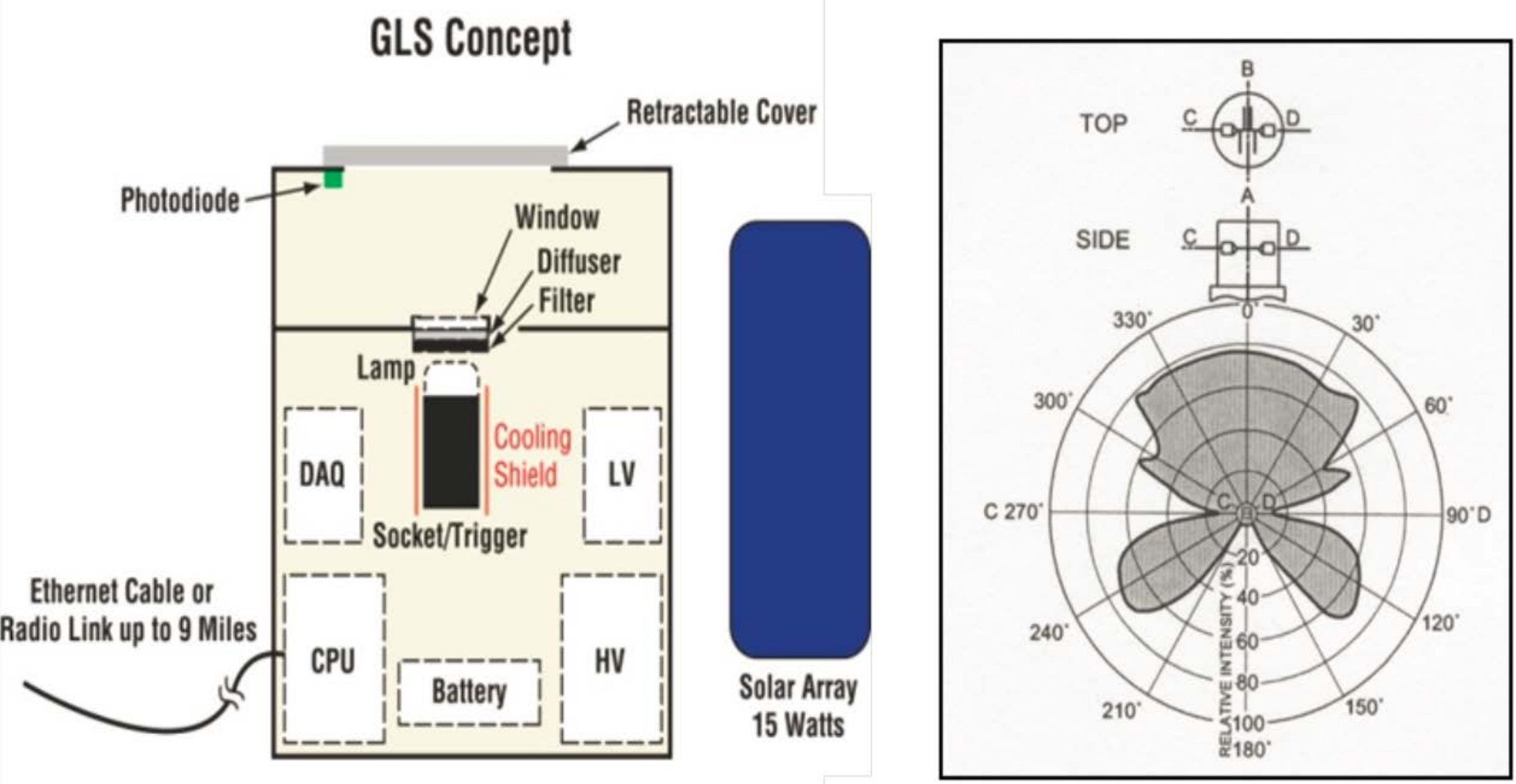}
  \caption{The configuration of a GLS flasher. The intensity distribution of the flashlamp is shown on the right.}
  \label{fig:flasher}
 \end{figure}
\begin{figure}
  \centering
  \includegraphics[width=0.47\textwidth]{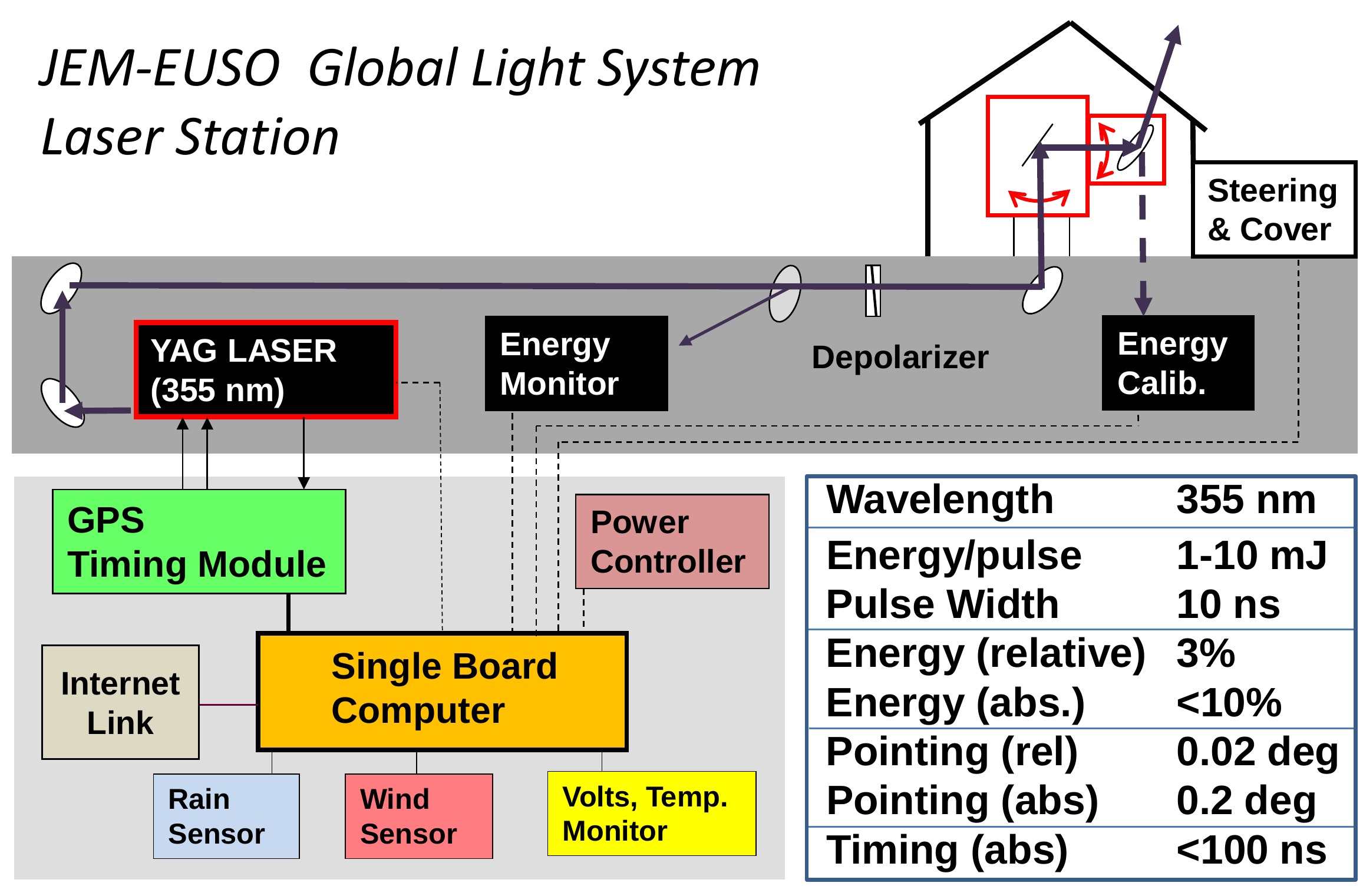}
  \caption{The configuration of a GLS laser system and table of performance targets.}
  \label{fig:GLSLaser}
 \end{figure}

\subsection {Laser Systems}
The design (Fig. \ref{fig:GLSLaser}), including component selection, draws on the design of two laser facilities \cite{mines_lasers} that have operated near the middle of the Pierre Auger Observatory  since 2004 and 2009 .  The
laser will be a frequency tripled YAG. Field proven models at HiRes and
Auger include the Quantel\cite{Quantel} Ultra, CFR and Centurion. The relative energy of each shot will be measured by a pyroelectric energy monitor probe. The net polarization of the beam will be
randomized so that for a given scattering angle, the atmosphere scatters the same amount of light
symmetrically about the beam axis. The steering mechanism will use a 
two orthogonal rotational stages with two steering mirrors (azimuth and elevation) 
mounted so that each mirror reflects the beam by a constant 90 degrees to minimize changes in the reflected
beam energy and polarization as a function of beam direction. In the parked position,
the beam will point down to a calibration probe that will measure the absolute beam energy downstream of all optics and calibrate the monitor energy probe. 
To facilitate identification of laser data within the JEM-EUSO data sample, the laser will
be triggered at precise times using a custom GPS timing module \cite{GPSY}. The time, direction, and energy of each shot will be recorded locally. The laser,energy monitors, and controls will be housed inside a temperature controlled shelter. The steering mechanism will be protected by an automated cover.

\subsection {Aircraft Systems}
A portable GLS system with flashers and a laser will be installed in a P3B airplane
managed by the NASA Airborne Science Program (ASP). The P3B 
has an upward viewing portal that is
available to install a flashlamp and a side port that will be fitted with
a fused silica window to transmit the horizontal UV laser pulses. The airplane will be deployed several times per year for under
flights of the ISS at night. The P3B will fly out 500~km from the
eastern seaboard to rendezvous with the ISS for a single under-flight.  
 (Since the earth rotates by some 22 degrees between each 90 minute ISS orbit
there will be one ISS overpass 
per P3B flight.) Over the length of the JEM-EUSO mission, these
flights will cover a range of altitudes, atmospheric and cloud
conditions, and moonlight.

\section {Testing EUSO-Balloon}
\begin{figure}
  \centering
  \includegraphics[width=0.30\textwidth]{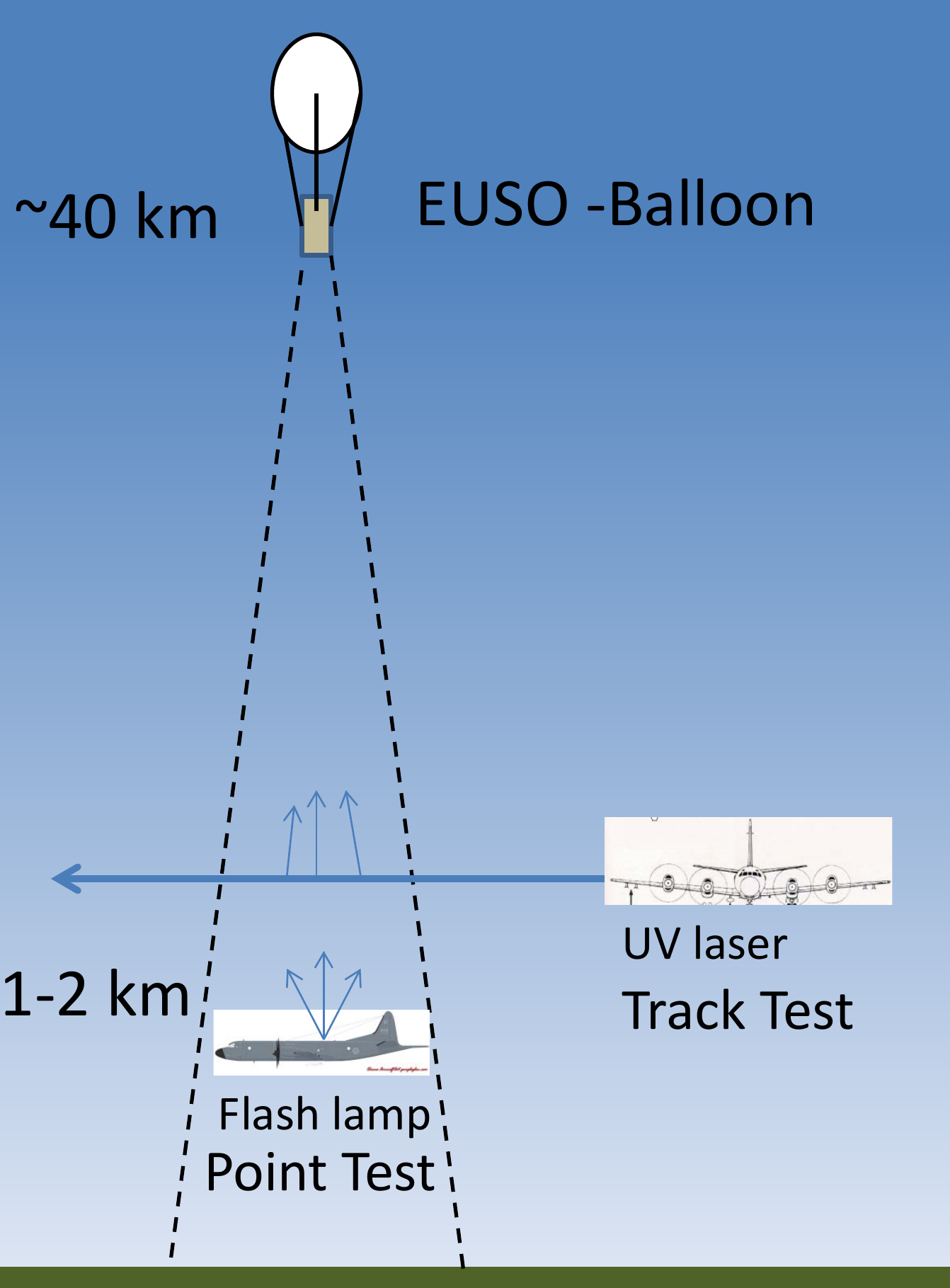}
  \caption{To test the EUSO-Balloon prototype detector an aircraft will fly under and next to the detector field of view with portable flasher and laser systems.}
  \label{fig:balloon}
 \end{figure}

\begin{figure}
  \includegraphics[width=0.45\textwidth]{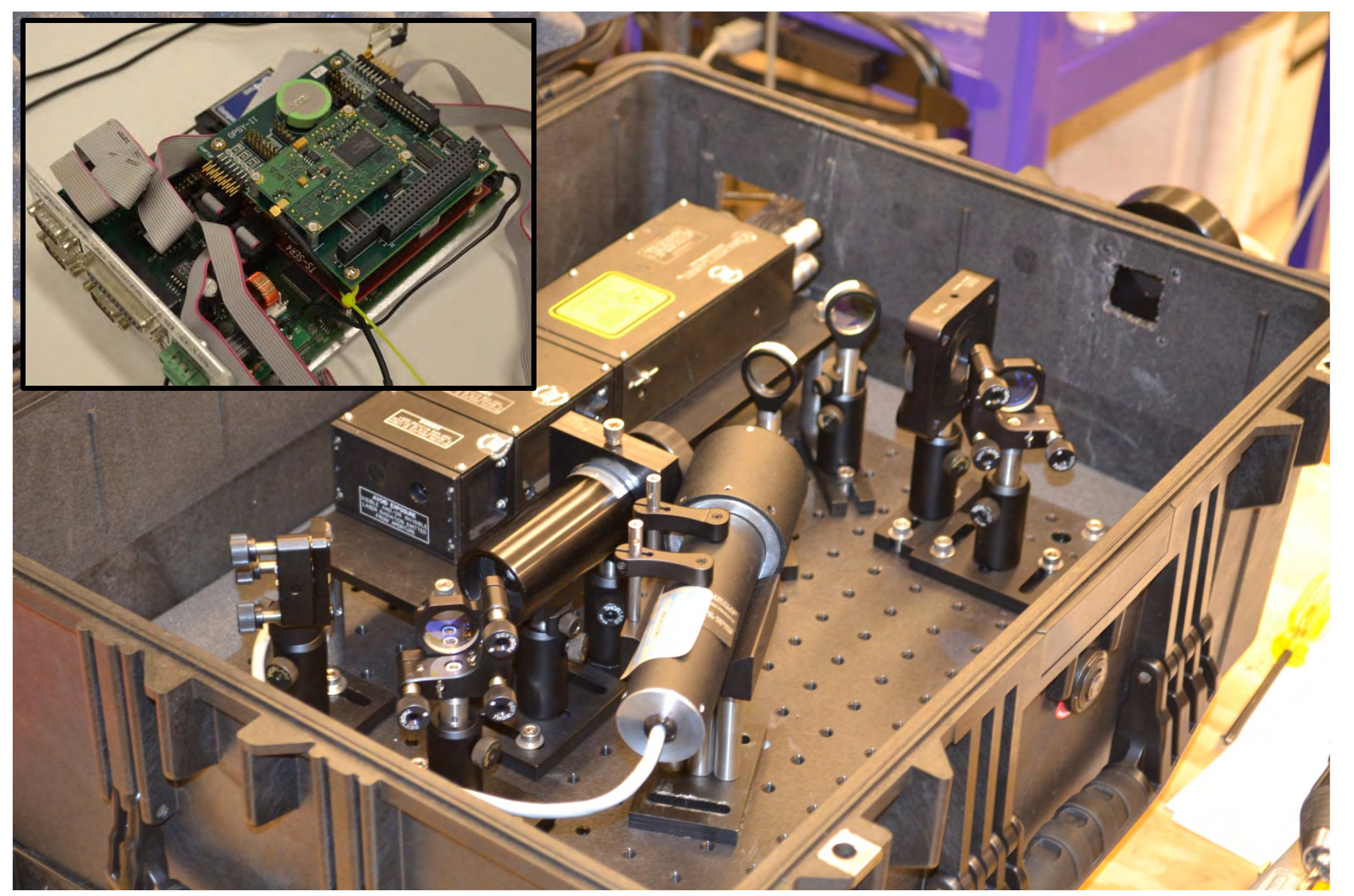}
  \caption{A portable laser system developed for the EUSO-Balloon tests. The insert shows the single board TS-5500 computer and GPSY timing unit.}
  \label{fig:gls-portable-photo}
 \end{figure}

\begin{figure}
  \centering
  \includegraphics[width=0.45\textwidth]{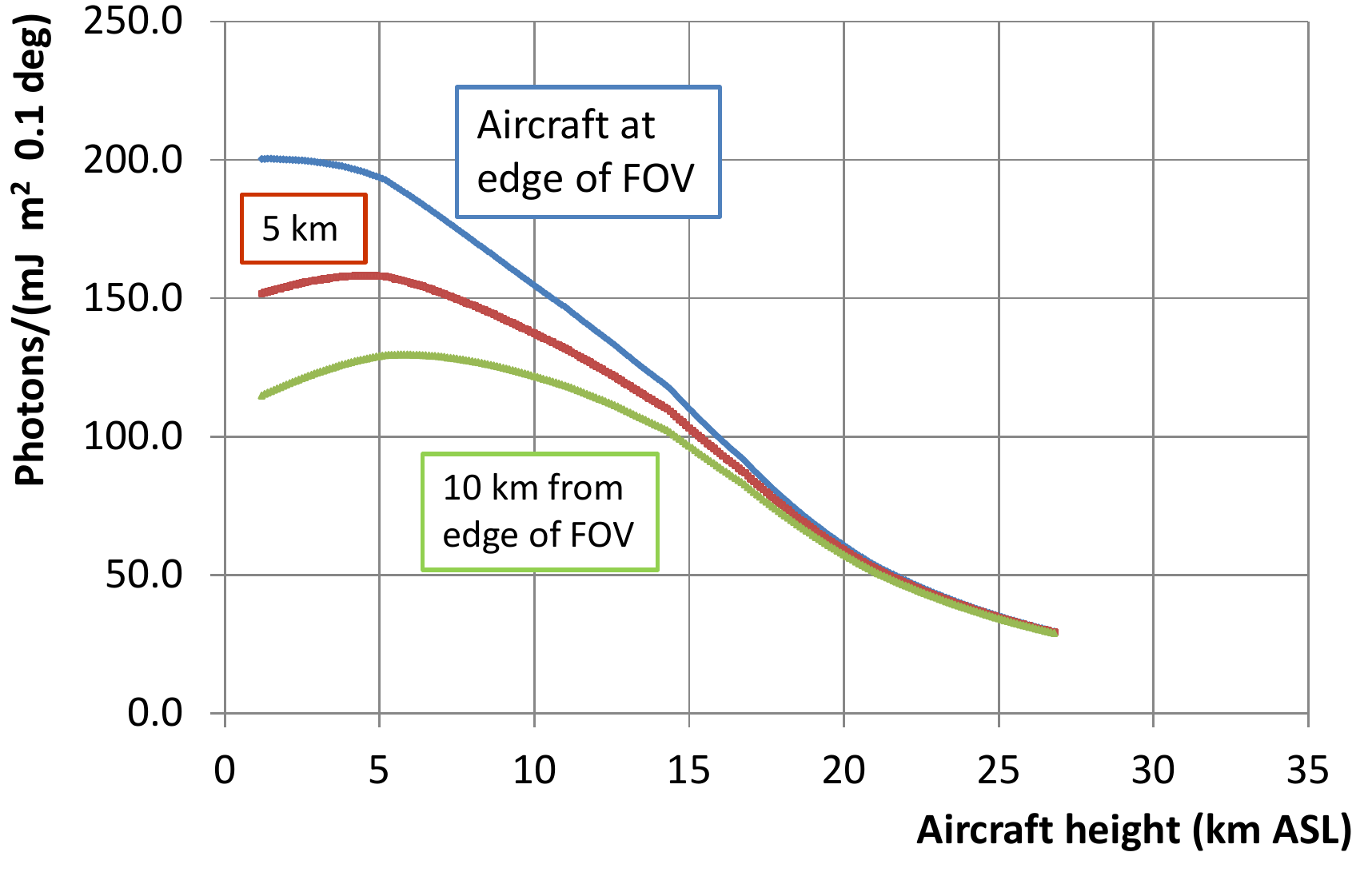}
  \caption{The light flux per mJ of laser energy reaching the expected 40 km altitude of EUSO-Balloon from a horizontal laser shot as a function of the height of the aircraft above sea level (ASL). The three curves correspond to different horizontal distances between the aircraft and the edge of the EUSO-Balloon field of view (FOV).}
  \label{fig:balloon-plot}
 \end{figure}

A prototype airborne GLS system will be deployed in an aircraft to support the suborbital
EUSO-Balloon mission \cite{EUSO-Balloon-Web, EUSO-Balloon-ICRC} (Fig. \ref{fig:balloon}).  The launch is planned for 2014 from Timmins Ontario. This mission sponsored by CNES is intended to be a full-scale end-to-end test of the JEM-EUSO proof of concept and technique, test the operation of key components, and measure the UV background below 40 km. Although the exposure will be limited to a flight of a few hours, EUSO-Balloon may image the first EAS looking down on the earth's atmosphere. To demonstrate the sensitivity to EASs, EUSO-Balloon will also measure flashes and tracks from the airborne GLS system. The laser system under development for this test is shown in figure \ref{fig:gls-portable-photo}. Since the balloon will  travel more slowly than the aircraft (unlike the ISS), 
the aircraft can fly multiple passes to test the instrument. The estimated light flux
for the planned horizontal laser shots reaching the 40 km elevation of the balloon is shown in figure \ref{fig:balloon-plot}. Due to a convenient compensation between scattering out of the beam and transmission 
between the beam and the balloon, the flux reaching the balloon is relatively insensitive to the altitude of the airplane when its flight
path is below 5 km.  

\section{Conclusions}
Understanding the high energy cosmic accelerators will require understanding the performance of the JEM-EUSO instrument while it orbits the earth measuring the cosmic messengers from these unknown sources. For this reason, the JEM-EUSO detector will also record a set of reference data interleaved with the cosmic measurements. This reference data will include UV flashes and tracks generated by the calibrated flashlamp and laser sources that will comprise the JEM-EUSO Global Light System. Design work and a search for sites is underway. A prototype portable GLS station is also being assembled to test the EUSO-Balloon JEM-EUSO prototype detector and will also be used for tests of the ground-based EUSO-TA prototype \cite{EUSO-TA-ICRC}.

\vspace*{0.3cm}
{
{\footnotesize{{\bf Acknowledgment: }{This work is supported by NASA grant \mbox{NNX13AH55G, NNX13AH53G} }

}}
\clearpage

%% file: icrc2013-1256.tex



\title{JEM-EUSO Design for Accommodation on the SpaceX Dragon Spacecraft}

\shorttitle{JEM EUSO on Dragon}

\authors{
J.H. Adams, Jr.$^{1}$,
R.M. Young$^{2}$,
A. Olinto$^{3}$
for the JEM-EUSO Collaboration.
}

\afiliations{
$^1$ University of Alabama in Huntsville \\
$^2$ NASA/Marshall Space Flight Center \\
$^3$ University of Chicago \\
}

\email{jim.adams@uah.edu} 

\abstract{The JEM-EUSO mission has been planned for launch on JAXAâ€™s H2 Launch Vehicle. Recently, the SpaceX Dragon spacecraft has emerged as an alternative payload carrier for JEM-EUSO. This paper discusses the accommodations that are available for JEM-EUSO in the Dragon Trunk and a concept for the re-design of JEM-EUSO so that it can be launched on Dragon.}


\maketitle

\section{Introduction}

To increase the launch opportunities for Extreme Universe Space Observatory on the Japanese Experiment Module (JEM-EUSO) \cite{bib:JEMEUSO}, we are designing a version of JEM-EUSO that can be accommodated in the Trunk section of the SpaceX Dragon Spacecraft. SpaceX began regular missions to deliver cargo to the International Space Station (ISS) in October 2012 \cite{bib:SpaceX}. In addition to delivering cargo to the ISS, NASA plans to use commercial services to deliver space crews as well. SpaceX's Dragon spacecraft is a contender for providing this service also.

On every launch Dragon carries a section, called the Trunk, which is an unpressurized cargo carrier. In this paper, the payload accommodations available for JEM-EUSO in the Trunk are described and a concept for how JEM-EUSO could be re-designed to fit in the Trunk is presented.

\section{The SpaceX Dragon System}

\begin{figure}
  \centering
  \includegraphics[width=0.48\textwidth]{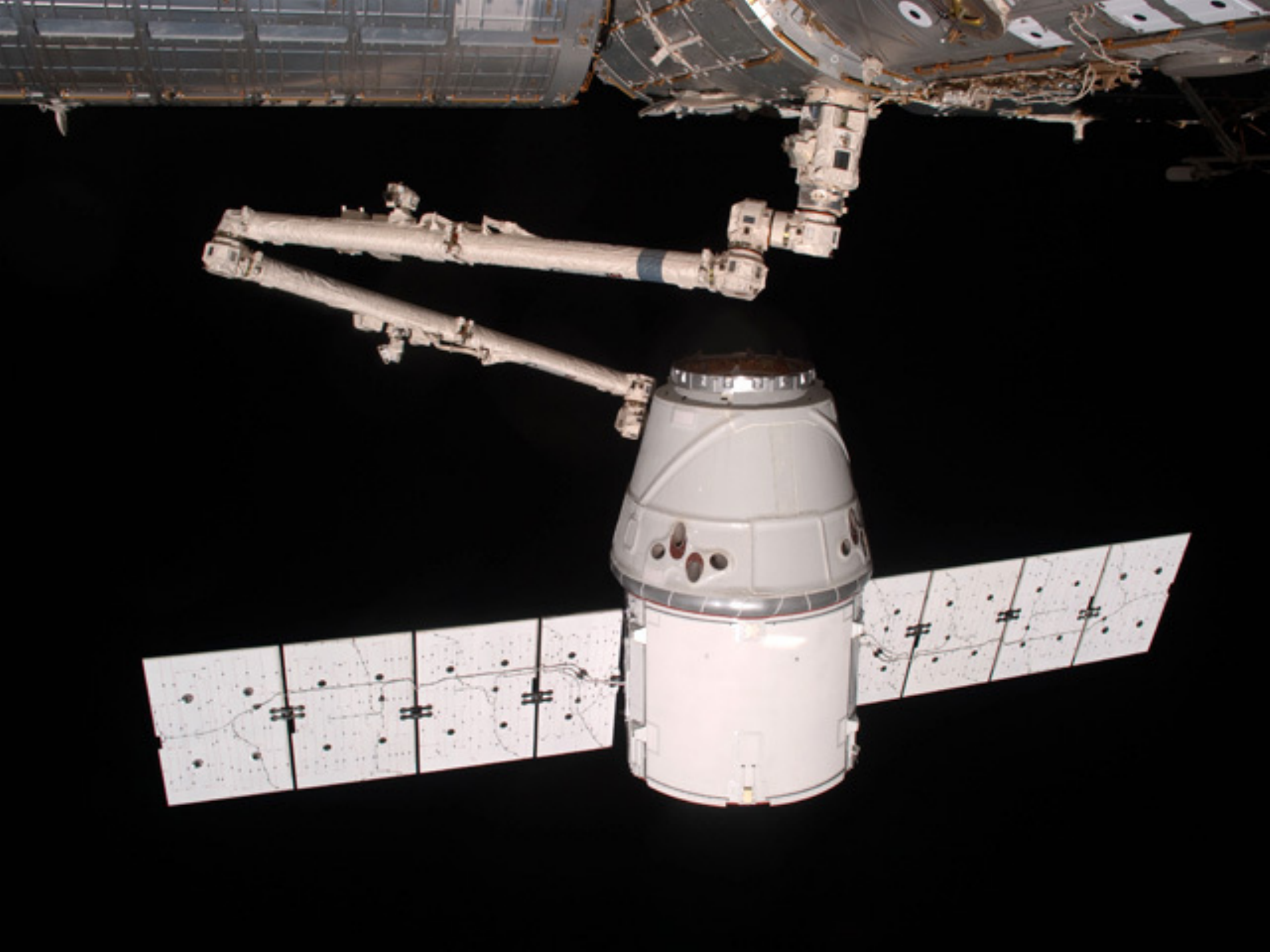}
  \caption{The SpaceX Dragon attached to the ISS Remote Manipulator Arm on May 25, 2012 (taken from \cite{bib:SpaceX}).}
  \label{Dragon_fig}
 \end{figure}

The Dragon Capsule is launched into orbit on SpaceX's Falcon 9. After achieving orbit, the Dragon (with the Trunk attached) proceeds to the ISS where it station-keeps until it is captured by the ISS Remote Manipulator Arm (RMS) as shown in figure \ref{Dragon_fig} which is taken from \cite{bib:SpaceX}. 

After docking, cargo intended for the interior of the ISS is removed from the pressurized Dragon module through the docking hatch. Instruments intended for mounting on the exterior of the ISS are carried in the unpressurized Trunk (the section with solar panels attached as shown in figure \ref{Dragon_fig}). The instrument is removed by the RMS and docked to a payload attachment point on the ISS. Old instruments may be installed in the trunk for disposal. After the mission to the ISS is complete, Dragon is undocked and it moves away from the ISS. Some time later it is deorbited. Before re-entering the atmosphere, the Trunk is jettisoned and it burns up in the atmosphere. The Dragon capsule returns, making a soft landing in the ocean.

The payload capacity of Dragon is 6000 kg which is divided between the pressurized and unpressurized cargo. The unpressurized cargo volume in the Trunk is 14 m$^{3}$.

\begin{figure}
  \centering
  \includegraphics[width=0.48\textwidth]{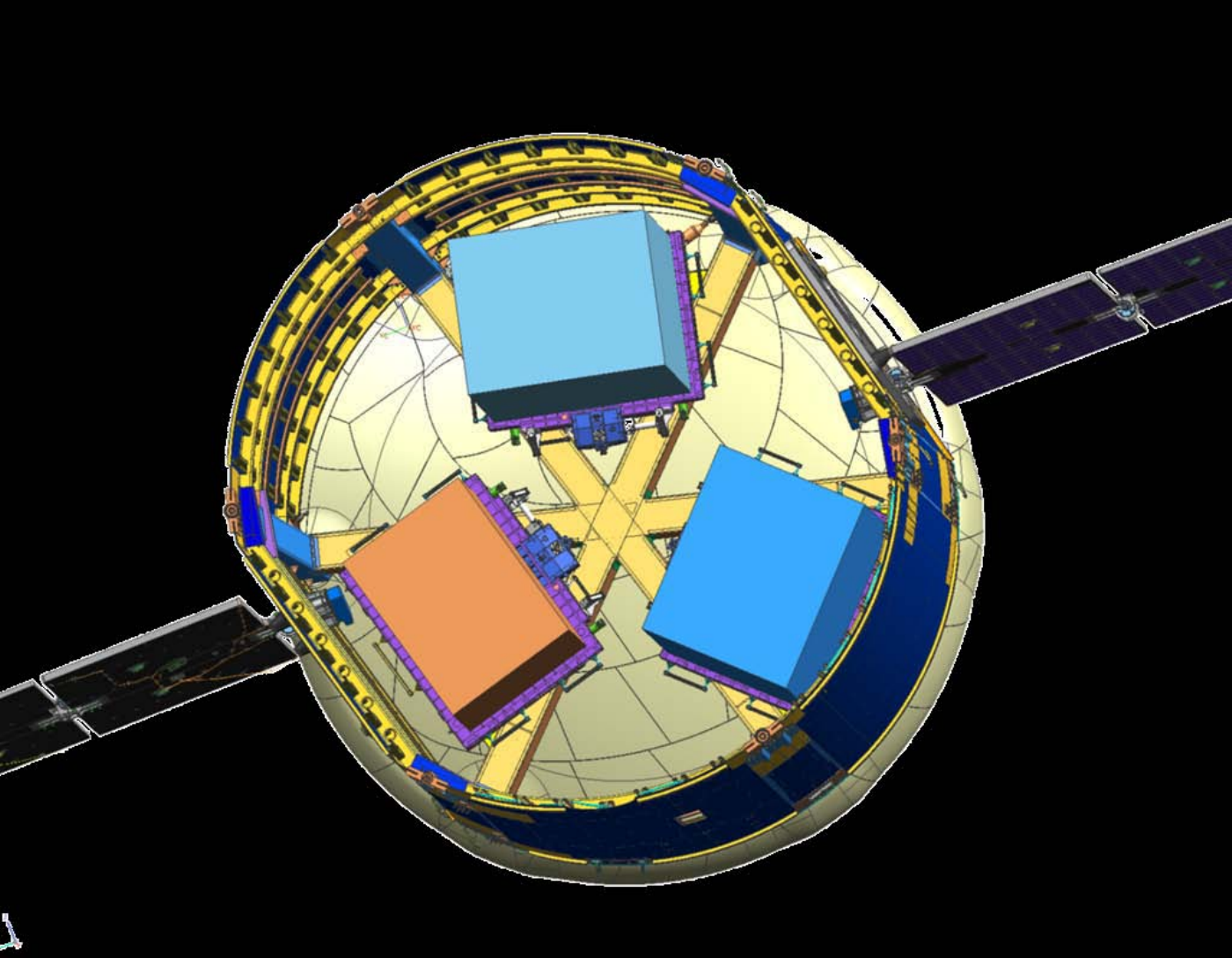}
  \caption{The Trunk showing three payloads attached in side (taken from \cite{bib:Trunk}).}
  \label{Trunk_fig}
\end{figure}

\section{Accommodations in the Trunk}

\begin{figure*}
  \centering
  \begin{minipage}{160mm}
  \includegraphics[width=160mm]{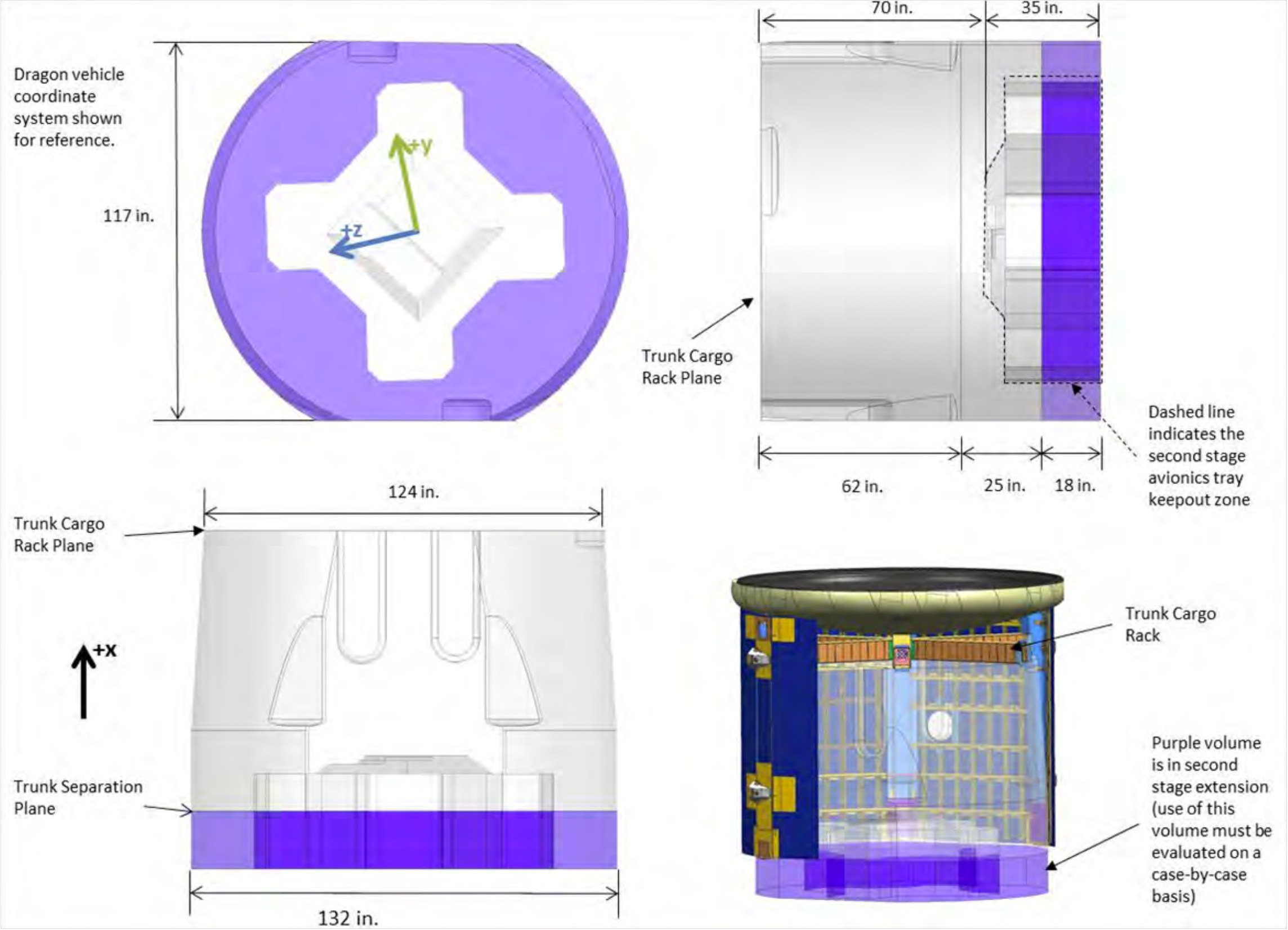}
  \caption{This figure shows the space available in the Dragon Trunk for payloads\cite{bib:IDD}.}
  \label{Accommodation_fig}
  \end{minipage}
 \end{figure*}

The space available in the trunk for payloads \cite{bib:IDD} is shown in figure \ref{Accommodation_fig}.

\subsection{Cargo Space in the Trunk}

As shown in figure \ref{Accommodation_fig}, the Trunk is cylindrical with tapered and straight sections. The diameter increases from 315 cm (124 inches) to 335 cm (132 inches) There are flat areas on the sides were the solar arrays are folded for launch. The resulting minimum diameter of the trunk is 297 cm (117 inches). The maximum length of the payload volume in the trunk is 221 cm (87 inches). The avionics tray on the second stage of the Falcon 9 extends up into the center of the trunk as shown in figure \ref{Accommodation_fig}. As a result, the length of the payload volume is reduced to 178 cm (70 inches). As shown, there may be some additional volume available in the in the second stage extension. This is an annular region 46 cm (18 inches) deep lying between the avionics tray and the inner surface of the second stage extension.

\subsection {Payload Attachment in the Trunk}

Payloads are attached under the Dragon module at the small end of the Trunk. The opposite end of the trunk is open once the second stage of the Falcon 9 separates after boosting the Dragon spacecraft into orbit. Figure \ref{Trunk_fig} shows an example of three payloads attached inside the Trunk.

\subsection {Release Mechanisms}

The payloads shown in Figure \ref{Trunk_fig} are attached using the Flight Releasable Attachment Mechanism (FRAM). A releasable attachment mechanism can provide data and power as well as thermal conduction and radiation pathways. JEM-EUSO is too heavy for the FRAM so a special releasable attachment mechanism must be designed. The interface definition for a releasable attachment mechanism in Dragon calls for the passive side of the mechanism to be attached to the framework on the Dragon end of the Trunk and the active side to be attached to the payload. This means that the signal to release JEM-EUSO from the Trunk must come through JEM-EUSO. This signal will be given through the ISS RMS as explained below.

\section{JEM-EUSO Redesign Concepts}

To accommodate JEM-EUSO in the Dragon Trunk several modifications to the JEM-EUSO design will be needed. In its design for HTV, JEM-EUSO attaches to HTVs Exposed Pallet at the entrance aperture end of the telescope. This means that the collapsing telescope tube structure must be strong enough to transmit the launch loads to the focal surface, which is the heaviest part of JEM-EUSO. Flying on Dragon makes it possible to attach JEM-EUSO inside the Trunk by its focal-surface end. This should result in a design for JEM-EUSO that saves weight.

In the HTV design it is necessary for the lenses and the focal surface to have flat sides. The maximum lens diameter had to be increased in this design to preserve the size of the entrance aperture. Because the Dragon Trunk has a nearly circular cross section, the lenses and focal surface in JEM-EUSO can be round and somewhat smaller.

\begin{figure}[h!]
  \centering
  \includegraphics[width=0.48\textwidth]{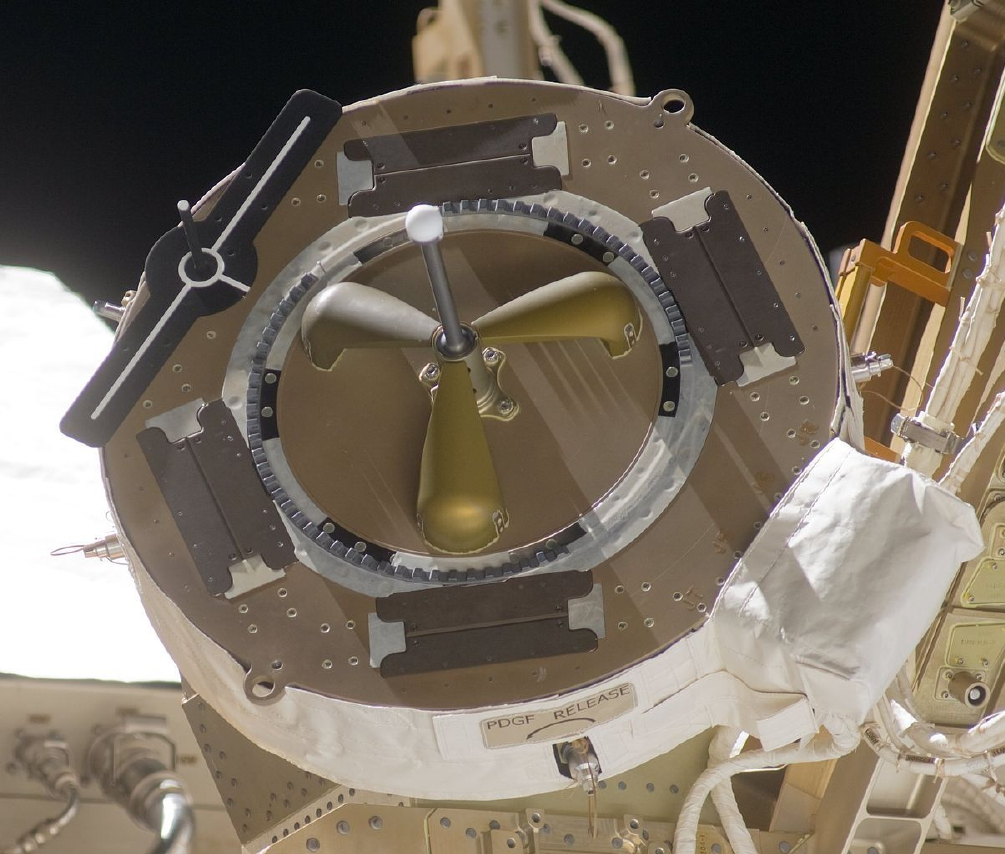}
  \caption{This figure shows a Power Video Grapple fixture which must be mounted on JEM-EUSO. It is used to release the payload from the Dragon Trunk and extract it from the Trunk (courtesy of NASA).}
  \label{PVGF_fig}
 \end{figure}

Because the active side of the releasable attachment mechanism is attached to JEM-EUSO, it will be necessary to use the Power Video Grapple fixture (PVGF) shown in Figure \ref{PVGF_fig}. This makes it possible to transmit the signal to release JEM-EUSO from the Dragon Trunk. This signal travels from the ISS RMS through the PVGF which is connected to the active side of releasable attachment mechanism on the other end of JEM-EUSO, releasing it from Dragon. The PVGF must be mounted at the open end of the Trunk so it is accessible to the ISS RMS.

\begin{figure}[h!]
  \centering
  \includegraphics[width=0.48\textwidth]{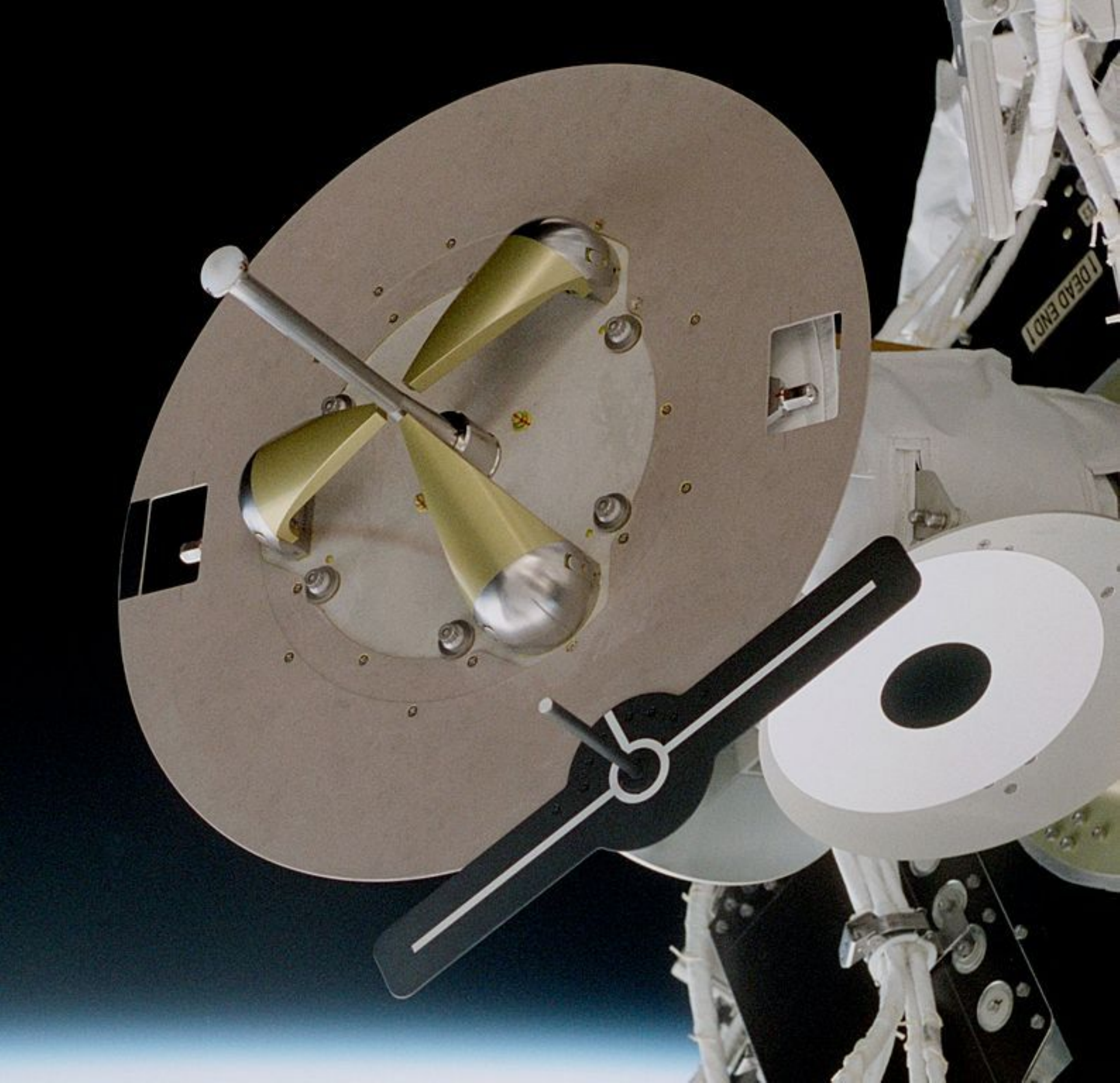}
  \caption{This figure shows a Flight Releasable Grapple fixture which must be mounted on the focal surface on JEM-EUSO. It is where the JEM RMS attaches during the handoff from the ISS RMS (courtesy of NASA).}
  \label{FRGF_fig}
 \end{figure}

Once removed from the Trunk, the ISS RMS will hand JEM-EUSO off to the RMS on the JEM. The JEM RMS will attach via a Flight Releasable Grapple Fixture (FRGF) shown in Figure \ref{FRGF_fig}. This fixture must be mounted on the focal surface end of JEM-EUSO near the Payload Interface Unit (PIU). The PIU is the device that attaches JEM-EUSO to the Exposed Facility (EF) on JEM. Mounting the FRGF near the PIU will permit the JEM RMS to exercise the control over JEM-EUSO that is needed to successfully mate the PIU with its matching part on the JEM EF.

\section{Summary}

We have presented a description of the accommodations available to JEM-EUSO in the Dragon Trunk. We conclude that JEM-EUSO can be accommodated on Dragon flights to the ISS. Using Dragon enables us to re-design JEM-EUSO in ways that should be beneficial. As pointed out above, there is reason to believe that JEM-EUSO, redesigned to fit in the Dragon Trunk, can be somewhat smaller and lighter while preserving its entrance aperture area and field of view.

\section{Acknowledgments}

The authors would like to thank Chi Min Chang, Peter Hasbrook and Steven Huning of NASA/JSC for their help with the accommodation study. We also want acknowledge support for our work from NASA Grant NNX13AH53G.

\clearpage

%% file: icrc2013-0343.tex



\title{Multi-Anode Photomultiplier Tube reliability analysis and radiation hardness assurance for the JEM-EUSO Space mission}

\shorttitle{JEM-EUSO ICRC 2013}

\authors{
H.Prieto-Alfonso$^{1}$,
K.Tsuno$^{2}$,
L.del Peral$^{1}$,
M.Casolino$^{2,3}$,
J.A. Morales de los Rios$^{1,2}$,
G. Saez-Cano$^{1}$,
T.Ebisuzaki$^{4}$ and 
M.D.Rodr\'iguez Fr\'ias $^{1}$
for the JEM-EUSO Collaboration.
}

\afiliations{
$^1$ SPace and AStroparticle (SPAS) Group. UAH, Madrid. Spain \\
$^2$ EUSO Team, RIKEN, Wako, Japan \\
$^3$ INFN Roma Tor Vergata, Roma, Italy\\
$^4$ Computational Astrophysics Laboratory. RIKEN, Wako, Japan\\
}

\email{hector.prietoa@uah.es} 

\abstract{Reliability assessment is concerned with the analysis of devices and systems whose individual components are prone to fail. This analysis documents the process and results of reliability determination of the JEM-EUSO PhotoMultiplier Tube (PMT) component under the Total Ionizing Dose (TIDs). In terms of TIDs, the PMTs that may fail due to this type of radiation is of the order of 246 PMT from a total amount of 4932 PMT, which cover the focal surface of the telescope. This means a reliability of around 95\%. However, the calculations show that the reliability of the "failing components", the remaining 5\% of the PMTs, is around 80\% in five years of operation of the JEM-EUSO Space Mission. Therefore, it can be concluded that around 99\% of the PMT's in terms of TIDs will complete their operation without failure, ensuring the success of the mission as far as radiation TIDs is concerned.}


\maketitle

\section{Introduction}

JEM-EUSO \cite{ijemeuso} is a large imaging telescope designed to study the Ultra-High Energy Cosmic Rays (UHECR) at energies above \textbf{$10^{20}$} eV. Looking downward the Earth from the International Space Station (ISS) it will detect such particles observing the UV light generated by Extensive Air Showers (EAS) the UHECRs develop in the atmosphere. The scientific objectives of the mission include charged particle astronomy and astrophysics, with the aim at extending the measurement of the energy spectra of the cosmic radiation beyond the Greisen-Zatsepin-Kuzmin (GZK) effect \cite{greisen}, together with the detection of Extremely High Energy Gamma Rays (EHEGR) and of Extremely High Energy Neutrinos (EHEN). 

JEM-EUSO is being designed to operate for 5 years on board the ISS orbiting in a Low Elevation Orbit (LEO) around the Earth at altitude of about 400 km. As for any mission to be operated in space, JEM-EUSO must comply to specific requirements, i.e. high radiation doses, unaccessibility and remote controlled operation. That is why the reliability analysis and radiation hardness assurance is extremely important in order to determine the tolerance and redundacy requirements within the system as previous studies \cite{Prieto, Prieto2} show in case of FPGA which also applies to PMTs.

The design and the construction of the JEM-EUSO telescope is a real technical challenge, as it involves the use of new technologies from the laboratories of both industrial and research in areas as diverse as large optical and accurate Fresnel lenses, a technique of photo detection highly sensitive with very accurate resolution, and very innovative analog and digital electronics as well.

\section{Objectives}

The main aim of this work is to determine the radiation hardness assurance to evaluate its present and potential reliability of the JEM-EUSO Photomultiplier Tube (PMT’s) implemented on the focal surface of the Space telescope, in order to ascertain the viability for this mission, since the PMT is a critical part of the instrumentation.

\section{Analysis}

PMTs have been used in the past from UV to near-IR photon detection. When used in combination with scintillation or Cherenkov materials, they can also detect more energetic ionising radiations. PMTs consist of a vacuum tube containing a cathode with a high photoelectric yield, and a series of dynodes with high secondary electron yield, each dynode biased to a steadily increasing potential before the anode is reached. The potential gradient ensures amplification with the multiplication of the number of electrons so that a single particle can release typically $10^6$ electrons which can be detected electronically \cite{ESAStandard,Hipparcos2}.

Background events can be induced in a PMT by one or more of the following mechanisms:

\begin{itemize}
\item Direct ionisation of the cathode or dynode by a particle producing secondary electrons.
\item Fluorescence, or more generally scintillation, in any optical components of the PMT (or instrument which are in line-of-sight of the photocathode) induced as a result of ionisation by an incident particle.
\item Cherenkov radiation induced in any optical components of the PMT (or instrument) from particles above the Cherenkov threshold for the material.
\end{itemize}

Previous space missions like HIPPARCOS measured background effects on PMTs due to Cherenkov and fluorescence processes from radiation- belt electrons, magnetospheric electron events and solar proton events \cite{Hipparcos1, Hipparcos2}

\section{Radiation Hardness Assurance}

The Radiation Hardness assurance of PMT's for their qualification requires meeting stringent radiation tolerance levels. The majority of radiation hardness assurance have so far focused on laboratory test, therefore, as a first step towards understanding the long-term reliability of PMT's in hostile radiation environments, it is required to perform an analytical and theoretical, probabilistic estimation which predicts the reliability of the PMT’s in space environments and radiation conditions for JEM-EUSO. Its focal surface is a spherical curved surface, of area 4.5 m$^2$ and it is covered with about 5,000 Multi-Anode PhotoMultiplier Tubes (MAPMT) Hamamatsu R11265-03-M64 MOD: MAPMT. The focal surface detector consists of Photo-Detector Modules (PDMs), each of which consists of 9 Elementary Cells (ECs). The EC implements 4 units of MAPMTs. Therefore, about 1,233 ECs or about 137 PDMs are arranged on the whole focal surface with 384,000 pixels \cite{purplebook}.

\subsection{Total Ionizing Dose Radiation Hardness Assurance Model}

TID is defined as the amount of energy deposited by ionisation or excitation in a material per unit mass of material. Since the dose is dependent on the target material, the dose is expressed in rad(Si). The components most sensitive to TID are active electronic devices such as transistors and integrated circuits (ICs). Their sensitivity thresholds tipically range from 1 krad(Si) to 1Mrad(Si) depending on the technologies used. Total Ionising dose (TID) degradation in microelectronics results from the build up of charge in insulating layers, and has a cumulative effect on electronics, resulting in a gradual loss of performance and eventual failure \cite{ESAStandard}.

Provided that the particle intensity and spectrum does not change significantly travelling through the material, TID can be determined from the charged particle fluence at the surface of the material, and the electronic stopping power of the particle based on the approximate formula:

\begin{normalsize}
\begin{equation}
 D = \frac{1}{\rho}\int^{E_2}_{E_1}\psi(E)\frac{dE}{dx}(E)\mathrm{dE}
\end{equation}
\end{normalsize}

where $\rho$ is the mass density of the material, $\psi(E)$ is the differential energy espectrum defined between E$_1$ and E$_2$, and dE/dx is the stopping power in units of energy loss per unit particle pathlength.

\begin{figure}[htb!]
 \begin{center}
\includegraphics[scale=0.41]{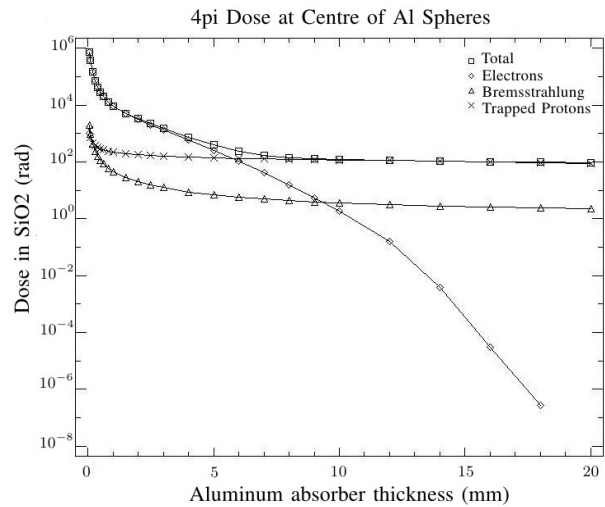} 
  \caption{Dose as a function of shielding \cite{spenvis}}
  \label{pmtTID}
   \end{center}
\end{figure}

The TID for the JEM-EUSO Mission was calculated using the SPENVIS computer software \cite{spenvis} as well as the orbital parameters of the international space station (Table \ref{tab:The International Space Station in Orbit (ISS)}). Appropriate parameter values for JEM-EUSO were collected and then used as input for SPENVIS. The basic parameters for the mission were the type of trajectory path, Mission Duration, Start data and mission Space Segments. According to the results shown by SPENVIS, and taking a 3 mm shielding for JEM-EUSO, the dose rate will be mostly due to trapped protons, and bremsstrahlung (Fig. \ref{PMTtmd}) for a total dose estimate of 10 krad (Fig. \ref{pmtTID}), which leads to conclude that the type of radiation and shielding is the appropriate.

\begin{table}[htb!]
\caption{the international Space station in orbit (ISS)\cite{iISS}}
\begin{small}
\begin{center}
\begin{tabular}{lc}
\hline
\textbf{Specs}&\textbf{Value}\\
\hline
Brightness &Approximately -4 (less than Venus)\\
\hline
Launch Window&5-10 min\\
\hline
Orbital Altitude& {361 km at Perigee - 437 km at apogee} \\ 
\hline
Mass & approx. 420000 kg\\
\hline
Dimensions & 111,08m by 89,2 m\\
\hline
Speed& approx. 2760 km/h\\
\hline
Orbital Inclination & $51,5947^{\circ}$\\
\hline
Orbital Period& approx. 90min\\
\hline
Observational Visibility & 60 N $\&$ 60 S\\
\hline
Orbital Type& Elliptical\\
\hline
\end{tabular}
\label{tab:The International Space Station in Orbit (ISS)}
\end{center}
\end{small}
\end{table} 

\begin{figure}[htb!]
 \begin{center}
\includegraphics[scale=0.40]{icrc2013-0343-PMTtmdrad.pdf} 
  \caption{SPENVIS PMT total mission dose \cite{spenvis}}
  \label{PMTtmd}
   \end{center}
\end{figure}

The reliability of a PMT under TID may be evaluated, according to \cite{TIDreliability}, taking into account that the probability of failure is equal to the probability that its TID exceeds its hardness. Then, when a component with a radiation hardness $H$, received a TID $x_D$ the reliability R(t) of the component may be calculated by the joint distribution function as:

\begin{equation}
R(t)=\int^{\infty}_0 \int^{\infty}_{x_H} f(x_D,x_H,t)\mathrm{dx}_H\mathrm{dx}_D
\end{equation}

where $f(x_D,x_H)$ is the joint distribution function, $x_D$ is the TID received by PMT in one year; $x_H=\frac{x_D}{H} \cdot t$, is the ratio between the TID received and the radiation hardness of the PMT; and, t is the exposure time to the radiation in years.

LEO refers to orbits in the 100-1,000 km altitude range, which includes Earth-Observing Satellites (EOS). A special case is the Space Station (ISS) at $\sim400$ km. The environment in LEO is fairly benign, with a typical dose rate of $x_D=0.1$ krad/year.

For a mission with a typical duration of 3-5 years, the total dose is $<$ 0.5 krad \cite{leo}. Hence, taking into account a radiation hardness for JEM-EUSO of $H=10$ krad \cite{purplebook}, the PMT time-dependent reliability is as follows:
   
\begin{eqnarray}
R(t) & = \int^{\infty}_0 e^{-x_D}\int^{\infty}_{x_H} e^{-x_H}\mathrm{dx}_H\mathrm{dx}_D=e^{-x_H} \nonumber \\
& = e ^{\left(-\frac{0.1}{10krad}\cdot t [yr]\right)}
\label{tid}
\end{eqnarray} 

\begin{center}
\begin{figure}[htb!]
  \centering
\includegraphics[scale=0.24]{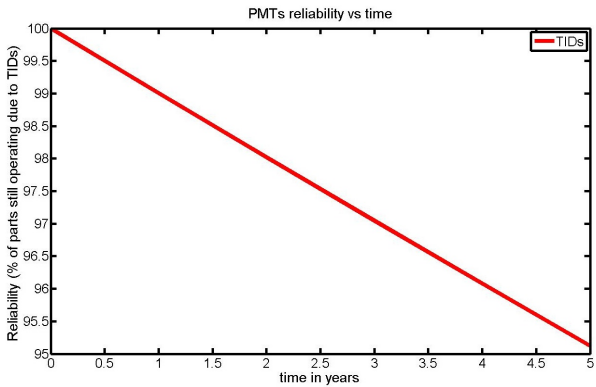}  
  \caption{Reliability vs time of JEM-EUSO PMTs due to TIDs according to the TID radiation hardness assurance model}
  \label{relitid1}
\end{figure}
\end{center}

An approach to comprise the reliability estimation due to TIDs is analyzing its behavior over time. In any case, considering JEM-EUSO time mission is 5 years and according to equation \ref{tid}, the PMTs reliability is around 95\% in case of one PMT. This behaviour is shown in Figure \ref{relitid1}. 

\begin{center}
\begin{figure}[htb!]
  \centering
\includegraphics[scale=0.27]{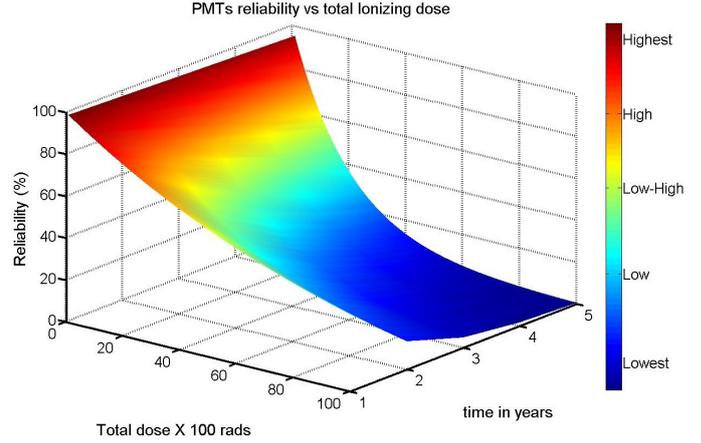} 
  \caption{3D view of the obtained reliability vs time due to TIDs of JEM-EUSO PMTs }
  \label{relitid2}
\end{figure}
\end{center}

If we assume higher (TIDs) radiation levels, obviously, the reliability of the PMTs will have a considerable decrease. Taking a look at  Figure \ref{relitid2} assuming that the radiation be now $H=20$ krad total dose, the reliability range would roughly fall between a maximum of 80\% and a minimum of 70\%.

\begin{center}
\begin{figure}[htb!]
  \centering
\includegraphics[scale=0.26]{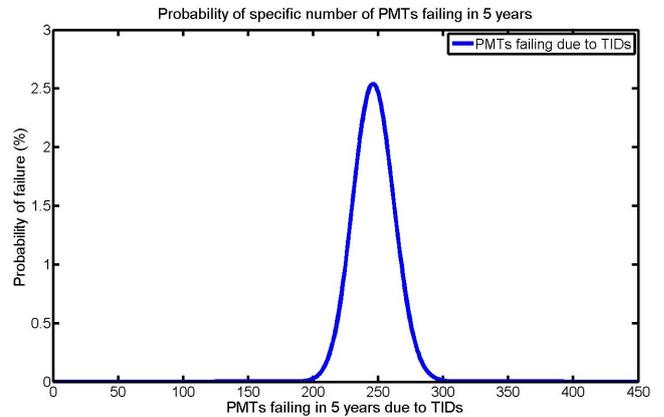}
  \caption{Probability of exactly PMT failures in 5 years}
  \label{relitid3}
\end{figure}
\end{center}

The estimation obtained by applying the total ionizing dose radiation hardness assurance model provides the reliability for a single PMT. Hence, to determine the reliability of all PMTs that will be used in the JEM-EUSO telescope focal surface, it is necessary to apply the Poisson distribution.This is shown in Figure \ref{relitid3}. In this case, for a total amount of 4932 PMTs, 246 are expected to fail with a probability of 2.5\% of getting that specific number of failure. It does suggest that the PMT designed for JEM-EUSO is robust and highly reliable against the influence of TIDs.

\begin{center}
\begin{figure}[htb!]
  \centering
\includegraphics[scale=0.26]{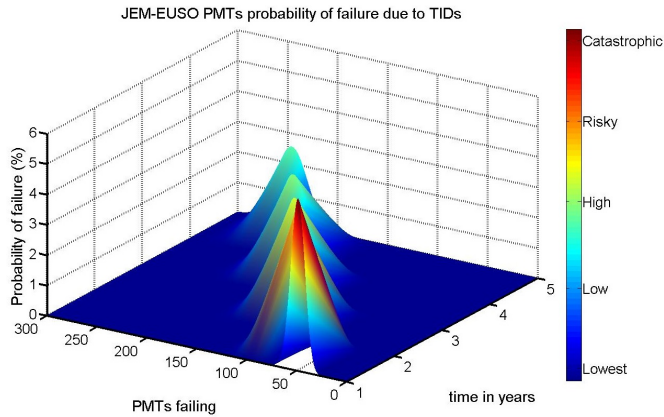} 
  \caption{3D view of reliability vs time due to TIDs according to the Total Ionizing Dose Radiation hardness assurance model}
  \label{relitid4}
\end{figure}
\end{center}

Figure \ref{relitid4}, shows the exact probability number of failures over time.The behavior of the PMTs under TIDs is quite similar during the mission duration. The probability of having 246 PMTs (which are prone to fail) failing over time is significantly low.

\begin{center}
\begin{figure}[htb!]
  \centering
\includegraphics[scale=0.25]{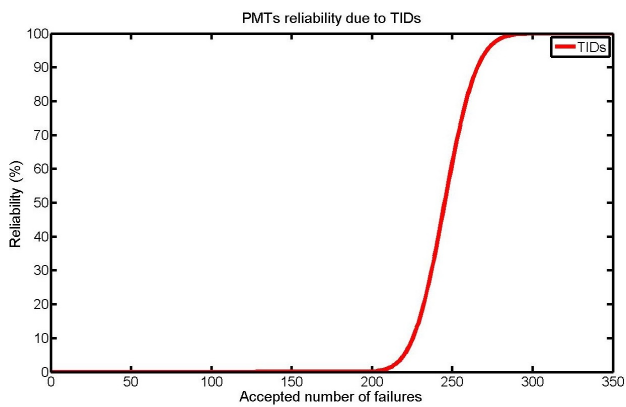}
  \caption{Accepted number of failures of PMTs under the influence of TIDs} 
  \label{reli}
\end{figure}
\end{center}

As a final step in determining the reliability of the JEM-EUSO PMTs under TIDs influence, it is necessary to know the reliability of the components that are expected to fail. Following the Poisson cumulative  distribution, this analysis has been carried out and the result is shown in Figure \ref{reli}. This plot can be considered as a further confirmation that JEM-EUSO PMTs meet specific requirements for Space environment. It also shows that concerning the 246 PMTs that are "expected" to fail, their reliability is around 80\%. 

\section{Conclusions}

The reliability of a PMT to be operated for space applications is closely related to the exposure it may have against Total Ionizing Dose. A method to calculate the reliability and radiation hardness assurance of PMTs under the effects of TIDs has been presented. This technique introduced a model which estimates the effects of the accumulation of TIDs in the PMTs during the time in which it is operating in Space.

In terms of Total Ionizing Dose (TIDs), the number of PMTs that may fail due to this type of radiation is of the order of 246. However, the calculations show that the reliability of these components that could fail is around 80\% in five years of operation. Therefore, it is reasonable to conclude that around 99\% of the PMT's in terms of radiation TIDs will complete their operation without interruption, ensuring the success of the mission as far as regards radiation TIDs.

\section*{Acknowledgments}

The JEM-EUSO Space Mission is funded in Spain under projects AYA2009-06037-E/AYA, AYA-ESP2010-19082 and the coordinated projects AYA-ESP2011-29489-C03 and AYA-ESP2012-39115-C03. We acknowledge CAM ASTROMADRID S2009/ESP-1496 and MICINN MULTIDARK CSD2009-00064 Projects for their finantial support to JEM-EUSO as well. H.Prieto-Alfonso wants to acknowledge the support from International Program Associates (IPAs),which granted his one year stay at the RIKEN institute, Wako-shi, Japan.

\clearpage

%% file: icrc2013-0432.tex



\title{Second level trigger and Cluster Control Board for the JEM-EUSO mission}

\shorttitle{J. Bayer \etal The Cluster Control Board for JEM-EUSO}

\authors{
	J\"org Bayer$^{1}$,
	Giuseppe Distratis$^{1}$,
	Daniel Gottschall$^{1}$,
	Andrea Santangelo$^{1}$,
	Christoph Tenzer$^{1}$
	Mario Bertaina$^{2,3}$,
	Marco Casolino$^{4,5,6}$,
	Giuseppe Osteria$^{7}$
	for the JEM-EUSO Collaboration.
	}

\afiliations{
$^1$ Institute for Astronomy and Astrophysics, Kepler Center, University of T\"ubingen, Germany\\
$^2$ Dipartimento di Fisica, Universita' di Torino, Italy\\
$^4$ RIKEN Advanced Science Institute, Wako, Japan\\
$^7$ Istituto Nazionale di Fisica Nucleare - Sezione di Napoli, Italy\\
\scriptsize{
$^{3}$ Istituto Nazionale di Fisica Nucleare - Sezione di Torino, Italy\\
$^{5}$ Istituto Nazionale di Fisica Nucleare - Sezione di Roma Tor Vergata, Italy\\
$^{6}$ Universita' di Roma Tor Vergata - Dipartimento di Fisica, Roma, Italy
}
}

\email{bayer@astro.uni-tuebingen.de}

\abstract{The Cluster Control Board (CCB) is one of the key elements of the JEM-EUSO read-out
electronics and manages the data received from nine Photo Detector
Modules (PDM). To reduce the large amount of data produced at the detector level
and to discriminate good events associated to Extensive Air Shower (EAS) from the spurious
events, a hierarchical trigger scheme over two levels has been developed.
The first trigger level 'L1' is implemented in the PDM electronics and the
second trigger level 'L2' in the CCB electronics. After the
processing of the data, potentially good events are transmitted to the
onboard CPU. In this paper, we will first present the algorithm developed, focusing
on its implementation in hardware. The algorithm aims at
distinguishing the unique patterns produced on the focal surface by the EAS from the
ones produced by background events. It is based on the scan of a predefined set of
directions, which covers the complete parameter space. To fulfill the requirement on the
processing time, the algorithm was optimized and implemented in a Field Programmable Gate Array (FPGA) in
order to make use of its parallel processing capabilities.
A prototype board has been produced and its functionality was validated with a laboratory
test setup. Furthermore, a dedicated version of the CCB was developed and produced for the
JEM-EUSO pathfinder missions TA-EUSO and EUSO-Balloon. Again, the CCB proved its
functionality during several integration campaigns with the other parts of the read-out
electronics and it was possible to set up the complete read-out chain of the detector.

After presenting the current architecture of the CCB and discussing the complex interfaces
with the other elements of the read-out electronics, we will report on the performance of
the prototype boards.}


\maketitle

\section{Introduction}
\label{sec:Introduction}
The planned Extreme Universe Space Observatory (EUSO) - attached to the Japanese
Experiment Module (JEM) of the International Space Station (ISS) -
is a large Ultra Violet (UV) telescope to investigate the nature and origin of
the Ultra High Energy Cosmic Rays (UHECRs) by observing the fluorescence light
produced in Extensive Air Showers (EAS).

The main instrument of JEM-EUSO is a super-wide $\pm 30^{\circ}$ Field of View (FoV)
telescope, which will be able to trace the fluorescence tracks generated by the
primary particles with a timing resolution of 2.5 $\mu$s and a spatial resolution
of $0.07^{\circ}$ (corresponding to about 550 m on ground). This allows to reconstruct the
incoming direction of the UHECR with an accuracy better than a few degrees \cite{takahashi}.

As the electronics has to handle over $3.15\cdot 10^5$ pixels, the Focal Surface (FS) has
been partitioned into subsections - the Photo-Detector Modules (PDMs) - and a
multi-level trigger scheme has been developed.
The first trigger level (L1) consists of three sub-levels implemented within
the front-end Application Specific Integrated Circuit (ASIC) \cite{ahmmad} and the PDM electronics.
The 'second level' trigger (L2) is implemented in the CCB electronics and will be described
in the following section.

To demonstrate the technology and to prove the detection technique, two pathfinder
instruments are being developed by the JEM-EUSO collaboration; TA-EUSO and EUSO-Balloon \cite{gcasolino, ballmoos, moretto}.
Both instruments are scaled down versions of the JEM-EUSO detector but inherit all key
elements of the electronics - e.g. one PDM and one CCB. For this purpose, 
a dedicated version of the CCB was developed and produced.

\section{L2 Trigger Algorithm}
\label{sec:L2TriggerAlgorithm}

An EAS will produce fluorescence light by exciting nitrogen molecules on its
ultra relativistic journey through the atmosphere. 
The amount of light is proportional to the number of secondary particles produced
and increases until the shower reaches its maximum.
Projected onto the focal surface of the detector, this corresponds to a
spot moving on a straight line from frame to frame.
Due to the projection, its speed and direction 
depends on the incoming direction of the UHECR
and the exposure time of the detector - called Gate Time Unit (GTU).
The principle of the L2 trigger algorithm is therefore trying to follow the spot
to distinguish these events from the background.

Technically, this is done by integrating the photon counting values along the
track over some predefined time. This value is then compared to some threshold
above the background and an L2 trigger is generated if the threshold is exceeded.
As the incoming direction of the UHECR is unknown, the approach is to use a set
of directions for the integration which covers the complete parameter space.
The number of directions was optimized to comply with constraints
in the available computing power - e.g. power-, weight-, and size-requirements
and the processing time of the L2 algorithm.

Currently implemented is a total of 375 starting points for the integration,
which are distributed equally over time and position around the trigger seed
(L1 trigger from the PDM, see also \cite{park2011}).
For each starting point, the integration will be performed for a box of 3x3 pixels,
over $\pm$7 GTUs and for 67 directions. Details on the optimization process
and the L2 algorithm can be found in \cite{thesis_bayer, bayer2011}.

\section{Cluster Control Board}
\label{sec:ClusterControlBoard}

\subsection{Overview}
\label{subsec:CCBOverview}

\begin{figure}
	\centering
	\includegraphics[width=0.44\textwidth, keepaspectratio=true]{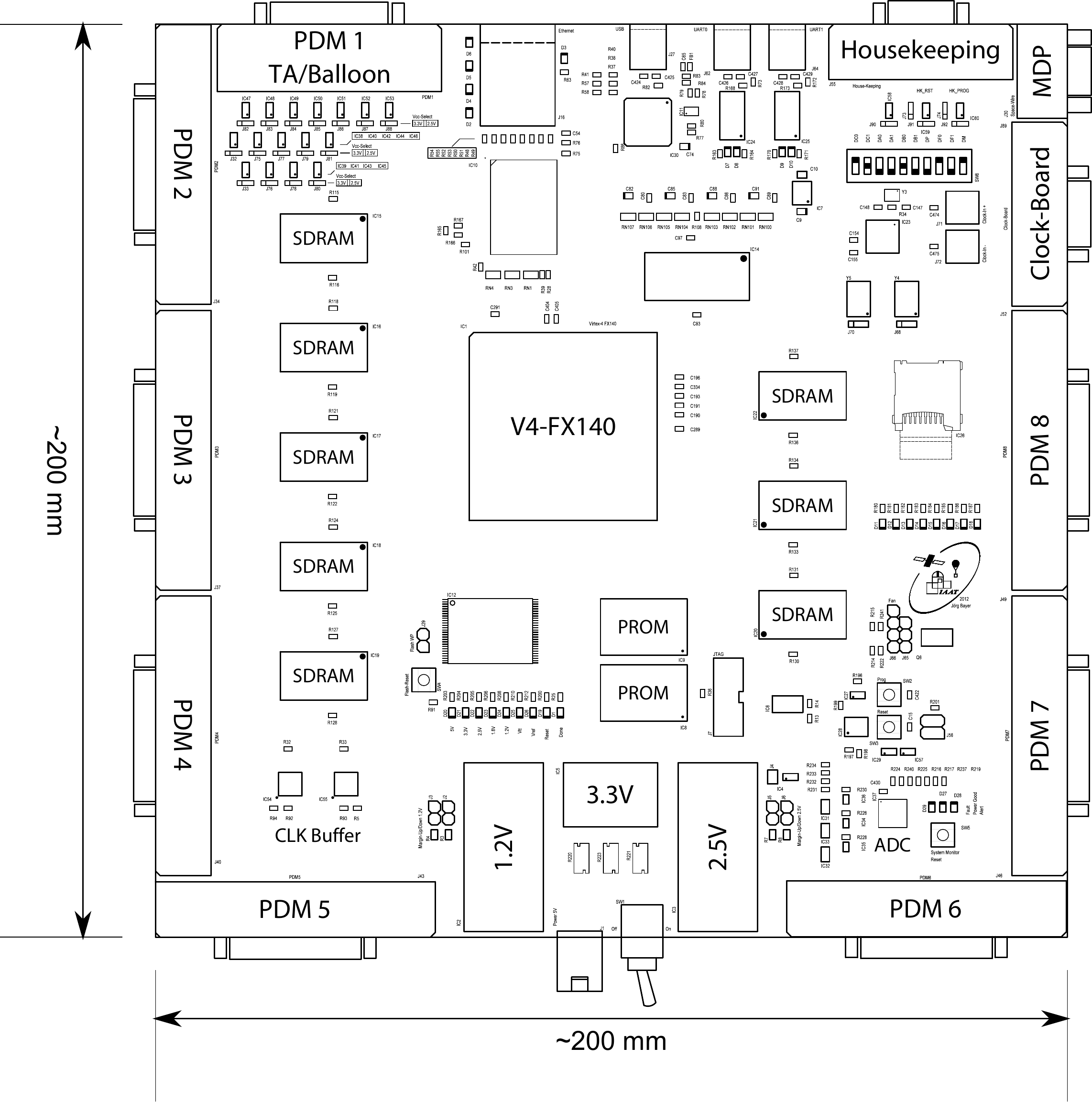}
	\caption{Layout of the JEM-EUSO prototype CCB.
		The 'heart' of the CCB is a V4-FX140 FPGA from Xilinx which is surrounded
		by Static Random-Access Memory (SRAM) modules and the connectors for the various
		interfaces.}
	\label{pic:CCB_schematic}
\end{figure}

\begin{figure}
	\centering
	\includegraphics[width=0.42\textwidth, keepaspectratio=true]{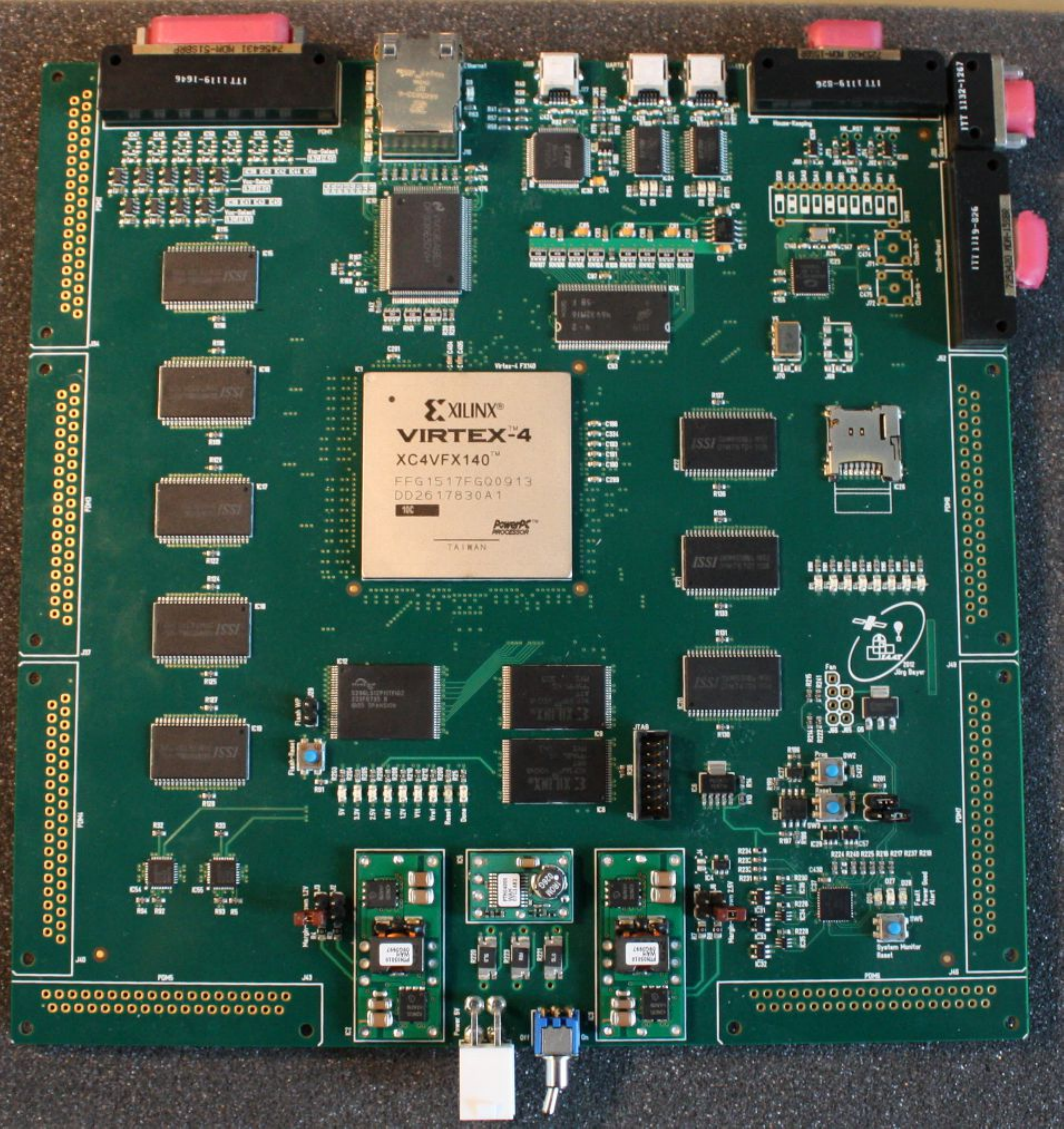}
	\caption{Picture of the produced and assembled JEM-EUSO prototype CCB (only one PDM
						connector is mounted).}
	\label{pic:CCB_PCB}
\end{figure}

A schematic drawing of the layout of the CCB prototype is given in Figure \ref{pic:CCB_schematic}
and a picture of the produced board in Figure \ref{pic:CCB_PCB}.
The main parts of the CCB are the FPGA, the external memories, and the connectors
for the various interfaces.
To comply with the requirement for space-qualified electronics, a Virtex 4-FX140 was
selected as it is available in a radiation tolerant version (see \cite{xilinx}) and due to its high number
of input-/output-pins. This high number is needed to interface nine PDMs, the Mission Data Processor (MDP),
the Housekeeping-System (HK) and the Clock-Board (see also Section \ref{subsec:CCBInterfaces}).
The external memory is necessary to buffer the large amount of PDM data while
the L2 calculation is performed. The prototype board contains a few changes with respect
to the baseline flight-model. One PDM connector was removed to free up some I/O pins which
are used for developing and debugging purposes. Furthermore, another PDM connector
was extended in the number of communication lines to have a few additional 'spare' lines
during the TA-EUSO and EUSO-Balloon integration phase. Other components to be mentioned
are two clock buffers which distribute the system clocks to the PDMs, an Analog-to-Digital-Converter (ADC)
for the Housekeeping, two Programmable Read-Only Memories (PROMs) which hold the
configuration data of the FPGA and the power-supply.

\begin{figure}
	\centering
	\includegraphics[width=0.45\textwidth, keepaspectratio=true]{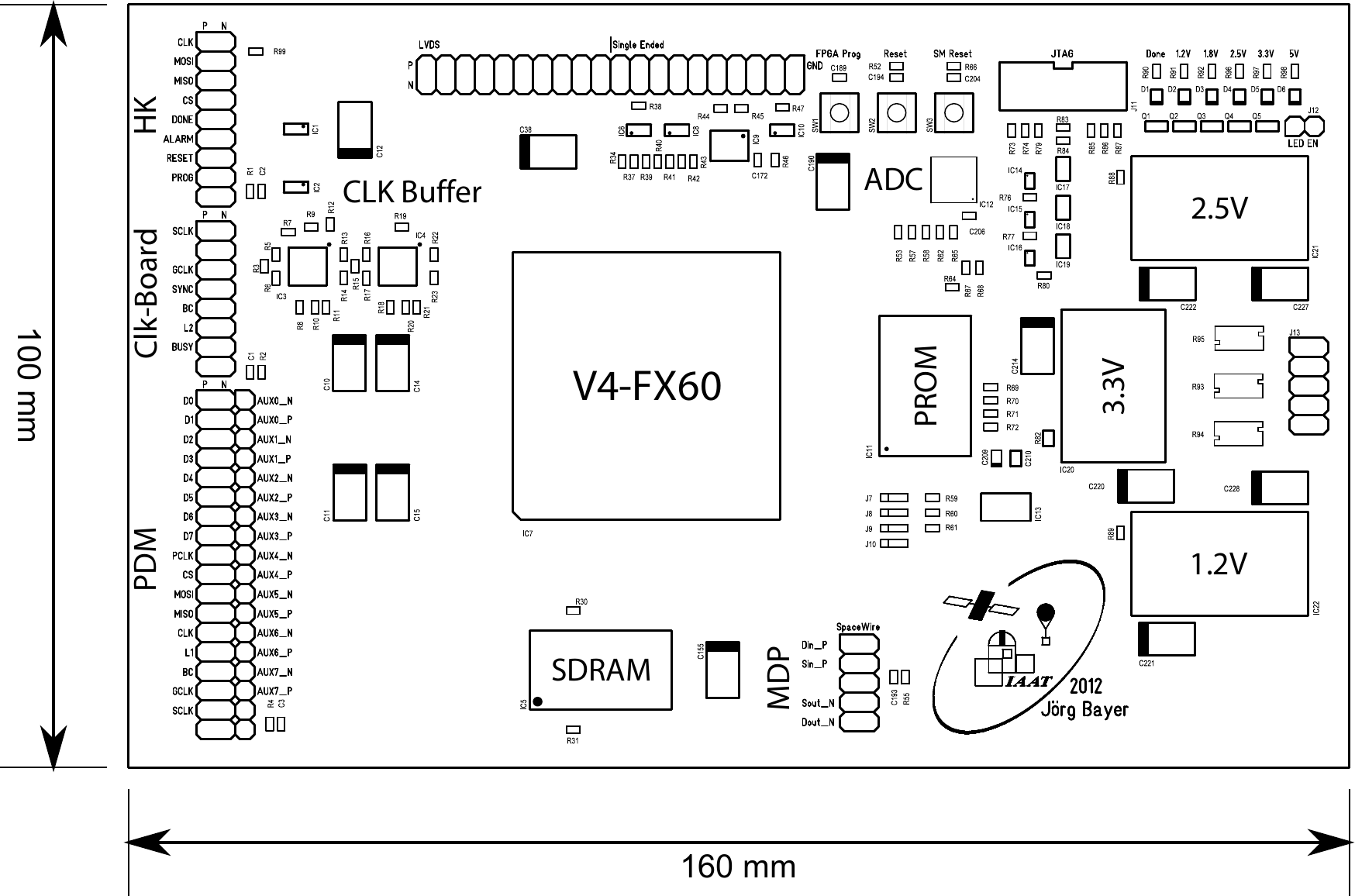}
	\caption{Layout of the TA-EUSO/EUSO-Balloon CCB.
		This is a smaller version of the JEM-EUSO CCB and designed to interface only one PDM.
		Therefore, a smaller member of the Virtex-4 family is used - a V4-FX60 - and
		only one SRAM.
	}
	\label{pic:CCB_Balloon_schematic}
\end{figure}

The layout of the CCB for TA-EUSO and EUSO-Balloon is given in Figure
\ref{pic:CCB_Balloon_schematic}. Since it interfaces only one PDM,
this board can bee seen as a scaled down version of the JEM-EUSO CCB.
To reduce power, weight, and size, it is built around a smaller FPGA of the
same family (V4-FX60), as these projects will act as a proof of technology for JEM-EUSO.
This applies also to the other components: the same SRAM and PROM (only one), ADC,
clock buffers, and a smaller power supply with the same technology are used
on this version of the CCB.

\subsection{Interfaces}
\label{subsec:CCBInterfaces}
The following section will give a brief overview on the various interfaces of
the CCB and their purpose. All interfaces are realized with the Low Voltage Differential Signaling
(LVDS) standard which adds more reliability, but doubles the number of lines needed.

\paragraph{PDM}
\label{sec:PDM_interface}
The interface between PDM and CCB for the scientific data from the detector
and for controlling/monitoring the PDMs is a critical part in the processing
chain as many other parts rely on the performance of this interface (e.g. the
ring buffer on the PDMs, the implementation of the L2 algorithm and the dead-time
of the instrument). In order to reduce the dead-time of the instrument, the event
data (around 2.7 Mbit per PDM and event) has to be transferred as fast as possible
within the hardware constraints.

The current baseline is a 8-bit wide, source synchronous data bus working at 40 MHz
for the event data and a standard Serial Peripheral Interface (SPI) for commands, configuration
and to retrieve the status of the PDM.

After receiving an L1- or an external-trigger, the CCB will broadcast the
trigger to all connected PDMs and their ring buffers will be frozen.
If a particular PDM is ready for sending the data, it will change its status register and the download
will be started by the CCB by sending a 'data request' command over the command interface.

The data package is composed of an 'Event Summary' (which contains the trigger seed for the L2)
followed by 128 GTU data frames. Each GTU data frame contains the photon counting values from the
whole PDM (48x48 pixels) and the data from the charge integration module of the ASIC (KI data).
After all data is sent to the CCB, the PDM will arm its ring-buffers again for the next run.

\paragraph{MDP}
\label{sec:MDP_interface}
The communication with the MDP has been realized with a SpaceWire interface due to its
approved reliability. In the current baseline, the interface runs at 200 MHz, which
translates into a data rate of 200 Mbit/sec. 
The complete data traffic between the MDP and the CCB takes place over this interface
in form of standardized data packets called 'Messages'. All messages leaving the CCB
are provided with a 4-byte large CRC in order to allow the CPU to assure their data
integrity.

\paragraph{HK}
\label{sec:HK_interface}
The communication to the HK-System is established over an SPI, where the CCB acts as slave.
The CCB provides several registers which will be read by the HK-System on a regular basis.
Data provided by the CCB includes voltages, currents and temperatures as well as status
registers from the CCB and the PDM.

\paragraph{Clock-Board}
\label{sec:Clock_Board_interface}
The Clock-Board supplies the CCB with the system clock and the GTU clock which
are both distributed to the PDM. In addition the external trigger will be received
from, and the L2 trigger sent to the Clock-Board. A synchronization signal will be
used to assign an absolute time-stamp to the events.

\subsection{Architecture}
\label{subsec:CCBArchitecture}

\begin{figure*}
	\centering
	\includegraphics[width=0.9\textwidth, keepaspectratio=true]{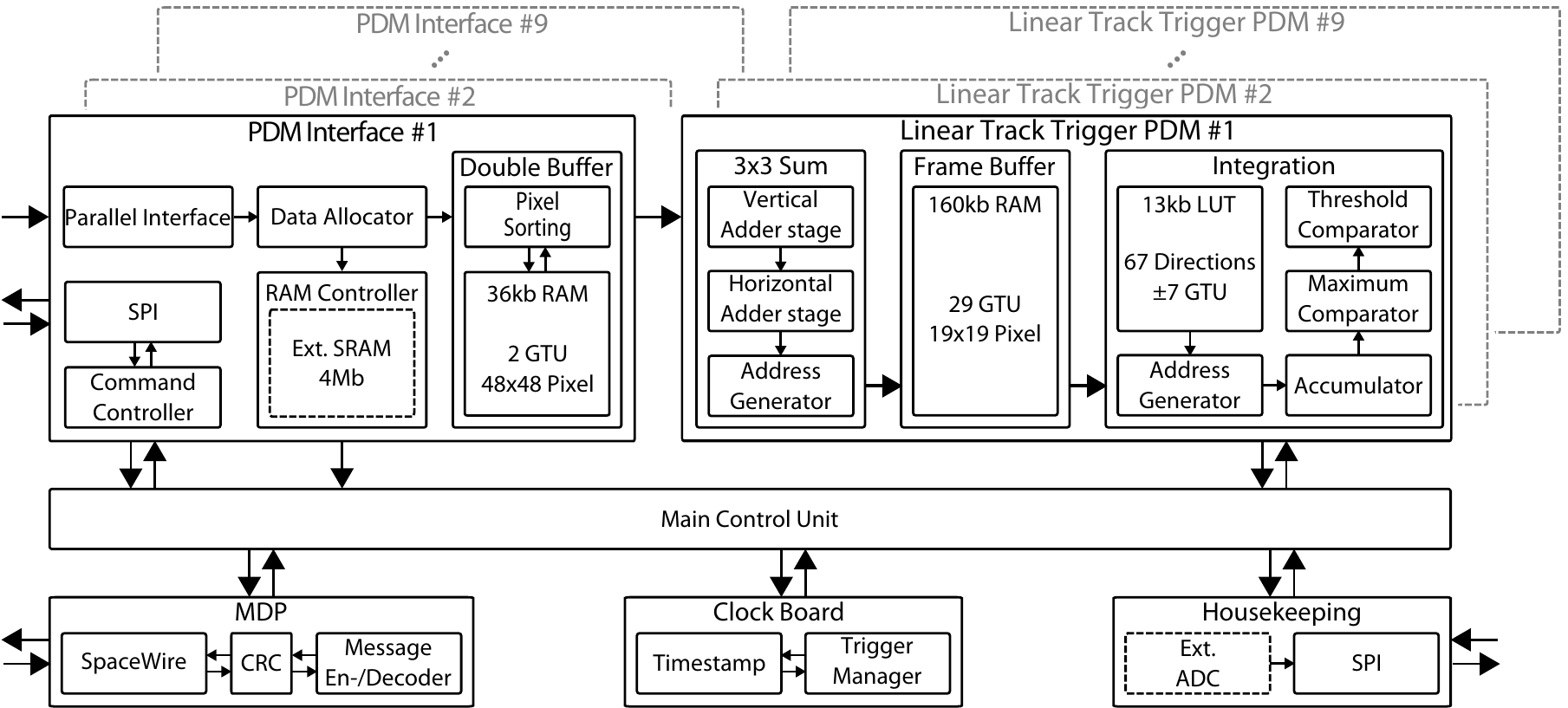}
	\caption{The current block diagram of the CCB FPGA.
		Most of the logic resources are needed for the nine 'Linear Track Trigger'
		modules which perform the L2 trigger algorithm and handle the data transfer
		from the PDM. As it is not possible to hold the data of all nine PDMs
		inside the FPGA RAM an interface to an external memory is needed.}
	\label{pic:FPGA_block}
\end{figure*}

The current block diagram for the FPGA is given in Figure \ref{pic:FPGA_block}.
Most of the FPGA's logic resources are consumed by the PDM interfaces and the
Linear Track Trigger (LTT) modules which perform the L2 calculation. To cope with the
requirements on the trigger evaluation time, which impacts the overall dead time of the
instrument, a highly parallel and pipelined architecture was chosen.
Therefore, data from all connected PDMs are processed independently and in parallel,
which reduces the time needed and adds reliability due to the fact that the
CCB is still operable if one or more PDMs will fail.

When the scientific data arrive at the CCB, the data stream will take two
different 'routes'. First, the complete raw data will be written to an external memory
where it resides until the L2 evaluation is done. In case of an L2 trigger it will be read
back and sent to the CPU - otherwise discarded.
Second, the photon counting data will be handed over to a 'Double Buffer' from where
it will be accessed by the LTT module. This buffer is necessary to reassemble the
frames from the single pixels as the data stream is disordered due to the channel
mapping of the ASICs. While the second frame will be reassembled, the first frame will be
fed into the '3x3 Sum' module which performs the summation of the 9-pixels blocks in two
stages (vertical and horizontal). As this is done for the whole frame, the redundant
3x3 summations will be reduced to a minimum. In that way, a new frame is generated where
each pixel contains the sum of the surrounding ones. This new frame is then trimmed to
a 19x19 pixel frame around the trigger seed and stored inside the 'Frame Buffer'.

When the 'Frame Buffer' contains all data for the direction integration (a total
of $\pm$14 GTUs around the trigger seed), the 'Address Generator' allocates the
different 3x3 sums to the 'Accumulator', which finally makes the direction integration.
The necessary 3x3 sum depends on the current starting point for the integration and the
offsets for the various directions which are stored in a Look Up Table (LUT).
It should be clear that the previous 3x3 summation now reduces the RAM read access
drastically - instead of 9 pixels times 15 GTUs per direction we only need to
allocate the pixels from the 15 GTUs.

Finally, the 'Maximum Comparator' selects the maximum integration value which is
then compared to the trigger threshold after all directions are completed.
In case the threshold is exceeded, the CCB sends a trigger to the Clock-Board
and the raw event data from the external memory to the CPU via the SpaceWire
interface.

\subsection{Test-bench and performance results}
\label{subsec:CCBResults}
To verify the correct behavior of the CCB, several tests were developed to prove
its functionality during the different stages of the development.

Starting with basic simulations of the developed FPGA logic, a first
hardware test of the LTT module was successfully performed by processing simulated PDM data
within the FPGA (see \cite{thesis_bayer, bayer2011}). This test already showed that the
module is working correctly and that the L2 processing time is less than 5 ms.
Additionally, the performance of the implemented set of directions was compared to
software simulations. This evaluation showed only minor differences in detecting the events,
which did not affect the overall trigger efficiency.

To validate the functionality of the CCB as a whole, including the interfaces and the
control-logic of the different modules, a stand-alone 'PDM-Simulator' was developed
(see \cite{thesis_gottschall}). The main purpose of the PDM-Simulator is to mimic
the behavior of the PDM which basically means to provide the data- and command-interfaces
and to generate an L1 trigger.
Designed around a Spartan 3 FPGA, the PDM-Simulator contains a large Synchronous
Dynamic Random Access Memory (SDRAM) and a Universal Serial Bus (USB) connection.
Over the USB connection, the PDM-Simulator receives simulated events generated
by the EUSO Simulation and Analysis Framework (ESAF) and stores them inside the SDRAM.
After a complete event is received, an L1 trigger is generated and the data
are sent to the CCB. Additionally, the PDM-Simulator is also acting as a
'Clock-Board-Simulator' by providing the necessary system- and GTU-clock to the CCB.

The processed data from the CCB is then sent to a computer, which is equipped
with a commercial SpaceWire card and the so called 'Near Real Time Analysis' (NRTA) software
provides a quick look to the received data.
To complete the hardware test-bench, a small micro-controller reads the various
Houskeeping registers and sends the data to the same computer.

By extensive use of this setup, the functionality of the hardware was validated
and optimized. Based on the current implementation and the assumption that the
L1 trigger is located in the middle of the event data (at GTU \#64), the L2 trigger
calculation will be finished in less than 1 ms after all data is received from the PDM.
Furthermore, if the L1 trigger will be adjusted to occur a few GTUs earlier, the
LTT will not introduce any additional dead-time to the system.

In addition to the laboratory setup, the TA-EUSO CCB was extensively
tested during several integration campaigns which were finished successfully with
the CCB working as intended.

\section{Conclusion \& Outlook}
\label{sec:ConclusionOutlook}
The design of the JEM-EUSO prototype CCB is finalized, schematic and layout is done and
a first board has been produced and tested. The internal logic of the
FPGA is working as intended and all requirements are met. Especially the current hardware
implementation of the L2 algorithm complies to the requirements which could be
achieved with a high grade of parallelization for the calculation process.
Nevertheless, it should be stressed that the hardware is still in the development phase
and a long way is necessary to reach the final space-qualified hardware.

The TA-EUSO CCB has already been successfully integrated with the other components
and it was possible to setup the complete read-out chain. It will start operation
in the next months (see \cite{gcasolino}).

Finally, the EUSO-Balloon CCB is in Phase C/D and the flight-models have been
produced. The next major step is to finalize the thermal management
and perform the qualification tests inside a climate-chamber in low-pressure
environment.

\vspace*{0.5cm}
{
\footnotesize{{\bf Acknowledgment:}{
This work is partially supported by the Bundesministerium f\"ur Wirtschaft und Technologie through the Deutsches Zentrum f\"ur Luft und Raumfahrt (Grant FKZ 50 QT 1101).
We wish to thank the European Space Agency for the support through a Topical Team grant.}}


\clearpage
}

%% file: icrc2013-1089.tex



\title{Performance of the SPACIROC front-end ASIC for JEM-EUSO}

\shorttitle{FE ASIC for JEM-EUSO}

\authors{
H. Miyamoto$^{1}$,
K. Yoshida$^{2}$,
F. Kajino$^{2}$,\\
S. Ahmad$^{3}$,
P. Barrillon$^{1}$,
S. Blin-Bondil$^{1,4}$,
S. Dagoret-Campagne$^{1}$,\\
C. de La Taille$^{4}$,
F. Dulucq$^{4}$,
P. Gorodetzky$^{5}$,
T. Iguchi$^{2}$,
H. Ikeda$^{6}$,\\
Y. Kawasaki$^{7}$,
G. Martin-Chassard$^{4}$
for the JEM-EUSO Collaboration.
}

\afiliations{
$^1$ Laboratoire de $l'Acc\acute{e}l\acute{e}rateur\ Lin\acute{e}aire$,
Univ Paris Sud-11,CNRS/IN2P3,
UMR8607 du CNRS, Centre Scirentifique d'Orsay,
Bat 200 - BP 34, 91898 Orsay Cedex, France \\
$^2$ Department of Physics, Konan University,
Okamoto 8-9-1, Higashinada, Kobe, Hyogo 658-8501, Japan \\
$^3$ Weeroc SAS,
86 rue de Paris, 91400 Orsay, France \\
$^4$ OMEGA Micro, Ecole Polytechnique, CNRS/IN2P3, LLR Aile 4, 91128
Palaiseau Cedex, France\\
$^5$ Astro Particle et Cosmologie, Univ Paris Diderot, CNRS/IN2P3,
10 rue A. Domon et L. Duquet, 75013 Paris, France \\
$^6$ ISAS/JAXA, 3-1-1 Yoshinodai, Sagamihara, Kanagawa 252-5210, Japan \\
$^7$ RIKEN 2-1 Hirosawa, Wako, Saitama 351-0198, Japan \\
}

\email{miyamoto@lal.in2p3.fr}

\abstract{
SPACIROC (Spatial Photomultiplier Array Counting and Integrating
ReadOut Chip) is a front-end (FE) ASIC designed for the space-borne
fluorescence telescope JEM-EUSO (Extreme Universe Space Observatory
on board Japanese Experiment Module).
This device performs single photon counting in a dynamic range
of 1 photoelectron (PE) to 300 PEs/pixel/2.5 $\mu$s, with double
pulse resolution of 30 $ns$, and low power consumption ($<$1 mW/ch).
Input photons are measured with two modes: Photon Counting (PC)
mode and Charge-to-Time conversion, so called KI,
mode for the multiplexed channels.
Combination of these two features enables the large dynamic range
as described above.
After successful testing phase of the first prototype of SPACIROC
(SPACIROC1), the second prototype (SPACIROC2) was developed and
tested since May 2012.
The main improvements are the following:
lower power consumption due to better power management,
enhancement in Photon Counting time resolution and extension of
the KI maximum input rate.
SPACIROC1 chips were integrated into the front-end electronics
(FEE) of an instrument pathfinder for detecting gamma ray
bursts - the Ultra Fast Flash Observatory (UFFO) which is foreseen
to be launched in 2013.
Towards the end of 2012, the FE board designed around SPACIROC1
chips have been fabricated for the EUSO-BALLOON \cite{jemeuso6}
\cite{jemeuso7} and TA-EUSO \cite{jemeuso8} projects.
We report here on the performance of SPACIROC1 and SPACIROC2 such
as single photon counting ability, double pulse resolution, dynamic
range, linearity and power consumption.
}


\maketitle

\section{Introduction}
JEM-EUSO is a mission designed to observe
Extreme Energy Cosmic Rays with a space-borne fluorescence
telescope on the International Space Station (ISS). 
The detector will consist of 5,000 1-inch-square Multi-Anode
Photomultiplier Tubes (MAPMTs),
and will allow an area of about $10^5 km^2$
of Earth's atmosphere to be imaged in the field of view.
The currently targetted launch is in 2017 in the framework
of the second phase
of JEM/EF (Japanese Experiment Module/Exposure Facility)
utilization.
The JEM-EUSO telescope will determine the energies and
directions of extreme energy primary particles by recording the tracks
of Extensive Air Shower (EAS) with a time resolution of about
$2.5{\mu}s$ and a spatial resolution of 0.1$^\circ$.
About the JEM-EUSO status and general project information, see also
\cite{jemeuso2},\cite{jemeuso3} and \cite{jemeuso4} in this conference.\\\\\\
\subsection{JEM-EUSO Focal Surface}
 \begin{figure}[h]
  \centering
  \includegraphics[width=0.85\hsize]{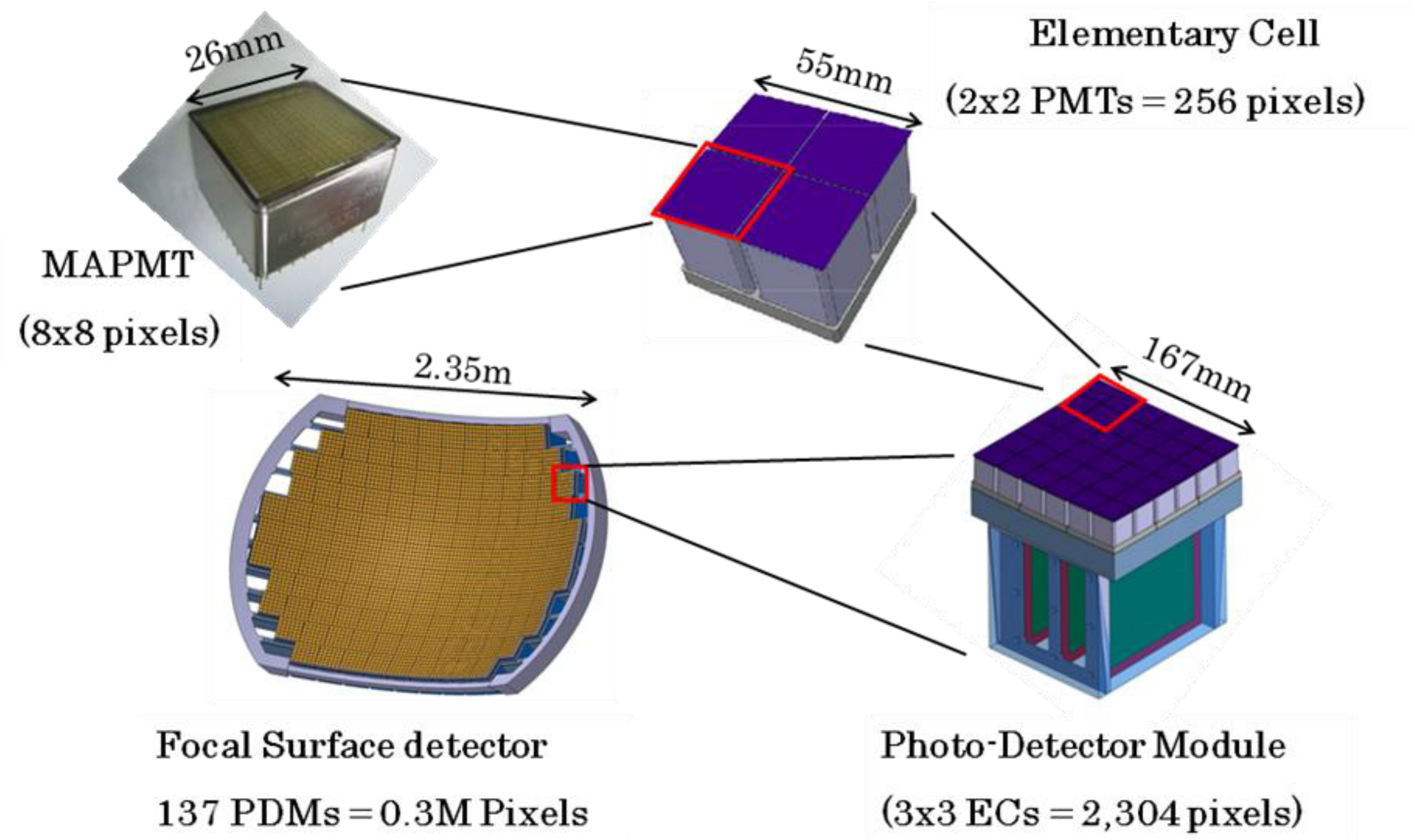}\vspace{-0.3cm}
  \caption{FS detector modules.}
\vspace{-0.5cm}
  \label{FS_DM}
 \end{figure}
The Focal Surface (FS) of the JEM-EUSO telescope has a structure
of curved surface of 2.35 m in diameter.
It is covered with 5,000 64-channel MAPMTs (Hamamatsu R11265-M64)
arranged in about 137 of JEM-EUSO Photo-Detector Modules (PDMs).
A PDM consists of an array of 3$\times$3 Elementary Cells (ECs),
each of which consist of 2$\times$2 MAPMTs (See the Fig.\ref{FS_DM}).
In the JEM-EUSO DAQ chain, an MAPMT captures single photons,
converts them in its photocathode into photoelectrons
and induces pulses from the charges on their anodes and dynode output.
The FE ASIC transforms the charges from MAPMTs into digital
numbers to be processed in the next stages of digital electronics.
Similarly the trigger stages process digitally those charges which
have been previously converted into numbers.
About the JEM-EUSO focal surface, see also the contributions
\cite{jemeuso5} and \cite{jemeuso6} in this conference.
\vspace{-0.4cm}
\section{Front-End ASIC: SPACIROC}
In terms of functionality, our FE ASIC must be able to deal with
different types of anode signals.
The anode pulses of the MAPMT will be in discrete mode
(Photon Counting) and also in DC mode (integration) for the wide
range of signals originated by the different types of event which
could be observed by JEM-EUSO.
Both Photon Counting and integration functions are the basis of
the analog part of this ASIC.
To complete the readout solution, the digital part is also
implemented for the analog-to-digital conversion for each acquisition
window, called the Gate Time Unit (GTU$ = 2.5\mu$s).
\subsection*{Requirement}
For the JEM-EUSO FE ASIC, some precautions are taken into account
for SPACIROC as it is required to work under the space
environment.
Due to the limited power budget provided by the ISS for JEM-EUSO,
the low power consumption is essential for the FE ASIC.
Radiation hardness is another strong requirement.
The operation requirements for the JEM-EUSO FE ASIC are summarized
as follows:
\begin{itemize}
\setlength{\itemsep}{-0.5mm}
\setlength{\itemindent}{-3mm}
\item Power consumption $<$ 1 mW/ch
\item 100\% trigger efficiency in Photon Counting at 50 fC,
equivalent to 1/3 PE with an MAPMT gain of 10$^6$
\item Dynamic range in charge measurement 1.5 PE to 150 PEs/GTU/pixel
or a sensitivity of factor 100
\item Radiation hardness. Expected accumulated radiation dose for
5 year operation: $\sim$30 krad
\item Data sampling: GTU = 2.5 $\mu$s (400 kHz).
\item Linearity in Photon Counting: $\geq$ 30 PEs/GTU
\item Double pulse resolution: $\leq$ 30 ns
\end{itemize}
\subsection{SPACIROC}
\vspace{-0.3cm}
\begin{figure}[h]
  \centering
  \includegraphics[width=3.0in]{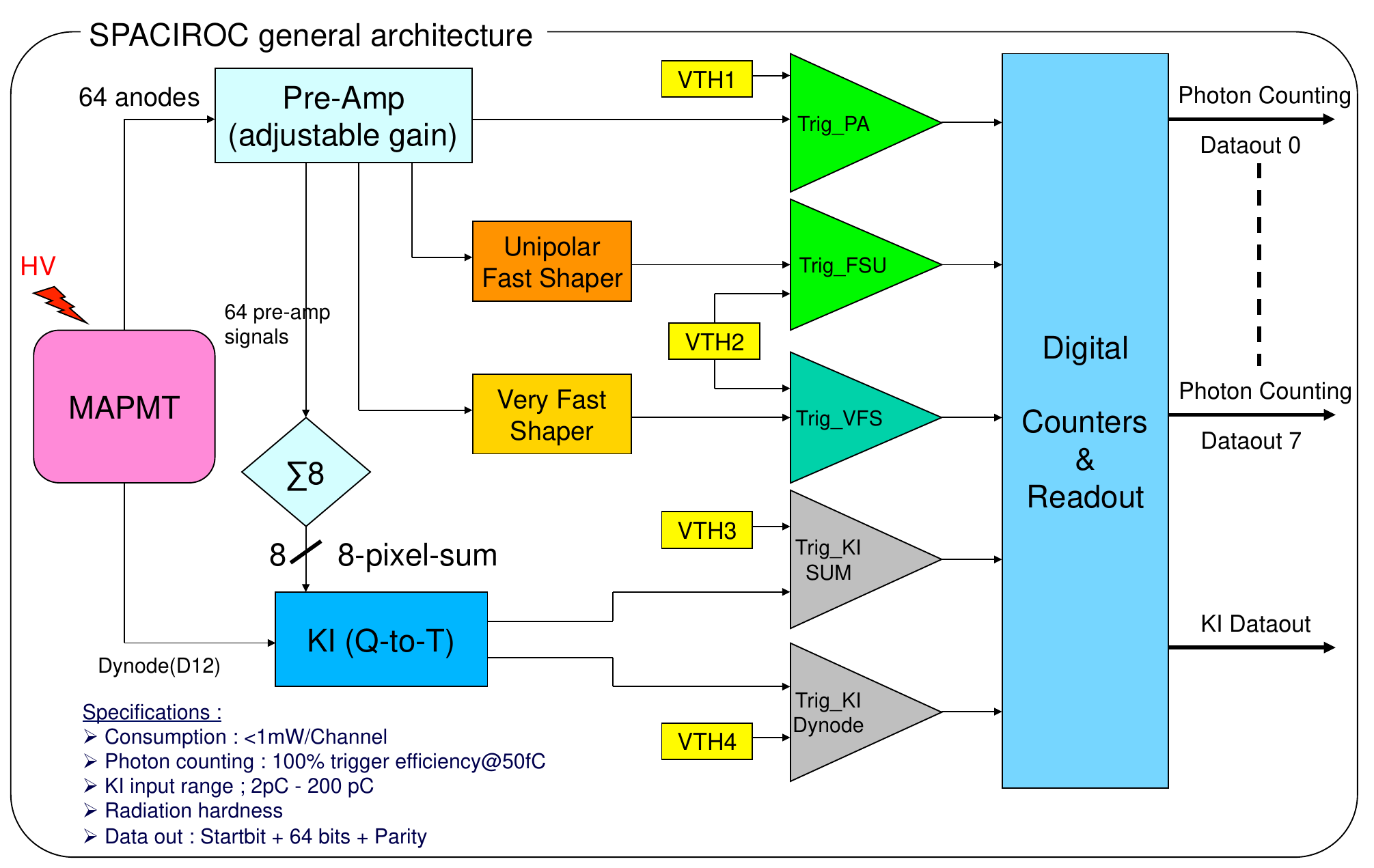}\vspace{-0.3cm}
  \caption{SPACIROC general architecture.}\vspace{-0.2cm}
  \label{SPACIROCschem}
\end{figure}
Fig. \ref{SPACIROCschem} shows the general architecture of the
SPACIROC1.
It consists of two analog blocks and one digital part.
One of the analog parts is dedicated to the Photon Counting, and
the other is dedicated to the KI converter.
The digital part is built to count the detected photons. 
The 64-channel Photon Counting block discriminates the preamplifier
signal into trigger pulses.
Each of the 64 channels of the Photon Counting block consist of a
preamplifier, two shapers and three comparators (trigger
discriminators) as below:
\begin{itemize}
\setlength{\itemsep}{-0.5mm}
\setlength{\itemindent}{-3mm}
\item Trig$\_$pa $\ \ \ :$ Trigger of the signal coming directly
from preamplifiers
\item Trig$\_$FSU : Trigger of signal from Unipolar Fast Shaper
(FSU)
\item Trig$\_$VFS : Trigger of signal from Very Fast Shaper (VFS).
\end{itemize}
The charge signals from the 64 anodes of a PMT are first fed into
preamplifiers before being sent to various shapers and discriminators
in the latter part of ASIC.
This is then sent to the above 3 Photon Counting outputs: preamplifier,
FSU, VFS.
At the end of a GTU, the counter values are readout through 8 serial
links in order to reduce overhead.
The first 8 inputs of the Charge-to-Time converter (KI) take the
pre-amplified signals from the Photon Counting (sum of every 8
channels), while the 9th input takes a signal coming directly from
the last dynode of the MAPMT.
In a similar manner to the Photon Counting readout, the counter
data are sent through a serial link at the end of each GTU.
\subsubsection{SPACIROC2}
A new version of the SPACIROC chip has been developed in order
to improve the performance of the first prototype.
Based on the ASIC characterization results and the feedbacks
from the UFFO pathfinder project, the design improvements are
mostly applied to the analog part of the ASIC.
The digital design was untouched for this ASIC version.
In terms of performance enhancement, the targets for SPACIROC2
are the following:
\begin{itemize}
\setlength{\itemsep}{-0.6mm}
\setlength{\itemindent}{-3mm}
\item Reduction on the power consumption by 30\%
\item Improvement on the double pulse resolution for the Photon
Counting ($<$30 ns)
\item Improvement on the KI converter: Dynamic range
extension and Reset implementation
\end{itemize}
The main improvement of the Photon Counting part was to achieve
a better time
resolution while maintaining lower power consumption.
As the FSU trigger design is considered as the baseline for
the Photon Counting part, its design is untouched.
Only minor modifications related to the power consumption have been
carried out for FSU trigger.
The main modifications of this part were done for the preamplifier
trigger which exhibits the lowest power consumption compared to the
other trigger design.\\
\begin{figure}[h]
  \centering
  \includegraphics[width=3.2in]{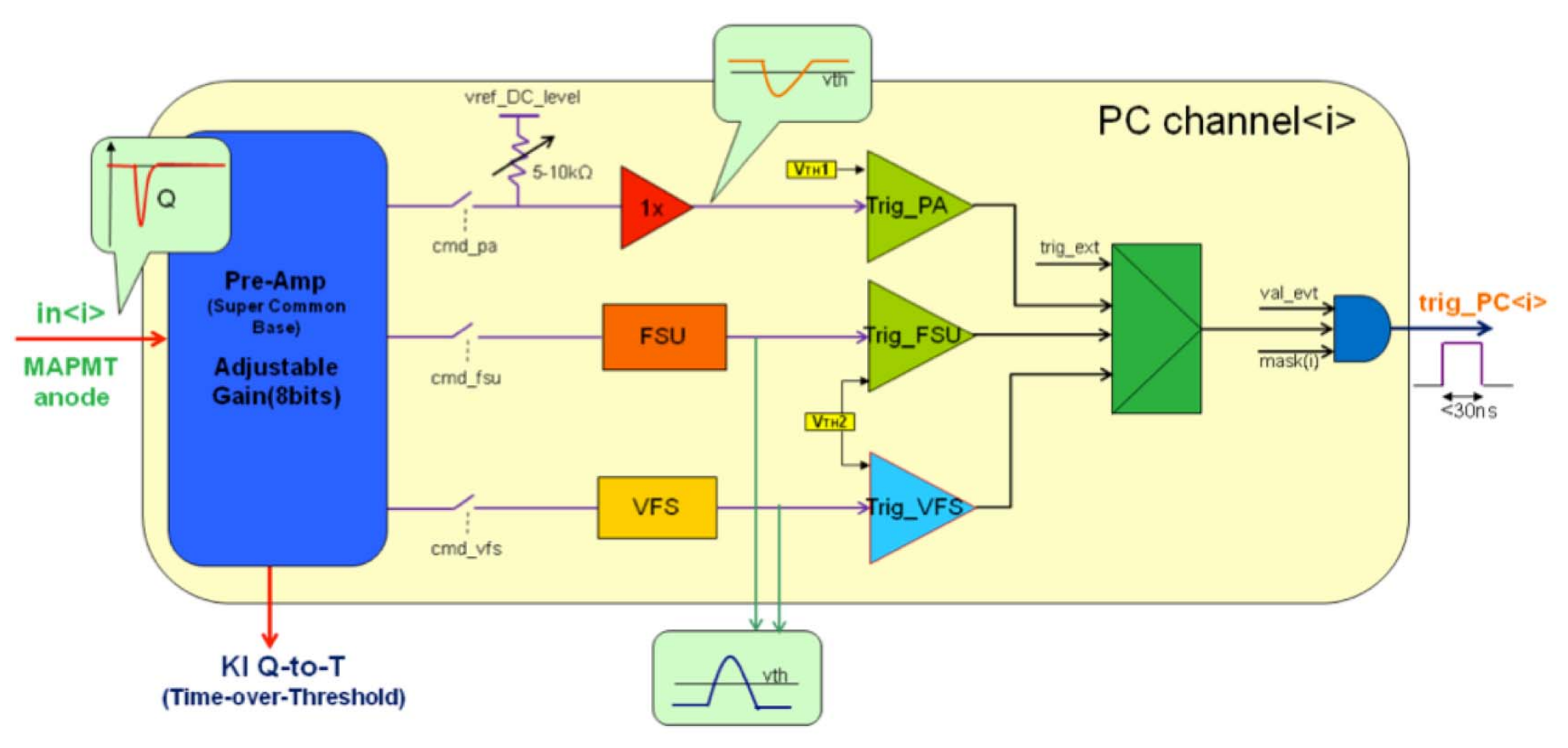}\vspace{-0.3cm}
  \caption{SPACIROC2 Photon Counting general architecture.}\vspace{-0.3cm}
  \label{S2analogHL}
\end{figure}

The architecture of the Photon Counting part is illustrated in
figure \ref{S2analogHL}.
The general architecture of SPACIROC2 is nearly identical to
the one shown previously in the Fig.\ref{SPACIROCschem}.
The modification on the architecture is visible only for the
preamplifier trigger (Trig\_PA) where a variable resistance
and a buffer have been introduced.
SPACIROC2 ASIC was submitted for prodution in November 2011,
The packaged chips were then received in March 2012.
\subsubsection{Characterization measurements}
\begin{figure}[h]
  \centering
  \includegraphics[width=3.2in]{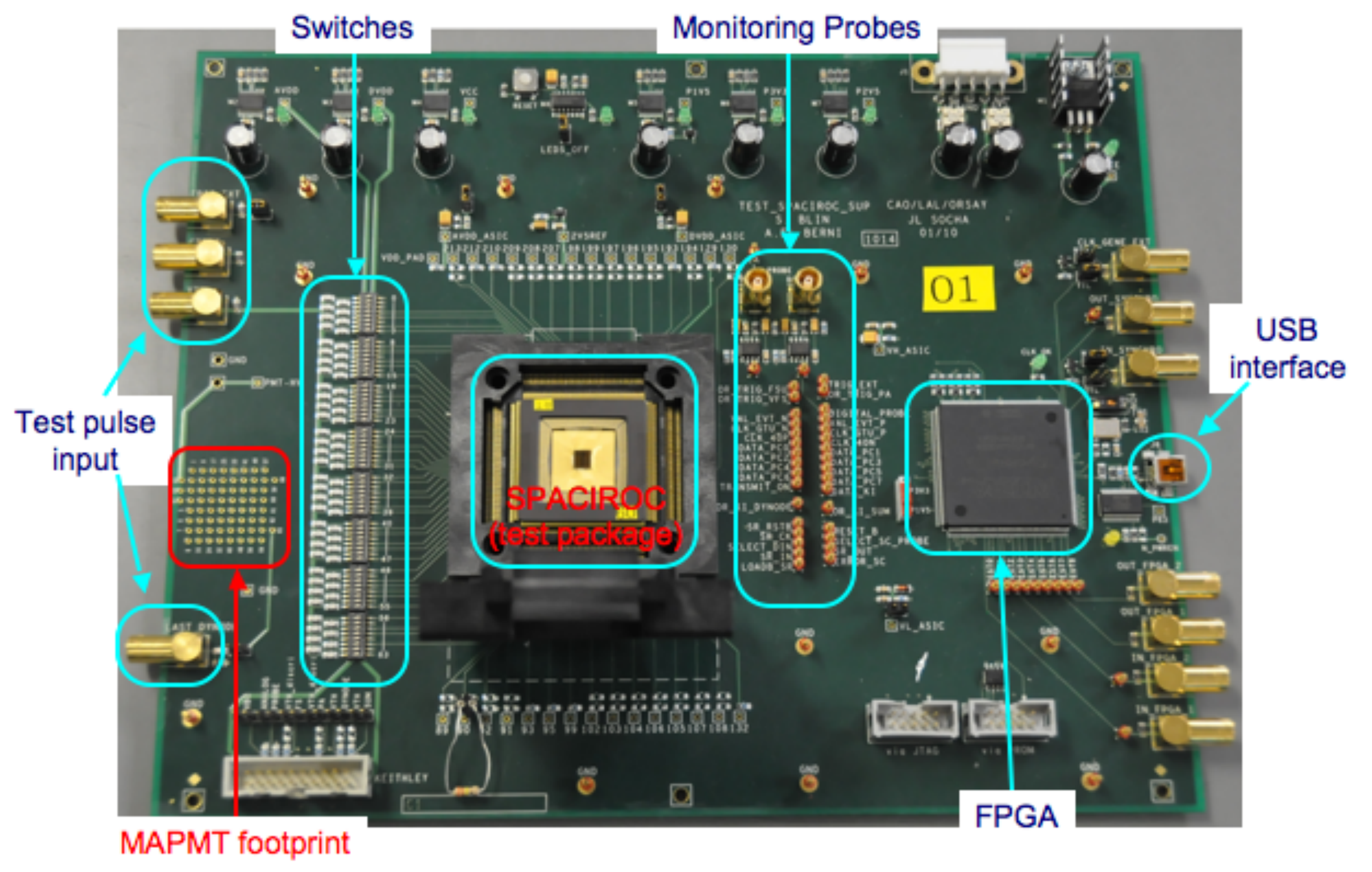}\vspace{-0.4cm}
  \caption{Test board for SPACIROC.}\vspace{-0.3cm}
  \label{testBoard}
 \end{figure}
For testing and characterizing the ASIC, a dedicated test board was
developed as shown in figure \ref{testBoard}.
It consists of a socket for ASIC, an FPGA and a USB connection
to PC, MAPMT footprint, and various test points.
The following features are available for the tests:
  \begin{itemize}
    \setlength{\itemsep}{-0.5mm}
    \setlength{\itemindent}{-5mm}
  \item Various types of ASIC input:
    \vspace{-0.2cm}
    \begin{itemize}
      \setlength{\itemsep}{-0.5mm}
      \setlength{\itemindent}{-11mm}
    \item ASIC internal charge injection via 2 pF capacitance,
    \item Onboard charge injection via 10 pF capacitance,
    \item Footprint for 64-pin Hamamatsu R11265-M64\\ \hspace{-10mm}MAPMT,
    \item External trigger for digital part.
    \end{itemize}
  \item Various points for monitoring signals of:
    \vspace{-0.3cm}
    \begin{itemize}
      \setlength{\itemsep}{-0.5mm}
      \setlength{\itemindent}{-11mm}
      \item Discriminator outputs,
      \item Discriminator OR outputs,
      \item Clocks \& Digital signals,
      \item Probe \& Slow Control signals
    \end{itemize}
  \end{itemize}
\vspace{-0.1cm}
For the characteristics tests of the ASICs, test pulses were fed into
the board from a pulse generator.
The registers inside the ASIC were controlled by a PC
using a LabView software via an FPGA and USB connection.
Typical measurements such as noise, S-curves and charge injection
have been carried out in order to check the performance of the
ASIC, both for SPACIORC1 and SPACIROC2 with the test board as described
in the following sections.

\subsubsection*{S-curves}
\begin{figure}[h]
\vspace{-0.3cm}
  \centering
  \includegraphics[width=3.2in,height=1.7in]{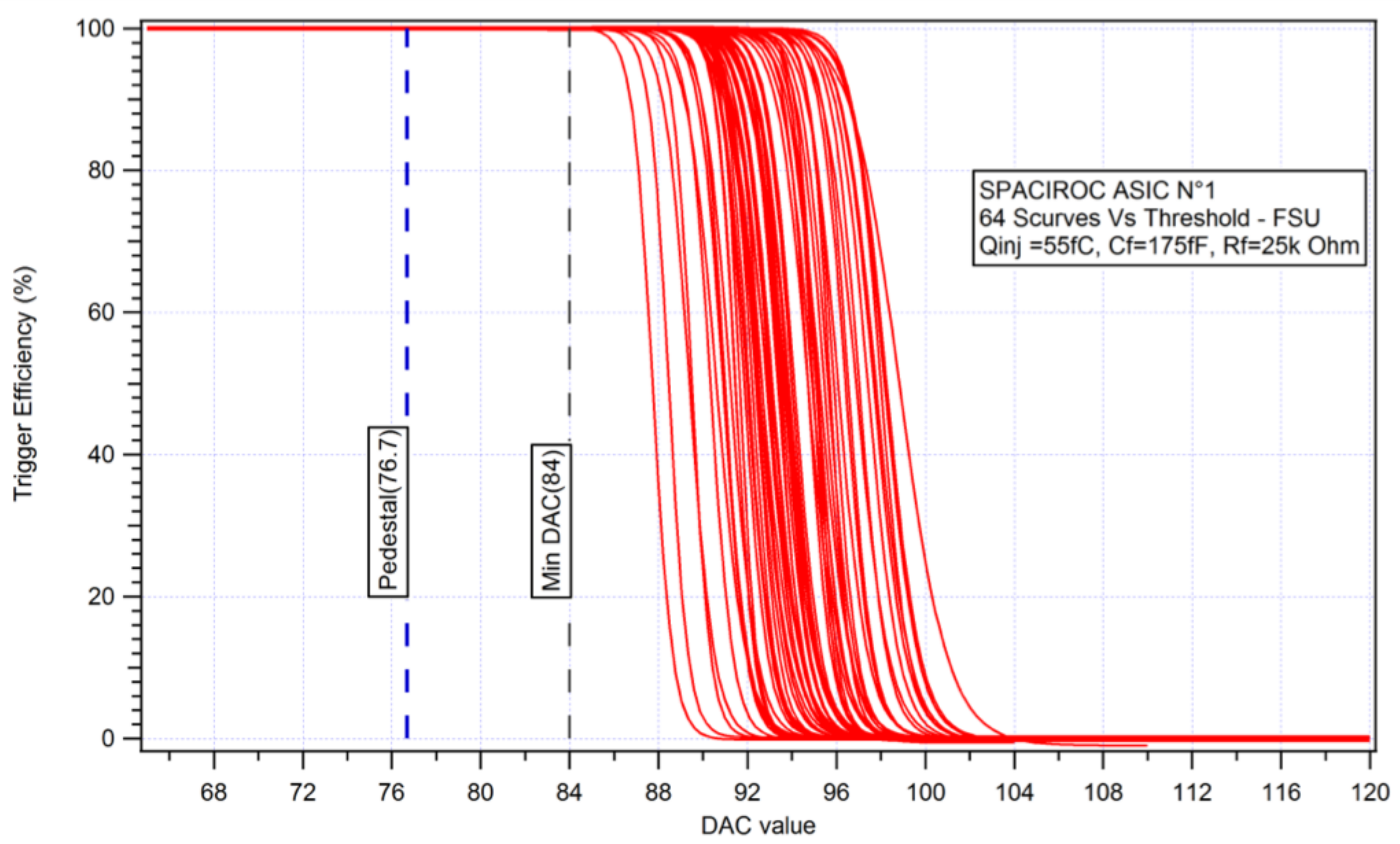}\vspace{-0.4cm}
  \caption{S-curves of FSU}\vspace{-0.3cm}
  \label{ScurvesFSUS1}
\end{figure}

One of the most important aspects of Photon Counting is the
trigger efficiency.
Typically it is done by scanning the threshold (DAC) for a
given injected charges.
The resulting plot is known as S-curves which is in fact the
cumulative distribution function of the probability to generate
a trigger.
For the Photon Counting, the minimum charge required by JEM-EUSO
in order to achieve 100\% triggering efficiency is 50fC
(1/3 PE for an MAPMT gain of $10^6$).\\
Fig. \ref{ScurvesFSUS1} shows S-curves for
all the 64ch of a SPACIROC chip, obtained with
Trig\_FSU response for an input charge of 50 fC.
We use the S-curves for the gain estimation as described in
the following section.
\subsubsection*{Linearity}
\begin{figure}[h]
  \centering
  \vspace{-0.2cm}
  \includegraphics[width=3in,height=1.3in]{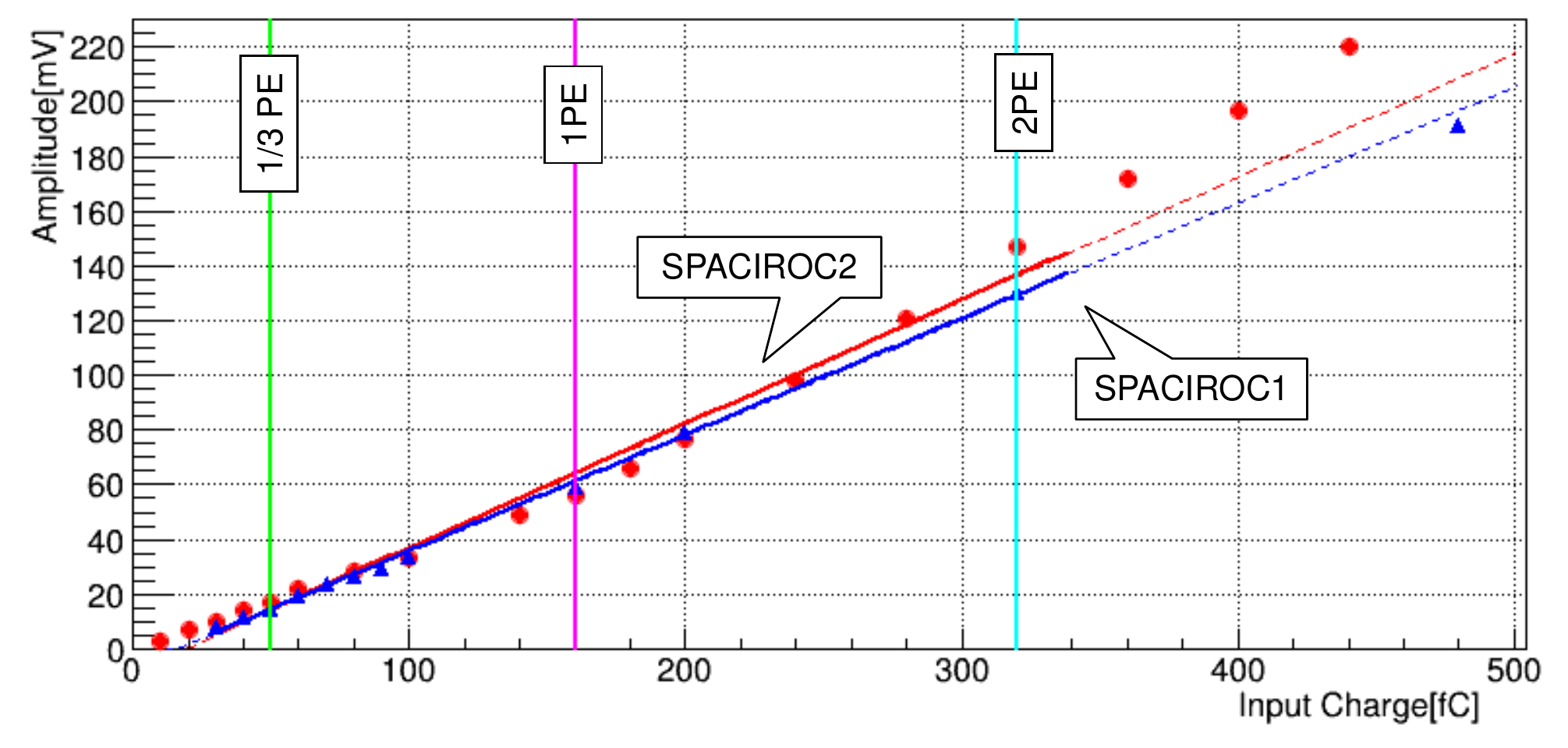}\\
  \includegraphics[width=3in,height=1.3in]{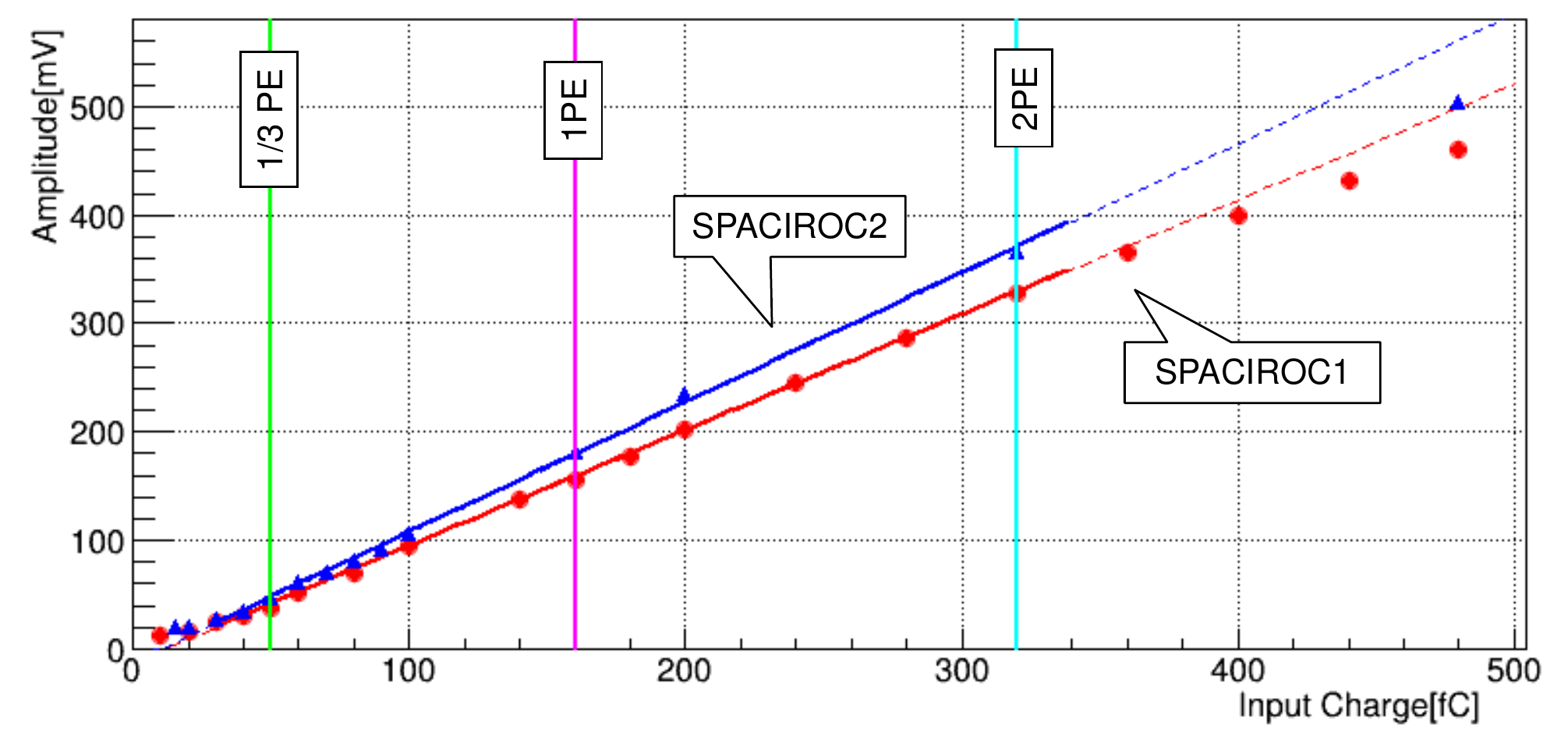}\vspace{-0.4cm}
  \caption{
    Top: Linearity of Trig\_PA analog pulse height of SPACIROC1 (red
    lines and circles) and SPACIROC2 (blue, triangle) as a function
    of injected charge.
    Bottom: Linearity of Trig\_FSU analog pulse height of SPACIROC1
    (red, circle) and SPACIROC2 (blue lines and triangles).}
  \label{linearity}
  \includegraphics[width=3in,height=1.4in]{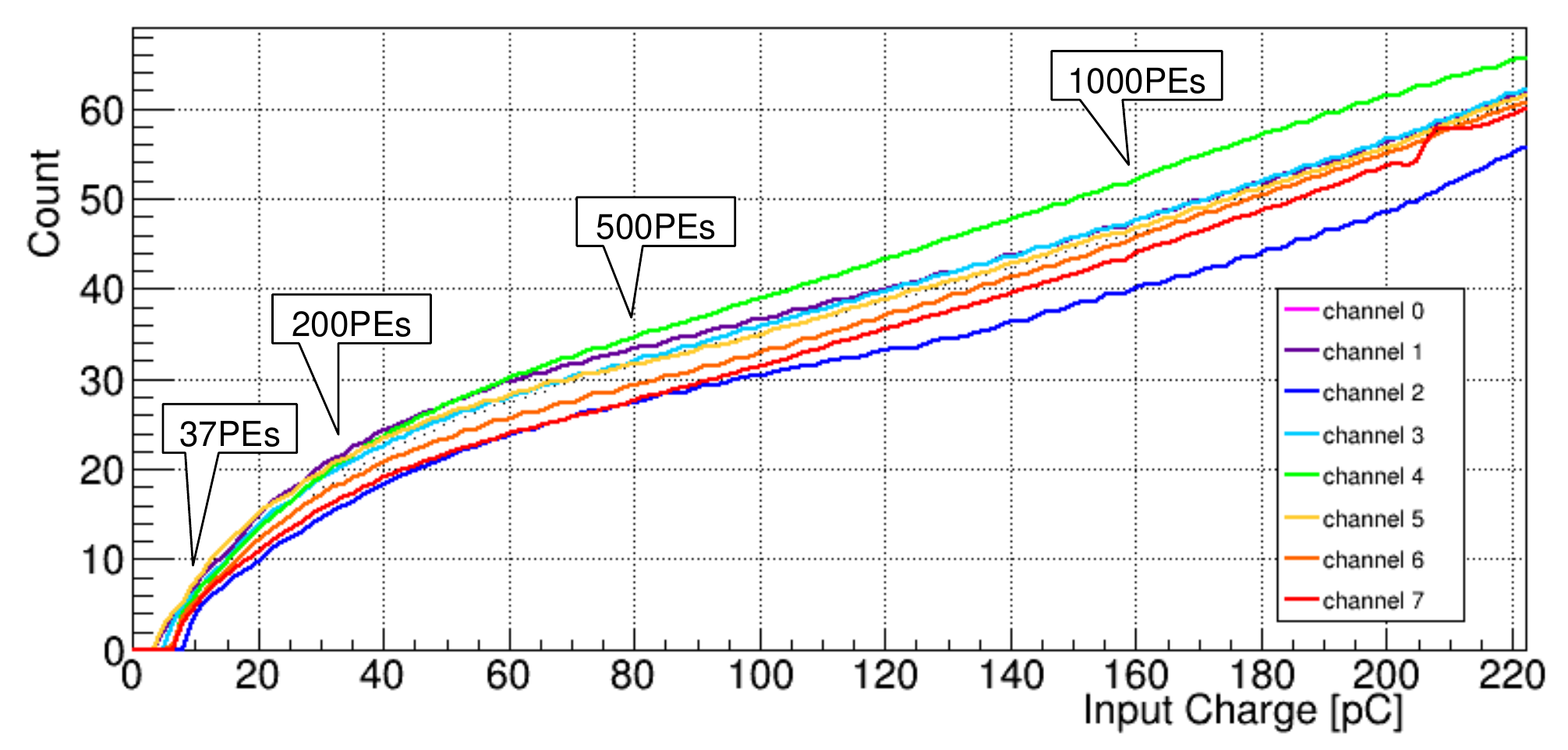}\vspace{-0.5cm}
  \caption{KI 8-Pixel-Sum (SPACIROC2) count as a function
    of input charge.
    The measurement was done from 6 pC to 220 pC.}\vspace{-0.3cm}
  \label{linearityKI}
\end{figure}

Data from S-curves can be used to verify the gain and determine
the minimum detectable signal from the Photon Counting mode.
By estimating the 50\% triggering efficiency as a function
of the injected input charge, the gain can be deduced from
the slope of the curve.
It also gives a good indication of linearity of each triggering
chain as a function of the input charge.
The top panel of Fig. \ref{linearity} shows the linearity of the
Trig\_PA analog pulse height of SPACIROC1 (red lines and circles)
and SPACIORC2
(blue lines and triangles) as a function of input charge,
and the bottom panel of Fig. \ref{linearity} shows the same for
Trig\_FSU analog pulse height.
As shown in the figures we confirmed that the ASIC responds linearly
up to at least 320 fC, which corresponds to the charge of 2 PEs.
In addition, Fig. \ref{linearityKI} shows the measurement results of
the KI converter of SPACIROC2 as a function of injected charge.
The result indicates that the KI Charge-to-Time module is capable
of integrating input charges up to 220 pC.
\subsubsection*{Double Pulse Resolution}
The Double Pulse Resolution has been measured for the Photon
Counting part.
It is done by injecting two pulses of 1PE each, separated
by a given delay.
The threshold is set around 1/3 PE in order to trigger both inputs.
The delay between both inputs was slowly decreased until there is
no clear trigger generated separately for both inputs.
In this measurement, 2 input pulses of a little less than
1 PE (132fC) were injected.\\
\vspace{-0.5cm}
\begin{figure}[h]
  \centering
  \includegraphics[width=3.2in]{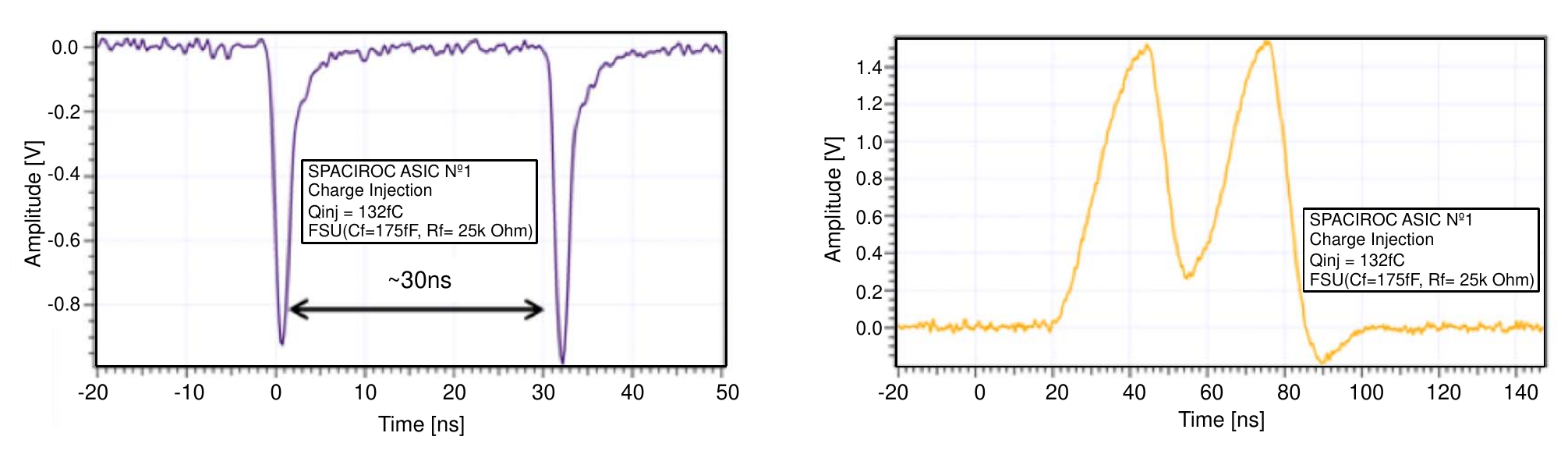}\vspace{-0.3cm}
  \caption{Left: Input pulses separated by 30 ns.
  Right: FSU waveforms for input charge separated by 30 ns.}\vspace{-0.3cm}
  \label{DPresInputPulse}
\end{figure}
\vspace{-0.2cm}

As a result, the double pulse resolutions for the cases using
each descriminators are summarized below:\\
\vspace{-0.5cm}
\begin{table}[h]
\begin{tabular}{|c|c|c|} \hline
\multicolumn{1}{|c}{}{Photon Counting} &
\multicolumn{2}{|c|}{Double Pulse Resolution [ns]} \\ \cline{2-3}
\multicolumn{1}{|c|}{Trigger} & {SPACIROC2} & {SPACIROC1}\\ \hline
{Trig\_PA}&{28.5}&{36}\\ \hline
{Trig\_FSU}&{30}&{30}\\ \hline
{Trig\_VFS}&{22}&{20}\\ \hline
\end{tabular}
\end{table}
\vspace{-0.5cm}

\hspace{-0.5cm}
The FSU is considered as the baseline for the Photon Counting
mostly because of the noise performance and the reasonable pulse
separation as shown by the result.
\subsubsection*{Power Consumption}
For SPACIROC2, a maximum of 0.8 mW per detection
channel has been allocated.
For SPACIROC1, the measured power consumption is 1.1 mW/channel.
This is partially due to design bugs which make some unused
components always on.
Since these bugs were corrected in the next version of the ASIC,
a reduction of the power consumption was expected.
Indeed, 0.83 mW/ch was measured for SPACIROC2, showing that the
requirements were nearly matched.
\subsection*{Radiation tolerance experiment at HIMAC}
As it will be operated in space, radiation hardness
is required for the JEM-EUSO FE ASIC.
We performed the radiation tolerance experiment at HIMAC with
collaborators of RIKEN and NIRS (National Instrument of Radiological
Science).
We irradiated Fe-ion beam with the energy of 500 MeV amu$^{-1}$ at
the beam exit, while the beam energy loss is 1.53 GeV/(g/cm$^2$)
and the spill interval is 3.3 sec.\\
We performed measurements of thirty minutes three times.
As the beam size is about 0.8 cm$^2$ and the size of ASIC is about
0.2 cm$^2$, after considering the amont of radiation on the ISS on
the orbit and the readout efficiency of the ASIC counter,
the number of particles radiated on the ASIC (3.78 $\times$ 10$^8$) is
equivalent to the amount of 6.4 years radiation in the ISS environment.
We checked the behavior of the ASIC after the irradiation experiment,
and we found no effects of ``Single Event Latch up'' or
``Single Event Upset''.
Thus, we conclude that the radiation tolerance of the SPACIROC is
suitable for operation in the space.\\
\vspace{-0.3cm}
\subsection{MAPMT measurement}
\begin{figure}[h]
\vspace{-0.3cm}
  \centering
  \includegraphics[width=2.8in,height=1.4in]{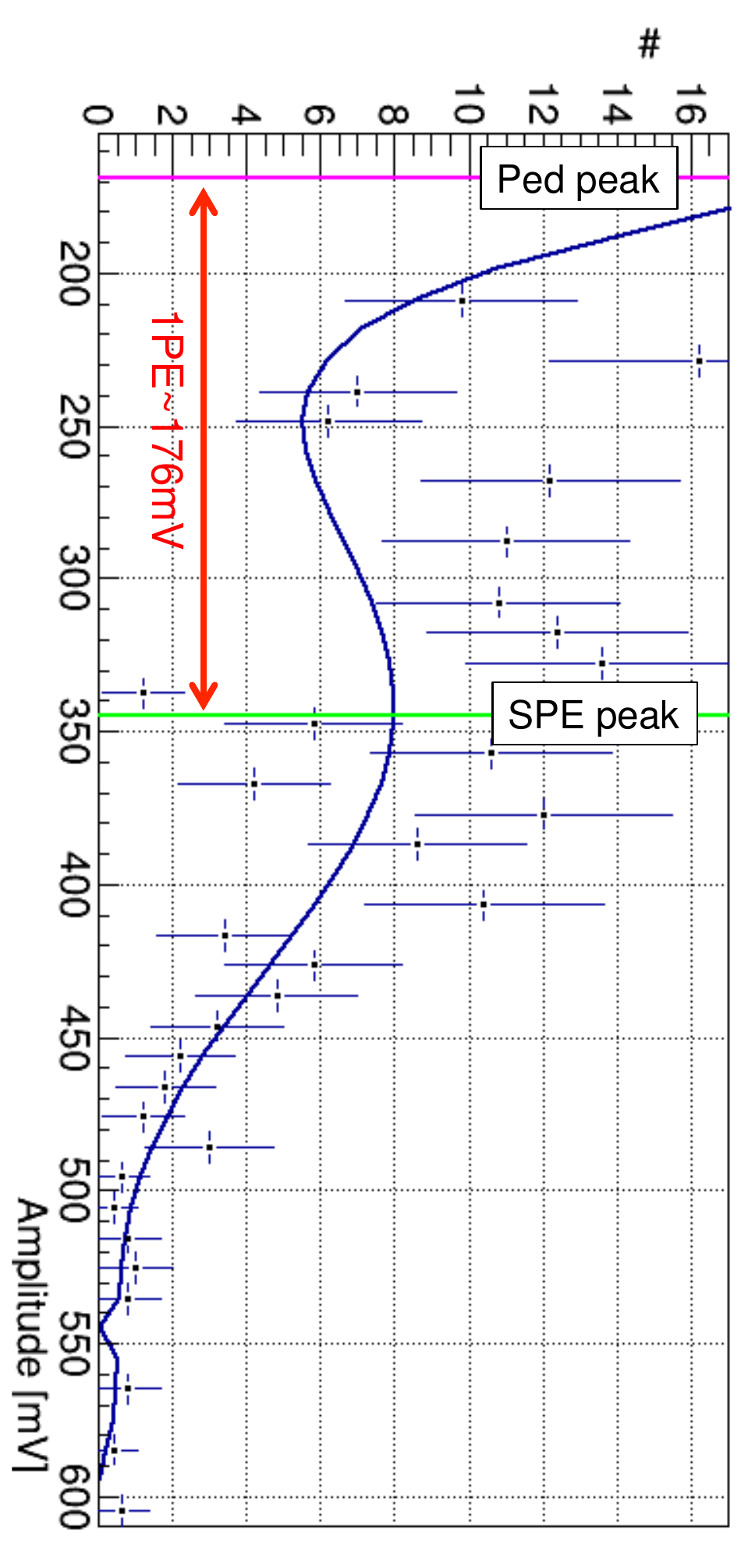}\vspace{-0.3cm}
  \caption{HAMAMATSU R11265-M64 pixel 36 SPE spectrum
  obtained from the ASIC S-curves derivation}\vspace{-0.4cm}
  \label{SPE}
\end{figure}

A series of tests using MAPMT have been carried out by our group
in France and in Japan.
We included an MAPMT onto the MAPMT footprint on the SPACIROC1
test board, and supplied a voltage of 900V to the MAPMT cathode.
We obtained a gain of about 10$^6$ at this voltage.
In our setup, the test board triggers with a frequency of
GTU (400 kHz), which is synhcronized with a pulse generator
driving a UV LED ($\lambda$=370 nm).
An integrating sphere splits the light uniformly between the MAPMT
and the NIST photodiode.
The MAPMT anodes signals are fed directly into the ASIC and the data
acquisition is done through LabVIEW.\\
The S-curve is the cumulative distribution function of the analog pulse
output spectrum, so the spectrum can be obtained from the observation
of S-curves.
The light intensity has to be adjusted very low in order to have
1\% of the Single Photoelectron (SPE) signal rate over the pedestal.\\
Fig. \ref{SPE} shows an SPE spectrum obtained with the FSU on one of
the 64 pixels of MAPMT.
It is obtained as the differential curve of the S-curve.
The blue points show the calculated data with statistical error, 
and the blue line is obtained by differentiating the smoothed S-curve.
The green line shows the SPE peak and the magenta shows the
peak position of pedestal.
The SPE peak in amplitude [mV] is consistent with the expected
value for a PMT gain of 10$^6$.
This gain is estimated around 176 mV with Trig\_FSU using test
pulses as shown in the bottom panel of Fig. \ref{linearity}.
\vspace{-0.2cm}
\section{Conclusions}
A series of tests have been carried out for the first and second version
of FE ASIC for the FS of JEM-EUSO mission.
The results showed that the fundamental functions of the ASICs
work very well without any critical problems.
Some minor problems of the first version were solved and the
functions have been satisfactorily improved in the second version of
the ASICs.
The third version of the ASIC (SPACIROC3) is planned to be submitted
this year. We aim to improve the double pulse resolution to 10 ns
as well as power consumption to around 0.6 mW/ch.\vspace{0.2cm}
\\
%
{
\footnotesize{{\bf Acknowledgment:}{
This work is mainly supported by CNES and IN2P3.
It was also partially supported by Basic Science Interdisciplinary 
Research Projects of RIKEN and JSPS KAKENHI Grant (22340063, 23340081
and 24244042).
}}

}
\clearpage

%% file: icrc2013-0832.tex




\title{The TA-EUSO and EUSO-Balloon optics designs}

\shorttitle{The TA-EUSO and EUSO-Balloon optics designs}

\authors{
Yoshiyuki Takizawa$^{1}$,
Alessandro Zuccaro Marchi$^{1}$,
Toshikazu Ebisuzaki$^{1}$
for the JEM-EUSO Collaboration.
}

\afiliations{
$^1$ RIKEN, Japan \\
}

\email{takky@riken.jp} 

\abstract{In this paper we describe the details of the design of the optics being developed for the TA-EUSO and the EUSO-Balloon experiments. These are pathfinders experiments of the JEM-EUSO mission. TA-EUSO (Telescope Array EUSO) observes the fluorescence light from extensive air showers, generated by ultra high energy cosmic rays, detected at the Telescope Array site in Utah. The EUSO-Balloon will observe extensive air showers from a gondola of a stratospheric balloon. 
We have developed feasible optics designs for these experiments based on the JEM- EUSO optics development. These designs are simplified version of the JEM-EUSO baseline optics design. The optics of TA-EUSO consists of two 1m square flat Fresnel PMMA lenses, while the EUSO-Balloon optics include, beside the two 1m square flat Fresnel PMMA lenses, an additional 1m square flat diffractive PMMA lens. }


\maketitle

\section{Introduction}
The Extreme Universe Space Observatory on the Japanese Experiment Module (JEM-EUSO) of the International Space Station (ISS) will be attached to the Exposure Facility on the JEM located of the ISS. Its main goal is to observe UV fluorescence images of UHECR air shower in the Earth atmosphere with a field of view of 60${}^\circ$ \cite{rbib:EUSOperf} . 

The TA-EUSO and EUSO-Balloon projects are pathfinder experiments for the JEM-EUSO, in which the manufacturing of several key components of the telescope will be tested together with the observational technique to detect the cosmic ray air shower events. The details of both missions are described in \cite{rbib:TA-EUSO} and \cite{rbib:Balloon}. In this paper, we focus on the designs of the optics of the two pathfinders experiments.

To maximize the sharing between the two projects of the machine time required by their manufacturing, it was decided that the two projects would have used the same optical design for the front and the rear lenses. 
Therefore, the EUSO-Balloon optics consists of two TA-EUSO lenses and of an additional flat diffractive lens. The diffractive lens is placed between the front lens and the rear lens to produce the fine RMS spot size ($<$ 2.8mm pixels size). The lens material is the UV transmittance grade PMMA (PMMA-000, Mitsubishi Rayon CO., LTD.). All lenses have a size of 1m by 1m shape and the thickness is 8 mm. This size was decided by the observational performance and commercial availability of PMMA base material. The details of the manufacturing of TA-EUSO/EUSO-Balloon lenses are described in \cite{rbib:Manufacturing}. 

\section{The concept of the TA-EUSO and EUSO-Balloon optics}
The TA-EUSO and the EUSO-Balloon optics is designed taking into account the optics performance and verifying the lens manufacturing technology. Since the JEM-EUSO lenses have a side cuts form, TA-EUSO and EUSO-Balloon lenses manufacturing is a good opportunity to verify how to manufacture non-circular shape lens.
As far as the optics design is concerned, the front and the rear lens of both experiments have a fresnel surface, while the middle lens of the EUSO-Balloon needs a diffractive surface to obtain a small spot size ($<$ 2.8mm pixels size). The diffractive lens counteracts the dispersion of the lens material refractive index, so it is able to reduce the color aberration. The surface roughness of the lenses should be smaller than 20 nm RMS. The photon collection efficiency (PCE) of both optics should have $\sim$40\% for the field angle 0$^\circ$. These values come from the JEM-EUSO optics requirements. PCE is calculated with a formula in our developed raytrace simulation:
\begin{equation}
PCE = \frac{Photon\ counts\ in\ a\ pixel\ of\ the\ detector}{Photon\ count\ hitting\ the\ front\ lens}\label{eqn:effi}
\end{equation}

To meet the observational requirements, the RMS spot size should be smaller than the pixel size of the detector for EUSO-Balloon. This requirement is the same for the JEM-EUSO optics. the Field of view (FOV) of the Photo Detector Module (PDM) of EUSO-Balloon is larger than 12$^\circ$. This value comes from sciences requirements \cite{rbib:Balloon}.

For TA-EUSO, the RMS spot size requirements are more relaxed since FOV of 5$\times$5 pixels of Telescope Array is nearly equal to the full FOV of TA-EUSO with one PDM. PDM has 48$\times$48 pixels. There are $\sim$9.6$\times$9.6 pixels of PDM in a pixel FOV of Telescope Array. In that case, the RMS spot size should be smaller enough than $\sim$9.6$\times$9.6 pixels, TA-EUSO is able to observe a resolving image in a Telescope Array pixel FOV. 
The FOV of the PDM is larger than 8$^\circ$. This value comes from sciences requirements \cite{rbib:TA-EUSO}. \\

The entrance pupil area of both optics should be larger than an area of 1m diameter. This requirement comes from \cite{rbib:TA-EUSO, rbib:Balloon}.

\section{Optics design and performance}
In this section, we describe in details the TA-EUSO and EUSO-Balloon optics respectively. \\

\subsection{Lens material}
There are two candidates for the material of the lenses: the PMMA-000 and CYTOP.
PMMA-000 (Mitsubishi Rayon Co, Ltd. product) is a special-grade UV transmittance Poly-methyl methacrylate \cite{rbib:ESA-PA}.
The seconde candidate is CYTOP (Asahi Glass Co, Ltd. product), an amorphous, soluble perfluoropolymer.
CYTOP has 95\% transmittance between UV and near-IR and high resistance against space environment (bombardment with atomic oxygen and radiation dose) because of the strong chemical stability by fluoro-bond.

PMMA-000 has been selected for the TA-EUSO and the EUSO-Balloon experiments mainly for two reasons: 1) non-space mission and 2) commercial availability of the lens material. 
\subsubsection{Characteristic of PMMA-000}
\begin{figure}[!h]
 \centering
 \includegraphics[width=0.48\textwidth]{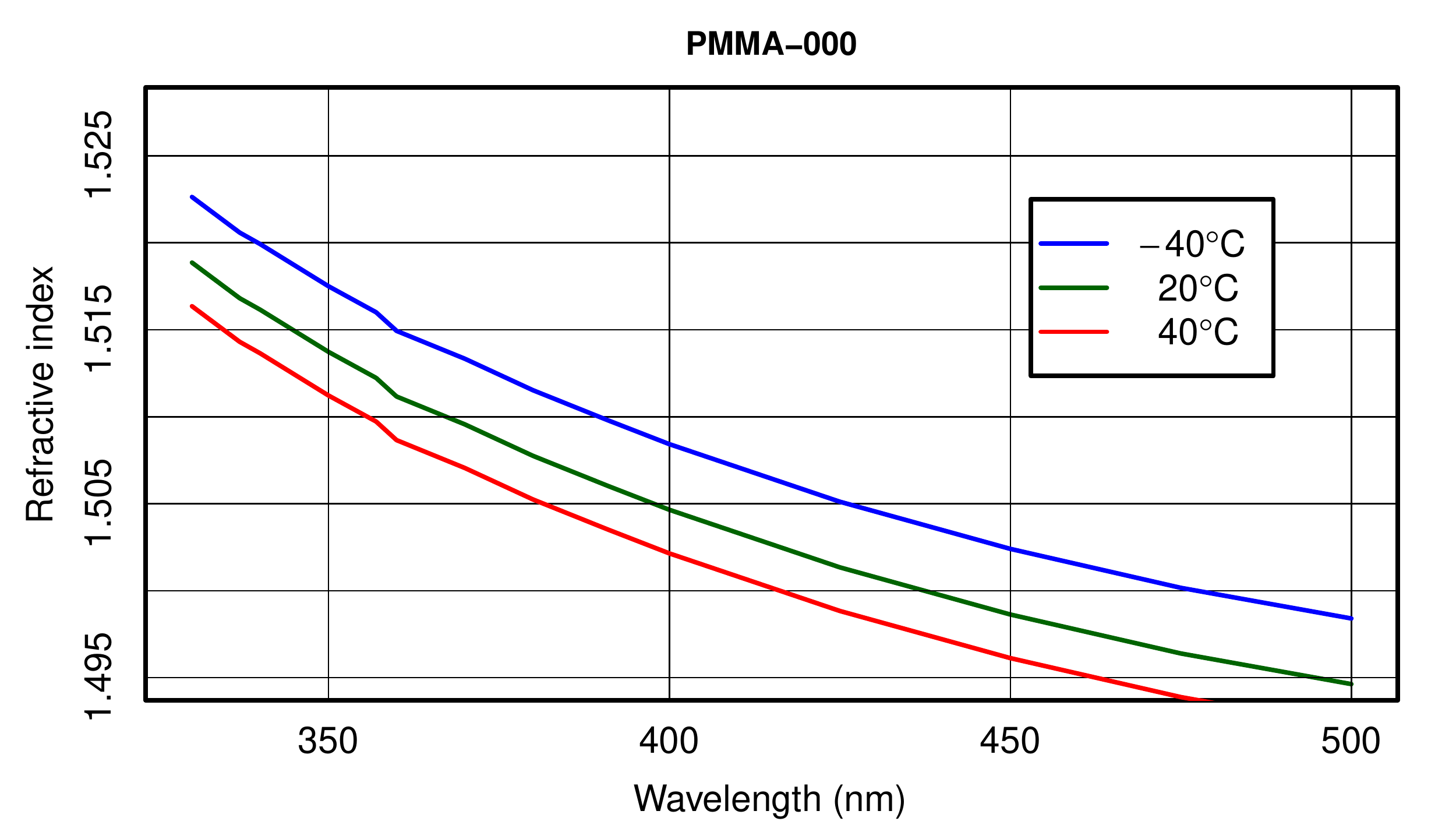}
 \caption{PMMA-000 refractive index in the near UV region with different temperatures ($-$40$^\circ$C, 20$^\circ$C and 40$^\circ$C).}
 \label{refra_fig}
\end{figure}

The refractive index in the near UV region for three different temperatures ($-$40$^\circ$C, 20$^\circ$C and 40$^\circ$C) is shown in Fig. \ref{refra_fig}. The 20$^\circ$C curve has been measured, while the $-$40$^\circ$C and 40$^\circ$C are calculated according to \cite{rbib:PMMA-formula}:
\[n(t) = n_0 + at + bt^2; \]
\[a = -0.000115,\]
\[b = -5.17358\cdot 10^{-7},\]
\[n_0\;is\;a\;reference\;refractive\;index.\]
\[Our\;measured\;data\;of\;20^{\circ}C\;is\;used\;as\;n_0. \] 
The TA-EUSO and EUSO-Balloon experiments will be operated in various temperature environments. TA-EUSO is deployed at the Telescope Array site in Utah, USA. The temperature of TA-EUSO will be changed between $-$30$^\circ$C $\sim$ 40$^\circ$C by the seasonal variation. On the other hand, the temperature of the EUSO-Balloon lenses will change with balloon flight conditions (trajectory, altitude, etc). Each lens has to withstand different temperatures. The front lens temperature will be in the range $-$40$^\circ$C $\sim$ 0$^\circ$C, depending on flight conditions. The inner lenses' temperatures will be higher than the front lens, and with smaller variations, due to the heating by electronics components. 

\begin{figure}[!h]
 \centering
 \includegraphics[width=0.45\textwidth]{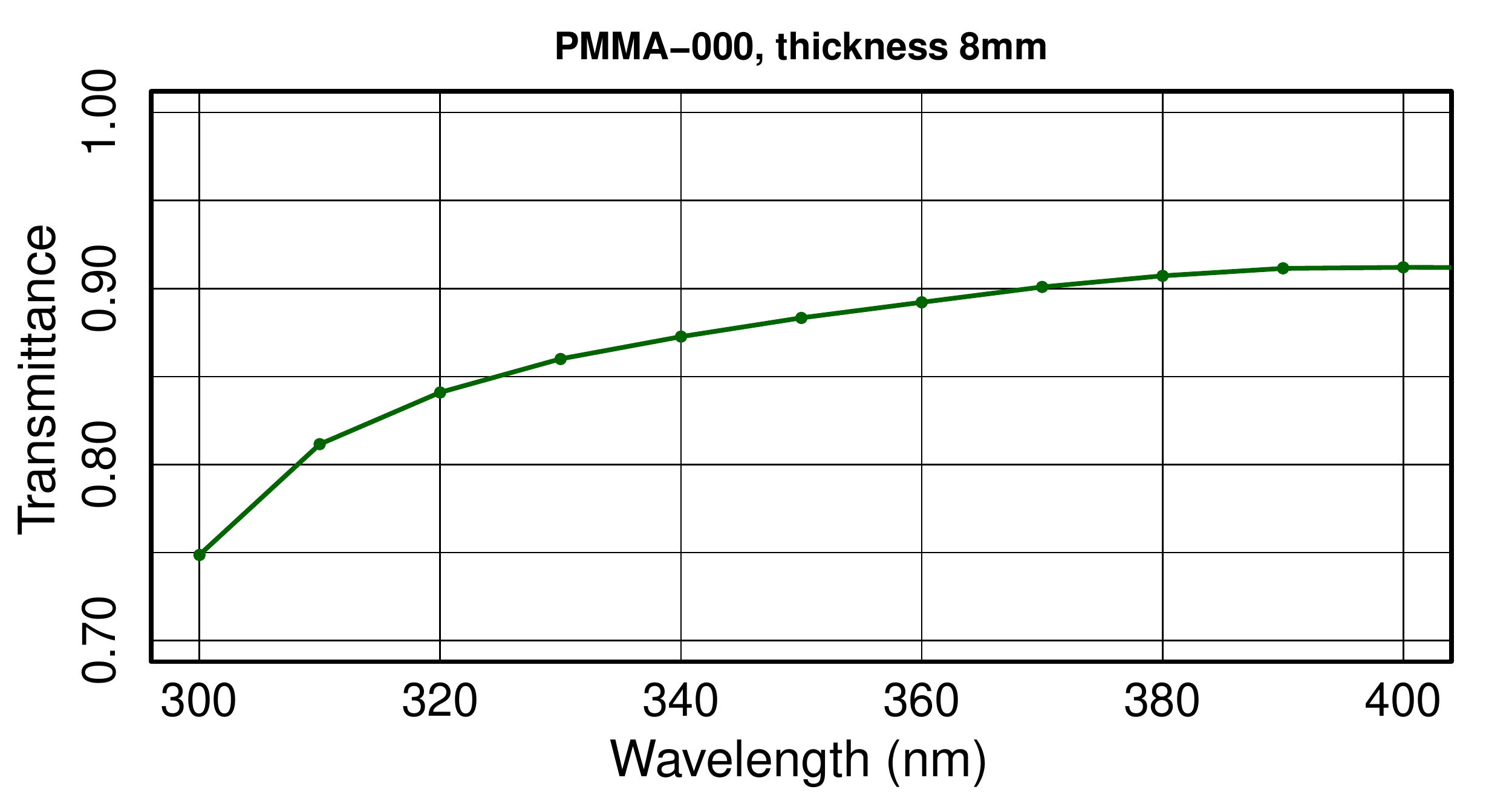}
 \caption{The 8mm thick PMMA transmittance in the near UV region.}
 \label{tra_fig}
\end{figure}

The PMMA transmittance with 8mm thickness in near UV region is shown in Fig. \ref{tra_fig}. 

\subsection{The TA-EUSO optics design and its performance}
\begin{figure}[!h]
  \centering
  \includegraphics[width=0.48\textwidth]{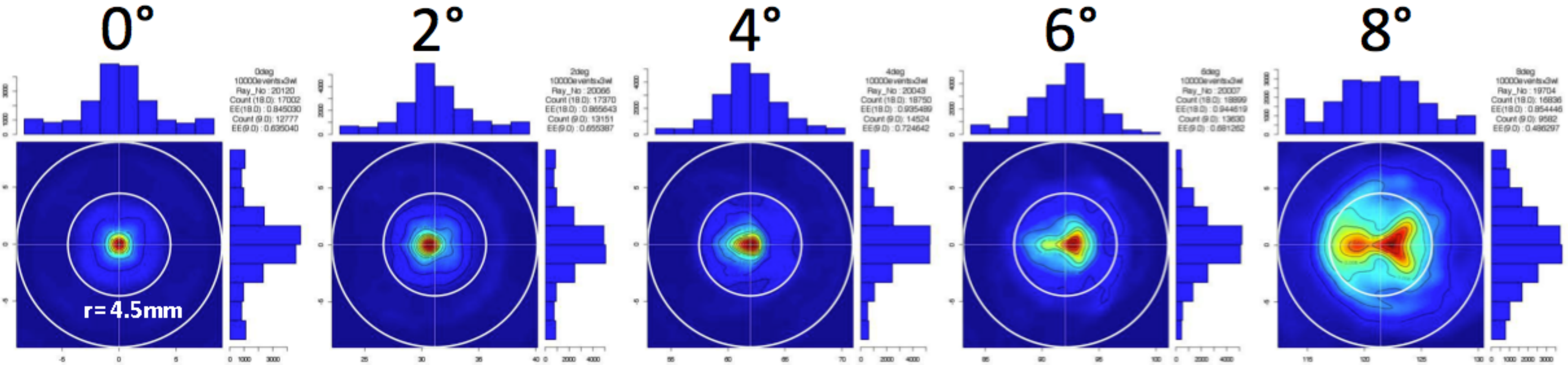}
  \caption{Spot diagrams of the TA-EUSO between FOV 0$^\circ$ and 8$^\circ$. Inner circle radius is 4.5 mm. Outer circle radius is 9 mm. }
  \label{TA-EUSO_spot}
\end{figure}

The requirements and the designed values of the TA-EUSO optics are shown in Table \ref{table_TA-EUSO_req}. The angular resolution has to be better than the Telescope Array Fluorescence Detector (TA-FD). The field of view of a TA-FD pixel is 1$^\circ$. It corresponds to $\sim$9.6$\times$9.6 pixels for the TA-EUSO. The RMS spot size of TA-EUSO should therefore be smaller than $\sim$9.6$\times$9.6 pixels to resolve a TA-FD pixel. Our best design results in a spot size, in case of the two lenses system, of 9 mm RMS, which corresponds to $\sim$3$\times$3 pixels of PDM. Specifications of the the TA-EUSO optics design is shown in Table \ref{table_TA-EUSO_spe}.\\
The front side of the front lens is a plane surface to clean dust from the surface easily. This is particularly important since TA-EUSO is going to be deployed at the Telescope Array site, where sand and dust might accumulate on the front surface of the front lens. The back side of the front lens has a Fresnel structure. The back side of the rear lens is again a plane surface that can be easily cleaned. The front side of the rear lens has a Fresnel structure.\\
The design was implemented for temperatures of front and rear lenses of 20$^\circ$C. This optics is insensitive to changes of temperature between 0$^\circ$C and 40$^\circ$C.
Full FOV of TA-EUSO is $\pm$8$^\circ$ for two PDMs. The spot diagrams of the TA-EUSO between FOV 0$^\circ$ and 8$^\circ$ are shown in Fig. \ref{TA-EUSO_spot}.\\
The loss factors of the optics system represent the losses of light on-axis (Table \ref{table_TA-EUSO_los}). The obscurations due to the back-cuts at off-axis angles are implemented in the efficiency calculations. Photon collection efficiency of the TA-EUSO is shown in Table \ref{table_TA-EUSO_pce}.


\begin{table}[!h]
\begin{center}
\begin{tabular}{|l|c|c|}
\hline              &Requirements        & Design result \\ \hline
Optical system &  2 or 3 lenses sys.&   2 lenses sys.\\ \hline
Focal length & - & 1562.18mm \\ \hline
FOV for a PDM & $> \pm$4$^\circ$ & $\pm$4$^\circ$ \\ \hline
RMS spot size & $< $ PMT size & 9 mm @ 0$^\circ$  \\ \hline
Entrance pupile& $>$ 0.785 m$^2$ &  0.95 m$^2$  \\ \hline
Base shape of lens & Flat type & Flat type \\ \hline
Lens material & PMMA-000 & PMMA-000 \\ \hline
Lens thickness & $< $ 10 mm & 8mm \\ \hline
FS curvature &2505 mm&2505 mm  \\ \hline
\end{tabular}
\caption{The TA-EUSO optics design requirement parameters and designed values.}
\label{table_TA-EUSO_req}
\end{center}
\end{table}


\begin{figure}[!h]
  \centering
  \includegraphics[width=0.48\textwidth]{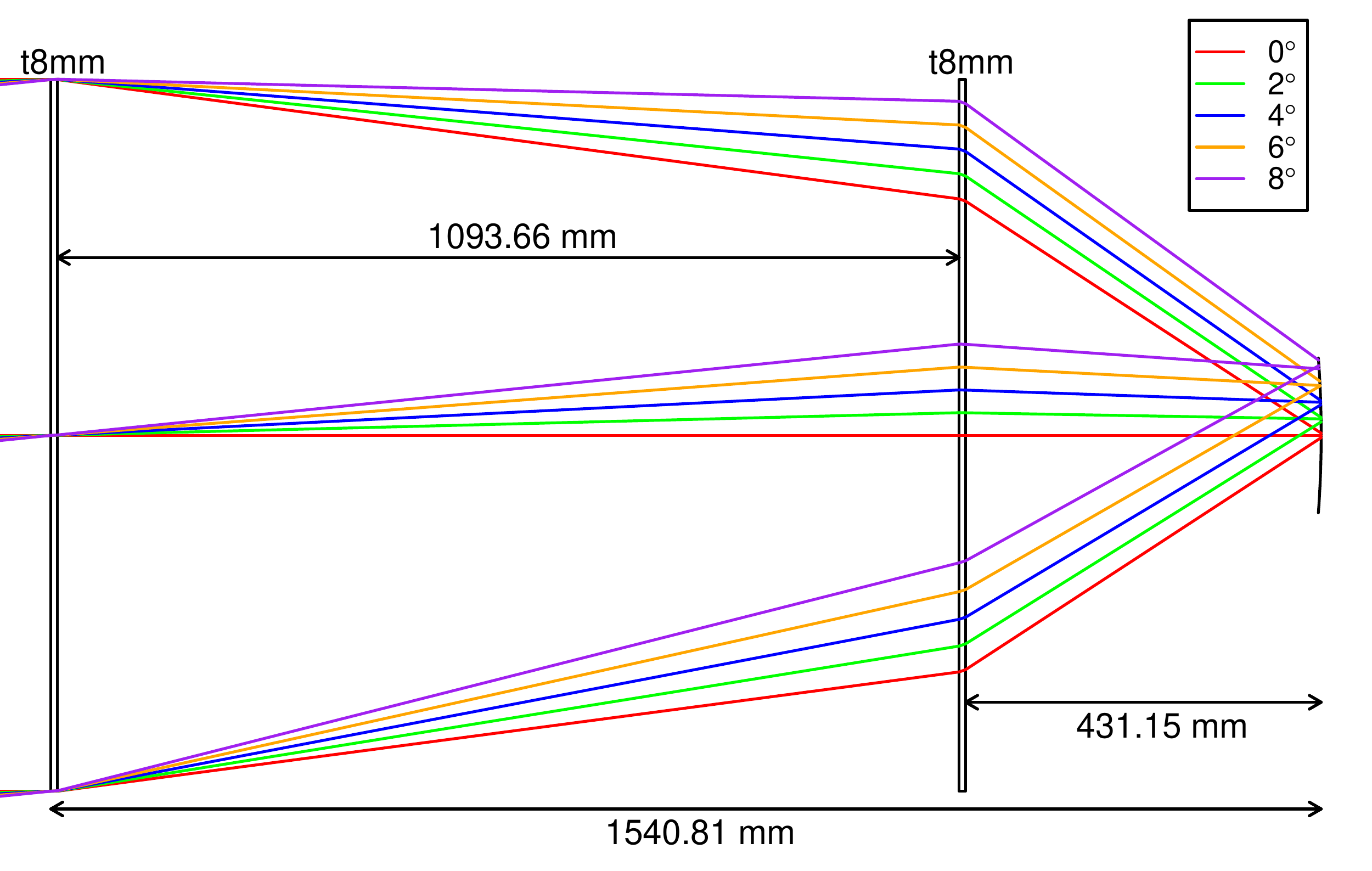}
  \caption{The optics design of the TA-EUSO.}
  \label{TA-EUSO_optics}
\end{figure}


\begin{table}[h]
\centering
\begin{tabular}{|l|c|c|c|c|}
\hline
& \multicolumn{2}{|c|}{ Front lens} & \multicolumn{2}{|c|}{ Rear lens} \\ \hline 
Material  &\multicolumn{2}{|c|}{ PMMA-000} & \multicolumn{2}{|c|}{ PMMA-000} \\ \hline 
Lens shape  &\multicolumn{2}{|c|}{ 1m square} & \multicolumn{2}{|c|}{ 1m square} \\ \hline 
Thickness  &\multicolumn{2}{|c|}{ 8 mm } & \multicolumn{2}{|c|}{ 8 mm } \\ \hline 
Weight [kg]  &\multicolumn{2}{|c|}{ 9.6 } & \multicolumn{2}{|c|}{9.6 } \\ \hline 
\multirow{2}{*}{Surface type}  & Front& Back& Front& Back\\ \cline{2-5} 
 & Plane & Fresnel  & Fresnel & Plane\\ \hline 
\end{tabular}
\caption{Specifications of the TA-EUSO optics.}
\label{table_TA-EUSO_spe}
\end{table}


\begin{table}
\begin{center}
\begin{tabular}{|l|c|c|}
\hline
Item & Loss factor (\%) \\ \hline
Suface roughness & 2\\ \hline
Diffractive structure depth error & 0\\ \hline
Fresnel facet back-cuts error& 4\\ \hline
Support structure obscuration  & 1\\ \hline
Total & 7\\ \hline
\end{tabular}
\caption{Loss factors of the TA-EUSO. }
\label{table_TA-EUSO_los}
\end{center}
\end{table}


\begin{table}
\begin{center}
\begin{tabular}{|c|c|}
\hline
Field angle & Efficiency \\ \hline
0$^\circ$ & 40\% \\ \hline
2$^\circ$ & 41\% \\ \hline
4$^\circ$ & 46\% \\ \hline
6$^\circ$ & 43\% \\ \hline
8$^\circ$ & 30\% \\ \hline
\end{tabular}
\caption{The photon collection efficiency of the TA-EUSO for three wavelengths (337, 357 391nm). These efficiencies are taken into account of the surface reflection, the material absorption and the loss factors of Table \ref{table_TA-EUSO_los}, except for the support structure obscuration.}
\label{table_TA-EUSO_pce}
\end{center}
\end{table}

\subsection{The EUSO-Balloon optics design and its performance}
\begin{figure}[!h]
  \centering
  \includegraphics[width=0.48\textwidth]{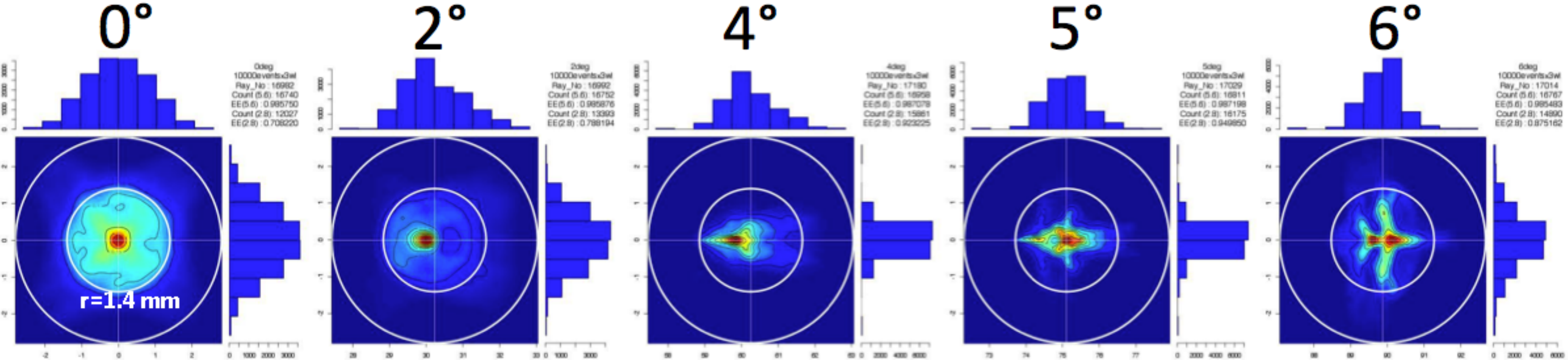}
  \caption{Spot diagrams of the EUSO-Balloon between FOV 0$^\circ$ and 6$^\circ$. Inner circle radius is 1.4 mm. Outer circle radius is 2.8 mm. }
  \label{Bal_spot}
\end{figure}

The requirements and the designed values of the EUSO-Balloon optics are shown in Table \ref{table_EUSO-Balloon_req}. The RMS spot size must be smaller than the PMT pixel size (= 2.8 mm). The EUSO-Balloon optics consists of two TA-EUSO lenses and an additional flat diffractive lens. The diffractive lens is placed between the front lens and the rear lens to correct for chromatic aberration and to obtain a small RMS spot size ($<$ 2.8mm pixel size). The specifications of the EUSO-Balloon optics design are shown in Table \ref{table_EUSO-Balloon_sep}. 
The design was done for operating temperatures of the front lens equal to $-$40$^\circ$C, for the middle lens and the rear lens equal to 10$^\circ$C. The optics design takes into account only changing the temperature of the rear lens, since the focusing power of the first lens is weak and the diffractive lens is not sensitive to changes of temperature. The rear lens is connected with the PDM electronics components by thermal radiation heating. We should study the thermal condition between the rear lens and the PDM components, and then, adjust the focusing point moving the PDM.
The Full FOV of EUSO-Balloon is $\pm$6$^\circ$ for the PDM. The spot diagrams of the EUSO-Balloon, between FOV 0$^\circ$ and 6$^\circ$, are shown in Fig \ref{Bal_spot}. The loss factor of the optics system is shown in Table \ref{table_TEUSO-Balloon_los}. PCE of the EUSO-Balloon is shown in Table \ref{table_EUSO-Balloon_pce}. 

\begin{table}[!h]
\begin{center}
\begin{tabular}{|l|c|c|}
\hline                  &Requirements        & Design result \\ \hline
Optical system &  3 lenses sys.&   3 lenses sys.\\ \hline
Focal length & - & 1620.717mm  \\ \hline
FOV for a PDM & $> \pm$6$^\circ$ & $\pm$6$^\circ$ \\ \hline
RMS spot size & $< $ pixel size & 1.6 mm @ 0$^\circ$  \\ \hline
Entrance pupile& $>$ 0.785 m$^2$ &  0.95 m$^2$  \\ \hline
Base shape of lens & Flat type & Flat type \\ \hline
Lens material & PMMA-000 & PMMA-000 \\ \hline
Lens thickness & $< $ 10 mm & 8mm \\ \hline
FS curvature &2505 mm&2505 mm  \\ \hline
\end{tabular}
\caption{The EUSO-Balloon optics design requirement parameters and designed values.}
\label{table_EUSO-Balloon_req}. 
\end{center}
\end{table}

\begin{figure}[!h]
  \centering
  \includegraphics[width=0.49\textwidth]{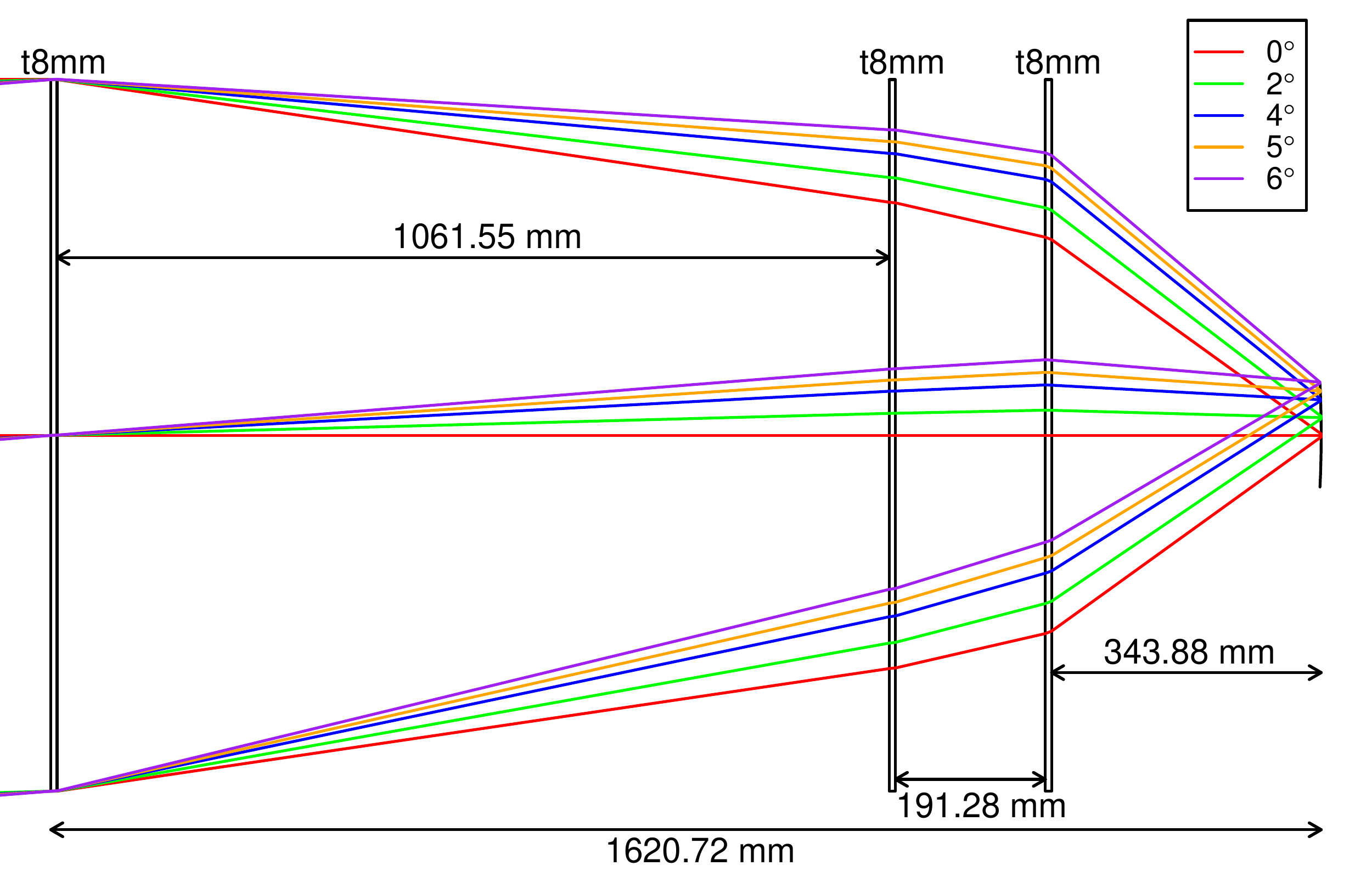}
  \caption{The optics design of the EUSO-Balloon.}
  \label{balloo_optics}
\end{figure}

\begin{table*}
\centering
\begin{tabular}{|l|c|c|c|c|c|c|}
\hline
& \multicolumn{2}{|c|}{ Front lens} & \multicolumn{2}{|c|}{ Middle lens}  & \multicolumn{2}{|c|}{ Rear lens} \\ \hline 
Material  &\multicolumn{2}{|c|}{ PMMA-000} & \multicolumn{2}{|c|}{ PMMA-000} & \multicolumn{2}{|c|}{ PMMA-000} \\ \hline 
Lens shape  &\multicolumn{2}{|c|}{ 1m square} & \multicolumn{2}{|c|}{ 1m square}& \multicolumn{2}{|c|}{ 1m square} \\ \hline 
Thickness  &\multicolumn{2}{|c|}{ 8 mm } & \multicolumn{2}{|c|}{ 8 mm } & \multicolumn{2}{|c|}{ 8 mm } \\ \hline 
Weight [kg]  &\multicolumn{2}{|c|}{ 9.6 } & \multicolumn{2}{|c|}{9.6 } & \multicolumn{2}{|c|}{9.6 } \\ \hline 
\multirow{2}{*}{Surface type}  & Front & Back & Front& Back& Front& Back\\ \cline{2-7} 
　& Plane & Fresnel  & Plane & Diffractive & Fresnel & Plane\\ \hline 
\end{tabular}
\caption{Specifications of the EUSO-Balloon optics}
\label{table_EUSO-Balloon_sep}
\end{table*}

\begin{table}
\begin{center}
\begin{tabular}{|l|c|c|}
\hline
Items & Loss factor (\%) \\ \hline
Suface roughness & 3\\ \hline
Diffractive structure depth error & 7\\ \hline
Fresnel facet back-cuts error& 4\\ \hline
Support structure obscuration  & 4\\ \hline
Total & 18\\ \hline
\end{tabular}
\caption{Loss factors of the EUSO-Balloon. }
\label{table_TEUSO-Balloon_los}
\end{center}
\end{table}


\begin{table}[!h]
\begin{center}
\begin{tabular}{|c|c|}
\hline
Field angle & Efficiency \\ \hline
0$^\circ$ & 35\% \\ \hline
2$^\circ$ & 39\% \\ \hline
4$^\circ$ & 46\% \\ \hline
5$^\circ$ & 47\% \\ \hline
6$^\circ$ & 43\% \\ \hline
\end{tabular}
\caption{The photon collection efficiency of the EUSO-Balloon for three wavelengths (337, 357 391nm). These efficiencies are taken into account of the surface reflection, the material absorption and the loss factors of Table \ref{table_TEUSO-Balloon_los}, .except for the support structure obscuration.}
\label{table_EUSO-Balloon_pce}
\end{center}
\end{table}

\section{Conclusions}
We have developed feasible optics designs for the TA-EUSO and EUSO-Balloon projects based on the JEM-EUSO optics development. 
These designs are simplified version of the JEM-EUSO baseline optics design, and are very similar to the central portion of the JEM-EUSO lenses. These designs are taking into account the manufacturability of the lenses under strong time constraint. 
The TA-EUSO and the EUSO-Balloon shared the design of the front lens and the rear lens. The EUSO-Balloon optics is a suitable design to demonstrate and to test a representative of the whole JEM-EUSO instrument. If we manufacture the middle lens (the diffractive lens) for TA-EUSO, TA-EUSO will acquire the same focusing power of EUSO-Balloon. Its optical system is able to observe the lateral distribution image of the inside of a TA-FD pixel with high resolution. 

The optics design parameters for the main JEM-EUSO missions will be improved based on the real production of the optics of these two pathfinder experiments.

\vspace*{0.5cm}
{
\footnotesize{{\bf Acknowledgment:}{This work was partially supported by Basic Science Interdisciplinary 
Research Projects of RIKEN and JSPS KAKENHI Grant (22340063, 23340081, and 
24244042), by the Italian Ministry of Foreign Affairs, General Direction 
for the Cultural Promotion and Cooperation, by the 'Helmholtz Alliance 
for Astroparticle Physics HAP' funded by the Initiative and Networking Fund 
of the Helmholtz Association, Germany, and by Slovak Academy 
of Sciences MVTS JEM-EUSO as well as VEGA grant agency project 2/0081/10.
The Spanish Consortium involved in the JEM-EUSO Space
Mission is funded by MICINN under projects AYA2009-
06037-E/ESP, AYA-ESP 2010-19082, AYA2011-29489-C03-
01, AYA2012-39115-C03-01, CSD2009-00064 (Consolider MULTIDARK)
and by Comunidad de Madrid (CAM) under project S2009/ESP-1496.
}}

}
\clearpage

%% file: icrc2013-1040.tex




\title{Manufacturing of the TA-EUSO and the EUSO-Balloon lenses}

\shorttitle{Manufacturing of the TA-EUSO and the EUSO-Balloon lenses}

\authors{
Yousuke Hachisu$^{1}$,
Yoshihiro Uehara$^{1}$,
Hitoshi Ohomori$^{1}$,
Yoshiyuki Takizawa$^{1}$,
Alessandro Zuccaro Marchi$^{1}$,
Toshikazu Ebisuzaki$^{1}$
for the JEM-EUSO Collaboration.
}

\afiliations{
$^1$ RIKEN, Japan \\
}

\email{hachisu@optics.gr.jp} 

\abstract{The TA-EUSO and EUSO-Balloon are the pathfinder experiments for the JEM-EUSO mission. The TA-EUSO observes fluorescence light from the cosmic ray air showers at the Telescope Array (TA) site with better resolution than the fluorescence detector of TA. The EUSO-Balloon is designed to observe extensive air shower from a stratospheric balloon. The optics team has developed the Bread Board Model (BBM) of the JEM-EUSO optics between 2007 and 2011. We have reached the fundamental technique of the meter scale fresnel lens manufacturing through the BBM lens manufacturing. The TA-EUSO optics consists of two 1m square flat Fresnel lenses, built in UV grade Poly Methyl Meth-Acrylate (PMMA). The TA-EUSO lenses and the EUSO-Balloon middle lens have been manufactured successfully. The manufacturing of  the EUSO-Balloon front lens and rear lens will be completed within summer 2013. The EUSO-Balloon optics consists of two 1m square flat Fresnel PMMA lenses and of a 1m square flat diffractive PMMA lens. This paper describes the details of the manufacturing the lenses focusing on the quality of surface roughness.}


\maketitle

\section{Introduction}

The Extreme Universe Space Observatory on the Japanese Experiment Module (JEM-EUSO) of the International Space Station (ISS) is being designed to be attached to the Exposure Facility on the JEM of ISS, and observe the earth atmosphere with a field of view of 60$^\circ$ \cite{biba:EUSOperf}. JEM-EUSO is designed to acquire the UV images produced by extensive air showers propagating in atmosphere and exciting nitrogen atoms. This allows determining the energy and trajectory of the incoming primary cosmic ray. The mission will help to discover the origin of cosmic rays and the acceleration processes responsible for producing these extremely energetic particles while potentially opening a door to new physics studies. 

\begin{figure}[!h]
  \centering
  \includegraphics[width=0.43\textwidth]{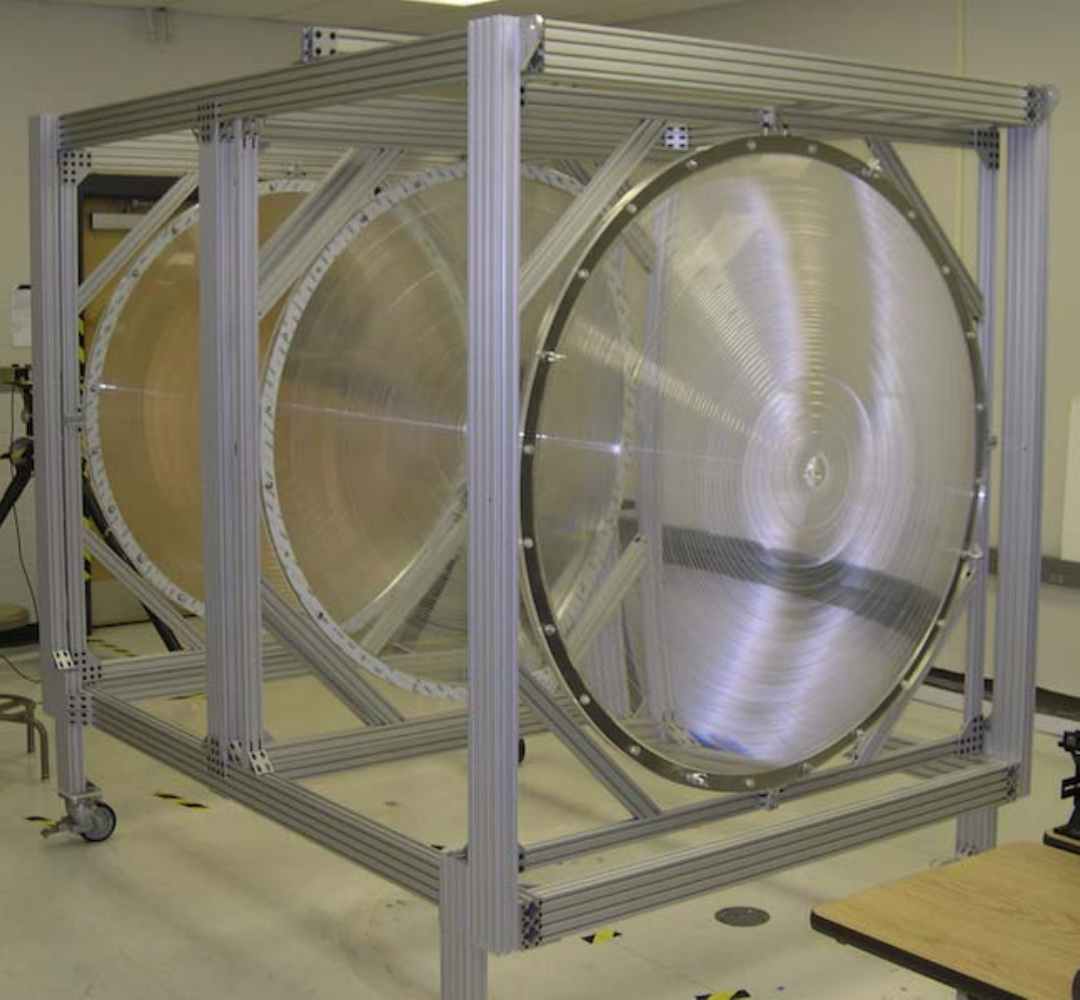}
  \caption{The bread board model (BBM) of the JEM-EUSO optics. The lens diameter is 1.5 m.}
  \label{BBM_optics}
 \end{figure}

The lens manufacturing for the JEM-EUSO optics is one of key technologies to observe the extensive air shower emission from space. The optics team has developed the Bread Board Model (BBM) of the JEM-EUSO optics between 2007 and 2011 [Fig  \ref{BBM_optics}]. 
The BBM optics consists of three lenses. Each lens replicates the central portion of the corresponding JEM-EUSO lens. Its size is 1.5 m in diameter. We have reached the fundamental technique of the meter scale fresnel lens manufacturing through the BBM lens manufacturing. The TA-EUSO and EUSO-Balloon projects are the pathfinder experiments for the JEM-EUSO mission, essential to test from the manufacturing of several key components of the telescope and the observational technique. The details of both experiments are described in \cite{biba:TA-EUSO} and \cite{biba:Balloon}. In this paper, we focus on the lens manufacturing. These lenses are manufactured by the same machine and technique which will be used for the JEM-EUSO development. We can verify the manufacturing quality and develop new manufacturing technology through the TA-EUSO and EUSO-Balloon lens fabrications. We have completed the manufacturing of the TA-EUSO lenses and the EUSO-Balloon middle lens in early May 2012. \\
 \section{Optics designs}
In order to optimize the manufacturing machine time, avoiding a conflict schedule between the two project it was decided that the final design of both optics would share the front and rear lens design. The EUSO-Balloon optics consists  of two TA-EUSO lenses and an additional flat diffractive lens. The diffractive lens is placed between the front lens and the rear lens to obtain a small RMS spot size ($<$ 2.8mm pixel size). Lens material is the UV transmittance grade PMMA (PMMA-000, Mitsubishi Rayon Co., LTD.). All lenses  are 1m by 1m square shape and their thickness is 8 mm. This size was decided with the observational performance and commercial availability of PMMA base material. The details of both optics design is described in \cite{biba:optics}. \\
\\

\subsection{The TA-EUSO optics design}
\begin{figure}[!h]
  \centering
  \includegraphics[width=0.45\textwidth]{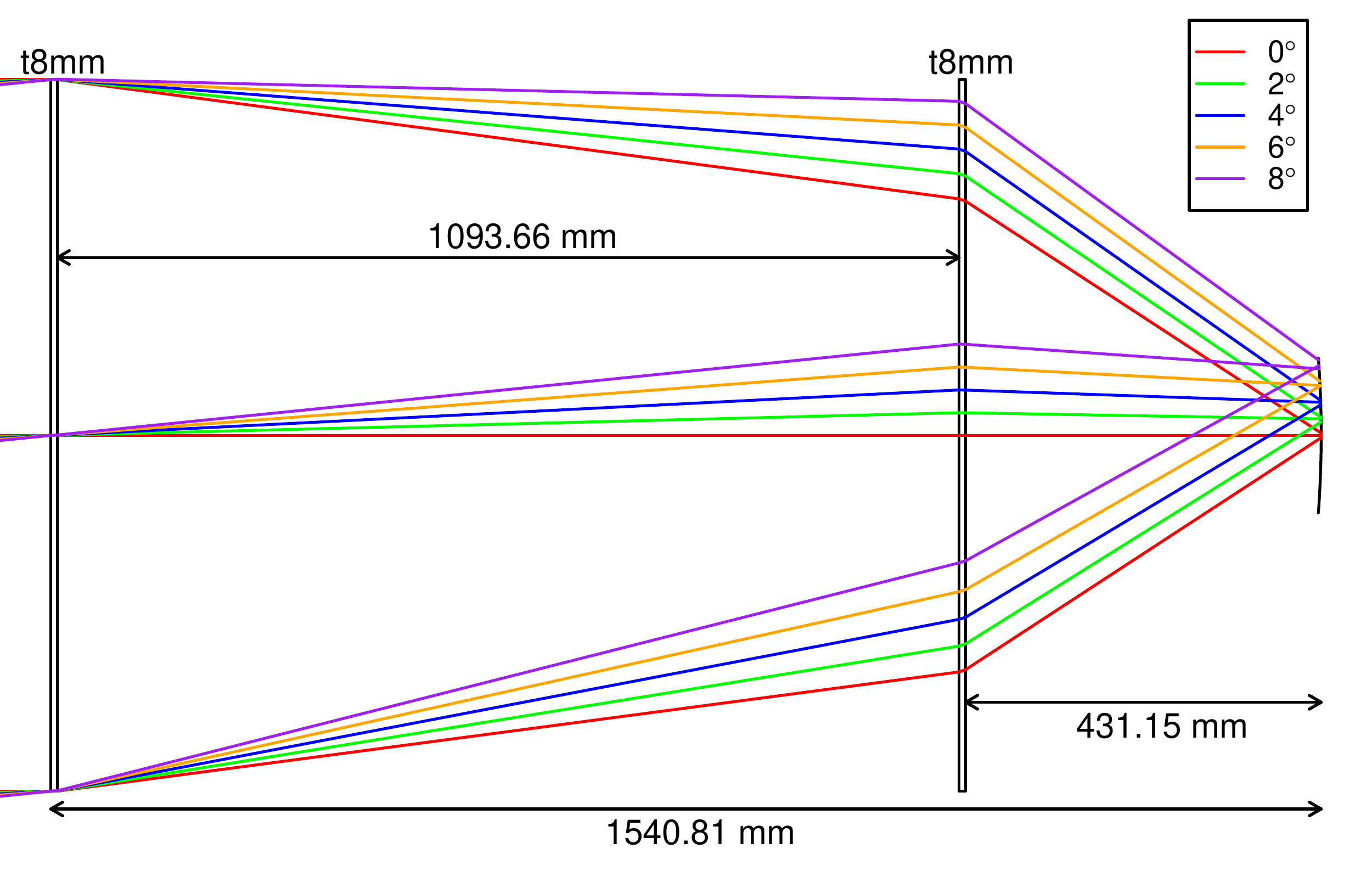}
  \caption{The optics design of the TA-EUSO.}
  \label{dTA-EUSO_optics}
\end{figure}

The TA-EUSO optics consists of two flat fresnel lenses, a front and a rear lens. As we already said, the EUSO-Balloon optics shares the same design for the front and rear lenses. The optics design is shown in Fig \ref{dTA-EUSO_optics} , while details on the surface features are shown in Table \ref{table_TA_lens}. 

\begin{table*}[!t]
\centering
\begin{tabular}{|c|c|c|c|c|}            \hline
\multirow{2}{*}{}& \multicolumn{2}{|c|}{Front lens}     & \multicolumn{2}{|c|}{Rear lens} \\ \cline{2-5} 
&Front side& Back side &Front side& Back side\\ \hline
Material & \multicolumn{2}{|c|}{PMMA-000} & \multicolumn{2}{|c|}{PMMA-000} \\ \hline
External form & \multicolumn{2}{|c|}{1m x 1m} & \multicolumn{2}{|c|}{1m x 1m} \\ \hline
Groove height & Plane&1mm & 2mm &  Plane\\ \hline
Groove width   & Plane& 0.5 mm - 47.79 mm & 1.65 mm - 36.71mm&  Plane\\ \hline
\end{tabular}
\caption{Characteristics of lens surfaces of the TA-EUSO optics.}
\label{table_TA_lens}
\end{table*}

\subsection{The EUSO-Balloon optics design}
\begin{figure}[!h]
  \centering
  \includegraphics[width=0.45\textwidth]{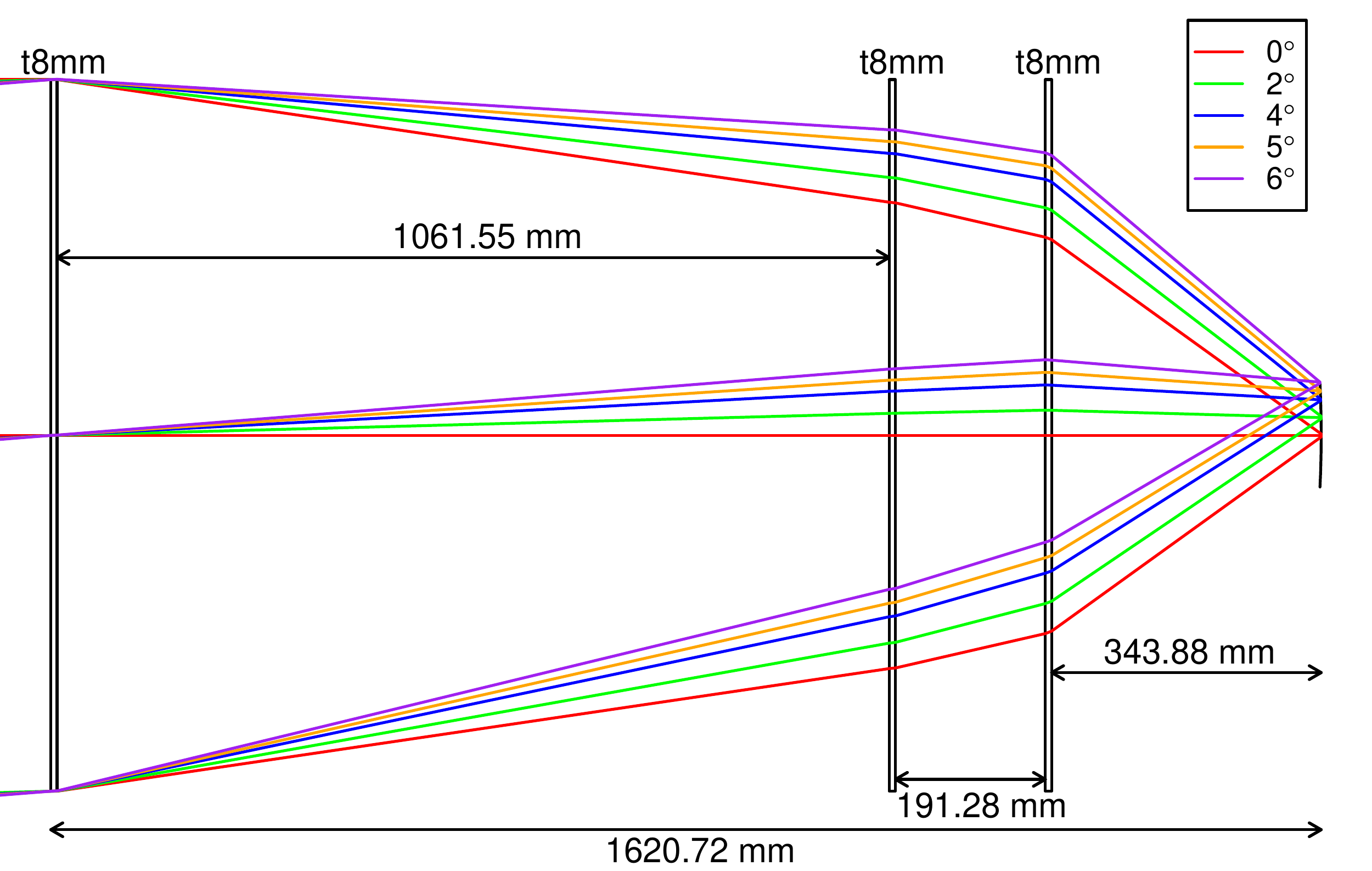}
  \caption{The optics design of theEUSO-Balloon.}
  \label{dballoo_optics}
\end{figure}

The optics design of the EUSO-Balloon optics is shown in Fig \ref{dballoo_optics}. A third middle diffractive lens is now included to correct achromatic aberration. The main features of lens surfaces are shown in Table \ref{table_balloon_lens}. 

\begin{table*}[!t]
\centering
\begin{tabular}{|c|c|c|c|c|c|c|}            \hline
\multirow{2}{*}{}& \multicolumn{2}{|c|}{Front lens}     & \multicolumn{2}{|c|}{Middle lens}    & \multicolumn{2}{|c|}{Rear lens} \\  \cline{2-7} 
&Front side& Back side &Front side& Back side&Front side& Back side\\ \hline
Material & \multicolumn{2}{|c|}{PMMA-000} & \multicolumn{2}{|c|}{PMMA-000} & \multicolumn{2}{|c|}{PMMA-000} \\ \hline
External form & \multicolumn{2}{|c|}{1m x 1m} & \multicolumn{2}{|c|}{1m x 1m} & \multicolumn{2}{|c|}{1m x 1m}\\ \hline
Groove height & Plane&1mm & Plane & 0.7$\mu$m& 2mm &  Plane \\ \hline
Groove width   & Plane& 0.003mm-2.63mm & Plane& 0.5mm-47.79mm & 1.65mm-36.71mm&  Plane\\ \hline
\end{tabular}
\caption{Characteristics of lens surfaces of the EUSO-Balloon optics.}
\label{table_balloon_lens}
\end{table*}

\section{Manufacturing lenses}
\begin{figure}[!h]
  \centering
  \includegraphics[width=0.45\textwidth]{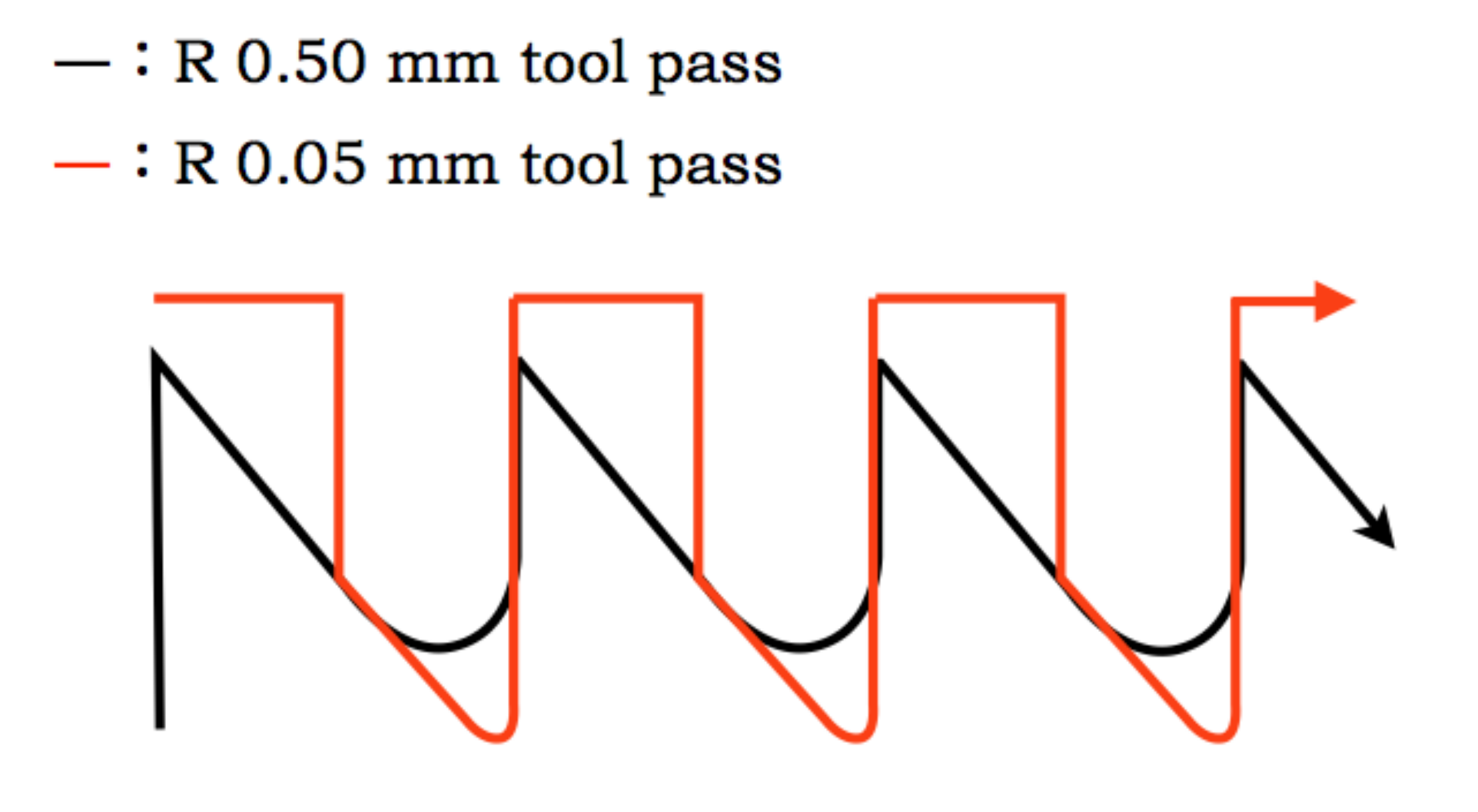}
  \caption{A schematic picture of the manufacturing of the groove edge with the use of the 0.05 mm radius tool. Black line shows the 0.5 mm radius tool pass. Red line is a pass of the 0.05mm tool pass. }
  \label{finecut}
\end{figure}

According to the requirement, the surface roughness of all lenses should be smaller than 20 nm RMS. This is the most important requirement to meet the observation performance. In fact, the surface roughness has a direct impact on to overall transmission. We have developed a technology to obtain smooth surfaces through the JEM-EUSO BBM lenses manufacturing. In the manufacturing process of the TA-EUSO and EUSO-Balloon lenses, we introduced a new technique to obtain better transmission than the one achieved for the BBM lenses. The BBM lenses manufacturing used only 0.5 mm radius diamond tool bite which can produce a smooth surface, with roughness smaller than 20 nm RMS. However, at large manufacturing radii, problem arose with the sharpness of the edge of the grooves, reducing severely the transmission of the BBM. This would have implied, for the rear lens of TA-EUSO, at radius about 550 mm, a fresnel groove slope angle of about 45${}^\circ$  and a groove height of 2 mm. In this case, the loss factor due to the 0.5 mm radius tool would have been 42\%. This is why, to eliminate this problem, we have developed a long-life 0.05 mm radius diamond tool. This sharp tool makes sharp edge of the groove [Fig \ref{finecut}].
The loss factor improves from $\sim$ 42\% to $\sim$ 4\% for the groove slope angle of 45${}^\circ$ and groove height of 2 mm. The surface roughness is measured by a mobile Atomic Force Microscope (AFM). Due to the size of the lens, the measurement values could only be obtained in a few points. In the past, we used optical interferometer measurements as evaluation criteria.  However, a recent study established a correlation with the AFM measurement result. We concluded that Rq 35 nm measured by AFM corresponds to RMS 20 nm measured by  optical interferometer.

\subsection{Manufacturing the TA-EUSO front lens}
\begin{figure}[!h]
  \centering
  \includegraphics[width=0.35\textwidth]{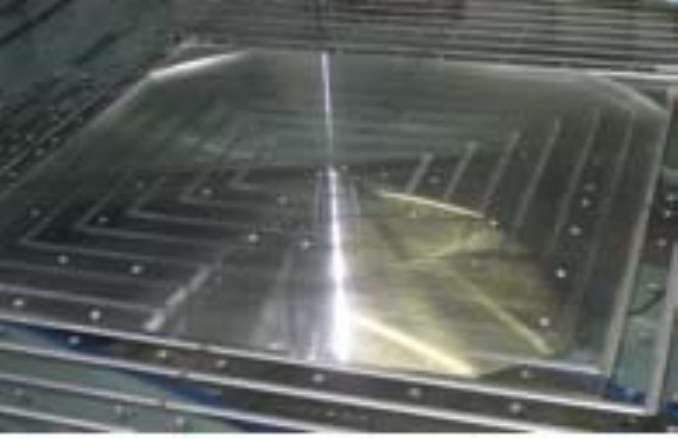}
  \caption{A photo of the TA-EUSO front lens on the machine after manufacturing.}
  \label{TA-front-lens-pict}
\end{figure}

The TA-EUSO front lens was manufactured using the machine parameters shown in Table \ref{table_TA_front_lens}. The front side of the front lens is a plane surface to clean dust from the surface easily. This is particularly important since TA-EUSO is going to be deployed at the Telescope Array site, where sand and dust might accumulate on the front surface of the front lens. The back side of the front lens has a Fresnel structure. The depth of groove is 1 mm. A photo of the TA-EUSO front lens on the machine after manufacturing is shown in Fig \ref{TA-front-lens-pict}. For both front and rear side, typical surface roughness as low as 12mm RMS has been achieved.

\begin{figure}[!h]
  \centering
  \includegraphics[width=0.38\textwidth]{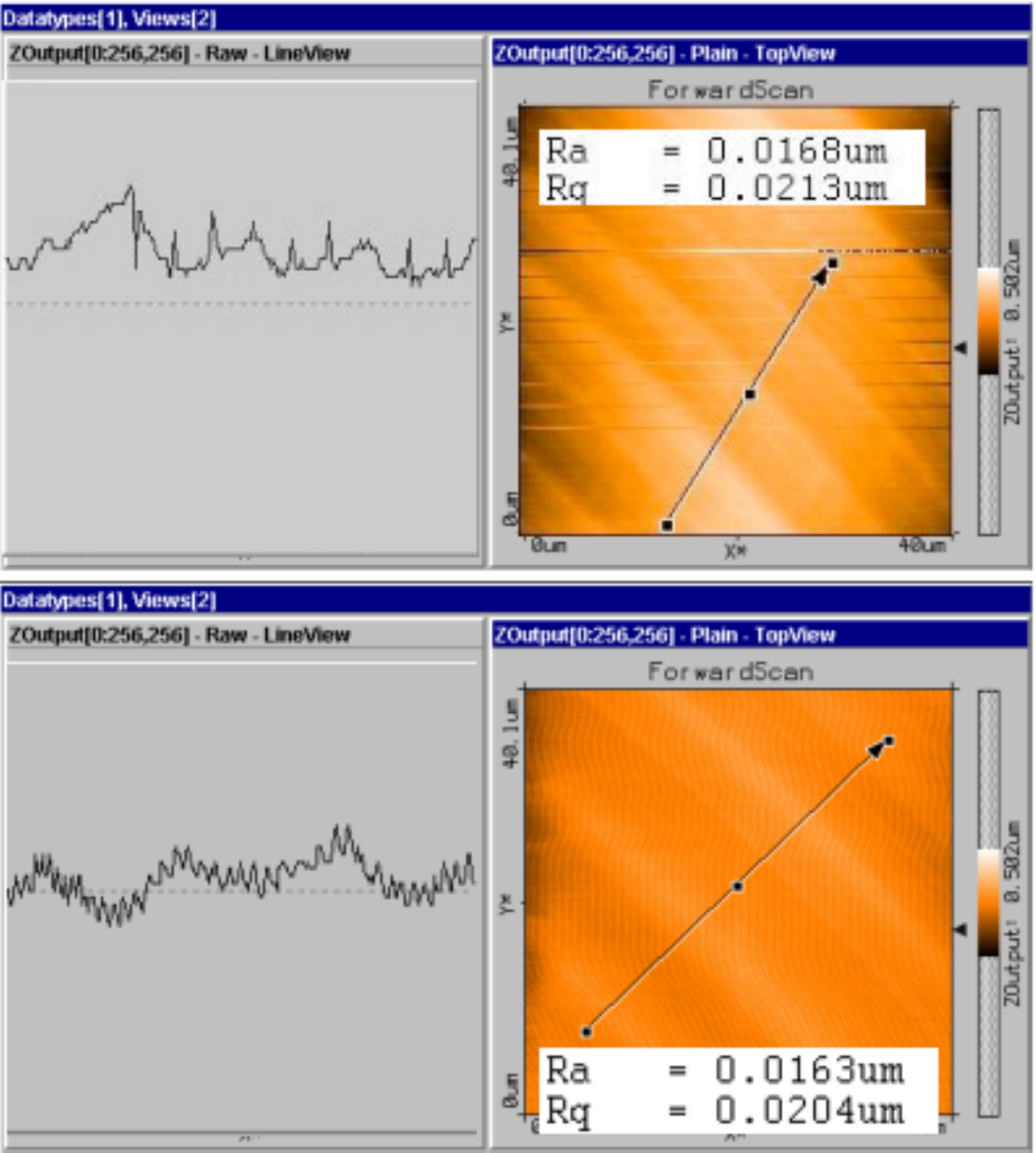}
  \caption{Surface roughness measurement by using a mobile AFM. Top panel: the front side of the front lens, Bottom  panel: the back side of the front lens}
  \label{TA-front-mes}
\end{figure}

\begin{table}[!h]
\centering
\begin{tabular}{|c|c|}            \hline
Tip radius of tool bit [mm] & 0.5, 0.05\\ \hline
Tool feeding speed [mm/min] & 0.15 $\sim$ 20\\ \hline
Depth of cut [mm/pass] & 0.01 $\sim$ 0.050\\ \hline
\end{tabular}
\caption{Manufacturing parameters for the TA-EUSO front lens.}
\label{table_TA_front_lens}
\end{table}

\subsection{Manufacturing the TA-EUSO rear lens}
\begin{figure}[!h]
  \centering
  \includegraphics[width=0.35\textwidth]{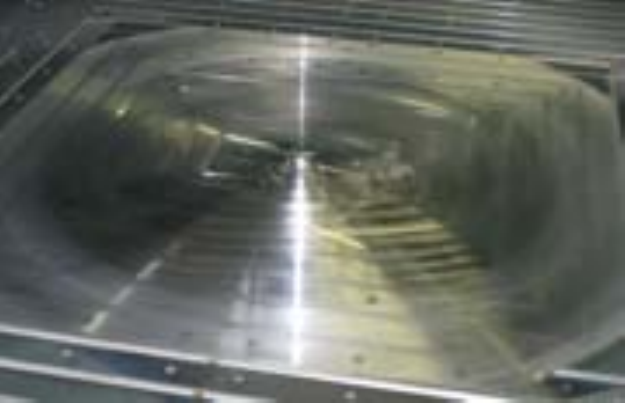}
  \caption{The TA-EUSO rear lens on the machine after manufacturing.}
  \label{TA-rear-lens-pict}
\end{figure}

TA-EUSO rear lens was manufactured using the machine parameters shown in Table \ref{table_TA_rear_lens}. The back side of the rear lens is again a plane surface that can be easily cleaned. The front side of the rear lens has a Fresnel structure. The depth of groove is 2 mm. A photo of the TA-EUSO rear lens on the machine after manufacturing is shown in Fig \ref{TA-rear-lens-pict}. Typical surface roughness of front side and back side are RMS $\sim$16 nm and RMS $\sim$11 nm.\\

\begin{figure}[!h]
  \centering
  \includegraphics[width=0.33\textwidth]{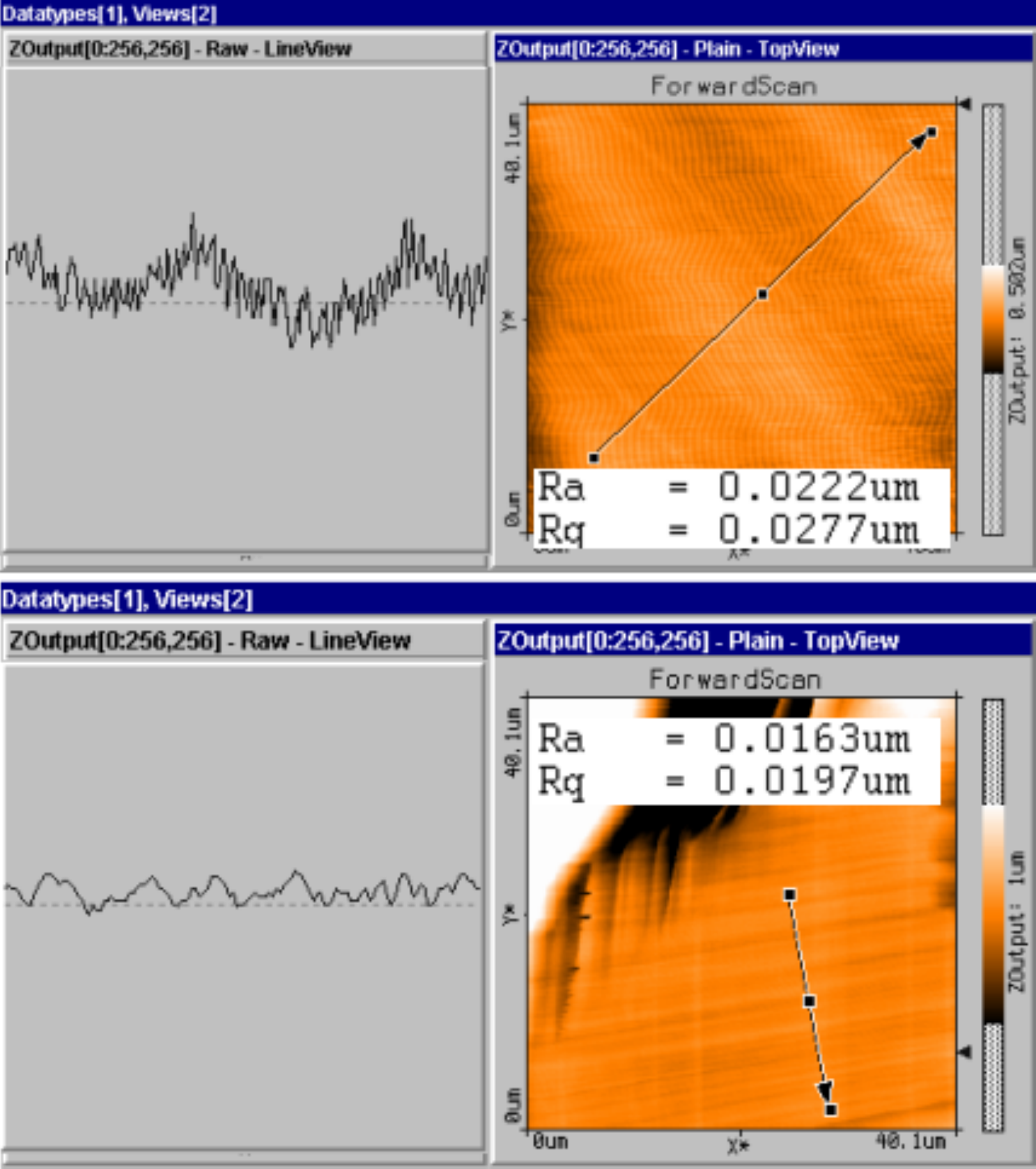}
  \caption{Surface roughness measurement using a mobile AFM. Top panel: the front side of the rear lens, bottom  panel: the back side of the rear lens.}
  \label{TA-rear-mes}
\end{figure}

\begin{table}[!h]
 \centering
\begin{tabular}{|c|c|}            \hline
Tip radius of tool bit [mm] &  0.5, 0.05\\ \hline
Tool feeding speed [mm/min] & 0.10 $\sim$ 15\\ \hline
Depth of cut [mm/pass] & 0.005 $\sim$ 0.030\\ \hline
\end{tabular}
\caption{Manufacturing parameters for the TA-EUSO rear lens.}
\label{table_TA_rear_lens}
\end{table}

\subsection{Manufacturing the EUSO-Balloon middle lens}
\begin{figure}[!h]
  \centering
  \includegraphics[width=0.4\textwidth]{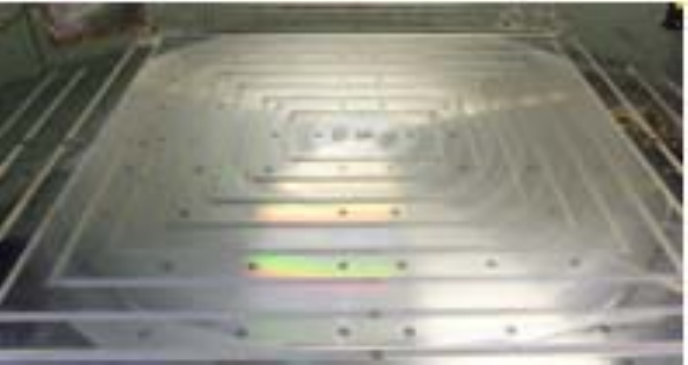}
  \caption{A photo of the EUSO-Balloon middle lens on the machine after manufacturing. Rainbows patterns due to diffraction structure are visible in this lens area.  }
  \label{Bal-mid-lens-pict}
\end{figure}

The EUSO-Balloon middle lens was manufactured using the machine parameters shown in Table \ref{table_Bal_mid_lens}. The front side of the middle lens is a plane surface. The back side of the middle lens has a diffractive structure. The depth of groove is 0.7 $\mu$m, which corresponds to the PMMA refractive index at wavelength 357 nm.  We can accept a manufacturing error of the depth of the diffractive groove around $\pm$10\%, that is  0.7 $\pm$ 0.07$\mu$m. A picture of the EUSO-Balloon middle lens  on the machine after manufacturing is shown in Fig \ref{Bal-mid-lens-pict}. Typical surface roughness of front side is RMS $\sim$11 nm. Typical diffractive groove hight is $\sim$ 0.74$\mu$m.

\begin{figure}[!h]
  \centering
  \includegraphics[width=0.38\textwidth]{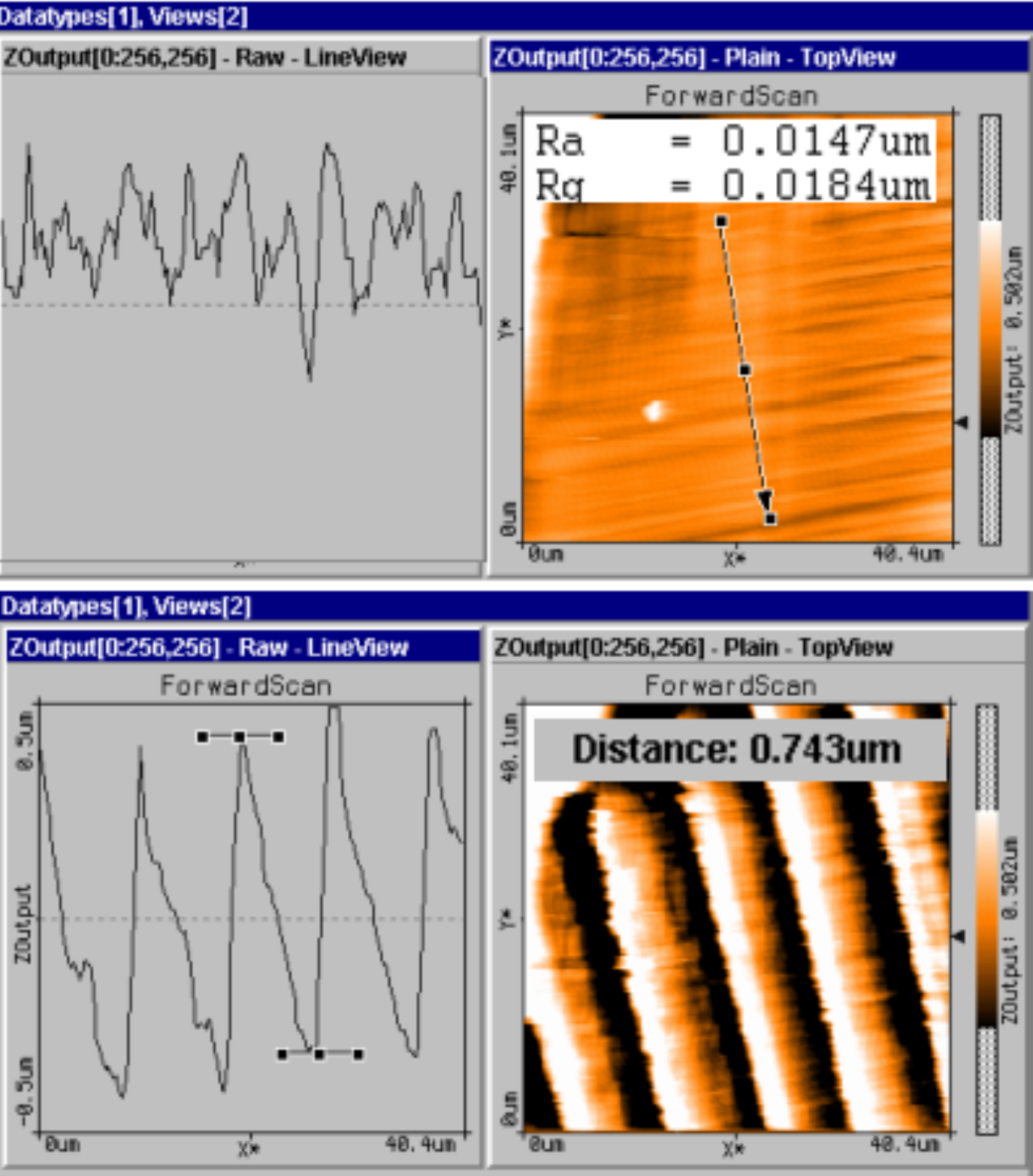}
  \caption{Surface roughness and diffractive structure measurement using a mobile AFM. Top panel: the front side of the middle lens, bottom  panel: the back side of the middle lens (diffractive surface).}
  \label{Bal-mid-mes}
\end{figure}

\begin{table}[!h]
\begin{tabular}{|c|c|}            \hline
Tip radius of tool bit [mm] & 0.05\\ \hline
Tool feeding speed [mm/min] & 0.5 $\sim$ 1.0\\ \hline
Depth of cut [mm/pass] & 0.005 $\sim$ 0.020\\ \hline
\end{tabular}
\caption{Manufacturing parameters of diffractive lens for the EUSO-Balloon middle lens.}
\label{table_Bal_mid_lens}
\end{table}

\subsection{Manufacturing the EUSO-Balloon front lens and rear lens}
We have started manufacturing the EUSO Balloon rear lens at the beginning of May 2013. The Balloon front lens and rear lens will be completed within summer 2013. EUSO-Balloon and TA-EUSO optics share the front lens and the rear lens design as mentioned in section 2. We  already have experienced manufacturing the front and rear lenses through the TA-EUSO lenses fabrication. We therefore expect that both lenses will be manufactured with much better quality than the TA-EUSO lenses.

\section{Conclusions}
We have completed the manufacturing of the TA-EUSO lenses and the EUSO-Balloon middle lens with better surface roughness than 20 nm RMS. The manufacturing of  the EUSO-Balloon front lens and rear lens will be completed within summer 2013.

The present study shows that it is possible to produce large PMMA fresnel lenses and  a large diffractive lens for the actual JEM-EUSO mission pathfinders. During the manufacturing process of the TA-EUSO and the EUSO-Balloon lenses, several improvements were introduced in the technique.

We will, of course, continue our study in optimizing the machining parameters and understanding effects of the manufacturing environments ( ambient temperature, ambient vibration, etc) to continue developing the technique that will be essential, in the near time, for the production of the JEM-EUSO flight model lenses.

\vspace*{0.5cm}
{
\footnotesize{{\bf Acknowledgment:}{This work was partially supported by Basic Science Interdisciplinary 
Research Projects of RIKEN and JSPS KAKENHI Grant (22340063, 23340081, and 
24244042), by the Italian Ministry of Foreign Affairs, General Direction 
for the Cultural Promotion and Cooperation, by the 'Helmholtz Alliance 
for Astroparticle Physics HAP' funded by the Initiative and Networking Fund 
of the Helmholtz Association, Germany, and by Slovak Academy  
of Sciences MVTS JEM-EUSO as well as VEGA grant agency project 2/0081/10.
The Spanish Consortium involved in the JEM-EUSO Space
Mission is funded by MICINN under projects AYA2009-
06037-E/ESP, AYA-ESP 2010-19082, AYA2011-29489-C03-
01, AYA2012-39115-C03-01, CSD2009-00064 (Consolider MULTIDARK)
and by Comunidad de Madrid (CAM) under project S2009/ESP-1496.
}}

}

\clearpage

%% file: icrc2013-0765.tex



\title{The Electronics of the EUSO-Balloon UV camera}

\shorttitle{EUSO-Balloon front-end}

\authors{
H. Miyamoto$^{1}$, P. Barrillon$^{1}$, P. von Ballmoos$^{2}$, S. Blin-Bondil$^{1}$, M. Casolino$^{3,4,5}$, S. Dagoret-Campagne$^{1}$, M. Dupieux$^{2}$, A. Ebersoldt$^{6}$, Ph. Gorodetzky$^{7}$, A. Haungs$^{6}$, Hirokazu Ikeda$^{8}$, A. Jung$^{9}$, Fumiyoshi Kajino$^{10}$, Y. Kawasaki$^{3}$, H. Lim$^{9}$, C. Moretto$^{2}$,  E. Parizot$^{3}$, I.H. Park$^{11}$, P. Picozza$^{3,4,5}$, P. Prat$^{3}$, G. Pr\'ev\^{o}t$^{3}$, J. Rabanal$^{1}$, M. Ricci$^{12}$, A. Santangelo$^{13}$, J. Szabelski$^{14}$, K. Tsuno$^{3}$ for the JEM-EUSO Collaboration$^{15}$.
}

\afiliations{
$^1$ Laboratoire de l'Acc\'el\'erateur Lin\'eaire, Univ Paris Sud-11, CNRS/IN2P3, Orsay, France\\
$^2$ Institut de Recherche en Astrophysique et Planétologie, Toulouse, France\\
$^3$ RIKEN Advanced Science Institute, Wako, Japan\\
$^4$ Istituto Nazionale di Fisica Nucleare - Sezione di Roma Tor Vergata, Italy\\
$^5$ Universita' di Roma Tor Vergata - Dipartimento di Fisica, Roma, Italy\\
$^6$ Karlsruhe Institute of Technology (KIT), Germany\\
$^7$ AstroParticule et Cosmologie, Univ Paris Diderot, CNRS/IN2P3, Paris, France\\
$^8$ Institute of Space and Astronautical Science, JAXA, Japan\\
$^9$ Ewha Womans University, Seoul, Republic of Korea\\
$^{10}$ Department of Physics, Konan University, Japan\\
$^{11}$ Sungkyunkwan University, Suwon-si, Kyung-gi-do, Republic of Korea\\
$^{12}$ Istituto Nazionale di Fisica Nucleare - Laboratori Nazionali di Frascati, Italy\\
$^{13}$ Institute for Astronomy and Astrophysics, Kepler Center, University of Tübingen, Germany\\
$^{14}$ National Centre for Nuclear Research, Lodz, Poland\\
$^{15}$ http://jemeuso.riken.jp
}

\email{Corresponding authors: barrillo@lal.in2p3.fr, miyamoto@lal.in2p3.fr} 

\abstract{The JEM-EUSO collaboration is currently developing the EUSO-BALLOON instrument, a pathfinder of the JEM-EUSO mission. 
Such an effort is led by the CNES, the French space agency, and involves several French institutes as well as several key institutes of the JEM-EUSO collaboration. The EUSO-Balloon instrument consists of an UV telescope and of an Infrared Camera. The UV telescope will operate at an altitude of 40 km, collecting background and possibly signal photons in the (290-430 nm) fluorescence UV range, the one in which the UV tracks generated by high energetic cosmic rays propagating in the earth's atmosphere are observed. The balloon experiment will be equipped with electronics and acquisition systems, as close as possible to the ones designed for the UV telescope of the main JEM-EUSO instrument. The past year has been devoted to the design, the fabrication and the tests of the prototypes of the optics, of the Photo Detector Module (PDM), of the digital processor and of the IR Camera of the EUSO-Balloon. In this contribution we focus on the PDM, the core element of the JEM-EUSO focal surface. We first describe all key items of the PDM,  from the photo-detectors to the FPGA board of the first stage of the digital processing. We then report on the tests carried out on the prototypes to assess their functionality and their suitability for a balloon mission.}


\maketitle

\section{Introduction}

\begin{figure}[ht]
  \centering
  \includegraphics[width=0.3\textwidth]{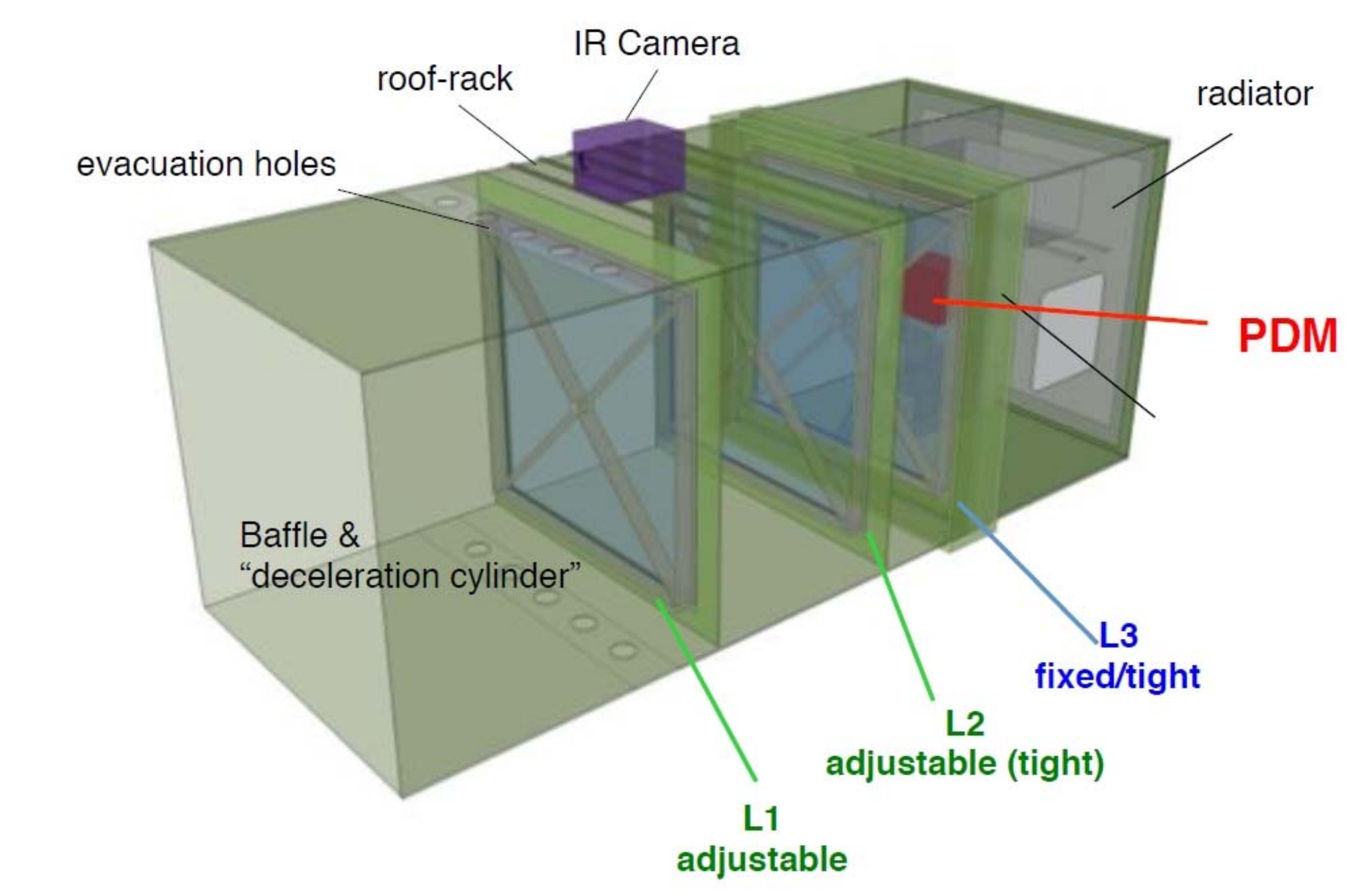}
  \caption{3D view of the EUSO-Balloon instrument. The 36 MAPMTs (bottom left part) and the ASIC boards (top right part) are visible.}
  \label{765-01}
 \end{figure}

The EUSO-Balloon experiment \cite{jbibEB-PVB,jbibEB-SD} is a pathfinder of the ISS mission JEM-EUSO \cite{jbibJEMEUSO} whose goal is to observe the Extensive Air Showers (EAS) produced in the atmosphere by the passage of the high energetic extraterrestrial particles with energy higher than $10^{19}$ eV. The particles of the showers generate a UV light track, observable in the 290-430 nm wavelength range, because of the fluorescence emission of Nitrogen molecules excited by collisions with secondary particles of the cascade. The detection of these UV photons is the base of the JEM-EUSO observational technique. In the EUSO-Balloon, a set of Fresnel lenses is used to focus the UV photons on the 2304 pixels of the Photo detector Module (PDM), the core element of the JEM-EUSO detection system and the focal surface of the Balloon instrument, as shown on figure \ref{765-01}. \\

 The main instrument of the EUSO-Ballon can be considered as an UV camera taking pictures every 2.5 $\mu$s (the size of a gate time unit, GTU). It will serve as a technogical demonstrator of the JEM-EUSO mission, since they share similar parts and concept elements, not only for the PDM electronics but also for the mechanics and the data processing system (acquisition and monitoring).\\
The EUSO-Balloon experiment will be also essential to test the technologies selected and developed in a severe environment. In fact the EUSO-Balloon will work at 40 km of altitude where the pressure is at 3 mbar, worse than in space conditions. \\
The science goals of the experiment are the study of the UV background below 40 km (the main contribution is known as nightglow) and the test of the trigger algorithms selectivity and rate.

\section{The PDM electronic}
\subsection{Overview}

\begin{figure}[ht]
  \centering
  \includegraphics[width=0.4\textwidth]{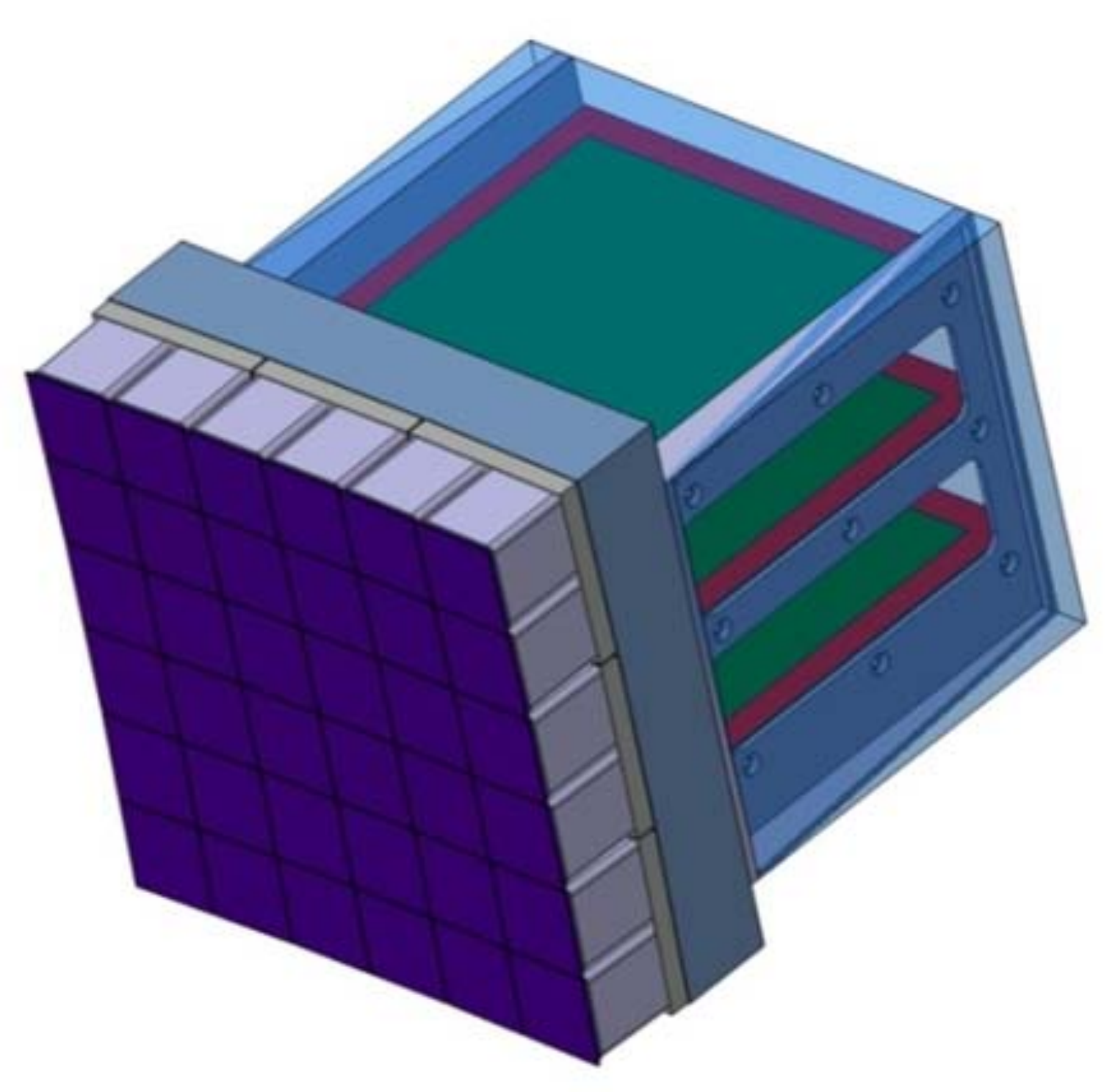}
  \caption{PDM 3D view.}
  \label{765-02}
 \end{figure}

The PDM (see figure \ref{765-02}) is composed by nine elementary cells (ECs) which are made of four Multi-Anode PhotoMulitpliers (MAPMT \cite{jbibMAPMT}, R11265-M64 from Hamamatsu) and their associated electronics. The EC unit electronics (see section \ref{sub_ec}) is very compact due to restrictive dimensions of the global mechanic. A set of small boards were designed to be placed in the shadow of the photomultipliers (MAPMT) used to detect the photons. Three different boards are used to supply 14 different high voltages and to collect the analog signals from the 64 channels of each MAPMT. \\
These signals are sent to a fourth board (see section \ref{sub_ec_asic}) containing the SPACIROC ASICs, which perform single photon counting and the estimate of the charge of the signal. The digitized data are sent to the PDM board (described in section \ref{sub_pdmb}) which controls up to six ASIC boards with an FPGA. The PDM contains the high voltage power supplies (see \ref{sub_hvps}) which provide the high voltages and is equipped with a system of switches to protect the MAPMTs from intense light flux.

\subsection{Elementary Cell unit}
\label{sub_ec}

\begin{figure}[ht]
  \centering
  \includegraphics[width=0.48\textwidth]{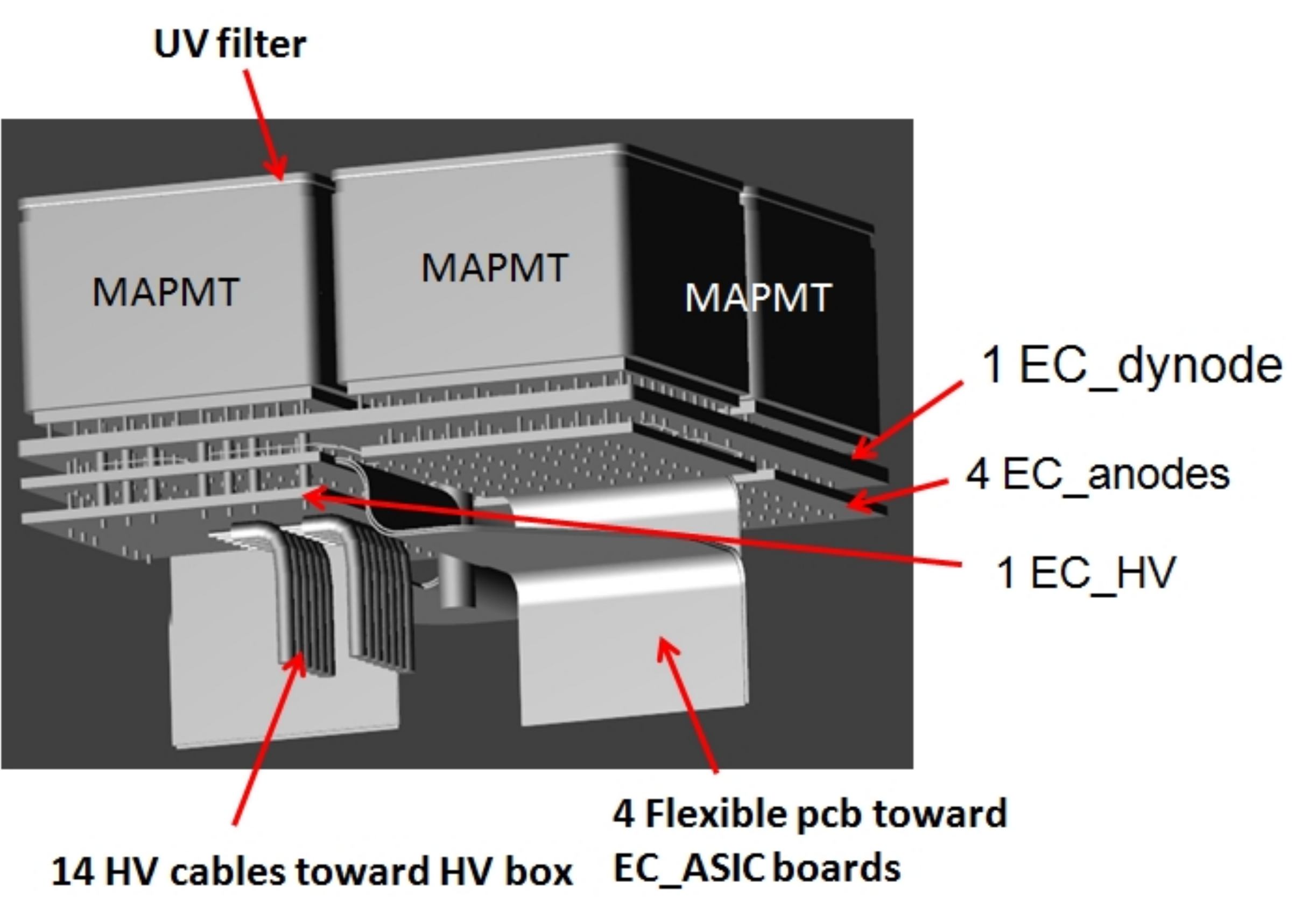}
  \caption{3D view of an EC unit.}
  \label{765-03}
 \end{figure}

Each EC unit is composed by four MAPMTs and a stack of three types of boards (see figure \ref{765-03}):\\
- One EC-DYNODE board, which distributes the high voltage to the four MAPMTs\\
- Four EC-ANODE boards, which collect the analog signals and transmit them to the next stage of the electronic chain \\
- One EC-HV, which interfaces the HVPS with the EC-dynode, transmitting the high voltage\\

\begin{figure}[ht]
  \centering
  \includegraphics[angle=270,width=0.48\textwidth]{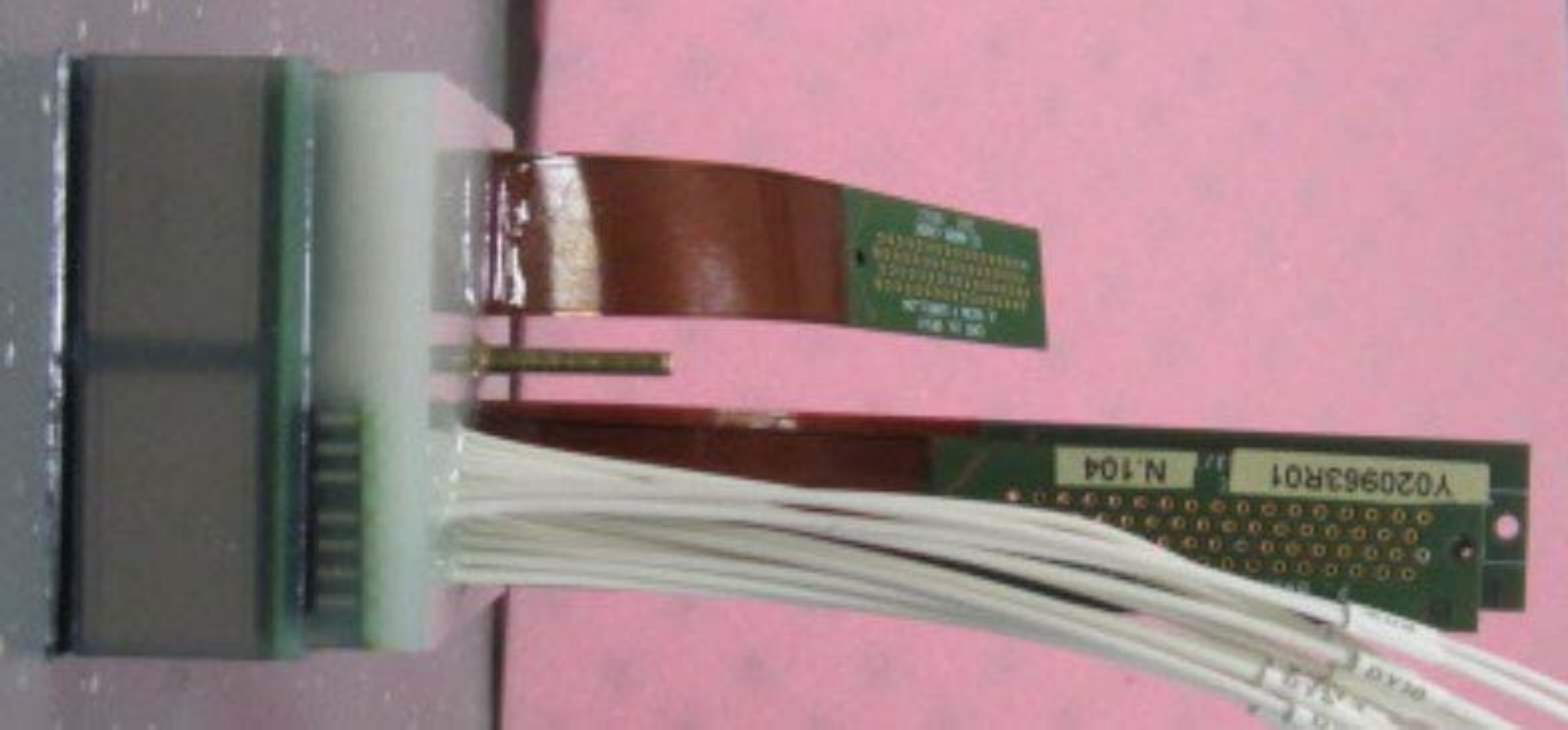}
  \caption{Picture of a potted EC unit.}
  \label{765-04}
 \end{figure}

Special care must be taken in the assembly of each EC unit to secure a firm montage in the mechanical structure. Because of the severe conditions of pressure, they have to be potted to protect them against destructive sparking induced possibly by high voltage. The figure \ref{765-04} shows a potted EC unit. Only the cables, a fixation screw and the UV filters are coming out of the potting.

\subsection{ASIC boards}
\label{sub_ec_asic}

\begin{figure}[ht]
  \centering
  \includegraphics[width=0.4\textwidth]{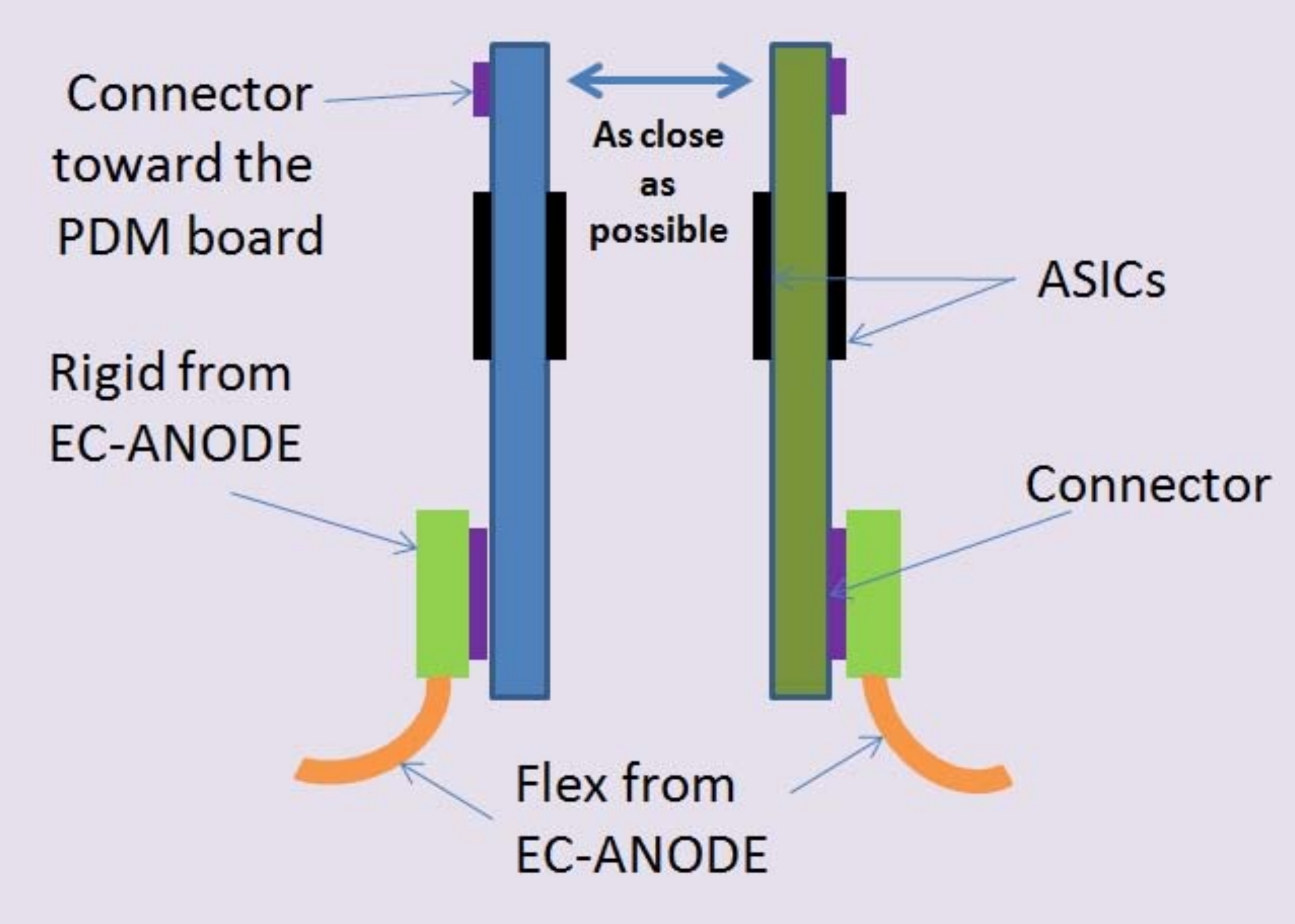}
  \caption{Connection between EC-ANODE and EC-ASIC boards.}
  \label{765-05}
 \end{figure}

To collect the analog signals coming out from the 36 MAPMTs (or EC-ANODE boards), six ASIC boards are fixed perpendicularly to the mechanical frame which welcomes the nine EC units. Figure \ref{765-05} illustrates this connection.

\begin{figure}[ht]
  \centering
  \includegraphics[width=0.5\textwidth]{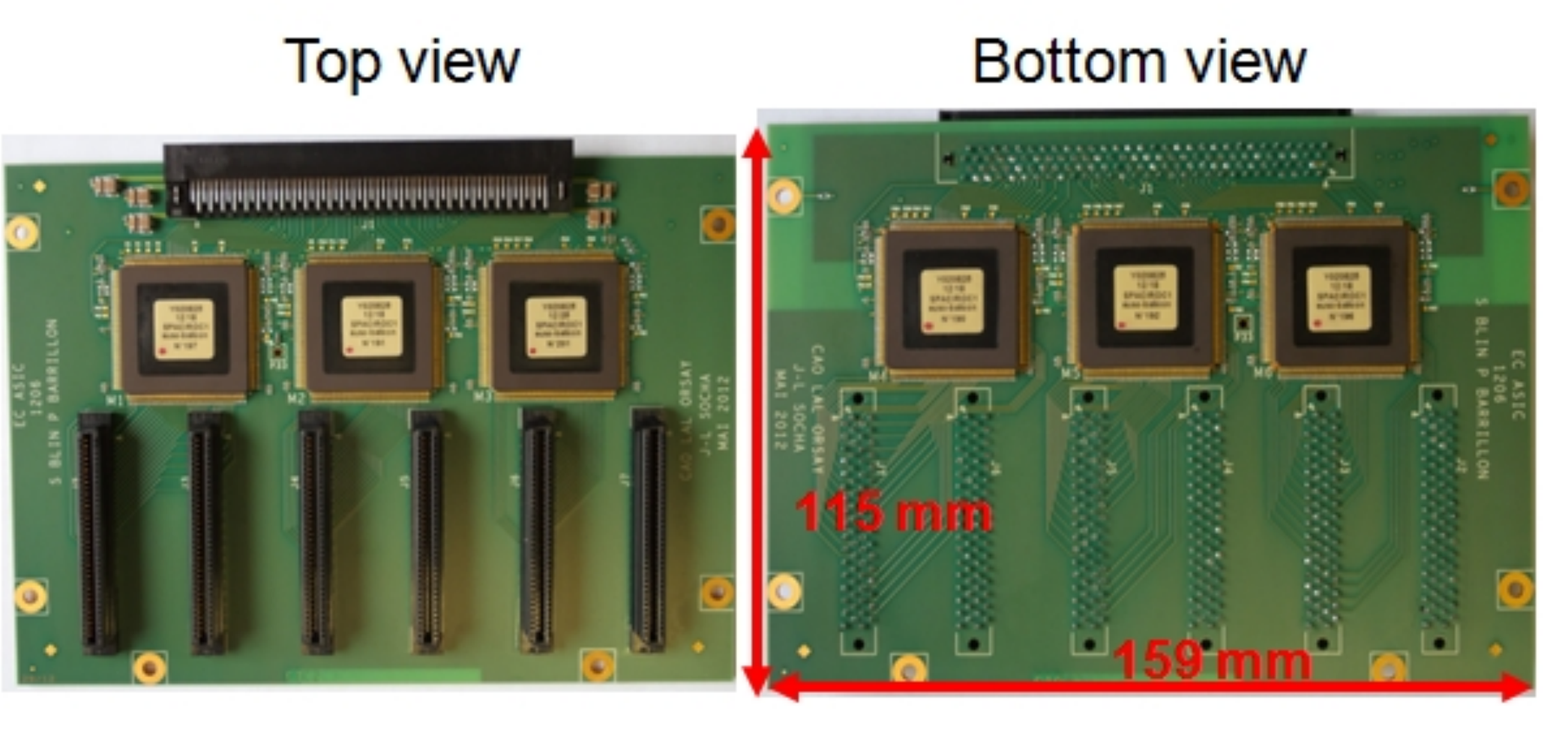}
  \caption{Top and bottom views of the prototype of the EC-ASIC board.}
  \label{765-06}
 \end{figure}

These six ASIC boards (also called EC-ASICs) welcome six ASICs, as well as six connectors toward the EC units and one connector toward the PDM board (section \ref{sub_pdmb}). Figure \ref{765-06} shows the top and bottom views of the first EC-ASIC produced and assembled. There are three packaged SPACIROC ASICs \cite{jbibSPACIROC} on each side. The ASIC performs photon couting for each pixel of the 36 MAPMTs, as well as an estimate of the charge of the gathering of eight pixels. Before integrating the ASIC on the board, its performances were checked successfully alone and coupled to a MAPMT.

The configuration (in a daisy chain mode) of the ASICs and the powers are sent through a connector by the PDM board, which also collects the digital data produced by the 36 ASICs.
 
\subsection{FPGA board}
\label{sub_pdmb}

\begin{figure}[ht]
  \centering
  \includegraphics[width=0.5\textwidth]{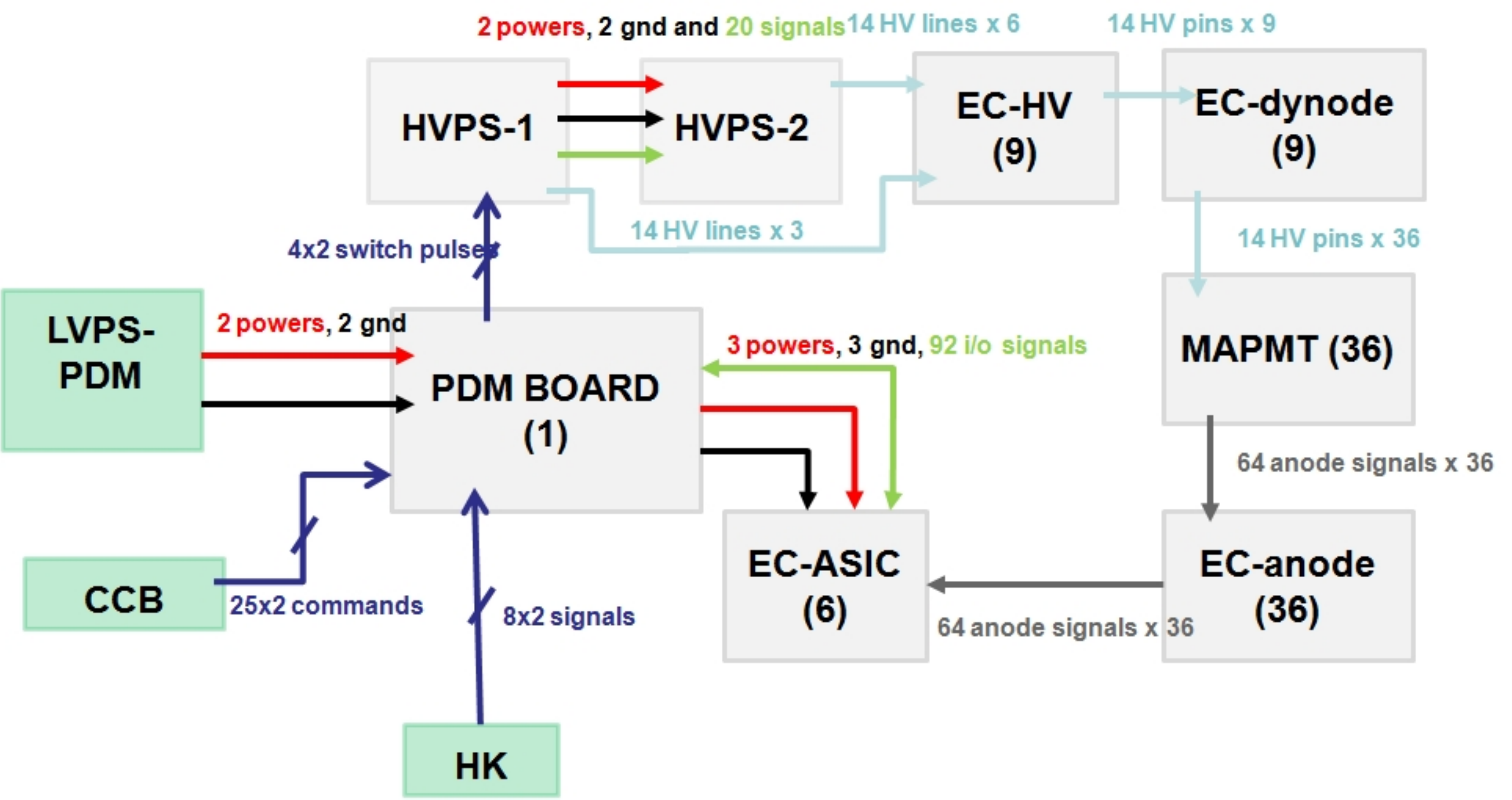}
  \caption{Electrical architecture of the PDM. CCB: Cluster Control Board. HK: Housekeeping. LVPS-PDM: Low voltage power supply.}
  \label{765-08}
 \end{figure}

As shown in figure \ref{765-08}, which presents the PDM electrical architecture, the PDM board is a central element which is connected to several components not only of the PDM but also of the Digital Processor system. It communicates with the LVPS-PDM (providing the power), and with the Cluster Control Board (the next stage of the data processing and trigger) \cite{jbibCCB}, the housekeeping board which distributes the necessary commands and collects relevant operational parameters, and one of the HVPS (high voltage power supply, see section \ref{sub_hvps}).

\begin{figure}[ht]
  \centering
  \includegraphics[width=0.5\textwidth]{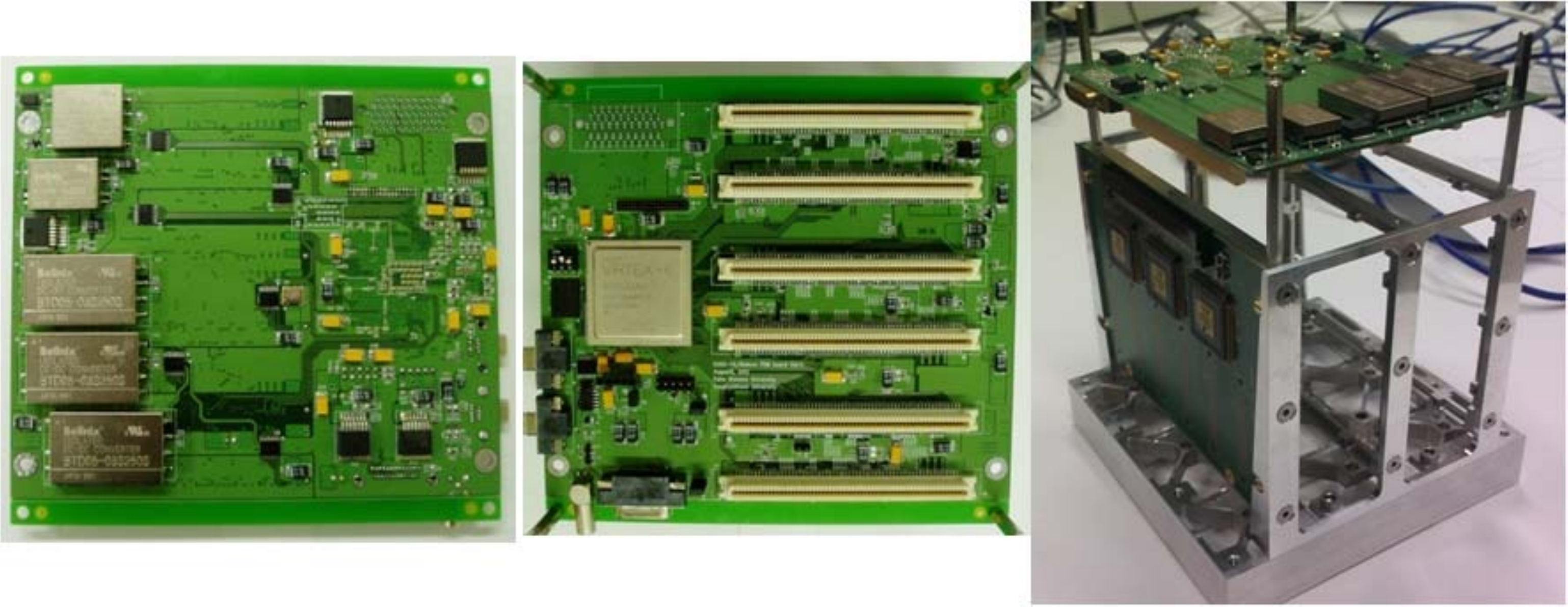}
  \caption{Top and bottom views of PDM board (left). PDM and EC-ASIC board fixed on the mechanical structure(right).}
  \label{765-07}
 \end{figure}

All the features needed for the interfaces but also for the data processing, such as the first level trigger algorithms, are integrated in the firmware of an FPGA (from Xilinx Virtex 6 family), the key element of the PDM board. The figure \ref{765-07} represents different views of this board, as well as its integration in the mechanical structure. In addition to the FPGA, this board is equipped with DC-DC converters and regulators to provide the different voltage sources required, the connectors toward the other boards and passive components.

\subsection{High voltage power supply}
\label{sub_hvps}

\begin{figure}[ht]
  \centering
  \includegraphics[width=0.5\textwidth]{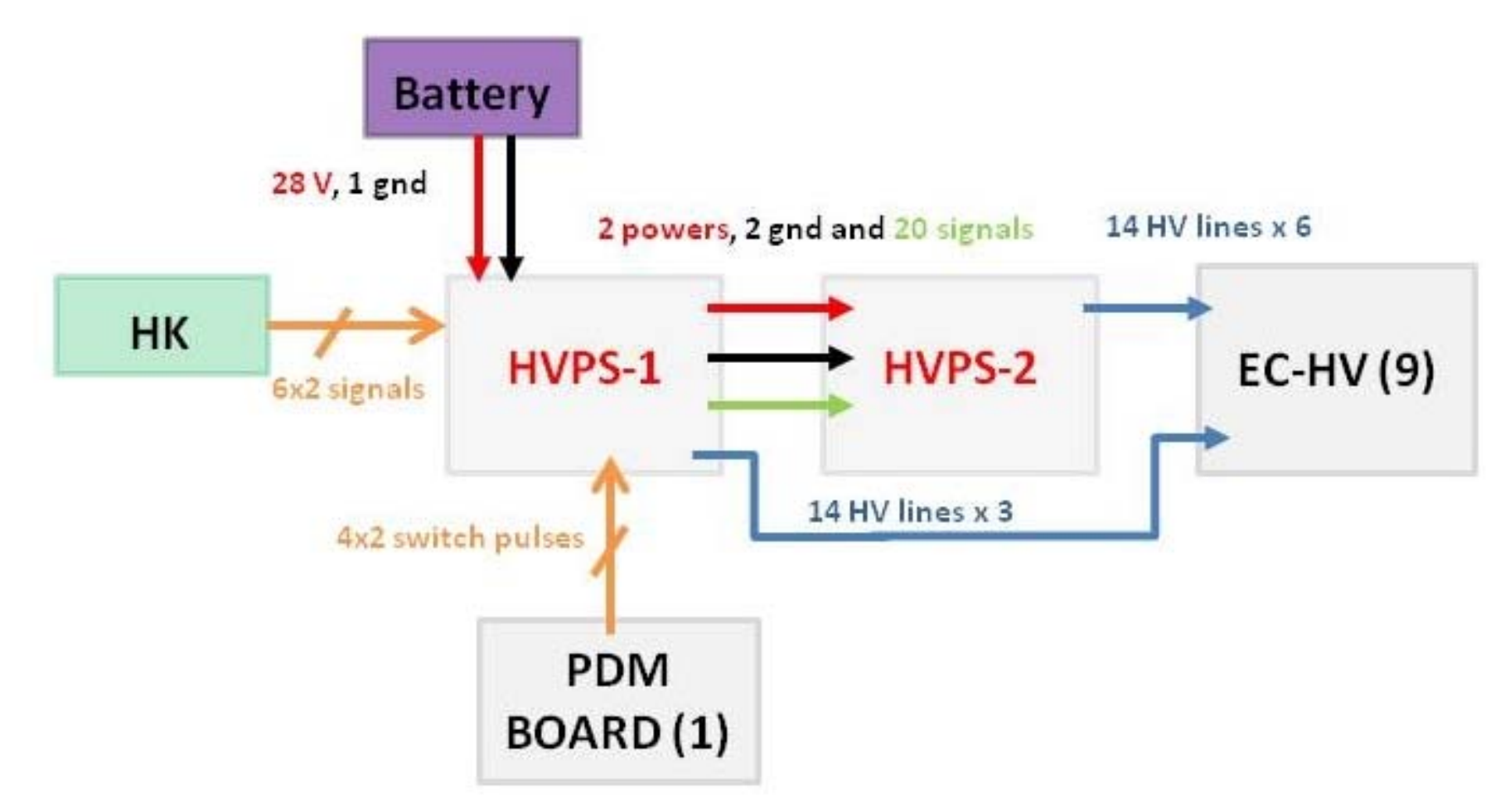}
  \caption{Schematic of the two HVPS with their signals and interfaces.}
  \label{765-09}
 \end{figure}

There are two HVPS (see figure \ref{765-09}), whose roles are to provide the 14 high voltage levels needed by each MAPMT. These voltages are generated by a Cockroft-Walton (CW) system from a 28V power source. This system was preferred to the classical voltage divider because of its very low power consumption (50 times less). The first and second HVPS produce the high voltage levels respectively for three and six EC units.\\
The HVPS also includes a system of switches which allows to cut the powering of all the MAPMTs in less than a GTU in case of strong light events.\\
Based on the charges estimated by the ASICs and FPGA logic, the PDM board produces a switch control signals sent to the first HVPS, which transmits it to the second one. A variant option is to check for a current increase on one of the last dynodes at the level of the HVPS.\\
The first HVPS interfaces the housekeeping board, which provides it with monitoring signals.

\section{Prototype tests}

\begin{figure}[ht]
  \centering
  \includegraphics[width=0.5\textwidth]{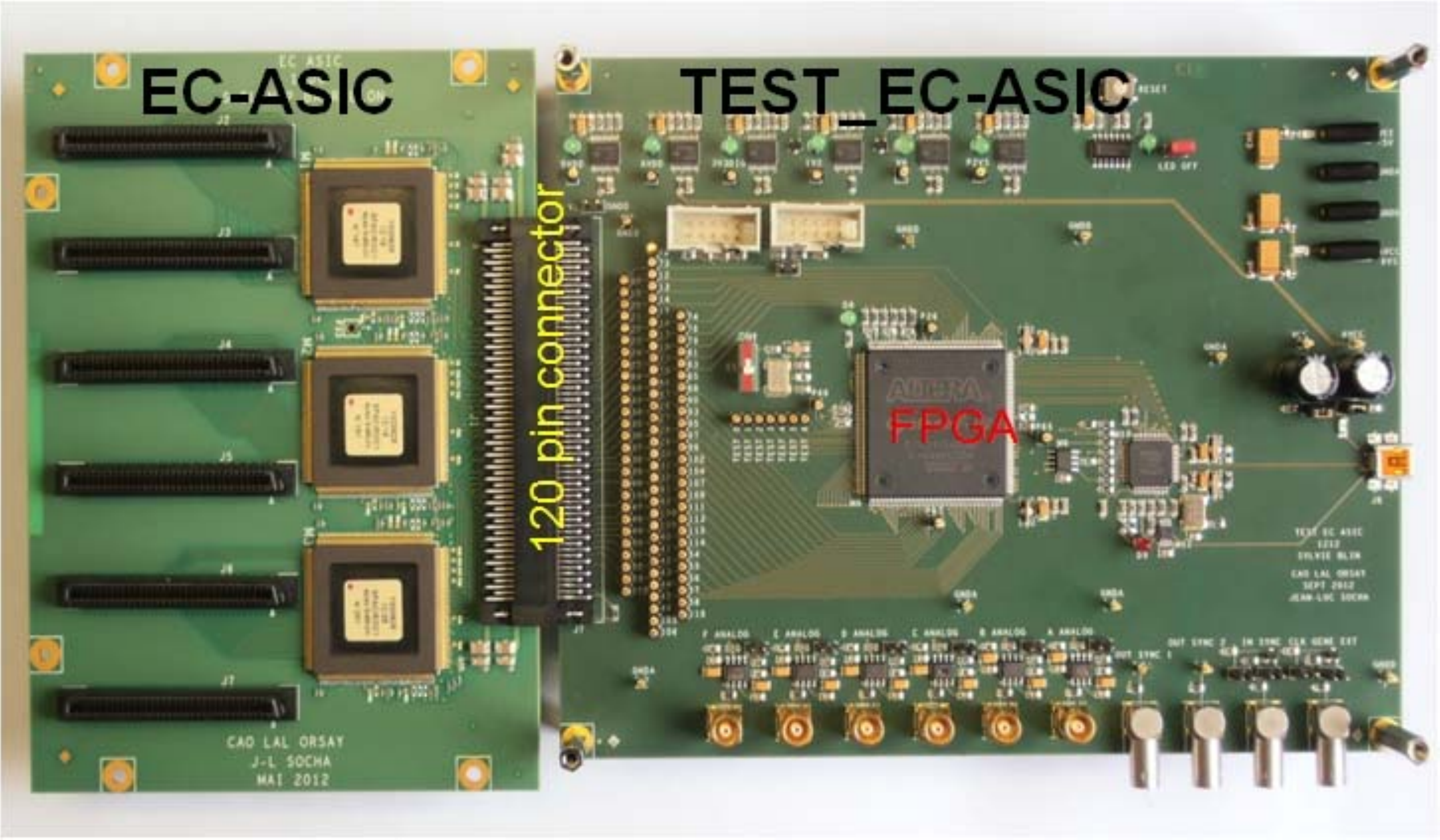}
  \caption{Picture of the ASIC prototype board connected to a dedicated test board.}
  \label{765-10}
 \end{figure}

In order to assess each board before starting the flight model production, prototypes were produced and tested in 2012.\\
Two prototypes of the EC units were assembled: a pure mechanical version and an electrical one. The first one was used to check the dimensions of the boards, the assembly sequence and the potting process. The second one (assembled later), while allowing finalising these aspects, permitted to check the performances. It was tested in a black box with a light source together with the prototype of the HVPS powering the EC unit up to 1100 V. The analog signals coming out from the MAPMTs were found as expected, proving that the system was working properly. Low pressure and temperature tests were carried out and no problem was noticed.\\
In parallel, a prototype of the ASIC board was tested with a dedicated test board (see figure \ref{765-10}). The latter has its own FPGA used to manage the ASICs configuration, data readout and communication protocole with a computer running Labview software. All the features were tested with a test bench that included multimeters, used to check the pedestal levels and the linearity of the threshold voltage, and a pulse generator which allowed injecting a MAPMT-like signal into the board. The measurements showed nice performances validating the design of the board as it was.\\

\begin{figure}[ht]
  \centering
  \includegraphics[width=0.5\textwidth]{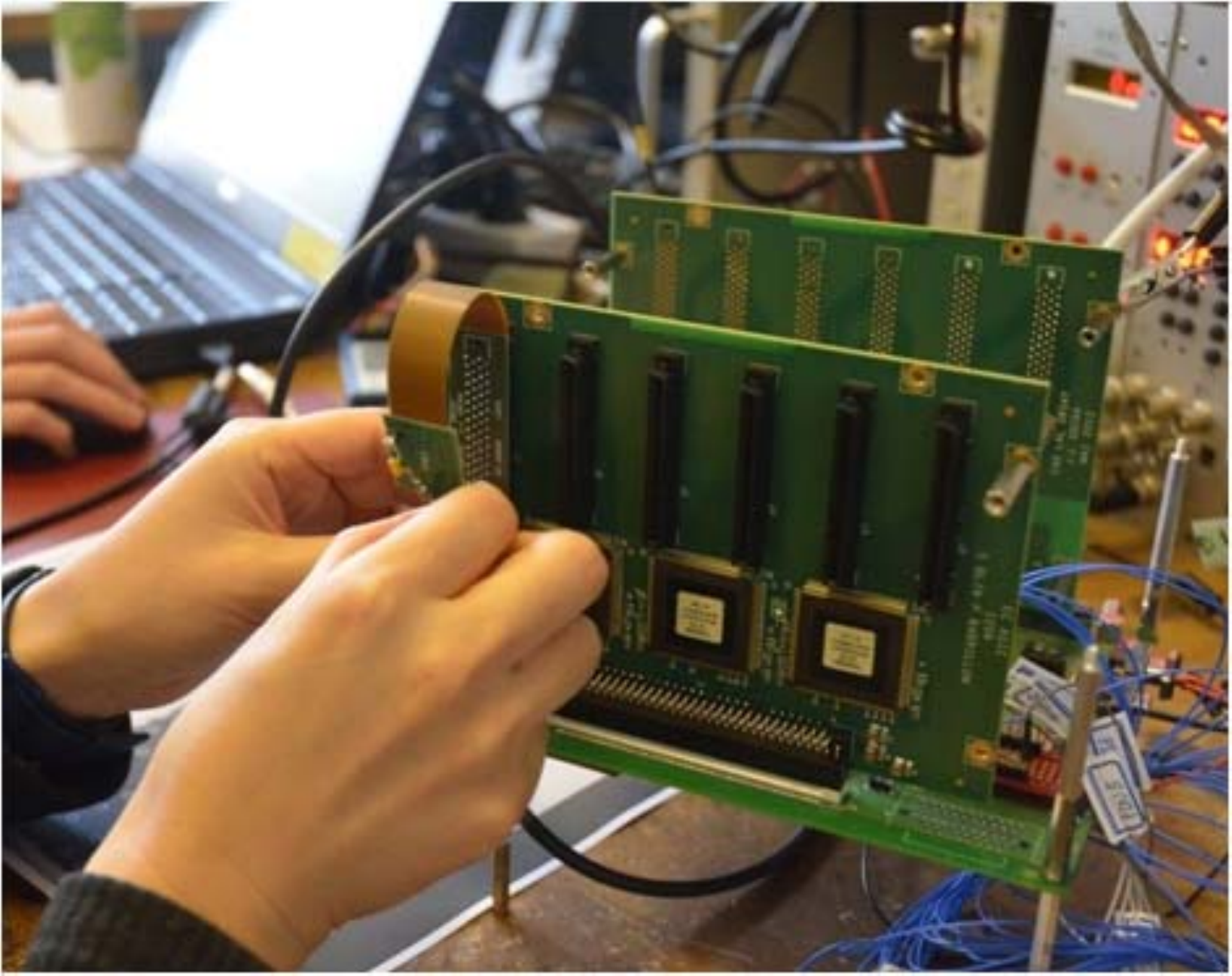}
  \caption{Picture of the PDM board with two ASIC boards connected during prototype tests.}
  \label{765-11}
 \end{figure}

After standard individual tests to check the consumption and the signal levels, the PDM board was tested together with ASIC boards with a set-up similar to the previous one (see figure \ref{765-11}). The interface between these two elements was checked by sending configuration to the ASICs and reading out data. Despite few minor adjustements in the firmware, all tests were successful proving that communications performed as expected.\\

In addition, the rest of the data processing chain was tested successfully with the PDM board, an ASIC board and an EC unit. These tests were carried out in Riken institute for the project TA-EUSO (Telescope Array) \cite{jbibTA-EUSO}, which also consists in a complete PDM that will be installed on the ground in Utah to perform similar and complementary measurements to the ones of EUSO-Balloon.

\section{Conclusion}

The 18th of December 2012, the results of all these prototype tests were presented to CNES during a review of the phase B of the EUSO-Balloon project. The conclusions from this review were that the flight model production of all the boards should start early 2013. The assembly of the EC units would follow on and then the individual tests of each element before the assembly and integration tests of the whole PDM. The goal is to provide a full instrument by the end of the year 2013 to be ready for the first balloon flight in 2014.

\section{Acknowledgment}

We wish to thank the CNES for the support to the EUSO Balloon studies. We also thank IN2P3, DLR, RIKEN, the Helmotz Alliance for Astroparticle Physics, INFN, which among other institutions have supported the development of the experiment.  We also acknowledge the support of ESA via the topical team on "JEM-EUSO".

\clearpage


%% file: icrc2013-0678.tex




\title{Global Description of EUSO-Balloon Instrument}

\shorttitle{Instrumentation for JEM-EUSO }

\authors{

C. Moretto$^{1}$, S. Dagoret-Campagne$^{1}$, J.H. Adams$^{19}$, P. von Ballmoos$^{2}$, P. Barrillon$^{1}$, J. Bayer$^{5}$, \mbox{M. Bertaina}$^{12}$, S. Blin-Bondil$^{1}$, F. Cafagna$^{7}$, M. Casolino$^{13,10,11}$, C. Catalano$^{2}$, P. Danto$^{4}$, \mbox{A. Ebersoldt}$^{6}$, T. Ebisuzaki$^{13}$, J. Evrard$^{4}$, Ph. Gorodetzky$^{3}$, A. Haungs$^{6}$, A. Jung$^{14}$, Y. Kawasaki$^{13}$, \mbox{H. Lim}$^{14}$, G. Medina-Tanco$^{15}$, H. Miyamoto$^{1}$, D. Monnier-Ragaigne$^{1}$, T. Omori$^{13}$, G. Osteria$^{9}$, \mbox{E. Parizot}$^3$, I.H. Park$^{14}$, P. Picozza$^{13,10,11}$, G. Pr\'ev\^ot$^{3}$,  H. Prieto$^{13,17}$, M. Ricci$^{8}$, M.D. Rodr\'iguez Fr\'ias$^{17}$, \mbox{A. Santangelo}$^{5}$, J. Szabelski$^{16}$, Y. Takizawa$^{13}$, K. Tsuno$^{13}$ 
for the JEM-EUSO Collaboration$^{19}$.
}

\afiliations{
$^1$ Laboratoire de l'Acc\'el\'erateur Lin\'eaire, Universit\'e Paris Sud-11, CNRS/IN2P3, Orsay, France\\
$^2$ Institut de Recherche en Astrophysique et Plan\'etologie, Toulouse, France\\
$^3$ AstroParticule et Cosmologie, Univ Paris Diderot, CNRS/IN2P3, Paris, France\\
$^4$ Centre National d'\'Etudes Spatiales, Centre Spatial de Toulouse, France\\
$^5$ Institute for Astronomy and Astrophysics, Kepler Center, University of T\"{u}bingen, Germany\\
$^6$ Karlsruhe Institute of Technology (KIT), Germany\\
$^7$ Istituto Nazionale di Fisica Nucleare - Sezione di Bari, Italy\\
$^8$ Istituto Nazionale di Fisica Nucleare - Laboratori Nazionali di Frascati, Italy\\
$^9$ Istituto Nazionale di Fisica Nucleare - Sezione di Napoli, Italy\\
$^{10}$ Istituto Nazionale di Fisica Nucleare - Sezione di Roma Tor Vergata, Italy\\
$^{11}$ UniversitaÕ di Roma Tor Vergata - Dipartimento di Fisica, Roma, Italy\\
$^{12}$ Dipartimento di Fisica dell'Universit\`a di Torino and INFN Torino, Torino, Italy\\
$^{13}$ RIKEN Advanced Science Institute, Wako, Japan\\
$^{14}$ Sungkyunkwan University, Suwon-si, Kyung-gi-do, Republic of Korea\\
$^{15}$ Universidad Nacional Aut\'onoma de M\'exico (UNAM), Mexico\\
$^{16}$ National Centre for Nuclear Research, Lodz, Poland\\
$^{17}$ Universidad de Alcal\'a (UAH), Madrid, Spain\\
$^{18}$ University of Alabama in Huntsville, Huntsville, USA\\
$^{19}$ http://jemeuso.riken.jp
}

\email{moretto@lal.in2p3.fr} 

\abstract{The EUSO-Balloon is a pathfinder of the JEM-EUSO mission, designed to be installed on-board the International Space Station before the end of this decade. 
The EUSO-Balloon instrument, conceived as a scaled-down version of the main mission, is currently developed as a payload of a stratospheric balloon operated by CNES, and will, most likely, be launched during the CNES flight campaign in 2014. Several key elements of JEM-EUSO have been implemented in the EUSO-Balloon. The instrument consists of an UV telescope, made of three Fresnel lenses, designed to focus the signal of the UV tracks, generated by highly energetic cosmic rays propagating in the earth's atmosphere, onto a finely pixelized UV camera. In this contribution, we review the main stages of the signal processing of the EUSO-Balloon instrument: the photodetection, the analog electronics, the trigger stages, which select events while rejecting random background, the acquisition system performing data storage and the monitoring, which allows the instrument control during operation. 
}



\maketitle

\section{Introduction}
The EUSO-Balloon is an experiment aiming at validating the conceptual design as well as the technologies foreseen for the ultra high energy cosmic ray space-based observatory  \mbox{JEM-EUSO} ~\cite{hbibEUSOperf}. The instrument, a scaled-down version of JEM-EUSO, includes several of the key components of the main space mission. Scientific and technical goals of the balloon instrument are reviewed in~\cite{hbibEBpath} and its simulation is presented in \cite{hbibEBSimulation}.
The EUSO-Balloon instrument, designed as payload of a stratospheric balloon operated by CNES,  will perform a series of night-flights, lasting from a few hours to tens of hours, at altitudes of $\sim40$ km, and at different locations. The proposed program requires payload recovery after landing either in water or ground, and the repairing of the instrument after each mission. Special atmospheric environmental conditions and recovery requirements imply a very careful design and dedicated tests on advanced prototypes. 

The paper is organised as follows. First, section~\ref{sec:OverviewInstrument} gives an overview of the instrument, including its particular mechanical design adapted to balloon flights.  Section~\ref{sec:Subsystems} provides details on the subsystems and highlights the main reasons for the chosen design. Section~\ref{sec:AssemblyTest} deals with all series of preliminary measurements and tests, mandatory before the acceptance of the instrument for the actual flight. Finally, control and analysis tasks to be performed during the operation, are mentioned in section ~\ref{sec:Operation}.

\section{The Instrument Overview}
\label{sec:OverviewInstrument}
\begin{figure} 
  \centering
  \includegraphics[width=0.48\textwidth]{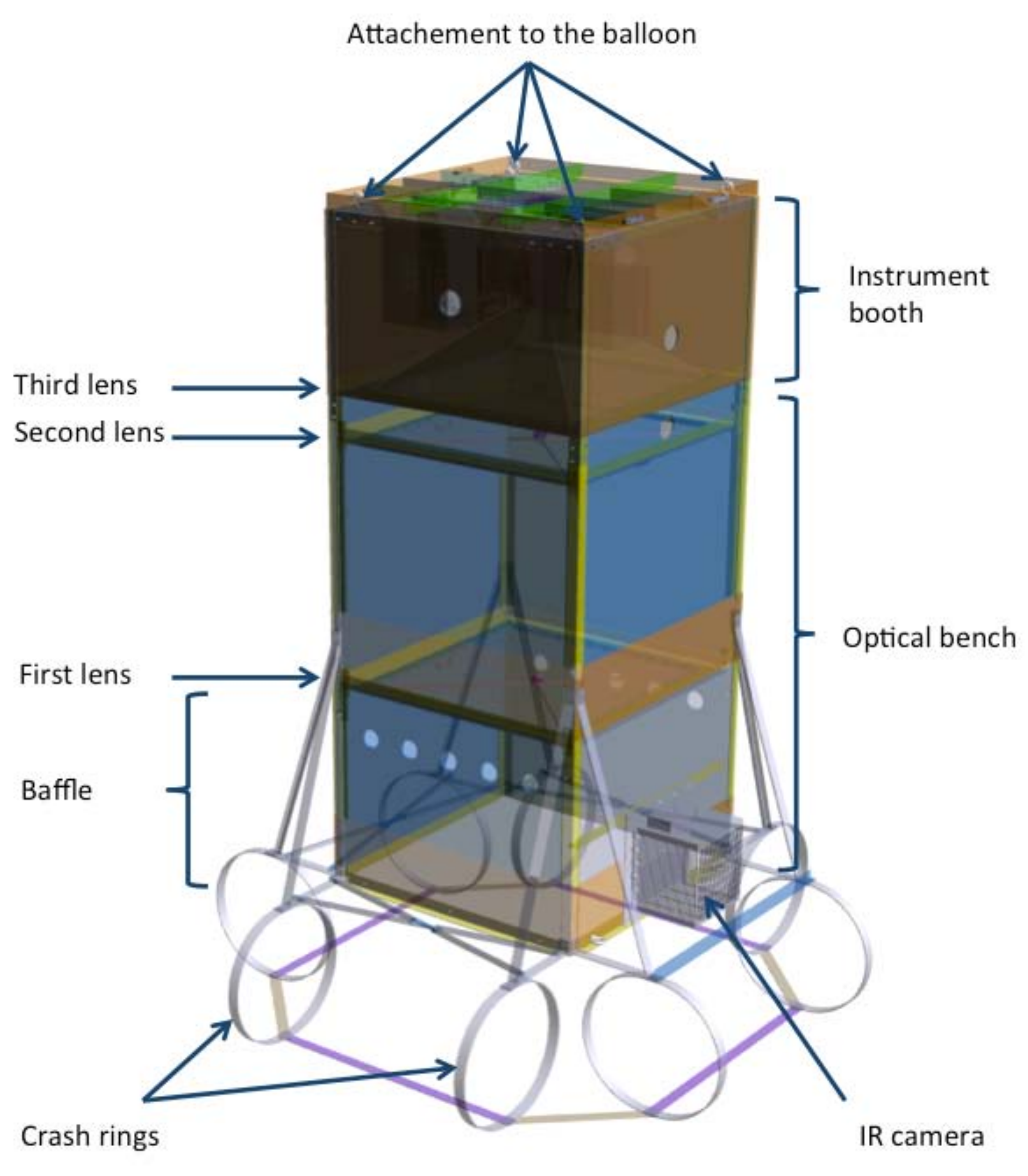}
  \caption{EUSO-Balloon Instrument Overview.}
  \label{fig:globalview}
 \end{figure}

The EUSO-Balloon instrument structure is shown in figure~\ref{fig:globalview} and its main characteristics are given in table~\ref{tab:properties}. The main mission parameters are justified in section~\ref{sec:Subsystems} devoted to the subsystems. The parallelepiped-shaped telescope
presents a wide field of view of 12$^\circ\times$12$^\circ$ for a collecting surface of  1 m$\times$1 m. During observations, it points to nadir observing the earth's atmosphere.
The main instrument basically consists of an optical bench associated to a detector module booth placed at the focal position. The optical bench encompasses two Fresnel lenses. The instrument booth includes the whole electronics inside a pressurized watertight box. The instrument booth is delimited by the third lens.

The main instrument includes an external roof-rack, which allows the accommodation of complementary instruments like an infra-red camera developed for atmosphere monitoring~\cite{hbibIR}.
 
\subsection{General Characteristics and Functions}
The optical subsystem includes the optical bench which focuses parallel light rays onto a finely pixelized focal surface, consisting of an array of  Multi-Anode Photomultipliers (MAPMTs) sensitive to UV light in the 290-430 \mbox{nm} range with very good photon-detection efficiency. The Focal Surface (FS) is completed by a complex electronics, which allows fast measurement within microsecond time scale, including photodetector protection against intense light flux by fast switches, auto-triggering capability, event filtering and event recording. 
The electronics records, for each triggered event, a sequence of 128 consecutive images with a time resolution of one Gate Time Unit (GTU) equal to 2.5 $\mu$s. 

\begin{table}[h!]
\centering
\includegraphics[width=0.48\textwidth]{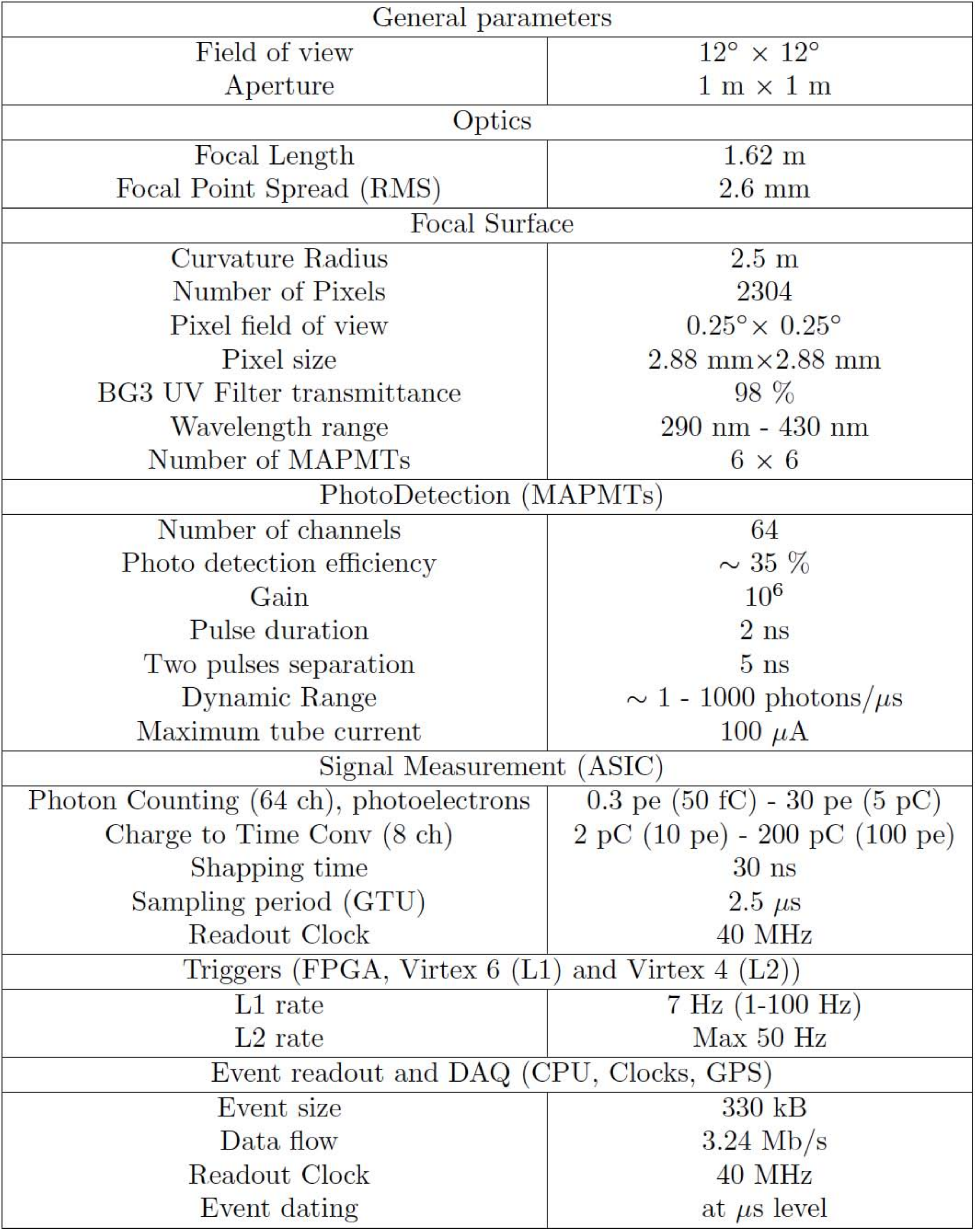}
\caption{Main features of the EUSO-Balloon instrument}
\label{tab:properties}
\end{table}

\subsection{Instrument Structure}
The mechanics of the instrument is made of Fibrelam\textregistered \ panels, arranged together through fiberglass sections.
The instrument is coated by an insulating cover to protect the components from fast temperature changes during balloon ascent and descent.
Special watertight valves, inserted in the optical bench, are used to enable pressure equilibrium with the external environment. Wherever the after-flight landing location occurs, the instrument must be recovered  with the smallest damages. The bottom part is therefore equipped with crash rings, which absorb strong deceleration (up to 15 g) when landing on ground. A baffle with special holes in the optical bench are used as a piston-effect to damp the shock for a fall over water.
The instrument booth, a totally watertight sealed box, consists of a central aluminium plate on which the various electronic boxes are fixed. One of its side is the third lens. The opposite side is an aluminium radiator used to dissipate the heat generated by the electronics. The instrument is surrounded by buoys to avoid sinking in case of splashdown and to raise straight up the instrument booth above the water level.

\section{The Instrument Subsystems}
\label{sec:Subsystems}
The main instrument is divided into the following main subsystems: 1) the Optics; 2) The Focal Surface (FS) which includes
the photodetector with the MAPMTs, the ASICs measuring signals, the Photo-Detector Module Board (PDMB) and the High Voltage Power Supplies (HVPS); 3) The Data Processing (DP) involving the Cluster Control Board (CCB) providing readout triggers and the Data Acquisition System (DAQ); 4) Utilities like the monitoring also called the housekeeping board (HK) and the low voltages power supplies (LVPS) associated to the batteries (PWP).
All these subsystems are all described below.

\subsection{Optics Subsystem}
The optics subsystem includes three lenses. Its goal is to provide the best focusing for the smallest focal distance. The focusing requirement is constrained by the pixel size of the photodetection system.
Due to the wide angular field of view, it is necessary to combine 3 flat lenses. The first and third lenses are one-sided focusing Fresnel lenses, while the lens in the middle is purely dispersive, necessary to correct for chromatic aberrations. They are made of PMMA material~\cite{hbibOptics}. Ray tracing calculations, including the temperature profile expected for flights in cold and warm cases, provide a focal length of 1.62 m and a focal point spread width of $\sim$2.6 mm, smaller than the pixel size. 

\subsection{Focal Surface Subsystem}
The FS is a slightly curved surface containing one Photo-Detector Module (PDM). The design matches the one of the JEM-EUSO central PDM. A PDM contains 36 MAPMTs and therefore it is an array of 48$\times$48 pixels of 2.88 mm $\times$ 2.88 mm size each, slightly larger than the focal point spread. The PDM is divided into a set of 9 identical Elementary Cells (ECs), which are matrices of 2$\times$2 MAPMTs. The photocathode is covered by a BG3 UV filter. Inside the PDM structure, the 9 ECs are disposed and tilted according to the appropriate shape required for the FS. We review in the following the main properties of this electronics.

\vspace*{-0.3cm}
\paragraph{MAPMTs} They are Hamamatsu photon detectors (R11265-M64) consisting of a matrix of 8$\times$8 pixels. Each pixel is associated to an anode generating a charge or a current in output. Their sensitivity is as low as a few tenths of photon and their dynamic range can extend up to few thousands photons per $\mu$s when working at their nominal high gain of $10^6$. The anode signals of the MAPMTs are measured and digitised by the ASICs and managed by an FPGA based PDM board, which performs also the first level trigger selection. More details on the PDM can be found in~\cite{hbibFrontEndEl}.

\vspace*{-0.3cm}
\paragraph{High voltage power supply}
MAPMTs are polarised with 14 high voltages. The latter are generated by a high voltage power supply (Crockoft-Walton, CW type, to limit power consumption). The nominal high voltage of the photocathode is -900 V for a MAPMT nominal gain at $10^6$. The effective dynamic range can be extended up to 10$^7$ photons/$\mu$s by reducing gradually the gain from $10^6$ by successive factors of $10^2$ down to a gain of 30.
Fast switches (SW) responsive at $\mu$s time-scale, adapt the voltage values to tune the MAPMT gain according to the intensity of photon flux. 
Because a large photon flux generating anode current above 100 $\mu$A would destroy the tube, this automatic control system can even switch off the voltage. The switching decision logic is implemented in the \mbox{FPGA}, which reads out the \mbox{ASICs}. In the PDM, there are 9 independent CW with their individual 9 SW, assembled into two separated HVPS boxes, each CW controlling independently the 9 ECs high voltages.

\vspace*{-0.3cm}
\paragraph{ASICs} 36 SPACIROC~\cite{hbibASIC} ASICs are used to perform anode signals measurement and digitisation. These ASICs have 64 channels. Their analog inputs are DC-coupled to the MAPMT anodes. They process in parallel the 64 analog signals in two modes: 1) photoelectron counting mode, in a range from 1/3 of photoelectrons up to 100 photoelectrons, by discriminating over a programmable threshold each of the channels; 
2) Integrating mode, estimating with a range from 20~pC to 200~pC, a 8 anodes current sum by time over threshold determination.
The 64 analog channels are balanced each-other relatively by gain matching over 8-bits.
The discrimination voltage level used in the photon-counting is provided by a 10-bit DAC (Digital to Amplitude Converter).
In both cases the digitisation is performed by 8-bits counters every \mbox{GTU}. There is no data buffering on the ASIC. The data are transferred, for each GTU, to the FPGA based PDMB at 40MHz.

\vspace*{-0.3cm}
\paragraph{PDMB} The instrument includes two trigger stages. The first level trigger (L1) is implemented in the FPGA (Xilinks Virtex 6) of the PDM-Board (PDMB), integrated in the PDM. The PDMB readouts the data from the 36 ASICs into its internal memory (the event buffer) each GTU to compute the L1 trigger. The L1 principle consists in counting an excess of signals over background in groups of 3$\times$3 pixels lasting more than a preset persistence time. The background rate is monitored continuously to adjust in real-time the trigger threshold keeping the L1 trigger rate compatible with the DAQ recording rate (a few Hz). The trigger is evaluated each GTU. Because Air-Showers may extend over 100 GTU, this trigger has the buffering capability over 128 consecutive GTU. To reduce the dead-time induced by event readout, the event buffer is doubled.

\subsection{Data Processing Subsystem}
Data acquisition and storage is the task of the Data Processing subsystem (DP).
The DP includes the CCB designed to perform the second level trigger L2, described in~\cite{hbibCCB}. For each generated L1 trigger, the CCB reads data corresponding to the 128 consecutive GTU from the PDMB buffer. 
In JEM-EUSO, the CCB combines information from 9 PDMs, to reduce the trigger rate to about a few Hz or less compatibly with the data storage capabilities of the DAQ. In the case of the EUSO-Balloon, the CCB, based on a Xilinx Virtex-4 FX-60, serves only one PDM and therefore the L2 trigger is not essential. However the L2 functionality will be tested. In addition to perform the second trigger stage, the CCB reads events from PDMB and passes them to the CPU. It also passes clock signals and configuration data to the PDM. The Clock-Board (CLKB), based on a Xilinks Virtex5 FPGA is part of the DP. It generates and distributes the system clock (40 MHz) and the GTU clock (400 kHz, 98\% duty cycle) to all devices. A GPS-Board provides information to perform event time tagging data with an accuracy of a few microseconds. The CPU (Motherboard iTX-i2705 model, processor  Atom N270 1.6 GHz) merges the event data with the time tagging data to build an event of a size of 330 kB. This implies a data flow of 3MB/s for a 10 Hz L1-L2 trigger. The CPU writes all data on disks (1 TB CZ Octane SATA II 2.5Ó SSD) and may also send to the balloon telemetry a subset of events flagged by the CCB, to allow monitoring.

\vspace*{-0.3cm}
\paragraph{Monitoring}
The instrument behaviour is controlled at low frequency by the Housekeeping system (HK) which is a part of the DP. It is based on a commercial micro controller board (Arduino Mega 2560) designed to control temperatures, voltages, and alarms raised by several boards. The CPU polls from time to time the alarms and initiates the corresponding foreseen actions. The HK is connected to the telemetry system to receive basic commands, namely those that allow to turn on-off most of the boards power supplies through relays.

\vspace*{-0.3cm}
\paragraph{Power Supply and Electrical Architecture}
The instrument runs autonomously thanks to a set of 60 battery cells providing 28 V (225 W during 24 H) to a set of low-voltage boards generating isolated-decoupled lower voltages to the PDM (HVPS and PDMB), DP (CPU,CLKB,GPSB,CCB and HK). The electrical architecture follows the EMC rules to prevent floating reference voltages induced by bad grounding (current ground loop effect).

\section{Assembly and Tests}
\label{sec:AssemblyTest}
After fabrication, the instrument has to be calibrated with great accuracy. 
The key goal of the calibration is to relate a measured digitised signal into the true number of photons impinging on the focal surface or on the first lens.
Thus the optics and the MAPMTs will be calibrated very accurately. Other subsystems like the trigger have to be tested once the instrument is close to final assembly.
Each of the subsystems of the instrument are calibrated if necessary and tested before the full integration. We also plan to test the integrated instrument before delivering it to the launch site.

\subsection{Optical Tests}
Even if the focal length of each lens and the combined focal length can be predicted by accurate ray tracing, error budgets are obviously associated to manufacturing. To achieve a resolution smaller than the pixel size, the optimum relative distance between the three lenses and the focal surface will be measured experimentally. This is performed by using a large parallel UV beam along the optical axis, sent over the first lens and measuring the focal length by adjusting the position of a CCD camera to get the narrowest focused spot.

\subsection{Measuring MAPMT Performances}
Each channel of the MAPMT is characterized by its photodetection efficiency and by the gain of the phototubes. Before assembling 4 MAPMTs into an EC, the MAPMTs are sorted according their gain. This allows to assembly ECs with homogeneous MAPMT gain for the same HV \cite{hbibPMT}. For this, gains have to be initially measured with sensitive, commercial multi-channels charge to digital converters imposing to work with a gain being a factor three above its nominal value (corresponding to HV around -1100~V). Once ECs are assembled, the gain and detection efficiency of each channel are measured with the ASICs at a nominal HV value of -900~V. Both types of measurements are done by illuminating the photocathode with a LED (monitored with a NIST-photodiode). MAPMTs are operated in single photoelectron mode~\cite{hbibCalib} to measure the single photoelectron spectrum for each of the 2304 pixels of the instrument camera.
This procedure allows to determine the exact high voltage to be applied to the MAPMTs photocathodes of each EC Units.

\subsection{ASIC Settings}
The ASICs measure the single photoelectron spectra at nominal high voltage for each of the channels by performing a series of runs ramping the discriminator voltage.
Since the relative gain of channels inside an EC-Unit can be slightly different, the ASICs allow balancing the discrepancies. This is done once the PDM is mounted and each MAPMT is associated to an ASIC. Then the nominal discriminator threshold, at 1/3 of a photoelectron, to be applied to each ASIC is established.

\subsection{Trigger Tests}
Once the PDM is mounted, including the PDMB, the L1 trigger algorithm performance is checked by illuminating the focal surface by a light spot moving closely to speed-of-light, generated by an "old" persistent-screen scope. 

\subsection{Instrument Tests}
Final tests will be performed after integration of all subsystems inside the instrument. A check of the correct final position of lenses and focal surface will be done by lighting up the first lens by a parallel UV beam along the optical axis. The size of the focused point on the focal surface will be minimised by finely adjusting the position of the PDM at the sub-millimetre scale. At the end of the integration and at launch site, basic health tests on the electronics will be performed by directly and uniformly illuminating the focal surface or the first lens by a LED-controlled, in single photon mode, as described in~\cite{hbibCalib}.

\section{Operation and Analysis}
\label{sec:Operation}
During the balloon flight operation, the instrument will be controlled from ground by an operator using a control program~\cite{hbibOffOnLineAna} interfaced to the TC/TM system (Telecommand and Telemetry) NOSYCA of CNES. At a given altitude reached by the balloon, a command will be issued to turn on the instrument. The HK system will turn on one by one each of the subsystems while monitoring parameters will be downloaded at ground. After the nominal set-up is reached, after having chosen the convenient configuration parameters for the ASICs and the triggers, the balloon operator will launch the DAQ program running on the CPU. Operators will control the basic run parameters, namely the background rate calculated by the PDMB. Conventionally thresholds auto-adapt to the required L1-L2 rates unless the operator forces another mode of trigger settings. At any moment, the instrument can be shut down. This will be certainly done during the descent phase.

\section{Conclusion}
\label{sec:Conclusion}
The EUSO-Balloon, a scaled-down pathfinder for the JEM-EUSO mission is being currently built to fly as a payload of a stratospheric balloon launched by CNES. It is by itself a complete autonomous instrument capable to validate the technique at the core of the JEM-EUSO mission. This first flight is expected during the CNES balloon campaign in 2014.

\vspace*{0.2cm}
{
\footnotesize{{\bf Acknowledgment:}{ This work was strongly, technically as well as financially supported  by \mbox{CNES} and by the JEM-EUSO collaboration. We thank ESA for supporting the JEM-EUSO team with the activities of the ESA Topical Team on "JEM-EUSO".}

}
}
\clearpage


%% file: icrc2013-0874.tex


\title{UV night background estimation in South Atlantic Anomaly}

\shorttitle{UV night background estimation in South Atlantic Anomaly}

\authors{
Pavol Bobik $^{1}$,
Marian Putis $^{1}$,
Mario Bertaina $^{2}$,
Svetlana Biktemerova$^{3}$,
Donatella Campana $^{4}$,
Francesco Fenu $^{5,6}$,
Fausto Guarino $^{4}$,
Karel Kudela $^{1}$,
Thomas Mernik $^{5}$,
Blahoslav Pastircak $^{1}$,
Kenji Shinozaki $^{5}$
for the JEM-EUSO Collaboration.
}

\afiliations{

$^1$ Institute of Experimental Physics Slovak Academy of Sciences, Kosice, Slovak Republic \\
$^2$ Department of General Physics, University of Torino, Torino, Italy \\
$^3$ Joint Institute for Nuclear Research, Joliot-Curie 6, 141980 Dubna, Russia \\
$^4$ Istituto Nazionale di Fisica Nucleare - Sezione di Napoli, Italy \\
$^5$ Institute for Astronomy and Astrophysics, University of T\"ubingen, Tubingen, Germany \\
$^6$ RIKEN Advanced Science Institute, Wako, Japan

}

\email{mernik@astro.uni-tuebingen.de} 

\abstract{The JEM-EUSO experiment at the International Space Station will detect the Earth night side UV light produced by UHECR due to their interaction with the atmosphere. The estimation of UV background in different conditions is necessary to precise the estimation of the experiment's operational efficiency.
In this article, we estimate an intensity of UV light during night inside the South Atlantic Anomaly and discuss its influence to the JEM-EUSO operational efficiency. Three sources of UV ligth were considered, galactic cosmic rays, airglow and Cherenkov light produced by trapped electrons in the detector lenses. For galactic cosmic rays a model based on simulation of cosmic rays trajectories in the geomagnetic field and secondary particle production in the atmosphere was used. Airglow production is evaluated by AURIC model \cite{bib:AURIC} estimations. The trapped electrons influence is evaluated using the GEANT 4 package \cite{bib:GEANT4_1,bib:GEANT4_2}.}


\maketitle

\section{Introduction}

Some physical mechanism can restrain measurements of the UV signal produced by extensive air showers created by ultra high energy cosmic rays (UHECR) in the South Atlantic Anomaly with the JEM-EUSO detector at the International Space Station. Geographical and time dependencies of UV light intensities at Earth night side in wavelength range 300 - 400 nm still need further investigation. Actual measurements suggest a latitudinal dependency of the UV intensity \cite{bib:Tatiana 2011}. 
There are not many experimental data on UV emissions and energetic electrons 
simultaneously measured at low altitudes, but contrary to high latitude UV nightglow 
production related to particle precipitation in the regions of aurora and outer radiation belt,
a mechanism of the mid-latitude UV enhancements is still unknown.
In this article we investigate the level of UV background in the South Atlantic Anomaly and its influence to the JEM-EUSO operational efficiency \cite{bib:Performance}. In this study we consider three possible sources of higher level of ultraviolet background (UV BG hereafter) in South Atlantic Anomaly (SAA). The first considered source is UV BG created by galactic cosmic ray (GCR) interactions with atmosphere. Second considered source is airglow production in the SAA. Third is interaction of electrons trapped in the Earth radiation belts with the JEM-EUSO optics, where relativistic electrons can create Cherenkov photons in the lenses.

\section{Galactic cosmic rays}

\begin{figure}[t]
  \centering
  \includegraphics[width=0.4\textwidth]{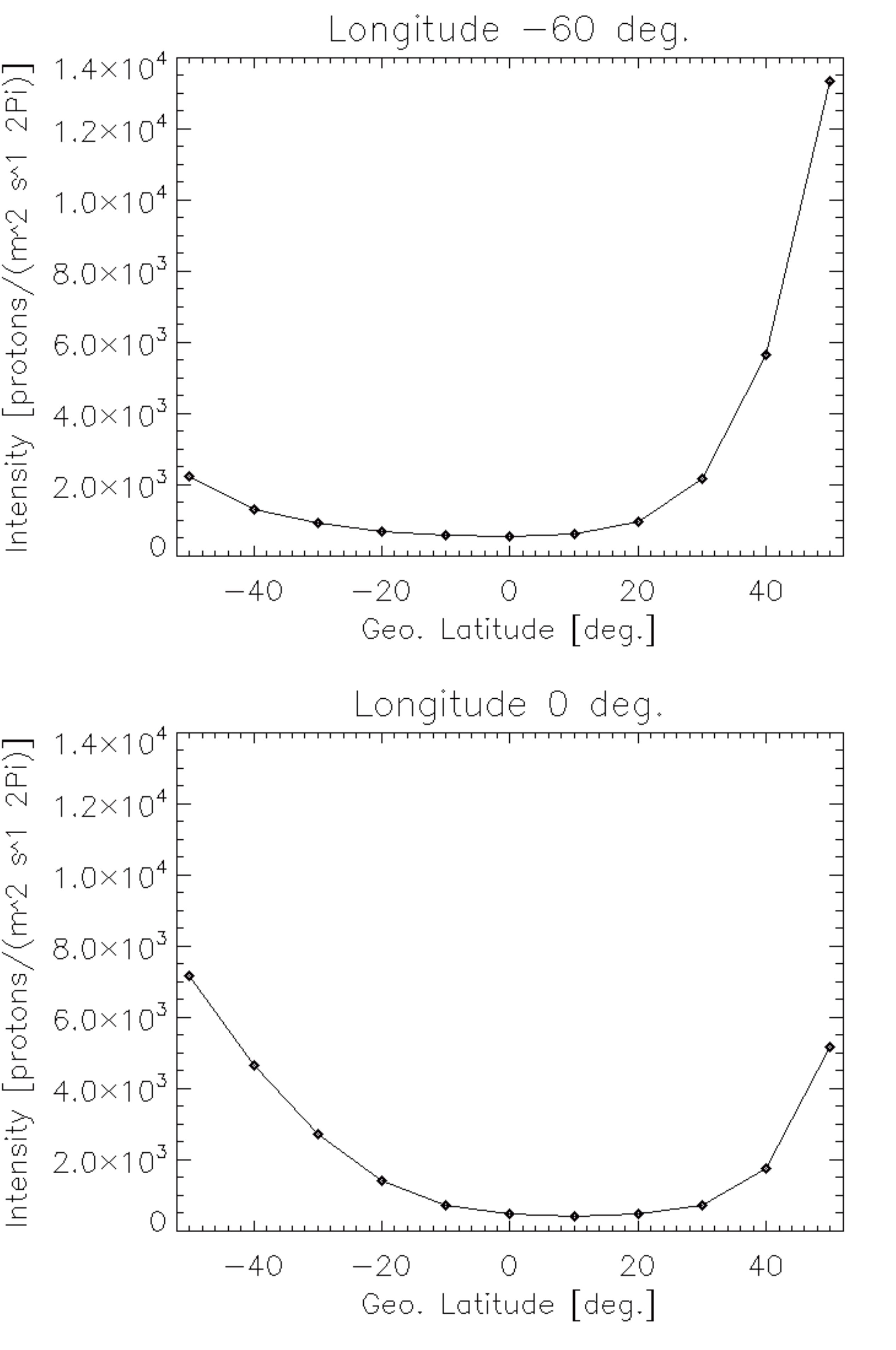}
  \caption{Intensities of cosmic rays proton component at altitude 382 km for two latitudinal profiles. For longitude 0$^\circ$ (bottom panel) and -60$^\circ$ (upper panel).}
  \label{GCrays_fig}
 \end{figure}

Galactic cosmic rays from interstellar space enter the heliosphere where they are modulated \cite{bib:HelMod ApJ} and a part of them reach 1 AU and enter the Earth's magnetosphere. The magnetosphere acts as filter to cosmic rays with variable transparency for different energies of cosmic rays.
We evaluate UV BG created by GCR over entire Earth surface in reference \cite{bib:AdvSR 2012} without taking SAA into account. We use measured AMS-01 proton and helium spectra from precursor fligth onboard the Space Shuttle Discovery mission STS-91 \cite{bib:Aguilar 2002, bib:Alcaraz 2000}. Because the published AMS spectra do not include the SAA region we need additional simulation to find magnetosphere transparency in SAA. To test a hypothesis that GCR in SAA can produce a significantly bigger amount of UV BG in comparison with other regions on Earth we evaluate a magnetosphere transparency for a set of points on the meridian line crossing SAA at International Space Station (ISS) altitude (382 km). Along the meridian line we set 11 points with latitudinal step of 10$^\circ$ from -50$^\circ$ to 50$^\circ$, covering latitudinal extension of ISS orbit with inclination 51.6$^\circ$. The used backtracing method for particles trajectories evaluation in geomagnetic field is described in \cite{bib:Geomag 1, bib:Geomag 2, bib:Geomag 3}. Specifically for this simulation we evaluate for every of the selected points 576 directions (one direction per solid angle of 0.0109 sr) of incoming cosmic rays covering half sphere directed outward to space. For every direction we simulated 20 thousands energies with incremental rigidity step 0.01 GV from 0.01 GV till 200 GV. As unmodulated spectrum of protons at 1 AU we take the spectrum from region 10 of the AMS-01 measurements \cite{bib:Alcaraz 2000}. However, this is the spectrum at 1AU from 1998 and we want to estimate a spectrum for the years 2017 to 2020, where situation in both periods can be similar because JEM-EUSO will measure during declining phase of solar activity with reaching similar solar minimum condition as was in 1998. 
The intensities of cosmic rays are evaluated from transmission functions constructed from allowed trajectories for used set of latitudinal points. The results for places at line with geographical longitude -60$^\circ$ crossing SAA and for comparison also for line with longitude 0$^\circ$ are presented in figure \ref{GCrays_fig}. The cosmic rays intensities decrease in equatorward direction, with a minimum close to the geomagnetic equator. Intensities in SAA region do not exceed numbers on similar latitudes at meridian line  with longitudes 0$^\circ$. Production of UV photons at longitudes with 0$^\circ$ was estimated \cite{bib:AdvSR 2012} as very small (i.e. less than 0.01 percent) in comparison to other sources of UV BG (nightglow, zodiacal light, integrated star light) at the Earth's night side. Therefore also the number of photons produced by GCR in SAA will be very small. We can conclude that GCR do not increase background in SAA significantly and consequently do not affect operational efficiency of the JEM-EUSO experiment. 

\section{Airglow production}

We present results of UV nightglow radiation in the wavelength range from 300 - 400 nm obtained by the AURIC model, which is computational tool for upper atmosphere radiation provided by Computational Physics, Inc. \cite{bib:AURIC}. AURIC is able to compute dayglow and nightglow radiation for many spectral features. We calculate radiation in wavelength range of 300 - 400 nm for Herzberg I,II and Chamberlain radiation. Purpose of this study was to estimate the prominence of South Atlantic Anomaly in global UV nightglow radiation.

Computation was provided in a range of latitude from $-85.5$$^\circ$ to $85.5$$^\circ$ and a longitude from $-180$$^\circ$ to $180$$^\circ$. 
This area was divided into 12 bins in longitude and 35 bins in latitude (grid with 420 cells). The center of each cell was used as input geoposition for the AURIC code. We calculate UV radiation for 4 months (March, June, September, December) from 1970 to 1994 (1994 is the last year in the AURIC parameters database). And for all months in years 1990 and 1994. The value of the nightglow radiation at given position for the period of interest (years in range 1970 - 1994 and separate years 1990, 1994) was obtained in the following way. In each month only one night was taken from day 20 to 21. We assume that radiation is not changing dramatically during one month. For this night, which is defined by solar zenith angle $> 110$$^\circ$, we obtain radiation for each full hour of local time. Final value was obtained as average of all values at given position for the period of interest. 

\begin{figure}[t]
  \centering
  \includegraphics[width=0.4\textwidth]{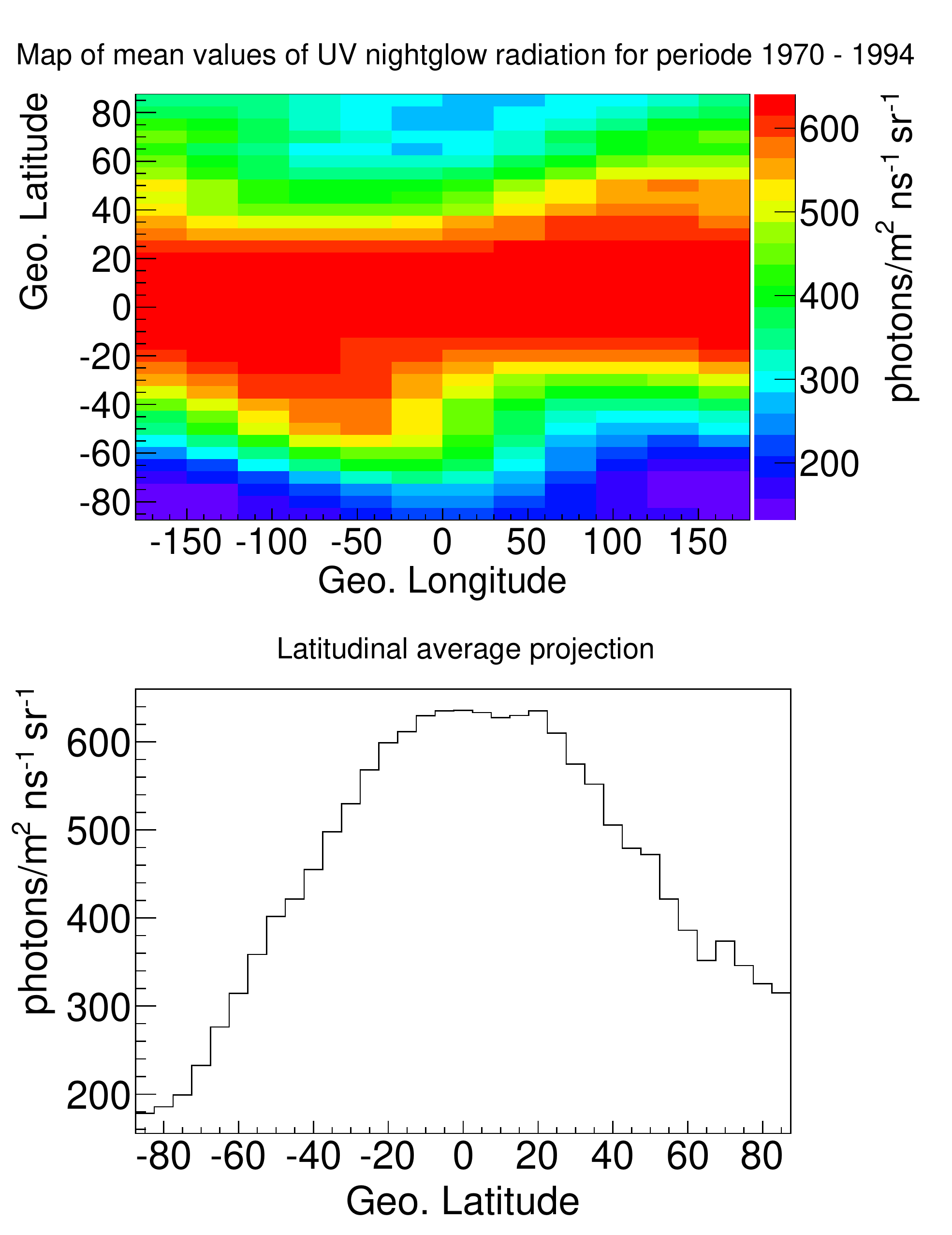}
  \caption{Top panel: Map of average values of UV nightglow radiation in March, June, September and December (together) for years in range from 1970 to 1994. Bottom panel: Latitudinal average projection of the map.}
  \label{All_year_4_months_fig}
 \end{figure}

We use the radiation obtained in this way to create a map of average values for the years from 1970 to 1994 (Fig. \ref{All_year_4_months_fig}). From this picture we can see an area with increased  values of radiation from $-20$$^\circ$ to $-50$$^\circ$ geographic latitude and from $-90$$^\circ$ to $0$$^\circ$ geographic longitude. This increase is not visible for example on east hemisphere (positive longitude) in same latitudes. However, this radiation is not bigger then other areas around equator (Fig. \ref{All_year_4_months_fig} bottom).

\begin{figure}[t]
  \centering
  \includegraphics[width=0.4\textwidth]{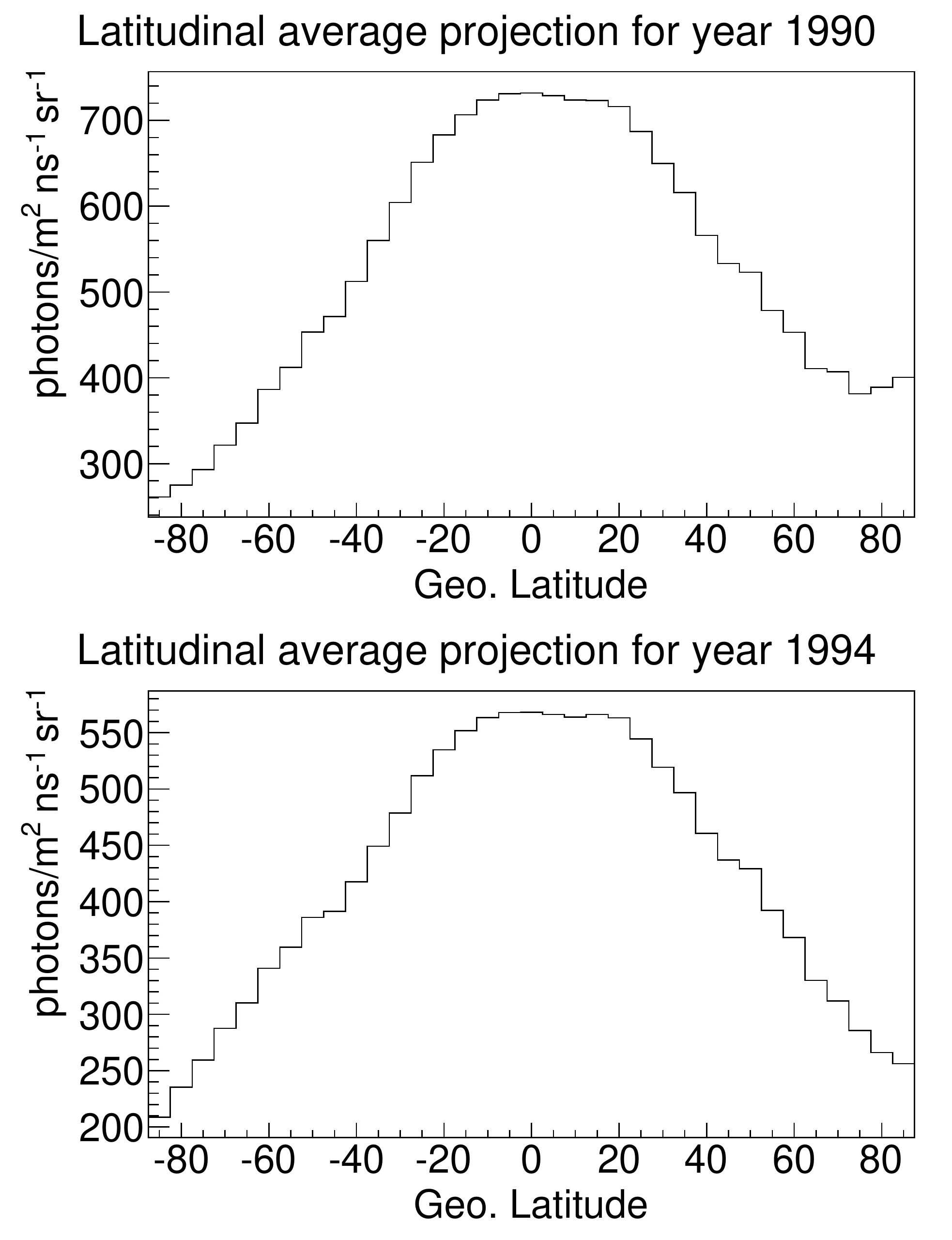}
  \caption{Top panel: Latitudinal average projection of the map of UV nightglow average values for all months in 1990. Bottom panel: Latitudinal average projection of the map of UV nightglow average values for all months in 1994.}
  \label{1990_1994_year_all_months_fig}
 \end{figure}

On figure \ref{1990_1994_year_all_months_fig} top panel the latitudinal average projection of map of mean values for year 1990 (close to solar maximum) and on figure \ref{1990_1994_year_all_months_fig} bottom panel the same picture for 1994 (close to solar minimum)is shown. From this figure and from figure \ref{All_year_4_months_fig} it is obvious that the average values of UV nightglow radiation during solar maximum are bigger than values of the nightglow radiation at the whole period (from 1970 to 1994) and the radiation during the year 1994 is lowest. There is no significant increase of radiation in SAA area for these two years with respect to other geopositions.

\begin{figure}[t]
  \centering
  \includegraphics[width=0.4\textwidth]{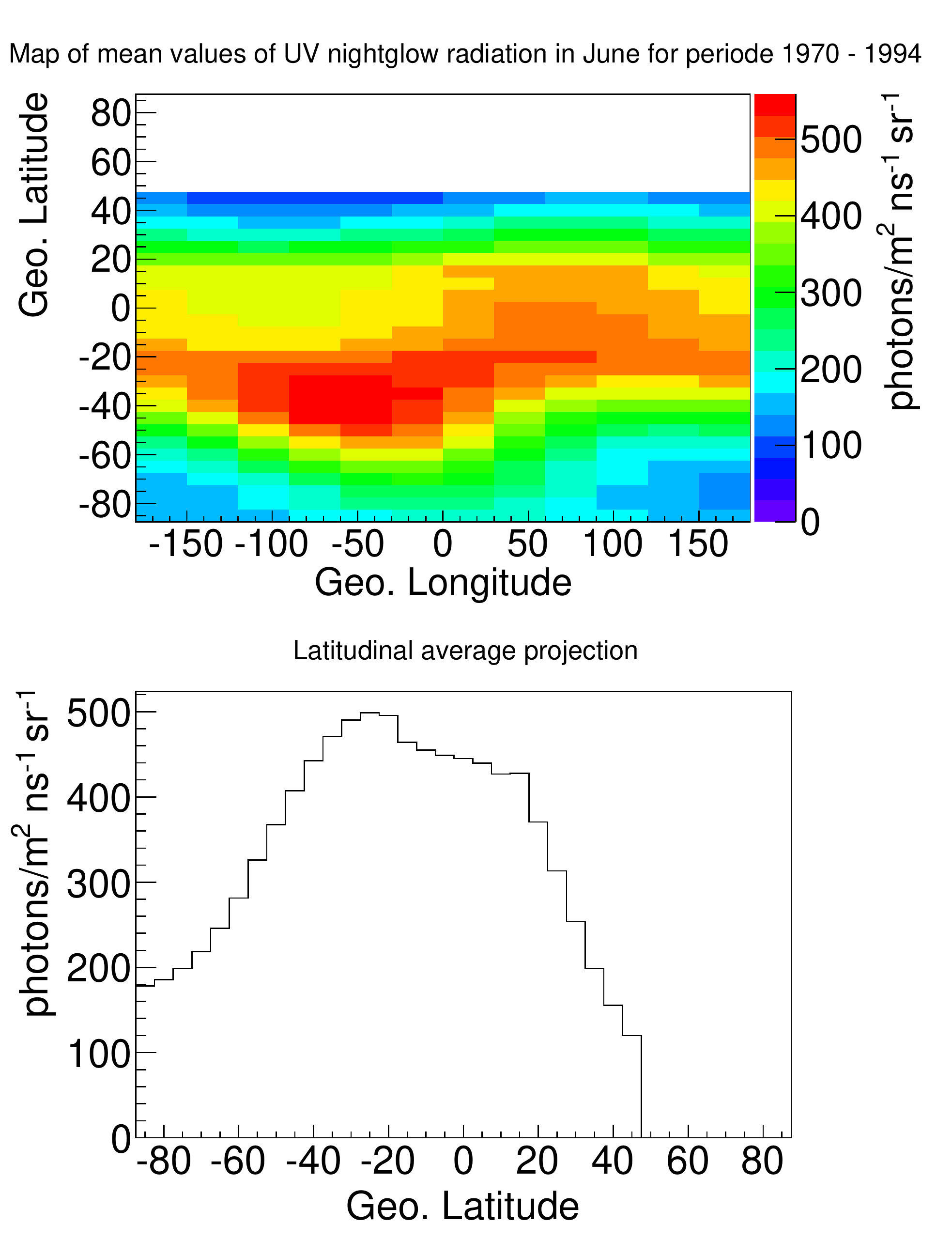}
  \caption{Top panel: Map of UV nightglow average values in June for years in range from 1970 to 1994. Bottom panel: Latitudinal average projection of the map.}
  \label{All_year_June_month_fig}
 \end{figure}

Figure \ref{All_year_June_month_fig} shows nightglow radiation for whole period from 1970 to 1994 only for June. In this case we can see that the area of SAA is most dominant, but still it does not reach bigger average values than average values of places with different positions on figure \ref{All_year_4_months_fig}.

However in some periods we can observe maximum of produced UV light in the South Atlantic Anomaly, but generally this maximum does not exceed the usual maxima on the Earth surface. Thus, the influence of UV BG in SAA to JEM-EUSO measurements must be part of a wider study devoted to the UV BG, generally.

\section{Effect of trapped electrons}

We use a SPENVIS AE-8 model \cite{bib:SPENVIS} to estimate the intensity of the trapped electrons along ISS trajectory. 
A visualisation of intensity distributions with energies over 40 keV is presented in figure \ref{Fig_5_el_1bin_Smax}. 
The higher intensities of electrons in the South Atlantic Anomaly and in the region over North America with high geomagnetic latitudes can be clearly seen. 
The maximum intensities in the center of SAA reaches values in order of millions of electrons per cm$^{2}$/s.
 The evaluated spectrum of trapped electrons from a solar maximum period in the center of the South Atlantic Anomaly was used as input for simulation of electrons in detector optics by the GEANT 4 package \cite{bib:GEANT4_1,bib:GEANT4_2}. The results show that one electron approximatelly produce 0.1 Cherenkov photons at the detectors focal surface. In the center of SAA, approximatelly to one square meter of detector lenses surface enter $10^{10}$ electrons per second. They produce $1*10^{9}$ photons per m$^{2}$ of FS per second, i.e. 1 photon per m$^{2}$ ns. 
This is approximatelly 1\% in comparison to photons which pass the detector and reach the FS from the standard UV BG of 500 ph/(m$^{2}$ ns sr). This leads to conclusion that electrons trapped in non disturbed magnetosphere do not affect the JEM-EUSO operational duty cycle significantly.

But even if production of photons in the detector lenses during the periods with nondisturbed magnetosphere is small, during disturbed periods intensities of trapped electrons can increase by two orders of magnitudes. 
Also regions with high intensity of trapped electrons is extended far beyond SAA borders during geomagnetic storms \cite{bib:AOS}. This effect should be considered and we will estimate it in future.

 \begin{figure}[t!]
  \centering
  \includegraphics[width=0.4\textwidth]{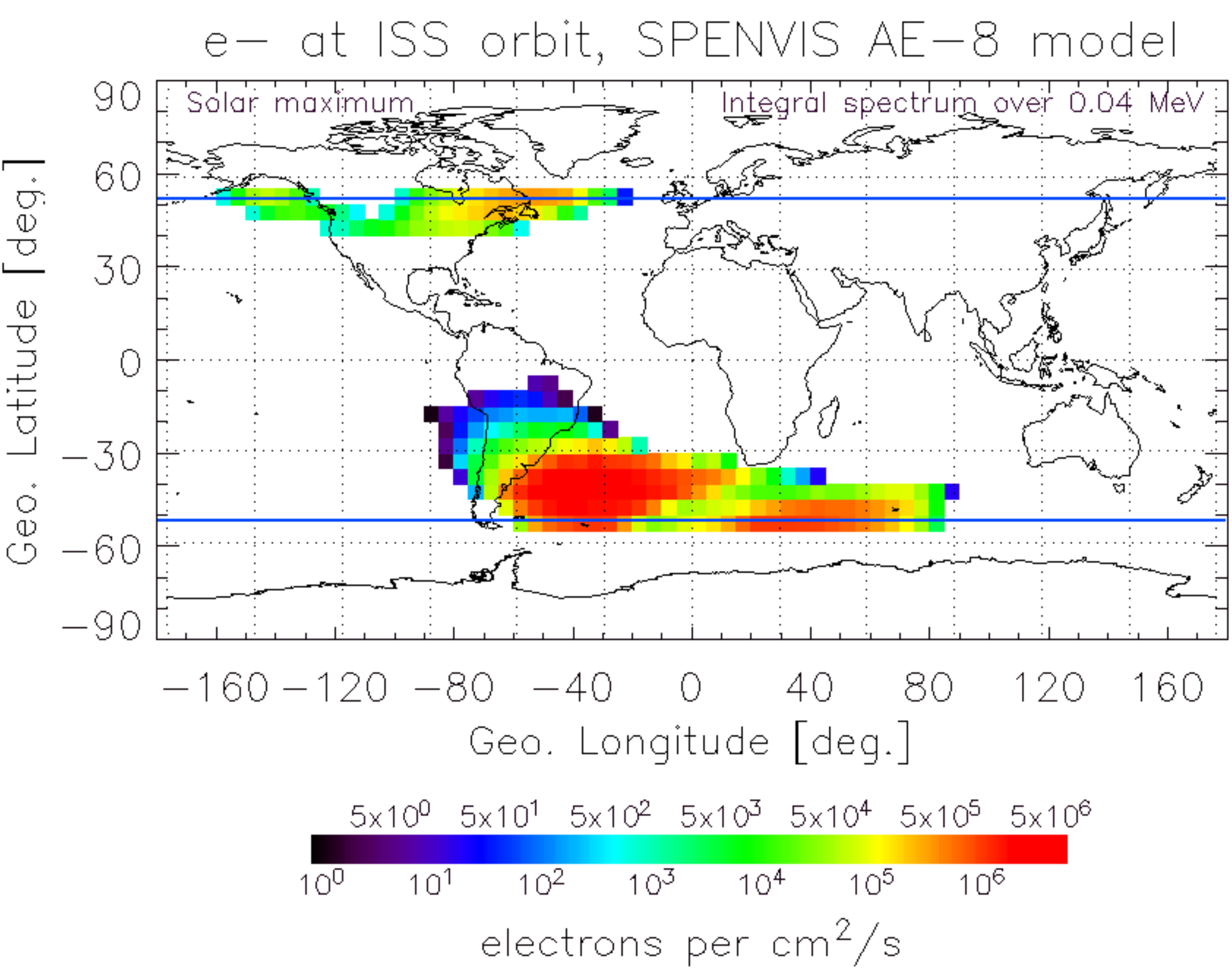}
  \caption{Intensities of the trapped electrons over areas covered by ISS trajectory evaluated from SPENVIS AE-8 model for solar maximum period. Blue lines show ISS trajectory bordes at geographical latitudes -51.6$^\circ$ and 51.6$^\circ$.}
  \label{Trapped_el_1bin_Smax_fig}
 \end{figure}

\section{Conclusions}

We investigate three possible mechanism to produce a UV light during night inside South Atlantic Anomaly 
and their influence on JEM-EUSO detector measurements. While UV background from GCR is very small to have 
any influence to JEM-EUSO measurements, UV light produced by airglow should be considered. 
However not as special contribution in SAA but as influence of UV BG produced by airglow globally where 
the amount of produced UV light in the wavelength range of 300-400 nm depends on geographical position with 
maximum production in average around geomagnetic equator and on time. The trapped electrons increase UV background 
registered by the detector in the center of SAA due to production of Cherenkov light in the detector lenses by approximately about one (few) percent and do not affect the operational duty cycle. Possible influence of trapped electrons should be considered during periods with disturbed magnetosphere (with higher Kp or Dst index) and will be evaluated in near future.

\vspace*{0.5cm}
{
\footnotesize{{\bf Acknowledgment:}{This work was supported by Slovak Academy of Sciences MVTS JEM-EUSO as well as VEGA grant agency project 2/0076/13.
}}

}
\clearpage

%% file: icrc2013-0919.tex



\title{Retrieving cloud top height in the JEM-EUSO cosmic-ray observation system}

\shorttitle{Retrieving cloud top height in the JEM-EUSO cosmic-ray observation system}

\authors{
A.Anzalone$^{1,5}$,
M.Bertaina$^{2}$,
S.Briz$^{3}$,
R.Cremonini$^{2}$,
F.Isgr\`o$^{4,5}$
for the JEM-EUSO Collaboration.
}

\afiliations{
$^1$ INAF-IASF, Istituto di Astrofisica Spaziale e Fisica Cosmica, Palermo, Italy \\
$^2$ Fisica, Universit\`a degli Studi di Torino, Italy  \\
$^3$ Universidad Carlos III de Madrid, Spain \\
$^4$ DIETI, Universit\`a degli Studi di Napoli Federico II, Italy \\
$^5$ Istituto Nazionale di Fisica Nucleare - Sezione di Catania and Napoli, Italy
}

\email{anna.anzalone@ifc.inaf.it}

\abstract{Clouds and atmospheric conditions affect the strength of the fluorescence light and the Cherenkov signal received from Extensive Air Showers. JEM-EUSO will observe the conditions of the atmosphere and clouds in the field of view of the telescope making use of a state-of-art atmospheric monitoring system. This work revises already existing and proposes new algorithms for cloud detection and height estimation. The algorithms have been checked by analysing scenes retrieved by operational atmospheric sensors similar to the IR camera on board JEM-EUSO atmospheric 
monitoring system (Meteosat, Aqua, Calipso). Results are very promising, although some algorithms have to be more extensively validated considering all type of clouds and scenarios. }


\maketitle

\section{Introduction }
The JEM-EUSO space observatory is foreseen to be launched and attached to the Japanese module of the International Space Station (ISS) in 2017. Its aim is to observe UV photon tracks produced by Ultra High Energy Cosmic Rays (UHECR) developing in the atmosphere and producing Extensive Air Showers (EAS)\cite{nbib:EUSOperf,bib:schoenberg}. \\
Clouds affect fluorescence and Cherenkov signal received from EAS, as well as its reconstruction efficiency. Estimation accuracy depends on cloud coverage and cloud top height (CTH). The strength of the fluorescent light also depends on the transparency of the atmosphere on the cloud coverage and on the height of the cloud top \cite{SaezCano2012}. 
There are many satellites that provide atmospheric information from multi-spectral observations with good spatial and temporal resolutions, for instance, geostationary satellites (GOES, MSG, etc), LEO satellites (Terra/Aqua, HIRS, etc.), CALIPSO, etc. However JEM-EUSO requires spatially and temporally simultaneous information of those atmospheric parameters.\\
In order to monitor the atmospheric conditions and the cloud coverage in the JEM-EUSO FoV a state-of-art Atmospheric Monitoring System (AMS) will be included in the system \cite{Neronov2011}. The AMS is crucial to estimate the effective exposure with high accuracy and to increase the confidence level in the UHECRs events just above the threshold energy of the telescope. The AMS consists of a LIDAR, an infrared (IR) camera and global atmospheric models. The LIDAR will measure the optical depth profiles of the atmosphere in selected directions. The IR camera will provide the cloud coverage and the cloud temperature. This work focuses on the algorithms to retrieve the CTH from the radiance measured by the IR camera.
CTH retrieval can be performed using either stereo vision algorithms or accurate radiometric information. The first methodology requires two different views of the same scene. The stereo technique has been applied on a stereo system composed by two Meteosat geostationary satellites of the new generation. The second one is based on the relationship between the cloud temperature and the cloud height. Some algorithms have been developed to consider atmospheric effects and to retrieve the cloud temperature from the brightness temperature (BT) which is calculated from the radiation measured by the IR camera after the calibration procedure. Data provided by the global atmospheric models will also be used to obtain the cloud height from the temperature data. The radiometric technique has been checked with MODIS images. However MODIS also can present some inaccuracies. A parallel inter-satellite comparison has been carried out to quantify possible discrepancies in satellite observations.

\section{Stereo Vision Algorithm }
\begin{figure}[t]
  \centering
  \includegraphics[width=0.4\textwidth]{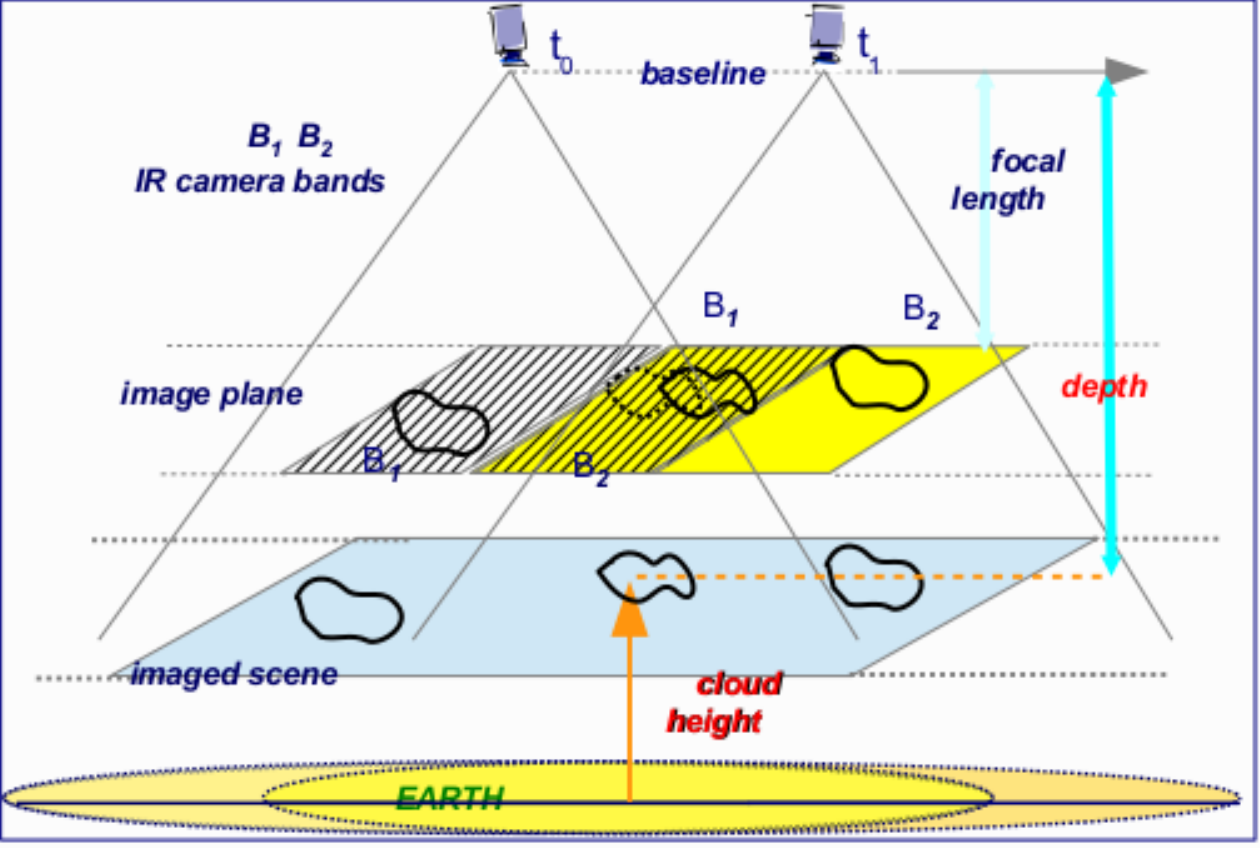}
  \caption{'JEM-EUSO Stereo System'.The scheme shows the overlapping needed for stereo be applied. As the IR camera moves towards right, consecutive images overlap in different bands.}
  \label{JSS}
 \end{figure}

The Stereo method is a different approach that is under test for estimating the CTH. The studies found in the recent literature \cite{Moroney2002,Muller2007} show that the improvements obtained with the new multiview instruments on board polar satellites and the better specs of the new geostationary satellites, provide good results in comparison with the past.\\
Stereo vision in general attempts to infer information on the 3D structure and distance of a scene, from two images taken by two spatially separated cameras. Each camera gets a different view of the same object, and the parallax effect, \textit{disparity} in the following, is used to reconstruct the \textit{depth} information, i.e. the distance of the object from the visual sensor.\\
In our project the 'JEM-EUSO Stereo System' is composed by the ISS, the IR camera and the ISS movement and it is constrained by the mission requirements. Instead of having two different cameras, the stereo imaging is accomplished by one camera moving along the observed scene, exploiting the ISS displacement. The scene results imaged from two different views and the intersection is processed to retrieve the distance from the IR device. Finally the CTH is obtained by subtracting the estimated \textit{depth} from the known ISS altitude. According to the specifications of the IR sensor, when it takes an image half of the scene is observed by the camera in the first band and the other half in the second band. In the next shot (after 17 s) the scene will appear displaced in such a way that the part of the scene seen in the first image by the band B1, will be seen in the second image by the band B2. Referring to the figure \ref{JSS}, the cloud in the middle of the imaged scene lies in the intersection of the views and it is shot both in B1 band and B2 band (see the image plane). In following shots the cloud on the right will be imaged in the same way and the whole FOV will be totally covered. In the figure \ref{JSS} it's also visible the parallax effect as an apparent motion of the cloud in the middle of the scene, highlighted  in the overlapped part of the two views in the image plane.

\begin{figure}[t]
  \centering
  \includegraphics[width=0.4\textwidth]{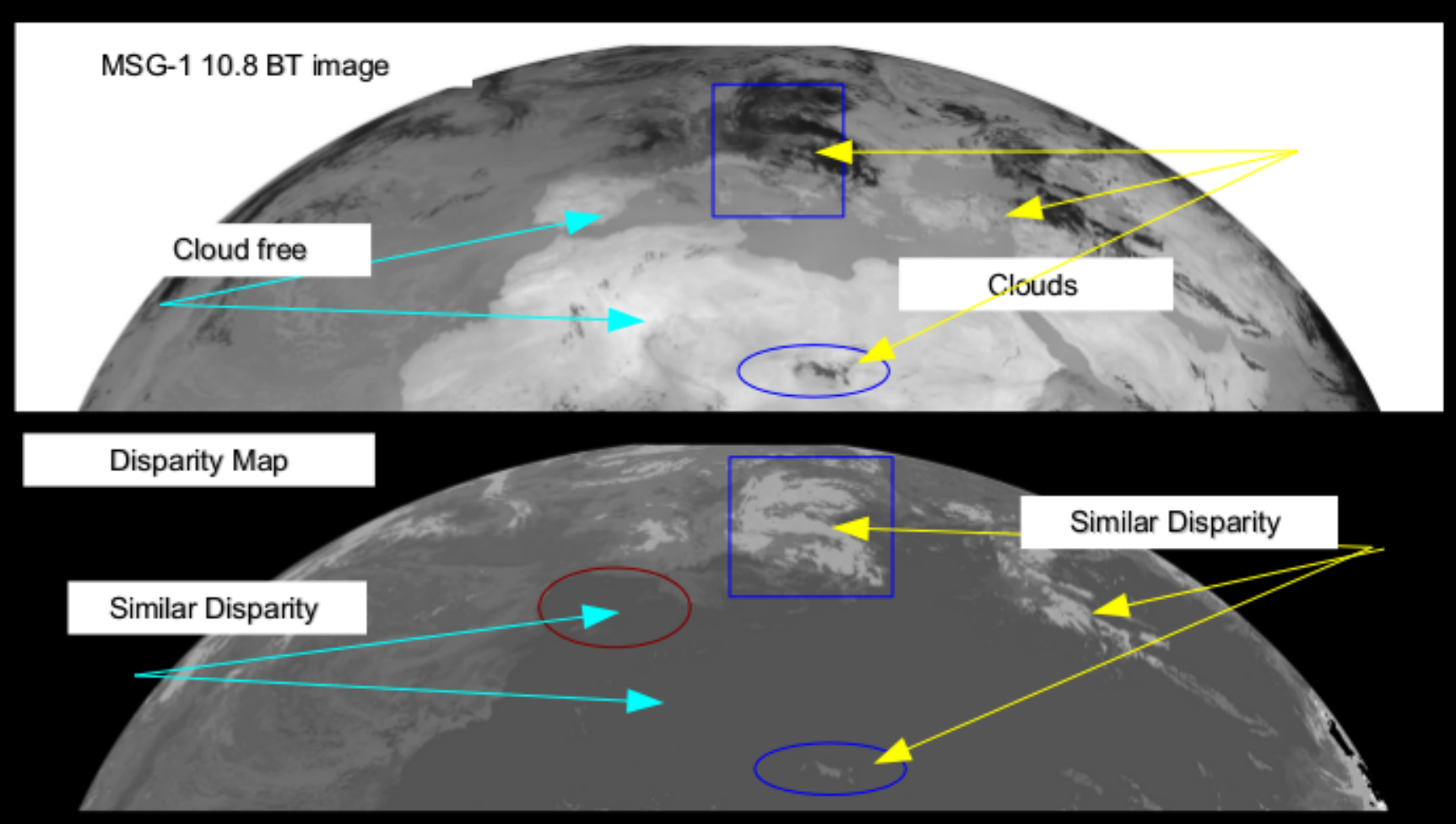}
  \caption{Disparity map.Top:One of the two MSG images used to estimate the map. Bottom: The corresponding disparity map where points having the same depth have the same gray level. The brightest points are the closest to the sensor and therefore the highest from the ground.}
  \label{Dsparity}
 \end{figure}

Multispectral stereo algorithms are currently under test. However, as a first step of the study a slightly modified version of the method in \cite{Manizade2006}, has been applied to mono band satellite stereo pairs. As in the common stereo methods applied to the satellite images, this method is not affected by the possible difference between the real temperature of a cloud and its BT, because the \textit{disparity} (i.e. the parallax effect) is estimated by geometric evaluations without any other extra information.\\
At first the method detects a fine segmentation of the BT layers in both members of the pair, creating a set of binary cloud masks, ROI masks in the following. The corresponding masks will be similar if the sensors of the stereo system are quite synchronous and images are in the same bands, as in our experiments. For the future application of this method to bi-spectral stereo pairs, it will be taken into account the possible discrepancy between values of the same image features in the two different bands.\\
This first step assumes that same BT layers have same heights. These values will not depend on the reliability of the measured BT for the reason mentioned before. Afterwards for each ROI mask the corresponding disparity is searched by looking for the best positional match with the other ROI mask of the pair. We have used the Euclidean distance for the similarity function.\\
The satellite images used for this work, are provided by the Meteosat Second Generation data base. The stereo system is composed of the two geostationary satellites MSG-8 and MSG-9 that although their baseline doesn't allow obtaining a good accuracy for the height reconstruction, they can be used as test for the disparity estimation that is a crucial step for the final height estimation. They are nearly synchronous, with a resolution worse than the one of the IR camera but in the same bands.  \\
An example of disparity map is shown on the bottom of the figure \ref{Dsparity}, obtained applying the method to the mono band MSG stereo pair (2011/07/20 at 12:00 UTC) and one of the input image is shown on the top of the figure. Areas having the same disparity grey level, represent points of the scene with the same distance from the IR camera, the more distant an object is, the smallest is the disparity and the darkest is the colour. \\ The preliminary results on the disparity estimation show a good agreement with what was expected. Further tests are in progress on different stereo satellite configuration and sensors.

\section{Temperature Retrieval Algorithms}
Due to atmospheric effects, the IR radiance emitted by the clouds is not the received one by the IR camera. Therefore the BT retrieved from the measured radiance is not the temperature of the cloud. To correct the atmospheric effects and obtain the cloud temperature some algorithms can be applied.\\
As the JEM EUSO IR camera will have two 1$\mu$m-width bands centered at 10.8 and 12 $\mu$m (hereafter referred as B1 and B2 respectively) we can develop a split window algorithm (SWA) to retrieve the cloud top temperature (CTT) of water clouds from BTs measured by the IR camera in these bands \cite{Inoue1987, Nauss2005}.\\
The retrieval algorithm is based on radiometric simulations of the
physical problem. The simulations were performed by means of the
radiative transfer equation and MODTRAN atmospheric simulation code
and considering only thick water clouds (emissivity 1). The eq. 1
shows this algorithm
\begin{equation}
T_{r}(B_{1}B_{2}) = -0.53819 + 2.6331T_{B_{1}} - 1.6305T_{B_{2}}
\end{equation}
Where $T_r$ is the cloud top temperature retrieved by the SWA, 
$T_{B_1}$ is the brightness temperature in the band centered in 10.8
$\mu$m, and
$T_{B_2}$ is the brightness temperature in the band centered in 12 $\mu$m.

\begin{figure}[tb]
\begin{center}
\begin{tabular}{cc}

\includegraphics[width=0.4\textwidth]{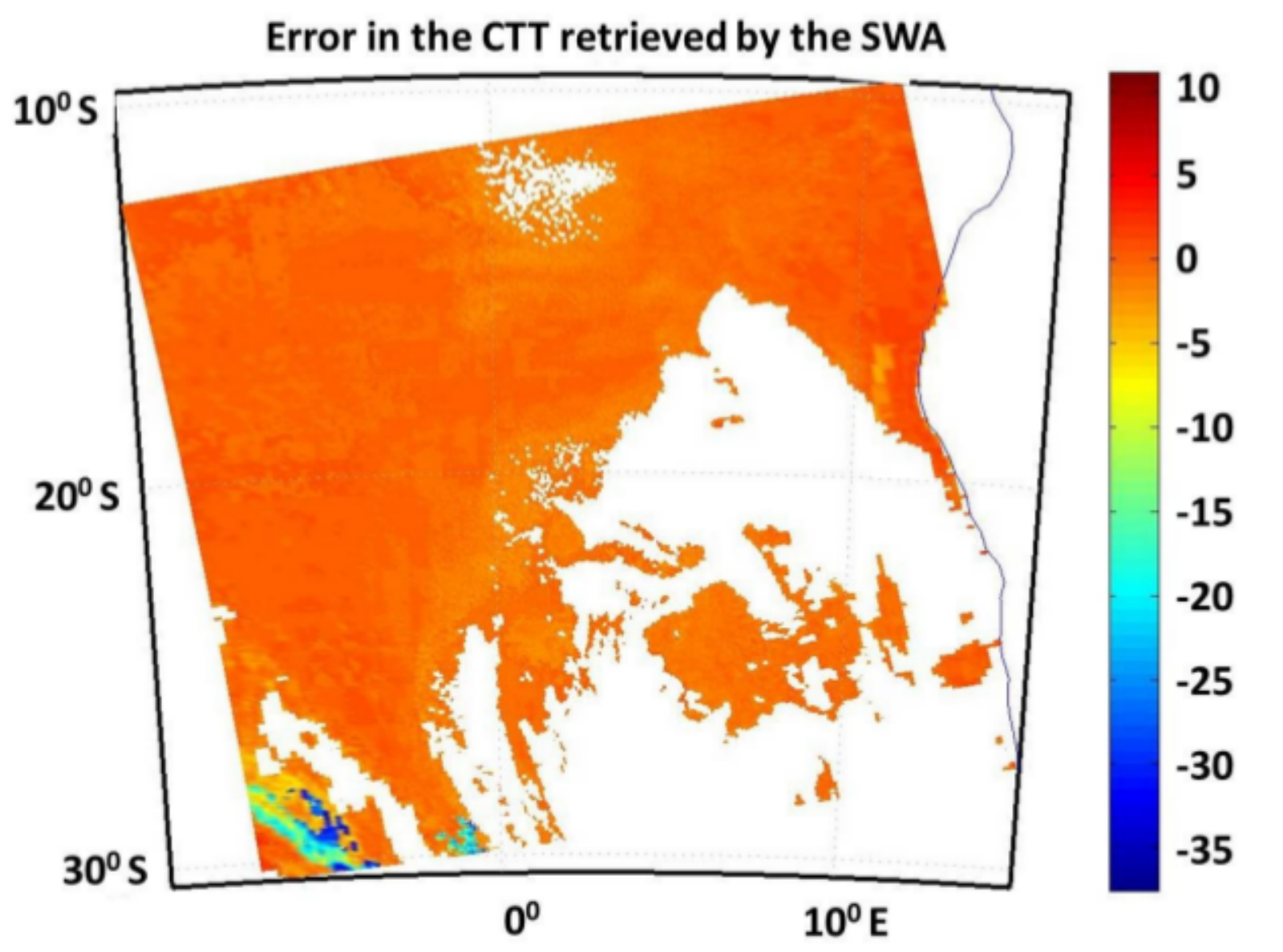} \\
\includegraphics[width=0.4\textwidth]{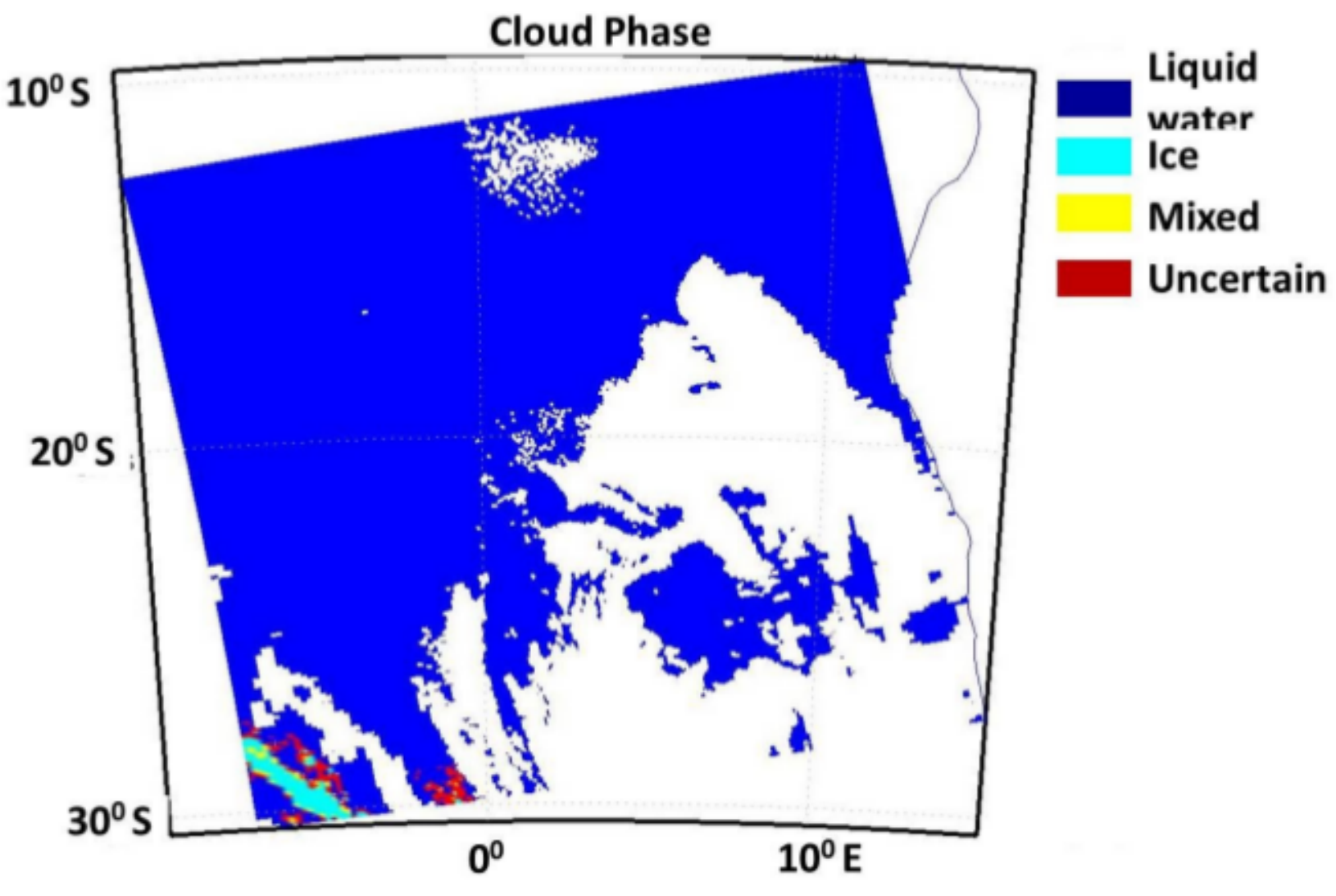}\\

\end{tabular}
\end{center}
\caption{Top: Temperature retrieval errors obtained when the SWA is applied to a MODIS image. Bottom: Cloud phase provided by MODIS }
\label{TempErrors}
\end{figure}

The SWA has been applied successfully to simulations of thick water
clouds in different scenarios. Errors are bigger for low clouds (0.5
km) and atmospheres with high water vapour content. However errors
keep below 0.3 K even for these worse situations, which is a very good
result. \\ 
In order to apply the SWA to real data MODIS images have been
used. From all the products that MODIS provides we have used only some
of them: BTs in bands 11 and 12 $\mu$m (IR camera bands), the cloud
temperature, the emissivity and the cloud phase.
The CTT retrieved by the IR camera and the CTT image provided by MODIS
are subtracted to calculate the retrieved temperature error. This
procedure has been applied to a MODIS image of South Hemisphere
(Atlantic Ocean) at 10S-30S latitude and 10W-15E longitude (07/10/2011
at 13:35 h UTC). In figure \ref{TempErrors} the corresponding error
image is shown (top image).

To clarify the error sources the error image has been compared with the emissivity and phase images. Although the correlation between the error and the emissivity is poor, the correlation with the phase is quite good.
The figure \ref{TempErrors} (bottom image) shows the cloud phase in which, blue denotes water, cyan is assigned to ice, yellow is used for pixels with mixture of water and ice and red corresponds to unknown phase. Errors for water phase keep below 1 K for the 95\% of the pixels and errors higher than 1 K can be related to pixels with effective emissivity lower than 1  (30\% aprox). \\
The conclusion is the following: the SWA is able to retrieve the temperature of thick water clouds with high accuracy, although it is not applicable to thin water clouds or ice clouds, as would be expected since the SWA was designed just for thick water clouds.
Other strategies are being developed to retrieve the temperature of thin clouds and to identify ice clouds.
The retrieval of the CTH will be carried out by using the vertical profile of temperature, as it will be shown in next section.

\section{Inter-Satellite Comparison of Cloud Height Retrieval}
\begin{figure}[tb]
\begin{center}
\begin{tabular}{cc}

\includegraphics[width=0.4\textwidth]{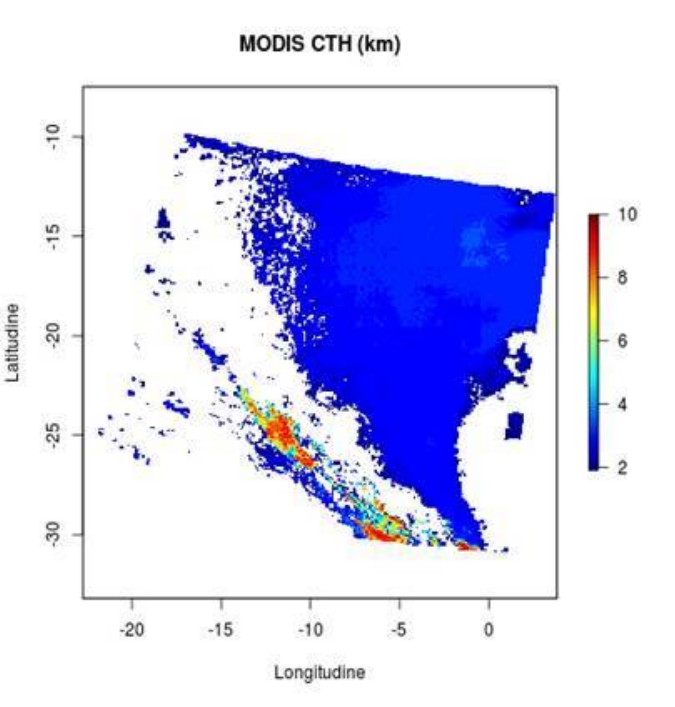} \\
\includegraphics[width=0.4\textwidth]{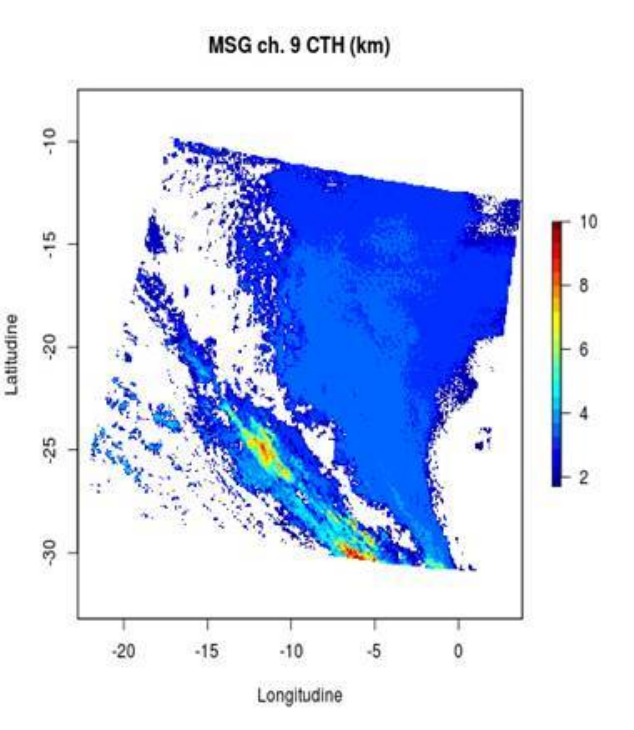}\\

\end{tabular}
\end{center}
\caption{Top: CTHs on July 10, 2011 from MODIS. Bottom: CTHs on July 10, 2011 from SEVIRI.}
\label{InterComp}
\end{figure}

From the perspective of a satellite-based cloud observation, inter-satellite comparison is a common way to quantify uncertainties in observations and reduce the effects of certain types of sampling biases. \\
To assess CTH uncertainties an IR scene obtained from SEVIRI (on-board Meteosat-9 satellite) has been analyzed and CTH has been compared with those ones derived from MODIS (on-board NASA’s Aqua satellite) and CALIOP (main instrument of CALIPSO satellite constellation). The two first sensors are infrared radiometers \cite{Menzel2006,Schmetz2002} and the third one is a three-channel Lidar that uses a Nd:YAG laser emitting linearly polarized pulses of light at 1064 nm and 532 nm.
The analyzed scene has been collected by Meteosat-9 (MSG-9) on July 10, 2011 over the Gulf of Guinea, close to sub-satellite point. Clouds top height estimated by MODIS has been obtained by MYD06\_L2 cloud product relative to Aqua acquisition on 12:05 UTC. CTHs derived from MSG-9 have been obtained with two different estimations of CTT:\\
a) Tcloud = BT @ 10.8 $\mu$m without any correction\\
b) Tcloud  = 1.0178 * T$_{CH9}$ - 4.149\\
The correction has been derived from radiative simulations of different atmospheres and thick clouds at different levels, as for the SWA. \\
Warm sea surface improves cloud detection due to high differences with cloud temperature. For this reason, pixels with BT greater than 289.15 K have been considered cloudy. Meteosat-9 BT data for cloudy pixels has been converted in CTH (meters above sea level), deriving the correspondent height from the atmospheric sounding performed in St. Elena (latitude -15.93 N and longitude -5.66 E) on the same day at 12 UTC.\\
To compare different observations, all satellite-based data have been re-sampled to the MSG ground resolution and georeferred to WGS84 geographical coordinate systems. The figure \ref{InterComp} shows CTHs from MSG-9 and MODIS for the analyzed scene.

Considering only middle and low clouds in the comparison, both algorithms show a mean overestimation of SEVIRI-derived estimations respect to MODIS-derived ones. Values reduce from BT without any correction show a mean bias equal to 206 m, and with the correction the mean bias is equal to 59 m. The standard deviation is about constant for both algorithms ranging (339 and 369 m). When we consider high irregular clouds both mean bias and standard deviation increase respectively to 300 m and to 1,500 m. Part of this change is due to different sampling area of the sensors. The discrepancy increment can be also attributed to an improper application of the algorithm. This algorithm was designed to retrieve water thick clouds and  it cannot be applied with enough accuracy to ice or thin/broken water clouds as can be found in high or clouds boundaries respectively.
For this reason, a Sobel Edge operator \cite{Vincent2009} has been applied to the scene to detect cloud boundaries and to select uniformly covered regions.
When we consider uniformly covered area, mean bias and standard deviation between SEVIRI and MODIS reduce. CTHs derived from Meteosat-9 have been also compared with CTHs provided by the LIDAR CALIOP. Unfortunately the satellite passing over the Gulf of Guinea was on 01:12 UTC, several hours before MSG-9 scene on 10:59 UTC. Considering observation time, the sounding retrieved in Cape Town (latitude -33.96N and longitude 18.6E) on July 10, 2011 at 00 UTC has been considered to estimate CHTs. Comparison between CTHs derived from MSG-9 and CALIOP reveals uncertainties in CTH ranging from 500 m for low continuous clouds and about 900 m for high clouds, in good agreement with previous comparison with MODIS estimates. 

\section{Conclusions}
The correct interpretation of the main telescope data, and therefore the JEM-EUSO Mission success, depends largely on the knowledge of the atmosphere status. The IR camera, as a part of the AMS, will provide the coverage and height of the clouds in the JEM-EUSO FoV. In this work two different and complementary methodologies to retrieve the CTH from the IR images provided by the IR camera have been explained. \\
The first methodology is based on a stereo vision technique and it will provide the height of the cloud directly from two consecutive images of the IR camera. The preliminary results tested on mono band stereo pairs are satisfying and must be confirmed even on bi-band stereo images. The second one, based on radiative measurements, will retrieve first the CTT and the height will be inferred by using atmospheric vertical profiles from sounding or global atmospheric models. The SWA designed to retrieve thick water clouds has been checked with simulations and real images from MODIS with very good results. Algorithms for ice and thin clouds are being tested.\\
Nevertheless, the comparison with MODIS results is not definitive; therefore an inter-satellite comparison of cloud height retrievals has been carried out. Low clouds decks CTHs derived from MODIS and SEVIRI are in good agreement, meanwhile the agreement decreases with less extended clouds and high clouds (ice clouds). Analysis shows the same behaviour of SEVIRI-derived estimations towards MODIS and CALIOP for high clouds and in partially cloudy regions. The edge detection through gradient calculation can be a pre-selecting method to exclude the critical areas (broken or thin clouds). These conclusions are consistent with those ones of the radiative retrieval algorithms.
We can conclude that the methodologies proposed to retrieve the cloud top height from the IR camera are providing very good results, even though some parts are still being tested.

\vspace*{0.2cm}
{
\footnotesize{{\bf Acknowledgment:}{This work was partially supported by Basic Science Interdisciplinary Research Projects of RIKEN and JSPS KAKENHI Grant (22340063, 23340081, and 24244042), by the Italian Ministry of Foreign Affairs, General Direction for the Cultural Promotion and Cooperation, by the 'Helmholtz Alliance for Astroparticle Physics HAP' funded by the Initiative and Networking Fund of the Helmholtz Association, Germany, and by Slovak Academy of Sciences MVTS JEM-EUSO as well as VEGA grant agency project 2/0081/10. The Spanish Consortium involved in the JEM-EUSO Space Mission is funded by MICINN under projects AYA2009-06037-E/ESP, AYA-ESP 2010-19082, AYA2011-29489-C03-01, AYA2012-39115-C03-01,AYA-ESP2011-29489-C03,AYA-ESP2012-39115-C03-02, CSD2009-00064 (Consolider MULTIDARK) and by Comunidad de Madrid (CAM) under project S2009/ESP-1496. The authors thank the contribution of A.J.de Castro, I.Rodr\'{i}guez, I.Fern\'{a}ndez-G\'{o}mez. F.Cort\'{e}s, S.S\'{a}nchez and F.L\'{o}pez from UC3M and R.Bechini, A.Pagnone and M.Del Giudice from Universit\`a di Torino. 
}}

}

\clearpage

%% file: icrc2013-1281.tex


\title{Simulations of extensive air showers produced by UHECRs in cloudy sky to be detected by JEM-EUSO}

\shorttitle{Simulations of EAS in cloudy conditions}

\authors{
G. S\'AEZ CANO$^{1}$,
J.A. MORALES DE LOS R\'IOS$^{1,2}$,
K. SHINOZAKI$^{2}$,
L. DEL PERAL$^{1}$,
M. BERTAINA$^{3}$,
A. SANTANGELO$^{4}$ \&
M.D. RODR\'IGUEZ FR\'IAS$^{1}$
for the JEM-EUSO Collaboration.
}

\afiliations{
$^1$ SPace and AStroparticle (SPAS) Group, UAH, Ctra. Madrid-Barcelona, km. 33.7, E-28871, Madrid, Spain.\\
$^2$ RIKEN, 2-1 Hirosawa, Wako, Saitama 351-0198, Japan.\\
$^3$ Dipartimento di Fisica, Universit ́ di Torino $\&$ INFN Torino, Italy. \\
$^4$ Institute fuer Astronomie und Astrophysik Kepler Center for Astro and Particle Physics Eberhard Karls University
Tuebingen Germany. 
}

\email{lupe.saez@uah.es} 

\abstract{The origin of Ultra-high Energy Cosmic Rays (UHECRs) is still unidentified. Since, 
at such high energies, the cosmic ray flux is extremely small, a detector with huge observation area is needed.
JEM-EUSO is a novel observatory that will be located at the International Space Station to
observe Extensive Air Showers (EAS) produced by UHECRs in the Earth's atmosphere. An advantage of a space based telescope is that also 
observation is possible under certain cloudy conditions, where most of the shower develops above the cloud.
In the present work, we show how the EAS signal is modified in presence of uniform layer clouds. Also, a more realistic atmospheric model is
being implemented to properly account the photon propagation of EAS to the telescope.}


\maketitle

\section{Introduction}
Mechanisms that accelerate Ultra High Energy Cosmic Rays (UHECRs) to energies around and above $\sim10^{20}$ eV 
are still unknown \cite{klet}. 
Due to such high energies, direct detection of UHECRs is not feasible. However, they can be detected by measuring 
Extensive Air Showers (EAS), which are cascades of secondary cosmic rays particles that UHECRs produce when 
they interact with particles of the Earth's atmosphere. Due to this interaction, atmospheric nitrogen particles emit 
fluorescence light. Moreover, Cherenkov light is produced as a consequence of the ultrarelativistic velocity of the
particles. 

JEM-EUSO is a novel space-based experiment aiming to detect UHECRs with large statistics. It will be located at the International Space 
Station (ISS) in 2017 \cite{kadams}. Due to the ISS orbit, JEM-EUSO will observe different parts of the atmosphere, unlike ground
based telescopes which only observe cosmic rays from a certain region of the sky. To properly determine the energy,
arrival direction of the primary particle and its composition, a measurement of the light profile is needed.
However, this profile depends on atmospheric conditions and, at an altitude of 400 km and with a field of view (FoV) 
of $\pm$ 30$^{\circ}$, the telescope will traverse above different atmospheric conditions (such as clouds) within 
its large observation area. Therefore, an accurate monitoring of the atmosphere is needed. JEM-EUSO counts on an 
Atmospheric Monitoring system (AMS) consisting of a LIDAR (LIght Detection And Ranging) device and an infrared (IR) camera \cite{kMD}.
The main information the AMS will provide is the coverage of clouds, the cloud-top altitude distribution and the 
profile of the optical depths of the atmosphere w.r.t photoabsorption of UV light. Moreover, it will also take advantage
from the global atmospheric models like those 
generated by the Global Modeling and Assimilation Office (GMAO) or the European Center for Medium range Weather Forecasts (ECMWF)
\cite{kbertICRC}. Unlike ground-based telescopes, physical properties of some events in cloudy conditions may be
reconstructed. We assume that if the shower maximum is above the cloud top altitude \cite{kabu}, the reconstruction of
the shower parameters will be feasible. This not only depends on the altitude of the cloud, but also on the arrival 
direction of the shower and the type of the primary particle \cite{klupe}.\\

For shower simulation performed for this work, ESAF (Euso Simulation and Analysis Framework) was used \cite{kberat}. 
It is a software framework to simulate space-based cosmic observations \cite{kfenu}.
In this proceedings we study how the EAS features are influenced by the presence
of clouds, and we discuss our current work, which consists of improving the atmospheric model used by JEM-EUSO 
software.\\

\section{Photon propagation in the atmosphere}
Fluorescence and Cherenkov light produced by EAS are mainly emitted in the ultraviolet (UV) range. The Earth's atmosphere
absorbs and scatters this UV light, depending on the column density of the atmosphere particles between the location of
the produced EAS photons and the detector \cite{kneronov}. If light passes through matter, energy and frequency respectively may change due to absorption or
scattering of photons into and out of the beam \cite{krybi}. In the case of scattering and absorption in the
atmosphere, these processes depend on the type of atmospheric particles.
In clear sky condition, UV photon propagation through atmosphere mainly involves Rayleigh scattering
and absorption by Ozone in shorter wavelengths (320 nm). In ESAF, the transmittance of these processes are modeled by
LOWTRAN \cite{kberat}. In presence of water clouds, droplet size is comparable with UV light wavelength. Therefore, the Rayeleigh 
approximation is not valid, and Mie scattering must be considered. Moreover, for high clouds, such as cirrus-like clouds, 
where ice particles are present,
Mie scattering needs some correction factor since these particles can not be considered as spherical particles. \\

The fluorescence component from the EAS is emitted isotropically. Therefore, its main contribution to the detected
light by the JEM-EUSO space observatory will be the fluorescence light emitted in the direction of the telescope. The small 
scattered component which will arrive to the telescope, will have a time delay in respect to the direct signal, and
therefore will be treated as noise. Cherenkov light, on the other hand, is emitted very collimated along the path 
of the shower. Thus, the directly emitted Cherenkov component arriving to JEM-EUSO will be negligible. Main detected
Cherenkov component is scattered light, which will arrive at the same time as emitted photons. 
To correctly determine the energy deposit at a given level in the atmosphere (proportional to the fluorescence light
emitted in such level), it is necessary to know exactly how Cherenkov light is distributed over the light image, to
subtract it properly. \cite{kbib:cher}\\

\section{EAS simulation in different atmospheric conditions} 

EAS light coming to the telescope is observed as light spot moving with the velocity of light. This information is very
important to reconstruct the arrival direction of the UHECR. To reconstruct the energy, on the other hand, one needs to 
know the amount of produced fluorescence light. In presence of clouds, photons coming from below the cloud are attenuated
as a function of the cloud optical depth. If the cloud is optically thick enough, the signal after the cloud will be truncated 
\cite{klupe2}. Therefore, the shower track will be shorter than that for clear sky. The time duration of the signal will
be smaller as well. If these clouds are located at lower altitudes, still the measurement of the dominant part of the light
curve, which is the arrival time distribution of photons to the telescope pupil, is feasible. In addition, Cherenkov light will be highly
reflected on the top of the cloud, and will help to determine the arrival direction of landing location of the EAS
even better than in case of clear sky. If the cloud is optically thin enough, the signal after the cloud will be attenuated
but not truncated, and thus some signal from below the cloud will reach the telescope.

 \begin{figure}[h]
 \begin{center}
  \includegraphics[width=0.5\textwidth]{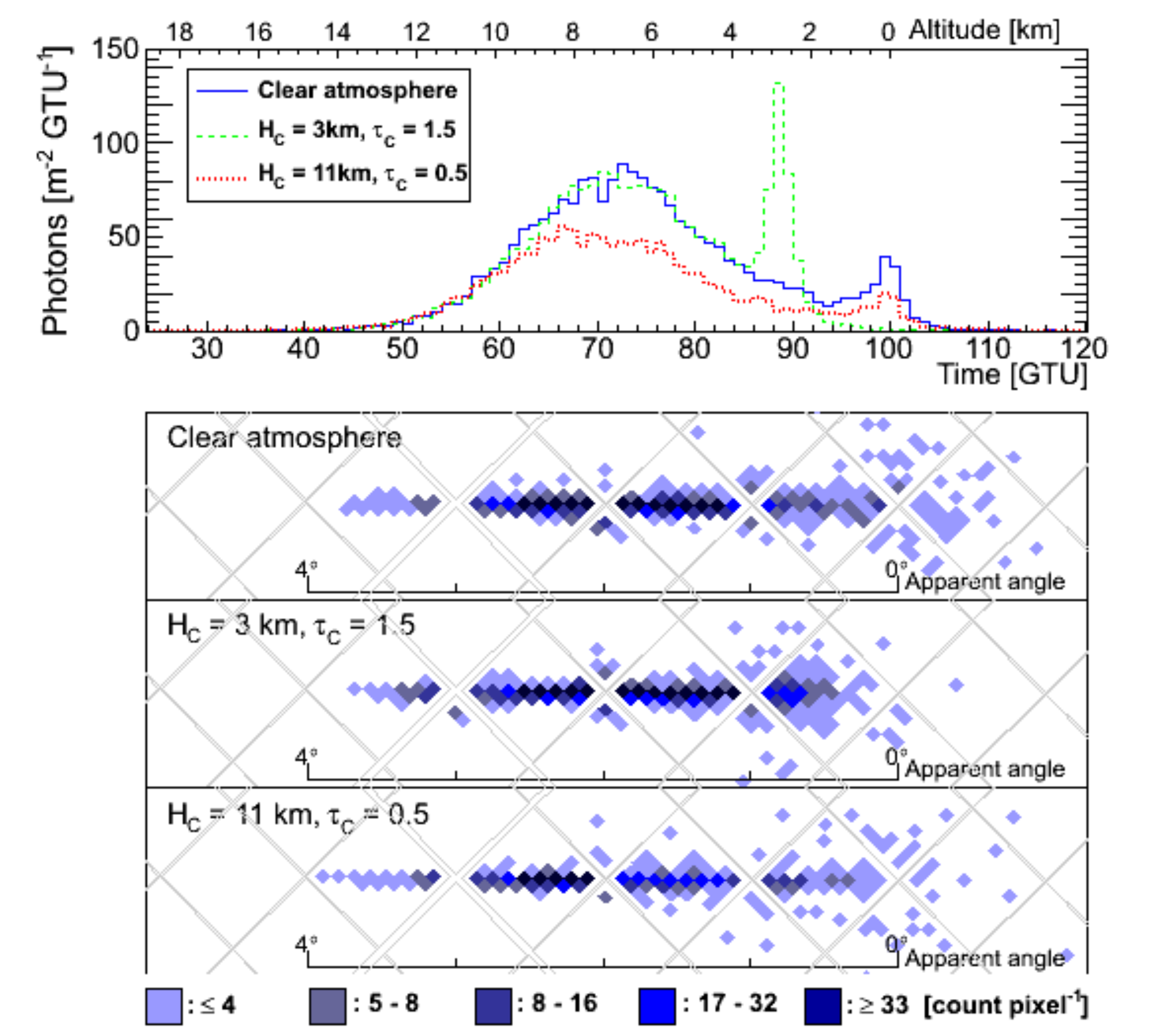}
  \end{center}
  \caption{The first pannel of the plot represents the arrival time distribution of photons (light curve) from 
  a typical EAS with an energy of 10$^{20}$ eV, a zenith angle of 60$^{\circ}$ and an azimuth angle of 45$^{\circ}$
  for three different atmospheric situations. The solid line shows the case for clear atmosphere. Dashed lines
  denote a case where a cloud with an optical depth ($\tau_{c}$) of 1 and an altitude (H$_{c}$) of 3 km is
  present. Dotted line represents another cloudy case ($\tau_{c}$=0.5 and H$_{c}$= 11). The top axis shows the
  altitude at which the photons have been 
  produced. The three bottom pannels show the EAS image in the focal surface detector for the three previous cases.
  The color scale indicates the number of signal counts per pixel. The position along the
EAS track corresponds to the arrival time on the top panel.}
    \label{lc_fs}
 \end{figure}

\begin{figure}[h]
\begin{center}
\label{figst}
\includegraphics[width=0.5\textwidth]{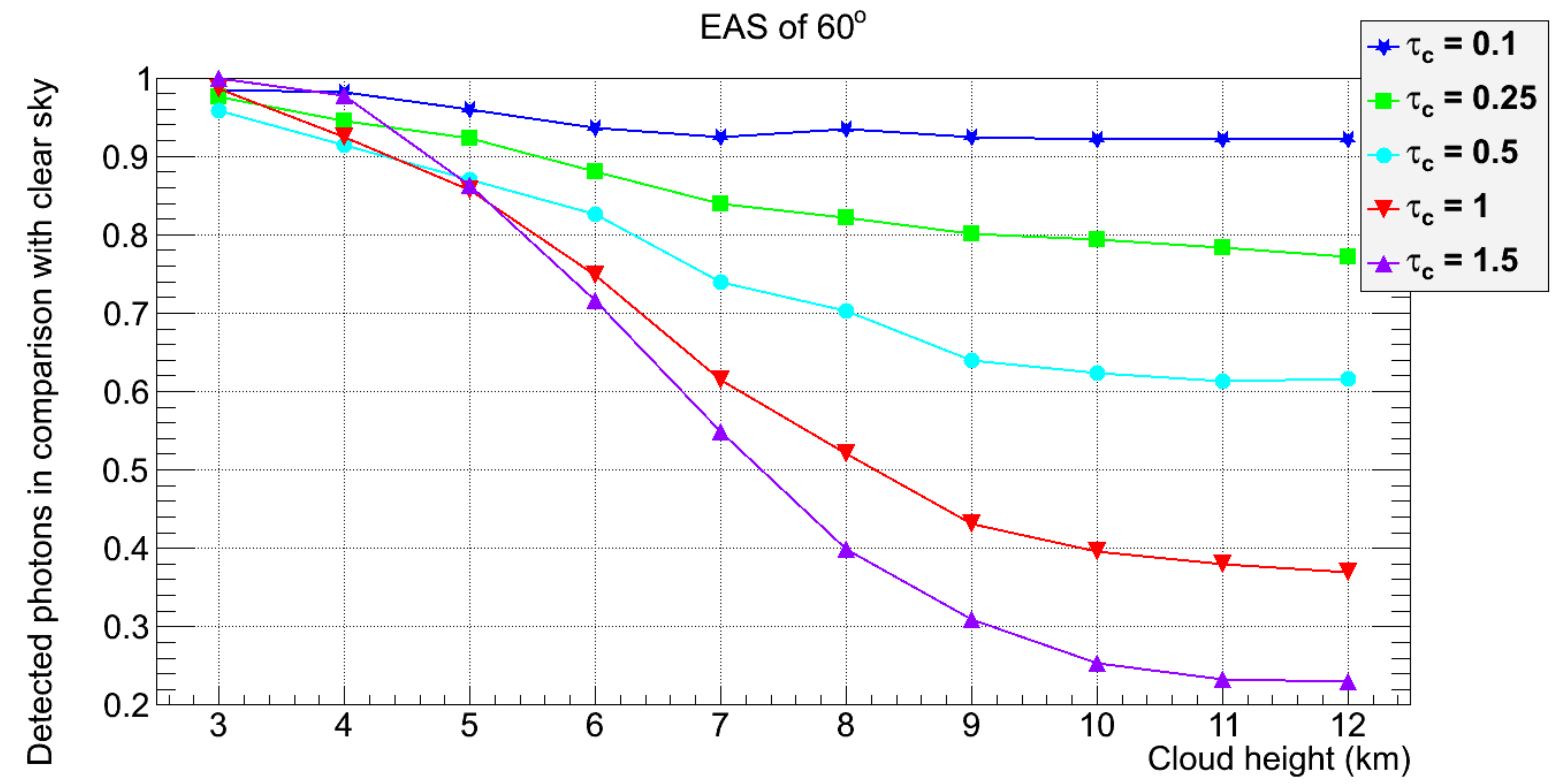} \\
\includegraphics[width=0.5\textwidth]{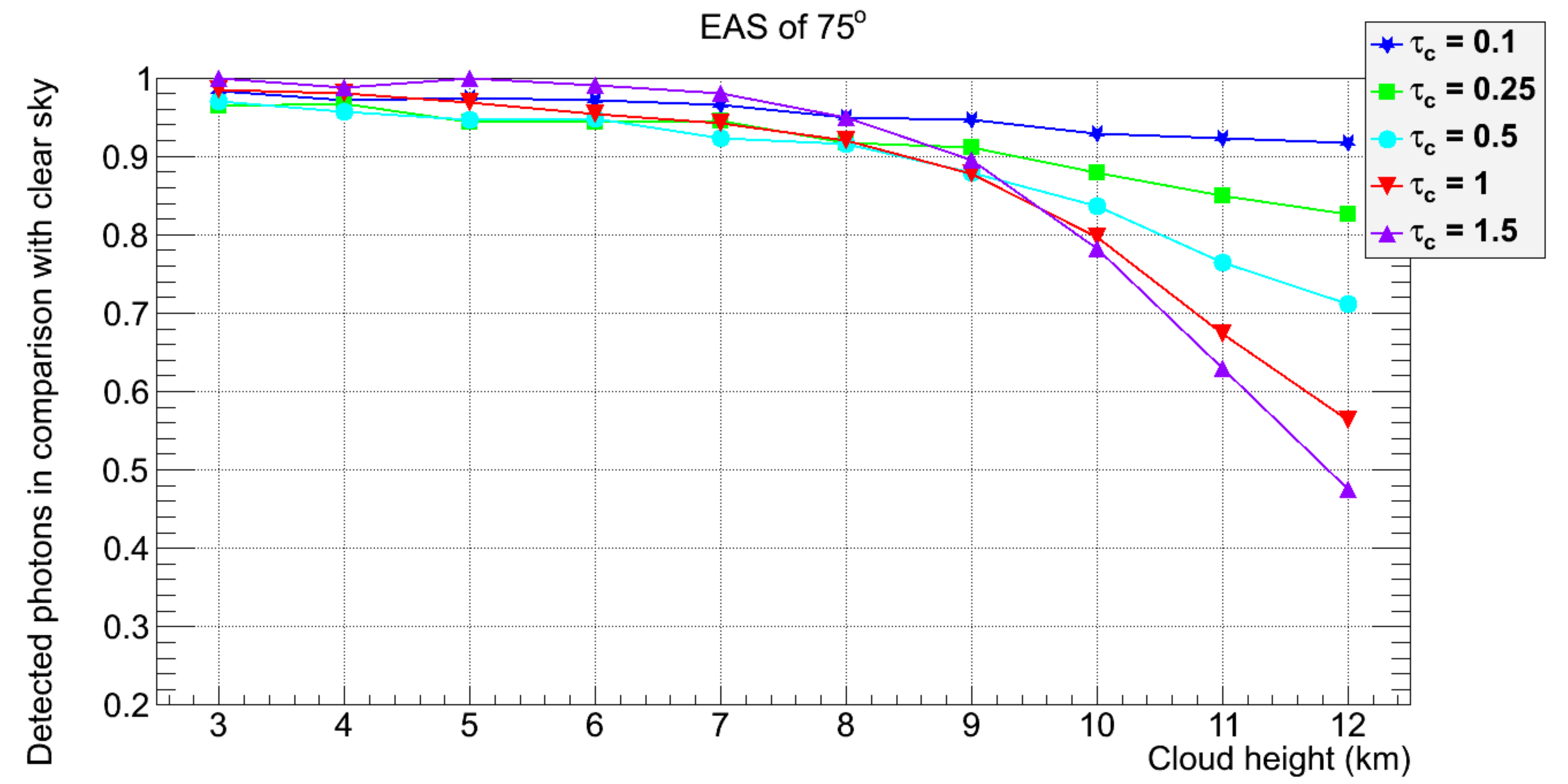}
\end{center}
\caption{Attenuation of photons in different clouds scenarios compared with clear atmosphere, as a function of cloud-top altitude
$H_{\rm C}$ along different optical depths $\tau_{\rm C}$ denoted by different symbols. Top and bottom panels indicate the cases of 
 $\theta=60^{\circ}$ and $\theta=75^{\circ}$, respectively. We observe how the attenuation is more pronounced as the altitude
 of the cloud is higher. Also, this attenuation has more impact in more vertical showers }
 \label{EAS_atten}
\end{figure}

An example of this difference in the light curve can be seen in the first pannel of Figure \ref{lc_fs}. We observe 
three light curves of typical EAS events with zenith angle of 60 $^{\circ}$ and energy of 10$^{20}$ eV in 
different atmospheric conditions. The X axis denotes the absolut time in GTUs (1GTU=2.5$\mu$s). The Y axis 
represents the number of photons reaching the telescope, normalized by the detector area. The apparent movement 
of these three examples are plotted in the next three pannels. For the clear atmosphere case (second pannel), 
the apparent movement extends $\sim3^{\circ}$and lasts $\sim$ 60 GTUs (=150 $\mu$s) \cite{kguz}. In third and fourth pannels
we observe how EAS signals are modified in the presence of clouds. If the optical depth of the shower is large
enough, as in the third pannel (case of a cloud of 3 km altitude and 1.5 optical depth), it is demonstrated
that no shower track is visible after the cloud. Apparent EAS image will last around 40 GTUs and will
extend a bit more than 2$^{\circ}$. However, since the cloud is lower than the depth of maximum development of the shower and there
is enough shower track, one can apply the reconstruction techniques similar to the used for clear atmosphere,
for the data measured from above the cloud. For a cloud with an small enough optical depth, as the case
represented in the fourth pannel, photons originated below the cloud will be attenuated but there will be 
still some contribution to the EAS signal. Therefore, angular reconstruction will be similar to that one
for a clear atmosphere. However, due to the attenuation, the energy might be understimated
if no correction is applied during the reconstruction with AMS information. Optically thin clouds that have more impact  
are those with highest altitudes, because most part of the shower will be located below the cloud and therefore, the EAS signal will 
suffer more attenuation. Figure \ref{EAS_atten} shows the attenuation produced by different optically thin clouds for a $60^{\circ}$ and a $75^{\circ}$ EAS.

\section {EAS and End to End IR camera simulations}

\begin{figure}[t]
  \centering
  \includegraphics[width=0.5\textwidth]{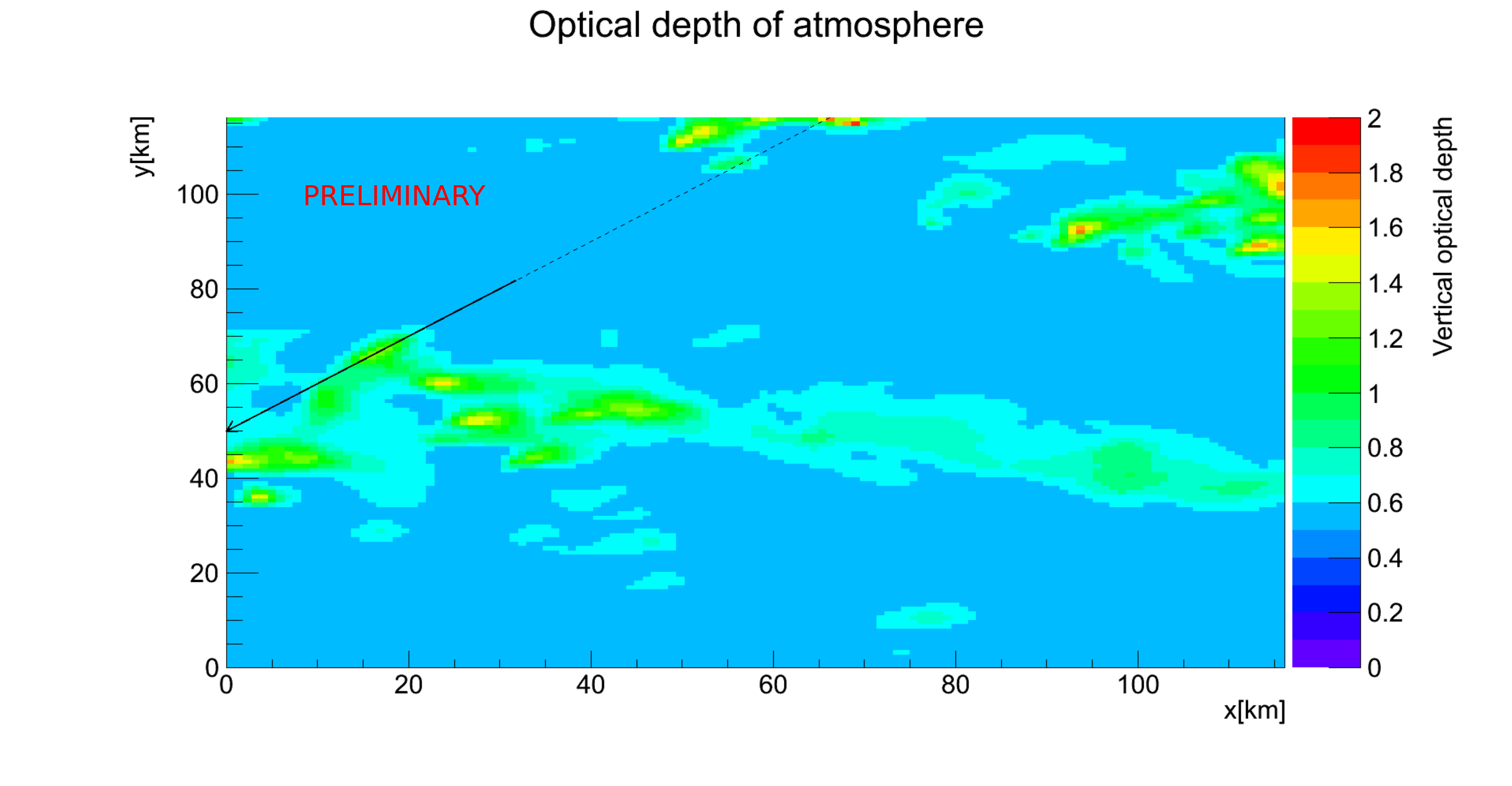}
 \caption{60$^{\circ}$ shower track in presence of a cloudy scenario from SCSMEX}
  \label{OD_ICRC}
 \end{figure}

Up to now ESAF has been using a simple module to include clouds in the atmospheric conditions to simulate EAS. It consists of
a test cloud (uniform and homogeneus layer) either chosen manually or from TOVS database \cite{kTOVS}, whose physics parameters
are the cloud top altitude, its optical depth and its physical thickness. Since months ago, the SPAS group from Alcal\'a
University (Spain) is working on the simulation of different atmospheric and cloudy conditions, by using a more reallistic
model such as atmospheric simulations from the Satellite Data
Simulator Unit (SDSU) software \cite{kmorICRC2}. This model is based on the images we expect to obtain from the JEM-EUSO infrared camera.
It considers simulated radiation produced by the Earth's surface and atmosphere, the effect of the optics, the detector, 
the electronics and the image compression algorithm \cite{kmorICRC}. To perform the simulations, the atmosphere 
is divided in three dimensional cells, filled with different atmospheric properties, including cloud properties.
In this work, we use a cloudy scenario from the South China Sea Monsoon Experiment (SCSMEX) \ref{OD_ICRC}.

\subsection {EAS simulation in ESAF}

The number of photons produced by a typical shower of 10$^{20}$eV is  $\sim10^{15}$. Therefore, raytracing each photon is not
feasible. For this reason, in ESAF the light simulation is done by introducing the concept of bunch. The shower longitudinal
distribution is split in length steps dL (we choose dL=10$g/cm^{2}$). At each step, one bunch for the fluorescence 
emission and another for the emitted Cherenkov light are produced \cite{kberat}.


In this work, we calculate where every bunch of photons has been created (in this case, for a 60$^{\circ}$ shower). 
However, for this preliminary study, only bunches of fluorescence photons have been considered (due to their isotropic emission).
We need to calculate the optical depth between the emission point and the telescope (\ref{eq3})

\begin{equation}
\label{eq3}
OD= \frac{\rho \times L}{\Lambda}=\frac{X}{\Lambda}=\alpha \times L
\end{equation}

Where $\rho$ is the atmospheric density, $\Lambda$ is the attenuation length, L is the path, $\alpha$ is the attenuation 
coefficient and X is the slant depth.


We calculate the optical depth between the location of each bunch emission and JEM-EUSO for a 60$^{\circ}$ shower
from the atmospheric properties of this cloudy scenario. Considering also the amount of photons produced in each bunch, 
we calculate the number of photons reaching the telescope:

\begin{equation}
\label{eq2}
I_{det}=I_{0}\times e^{-OD} \times Eff \times \frac{Area_{JE}}{(4  \pi  d^{2})}
\end{equation}

\begin{figure}[h]
\begin{center}
\label{yfigst}
\includegraphics[width=0.5\textwidth]{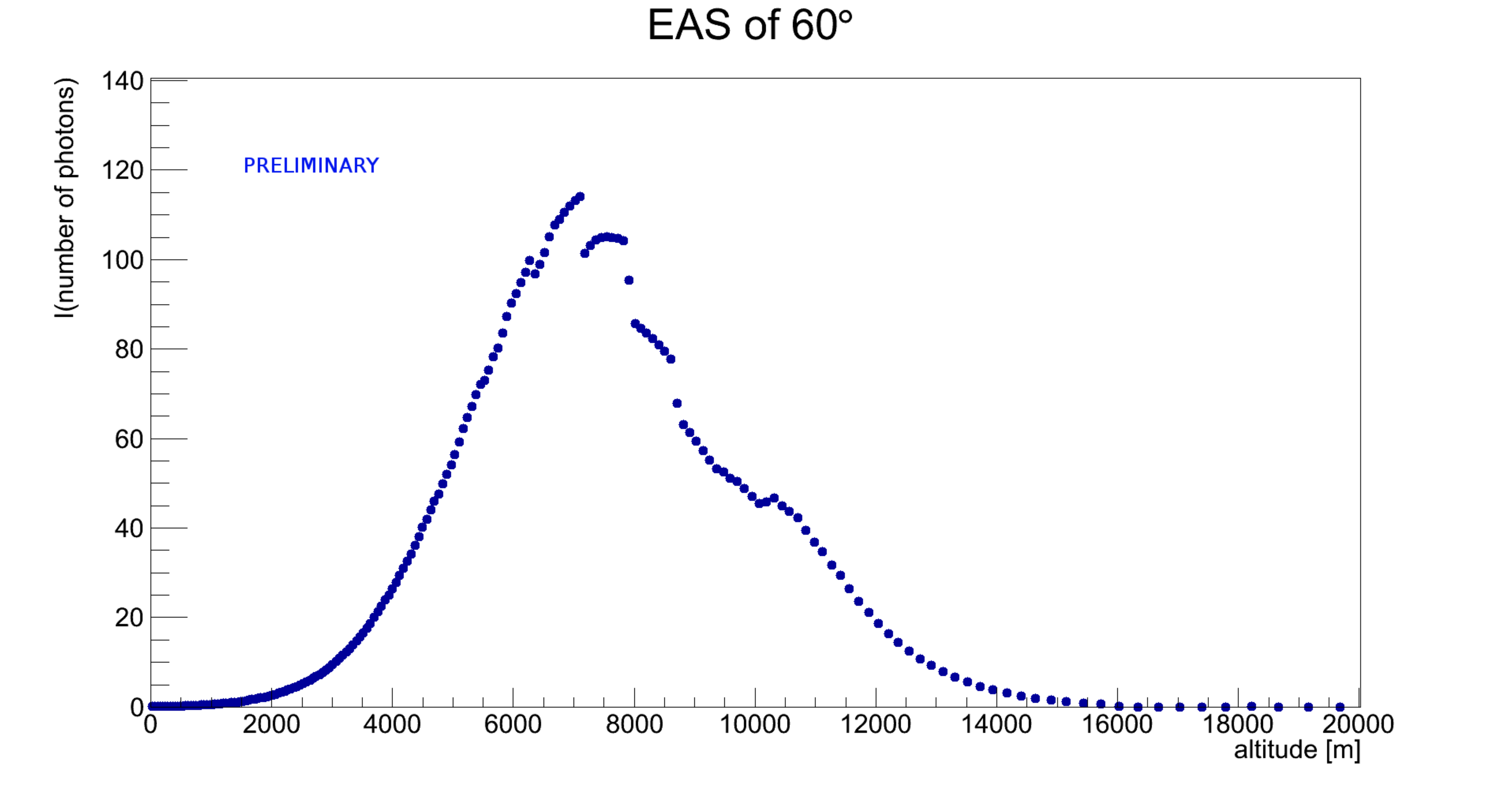} \\
\includegraphics[width=0.5\textwidth]{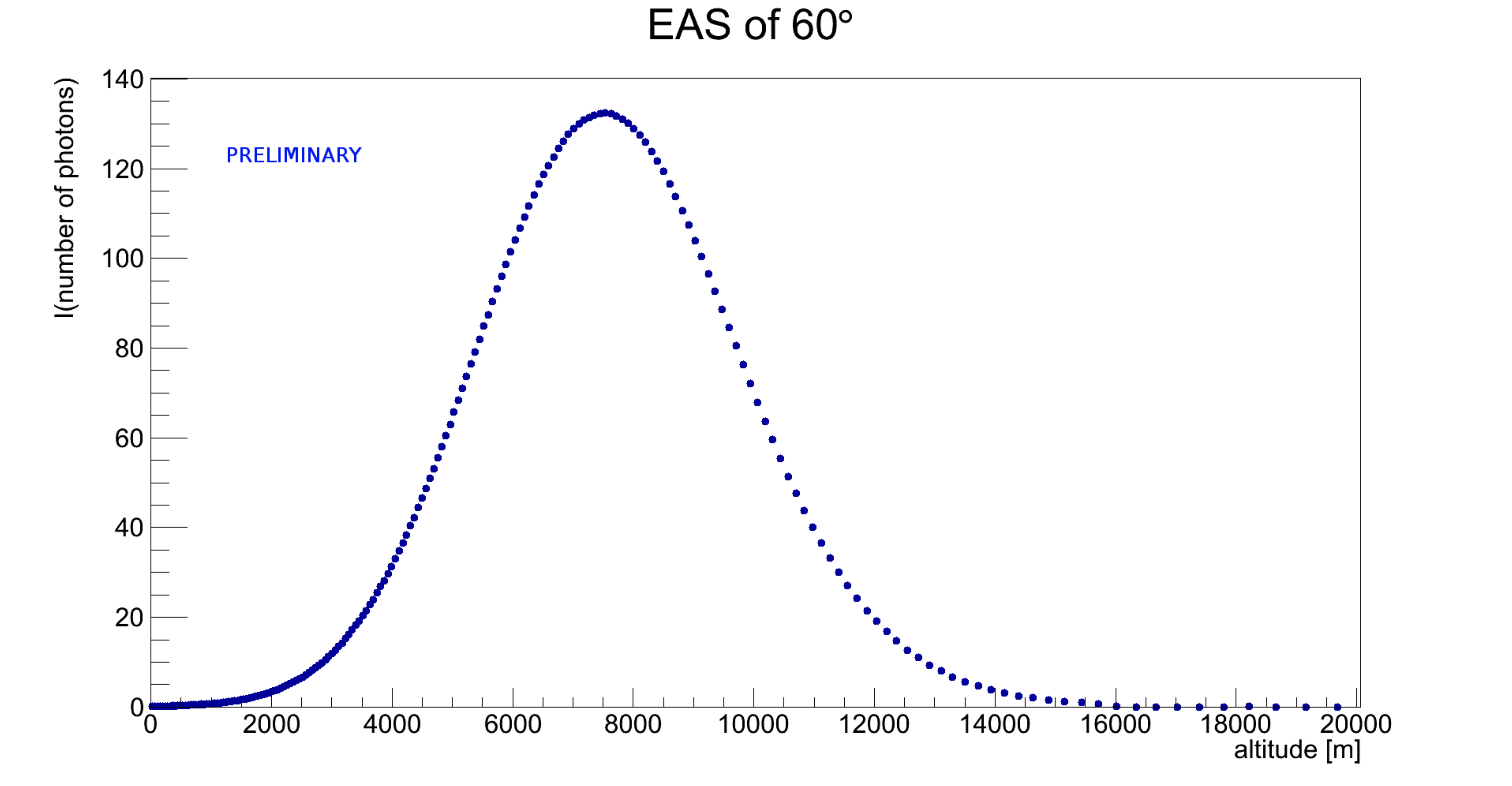}
\end{center}
\caption{Number of photons reaching the telescope for a standard shower in clear atmosphere (bottom pannel) and
in the presence of the previous cloudy scenario from the South China Sea Monsoon Experiment}
 \label{I_I0_test}
\end{figure}

Where I$_{det}$ is the number of photons reaching JEM-EUSO, I$_{0}$ is the produced photons in each bunch, OD is the 
optical depth from the emission location to the JEM-EUSO location, Eff is the detector
efficiency, Area$_{JE}$ is the detector area, and d is the distance between
the bunch and the telescope. \\ 
 In Figure \ref{I_I0_test} we observe the number of photons reaching the telescope for a standard shower in clear
atmosphere (bottom pannel) and in the presence of a cloudy SCSMEX scenario (top pannel). In the top pannel, photons produced above 
11km are propagated in clear sky. Thus, the light curve from 11 km 
is the same in both figures (top and bottom). Between 7km and 11km, photons emitted in the direction of the
telescope will suffer an attenuation due to a higher optical depth produced by the presence of a cloud. The cloud optical depth varies;
for bunches located at 10 km it has a value of $\simeq$ 0.4, which is almost 3 times higher than the expected without the cloud presence and,
thus, the number of photons reaching JEM-EUSO will be around 75$\%$ comparing with clear sky. The optical depth in the cloudy scenario
for bunches located at 9km will be $\simeq$ 0.7, and only 60$\%$ of photons comparing with clear sky will reach JEM-EUSO. This results
are preliminary.

For bunches located at the end of the shower, where very few photons are produced, the influence of a cloud is small when physical parameters of the shower 
are reconstructed.
 \begin{figure}[t]
  \centering
  \includegraphics[width=0.5\textwidth]{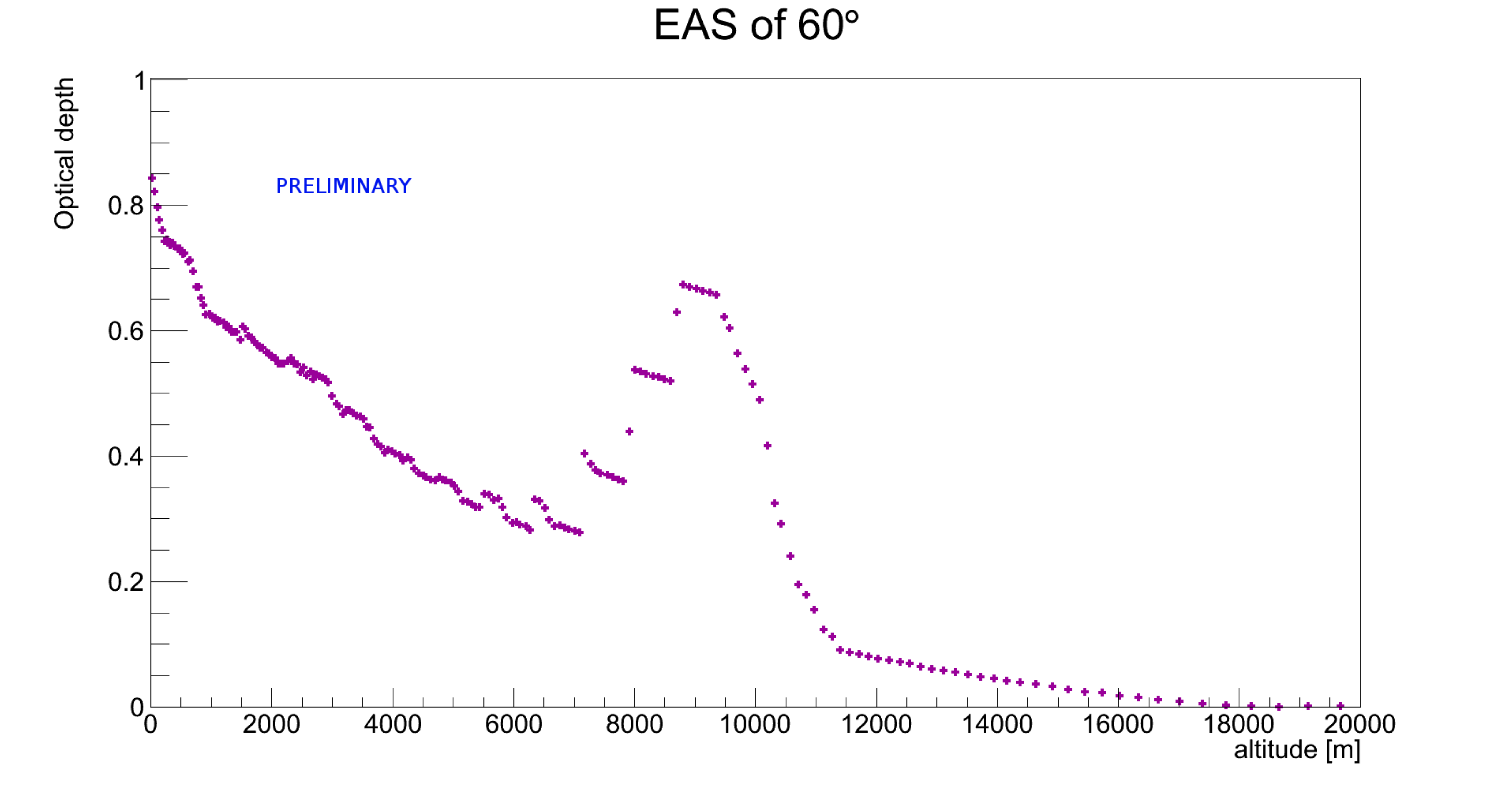}
  \caption{Optical depth from each fluorescence bunch to the telescope, for the case of a 60$^{\circ}$ shower 
   in the presence of a monsoon cloud taken from SCSMEX.)}
  \label{OD_clearsky}
 \end{figure}

With this ongoing work we want to clarify how the EAS looks like in the presence of inhomogeneus and therefore more
reallistic clouds, as well as to have more flexibility for the implementation of different atmospheric profiles.  

\section{Summary and discussion}

An advantage of the space-based observation of EAS produced by UHECRs is that measurements
are also possible if the cloud top altitude is below the shower maximum. The influence
of the clouds not only depends on the height of the cloud and its optical depth, but also
on the inclination of the shower.

Low clouds influence nos significantly the EAS development since they only affect the last part of 
the shower development, and therefore, one may use the light curve up to the cloud to
reconstruct the energy and arrival direction of the shower. As long as the altitude of the cloud
increases, clouds will have more impact on the 
EAS signal. However, if the cloud is optically thin enough, signal after the cloud
will be attenuated but still visible, and therefore, also the angular reconstruction is 
little affected since it is based on the EAS apparent movement. The estimated energy
will be affected, since the EAS will look like a lower energy EAS for clear sky.
For these cases, the presence of AMS is more important. For optically thick clouds 
whose shower maximum is located above the cloud,
a bright Cherenkov reflected light from the top of such a cloud will be detected.
This may have a positive effect since the location of a point of the
shower track that helps to reconstruct geometrical and physical parameters of the
EAS is given. Nevertheless, since the shower track will look shorter, the quality
of the reconstruction must be compared to that one for more vertical showers
in clear atmosphere.

In order to have a deeper knowledge about how EAS signal will be distorted by
the presence of clouds in the JEM-EUSO detector, currently we are propagating
photons from different EAS inside clouds which have been simulated according
to the JEM-EUSO infrared camera response.\\

{
\footnotesize{{\bf Acknowledgment:}{This work was partially supported by Basic Science Interdisciplinary 
Research Projects of RIKEN and JSPS KAKENHI Grant (22340063, 23340081, and 
24244042), by the Italian Ministry of Foreign Affairs, General Direction 
for the Cultural Promotion and Cooperation, by the 'Helmholtz Alliance 
for Astroparticle Physics HAP' funded by the Initiative and Networking Fund 
of the Helmholtz Association, Germany, and by Slovak Academy  
of Sciences MVTS JEM-EUSO as well as VEGA grant agency project 2/0081/10.
The Spanish Consortium involved in the JEM-EUSO Space
Mission is funded by MICINN under projects AYA2009-
06037-E/ESP, AYA-ESP 2010-19082, AYA-ESP 2011-29489-C03-
01, AYA-ESP 2012-39115-C03-01, CSD2009-00064 (Consolider MULTIDARK)
and by Comunidad de Madrid (CAM) under project S2009/ESP-1496.
}}

}
\clearpage

%% file: icrc2013-0449.tex


\title{Absolute Fluorescence Spectrum and Yield Measurements for a wide range of experimental conditions }

\shorttitle{Absolute Fluorescence Spectrum and Yield Measurements for a wide range of experimental conditions }

\authors{
D. Monnier Ragaigne$^{1}$,
P. Gorodetzky$^{2}$,
C. Moretto$^{1}$,
 C. Blaksley$^{2}$,
 S. Dagoret-Campagne$^{1}$,
 A. Gonnin$^{1}$,
 H. Miyamoto$^{1}$,
 H. Monard$^{1}$
 F. Wicek$^{1}$
for the JEM-EUSO Collaboration.
}

\afiliations{
$^1$ Laboratoire de l'Acc\'el\'erateur Lin\'eaire, Univ Paris-Sud, CNRS/IN2P3, Orsay, France \\
$^2$ Laboratoire Astroparticule et Cosmologie, APC, Paris, France\\

}

\email{monnier@lal.in2p3.fr}

\abstract{
The fluorescence yield is the key parameter that allows fluorescence telescope experiments to estimate the air shower energies from the number of UV-fluorescence photons detected. This fluorescence emission is induced by the energy loss in the air medium of the secondary charged particles dominated by the electron/positron component.
The fluorescence emission is a line spectrum, of which the line at 337 nm is the strongest. Actual fluorescence telescopes measure the integral of this spectrum in the UV range.
The process of photo-deexcitation of the Nitrogen involved in the UV emission is dependent on the pressure, temperature and humidity and requires dedicated measurements at various thermodynamic states of air representative of the atmospheric conditions.
Most of the previous experiments have performed measurements on a subset of the dominant spectrum peaks and at standard pressure and temperature conditions with the statistical/systematic errors being at best at the 10\% level. For practical fluorescence telescope, these yield measurements have to be extrapolated to account for the whole spectrum and  for any atmospheric conditions with poorly known systematic errors.
We propose an experiment based on a few MeV electron beam produced by an accelerator, targeting a fiducial volume collecting all the fluorescence light emitted.
The goal is to measure the yield absolutely for each spectrum-line and also the integral spectrum by varying the thermodynamic conditions with a statistical accuracy better than 5\%.
This accuracy can be achieved through  control of the electron source intensity, knowledge of primary electron geometrical path and the fiducial volume where the fluorescence light is generated and collected containing more than 95\% of the deposited energy and the calibrated detection without relying on any Monte Carlo simulation.
}

\maketitle

\section{Introduction}

An extensive air shower is an hadronic shower consisting mostly of an electromagnetic component carrying a very large fraction of the total energy of the shower.
This fraction is constant over a wide range of incident energies and this is why the measurement of this electromagnetic component provides a good estimation of the energy of the primary cosmic ray particle.
The electromagnetic energy is dissipated by the secondary electrons/positrons from the particle cascade undergoing inelastic collisions loss with the air atoms. This dissipative effect is detectable because the Nitrogen deexcites by UV fluorescence emission into a spectrum including about 30 lines (270 nm-430 nm). 
Hence, this measurement is a homogeneous calorimetric determination of the shower energy. 
It is well known that the fluorescence light is proportional to the energy dissipated.

Whereas the global process of energy deposit by charged particles is well known and given by the Bethe-Bloch formula, the details of the effective geometrical energy deposition are not obvious due to the production of secondary delta-rays.
 A small fraction of them carrying a sizeable fraction of the primary electron energy may deposit their energy at far distances from the electron trajectory. This point raises the critical question of the volume size around the primary trajectory where most of the dissipated energy is contained. An underestimate of the emission volume would lead to an underestimate of the yield due to the energy leakage outside the volume, thus inducing a systematic overestimation of the shower energies.

In a laboratory measurement of the fluorescence yield, scientist controled electron  beam replaces favourably the shower electrons (the relationship between the fluorescence yield and the energy loss is well known).

The necessary containment volume is  proportional to the logarithm of the energy, and inversely proportional to the pressure. Its exact size for a given pressure, given by the range of the most energetic deltas (electrons) has to be evaluated by a Monte Carlo simulation.

 The actual fluorescence telescopes operate by counting the fluorescence photons integrating over the whole UV spectrum (\cite{lab1}, \cite{lab2},  \cite{lab3},  \cite{lab4}, \cite{lab5}) .

 However, the intensity of each emission line varies relatively one to another with the atmospheric conditions (\cite{lab8}, \cite{lab9},  \cite{lab10}) in a way which is not precisely  predictable. 

This means one cannot count on the knowledge of a single emission line to predict the behaviour of other lines and naturally that of the integrated spectrum.

Fluorescence telescopes practically measure the total number of photons emitted along a shower track in each of their pixels. From the pixel angular size and the shower distance one can calculate the shower-track-length. Defining the linear fluorescence yield as the ratio of the number of photons per unit length, one can then estimate the number of electrons crossing the field of view of the pixel.

From the number of electrons, one deduces the dissipated energy of the shower according to the Bethe-Bloch formula.

The essential factor for a precise absolute energy measurement of the shower is the fluorescence yield.

\section{Status of the Previous Measurements}
Up to now, there have been two kinds of experiments: 1) those with radioactive sources, simple table-top experiments, hence the first ones, but not very precise (about 15 \% per line, due to low counting rate); 2) beam experiments, more complex but giving more precise results.
The main one is the AirFly project (\cite{lab6}), which measured the properties of the most intense line (337 nm) in all atmospheric conditions. They also measured the different line yields relative to the 337nm line, but at one pressure value only (1 atm).
If their precision on the 337 nm line is good (4\%), the extrapolation to the integral spectrum at any atmospheric condition needs to be improved.

For all the reasons mentioned above, we propose an experiment where we will measure the properties of each individual line in all atmospheric conditions using an experimental setup common to all lines at the same time.

\section{Fluorescence Yield measurement issues}

\subsection{Control of the volume where energy deposit and light emission occurs}

The goal is to collect the fluorescence light emitted anywhere in the target volume, independently of the location of the emission. This requires to use an integrating sphere. 
The fraction of the light detected on a port on the surface of the sphere is proportional to the ratio of the area of this port to the total inner surface of the sphere.
The size of this sphere can be estimated at the lowest pressure : 0.1 atm corresponding to an altitude of 10 km. This corresponds to a radius of 20 cm for a 85\% energy containment for an electron energy of 4 MeV ~\cite{lab8}. 
At 1 atm, for the same radius, the contained energy  is 91\%, i.e. very close to the 0.1 atm value. This is due to the fact that only a very small numbers of deltas have energies such large that their ranges exceed the sphere radius.  
This is an acceptable compromise  between the total energy containment and the size of this sphere.  This missing energy will be treated in section~\ref{sec:press}.

\vspace{0.5cm}
\subsection{Temperature, pressure, humidity dependence}
The effective atmospheric conditions range from 1 atm to 0.1 atm in pressure, from 20$^\circ$C to -60$^\circ$C and the humidity varies from saturation to 1\% of the saturation level. Each fluorescence line is affected differently by these parameters. The weighted sum of the yield measurements of individual lines must be checked by a specific integrated measurement through a PMT equipped with  a BG3 filter identical to the one used in Telescope Array and JEM-EUSO. The Pierre Auger Observatory uses a similar filter.  

\section{Experimental Set-Up}

\begin{figure*}[!h]
  \centering
  \includegraphics[width=8cm]{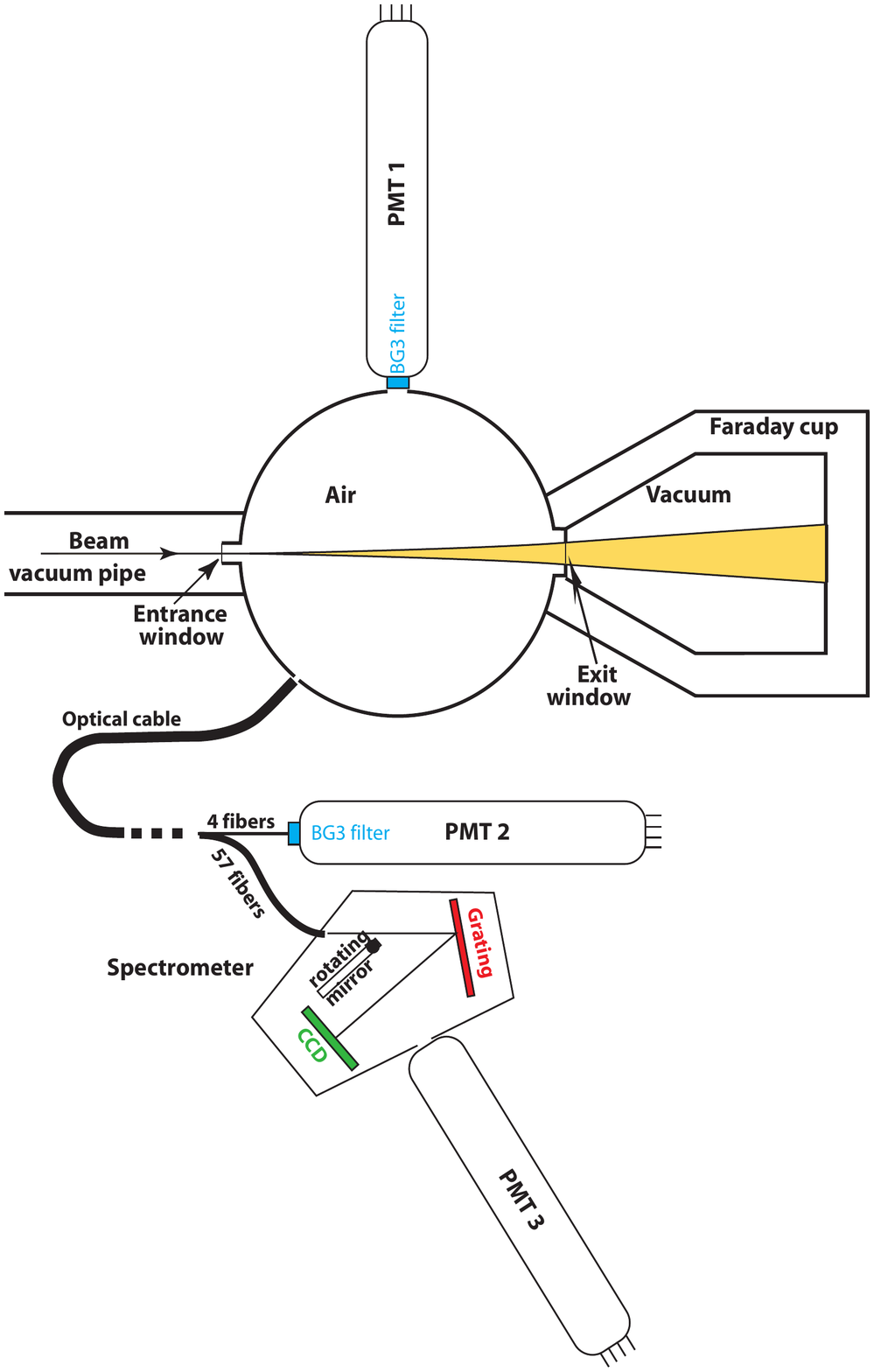}
 \caption{Design of experiment.}
  \label{wide_fig}
 \end{figure*}

The scheme in figure~\ref{wide_fig} explains the experimental setup proposed for the yields measurements.  
The electron beam arrives from the left, goes inside the sphere, being diffused by multiple scattering and exits the sphere towards a precise Faraday cup under vacuum.
This beam is bunched at 5 Hz. Each bunch has a time length of 8 ps and a total charge of 100 pC (PHIL accelerator ~\cite{lab12},~\cite{lab13}) . The energy will be 3-5 MeV with a maximum transverse size of 1mm (sigma), to be reduced in the future. 
The PMT1, equipped with a BG3 filter is used to monitor the emitted light. This tube has been calibrated before and with a 2\% accuracy. It then becomes a NIST PMT.

A small fraction of the fluorescence light is collected by an optical cable made of 61 silica step-index fibers of 0.1 mm diameter. 
Four of these are used to measure accurately the integrated light with the PMT2 also equipped with the BG3 filter.
57 fibers, arranged  vertically as a mono-fiber layer of 5.7 mm height and 0.1 mm thickness, enter a grating spectrometer (the grating being naturally vertical). The entrance slit of this spectrometer  allows to restrict the angular distribution of the light at the fiber exit.
Choosing a slit width of 0.1mm provides a resolution of 0.1 nm. 
The grating has 600 grooves per mm to cover 100 nm.  
The light is collected on a CCD (1024 horizontal pixels, along wavelength axis and 256 vertical pixels parallel to the slit), which is LN2 cooled to achieve a background noise of 1 electron per pixel per hour.
The whole wavelength range from 300 nm to 400 nm is measured only once. Even if the PMT are not the same for the experiments and this measurement, the response of the photocathode is well known and taken into account (weighted with respect to the lines strenghts).
All used PMTs have been calibrated  (see section~\ref{sec:cal}).

Notice that a movable mirror located after the grating can direct the light toward an external output slit. This output light is detected by the PMT3, which will be used for the calibration of the CCD.

\vspace{1cm}
\subsection{Temperature and Humidity effects}
In order to study effects of temperatures down to -60$^\circ$, a Dewar will be set around the sphere. Care will be taken to protect all the ports in this Dewar from freezing.
Humidity will be set by introducing known partial pressure of water vapour, according to the pressure and temperature conditions.

\section{Measurement}

The goal of the measurement is to determine precisely the relative shape of the fluorescence spectrum such that the statistical accuracy of the contribution of each line is better than 1\%.
This measurement is carried out by the CCD. It requires to have detected at least 10$^4$ photons on the lowest intensity emission line.
But the CCD is not a photo-detector representative of the actual measurement in fluorescence telescope.
At the same time, the integral spectrum is recorded in the calibrated PMT2, similar to those used in Telescopes.
In this way, we can relate the whole spectrum of the CCD measurement to the PMT photon units.
Then, the calibration procedure will establish the relation between this PMT photon units to the amount of photons emitted inside the sphere.
Notice, the number of primary electrons is given by a Faraday cup acting as a beam dump.

The PMT1 monitors the number of photons inside the sphere to get a first estimate on the ratio of the number of photons in the sphere to the number of photons to the PMT2.

\section{Calibration}
\label{sec:cal}

\begin{figure*}[t]
  \centering
  \includegraphics[width=8cm]{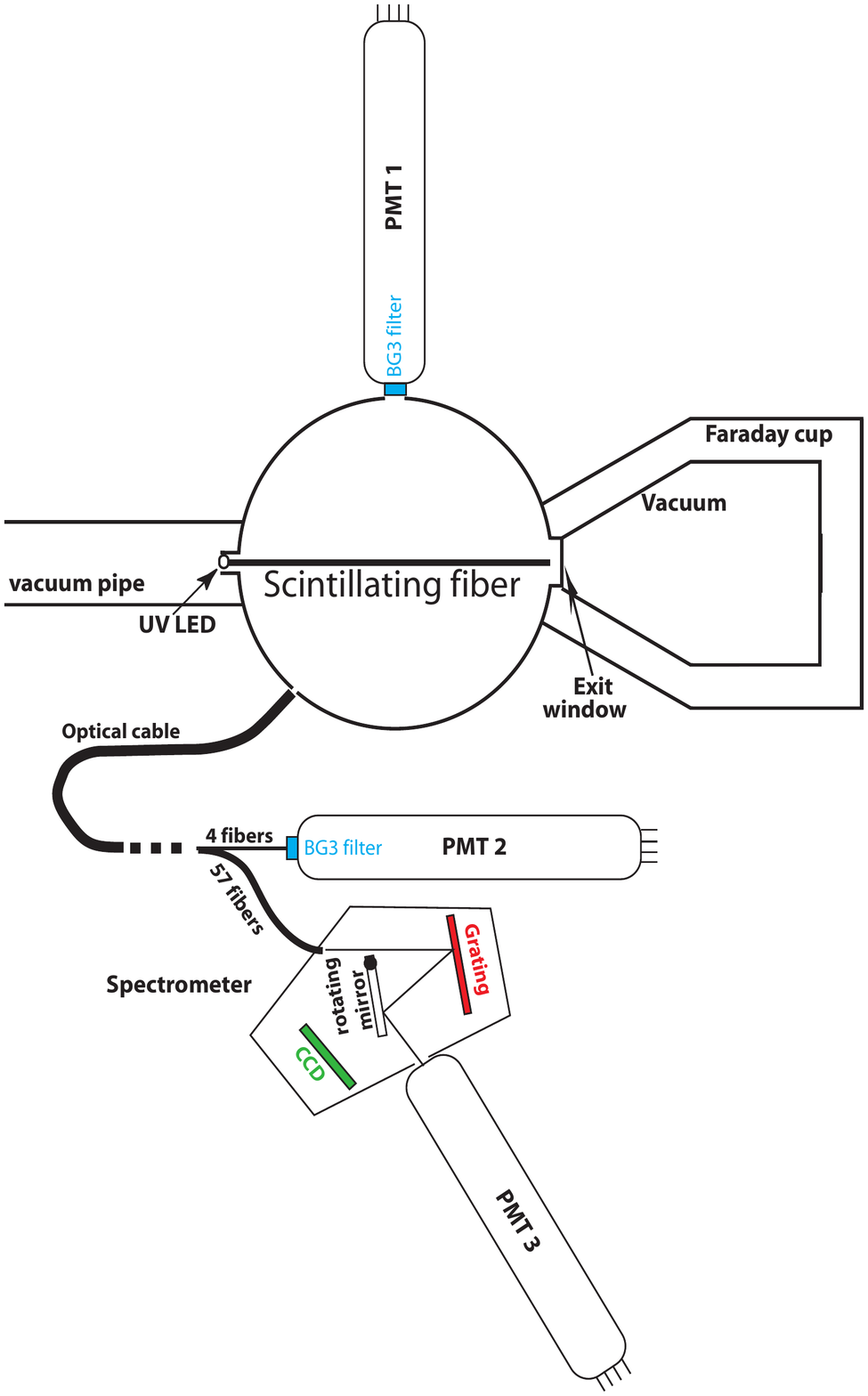}
  \caption{Calibration of the experiment.}
  \label{wide_fig2}
 \end{figure*}

The scheme in figure~\ref{wide_fig2} gives the calibration procedure. 
The purpose of the calibration is to estimate in a more precise and absolute way the light ratio between the total light generated in the sphere to the light detected in the CCD or reaching the PMT2.
To mimic the beam geometry, the light source will be a 1 mm scintillating fiber illuminated from outside the sphere by a UV LED.  The light is monochromatic within 10 nm.

All the PMTs are calibrated in gain and their efficiency is absolutely determined at 2 \% level by a patented method developped by Lefeuvre et al~\cite{lab7},~\cite{lab14}.
This method is based on the comparison between the PMT and a NIST-photodiode precise to 1.5 \%~\cite{lab15}. 
The variation of the ratio of photoelectrons produced in the PMT to the photons hitting this tube (in units of  Ampere per Watt) is known.
With this method, the PMT becomes a "NIST-PMT".
This calibration method is precise because it is made in single photoelectron mode where gain and efficiency are totally de-correlated. 

The PMT3 is calibrated with the same accuracy, the mirror (reflectivity 99\%) inside the spectrometer is tilted such that the light from the grating goes through a slit of the same width than one horizontal pixel of the CCD.
Scanning the emission wavelength profile of the scintillating fiber, we can compare the signal units of the CCD to the signal units of the PMT3.  
The attenuation of the number of photons from the sphere to the grating induced by reflection on interface, numerical aperture, fibers transmission is similar for the fluorescence yield measurement and the instrument calibration.
Their contributions are then compensated by the measurement of the calibration ratio.   

\vspace{0.5cm}
\subsection{High pressure}
\label{sec:press}

We saw that 15\% energy is missing at 0.1 atm and 9\% at 1 atm.
In the yield measurement the remaining uncertainty will be the fraction of high energy deltas escaping the sphere. This contributes to  the order of 5\% with an error  of 10\% on the energy leaked out.
For a 4 MeV beam, the highest energy delta is 2 MeV with a range of around 8 m at 1 atm. It is impracticable to make a sphere and a Dewar that big. 
We plan to get around this problem by increasing the pressure and maybe decreasing the beam energy until
all the deltas are stopped in the 20 cm sphere. Then a direct comparison of the energy loss with the Bethe -Bloch formula will provide the correction for the escaped energy.

\section{Conclusion}

Up to now the fluorescence yield of the brightest line at 337 nm has been measured in an absolute way in one set of conditions, 
 whereas fluorescence yields at the other wavelengths have been relatively measured for different conditions. 

This experiment will provide both the integrated measurement and fluorescence yields for each line with high
accuracy better than 5\% other a wide range of atmospheric conditions.

An current set-up with a small integrating sphere (6 cm diameter) is tested at PHIL accelerator in order to valid the whole system.

\vspace*{0.5cm}
{
\footnotesize{{\bf Acknowledgements}

{We thank F. Arqueros and J. Rosado for
their advices and  simulations.
We also thank the mechanics, PHIL, and vacuum team at LAL for the construction of the fluorescence bench.}}
This work has been financially supported by the GDR PCHE in France, APC laboratory, and LAL.

}

\clearpage

%% file: icrc2013-0900.tex



\title{Towards the Preliminary Design Review of the Infrared Camera of the JEM-EUSO Space Mission.}

\shorttitle{JEM-EUSO IR camera}

\authors{
M. D. Rodriguez Frias$^{1}$,
J. Licandro$^{2}$,
M. D. Sabau$^{3}$,
M. Reyes$^{2}$,
T. Belenguer$^{3,}$,
M. C. Gonzalez-Alvarado$^{3}$,
E. Joven$^{2}$,
J. A. Morales de los Rios$^{1,4}$,
M. Saez-Palomino$^{3}$,
H. Prieto-Alfonso$^{1}$,
G. Saez Cano$^{1}$,
J. H-Carretero$^{1}$,
S. Perez Cano$^{1}$ \&
L. del Peral$^{1}$
for the JEM-EUSO Collaboration.
}
\afiliations{
$^1$ SPace \& AStroparticle (SPAS) Group, UAH, Madrid, Spain \\
$^2$ Instituto de Astrofisica de Canarias (IAC), Tenerife, Spain. \\
$^3$ LINES laboratory, Instituto Nacional de Tecnica Aeroespacial (INTA), Madrid, Spain. \\
\scriptsize{
$^{4}$ now at: RIKEN, 2-1 Hirosawa, Wako, Saitama 351-0198, Japan.
}
}
\email{jose.moralesdelosrios@riken.jp} 

\abstract{An Atmospheric Monitoring System is a key element of a Space-based mission which aims to detect Ultra-High Energy Cosmic Rays (UHECR). The JEM-EUSO Space Mission has a dedicated Atmospheric Monitoring System that plays a fundamental role in our understanding of the atmospheric conditions in the Field of View of the telescope. Our Atmospheric Monitoring System consists of an infrared camera and a LIDAR. The full design, prototyping, construction under space qualification, assembly, integration and verification of the Infrared Camera is under responsibility of the Spanish Consortium within JEM-EUSO. The Infrared Camera Scientific Requirements Review (SRR) was achieved in December 2011 and the System Preliminary Design Review (SPDR) is forseen for 2013. The Infrared Camera of JEM-EUSO will contribute to ensure that the energy of the primary UHECR and the depth of maximum development of the Extensive Air Shower (EAS) are measured with an accuracy better than 30 \% and 120 $g/cm^2$ respectively.} 


\maketitle

\section{Introduction}

The Extreme Universe Space Observatory on the Japanese Experiment Module (JEM-EUSO) \cite{sbib:EUSOperf},\cite{bib:MDRodriguezFrias-Vulcano} of the International Space Station (ISS) is the first space-based mission worldwide in the field of Ultra High-Energy Cosmic Rays (UHECR) and will provide a real breakthrough toward the understanding of the Extreme Universe at the highest energies never detected from Space so far. JEM-EUSO from Space will pioneer the observation of cosmic rays at the extreme high energy range. Moreover, JEM-EUSO will use our atmosphere as a huge calorimeter to detect the electromagnetic and hadronic components of the Extensive Air Shower (EAS) produced by the primary UHECR.

\begin{figure}[t]
 \includegraphics[width=0.5\textwidth]{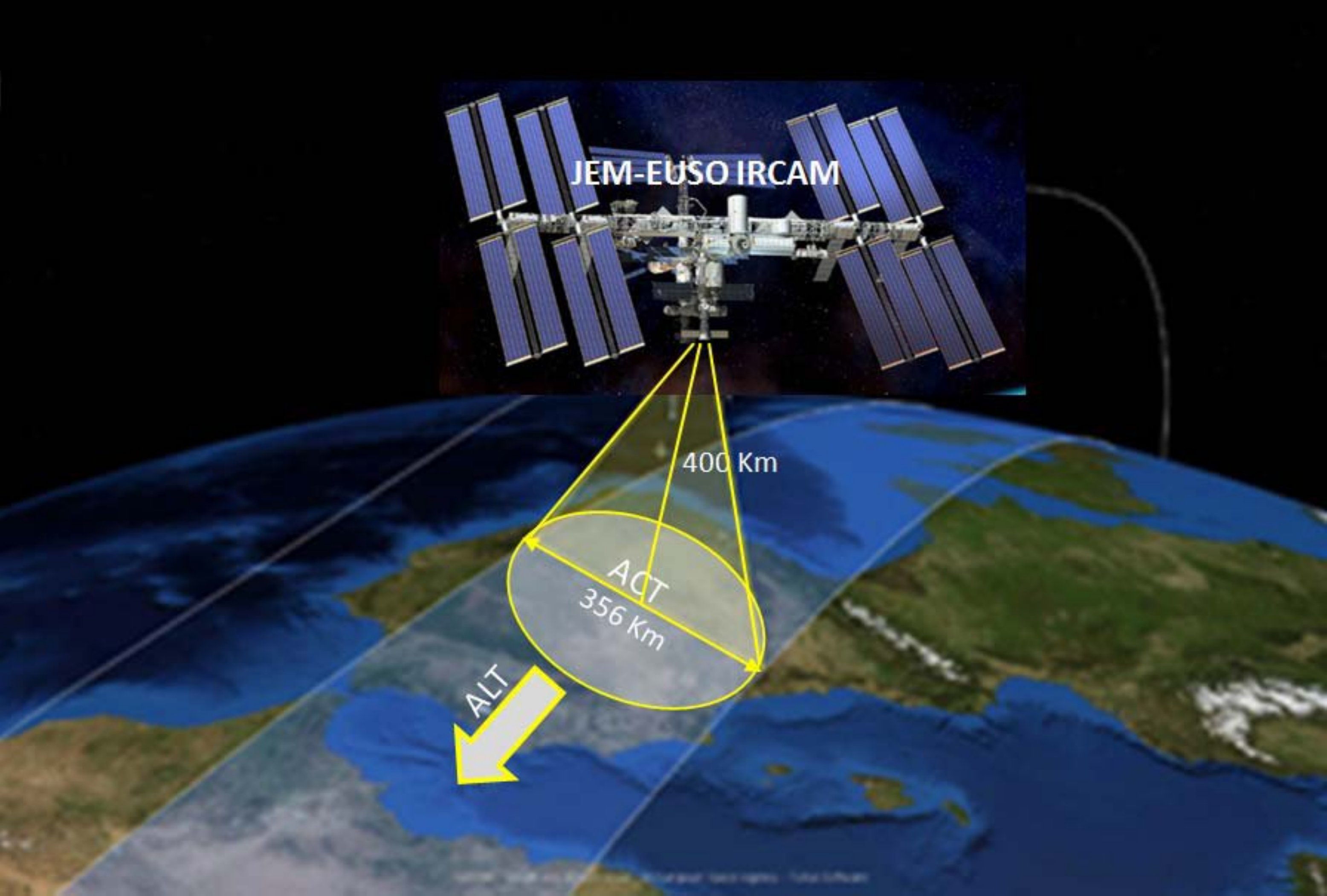}
\caption{Schematic view of the IR-Camera observation concept along the International Space Station track.}
\label{IRConcept:1}       
\end{figure}

At the UHECR regime observed by JEM-EUSO, above 10$^{19}$ eV, the existence of clouds will blur the observation of UHECRs \cite{bib:Guadalupe}. Therefore, the monitoring of the cloud coverage by the JEM-EUSO Atmospheric Monitor System (AMS) (\cite{bib:swada},\cite{bib:MDRodriguezFrias-CERN},\cite{bib:andriiICRC}) is crucial to estimate the effective exposure with high accuracy and to calibrate the UHECRs and EHECRs events just above the threshold energy of the telescope. The IR-Camera will be the instrument devoted to detect clouds and determine their top height in the FoV of the JEM-EUSO main instrument. The camera will provide a 2D image of the cloud top temperature, and using this image, with the LIDAR and the global models, the cloud top height under investigation will be achieved with an accuracy better than 500m during the observation period of the JEM-EUSO main instrument. The IR-Camera full design, prototyping, space qualified construction, assembly, verification and integration is under responsibility of the Spanish Consortium involved in JEM-EUSO. The scientific and technical requirements for the IR-Camera are far for being undemanding, and are summarized in Table \ref{aatab_req:1}.  

\begin{table}
\caption{Requirements for the IR-Camera of the JEM-EUSO Space Mission.}
\begin{tabular}{|l|ll|}
\hline\noalign{\smallskip}
Parameter & Target value &  Comments  \\
\noalign{\smallskip}\hline\noalign{\smallskip}
Measurement &  & Annual variation \\ range & 220 K - 320 K &  of cloud  \\ &  & temperature plus \\ & & 20 K margin    \\
\hline
  &  & Two atmospheric \\ Wavelength & 10-12 $\mu$m  & windows available: \\ & & 10.3-11.3 $\mu$m \\ & & and 11.5-12.5 $\mu$m    \\
\hline
FoV     & 48$^o$ &  Same as \\  & & main instrument \\
\hline
Spatial & 0.1$^o$ (Goal) & @FoV center \\ resolution &  0.2$^o$ (Threshold) &  \\
\hline
Absolute &  & 500 m in cloud \\ temperature & 3 K & top altitude \\ accuracy &  &  \\
\hline
Mass  & $\leq11$ kg & Inc 20$\%$ margin.    \\
\hline
Dimensions     & $400\times400\times370$ & w/o Insulation and \\ 
& & mounting bracket. \\
\hline
Power    & $\leq15$ W  &  Inc 20$\%$ margin.       \\
\hline
Lifetime    & 5 years In-orbit  &  +2 years On-ground    \\
\noalign{\smallskip}\hline
\end{tabular}
\label{aatab_req:1}  
\end{table}

Moreover, a dedicated End to End (E2E) simulation for the IR-Camera is under development \cite{aasdsu}. This work gives us the capabilities to study the impact of several scenarios of the atmosphere, in terms of retrieval temperature accuracy, detector capabilities, calibration procedures and correction factors to be taken into account for the final data products of the AMS system of the JEM-EUSO Space Mission. At this design state of the IR Camera, this E2E similator will give us answers in key points of the design, like the compression algorithms evaluation and estimation of the expected accuracy of the calibration options foreseen \cite{bib:joseICRC2013}.

\section{The IR-Camera System Preliminary Design}
\label{ircam:M}

The IR-Camera \cite{bib:joseICRC2011} is a microbolometer based infrared imaging system aimed to obtain the cloud coverage and cloud top altitude during the observation period of the JEM-EUSO main instrument. Its preliminary design can be divided into three main blocks: the Telescope Assembly, the Electronic Assembly and the Calibration Unit. The main function of the Telescope Assembly is to acquire the infrared radiation by means of an uncooled microbolometer and to convert it into digital counts. A dedicated optical design has been developed as well, with a huge angular field to complain with the wide FoV of the JEM-EUSO main telescope. Meanwhile the Electronic Assembly provides mechanisms to process and transmit the obtained images, the electrical system, the thermal control and to secure the communication with the platform computer. To assure the high demanding accuracy, a dedicated on-board calibration system is foreseen. Moreover, this System Preliminary Design is complemented by a challenging Mechanical and Thermal design to secure that the IR-Camera will be completely isolated.

\subsection{The Telescope Assembly; detector and FEE}
\label{detector:M}

\begin{figure}[t]
 \includegraphics[width=0.5\textwidth]{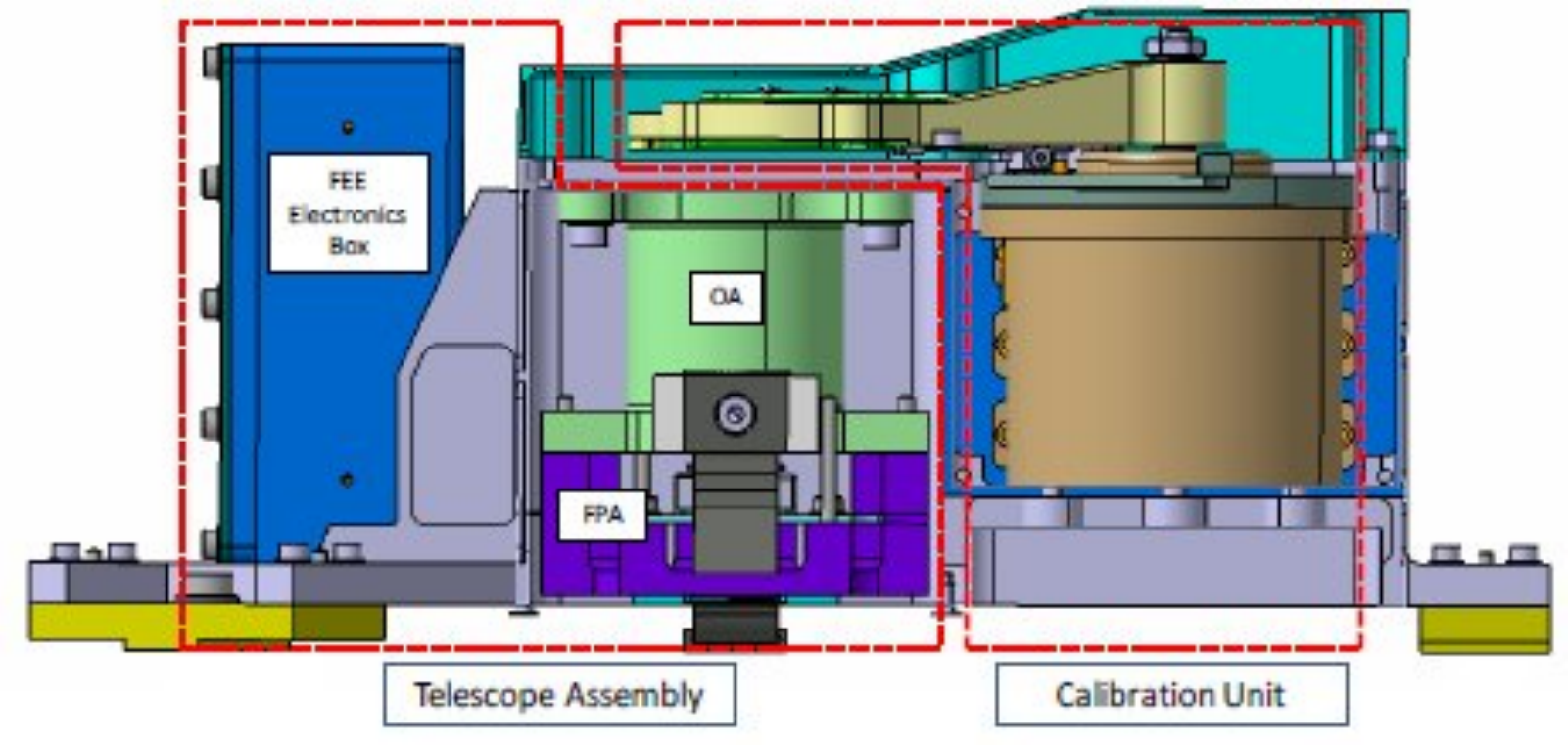}
\caption{Illustration of the preliminary design of the infrared camera telescope assembly.}
\label{TelescopeAssembly:1}       
\end{figure}

The IR-Camera Telescope assembly includes the Infrared detector ($\mu$Bolometer), the FEE (Front End electronic) and the Optical lens assembly (Figure \ref{TelescopeAssembly:1}).

The infrared detector that has been selected for the JEM-EUSO-IR camera is the UL04171 from the ULIS Company \cite{aaulis_datasheet}. The UL04171 is an infrared opto-electronic device comprised by a $\mu$bolometer Focal Plane Array (FPA); two dimensional detector array made from amorphous Silicon. The working operative temperature is around 30$^o$C and a dedicated TEC (Thermo-Electrical Cooler) has been implemented to guarantee a very stable temperature. The $\mu$Bolometer is supplied by the manufacturer in a vacuum sealed package with the readout electronics, peltier and temperature sensor integrated. Moreover, a protective window of Germanium glass has been implemented in the optical design.
 
The FEE (Front End Electronics) manages and drives the $\mu$Bolometer; It provides the bias and the sequencer and manages the images acquisition modes. The FEE communicates with the ICU and provides it the uncompressed raw images. The core of the FEE shall be a FPGA, VIRTEX family, in charge of implementing the main FEE functions. This includes the control of the UL04171, the generation of all the synchronism including the clocks generation and the interface with the sequencer. The polarization of the detector (bias, gain, offset generation and control) will be also controlled by the FPGA. The data acquisition will be implemented with an Analog Digital Converter (ADC) in each detector output channel previous to the FPGA input. The ADC number of bits will be chosen according to the pixel data resolution required by the IR-Camera.

\begin{figure}[b]
 \includegraphics[width=0.5\textwidth]{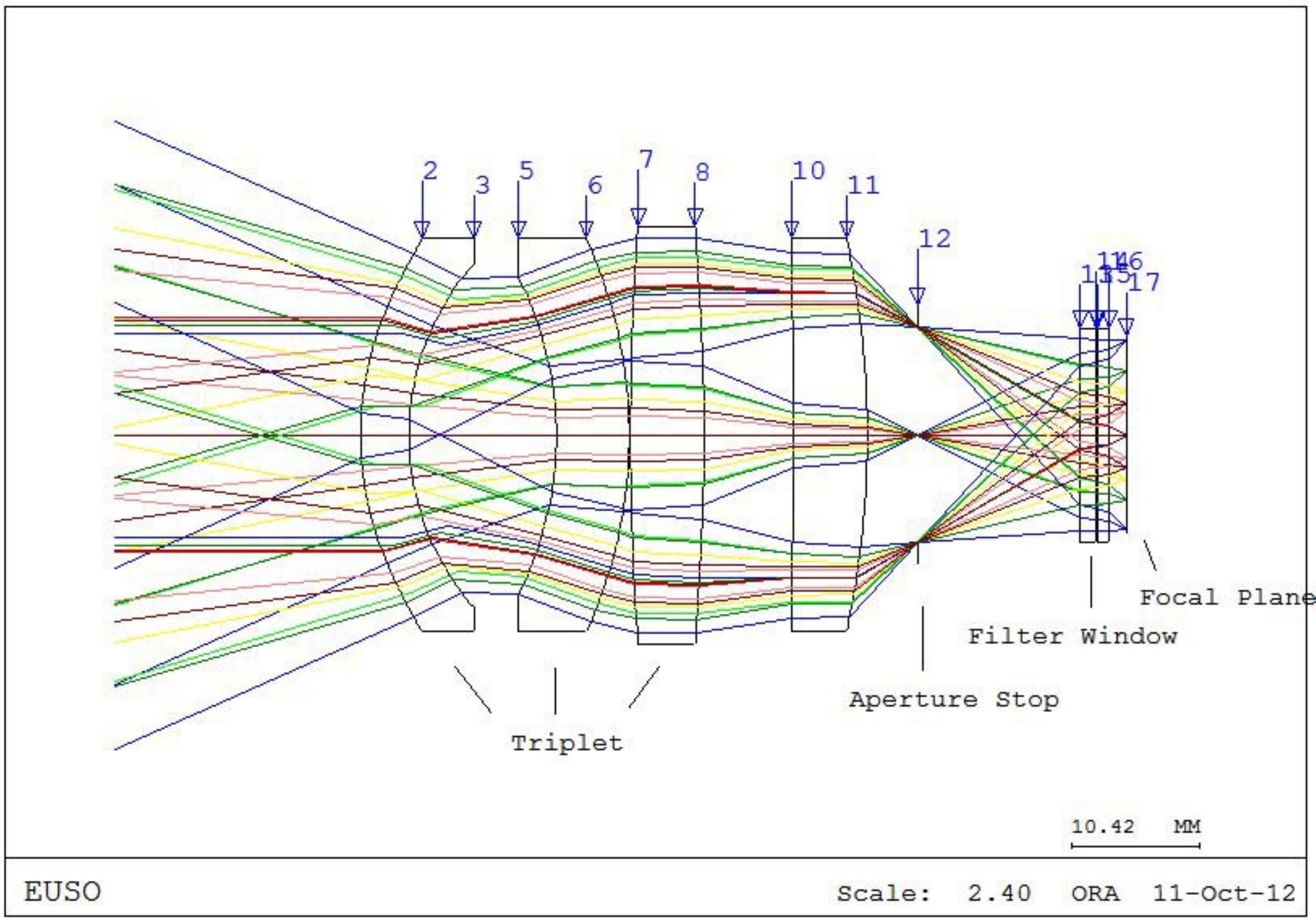}
\caption{Schematic illustration of the optic preliminary design of the Infrared Camera.}
\label{optics:F1}       
\end{figure}

\subsection{The Optical subsystem Preliminary Design}
\label{optics:M}

The Optical Assembly is one of the most critical sub-assemblies of the IR-Camera. For an optimal operation, the design of the Optical subsystem has to fulfill the following technical requirements: To acquire radiation at the mid infrared wavelength band (10-12.5) $\mu$m; To guarantee the requested (48$^o$) FoV; To be very fast in terms of $F\#$; To secure an optimal operative temperature for the ULIS (~29$^o$C) detector, for both, the cold operative case (-15$^o$C) and the hot operative case (15$^o$C); The thermal excursion of the lenses has to be less than ~20$^o$C and finally, to keep the Cold Stop temperature 15$^o$C below the ULIS $\mu$blometer temperature.

Presently, the optical system design (Figure \ref{optics:F1}). is a refractive objective based in a triplet with one more lens close to the stop and a window for the filters close to the focal plane. The first surface of the first lens and the second surface of the third lens are aspheric that allow a better quality of the complete system. The aperture stop is situated at 0.40 mm behind the fourth lens, in order to separate the optical system to the detection module. The system, consisting of four lenses, has a focal length of 19.10 mm, and a f-number of 1, and it shall work with a total FoV of 48$^o$. The overall length between the first surface to the focal plane is 62.30 mm. All the data shown below are related to extreme fields ($\pm$24$^o$), although better response is obtained for intermediate fields. The full system has been designed only with one optical material, Germanium with a refraction index of 4.003118.

\begin{figure}[h]
 \includegraphics[width=0.5\textwidth]{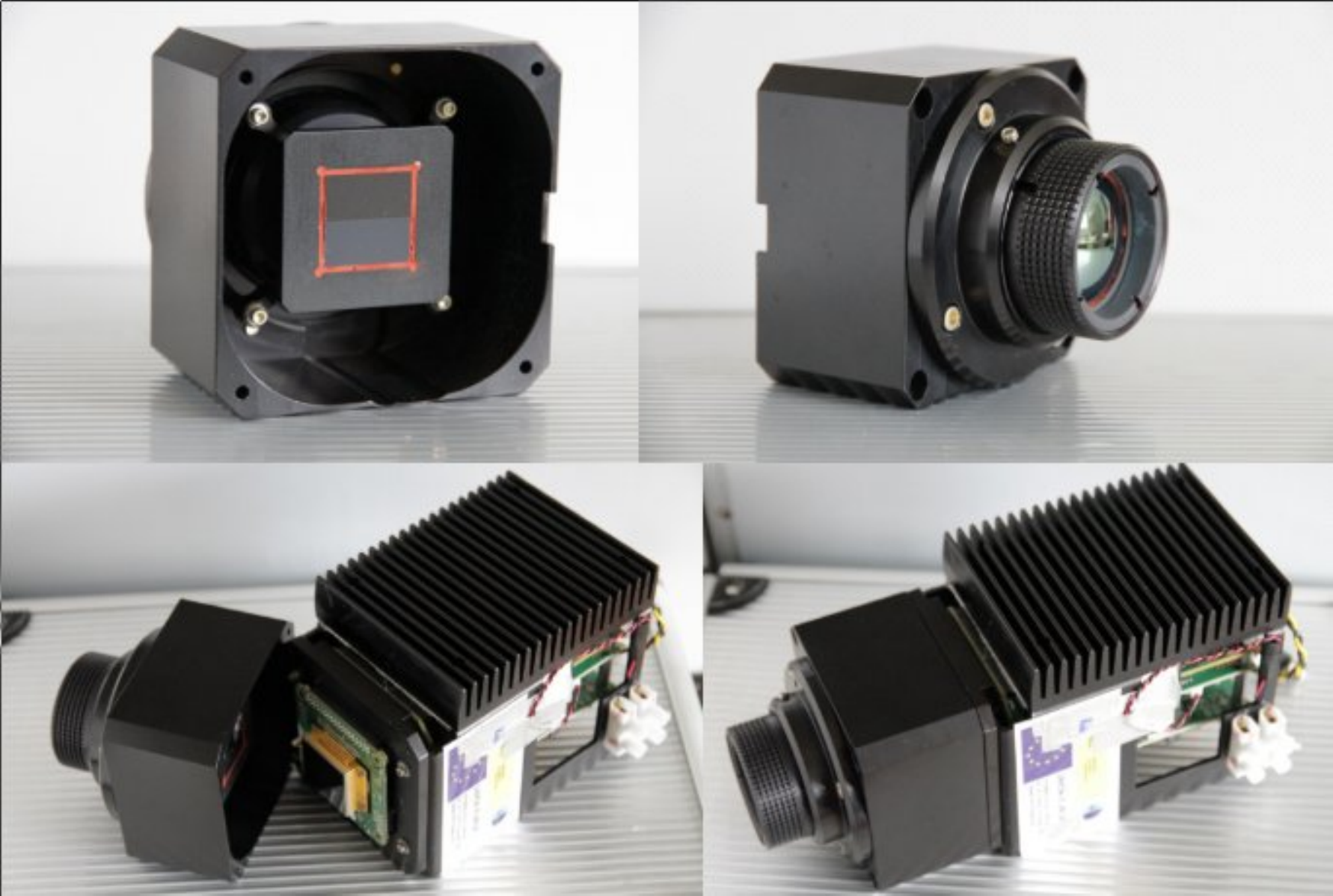}
\caption{Breadboard model manufactured at INTA facilities for the Infrared Camera of JEM-EUSO.}
\label{optics:F3}       
\end{figure}

A breadboard model (Figure \ref{optics:F3}) has been manufactured to test the optical performances of the system. The breadboard lenses have been mounted in the same way that will be assembled in the flight model. The tolerances and optomechanical process have been successfully tested and verified at this stage of development.

\subsection{The electronic assembly}
\label{electronics:M}

The Electronic Assembly is composed of two main sections: the Instrument Control Unit (ICU), and the Power Supply Unit (PSU). Both blocks follow cold redundancy architecture and are placed on individual PCBs so that four boards are defined: ICU Main, ICU Redundant, PSU Main, and PSU Redundant. The ICU controls and manages the overall system behavior, including the data management (compression, format), the power drivers and the mechanisms (shutter, blackbodies etc.) controller FPGA. The IR-Camera electronics shall provide mechanisms to process and transmit images obtained from an IR detector controlled by a dedicated FEE board, a Firmware (FW) solution is considered as baseline for this proposal.  

Data generated by the FEE is then processed by the Instrument Control Unit (ICU), which is in charge of controlling several aspects of the system management such as the electrical system, the thermal control and the communication with the platform computer. The Power Supply Unit (PSU) receives the main power bus from JEM-EUSO main telescope and it provides the required power regulation to the system and the sub-systems. The actuator will be managed by the ICU, providing control to a stepper motor and acquiring its position by means of micro-switches placed in the stable positions.

\subsection{The calibration subsystem}
\label{calibration:M}

\begin{figure}[h]
 \includegraphics[width=0.5\textwidth]{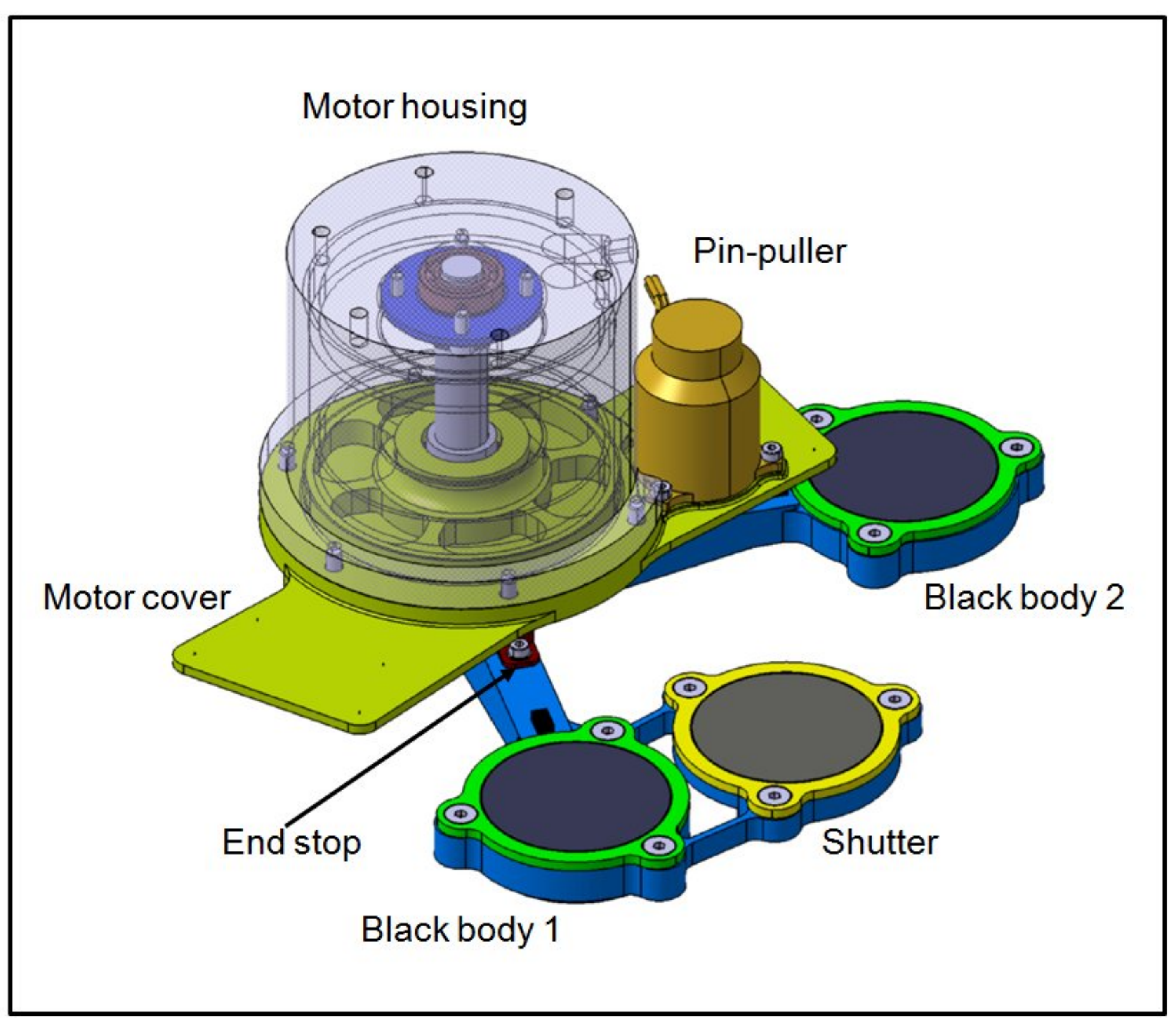}
\caption{Ilustration of the calibration subsystem.}
\label{Calibration:F1}       
\end{figure}

The calibration unit (Figure \ref{Calibration:F1}) is dedicated to manage and control the IR calibration operation. This unit has to guarantee a reference internal temperature to ensure the calibration of the data coming out of the FEE. Following the strategy of operational modes, four positions are provided from this unit: Acquire, Shutter (offset correction), Calibration Hot point, and Calibration Cold point. 

The position of shutter will be used to close the optic in the safe and off operation mode. Calibration unit is mainly composed of two Black Bodies with a temperature controlled Shutter, a moving mechanism and the motor, the positioning system and a calibration Thermal control.

\subsection{The thermal and mechanical design}
\label{thermal:M}

The mechanical structure contains and protects the Telescope Assembly and the Calibration Unit. It is attached to the bench of the JEM-EUSO Telescope by means of three flexure-pads. The Main Housing is an aluminium Al6082 monocoque body-shell. It has three different compartments to accommodate the required subsystems and provide overall stiffness and thermal isolation of the Optical Assembly and FPA from the Calibration Unit and the FEE. Stiffeners have been used to optimize the mass of the Main Housing structure. This Housing contains a stiff baseplate, which supports the Calibration Unit and the FEE Electronics Box, both contained in the IF plane to minimize the loads on this plane maintaining a low CoG. 

The Lenses Barrel has the mission to enclose and support the lenses, which are positioned with Spacers, and they are bonded with optical adhesive EPO-TEK 301-2. The Cold Stop is a sort of diaphragm between the last lens and the microbolometer. It is necessary for optical purposes, and its temperature must be around 15$^o$C lower than the ULIS temperature. This is achieved by means of a passive thermal control, and two (main and redundant) thermal sensors.

\begin{figure}[h]
 \includegraphics[width=0.5\textwidth]{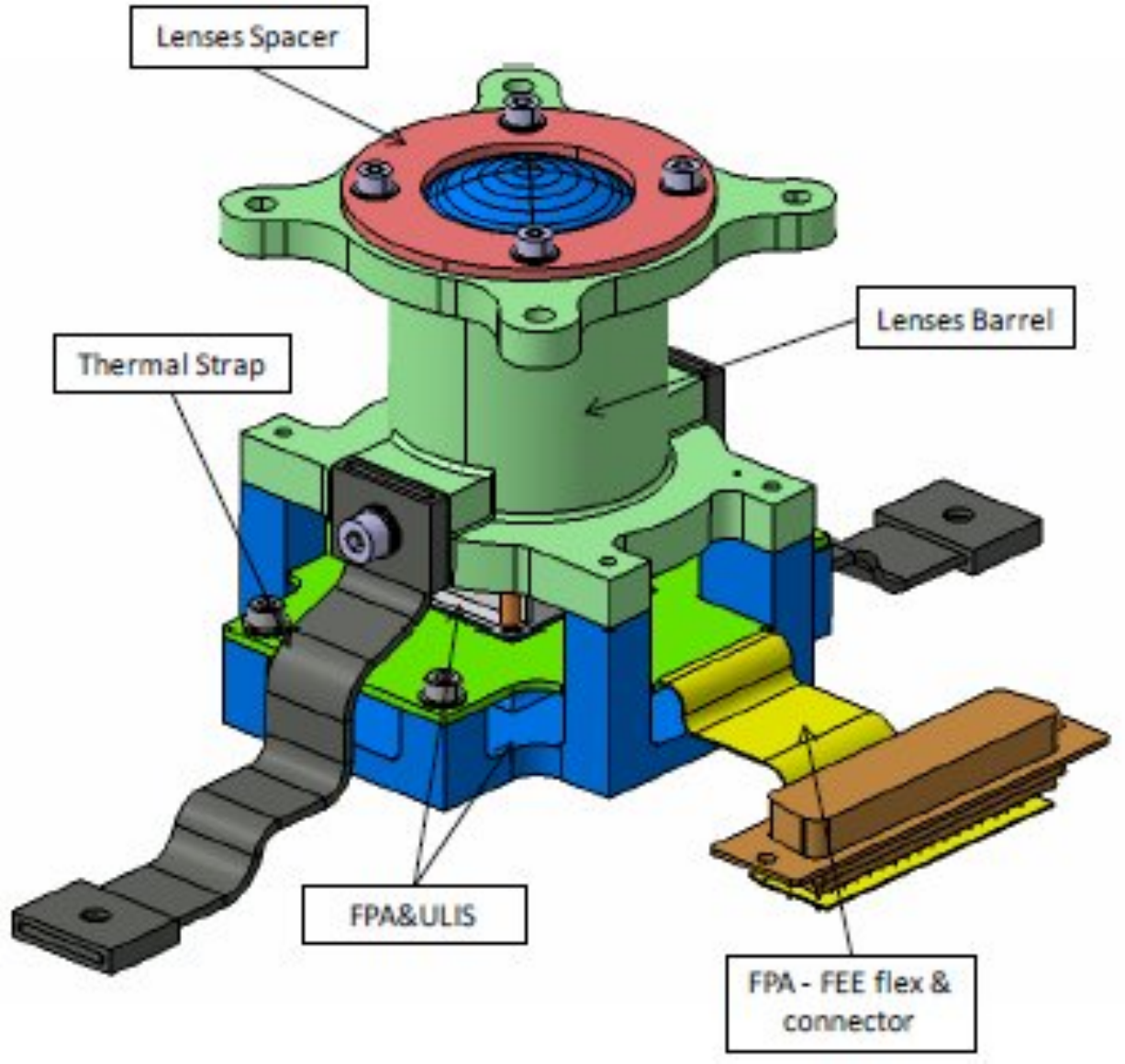}
\caption{Optic Assembly where the lenses barrel is shown.}
\label{thermal:F2}       
\end{figure}

\section{Conclusions}

Cosmic Rays Physics is one of the Fundamental Physics key issues and an essential branch of Astroparticle Physics. It aims, in an unique way to address many fundamental questions of the extreme and non-thermal Universe in the Astroparticle Physics domain at the highest energies never detected so far. Moreover, UHECR has witnessed a major breakthrough with the Pierre Auger Observatory (PAO) and the Telescope Array (TA) success. The results on UHECR by PAO and TA have pointed out the huge physics potential of this field that can be achieved by an upgrade of the performances of current ground-based instruments and with new space-based missions. To reach the largest exposures, space observatories are likely to be essential. The JEM-EUSO space observatory is aimed to achieve one of our main goals, reach the so called "Particle Astronomy Era".

The IR-Camera onboard JEM-EUSO will consist of a refractive optics made of germanium and an uncooled $\mu$bolometer array detector. The FoV of the IR-Camera is $48^\circ$, totally matching the FoV of the main JEM-EUSO telescope. The angular resolution, which corresponds to one pixel, is about $0.1^\circ$. A temperature-controlled shutter in the camera and blackbodies are used to calibrate background noise and gains of the detector to achieve an absolute temperature accuracy of $\sim 3$ K. Though the IR-Camera takes images continuously every 17s, in which the ISS moves 1/4 of the FoV of the JEM-EUSO telescope. In this paper an overall description of the present stage of design and development of the IR Camera of JEM-EUSO has been reviewed.

\vspace*{0.5cm}
{
\footnotesize{{\bf Acknowledgment:}{
The Spanish Consortium involved in the JEM-EUSO Space
Mission is funded by MICINN under projects AYA2009-
06037-E/ESP, AYA-ESP 2010-19082, AYA-ESP 2011-29489-C03-
01, AYA-ESP 2011-29489-C03-
02, AYA-ESP 2012-39115-C03-01, AYA-ESP 2012-39115-C03-03, CSD2009-00064 (Consolider MULTIDARK)
and by Comunidad de Madrid (CAM) under project S2009/ESP-1496. This work was partially supported by Basic Science Interdisciplinary 
Research Projects of RIKEN and JSPS KAKENHI Grant (22340063, 23340081, and 
24244042), by the Italian Ministry of Foreign Affairs, General Direction 
for the Cultural Promotion and Cooperation, by the 'Helmholtz Alliance 
for Astroparticle Physics HAP' funded by the Initiative and Networking Fund 
of the Helmholtz Association, Germany, and by Slovak Academy  
of Sciences MVTS JEM-EUSO as well as VEGA grant agency project 2/0081/10.
}}

}

\clearpage

%% file: icrc2013-0530.tex


\title{LIDAR treatment inside the ESAF Simulation Framework for the JEM-EUSO mission}

\shorttitle{LIDAR simulation for JEM-EUSO}

\authors{
S. Toscano$^{1}$,
L. Valore$^{2}$,
A. Neronov$^{1}$,
F. Guarino$^{2}$
for the JEM-EUSO Collaboration.
}

\afiliations{
$^1$ ISDC Data Centre for Astrophysics, Versoix, Switzerland \\
$^2$ Istituto Nazionale di Fisica Nucleare - Sezione di Napoli, Italy \\
}

\email{Simona.Toscano@unige.ch}

\abstract{JEM-EUSO is a next-generation fluorescence telescope which will detect Ultra High Energy Cosmic Rays (UHECR, cosmic rays with energies above $5\cdot10^{19}$ eV) from the International Space Station (ISS), by using the whole Earth as a detector. Being in such a peculiar location, JEM-EUSO will orbit the Earth and it will experience all possible weather conditions. The JEM-EUSO telescope will detect fluorescence UV emission from Extensive Air Showers (EAS) produced by UHECR penetrating in the atmosphere. To achieve a correct reconstruction of  UHECR energy and of the type of the primary cosmic ray particle, information about absorption and scattering properties of the atmosphere is required.\\
A LIght Detection And Ranging (LIDAR) device is foreseen as a part of the Atmospheric Monitoring system for the JEM-EUSO mission. The goal of the LIDAR is to provide measurements of extinction and scattering properties of the atmosphere along the EAS development path and between the EAS and JEM-EUSO. In order to test the capabilities of the LIDAR a simulation of this device has been implemented inside the ESAF Simulation Framework used for the JEM-EUSO mission.\\ 
In this contribution we will review the LIDAR simulation chain, focusing on the generation and propagation of photons in the atmosphere. First results from simulations will be shown for a laser beam propagating in different atmospheric conditions. }


\maketitle

\section{Introduction}

JEM-EUSO is a next-generation fluorescence telescope which will observe UV emission from UHECR induced Extensive Air Showers (EAS) from space, experiencing all possible weather conditions. It has been estimated that  $\sim$70\% of EAS detected by JEM-EUSO will be affected by scattering and absorption in the clouds and aerosol layers \cite{Mario_ICRC2011, Lupe2012}. Proper interpretation of the EAS signal, including the reconstruction of the energy, direction and identity of the UHECR particle requires a detailed knowledge of the influence of the scattering of UV light in clouds and aerosols on the detected fluorescence signal. Cloud- and aerosol-induced variations of the scattering and absorption properties at the location of EAS events distort the UV signal from EAS detected by JEM-EUSO. In the absence of detailed information on the presence and physical properties of the cloud and aerosol layers in the JEM-EUSO Field of View (FoV), distortions of the UV signal from EAS lead to systematic errors on the determination of the properties of UHECR from the UV light profiles.
The distortion of the EAS profiles could be corrected if detailed information on distribution and optical properties of the cloud/aerosol layers in the JEM-EUSO FoV is known. This information will be provided by the Atmospheric Monitoring (AM) system of JEM-EUSO. \\
The most relevant information about the absorption and scattering properties of clouds and aerosols is at the location around the EAS events and it will be provided by the LIDAR. The laser beam will be shot several times in the direction in which the EAS trigger occurred directly after the trigger is generated. With this pointing capability, the LIDAR device will be able to measure the backscattered signal in several directions around the supposed EAS maximum. \\
In order to study the system capabilities and to ensure that it fulfills all the requirements established for the JEM-EUSO instrument \cite{Picozza_ICRC2013} (energy resolution of 30\% and X$_{max}$ resolution of 120 g/cm$^2$), a simulation of the LIDAR device has been implemented as a part of the EUSO Simulation and Analysis Framework (ESAF) \cite{ESAFpaper} currently in use in the collaboration. \\
In this contribution we describe the generation of the laser track and the propagation of photons to the detector focal surface. The LIDAR simulated signal is shown in the case of clear sky and in the presence of clouds as an example.     

\section{The ESAF Simulation Framework for LIDAR}
\begin{figure}[t]
  \centering
  \includegraphics[width=0.5\textwidth]{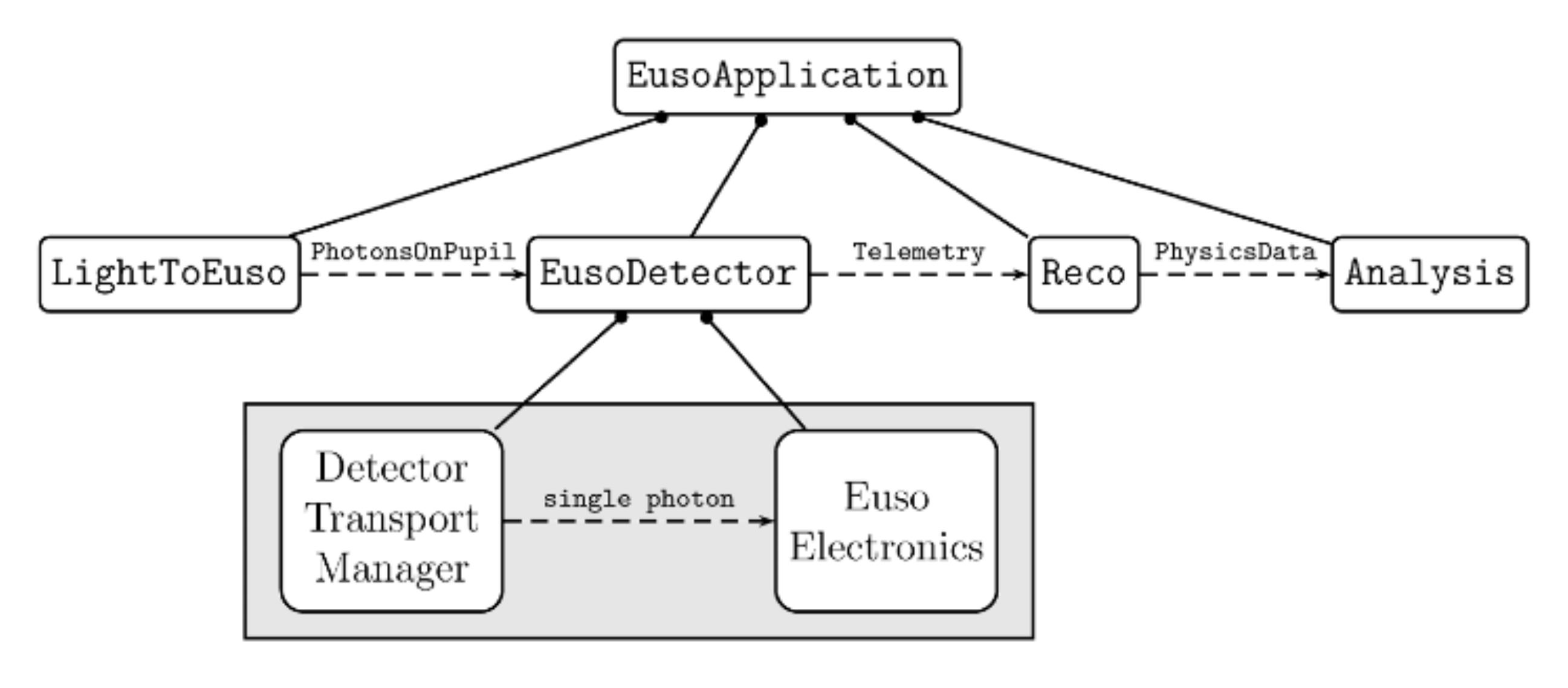}
  \caption{ESAF flux diagram from \cite{Fenu_ICRC2011}}
  \label{fig:ESAFScheme}
 \end{figure}
ESAF is a C++, Object Oriented, ROOT based, modular software designed to simulate space based UHECR detectors and currently used for the JEM-EUSO mission. It consists of several independent modules (LightToEuso, EusoDetector, Reco, Analysis) that take care of EAS simulation and reconstruction. The simulation is carried out by the first two modules: LightToEuso  simulates the shower development and the propagation of light through the atmosphere to the telescope; while EusoDetector allows the simulation of all the detector components up to telemetry. The reconstruction of the shower properties (direction, energy and primary particle) is performed inside the Reco framework. A flux diagram of the ESAF structure is shown in Fig.~\ref{fig:ESAFScheme}. In order to simulate the LIDAR backscattered signal the simulation framework has to be adapted. In the following sections we describe the changes needed to simulated the geometry of the laser beam and the propagation of photons. 

\subsection{Track geometry definition and generation of light}
\begin{figure}[t]
  \centering
  \includegraphics[width=0.4\textwidth]{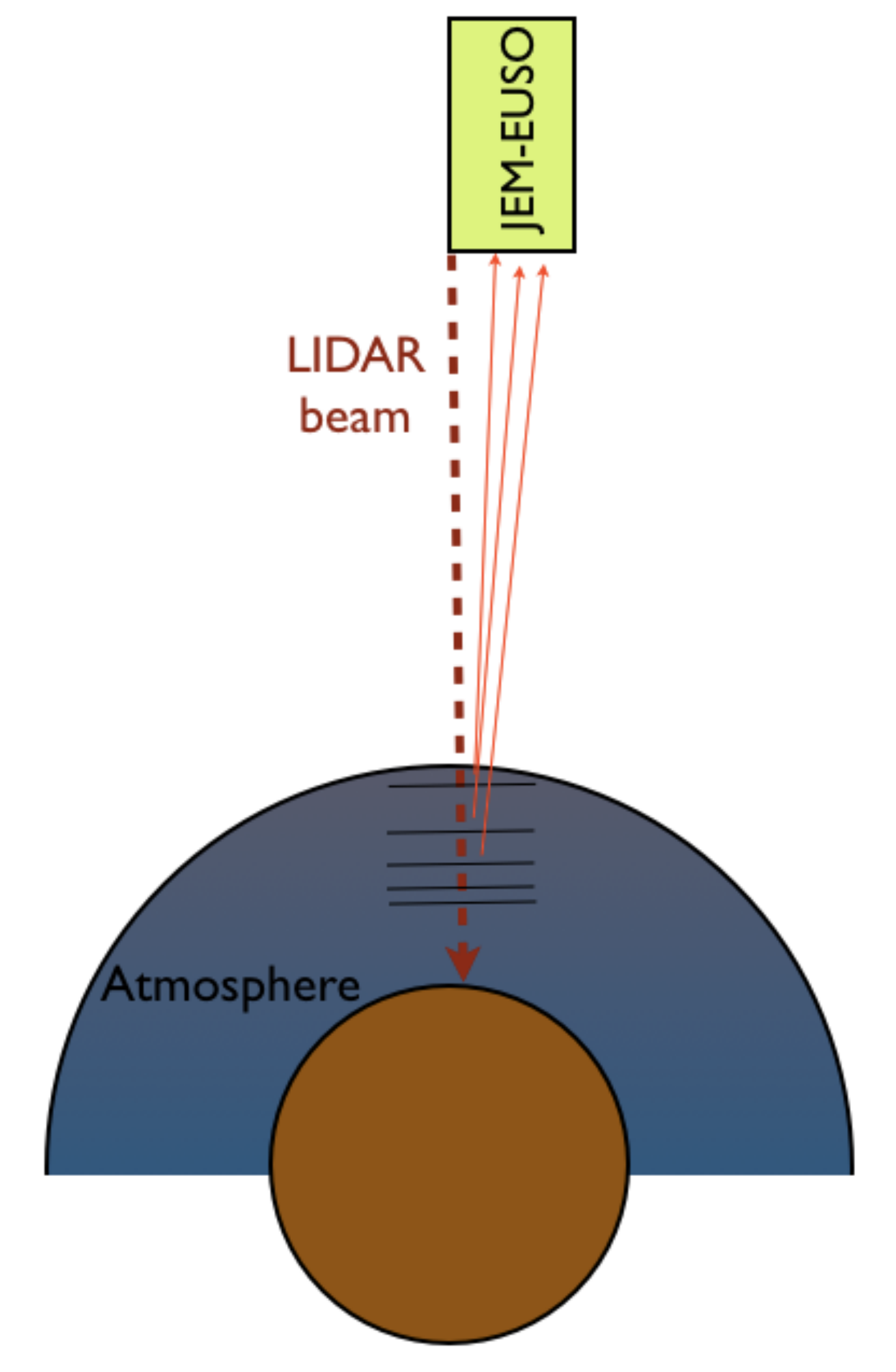}
  \caption{Illustrative picture of the LIDAR simulation scheme. The LIDAR beam interacts with the Earth atmosphere and the backscattered signal calculated in several steps along the beam track is transferred to the detector focal surface.}
  \label{fig:LaserSimuScheme}
 \end{figure}

The original ESAF code takes into account both Fluorescence and Cherenkov production from the shower but there is no treatment of the laser beam. For the implementation of the LIDAR device a new class of photons (\emph{Lidar}) has been introduced together with the existing ones (\emph{Fluo} and \emph{Cherenkov}). The Lidar photons are monochromatic and directed along the beam track; no lateral or wavelength distribution is needed as in the case of the photons from showers.  \\
In order to describe the laser beam a new class has been created inside the LightToEuso module. The number of initial photons in the beam is calculated from energy and wavelength, set when configuring ESAF, using the following formula: 
\begin{equation}
N_\gamma = E_{laser} / (c \cdot h /\lambda)
\label{eq:laser}
\end{equation}
where $E_{laser}$ is the energy per pulse in Joule, $\lambda$ is the laser wavelength in nm, $c$ the speed of light in m/s and $h$ the Planck constant in J$\cdot$s. \\
At this step the geometry of the track is defined. In the simplest case in which photons are propagated in bunches (see next section), position and time of new bunches of photons are generated along the track of the laser beam. Unlike for the case of the shower, in the case of laser there is no creation of new photons along the track. Photons are generated only at the initial step and then propagated through the atmosphere along the laser beam track. For this reason only the first bunch is filled with the initial number of photons $N_\gamma$, while the others are created but empty. An illustrative picture representing the scheme of the laser beam track simulation is reported in Fig.~\ref{fig:LaserSimuScheme}: photons from the laser beam interact at different altitudes in the atmosphere and they are backscattered toward the detector. 

\subsection{Light propagation in the atmosphere}
Once the photons are generated at first they need to be transported through the atmosphere down to the ground and then back to the detector pupil entrance. In ESAF there are two modes of propagating photons currently working in the simulation framework: 1) propagation of bunches and 2) Monte-Carlo code. In the first case, the shower simulation module provides for each ``Shower Step" the number of electrons, the electron energy distribution and the lateral distribution for both Fluorescence and Cherenkov emission. The light generation module uses them to compute \emph{bunches} of photons characterized by a mean position value, a mean direction according to the computed angular distribution, and a creation time. This new concept allows for fast simulations with no need to follow the fate of each individual photon\footnote{The typical number of photons produced by a shower of the energy of $10^{20}$ eV and $\theta = 60^\circ$is of the order of $10^{16}$}. The second is a ray-tracing algorithm in which a reducing factor is applied in oder to limit the number of photons to be simulated. Detailed explanations together with advantages and drawbacks of the two algorithms can be found in \cite{ESAFpaper}. \\
The bunch algorithm represents the fastest approach to the problem and it is currently the default choice for all the simulations carried out in JEM-EUSO. For this reason the light transport for LIDAR has been implemented at the moment only inside the bunch propagation.  A dedicated propagator is present in ESAF in order to simulate different conditions, from clear sky to clouds. In case of clear sky the propagation of photons is done calling the \emph{ClearSkyPropagator} module while the \emph{TestCloudsPropagator} is used to simulate the propagation in presence of clouds. In this context new methods have been introduced specifically for LIDAR inside the two propagators. In both case the \emph{LidarPropagate} method has been created. The method propagates bunches of photons in clear sky and cloudy conditions, creating single photons all along the path to be transferred to the detector pupil. In presence of clouds, the bunch algorithm takes into account the effects on the transmission values (depending on the cloud optical depth $\tau$ ), considering the photons scattered on this medium as lost. Once photons arrive at the focal surface the rest of the simulation is performed in the same way as for the shower case \cite{Fenu_ICRC2011}. 

\section{Results}
\begin{figure}[t]
  \centering
  \includegraphics[width=0.5\textwidth]{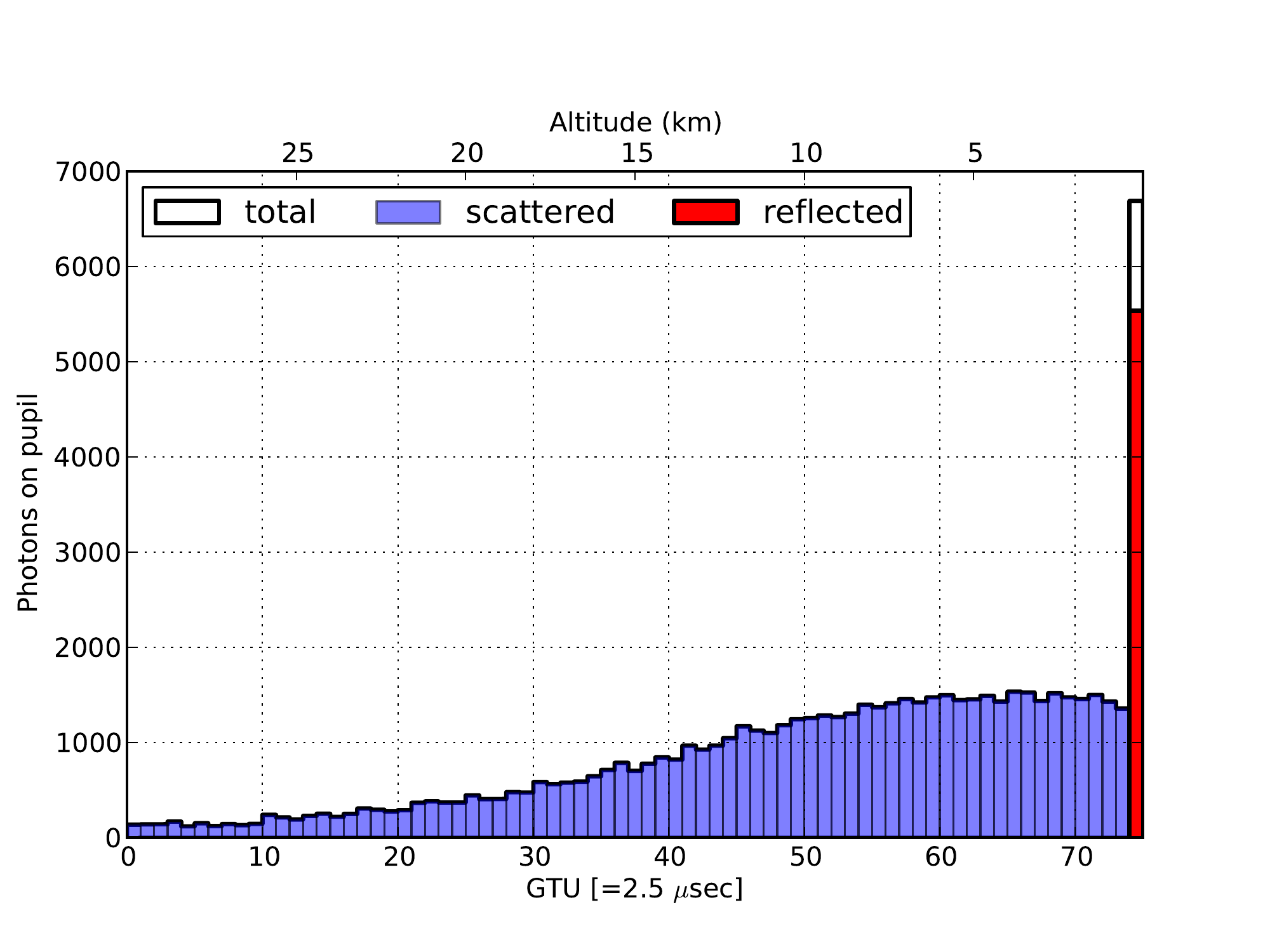}
  \caption{Distribution of the backscattered photons on pupil as a function of GTU in case of clear sky.}
  \label{fig:LIDARclear}
 \end{figure}

In this section we present the results obtained from the simulation code developed in ESAF to describe the LIDAR system for the clear sky case and in presence of clouds. Following the specification from the LIDAR design for the AM system \cite{AMS_ICRC2013}, LIDAR simulations have been carried out for a laser with $E_{laser} = $20 mJ and $\lambda = $355 nm shooting in nadir position (vertical in the centre of the telescope FoV). \\
In the case of clear sky condition the atmosphere has been considered purely molecular (no aerosol layers), where only Rayleigh scattering is considered. Fig~\ref{fig:LIDARclear} shows the distribution of photons at the pupil entrance of the JEM-EUSO telescope as a function of the time unit of the focal surface detector (GTU = 2.5 $\mu$s). The correspondence between time of the signal and altitude is also shown. The blue histogram is the air scattered component, while the red histogram represents the ``ground mark", generated by photons reaching ground and reflected back to the detector. The presence of the ground mark is a valuable information. In the case of an EAS the timing of the ground mark is used to reconstruct the geometry and the energy of the shower \cite{Reco_ICRC2011}. For LIDAR it represents a reference time that allows to correctly retrieve the properties of the atmosphere from the analysis of the received signal. 

\begin{figure}[t]
  \centering
\includegraphics[width=0.5\textwidth]{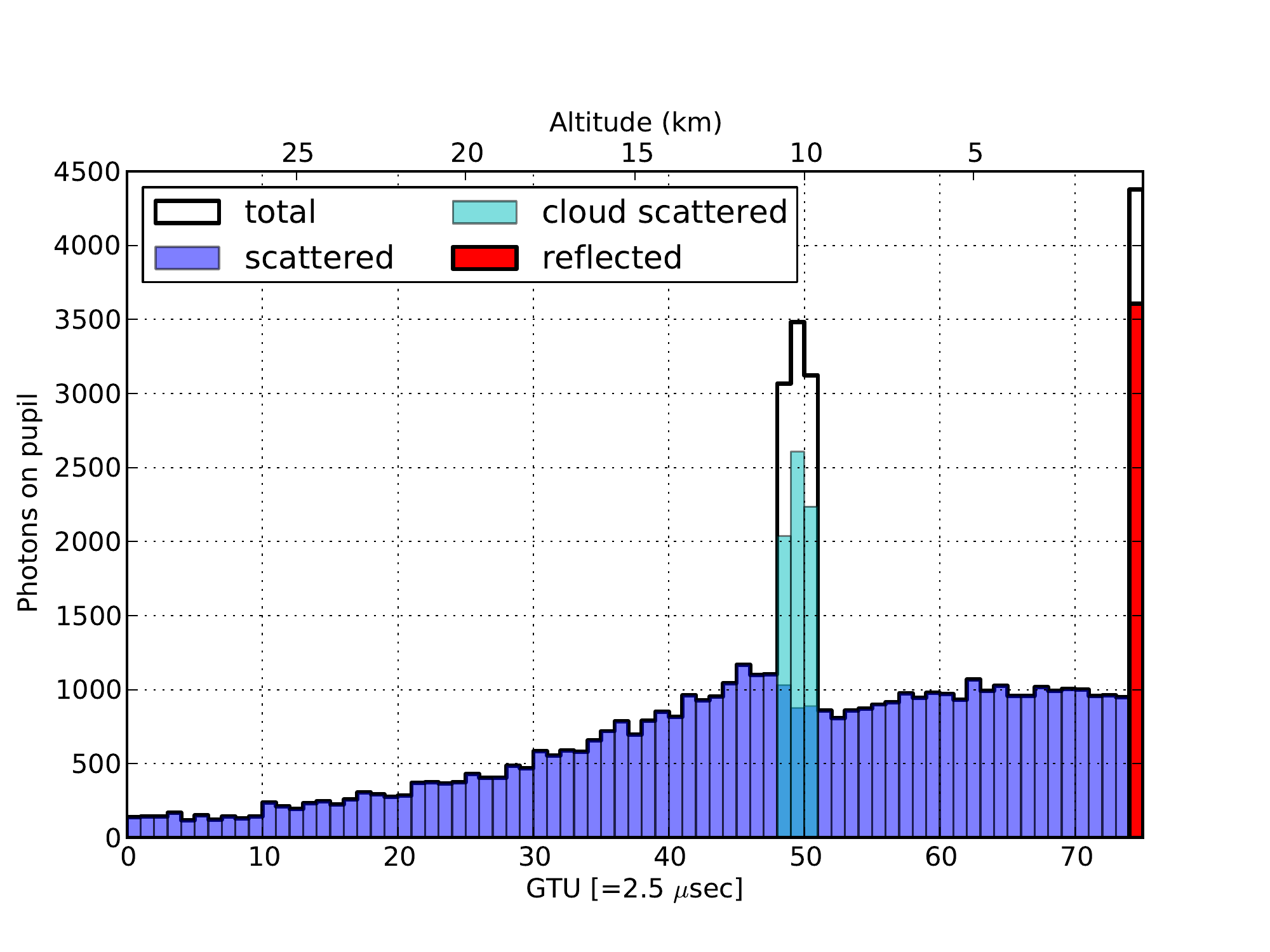} \includegraphics[width=0.5\textwidth]{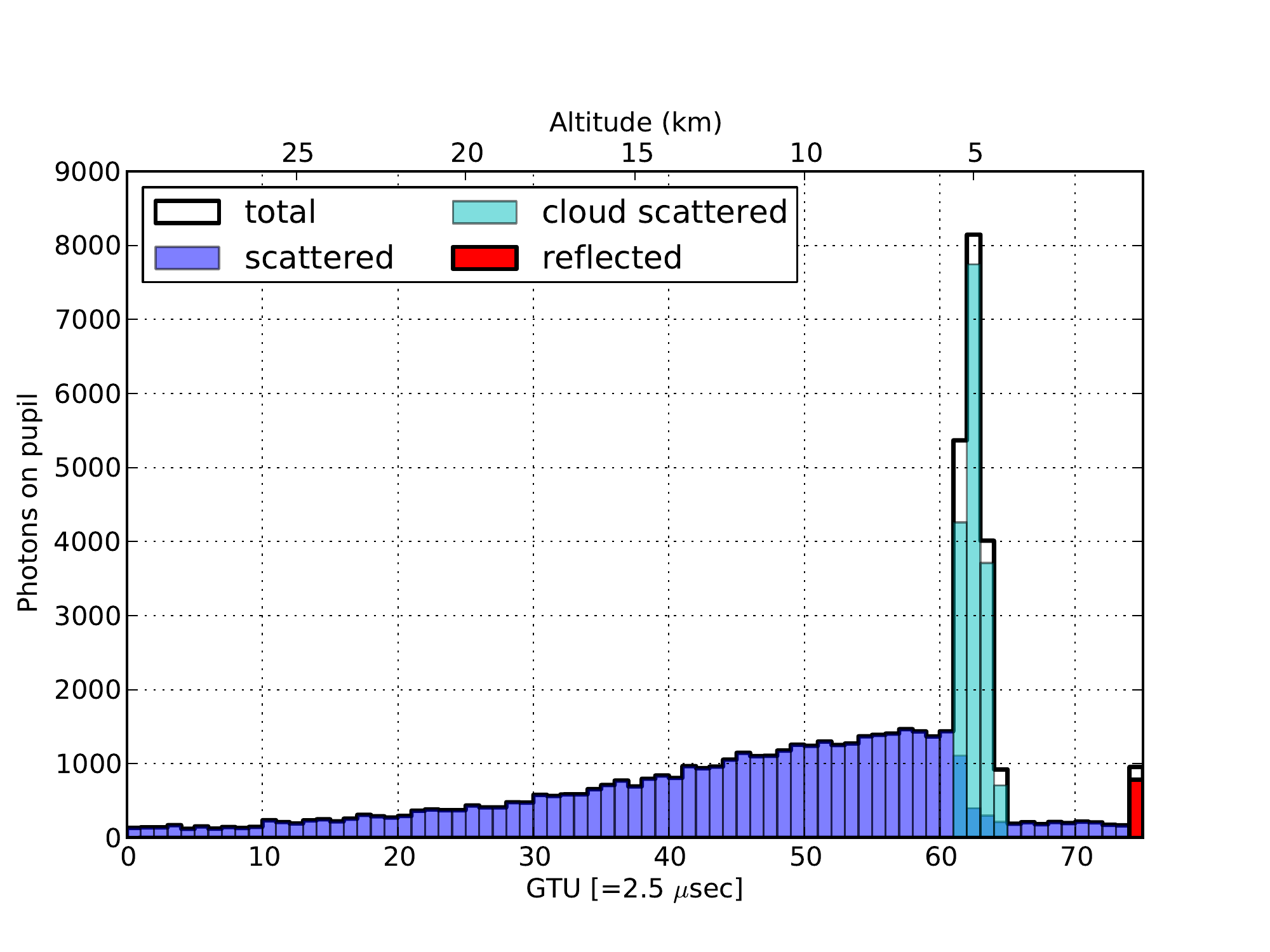}
  \caption{Distribution of the backscattered photons on pupil as a function of GTU in presence of an optically thin (\emph{top}) and optically thick (\emph{bottom}) cloud.}
  \label{fig:LIDARcloud}
 \end{figure}

\noindent 
Simulations show that the ground mark is attenuated when a cloud is present in the FoV of the JEM-EUSO telescope. Eventually it may even disappear if the cloud is optically thick. In this extreme case a reference time can be obtained from the mark generated by the photons reflected from the cloud top layers (``cloud mark"). Fig.~\ref{fig:LIDARcloud} shows the received signal as a function of time and altitude in presence of an optically thin and optically thick cloud. The optically thin ($\tau$ = 0.2) cloud has been simulated at an altitude of 10 km (top panel). The signal here is characterised by two features: the reflected component from the ground (in red) and the reflected component from cloud (in cyan). Because of the presence of the cloud, part of the laser photons are reflected from the cloud while part of them is transmitted. As a result, the entire signal is reduced by a factor of $e^{-\tau}$, and the ground mark is attenuated but still clearly visible.\\
A simulation has been performed also locating an optically thick ($\tau$ = 1) cloud at an altitude of 5 km (bottom panel). The strong mark generated by photons reflected from the cloud is the main characteristic of the time profile of the received signal. Unlike the first case, the ground mark is strongly suppressed because of the high optical depth of this cloud. Photons here are mostly reflected from the cloud top layers or trapped inside the cloud (due to multiple scattering processes). The faint ground mark visible in the distribution of photons at the pupil entrance may not be detectable and thus not suitable for the analysis of data. 

\section{General discussion and conclusions}
In this contribution we described the implementation done to the ESAF simulation framework to simulate the LIDAR device of the Atmospheric Monitoring system of the JEM-EUSO telescope. After a brief introduction on the general structure of the framework, we discussed the changes needed to simulate the geometry of the laser beam and the propagation of laser photons through the atmosphere. Examples of a simulated laser beam with $E_{laser}$ = 20 mJ and $\lambda$ = 355 nm have been shown in clear sky conditions and in the presence of clouds with different characteristics (optical depth and altitude). The signal received at the pupil entrance of the telescope is characterised by a couple of interesting features (ground and cloud mark). In a case in which a cloud is crossing the telescope FoV the information obtained from the LIDAR backscattered signal can be used to   retrieve the cloud parameters. From the time of the cloud mark it is possible to estimate the cloud top altitude, an information complementary to the measurement done with the Infrared Camera \cite{CloudTop_ICRC2013, sIRCamera_ICRC2013}. By fitting the scattering ratio (the ratio between the backscattered signal detected in the real condition and a reference profile which represents the backscattered signal in clear sky) below the cloud region it is possible to retrieve the cloud optical depth. A first analysis of LIDAR data has been performed using this simulation chain and it shows the capability of the system to measure the cloud parameters. A discussion on LIDAR data analysis and the possibility of correcting the shower profiles affected by the presence of clouds can be found in \cite{AMS_ICRC2013}.


\clearpage

%% file: icrc2013-0514.tex



\title{An End to End Simulation code for the IR-Camera of the JEM-EUSO Space Observatory.}

\shorttitle{A simulation code for the IR-Camera of JEM-EUSO.}

\authors{
J.A. Morales de los Rios$^{1,2}$,
L. del Peral$^{2}$,
G. Saez-Cano$^{2}$,
H. Prieto$^{2}$,
J. H-Carretero$^{2}$,
M.D. Sabau$^{3}$,
T. Belenguer$^{3}$,
C. Gonzalez Alvarado$^{3}$,
M. Sanz Palomino$^{3}$,
J. Licandro$^{4,5}$,
E. Joven$^{4}$,
M. Reyes$^{4}$ and
M.D. Rodriguez Frias$^{2}$,
for the JEM-EUSO Collaboration.
}

\afiliations{
$^1$ RIKEN, 2-1 Hirosawa, Wako, Saitama 351-0198, Japan. \\
$^2$ SPace \& AStroparticle (SPAS) Group, UAH, Madrid, Spain \\
$^3$ LINES laboratory, Instituto Nacional de Técnica Aeroespacial (INTA), Madrid, Spain. \\
$^4$ Instituto de Astrofísica de Canarias (IAC), Tenerife, Spain. \\
$^5$ Departamento de Astrofísica, Universidad de La Laguna, Tenerife, Spain. 
}

\email{jose.moralesdelosrios@riken.jp} 

\abstract{The Extreme Universe Space Observatory on the Japanese Experiment Module (JEM-EUSO) of the International Space Station (ISS) is the first space-based mission worldwide in the field of Ultra High-Energy Cosmic Rays (UHECR). JEM-EUSO will use our atmosphere as a huge calorimeter, to detect the electromagnetic components of the Extensive Air Shower (EAS). Therefore, the atmosphere must be calibrated and has to be considered as input for the analysis of the fluorescence signals.  The JEM-EUSO space observatory is implementing an Atmospheric Monitoring System (AMS), to gather data of the atmosphere status during the UHECR observation period, it will include an IR-Camera and a LIDAR. The AMS IR-Camera is an infrared imaging system aimed to detect the presence of clouds. Our paper is focused on the End to End (E2E) simulation developed for the IR-Camera of the JEM-EUSO Space Mission. This work gives us the capabilities to study the impact of several scenarios of the atmosphere, in terms of retrieval temperature accuracy, detection capabilities, calibration procedures, and correction factors to be taken into account for the final data products of the AMS system of the JEM-EUSO Space Mission.}


\maketitle

\section{Introduction}

The JEM-EUSO space observatory is foreseen to be launched and attached to the Japanese module of the International Space Station (ISS) in 2017 \cite{Adams}, \cite{MDRodriguezFrias-Vulcano}. It aims to observe UV photon tracks produced by Ultra High Energy Cosmic Rays (UHECR) developing in the atmosphere and producing Extensive Air Showers (EAS). However the atmospheric clouds blurs the UV radiation produced by the EAS \cite{lupeTAUP}. 

In order to monitor the atmosphere, and more important to obtain the cloud coverage in the JEM-EUSO FoV an Atmospheric Monitoring System (AMS) will be included in the telescope \cite{swada}. 
The AMS consists of a LIDAR, an infrared (IR) camera and global atmospheric models will be used as well. The LIDAR will measure the optical depth profiles of the atmosphere in selected directions. The IR-Camera will provide the cloud coverage and the cloud top height \cite{MDRodriguezFrias-CERN}. The global atmospheric models will be used to retrieve the atmospheric parameters (temperature, pressure and humidity vertical profiles) in the monitored region \cite{andriiICRC}. 


In this publication we disclose the status of the IR-Camera End to End (E2E) simulation fully developed for the IR-Camera of the JEM-EUSO Space Mission. This work gives us the capabilities to study the impact of several scenarios of the atmosphere, in terms of retrieval temperature accuracy, detection capabilities, calibration procedures, and correction factors to be taken into account for the final data products of the AM system of the JEM-EUSO Space Mission.

At this design stage of the IR-Camera prototype, this E2E simulation is giving us some answers in key points of the design, like the compression algorithms evaluation presented here.

\section{The IR-Camera Preliminary Design}

The Atmospheric Monitoring System (AMS) IR-Camera \cite{joseICRC} is a microbolometer based infrared imaging system aimed to obtain the cloud coverage and cloud top altitude during the observation period of the JEM-EUSO main instrument. The scientific and technical requirements for the IR-Camera are far for being undemanding, and are summarized in Table \ref{tab_req:1}.  Its preliminary design \cite{lolyICRC} can be divided into three main blocks: the Telescope Assembly, the Electronic Assembly, and the Calibration Unit.

\begin{figure}
 \includegraphics[width=0.35\textwidth]{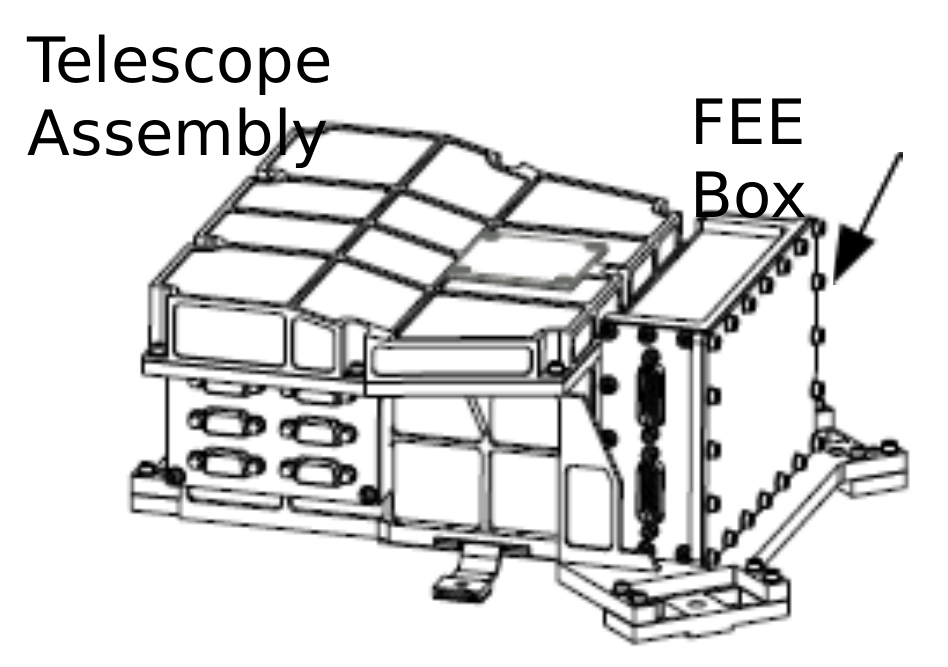}
\caption{IR-Camera Telescope Assembly Illustration.}
\label{telescope:F1}       
\end{figure}

The Telescope Assembly (Figure \ref{telescope:F1}) has to acquire the infrared radiation by means of an uncooled microbolometer and to convert it into digital counts. The IR-Camera Telescope assembly encompass the Infrared detector ($\mu$Bolometer), the FEE (Front End electronic) and the Optical lens assembly. The infrared detector that has been selected for the JEM-EUSO IR-Camera is the UL04171 from the ULIS Company \cite{ulis_datasheet}. The UL04171 is an infrared opto-electronic device comprised by a $\mu$bolometer Focal Plane Array (FPA). The FEE (Front End Electronic) manages and drives the $\mu$Bolometer; It provides the bias and the sequencer and manages the images acquisition modes.

Presently, the optical system design is a refractive objective based in a triplet with one more lens close to the stop and a window for the filters close to the focal plane. The first surface of the first lens and the second surface of the third lens are aspheric that allow a better quality of the complete system. The aperture stop is situated at 0.40mm behind the fourth lens, in order to separate the optical system to the detection module. The system, consisting of four lenses, has a focal length of 19.10mm, and a f$\#$ of 1, and it shall work with a total FoV of 48$^o$. The overall length between the first surface to the focal plane is 62.30mm.

The Electronic Assembly provides mechanisms to process and transmit the obtained images, the electrical system, the thermal control and to secure the communication with the platform computer. The Electronic Assembly is composed of two main sections: the Instrument Control Unit (ICU), and the Power Supply Unit (PSU). Data generated by the FEE is then processed by the Instrument Control Unit (ICU), which is in charge of controlling several aspects of the system management such as the electrical system, the thermal control and the communication with the platform computer. The Power Supply Unit (PSU) receives the main power bus from JEM-EUSO main telescope and it provides the required power regulation to the system and the sub-systems. A dedicated on-board calibration system is foreseen \cite{joseICRC},\cite{lolyICRC}. The calibration mechanism consists of a stepper motor governing the blackbodies and shutter. 


\begin{table}
\caption{Requirements for the IR-Camera of the JEM-EUSO Space Mission.}
\begin{tabular}{|l|ll|}
\hline\noalign{\smallskip}
Parameter & Target value &  Comments  \\
\noalign{\smallskip}\hline\noalign{\smallskip}
Measurement &  & Annual variation \\ range & 220 K - 320 K &  of cloud  \\ &  & temperature plus \\ & & 20 K margin    \\
\hline
  &  & Two atmospheric \\ Wavelength & 10-12 $\mu$m  & windows available: \\ & & 10.3-11.3 $\mu$m \\ & & and 11.5-12.5 $\mu$m    \\
\hline
FoV     & 48$^o$ &  Same as \\  & & main instrument \\
\hline
Spatial & 0.1$^o$ (Goal) & @FoV center \\ resolution &  0.2$^o$ (Threshold) &  \\
\hline
Absolute &  & 500 m in cloud \\ temperature & 3 K & top altitude \\ accuracy &  &  \\
\hline
Mass  & $\leq11$ kg & Inc 20$\%$ margin.    \\
\hline
Dimensions     & $400\times400\times370$ & w/o Insulation and \\ 
& & mounting bracket. \\
\hline
Power    & $\leq15$ W  &  Inc 20$\%$ margin.       \\
\hline
Lifetime    & 5 years On-orbit  &  +2 years On-ground    \\
\noalign{\smallskip}\hline
\end{tabular}
\label{tab_req:1}  
\end{table}

\section{IR-Camera Prototype Tests}
\label{ProtoTest:M} 

The main aim of these tests is to characterize the microbolometer detector to be used in the JEM-EUSO IR-Camera. This work made in the Astrophysics Institute of the Canary Islands (Instituto de Astrofisica de Canarias, IAC, Tenerife) has provided very useful information of the detector performances to be implemented in the IR-Camera. 

In order to acquire data and images from this FPA, a electronics prototype module developed by INO (Canada) has been used \cite{INO1}.  This electronics core is known as IRXCAM-640.  Although the chip architecture exploits a TEC less operation, the already integrated TEC and the control loop allow us 10 mK stability in temperature, keeping very low Noise Equivalent Temperature Difference (NETD) values. For the camera optics we decided to use a commercial unit, the Surnia lenses from Janos \cite{surnia}, capable to measure in the 7-14 $\mu m$ region. The main characteristics of this optics are: focal length = 25 mm and $f\#$=0.86, with a circular FoV of 45$^o$. Wavelength is limited in the 7 to 14$\mu m$ range.

The used infrared radiation source was a Black Body (model DCN-1000-L3) from HGH Systems Infrarouges (France) \cite{HGH}, with an emissive area equal to 75x75 mm and an absolute temperature range from -40$^o$C to 150$^o$C. The thermal uniformity is better than 0.01$^o$C with an stability of 0.002$^o$C. To complete the testing procedure a control system was built as well. The system consists of:  (a) 8 Pt-100 temperature sensors placed everywhere, (b) a commercial Lakeshore-218 8-channel temperature monitor and (c) a Proportional Integral Derivative (PID) control loop handled by a Lakeshore-331 temperature controller to keep the optics case in the 10 mK environment. The entire device has been synchronized and controlled by a friendly user-interface developed under NI-Labview, using a PC-platform. Most typical tests, as linearity, temperature stability, non-uniformity calibration or NETD, were fully automatized for these purposes.

\section{The End to End Simulation code.}

An End to End (E2E) dedicated simulation of the infrared camera will give us simulated infrared images of those we expect to obtain with the instrument. It provides us with the capabilities to study the impact of several scenarios of the atmosphere, in terms of retrieval temperature accuracy, to analyze the detection capabilities, calibration procedures, and correction factor to be taken into account for the final data products of the AMS system of the JEM-EUSO Space Mission. At this design stage of the IR-Camera prototype, this E2E simulation will give some answers in key points of the design, like the compression algorithms evaluation, and an estimation of the expected accuracy of various calibration options.

The simulation is a complex software, written in C++, and divided into several stages \cite{Jose2013}. It starts with the simulation of the IR scenario with atmospheric simulation software, like the Satellite Data Simulator Unit (SDSU) \cite{sdsu}. Instead of the simulator we can use real satellite IR images, taken by missions like MODIS \cite{modis} or CALIPSO \cite{calipso}. 
After the input scene is read by the simulator, an optics elements simulation takes place. Starting from the simulation of the diffraction, distortion and efficiency of the optics module, using the evaluated optics design with software Code-V \cite{codev}. The image is first blurred with the PSF (Point Spread Function) calculated for several regions of the optics. Then each pixel of the FoV image is transported to the position inside a distorted image, using a transport matrix calculated with the distortion data. Similar to the optics, the filters spectrum function made by the manufacturer is taken into account, and produces a 2-bands image which is later processed by the detector module. 

\begin{figure}
\begin{center}
 \includegraphics[width=0.35\textwidth]{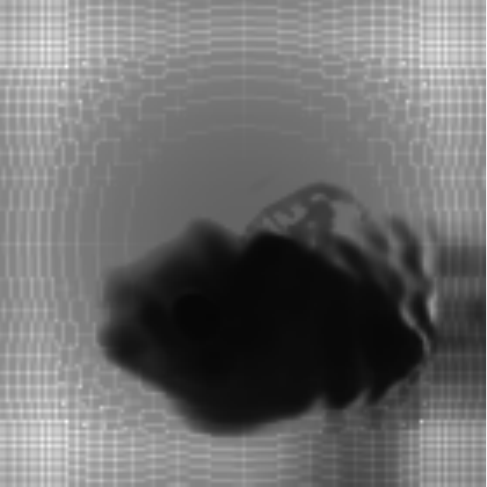}
 \end{center}
\caption{Grayscale image of a cloud brightness temperature simulated in SDSU + IR-Camera E2E. The white lines are produced by the non-continuous barrel distortion simulation.}
\label{E2E:1}       
\end{figure}

To create a model of a detector, we used results of the test described in section \ref{ProtoTest:M}. Therefore, we can translate the input values to analog voltage values that should be similar to the detector response. Moreover, we can apply the ADC (Analog to Digital Conversion) of 12bits, and its corresponding change to 10bits. As a last step, in the instrument simulation, we have compressed the image, using HP (Hewlet Packard) code LOCO-I/JPEG-LS algorithm \cite{HP-lab} with nearly loss-less code.
In Figure \ref{E2E:1} a simulated IR image is shown as an example. 


\begin{figure}
\includegraphics[width=0.45\textwidth]{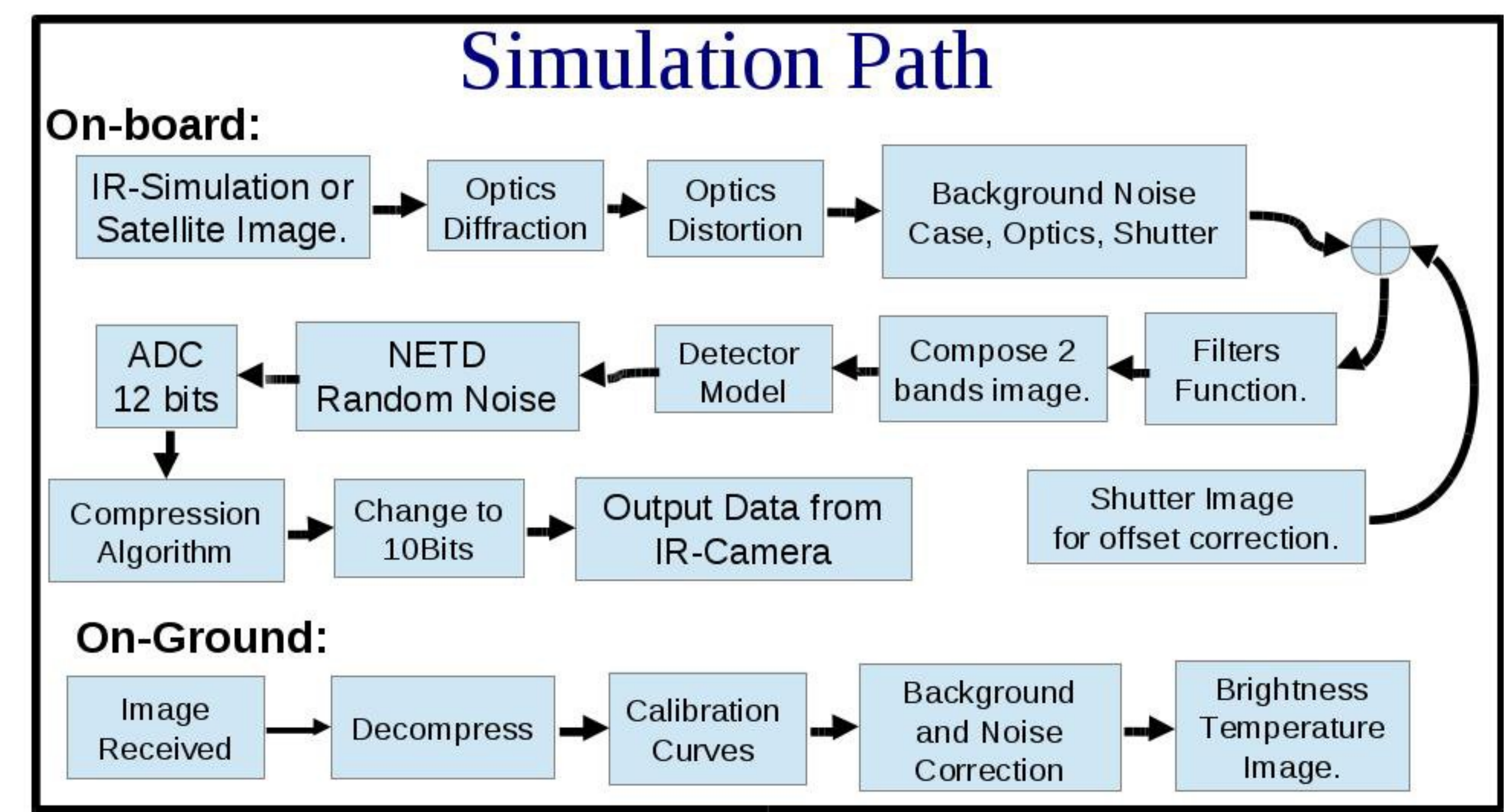}
\caption{ Block diagram of the simulation for the IR-Camera End to End simulation.}
\label{E2E:2}       
\end{figure}

In addition the simulation should include, at least, some on ground processing steps. Therefore, we have to perform decompression, implementation of calibration curves to convert digital values into temperature values, and background and noise reduction using feedback from housekeeping data. A data analysis module is foreseen to take the data from simulator, and real data from the IR-Camera to perform the analysis tasks with the algorithms for data retrieval. The output from this analysis module will be used as an input in the official codes for the performance analysis, and event reconstruction of the main telescope. A diagram of the simulation path explained before is shown in Figure \ref{E2E:2}.

\section{Compression Algorithm evaluation}
One of the key points to estimate the data rate bandwidth for the infrared camera requirements is the capability to compress the images to be sent. The evaluation of the impact of these algorithms is crucial to assure that the scientific data will not be compromised by the compression. To perform this trade off, we have used the simulator being developed for the instrument, and some test images created with SDSU. The compression algorithm under study is the HP Labs LOCO-I/JPEG-LS [4]. The procedure is very simple, we have used the simulated images of SDSU to create a control output image, and test these images with different compression factors to evaluate the compression ratio, and the impact on the data decompressed when compared to the control images. Figure \ref{E2E:2} shows the simulation path followed by the test and control images.

\begin{figure}
\begin{center}
\includegraphics[width=0.45\textwidth]{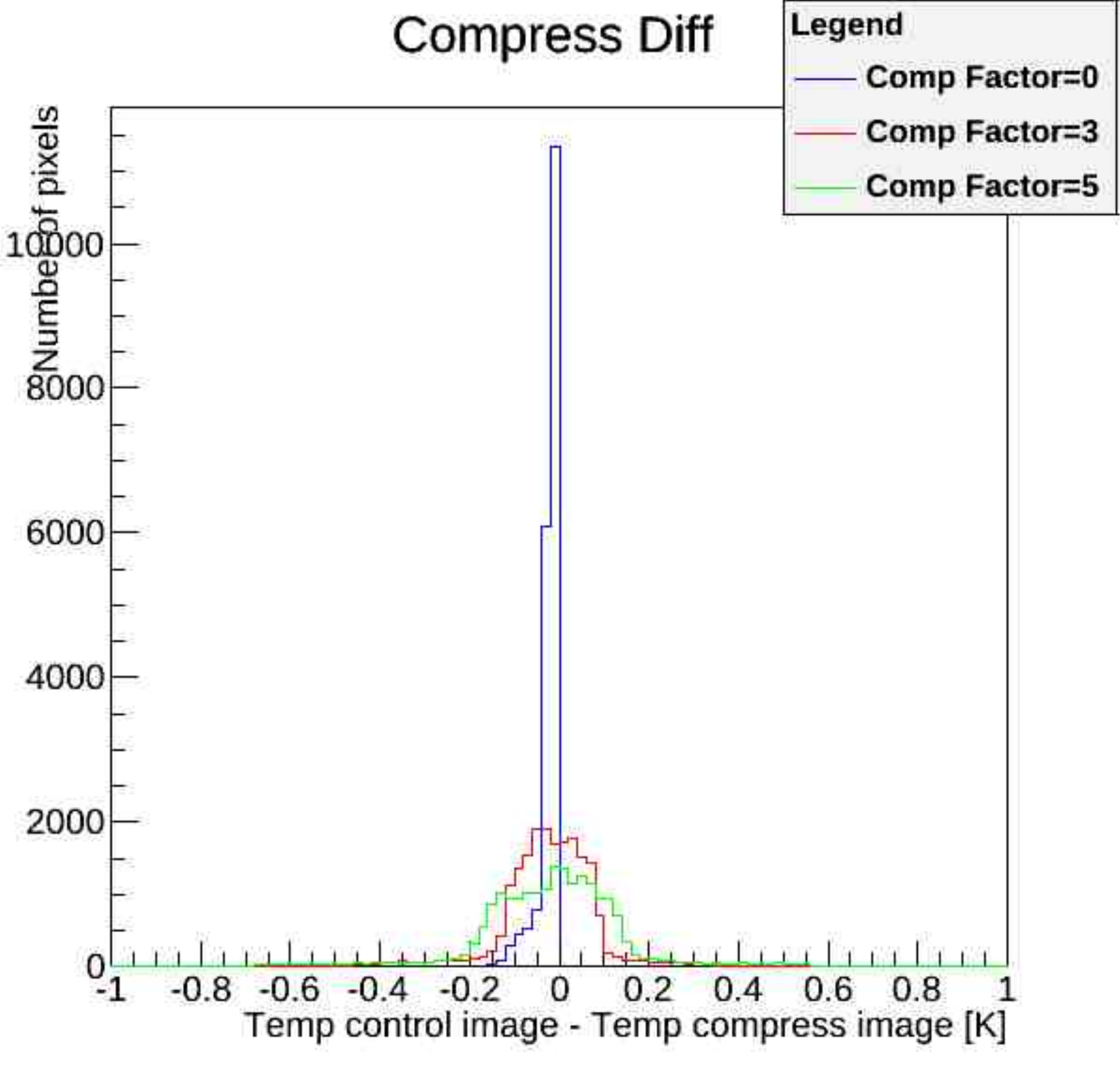}
\end{center}
\caption{Histogram with the pixel by pixel value comparison between the control image and the compressed-decompressed image.}
\label{Compress:F1}       
\end{figure}

For this first study we have evaluated compression factors of 0 (near-lossless), 3 and 5. Differences in the images cannot be appreciated by human eyes, therefore we have compared the values of each pixel of the test image, with the related pixel of the control image, and plot an histogram to evaluate the differences of the values. Results are presented in Table \ref{tab_comp:1}. The Output from pixels values comparison of test images with the control image are presented in Figure \ref{Compress:F1} for the compression factor of 0 (nearly loss-less), 3 and 5. 

\begin{table}
\caption{Analysis of one compression image.}
\begin{tabular}{|l|lll|}
\hline\noalign{\smallskip}
Comp Factor & 0 &  3 & 5  \\
\noalign{\smallskip}\hline\noalign{\smallskip}
\hline
In [kbits] & 235.2 & 235.2 & 235.2 \\
\noalign{\smallskip}
Out [kbits] & 65.232 & 34.788 & 29.288 \\
\noalign{\smallskip}
Comp Ratio & 3.6 & 6.76 & 7.92 \\
\noalign{\smallskip}
Typ Dispersion [$^o K$] & $\approx$0.10 & $\approx$0.15 & $\approx$0.2 \\
\noalign{\smallskip}
Max Dispersion [$^o K$] & 0.15 & 0.63 & 0.9 \\
\noalign{\smallskip}\hline
\end{tabular}
\label{tab_comp:1}  
\end{table}

\section{Conclusions}
 
At the UHECR regime observed by JEM-EUSO, above 10$^{19}$ eV, the existence of clouds will blur the observation of UHECRs. Therefore, the monitoring of the cloud coverage by the JEM-EUSO Atmospheric Monitor System (AMS) is crucial to estimate the effective exposure with high accuracy and to calibrate the UHECRs and EHECRs events just above the threshold energy of the telescope. The AMS IR-Camera of JEM-EUSO is an infrared imaging system aimed to detect the presence of clouds in the FoV of the JEM-EUSO main telescope and to obtain the cloud coverage and cloud top altitude during the observation period of the JEM-EUSO main instrument. Its full design, prototyping, space qualified construction, assembly, verification and integration is under responsability of the Spanish Consortium involved in JEM-EUSO. The observed radiation is basically related to the target temperature and emissivity and it can be used to get an estimate of the clouds top height.

 The development of the E2E simulation is an undergoing work, making the model more accurate, and covering each area of the infrared camera design deeply. The prototype test has given us the opportunity to acquire the knowledge to build the detector model. Moreover, the optics, designed by the INTA and characterized with CODE-V have been simulated with our code as well.

Our next objective is to try to address the impact of several design characteristics to have a very detailed study of the detection error, and to provide a plattform for the IR-Camera design engineers to test the changes neccesary in each step of the infrared camera development process. Moreover, from the compression algorithm trade off, we can conclude that the HP-LOCO algorithm is suitable for our IR-Camera and, a compression ratio of 3 is required to ensure a feasibly quality images with minor losses.

\vspace*{0.2cm}

{
\footnotesize{{\bf Acknowledgment:}{The Spanish Consortium involved in the JEM-EUSO Space
Mission is funded by MICINN \& MINECO under projects AYA2009-
06037-E/ESP, AYA-ESP 2010-19082, AYA-ESP 2011-29489-C03-
01, AYA-ESP 2011-29489-C03-02, AYA-ESP 2012-39115-C03-01, AYA-ESP 2012-39115-C03-03, CSD2009-00064 (Consolider MULTIDARK) and by Comunidad de Madrid (CAM) under project S2009/ESP-1496. This work was partially supported by Basic Science Interdisciplinary Research Projects of RIKEN and JSPS KAKENHI Grant (22340063, 23340081, and 
24244042), by the Italian Ministry of Foreign Affairs, General Direction 
for the Cultural Promotion and Cooperation, by the 'Helmholtz Alliance 
for Astroparticle Physics HAP' funded by the Initiative and Networking Fund of the Helmholtz Association, Germany, and by Slovak Academy  
of Sciences MVTS JEM-EUSO as well as VEGA grant agency project 2/0081/10.
The calculations were performed by using the RIKEN Integrated Cluster of Clusters (RICC) facility. J.A. Morales de los Rios wants to acknowledge the financial support from the UAH-FPI grant, and the RIKEN-IPA program.
}}

}
\clearpage